\documentclass[12pt,dvips,twoside,a4paper,footnoteintab]{book}

\usepackage[top=2cm,inner=3cm, includefoot,centering]{geometry}
\usepackage{setspace}
\usepackage{times,fancyhdr}
\usepackage{pdflscape}
\usepackage{rotating}
\usepackage{graphicx}
\usepackage{geometry}

\def\bx{{\bf x}}

\def\beq{\begin{equation}}
\def\eeq{\end{equation}}
\def\bea{\begin{eqnarray}}
\def\eea{\end{eqnarray}}
\def\bean{\begin{eqnarray*}}
\def\eean{\end{eqnarray*}}
\def\bx{{\bf x}}

\newcommand{\bear}{\begin{eqnarray}}

\newcommand{\eear}{\end{eqnarray}}

\newbox\pippobox

\def\na{\nabla}
\def\e{\epsilon}

\renewcommand{\[}{\left[}
\renewcommand{\]}{\right]}
\def\bx{{\bf x}}
\def\e{\mathbf{e}}
\def\sp{\sigma_+}

\def\y{\vartheta}

\def\l{\label}

\def\hsp5{\hspace{5mm}}
\def\lgth{[\,\mbox{length}\,]}

\PassOptionsToPackage{linktocpage}{hyperref}
\usepackage[draft=false]{hyperref}
\usepackage{threeparttable}

\newcommand{\sfrac}[2]{{\textstyle{#1\over#2}}}
\def\case#1/#2{\textstyle\frac{#1}{#2}}

\newcommand{\be}{\begin{equation}}
\newcommand{\ee}{\end{equation}}
\newcommand{\ben}{\begin{eqnarray}}
\newcommand{\een}{\end{eqnarray}}
\newtheorem{thm}{Theorem}
\newtheorem{lem}{Lemma}
\newtheorem{rem}{Remark}
\newtheorem{prop}{Proposition}
\newtheorem{defn}{Definition}

\def\PRTS{{Physics Reports} }

\newcommand{\PR}{{\it Phys. Rev.} }

\usepackage{epsfig}
\usepackage{bm}
\usepackage{amsfonts}
\usepackage{amsmath,amssymb}
\usepackage{dcolumn}
\usepackage{threeparttable}
\def\case#1/#2{\textstyle\frac{#1}{#2}}

\def\bea{\begin{eqnarray}}
\def\eea{\end{eqnarray}}
\newcommand{\s}{\sigma}
\newcommand{\la}{\lambda}

\newcommand{\vphi}{\varphi}

\onehalfspace

\begin{document}
\title{{
\vskip 0.45in \bfseries\scshape Cosmological dynamical systems }}
\author{\bfseries\itshape Genly Leon \footnote{Genly Leon is currently affiliated to Instituto de F\'isica, Pontificia Universidad Cat\'olica de Valpara\'iso, Casilla 4950, Valpara\'iso, Chile. E-mail address: genly.leon@ucv.cl}
 ~
and Carlos R. Fadragas\thanks{E-mail address: fadragas@uclv.edu.cu}\\
Department of Mathematics, \\
Universidad Central de Las Villas, \\Santa Clara \enskip CP 54830, Cuba and\\
Department of Physics, \\
Universidad Central de Las Villas, \\ Santa Clara \enskip CP
54830, Cuba}

\date{\today}

\maketitle

\renewcommand{\headrulewidth}{0pt}





\pagestyle{plain}

\pagenumbering{roman}

\tableofcontents

\listoffigures

\listoftables

\chapter*{Preface}
\addcontentsline{toc}{chapter}{Preface}

Over the last century, the investigations in Gravitation and
Cosmology have attracted the attention of thousands of human
minds. Apart this boom in popularity, the homework task for
cosmologists is to try to give an answer to the reasons why the
present Universe has the properties that are observed. In spite of
having a well-established theoretical framework like the
Einstein's gravity theory (GR), which is valid in an amplest rank
of energies and that has responded satisfactorily to numerous
empirical verifications, are numerous unanswered intriguing
questions.   Is the dark energy the cause of the acceleration of
the expansion? What is its equation of state nowadays? Why the
energy densities of matter and dark energy are of the same order
of magnitude nowadays? Will be the acceleration of the expansion
an indication that the action of Einstein-Hilbert must be
modified? Can be reached the isotropic degree of the universe that
is observed today, independently of the initial degree of
anisotropy? In this book are studied, from the perspective of the
dynamical systems, several Universe models that try to give
answers to some of these questions.

In chapter 1 we give a bird's eye view on cosmology and
cosmological problems. Cosmology is a particular way to embrace
the space-time, the gravity theory and the material content of the
Universe, in order to clarify the origins and the evolution of the
Universe as a whole. Within the most commonly accepted vision in
Cosmology, it is assumed that the geometry of the universe at the
large scale is described by General Relativity; although in this
book we are ascribed also to alternative theories. In this chapter
we put in a firmly theoretical setting the above questions.

Chapter 2 is devoted to a brief review on some results and useful
tools from the qualitative theory of dynamical systems. They
provide the theoretical basis for the qualitative study of
concrete cosmological models. This is the more mathematical in
character. However, to avoid a lengthly and tedious reading, we
omit several of the mathematical proofs. By this reason, it is
required that the reader will be managed with concepts from
Non-linear Differential Equations, particularly, for the stability
theory of singular points and periodic orbits, stable and unstable
manifolds; basic rudiments on Differential Geometry, and a firm
knowledge on Calculus. These skills, and a deeply understanding of
Chapter 1, will be required to read the book to the end.

Chapters 1 and 2 are a review of well-known results. Chapters 3,
4,  5 and 6 are devoted to our main results. In these chapters are
extended and settled in a substantially different,  more strict
mathematical language, several results obtained by one of us in
\cite{Leon:2008de,Leon2010,PLB2010,Leon:2009rc} (at these stage we
acknowledged to all the collaborators and to the Editorial houses;
that gave their permissions for using several of these previous
results). In chapter 6, we  provide a different approach to the
subject discussed in \cite{Lazkoz:2005bf}.

The aim of the chapter 3 is to extent several results related to
flat FRW models formulated in the conformal (Einstein) frame of
scalar-tensor gravitational theories, including $F(R)$ theories
through conformal transformation. We will focus mainly in a
particular era of the universe where matter and radiation
coexisted. The ordinary matter is described by a perfect fluid. We
consider models with and without radiation. We are interested in
investigating all possible scaling solutions in this regime, since
scaling late-time attractors provide a hint for avoiding or
alleviating the Coincidence Problem. Although we are mainly
interested in describing the early time dynamics, for completeness
we will focus also in the late-time dynamics of the models under
consideration. For flat FRW models we obtain sufficient conditions
under the potential for the asymptotic stability of the
non-negative local minima for the potential. Center manifold
theory is employed to analyze the stability solutions associated
to the local degenerated minimum and to the inflection points of
the potential. We prove, for arbitrary potentials and arbitrary
coupling functions of appropriate differentiable class, that the
scalar field almost always diverges into the past generalizing
previous results. It is designed a dynamical system well suited to
studying the stability of the singular points in that limit. We
obtain there: radiation-dominated cosmological solutions;
power-law scalar-field dominated inflationary cosmological
solutions; matter-kinetic-potential scaling solutions and
radiation-kinetic-potential scaling solutions. It is discussed, by
means of several worked examples, the link between our results and
the results obtained for specific $F(R)$ frameworks by using
appropriated conformal transformations. We prove an stability
theorem and two singularity theorems.

In chapter 4, we investigate several varying-mass dark-matter
particle models in the framework of phantom cosmology. We examine
whether there exist late-time cosmological solutions,
corresponding to an accelerating universe and possessing dark
energy and dark matter densities of the same order. Imposing
exponential or power-law potentials and exponential or power-law
mass dependence, we conclude that the coincidence problem cannot
be solved or even alleviated. We improve our previous analysis by
using the Center Manifold Theory to analyze the stability of the
nonhyperbolic fixed points in the phase space. Basically, we use
these cosmological models as examples of how to apply the Center
Manifold Theory in cosmology. Also, in this chapter we perform a
Poincar\'e compactification process allowing to construct a global
phase space containing all the cosmological information in both
finite and infinite regions. Are proved several stability
theorems.

Chapter 5 is devoted to a detailed phase-space analysis of
Ho\v{r}ava-Lifshitz cosmology, with and without the
detailed-balance condition. Under detailed-balance we find that
the universe can reach a bouncing-oscillatory state at late times,
in which dark-energy, behaving as a simple cosmological constant,
is dominant. In the case where the detailed-balance condition is
relaxed, we find that the universe reaches an eternally expanding,
dark-energy-dominated solution, with the oscillatory state
preserving also a small probability. To achieve the above results
we use the Center Manifold Theory to analyze the stability of the
nonhyperbolic fixed points in the phase space. Also, we perform a
Poincar\'e compactification process allowing to construct a global
phase space containing all the cosmological information in both
finite and infinite regions. We prove an stability theorem.

In chapter 6, we investigate the asymptotic behavior of Cardassian
cosmological models filled with a perfect fluid and a scalar field
with an exponential potential. Cardassian cosmologies arise from
modifications of the Friedmann equation, and among the different
proposals within that framework we will choose those of the form
$3 H^2-\rho \propto \rho^n$ with $n < 1$. We construct a three
dimensional dynamical system arising from the evolution equations.
Using standard dynamical systems techniques we find the fixed
points and characterize the solutions they represent. We pay
especial attention to the properties inherent to the modifications
and compare with the (standard) unmodified scenario. Among other
interesting results, we find there are no late-time scaling
attractors. We prove the stability (but not asymptotic stability)
of a cosmological solution representing a regime where the
Cardassian corrections dominates.

This book is intended to a wide audience, specially to cosmologist
and mathematicians working on the field of Gravitation and
Cosmology. Most parts of the text are at the level of a last year
math undergraduate student. Several parts of the book were used to
complement the course ``Dynamical Systems'' taught last Spring by
one of us (GL) to the 3rd and 4th year of the Mathematics degree
at Universidad Central de Las Villas (Santa Clara, Cuba).

{\bf Acknowledgements}

The authors wish to thank to LAP Lambert publishing for inviting
us to write this book. The MES of Cuba is acknowledged for partial
financial support of this investigation. This research was also
supported by the National Basic Science Program (PNCB) of Cuba and
Territorial CITMA Project (No. 1115). We also thank to Dr. E. N.
Saridakis for providing us with a nice review on
Ho\v{r}ava-Lifschitz cosmology. GL wishes to thanks to Dra. Ruth
Lazkoz and to Dr. Rolando Cardenas for their guidance on
cosmological studies. Professors Alan Coley and David Wands are
acknowledged for writing inspiring works. CRF thanks to
his son Carlos and his daughter Martica, without whose support and
encouragement they would not have completed this work.

Genly Leon and Carlos R. Fadragas

Santa Clara, November 25th, 2011.

\clearpage

\pagestyle{plain}

\pagenumbering{arabic} \setcounter{page}{1}

\chapter*{Introduction}
\addcontentsline{toc}{chapter}{Introduction}


Cosmology is a broad and promising area of research in applied
Mathematics and Physics. It is supported by reliable observed data
available in modern Astrophysical literature. As a mathematical
discipline, its methods come from two main sources: Differential
Geometry and Differential Equations. At the same time Cosmology
returns new problems not only to both fields, also to related
Physics.

There exist three essential elements whenever cosmological
modelling is concerned: the space-time, the gravity theory and the
material content of the Universe. Cosmology is a particular way to
embrace these basic elements for studying  the origins and the
evolution of the Universe as a whole.

There exists a basic hierarchy of space-times in descendant degree
of symmetry and, therefore, in an increasing degree of generality:
isotropic and homogeneous Friedmann-Lema\^itre-Robertson-Walker
(FLRW) space-time, homogenous space-time (Bianchi and
Kantowsky-Sachs), inhomogeneous space-time and generic space-time
(i.e. without symmetries).

Some of the theories of gravity are: General Relativity (GR),
higher order gravity (HOG) or extended gravity (EG), Scalar-tensor
theories (STT), String theory (ST), and others.

The matter fields have played important roles at different epochs
in the Universe evolution: vacuum, fluids (barionic/non-barionic
matter, dark matter, radiation), scalar fields, n-form fields,
etc.

There exist four standard ways of systematic investigation that
can be used to examining cosmological models:
    Obtaining and analyzing exact solutions;
    Heuristic approximation methods;
    Numerical simulation, and
    Qualitative analysis.
In the last case can be used three different approaches:
        (a) Piecewise approximation methods,
        (b) Hamiltonian methods,
        (c) Dynamical systems methods.
In the approach iv (a), the evolution of the model universe is
approximated through a sequence of epochs in which certain terms
in the governing differential equations can be neglected, leading
to a simpler system of equations. This heuristic approach is
firmly based in the existence of heteroclinic sequences, which is
a concept from iv (c). In the approach iv (b), the Einstein´s
equations are reduced to a Hamiltonian system dependent of time
for a particle (point universe) in two dimensions. This approach
has been used mainly for modelling and analyzing the dynamics of
the universe nearly the Big-Bang singularity. In the approach iv
(c), Einstein's equations for homogeneous cosmologies can be
described as an autonomous system of first order ordinary
differential equations plus certain algebraic constraints. In this
case the solution curves form a partition $\mathbb{R}^n$ in
orbits, defining a dynamical system in $\mathbb{R}^n$.  The
approach that is used is to start from local analysis and to be
extending, step by step, the regions of the state space and the
parameter space that are investigated. In the general case the
sets of the partition of the state space can be enumerated and
described. This study consists of several steps: the determination
of the singular points, the linealization in a vicinity of them,
the search of the eigenvalues of associated the Jacobian matrix,
the verification of the conditions of stability in a vicinity of
the singular points with physical sense, the finding of the their
sets of stability and instability and the determination of their
basin of attraction. Using this approach in \cite{WE} have been
obtained many results concerning the possible asymptotic
cosmological states in Bianchi and FRW models, whose material
content is a perfect fluid (usually modelling ``dark matter''
(MO), a component that plays an important role in the formation of
structures in the Universe, such as galaxies and clusters of
galaxies) with linear equation of state (with the possible
inclusion of a cosmological constant). Also several classes of
inhomogeneous models are examined comparing the results with those
obtained using numerical and Hamiltonian methods. This analysis is
extended in \cite{Coley:2003mj}, to other contexts, having
considered other material sources such as the scalar fields.

The investigations in Gravitation and Cosmology try to give an
answer to the reasons why the present Universe has the properties
that are observed. In spite of having a well-established
theoretical framework like the Einstein's gravity theory (GR), who
is valid in an amplest rank of energies and that has responded
satisfactorily to numerous empirical verifications, are numerous
unanswered intriguing questions.   Is the dark energy the cause of
the acceleration of the expansion? What is its equation of state
nowadays? Why the energy densities of matter and dark energy are
of the same order of magnitude nowadays? Will be the acceleration
of the expansion an indication that the action of Einstein-Hilbert
must be modified? Can be reached the isotropic degree of the
universe that is observed today, independently of the initial
degree of anisotropy?

In order to respond them the consensus says that it is necessary
to progress in the theoretical and/or phenomenological modelling
of the Universe on the basis of an increasing number of
observational data that inform to us into how it is the Universal
kinematics on great scales, and on the other hand, in the
deepening in the understanding of the fundamental theory
describing the gravitational interaction.

In this proposal are studied, from the perspective of the
dynamical systems, several models of the Universe as they try to
give answer to some of these questions. These techniques allow the
fine tuning of the initial conditions required to conciliate with
the observations.

There exists a huge amount of astrophysical data, collected since
1998 to the date, that are the basis of a new cosmological
paradigm, according to that, the universe is spatially flat and it
seems to be in an accelerated expansion phase. Strong evidences
for that coming from the Hubble diagram, the Supernovae type Ia
observations of velocity-luminosity, and the anisotropies observed
in the cosmic background radiation
\cite{Komatsu:2010fb,obs1,obs2,obs3,obs4,Perlmutter:1998np,Bennett:2003bz,Tegmark:2003ud,Allen:2004cd,Tegmark:2003uf,Komatsu:2003fd,Hinshaw:2003ex,Spergel:2003cb,Peiris:2003ff,Riess:1998cb,Riess:2004nr,Jassal:2004ej,Freedman:2000cf,Mould:1999ap,Choudhury:2003tj}.

It is convenient to measure the energy density of the different
species in terms of the critical density
$\rho_{c}(t_0)=3H^{2}_{0}/8\pi G$, where $H_{0}=(\dot{a}/a)_{0}$
is the current expansion rate of the universe. The critical
density is exactly the required density to make the universe flat.
If the energy density in the universe is lower than the critical
one, the universe will expand up to a maximum value and then it
will recollapsing; if it is lower, the universe would expand
forever. Following the recent measurements, the critical density
is given by $\rho_{c}(t_0)=1.88 h^2\times 10^{-26} kg\, m^{-3},$
where $h=0.72\pm 0.007$  \cite{Freedman:2000cf,Mould:1999ap}. The
dimensionless energy density parameter
${\Omega_{i}}={\rho_{i}}/\rho_{c},$ allows us to know the
contribution of the different energetic components in the Universe
to the total energy density (where $i$ is used to denote the
$i$-th component, e.g., matter (M), dark energy (DE), radiation
(R), etc.).

\begin{itemize}

\item The total energy density is bounded by
$0.98\lesssim\Omega_{total}\lesssim 1.08.$ This value is
determined by the angular anisotropy spectrum in cosmic microwave
background radiation (CMB). These observations, combined with the
reasonable hypothesis $h>0.5,$ shows that we live in a universe
with the critical density, that is a flat universe.

\item The observations of primordial deuterium originated in the
Big-Bang nucleosynthesis as well as the CMB observations shows
that baryons contribute around $\Omega_B=(0.024\pm 0.0012)h^{-2};$
since  $h=0.72\pm 0.007,$ $\Omega_{B}\cong0.04-0.06;$ thus, we
conclude that must of the matter content in the universe is
non-baryonic.

\item The observations related to large scale structures (LSS), as
well as they dynamics (rotation curves of galaxies, the estimate
of mass in clusters of galaxies, gravitational lensing, galactic
surveys ...), suggest that our universe contains a non-luminous
matter (dark matter; DM hereafter) composed by  weakly interacting
massive particles (WIMPs) which does cluster at galactic scales.
This matter source contributes about $\Omega_{DM}\cong 0.20-0.35$.

\item Combining the last observation with the first one, we
conclude that there exists at least one additional component in
the cosmic budget that contributes about $70\%$ of the critical
density. The initial analysis of some observations
\cite{Komatsu:2010fb,obs1,obs2,obs3,obs4,Perlmutter:1998np,Bennett:2003bz,Tegmark:2003ud,Allen:2004cd,Tegmark:2003uf,Komatsu:2003fd,Hinshaw:2003ex,Spergel:2003cb,Peiris:2003ff,Riess:1998cb,Riess:2004nr,Jassal:2004ej,Freedman:2000cf,Mould:1999ap,Choudhury:2003tj},
indicates that this energy-density source (called dark energy; DE
hereafter) is unclustered, has negative pressure and contributes
to the total content as $\Omega_{EO}\cong 0.60-0.75$. The
observations suggest that this component has the equation of state
(EoS) parameter $w=p/\rho\lesssim -0.78$

\item The universe also contains radiation contributing an energy
density $\Omega_R h^2=2.56\times 10^{-5}$ ($\Omega_{R}\cong
4.84-5.03\times 10^{-5}$). Today much of such radiation is due to
the photons in the CMB radiation. Its contribution is dramatically
irrelevant today but it would have been the dominant component in
the universe for redshifts larger than $z_{e}\cong
\Omega_{MO}/\Omega_R\cong 4\times 10^{4}\Omega_{MO} h^2.$

\item Thus, we conclude that:
$$\left(\Omega_{EO},\, \Omega_{MO},\, \Omega_{B},\,
\Omega_{R}\right)\cong \left(0.7,\, 0.26,\, 0.04,\, 5\times
10^{-5}\right).$$

\end{itemize}

In conclusion, we are facing the fact that the 96\% of the total
energetic content of our universe consist of energy/matter forms
whose nature is intriguingly unknown; thus, time and resources are
required to solve such an enigma. From this matter/energy, the
70\% counteracts gravity and it is responsible for the accelerated
expansion phase our universe is experiencing.  In the vast
literature on the subject, there are numerous descriptions which
have more and less arguments in favor, without achieving a
definitive answer.

These features of the universe led physicists to follow two
directions in order to explain the accelerated expansion. The
first is to introduce the concept of dark energy (see the reviews
\cite{Copeland:2006wr,Caldwell:2009ix,Turner:2007qg,Alam:2004jy,Leon:2009ce}
and references therein) in the right-hand-side of the field
equations, which could either be the simple cosmological constant
$\Lambda$ \footnote{This choice is seriously plagued by the well
known coincidence and fine tuning problems
\cite{Sahni:2002kh,Carroll:1991mt,Sahni:1999gb,Weinberg:1988cp}.}
or, one or several scalar fields
\cite{Tsujikawa:2006mw,Tsujikawa:2005ju,Amendola:2004be,Beck:2003wj,Maor:2002rd}.
The second is looking for alternative models
\cite{Sami:2009dk,Alcaniz:2006ay,Mannheim:2005bfa}.

Within the orthodox vision, the dynamics of the universe is well
described by the Einstein's equations for the gravitational field.
These relates the geometry of spacetime, to its matter content.
The conventional treatment in cosmology separates the study of the
universe in the large scales ($l\gtrsim 150 h^{-1} Mpc$) from the
study of structure formation in lower scales. The first is
modelled as an homogeneous and isotropic matter distribution
(Cosmological Principle) and the second is solved in terms of
gravitational instabilities which can amplify the small initial
density perturbations, leading to the formation of structures like
galaxies.

The Cosmological Principle allows for a substantial reduction of
the complexity of Einstein's field equations; particularly it is
reduced the number of independent equations. Exactly in such an
approximation, the universe expansion is described by a function
of time $a(t)$ (called scale factor, whose time derivative will be
denoted by $\dot a(t)$) and its evolution is governed by the
equations (we have used units in which $c=1$):

\be \frac{{\dot a}^2+k}{a^2}=\frac{8\pi G\rho}{3};\; d(\rho a^3)=-p d a^3.\ee

The first of them (called Friedmann equation) relates de expansion
rate with the energy density $\rho$ ($k=0, \pm 1$ is the parameter
characterizing the spatial curvature of the universe). The second
one, determines the evolution of the density $\rho=\rho(a)$ in
terms of the scale factor if it is considered an EoS $p=p(\rho)$.
Particularly, if $p=w \rho$ with $w$ (at least approximately) a
constant, then $\rho\propto a^{-3(1+w)}$ and, assuming $k=0$,
$a\propto t^{2/[3(1+w)]}.$

Additionally, the spatial part $\bf g$ of the geodesic
acceleration (this one measures the relative acceleration between
two geodesics in the spacetime) satisfies an exact equation in
general relativity given by \be \nabla\cdot {\bf g}=-4\pi G(\rho+3
p).\label{geodesica}\ee This shows that the source of the geodesic
acceleration is $(\rho+3p)$ and not $\rho.$ Thus, if $\rho+3p<0$
matter exerts a negative pressure that counteracts the action of
gravity driving the current accelerated expansion. This will be
achieved if matter satisfies $\rho>0$ and $w=p/\rho<-1/3.$ If we
consider a sufficiently large scale such that the Cosmological
Principle can be considered as valid, then, equation
(\ref{geodesica}) reduces to
\begin{equation} 3\frac{\ddot a}{a}=-4\pi G\left(\rho +3\,p
\right).
\end{equation}
As a consequence of the previous discussion, the acceleration
(deceleration) of the universe can be characterized by the sign of
$\ddot a$. If it is positive (negative) expansion is accelerated
(decelerated).

To characterize the current expansion, it is used the
``deceleration'' factor measured today, $q_0,$ which can be
identified in the Taylor expansion: \be
\frac{a(t)}{a(t_0)}=1+H_0(t-t_0)-\frac{q_0}{2}H_0(t-t_0)^2+\ldots
\ee This defines

\be q_0=-\frac{\ddot
a(t_0)}{a(t_0)}\frac{1}{H_0^2}=-\frac{a(t_0)\ddot a(t_0)}{\dot
a(t_0)^ 2},\ee where the overdot denotes derivative with respect
time; the subindex  $0$ refers to magnitudes measured at current
time ($a(t_0)$ is the value of the scale factor today).

Knowing the properties of matter in the universe, then $q_0$ it
not independent with respect to the parameters $H_0$ and
$\Omega_0.$ These two parameters are sufficient (at least in first
approximation) to describe all the possibilities. However, if we
do not know the properties of matter in the universe,  $q_0$
provides a complementary information. $q_0$ can be  measured
directly  by observing distant objects such as distant galaxies.

Perhaps the alternative to consider an scalar field to explain the
current acceleration should be a more interesting possibility
since this fields appears in Grand Unification theories (GUT); and
they are based in fundamental physics. However, they have its own
drawbacks \cite{Padmanbhan2005}:

\begin{itemize}

\item This models are degenerated in several senses:
\begin{itemize}
\item virtually any form of $a(t)$ can be modelled by a suitable
``designer'' $V(\phi);$ \item even when $w(a)$ is known, it is not
possible to univocally determine the nature of the scalar field
Lagrangian that gives it origin. At least two different forms of
the scalar field Lagrangian (corresponding to quintessence or
tachyonic field) could lead to the same $w(a)$ (see the explicit
construction in \cite{Padmanbhan2005}).
       \end{itemize}

\item All the scalar field potentials require fine tuning of the
parameters in order to be viable. This is obvious in the
quintessence models in which adding a constant to the potential is
the same as invoking a cosmological constant. So to make the
quintessence models to work, we first need to assume the
cosmological constant is zero.

\item By and large, the potentials used in the literature have no
natural field theoretical justification. All of them are
non-renormalizable in the conventional sense and has to be
interpreted as a low energy effective potential in an ad hoc
manner.

\item Although the large astrophysical observations are consistent
with the $\Lambda$-CDM model ($w=-1$) some of them favor the case
$w<-1.$ If that is true, the EoS parameter should crosses the
phantom barrier ($w=-1$).

\end{itemize}

Scalar fields have been also proposed to modelling dark matter in
the halo of spiral galaxies, where the centers predicted by such
models are scalar field condensates and its density profiles are
almost constants \cite{tona}. The scalar field can gravitationally
influence the galaxy causing that the orbital velocity of the
celestial objects within this region remains a constant,
explaining the flatness  of rotational curves of galaxies. This
fields can model also larger structures like clusters and
super-clusters of galaxies. This fact somewhat justify that we
consider an scalar field modelling dark matter in our Cardassian
model.

In Cardassian cosmologies \cite{Freese:2002sq} is enough to
consider a universe filled with cosmic dust (pressureless cold
dark matter) and perhaps radiation, to modelling an expanding
universe with geometric origin.

The dark energy can be described using scalar, vector or spinorial
fields. Within the scalar matter, the quintessence field is the
more investigated nowadays. Another possibility are the phantom
fields. This models implies an ``extraneous physics'' because its
kinetic energy is negative and there is also quantum instability.
However, the hybrid model called quintom gives a dynamics for its
equation of state favored by recent astrophysical observations
\cite{Leon:2009ce,Nesseris}.

Since 2004 to the date, researchers have deserved several journal
pages for the construction of dark energy models based in field
theory which allows for the phantom-barrier crossing. First than
all, a single scalar field with canonical kinetic energy
(quintessence) or non-conventional kinetic energy (phantom) do not
gives the desired result \cite{Melchiorri:2002ux,Vikman:2004dc}
(unless we consider non-minimally coupled scalar field models
based in Scalar-tensor theories
\cite{Curbelo:2005dh,Elizalde2004,Apostolopoulos:2006si,Bamba:2008xa,Bamba:2008hq,Setare:2008mb}).

The crossing of the phantom divide is a significant challenge for
theoretical physics. It was proved that the EoS of dark energy
cannot cross the phantom divide if 1.) the dark energy component
has an arbitrary scalar field Lagrangian, which has a general
dependence on the field itself and its first derivatives., 2.)
general relativity holds and 3.) the geometry of the universe is
the spatially flat Friedmann-Robertson-Walker
\cite{Vikman:2004dc}. Thus, realizing such a crossing is not a
trivial work.

To cross the phantom divide, we must break at least one of the
conditions enumerated above. The more simply way to do that is to
consider a Two-field model ({quintessence} and phantom). These
models have settled out explicitly and named quintom models
\cite{quintom,quintom1,quintom2,quintom3,quintom4,Guo:2004fq,Zhang:2005eg,Wei:2005fq,Wei:2005nw,Lazkoz:2006pa,stringinspired,stringinspired1,stringinspired2,Cai:2008gk,Saridakis:2009uu,Lazkoz2007,arbitrary,arbitrary1,arbitrary2}.
Quintom behavior (i.e., the $w=-1$ crossing) has been investigated
in the context of h-essence cosmologies
\cite{Wei:2005fq,Wei:2005nw}; in the context of holographic dark
energy
\cite{holographic,holographic1,holographic2,holographic3,holographic4};
inspired by string theory
\cite{stringinspired,stringinspired1,stringinspired2}; derived
from spinor matter \cite{Cai:2008gk}; for arbitrary potentials
\cite{Lazkoz2007,arbitrary,arbitrary1,arbitrary2}; using
isomorphic models consisting of three coupled oscillators, one of
which carries negative kinetic energy (particularly for
investigating the dynamical behavior of massless
quintom)\cite{setare1}. The crossing of the phantom divide is also
possible in the context of scalar tensor theories
\cite{Elizalde2004,Apostolopoulos:2006si,Bamba:2008xa,Bamba:2008hq,Setare:2008mb}
as well as in modified theories of gravity \cite{Nojiri:2006ri}.

Alternative approaches to dark energy are also the so-called
Extended Theory of Gravitation (ETG) and, in particular,
higher-order theories of gravity (HOG)
\cite{Carroll:2004de,Carroll:2003wy,Capozziello:2003gx,Capozziello:2003tk,Capozziello:2007ec,Wands:1993uu,Capozziello:2010wt,Capozziello:2009yj,Capozziello:2009nq,Capozziello:2007zz,Capozziello:2008gu,Capozziello:2002rd,Faraoni:2005vk,Ruggiero:2006qv,delaCruzDombriz:2006fj,Poplawski:2006kv,Brookfield:2006mq,Song:2006ej,Li:2006ag,Sotiriou:2006hs,Bertolami:2009ac,Bertolami:2009cd,Bertolami:2009ac,Bertolami:2007gv,Briscese:2006xu,Leon:2010pu}.
Such an approach can still be in the spirit of General Relativity
Theory (GRT) since the only request the Hilbert-Einstein action
should be generalized (by including non-linear terms in the Ricci
curvature $R$ and/or involving combinations of derivatives of $R$
\cite{Kerner:1982yg,Duruisseau:1986ga,Teyssandier:1989dw,Magnano:1987zz})
asking for a gravitational interaction acting, in principle, in
different ways in both cosmological
\cite{Capozziello:2003gx,Capozziello:2003tk,Capozziello:2006uv,Maeda:2004vm,Maeda:2004hu,Ohta:2004wk,Akune:2006dg}
and astrophysical \cite{Capozziello:2006uv,Capozziello:2006ph}
scales. In this case the field equations can be recast in a way
that the higher order corrections are written as an
energy-momentum tensor of geometrical origin describing an
``effective" source term on the right hand side of the standard
Einstein field equations
\cite{Capozziello:2003gx,Capozziello:2003tk}. These models have
been studied from the dynamical systems viewpoint in
\cite{Leon:2010pu,Carloni:2004kp,Carloni:2007eu,Carloni:2009nc,Carloni:2007br,Leach:2006br,Leach:2007ss,Goheer:2007wx,Goheer:2007wu,Goheer:2008tn,Goswami:2008fs,Miritzis:2009bi,Abdelwahab:2007jp}.

Other alternatives are the models based on extra-dimensional
theories, for example Randall-Sundrum  braneworlds of type 2
(RS2). Randall-Sundrum braneworls were first proposed in
\cite{rs1,rs2}. In these references was proved that for
non-factorizable geometries in five dimensions there exists a
single massless bound state confined in a domain wall or
three-brane. This bound state is the zero mode of the Kaluza-Klein
dimensional reduction and corresponds to the four-dimensional
graviton. The RS2 model, was proposed as an alternative mechanism
to the Kaluza-Klein compactifications \cite{rs1}, have been
intensively studied in the last years, among other reasons,
because its appreciable cosmological impact in the inflationary
scenario \cite{Hawkins:2000dq, Huey2001, Huey2002}. The setup of
the model start with the particles of the standard model confined
in a four dimensional hypersurface with positive tension embedded
in a 5-dimensional bulk with negative cosmological constant. It is
well-known that the cosmological field equations on the brane are
essentially different from the standard 4-dimensional cosmology.
Friedmann-Robertson branes with an scalar field trapped on it have
been investigated widely in the literature. In \cite{Gonzalez2009}
was investigated the dynamics of a scalar field with constant and
exponential potentials. These results were extended to a wider
class of self-interaction potential in \cite{Leyva2009} using a
method proposed in \cite{Fang2009} supporting the idea that this
scenario modifies gravity only at very high energy/short scales
(UV modifications only) having an appreciable impact on primordial
inflation but does not affecting the late-time dynamics of the
Universe unless if the energy density of the matter trapped in the
brane increase at late times \cite{Garcia-Salcedo2011a}.
Extra-dimensional alternatives are also the so-called
Ho\v{r}ava-Lifshitz cosmology which is a power-counting
renormalizable, ultra-violet (UV) complete theory of gravity
\cite{hor2,hor1,hor3,hor4}. This theory as an infrared (IR) fixed
point, namely General Relativity.

In \cite{Leon:2008de} were investigated coupled dark energy
models. There was proved that for coupling functions and
potentials of desired differentiability class, the scalar field is
unbounded to the past, but a set of zero-measure.  There was
devised a dynamical system appropriated to describe the past
asymptotic dynamics, allowing to classify scaling solutions. There
were presented asymptotic expansion rates for the cosmological
solutions near the initial singularity, extending previous
results.  In \cite{Leon2010} was investigated a cosmological model
based in STT (and, therefore, related by conformal transformations
with $F(R)$-theories) where it is considered a scalar field
coupled to matter, and radiation is included. There was proved
that the equilibrium points corresponding to the nonnegative local
minimums of the potential (associated with cosmological de Sitter
solutions) are asymptotically stable. In the same way as in
\cite{Leon:2008de}, in \cite{Leon2010} we prove that the scalar
field is unbounded towards the past, but a set of zero-measure. A
dynamical system was devised to investigate the dynamics near the
initial singularity, obtaining in that regime: radiation-dominated
scaling solutions; power-law inflationary scalar field dominated
solutions; matter-kinetic-radiation scaling solutions;
matter-potential-radiation scaling solutions. There were
investigated the important examples of modified gravity $F(R) = R
+ \alpha R^2$ (quadratic gravity) and $F(R) = R^n$ gravity. In
this book several of these results will be rigourously-established
and also improved from the mathematical viewpoint.

Concerning the cosmological evolution of quintom models, this
topic was investigated in \cite{Guo:2004fq} and
\cite{Zhang:2005eg,Lazkoz:2006pa}, from the dynamical systems
viewpoint, by considering exponential potentials. The difference
between \cite{Guo:2004fq} and \cite{Zhang:2005eg,Lazkoz:2006pa} is
that in the second case the potential function also accounts for
the interaction between the conventional scalar field and the
phantom field. In \cite{Zhang:2005eg} it had been proven that in
the absence of interactions, the phantom dominated solution should
be the attractor of the system and that the interaction does not
affect its attractor behavior. In \cite{Lazkoz:2006pa} the case in
which the interaction term dominates against the mixed terms of
the potential, was studied. It was proven there, that the
hypothesis in \cite{Zhang:2005eg} is correct only in the cases in
which the existence of the phantom phase excludes the existence of
scaling attractors (in which the energy density of the quintom
field and the energy density of DM are proportional). Some of this
results were extended in \cite{Lazkoz2007}, for arbitrary
potentials. There it was settled down under what conditions on the
potential it is possible to obtain scaling regimes. It was proved
there, that for arbitrary potentials having asymptotic exponential
behavior, scaling regimes are associated to the limit where the
scalar fields diverge.  Also it has been proven that the existence
of phantom attractors in this framework is not generic and
consequently the corresponding cosmological solutions lack the big
rip singularity. In the reference \cite{Leon:2009ce} it is
presented an exhaustive review of the quintom and quinstant
paradigms.

In the reference \cite{Escobar:2011cz} was investigated a scalar
field with arbitrary potential trapped in a RS2 model. There were
obtained sufficient conditions for the asymptotic stability of de
Sitter solution and  for the stability of scaling solutions as
well as for the stability of the scalar-field dominated solution
extending the results in \cite{Copeland1998a} to the higher
dimensional framework. In \cite{Escobar:2011cz} was proved, also,
the non-existence of late time attractors with 5D-modifications. A
fact that correlates with a transient primordial inflation. The
natural extension of the analysis in \cite{Escobar:2011cz} is to
consider a Bianchi I brane. Bianchi I models are the minimal
extension of the FRW metric to the anisotropic framework.
Homogenous but anisotropic geometries are well-known
\cite{Misner:1974qy,Peebles:1994xt}. Bianchi I, Bianchi III, and
Kantowski-Sachs can be a very good representation for the
homogeneous but anisotropic universe. They were investigated in
the framework of $f(R)$ cosmology from both numerical and
analytical viewpoint also incorporating the matter content (see
\cite{Leon:2010pu} and the references therein). The evolution of
cosmological braneworld models were investigated, for instance, in
\cite{Campos:2003bj,Campos:2001cn,Campos:2001pa,Campos:2000cn}. In
\cite{Campos:2001pa} it is presented a systematic analysis of FRW,
Bianchi I and Bianchi V  metrics in these scenarios. There it is
discussed the changes in the structure of the phase space with
respect the general-relativistic case. In \cite{Maartens2001} it
is studied the dynamics of a BI brane and it is showed that the
high energy effects from extra dimensional gravity removes the
anisotropic behavior near the initial singularity which is found
in general relativity. For a Bianchi I-RS2 brane, it is possible
to obtain isotropic late-time attractors compatibles with
accelerated expansion for general classes of self-interacting
potentials for a wide region in the parameter space (work in
progress). In this way, the universe isotropizes towards the
future, irrespectively of the initial anisotropy degree.

Following this line of reasoning, in \cite{Leon:2010pu}, are
investigated $F(R)=R^n$ theories in anisotropic Kantowski-Sachs
(KS) metrics. In this scenario the universe at late times can
result to a state of accelerating expansion, and additionally, for
a particular $n$-range ($2<n<3$) it exhibits phantom behavior.
Additionally, the universe has been isotropized, independently of
the anisotropy degree of the initial conditions, and it
asymptotically becomes flat. The fact that such features are in
agreement with observations
\cite{Komatsu:2010fb,obs1,obs2,obs3,obs4,Perlmutter:1998np,Bennett:2003bz,Tegmark:2003ud,Allen:2004cd,Tegmark:2003uf,Komatsu:2003fd,Hinshaw:2003ex,Spergel:2003cb,Peiris:2003ff,Riess:1998cb,Riess:2004nr,Jassal:2004ej,Freedman:2000cf,Mould:1999ap,Choudhury:2003tj}
is a significant advantage of the model. Moreover, in the case of
radiation ($n=2$, $w=1/3$) the aforementioned stable solution
corresponds to a de-Sitter expansion, and it can also describe the
inflationary epoch of the universe. Note that at first sight the
above behavior could be ascribed to the cosmic no-hair theorem
\cite{Wald}, which states that a solution of the cosmological
equations, with a positive cosmological constant and under the
perfect-fluid assumption for matter, converges to the de Sitter
solution at late times. However, we mention that such a theorem
holds for matter-fluids less stiff than radiation, but more
importantly it has been elaborated for General Relativity
\cite{Kitada-Maeda}, without a robust extension to higher order
gravitational theories \cite{Cotsakis:1993er}. In
\cite{Leon:2010pu} were extracted the results without relying at
all on the cosmic no-hair theorem, which is a significant
advantage of the analysis. Apart from the above behavior, in this
scenario the universe has a large probability to remain in a phase
of (isotropic or anisotropic) decelerating expansion for a long
time, before it will be attracted by the above global attractor at
late times, and this acts as an additional advantage of the model,
since it is in agreement with the observed cosmological behavior.
The Kantowski-Sachs anisotropic $R^n$-gravity can also lead to
contracting solutions, either accelerating or decelerating, which
are not globally stable. Thus, the universe can remain near these
states for a long time, before the dynamics remove it towards the
above expanding, accelerating, late-time attractors. One of the
most interesting behaviors is the possibility of the realization
of the transition between expanding and contracting solutions
during the evolution. That is, the scenario at hand can exhibit
the cosmological bounce or turnaround. Additionally, there can
also appear an eternal transition between expanding and
contracting phases, that is we can obtain cyclic cosmology. These
features can be of great significance for cosmology, since they
are desirable in order for a model to be free of past or future
singularities. In summary, anisotropic $R^n$-gravity has a very
rich cosmological behavior, and a large variety of evolutions and
late-time solutions, compatible with observations, that leads to
radically different implications comparing to the simple isotropic
scenarios.

In \cite{Leon:2009rc} was performed a detailed phase-space
analysis of Ho\v{r}ava-Lifshitz cosmology, with and without the
detailed-balance condition. Under detailed-balance the universe
can reach a bouncing-oscillatory state at late times, in which
dark-energy, behaving as a simple cosmological constant, is
dominant. In this book, we use the Center Manifold Theory to
investigate the stability of the de Sitter solution when
detailed-balance condition is relaxed. In this case (where the
detailed-balance condition is relaxed) the universe reaches an
eternally expanding, dark-energy-dominated solution, with the
oscillatory state preserving also a small probability. Since the
phase space of Ho\v{r}ava-Lifshitz  is, in general, non-compact,
in this book we complement the results in \cite{Leon:2009rc} by
performing a Poincar\'e compactification process  in order to
investigate the dynamics at infinity. This allows to construct a
global phase space containing all the cosmological information.

Finally, in \cite{PLB2010} was investigated several varying-mass
dark-matter particle models in the framework of phantom cosmology.
It is examined there whether there exist late-time cosmological
solutions, corresponding to an accelerating universe and
possessing dark energy and dark matter densities of the same
order. Imposing exponential or power-law potentials and
exponential or power-law mass dependence, the coincidence problem
cannot be solved or even alleviated. Thus, if dark energy is
attributed to the phantom paradigm, varying-mass dark matter
models cannot fulfill the basic requirement that led to their
construction. However, for completeness, in this book we use the
Center Manifold Theory to analyze the stability of the
non-hyperbolic fixed points in the phase space of dark-matter
particle models in the framework of phantom cosmology. Basically,
we use these cosmological models as examples of how to apply the
Center Manifold Theory in cosmology. Also, in this book we perform
a Poincar\'e compactification process allowing to construct a
global phase space containing all the cosmological information in
both finite and infinite regions.

Every one of these theoretical frameworks allows to explain
several features of our dynamical universe. However, in general
they are not valid to describe all the cosmic evolution. Following
one or another direction, the systematic way to examine all
possible cosmological behaviors of a particular  model is to use
dynamical systems techniques. Such an approach allows to by-pass
the high nonlinearities and order of the cosmological equations
(particularly, in the metric approach, $F(R)$ models gives fourth
order differential equations) which prevents a complete analytical
treatment, obtaining a good global qualitative dynamics of the
models under investigation.

Therefore, the proposition and validation of a cosmological model
to explain the accelerated expansion of the universe is an open
topic of discussion nowadays, which is verified in the variety of
models proposed, without arriving to a definitive proposal yet.
All these facts, explained before, justify the present
investigation.

According to that, our investigation problem is related to has far
the dynamical systems studies have contributed to the
understanding of the evolution of the early- and late-time
universe?; and more precisely, about if it is possible to apply
the theory of dynamical systems to select among the cosmological
paradigm, those models with proper past and future attractors.

This it is concreted in the following investigation question: How
can we determine the asymptotic behavior of typical cosmological
solutions with scalar fields, combining topological, analytical
and numerical methods of investigation?

Having examined the bibliography and having elaborated the
theoretical framework, we formulate as investigation hypothesis
that it is possible to obtain information about the local
properties of the flow associated to an autonomous system of
ordinary differential equations (ODEs), using qualitative
techniques and considering a proper normalization and a proper
parametrization only demanding good differentiability and
integrability conditions for the input functions of the models.

Our general objective will be analyze, using a combination of
topological, analytical and numerical techniques, the phase space
of our investigation objects.

In order to fulfill our general objective, we have traced the
specific objectives:
\begin{enumerate}
\item Obtaining all possible asymptotic behaviors for a
quintessence field non-minimally coupled to matter including or
not radiation, based on a STT.

\item Obtaining all possible asymptotic behaviors for mass-varying
dark matter-particles in the framework of phantom cosmologies; and
analyzing the viability of them in order to solve the coincidence
problem (why the energy densities of dark matter and of dark
energy are comparable in order the magnitude today?)

 \item Obtaining
all possible asymptotic behaviors for Ho\v{r}ava-Lifshitz
cosmologies with and without detailed-balance.

 \item Obtaining
all possible asymptotic behaviors for the so-called Cardassian
cosmologies.

\end{enumerate}

The book is organized as follows. In chapter 1 we give a bird's
eye view on cosmology and cosmological problems. Chapter 2 is
devoted to a brief review on some results and useful tools from
the qualitative theory of dynamical systems providing the
theoretical basis for the qualitative study of concrete
cosmological models. Chapters 3, 4,  5 and 6 are devoted to our
main results. In these chapters are extended and settled in a
substantially different,  more strict mathematical language,
several results obtained by one of us in
\cite{Leon:2008de,Leon2010,PLB2010,Leon:2009rc}. In chapter 6, we
provide a different approach to the subject discussed in
\cite{Lazkoz:2005bf}.

\chapter{A bird's eye view on cosmology and cosmological problems}

\section{Cosmological models}\label{modelos}

A cosmological model represents the Universe in a particular scale
\cite{Ellis:1998ct}. Within the most commonly accepted vision in
Cosmology, it is assumed that the geometry of the universe at the
large scale is described by General Relativity (see
\cite{Misner:1974qy,Hawking:1973uf,d'Inverno:1992rk,Wald:1984rg}).

Each cosmological model is defined specifying
(\cite{Ehlers:1993gf,Ellis:1971pg}):
\begin{enumerate}

\item[(i)] the  {\it spacetime geometry}, represented in an
averaged scale by the metric tensor $g_{\alpha \beta}(x^\gamma).$
Due the requirement of the compatibility with astrophysical
observations, the metric of a cosmological model should have as a
regular limit one of the Robertson- Walker (`RW') geometries; or
at least, they should have properties compatibles with the
observations inferred for the cosmological epoch that they are
intended to describe;

\item[(ii)] the {\it matter content}, represented in an averaged
scale, and its {\it physical behavior}. The physical behavior of
matter is described specifying the energy-momentum tensor of each
matter component, their governing equations, the thermodynamical
state equations and the interaction terms. Ideally, they should
have a plausible physical interest (explaining the matter content
of the universe from earlier epochs to the present, including the
majority of the physical interactions described up to date); and

\item[(iii)] the {\it interaction between matter and geometry}
--how the matter determines geometry, and at the same time how
geometry influences matter \cite{Wheeler}--. In general it is
assumed that this relation is given by the {\em Einstein's field
equations of the gravitational field\/} (EFEs) \footnote{Hereafter
we employ units in which $c = 1 = 8\pi G/c^{2}$. Thus, all
geometrical variables should have physical dimensions that are
integer powers of $\lgth$.}
\be \label{eq:efe} G_{\alpha\,\beta} \equiv R_{\alpha\,\beta} -
{\sfrac12}\,R\,g_{\alpha\,\beta} = T_{\alpha\,\beta} -
\Lambda\,g_{\alpha\,\beta} \ , \ee
(here we have  considered the possibility to introduce a non-null
cosmological constant $\Lambda$).  $R$ denotes the trace
$g^{\alpha\,\beta}R_{\alpha\,\beta}$. Assuming that the {\em
cosmological constant\/} $\Lambda$ satisfies the relation
$\nabla_{\alpha}\Lambda = 0$, i.e., if it is constant in time and
in space, EFEs guaranteed the conservation of the total
energy-momentum tensor through the {\it twice contracted Bianchi
identities}, \be \label{eq:cons} \nabla_{\beta}G^{\alpha\,\beta} =
0 \hsp5 \Rightarrow \hsp5 \nabla_{\beta}T^{\alpha\,\beta} = 0 \ ,
\ee

\end{enumerate}

These three ingredients determine the combined evolution of
geometry and the matter on it. The description should be
sufficiently complete to determine:

\begin{enumerate}

\item[(iv)] the  {\it observational relations} predicted by the
model for both discrete sources \footnote{Observations of discrete
sources (primarily galaxies, radio sources, infrared sources and
quasars) provide information about the structure of the universe
in the galactic epoch (say redshifts $z\lesssim 5$). These
observations can be grouped into three clases, {\it number count
surveys}, that provide direct evidence concerning isotropy about
our position, {\it redshift surveys}, that provide information
about the inhomogeneities in the distribution of galaxies, and
{\it peculiar velocity surveys}, that describe deviations from a
unifom Hubble  flow (see \cite{WE}, pp 65-67).}, the cosmic
microwave background radiation and the light-element abundances
from nucleosynthesis in the early universe; implying a
well-established theory for the {\it structure formation} in small
and large physical scales.

\end{enumerate}

Practically, as in any modelling of physical phenomena, we should
compromise the model complexity in order to obtain the desired
results. When referred to a cosmological model, this means
assuming the existence of some symmetries. The usual hypotheses to
describe the matter content are a combination of sources of any of
the following types:

\begin{enumerate}

\item[(v)] a fluid with a well-motivated physical state equation:
for example, a perfect fluid with an specific equation of state;

\item[(vi)] a mixture of fluids, usually with different
4-velocities;

\item[(vii)] a set of particles represented by a kinetic theory
description;

\item[(viii)] a scalar field $\phi$ (or several of them), with a
given self-interacting potential $V(\phi).$

\end{enumerate}

In the phenomenological fluid description of a matter source, the
standard decomposition of the energy-momentum tensor
$T_{\alpha\,\beta}$ with respect to a time-like vector field
$\vec{u}$ is given by: \be T_{\alpha\,\beta}=\rho u_\alpha
u_\beta+2 q_{(\alpha}u_{\beta)}+p(g_{\alpha\,\beta}+u_\alpha
u_\beta)+\pi_{\alpha\,\beta},\label{sourcedecomp}\ee  where $\rho$
denotes the relativistic energy density relative to $u^\alpha$,
$p$ is the isotropic pressure, $q^\alpha$ is the momentum density,
and $\pi_{\alpha\,\beta}$ is the trace-free anisotropic pressure.
We have  $q_\alpha u^\alpha=0,\; \pi_{\alpha\,\beta} u^\beta=0,\;
\pi_\alpha^{\alpha}=0,\;\pi_{\alpha\,\beta}=\pi_{\beta\,\alpha}.$
These quantities have to be related by appropriate thermodynamical
equations of state in order to close the system of equations.
These should provide a coherent representation of the underlying
physics to the geometrical spacetime fluid scenario.

In this book we restrict ourselves to a non-tilted perfect fluid
and to a scalar field. These matter sources admits, respectively,
the following decompositions.

\begin{enumerate}

\item A {\em non-tilted perfect fluid} is described by its
4-velocity $\vec{u}$, its energy density $\rho$ and pressure $p$,
with barotropic equation of state  $p=p(\rho).$ The
energy-momentum tensor is \be T_{\alpha\,\beta}=\rho u_\alpha
u_\beta+p(g_{\alpha\,\beta}+u_\alpha u_\beta),\; u_\alpha
u^\alpha=-1, \ee We will usually work with an equation of state
$p=(\gamma-1)\rho,$ where $\gamma$ is a constant. From a physical
point of view, the must important values are $\gamma=1$ (dust) and
$\gamma=4/3$ (radiation), specially for the early universe. The
value $\gamma=0$ corresponds to a cosmological constant and the
value $\gamma=2$ is occasionally considered corresponding to a
``stiff'' fluid. We assume that $\gamma$ satisfies
$0\leq\gamma\leq 2,$ in order to fulfill the energy requirements.
The values $\gamma>2$ corresponds to fluids with supra-luminous
propagation.

\item A canonical {\em scalar field} is described by the
energy-momentum tensor \be\label{sft}
T_{\alpha\,\beta}=\nabla_\alpha \phi\nabla_\beta\phi
-\left[\frac{1}{2}\nabla_\gamma\phi\nabla^\gamma\phi+V(\phi)\right]g_{\alpha\,
\beta},\ee where the potential $V(\phi)$ has to be specified. If
$\nabla_\alpha \phi$ is time-like, we can define a unit time-like
vector field  $\vec{u}$ normal to the surfaces
$\phi=\textstyle{const.}:$
$$u^\alpha=\frac{\nabla^\alpha
\phi}{(-\nabla_\gamma\phi\nabla^\gamma\phi)^{\frac{1}{2}}}.$$
Then, $T_{\alpha\,\beta}$ has the algebraic form of a perfect
fluid with
$$\rho=-\frac{1}{2} \nabla_\alpha\phi\nabla ^\alpha\phi+V(\phi),\;p=-\frac{1}{2}
\nabla_\alpha\phi\nabla ^\alpha\phi-V(\phi).$$ The equation of
motion for the scalar field is the Klein-Gordon equation
$$\nabla^\alpha\nabla _\alpha \phi-V'(\phi)=0,$$ which is a
consequence of \eqref{sft} and the conservation equation
\eqref{eq:cons}.

\end{enumerate}

\subsection{Homogeneity and isotropy: the Roberson-Walker metric}

The current astrophysical observations points that our observable
universe is homogeneous and isotropic at a great accuracy
\cite{Komatsu:2010fb,obs1,obs2,obs3,obs4,Perlmutter:1998np,Bennett:2003bz,Tegmark:2003ud,Allen:2004cd,Tegmark:2003uf,Komatsu:2003fd,Hinshaw:2003ex,Spergel:2003cb,Peiris:2003ff,Riess:1998cb,Riess:2004nr,Jassal:2004ej,Freedman:2000cf,Mould:1999ap,Choudhury:2003tj},
leading the large majority of cosmological works to focus on
homogeneous and isotropic geometries. Isotropy is the claim that
the universe looks the same in all directions. Direct evidence for
this comes from the smoothness of the Temperature of the CMB.
Homogeneity is the claim that the universe looks the same at every
point. We may therefore approximate the universe as a spatially
homogeneous and isotropic three-dimensional space which may expand
(or, in principle, contract) as a function of time. The metric on
such a spacetime in necessarily the Robertson-Walker form.

Spatial isotropy implies spherical symmetry. Choosing a point as
an origin, and using coordinates $(x,\vartheta,\varphi)$ around
this point, the spatial line element must take the form
\begin{equation}
d\sigma^2=dr^2+f^{2}(r)\left(d\theta^2+\sin^2\theta
d\varphi^2\right) \ ,
\end{equation}
where $f(r)$ is a real function, which, is the metric is to be
nonsingular at the origin , obeys the law $f(r)\sim r$ as
$r\rightarrow 0$. It is proved that the more general metric of a
spacetime consistent with homogeneity and isotropy is
\begin{equation}
\label{frwmetric} ds^2=-dt^2+a^2(t) \left[
d\rho^2+f^2(\rho)\left(d\theta^2+\sin^2\theta
d\phi^2\right)\right] \ ,
\end{equation}
where the three possibilities for  $f(\rho)$ are to be  \be
f(\rho)= \{\sin(\rho), ~~ \rho, ~~ \sinh(\rho)\} \ .  \ee This is
a purely geometric fact, independent of the details of general
relativity \cite{Trodden:2004st}. We have used spherical polar
coordinates $(\rho,\theta,\phi)$, since spatial isotropy implies
spherical symmetry about every point. The time coordinate $t$,
which is the proper time as measured by a comoving observer (one
at constant spatial coordinates), is referred to as cosmic time,
and the function $a(t)$ is called the scale factor.

There are two other useful forms for the RW metric. Fist, a simple
change of variables in the radial coordinate yields
\begin{equation}
\label{frwmetric2} ds^2=-dt^2 +a^2(t)\left[\frac{dx^2}{1-kx^2}
+x^2\left(d\vartheta^2+\sin^2\vartheta d\varphi^2\right)\right] \
,
\end{equation}
where
\begin{equation}
\label{curvature} k=\left\{\begin{array}{ll}
+1 &  \ \ \ \ \ \mbox{si $f(\rho)=\sin(\rho)$} \\
~~0 &  \ \ \ \ \ \mbox{si $f(\rho)=\rho$} \\
-1 &  \ \ \ \ \ \mbox{si $f(\rho)=\sinh(\rho)$}
\end{array}\right. \ .
\end{equation}

Geometrically, $k$ describes the curvature of the spatial sections
(slices at constant cosmic time). $k=+1$ corresponds to positively
curved spatial sections (locally isometric to 3-spheres); $k=0$
corresponds to local flatness, and $k=-1$ corresponds to
negatively curved (locally hyperbolic) spatial sections. These are
all local statements, which should be expected from a local theory
such as GR \cite{Trodden:2004st}.

A second coordinate change, which may be applied to either
~(\ref{frwmetric}) or  ~(\ref{frwmetric2}), is to transform to a
{\it conformal time}, $\tau$, via
\begin{equation}
\label{conformaltime} \tau(t)\equiv \int^t \frac{dt'}{a(t')} \ .
\end{equation}
Applying this to~(\ref{frwmetric2}) yields
\begin{equation}
\label{conffrwmetric2} ds^2= a^2(\tau)\left[-d\tau^2 +
\frac{dr^2}{1-kr^2}+r^2\left(d\theta^2+\sin^2\theta
d\phi^2\right)\right] \ ,
\end{equation}
where we have written $a(\tau)\equiv a[t(\tau)]$ as is
conventional. The conformal time does not measure the proper time
 for any particular observer, but it does simplify some calculations.

A particular useful quantity to define form the scale factor is
the  {\it Hubble parameter} (sometimes called Hubble constant),
given by
\begin{equation}
\label{Hubbleconstant} H\equiv \frac{{\dot a}}{a} \ .
\end{equation}
The Hubble parameter relates how fast the most distant galaxies
are receding form us to their distance from us via Hubble's law
$v\simeq H d.$

\subsection{Dynamics: the Friedmann equations}

As mentioned, the RW metric is a purely kinematic consequence of
requiring homogeneity and isotropy of our spatial sections. We
next turn to the dynamics, in form of differential equations
governing the evolution of the scale factor $a(t)$. These will
come from applying Einstein's equations , \be R_{\mu\nu}-{1\over
2}Rg_{\mu\nu} =8\pi G T_{\mu\nu}
  \label{einstein}
\ee to the RW metric.

It is often to adopt the perfect fluid form for the
energy-momentum tensor of cosmological matter. This form is
\begin{equation}
\label{perfectfluid} T_{\mu\nu} = (\rho + p)U_\mu U_\nu + p
g_{\mu\nu}\ ,
\end{equation}
where $U^{\mu}$ is the fluid 4-velocity, $\rho$ is the energy
density in the rest frame of the fluid and $p$ is the pressure in
that same frame. The pressure is necessarily isotropic for
consistency with RW matric. Similarly, fluid elements will be
comoving in the cosmological rest frame, so that the normalized
4-velocity in the coordinates of (\ref{frwmetric2}) will be \be
  U^\mu = (1,0,0,0)\ .
\ee Therefore, the energy-momentum tensor takes the form  \be
  T_{\mu\nu} = \left(%
\begin{array}{cc}
  \rho & 0 \\
  0 & p g_{ij} \\
\end{array}%
\right)\ee where $g_{ij}$ represents the spatial metric (including
the factor of $a^2$).

Armed with this simplified description for matter, we are now
ready to apply Einstein's equation (\ref{einstein}) to cosmology.
Using~(\ref{frwmetric2}) and~(\ref{perfectfluid}), we obtain two
equations. The first is known as the Friedmann equation,
\begin{equation}
\label{Friedmann} H^2 \equiv \left(\frac{{\dot
a}}{a}\right)^2=\frac{8\pi G}{3}\sum_i \rho_i -\frac{k}{a^2} \ ,
\end{equation}
where an overdot denotes a derivative with respect to cosmic time
$t$ and $i$ indexes all different types of energy in the universe.
This equation is a constraint equation, in the sense that we are
not allowed to freely specify the time derivative  $\dot{a}$; it
is determined in terms of the energy density and curvature. The
second equation, which is an evolution equation, is
\begin{equation}
\label{2ndeinsteineqn} \frac{\ddot a}{a}
+\frac{1}{2}\left(\frac{{\dot a}}{a}\right)^2=-4\pi G\sum_i p_i
-\frac{k}{2a^2} \ .
\end{equation}
It is often useful to combine ~(\ref{Friedmann})
and~(\ref{2ndeinsteineqn}) to obtain the  {\it acceleration
equation}
\begin{equation}
\label{acceleration} \frac{{\ddot a}}{a}=-\frac{4\pi G}{3}\sum_i
\left(\rho_i +3p_i \right) \ .
\end{equation}
In fact, if we know the magnitudes and evolutions of the different
energy density components  $\rho_i$, the Friedmann equation
(\ref{Friedmann}) is sufficient to solve for the evolution
uniquely. The acceleration equation is conceptually useful, but
rarely invoked in calculations \cite{Trodden:2004st}.

The Friedmann equation relates the rate of increase of the scale
factor, as encoded by the Hubble parameter, to the total energy
density of all matter in the universe. We may use the Friedmann
equation to define, at any given time, a critical energy density,
\begin{equation}
\label{criticaldensity} \rho_c\equiv \frac{3H^2}{8\pi G} \ ,
\end{equation}
for which the spatial sections must be precisely flat ($k=0$). We
then define the density parameter
\begin{equation}
\label{omega} \Omega_{\rm total} \equiv \frac{\rho}{\rho_c} \ ,
\end{equation}
which allows us to relate the total energy density in the universe
to its local geometry via
\begin{eqnarray}
\Omega_{\rm total}>1 & \Leftrightarrow & k=+1 \nonumber \\
\Omega_{\rm total}=1 & \Leftrightarrow & k=0 \\
\Omega_{\rm total}<1 & \Leftrightarrow & k=-1 \nonumber \ .
\end{eqnarray}
It is often convenient to define the fractions of the critical
energy density in each different component by
\begin{equation}
\Omega_i=\frac{\rho_i}{\rho_c} \ .
\end{equation}

Energy conservation is expressed in GR by the vanishing of the
covariant divergence of the energy-momentum tensor, \be
  \nabla_\mu T^{\mu\nu} = 0\ .
\ee Applying this to our assumptions --the RW metric
(\ref{frwmetric2}) and a perfect energy-momentum tensor
(\ref{perfectfluid})-- yields to a single energy-conservation
equation,
\begin{equation}
\label{energyconservation} {\dot \rho} + 3H(\rho+p)=0 \ .
\end{equation}
This equation is actually not independent of the Friedmann and
acceleration equations, but it is required for consistency. it
implies that the expansion of the universe (as specified by $H$)
can lead to local changes in the energy density. Note that there
is no notion of conservation of ``total energy'', as energy can be
interchanged between matter and the spacetime geometry.

To close the equations we need to specify the equation of state.
Within the fluid approximation we are using we may assume that the
pressure is a single-valued function of the energy density
$p=p(\rho)$. It is often convenient to define an equation of state
(EoS) parameter, $w$, by \be p= w\rho\ . \ee Several sources of
cosmological matter satisfies this relation with $w$ constant. For
example, $w=0$ corresponds to pressureless matter, or dust -- any
collection of massive non-relativistic particles would qualify.
Similarly, $w=1/3$ corresponds to a gas of radiation, whether it
be actual photons or other highly relativistic species.

A constant value of $w$ leads to a great simplification in solving
our equations. In particular, using~(\ref{energyconservation}), we
see that the energy density evolves with the scale factor
according to
\begin{equation}
\label{energydensity} \rho(a) \propto \frac{1}{a(t)^{3(1+w)}} \ .
\end{equation}

We have not included a cosmological constant $\Lambda$ in the
gravitational equations. This is because it is equivalent to treat
any cosmological constant as a component of the energy density in
the universe. In fact, adding a cosmological constant $\Lambda$ to
Einstein's equation is equivalent to including an energy-momentum
tensor of the form \be
  T_{\mu\nu} = -{\Lambda \over 8\pi G} g_{\mu\nu}\ .
\ee This is simply a perfect fluid with energy momentum
tensor~(\ref{perfectfluid}) with
\begin{eqnarray}
\rho_{\Lambda} & = & \frac{\Lambda}{8\pi G} \nonumber \\
p_{\Lambda} & = & -\rho_{\Lambda} \ ,
\end{eqnarray}
so that the EoS parameter is \be
  w_{\Lambda}=-1\ .
\ee This implies that the energy density is a constant, \be
 \rho_\Lambda = {\rm constant}\ .
\ee Thus, this energy is constant throughout spacetime; we say
that the cosmological constant is equivalent to  {\it vacuum
energy}.

Similarly, it is sometime useful to consider any non-null
curvature as yet another component of the cosmological energy
budget, obeying
\begin{eqnarray}
\rho_k & = & -\frac{3k}{8\pi Ga^2} \nonumber \\
p_k & = & \frac{k}{8\pi Ga^2}\ ,
\end{eqnarray}
so that \be w_k=-1/3 \ . \ee It is not an energy density, of
course; $\rho_k$ is simply a convenient way to keep track of how
mush energy density is lacking, in comparison to a flat universe.

\subsection{Flat universes}

It is much easier to find exact solution to the cosmological
equations for $k=0$. Fortunately to us, nowadays we can appeal to
mare than mathematical simplicity to make this choice. The modern
cosmological observations, particularly precision measurements of
the CMB, show the universe today to be extremely spatially flat.

In the case of flat spatial sections and for constant equation of
state $w$, we can exactly solve the Friedmann equation
(\ref{energydensity}) to obtain
\begin{equation}
\label{flatsolution} a(t)=a_0 \left(t\over t_0\right)^{2/3(1+w)} \
,
\end{equation}
where $a_0$ is the scale factor today, unless $w=-1$, in which
case we obtain $a(t) \propto e^{Ht}$. Applying this result to the
more commonly used matter sources we obtain the results showed in
table ~\ref{sourcestable}.

\begin{table*}[t]
\caption[A summary of the behaviors of the most important sources
of energy density in cosmology.]{A summary of the behaviors of the
most important sources of energy density in cosmology.  The
behavior of the scale factor applies to the case of a flat
universe; the behavior of the energy densities is perfectly
general.} \label{sourcestable}
\begin{center}
\begin{tabular}{l|l|l}
Type of energy & $\rho(a)$ & $a(t)$ \\ \hline
Dust & $a^{-3}$ & $t^{2/3}$ \\
Radiation & $a^{-4}$ & $t^{1/2}$ \\
Cosmological constant & constant & $e^{Ht}$ \\
\end{tabular}
\end{center}
\end{table*}

\subsection{Including curvature}
It is true that we know observationally that the universe today is
flat to a high degree of accuracy. However, it is instructive, and
useful when considering early cosmology, to consider how the
solutions we have already identified change when curvature is
included. We discuss some examples from \cite{Trodden:2004st} by
working in terms on the conformal time $\tau$.

Let us first consider models in which the energy density is
dominated by matter ($w=0$). In terms of conformal time, Einstein
equations become
\begin{eqnarray}
3(k+h^2) & = & 8\pi G\rho a^2 \nonumber \\
k+h^2+2h' & = & 0 \ ,
\end{eqnarray}
where a prime denotes derivative with respect to the conformal
time and $h(\tau) \equiv a'/a$. These equations are then easily
solved for $h(\tau)$ giving
\begin{equation}
\label{hoftaudust} h(\tau)=\left\{\begin{array}{ll}
\cot(\tau/2) & \ \ \ \ k=1 \\
2/\tau & \ \ \ \ k=0 \\
\coth(\tau/2) & \ \ \ \ k=-1
\end{array} \right. \ .
\end{equation}
This then yields
\begin{equation}
\label{aoftaudust} a(\tau)\propto \left\{\begin{array}{ll}
1-\cos(\tau) & \ \ \ \ k=1 \\
\tau^2/2 & \ \ \ \ k=0 \\
\cosh(\tau)-1 & \ \ \ \ k=-1
\end{array} \right. \ .
\end{equation}
One may use this to derive the connection between cosmic time and
conformal time, which here is
\begin{equation}
\label{toftaudust} t(\tau)\propto \left\{\begin{array}{ll}
\tau-\sin(\tau) & \ \ \ \ k=1 \\
\tau^3/6 & \ \ \ \ k=0 \\
\sinh(\tau)-\tau & \ \ \ \ k=-1
\end{array} \right. \ .
\end{equation}

Next we consider models dominated by radiation ($w=1/3$). In terms
of conformal time, the Einstein equations become
\begin{eqnarray}
3(k+h^2) & = & 8\pi G\rho a^2 \nonumber \\
k+h^2+2h' & = & -\frac{8\pi G \rho}{3}a^2 \ .
\end{eqnarray}
Solving as we did above yields
\begin{equation}
\label{hoftaurad} h(\tau)=\left\{\begin{array}{ll}
\cot(\tau) & \ \ \ \ k=1 \\
1/\tau & \ \ \ \ k=0 \\
\coth(\tau) & \ \ \ \ k=-1
\end{array} \right. \ ,
\end{equation}
\begin{equation}
\label{aoftaurad} a(\tau)\propto \left\{\begin{array}{ll}
\sin(\tau) & \ \ \ \ k=1 \\
\tau & \ \ \ \ k=0 \\
\sinh(\tau) & \ \ \ \ k=-1
\end{array} \right. \ ,
\end{equation}
and
\begin{equation}
\label{toftaurad} t(\tau)\propto \left\{\begin{array}{ll}
1-\cos(\tau) & \ \ \ \ k=1 \\
\tau^2/2 & \ \ \ \ k=0 \\
\cosh(\tau)-1 & \ \ \ \ k=-1
\end{array} \right. \ .
\end{equation}

Is is straightforward to interpret these solution by examining the
behavior of the scale factor $a(\tau)$; the qualitative features
are the same for matter- or radiation-domination. In both cases,
the universes with positive curvature ($k=+1$) expand from an
initial singularity with $a=0$, and latter recollapsing again. The
initial singularity is the Big Bang, and the final singularity is
sometimes called the Big-Crunch. The universes with zero or
negative curvature begin at the Big-Bang and expand forever.

\subsection{Geometry, destiny and dark energy}

We know that in matter- or radiation-dominated universes with
energy density greater than the critical would ultimately
collapse, while those with less than the critical would expand
forever, with flat universes lying in the border between the two.
For the case of purely dust-filled universes this is easily seen
form  ~(\ref{aoftaudust}) and~(\ref{aoftaurad}).

The connection between geometry and destiny was quite reasonably
as long as dust and radiation were the only types of energy
relevant in the late universe.

In recent years it has become clear that the dominant component of
energy density in the present universe is neither dust nor
radiation, but rather is dark energy. This component is
characterized by an equation of state parameter  $w<-1/3$.

For simplicity let us focus on what happens if the only energy
density in the universe is a cosmological constant, with $w=-1.$
In this case the Friedmann equation may be solve for any value of
the spatial curvature parameter. If $\Lambda
>0$ then the solutions are
\begin{equation}
\label{positiveLambdasolns}
\frac{a(t)}{a_0}=\left\{\begin{array}{ll}
\cosh\left(\sqrt{\frac{\Lambda}{3}} t\right) & \ \ \ \ k=+1  \\
\exp\left(\sqrt{\frac{\Lambda}{3}} t\right) & \ \ \ \ k=0  \\
\sinh\left(\sqrt{\frac{\Lambda}{3}} t\right) & \ \ \ \ k=-1
\end{array}\right. \ ,
\end{equation}
were we have encountered the case $k=0$ earlier. It is immediately
clear that in the limit $t \rightarrow\infty$, all the solutions
expand exponentially, independently of the spatial curvature. In
fact, these solutions are exactly the same spacetime {\it de
Sitter space}, just in different coordinate systems. The crucial
point is that the universe clearly expands forever in these
spacetimes, irrespective of the value of the spatial curvature.
Note, however, that no all of the solutions in
(\ref{positiveLambdasolns}) actually cover all the de Sitter
spacetime; the $k=0$ and $k=-1$ solutions represents coordinate
patches which only cover part of the manifold. For completeness,
let us complete the description of spacetimes with a cosmological
constant by considering the case $\Lambda<0$. This spacetime is
called {\it Anti-de Sitter space} (AdS) and it should be clear
from the Friedmann equation that such an spacetime can only exists
in a space with spatial curvature $k=-1$. The corresponding
solution for the scale factor is
\begin{equation}
\label{antidesitter} a(t)=a_0 \sin\left(\sqrt{-\frac{\Lambda}{3}}
t\right) \ .
\end{equation}
Once again, this solution does not cover all of AdS.

\subsection{Scalar fields and Dark Energy}\label{camposescalares}

Dark energy is one of the hottest topic in precision cosmology
(see the recent reviews
\cite{Copeland:2006wr,Caldwell:2009ix,Turner:2007qg,Alam:2004jy,Leon:2009ce,Sahni}),
with the cosmological constant being the more arguably and
economical candidate. This proposal suffers, however, from the
well-known problem of fine tuning. That is, there exists a (yet)
unexplained discrepancy (close to 120 orders of magnitude) between
the cosmological observed value and the value predicted by Quantum
Field Theory. This problem has been baptized as the Cosmological
Constant Problem.

Rather than dealing directly with the cosmological constant a
number of alternatives routes have been proposed which skirt
around this thorny issue. An incomplete list includes

\begin{enumerate}
\item[i)] Quiessence with $w\equiv p_{DE}/\rho_{DE}=const.,$ the
cosmological constant $\Lambda$ ($w=-1$) is a special member of
this class.

\item[ii)] Quintessence models which are inspired by the simplest
class of inflationary models of the early universe and employ a
scalar field rolling down a potential $V(\phi)$ to achieve
acceleration. Quintessence potentials with $V'' V/(V')^2\geq 1$
have the attractive property that the dark energy approaches a
common evolutionary ``tracker path'' from a wide range of initial
conditions.

\item[iii)] K-essence which is characterized by a scalar field
with a non-canonical kinetic energy. The most general scalar fiedl
action is a function of $\phi$ and $X\equiv
-\frac{1}{2}(\nabla\phi)^2.$

\item[iv)] Tachyon fields with the more general Lagrangian
$\mathcal{L}_\phi=V(\phi)\sqrt{-det(g_{\alpha \beta}+\partial
_\alpha \phi \partial_\beta\phi)}.$

 \item[v)] The Chaplygin gas model (CG) has the equation of
state $p\propto -1/\rho$ and evolves as $\rho=\sqrt{A+B(1+z)^6}$
where $z$ is the redshift, $z=a(t0)/a(t)-1.$ It therefore behaves
like dark matter at early times ($z\gg 1$) and like the
cosmological constant at late times.  CG appears to be the
simplest model attempting to unify DE and non-baryonic cold dark
matter.

\item[vi)] ``Phantom'' DE ($w<-1$).

\item[vii)] Oscillating DE.

\item[viii)] Models with interactions between DE and DM.

\item[ix)] Scalar-tensor DE models.

\item[x)] Modified Gravity DE models in which the gravitational
Lagrangian is changed from $R$ to $F(R)$ where $R$ is the
curvature scalar and $F$ is an arbitrary function.

\item[xi)] Dark energy driven by quantum effects.

\item[xii)] Higher dimensional ``braneworlds'' models in which
acceleration is caused by the leakage of gravity into extra
dimensions.

\item[xiii)] Holographic dark energy, etc.
\end{enumerate}

In table (\ref{tab:ModelosDeCamposEscalaresParaEO}) we enumerate
some of the Lagrangian densities of the most common models for DE.
Additionally we present the corresponding equations of motion for
the scalar field in the case of homogeneous cosmologies.

\begin{table}[htbp]
   \centering
   {\small
\caption{DE models based on scalar fields
($X\equiv-\frac{1}{2}g^{\alpha\,\beta}\partial_\alpha\phi\partial_\beta\phi$).
}    \label{tab:ModelosDeCamposEscalaresParaEO}
\begin{tabular}{@{\hspace{4pt}}c@{\hspace{14pt}}c@{\hspace{14pt}}c@{\hspace{2pt}}}
\hline
\hline\\[-0.3cm]
Model & Lagrangian Density& Motion Equations. \\[0.1cm]
\hline\\[-0.2cm]
Quintessence& $\mathcal{L}_\phi=-V(\phi)+X$& $\ddot\phi+3H\dot\phi+\frac{d V}{d\phi}=0$\\[0.2cm]
Tachyon & $\mathcal{L}_\phi=-V(\phi)\sqrt{1-2\,X}$&
$\frac{\ddot\phi}{1-\dot\phi^2}+3H\dot\phi+\frac{1}{V}\frac{d
V}{d\phi}=0$\\[0.2cm]
Phantom& $\mathcal{L}_\phi=-V(\phi)-X$& $\ddot\phi+3H\dot\phi-\frac{d V}{d\phi}=0$ \\[0.2cm]
K-essence& $\mathcal{L}_\phi=L(\phi,X)$ & \\ & $L$ no lineal en
$X$ & $\left(\frac{\partial L}{\partial X}+2 X \frac{\partial^2
L}{\partial X^2}\right)\ddot\phi+\frac{\partial L}{\partial
X}\left(3 H\dot \phi\right)+$
 \\ &  & $\frac{\partial^2 L}{\partial\phi\partial X}\dot\phi^2-\frac{\partial L}{\partial\phi}=0$ \\[0.4cm]
\hline \hline
       \end{tabular}}
   \end{table}

\underline{Quintessence}\\ The most popular scalar field DE models
are the quintessence field
\cite{Sahni:2002kh,Padmanabhan:2002ji,Kolda:1998wq}.  These models
are described  by a conventional scalar field $\phi$ minimally
coupled to gravity. An adequate choice of the interaction
potential gives the late time acceleration of the universe. The
action for quintessence is given by: \be S=\int d{}^4
x\sqrt{|g|}\mathcal{L}_\phi\label{eq1.1.2}\ee (see definition of
$\mathcal{L}_\phi$ in table
(\ref{tab:ModelosDeCamposEscalaresParaEO})).

The energy-momentum tensor of quintessence field (obtained by
varying the action (\ref{eq1.1.2}) with respect to
$g^{\alpha\,\beta}$ and having into account the identity
$\delta\sqrt{|g|}=-(1/2)\sqrt{|g|} g_{\alpha\,\beta}\delta
g^{\alpha\,\beta}$) is given by:

\be T_{\alpha\,\beta}\equiv-\frac{2}{\sqrt{|g|}}\frac{\delta
S}{\delta
g^{\alpha\,\beta}}=\partial_\alpha\phi\partial_\beta\phi-g_{\alpha\,\beta}
\left[\frac{1}{2}g^{\gamma\,\delta}\partial_\gamma\phi\partial_\delta\phi+V(\phi)\right]\label{1.1.5}.\ee

In the spatially homogeneous case, the energy density and
isotropic pressure are given respectively  by
$\rho=-T^0_0=\frac{1}{2} \dot\phi^2+V(\phi),$ $p=T^i_i=\frac{1}{2}
\dot\phi^2- V(\phi).$ We assume that
$w\equiv\frac{p}{\rho}\in[-1,-\frac{1}{3}].$

\underline{Phantom}\\ Recently it has been argued that (combined)
astrophysical observations (from SNIa and CMB) might favour a DE
component with EoS parameter $\omega=p/\rho<-1$
\cite{Riess:2004nr,Jassal:2004ej,Choudhury:2003tj,Alam:2004jy,Alam:2003fg,Feng:2004ad},
where $p$ is the pressure and $\rho$ is the nergy density of the
fluid. Sources sharing this property violate the null dominant
energy condition (NDEC) \footnote{That NDEC-violating sources can
occur has been argued decades ago, e.g., see references
\cite{Nilles:1983ge,Barrow:1988yc,Pollock:1988xe}.}.
NDEC-violating sources have been investigated as possible DE
candidates and have been called phantom components
\cite{Caldwell:1999ew,Nojiri:2003vn}.

Since NDEC prevents instability of the vacuum or propagation of
energy outside the light cone, phantom models are intrinsically
unstable \footnote{Another very strange property of phantom
universes is that their entropy is negative
\cite{Brevik:2004sd}.}. Nevertheless, if thought of as effective
field theories (valid up to a given momentum cutoff) these models
could be phenomenologically viable \cite{Carroll:2003st}.

Phantom DE can be described by an unconventional scalar field
$\phi$ minimally coupled to gravity. Its action is given by: \be
S=\int d{}^4 x\sqrt{|g|}\mathcal{L}_\phi\label{1.1.2new}\ee (see
definition of $\mathcal{L}_\phi$ in table
\ref{tab:ModelosDeCamposEscalaresParaEO}). The energy-momentum
tensor y very similar to (\ref{1.1.5}) with the identification
\begin{equation}\partial_\alpha\phi\partial_\beta\phi\rightarrow
-\partial_\alpha\phi\partial_\beta\phi.\end{equation}

Due to the unorthodox character of considering the wrong sign of
the kinetic term we want to comment further on this particular. As
already noted, the argument most used for considering phantom
matter at classical level is linked with the idea that, at long
distances, the field theory of the phantom particle is an
effective theory whose ultraviolet completion is well defined and
respects unitarity \cite{Cline:2003gs}. In effect, the divergent
nature of the instability caused by a massive ghost excitations
can be avoided only by imposing a Lorentz non-invariant momentum
space cutoff such that the vacuum decay rate would be slow enough
on cosmological time scales. In \cite{Cline:2003gs}, the magnitude
of the cutoff is estimated by requiring consistency with
observational constraints. The authors found that the
low-effective ghosts must originate from new physics far below the
TeV scale. String theory, in particular is ruled out as possible
source for effective ghosts \cite{Cline:2003gs}. A possibility,
also anticipated in \cite{Cline:2003gs}, is that the phantom field
massive ghost excitations might come from a low-energy sector that
is completely hidden from the standard model particles, except for
gravitational couplings.  Braneworlds, in particular
Randall-Sundrum type 2 (RS2) proposed in the reference
\cite{Randall:1999vf} could be such a source for effective ghosts
required to hold the phantom dark energy hypothesis
\cite{Curbelo:2005dh}. Besides, the effective phantom nature of
the DE can be reinterpreted as arising from dynamical-screening of
the brane cosmological constant in Dvali-Gabadadze-Porrati (DGP)
braneworld model (proposed in \cite{Dvali:2000hr}) with a standard
brane cosmological constant \cite{Lue:2004za}.

\underline{The Big-Rip singularity}\\ To the number of unwanted
properties of a phantom component with ``supernegative'' EOS
parameter $\omega_{ph}=p_{ph}/\rho_{ph}<-1$, we add the fact that
its energy density $\rho_{ph}$ increases in an expanding universe.
\footnote{Alternatives to phantom models to account for
supernegative EOS parameter have been considered also. See, for
instance, references \cite{Onemli:2002hr}.} This property
ultimately leads to a catastrophic (future) big rip singularity
that is characterized by divergences in the scale factor $a$, the
Hubble parameter $H$ and its time-derivative $\dot H$
\cite{Chimento:2004ps}. In fact, the Einstein's field equations
for a flat FRW universe with line element  $ds^2 = - dt^2 + a^2(t)
\left(dx^2 + d{\vartheta}^2 + \sin^2 {\vartheta}\,
d{\varphi}^2\right),$ filled with a barotopic perfect fluid with
EoS $w=p/\rho$ ($w$ const.) admits the solution
$H=\frac{2}{3(1+\omega)(t-t_0)},\; a(t) \propto
(t-t_0)^{\frac{2}{3(1+w)}},\; \rho\propto a^{-3(1+w)},$ where
$t_0$ is a constant and $w\neq -1.$ Observe that for $w<-1$
describe a contracting universe. However, these equations admits
an expanding solution given by $a(t)=(t_s-t)^{\frac{2}{3(1+w)}},$
where $t_s$ is a constant. This corresponds to a
super-inflationary solution where the Hubble scalar and curvature
increases with time respectively as $H=\frac{n}{t_s-t},\;
n=-\frac{2}{3(1+w)}>0,\; R=6\left(2 H^2+\dot H\right)=\frac{6
n(2n+1)}{(t_s-t)^2}$ .The scale factor diverge (and consequently,
matter density) as $t\rightarrow t_s$ (which corresponds to an
infinitely large quantity in a finite proper time towards the
future) provided $w<-1$. In that limit, the Hubble scalar and
curvature become infinity. This situation has been referred as
Big-Rip singularity. This is an inherent property of all generic
phantom DE model \cite{Caldwell:2003vq}. A detailed study of the
kinds of singularity might occur in phantom scenarios (including
the big rip) has been the target of references
\cite{Fernandez-Jambrina:2004yy,Nojiri:2005sx,Fernandez-Jambrina:2006hj}.

This singularity is at a finite amount of proper time into the
future but, before it is reached, the phantom energy will rip
apart all bound structures, including molecules, atoms and nuclei.
To avoid this catastrophic event (also called "cosmic doomsday"),
some models and/or mechanisms have been invoked. In the references
\cite{Nojiri:2004pf,Nojiri:2004ip,Elizalde:2004mq}, for instance,
it has been shown that this singularity in the future of the
cosmic evolution might be avoided or, at least, made milder if
quantum effects are taken into consideration. Instead, a suitable
perturbation of de Sitter equation of state can also lead to
classical evolution free of the big rip \cite{McInnes:2001zw}.
Gravitational back reaction effects \cite{Wu:2004ex} and scalar
fields with negative kinetic energy term with self interaction
potentials with maxima \cite{Carroll:2003st,Singh:2003vx} have
also been considered. Another way to avoid the unwanted big rip
singularity is to allow for a suitable interaction between the
phantom energy and the background (DM) fluid
\cite{Curbelo:2005dh,Guo:2004vg,Cai:2004dk}.

\underline{Interacting Dark energy}\\ Although experimental tests
in the solar system impose severe restrictions on the possibility
of non-minimal coupling between the DE and ordinary matter fluids
\cite{Will:1993ns}, due to the unknown nature of the dark matter
(DM) as part of the background, it is possible to have additional
(non gravitational) interactions between the DE and the DM
components, without conflict with the experimental data. Thus, if
there is transfer of energy from the phantom component to the
background fluid, it is possible to arrange the free parameters of
the model, in such a way that the energy densities of both
components decrease with time, thus avoiding the big rip
\cite{Curbelo:2005dh,Guo:2004vg}.

Since there is exchange of energy between the phantom and the
background fluids, the energy is not conserved separately for each
component. Instead, the continuity equation for each fluid shows a
source (interaction) term \cite{Curbelo:2005dh}:
\begin{equation}
\dot\rho_m+3H(\rho_M +p_m)=Q, \label{backconteq}
\end{equation}
\begin{equation}
\dot\rho_{DE}+3H(\rho_{DE}+p_{DE})=-Q, \label{phconteq}
\end{equation}
where the dot accounts for derivative with respect to the cosmic
time and $Q$ is the interaction term. Note that the total energy
density $\rho_T=\rho_M+\rho_{DE}$ ($p_T= p_M+p_{DE}$) is indeed
conserved: $\dot\rho_T+3H(\rho_T+p_T)=0$. To specify the general
form of the interaction term, the authors look at a scalar-tensor
theory of gravity where the matter degrees of freedom and the
scalar field are coupled in the action through the scalar-tensor
metric $\chi(\phi)^{-1}g_{ab}$
\cite{Curbelo:2005dh,Kaloper:1997sh}:
\begin{eqnarray}
&&S_{ST}=\int d^4x
\sqrt{|g|}\left\{\frac{R}{2}-\frac{\epsilon}{2}(\nabla\phi)^2-V(\phi)
+\right. \nonumber\\ &&  \ \ \ \ \ \ \ \ \ \ \ \ \ \ \ \ \ \
\left. \chi(\phi)^{-2}{\cal
L}_m(\mu,\nabla\mu,\chi^{-1}g_{ab})\right\}, \label{staction}
\end{eqnarray}
where $\epsilon=\pm 1$ ($\epsilon=-1$ for phantom models and
$\epsilon=1$ for quintessence models), $V(\phi)$ is the scalar
field self-interaction potential,  $\chi(\phi)^{-2}$ is the
coupling function, ${\cal L}_m$ is the matter Lagrangian and $\mu$
is the collective name for the matter degrees of freedom. It can
be shown that, in terms of the coupling function $\chi(\phi)$, the
interaction term $Q$ in equations (\ref{backconteq}) and
(\ref{phconteq}), can be written in the following form:
\begin{equation}
Q=\rho_M H\left[a\frac{d(\ln{\bar\chi})}{da}\right],
\label{interactionterm}
\end{equation}
where it is used the  ``reduced'' notation
$\bar\chi(a)\equiv\chi(a)^{(3\omega_M-1)/2}$ and it has been
assumed that the coupling can be written as a function of the
scale factor $\bar\chi=\bar\chi(a)$. Comparing this with other
interaction terms in the bibliography, one can obtain the
functional form of the coupling function $\bar\chi$ in each case.
In the reference \cite{Guo:2004vg}, for instance,
$Q=3Hc^2(\rho_{ph}+\rho_M)= 3c^2H\rho_M(r+1)/r$, where $c^2$
denotes the transfer strength and $r\equiv
\Omega_{m}/\Omega_{ph}$. If one compares this expression with
(\ref{interactionterm}) one obtains the following coupling
function:
\begin{equation}
\bar\chi(a)=\bar\chi_0\;e^{3\int\frac{da}{a}(\frac{r+1}{r})c^2},
\label{xizhang}
\end{equation}
where $\bar\chi_0$ is an arbitrary integration constant. If
$c^2=c_0^2=const.$ and $r=r_0=const.$, then
$\bar\chi=\bar\chi_0\;a^{3c_0^2(r_0+1)/r_0}$.

\underline{The coincidence problem} \\ Models with interaction
between the phantom and the DM components are also appealing since
the coincidence problem (why the energy densities of dark matter
and dark energy are of the same order precisely at present?) can
be solved or, at least, smoothed out by considering a
non-gravitational coupling between both dark sources
\cite{Curbelo:2005dh,Amendola:1999er,Amendola:1999qq,Bean:2000zm,Chimento:2003ie,Chimento:2000kq,Chimento:2003sb,Zimdahl:2001ar,Gonzalez:2006cj}.

It has been shown, in particular, that a suitable coupling, can
produce scaling solutions. The way in which the coupling is
approached is not unique. In reference
\cite{Amendola:1999er,Amendola:1999qq}, for instance, the coupling
is introduced by hand. In
\cite{Chimento:2003ie,Chimento:2000kq,Chimento:2003sb} the type of
coupling is not specified from the beginning. Instead, the form of
the interaction term is fixed by the requirement that the ratio of
the energy densities of DM and quintessence has an stable fixed
point during the evolution that solves the coincidence; in
\cite{Chimento:2003ie} a suitable interaction between the
quintessence field and DM leads to a transition from the
domination matter era to an accelerated expansion epoch in the
model proposed in
\cite{Chimento:2003ie,Chimento:2000kq,Chimento:2003sb}. A model
derived from the Dilaton is studied in \cite{Bean:2000zm}. In this
model the coupling function is chosen as a Fourier expansion
around some minimum of the scalar field.

It is interesting to discuss, in the general case, under which
conditions the coincidence problem might be avoided in models with
interaction among the components in the mixture. In this sense one
expects that a regime with simultaneous non zero values of the
density parameters of the interacting components is a singular
point of the corresponding dynamical system, so the system lives
in this state for a sufficiently long period of time and, hence,
the coincidence does not arises.

For this purpose it is recommended to study the dynamics of the
ratio function $r$:

\be
r=\frac{\rho_M}{\rho_\phi}=\frac{\Omega_m}{\Omega_\phi},\label{ratio}\ee
in respect to the time variable $\tau\equiv \ln a$ (it is related
to the cosmic time through $d\tau=H dt$). The following generic
evolution equation holds for $r$:

\be r'=\frac{\rho_M}{\rho_\phi}\left(\frac{\rho_M'}{\rho_M}-
\frac{\rho_\phi'}{\rho_\phi}\right)=f(r),\label{evolutionr}\ee
where the prime denotes derivative in respect to $\tau$, and $f$
is an arbitrary function (at least of class ${\cal C}^1$) of $r$.
One is then primarily interested in the equilibrium points of
equation (\ref{evolutionr}), i.e., those points $r_{ei}$ at which
$f(r_{ei})=0$. After that one expands $f$ in the neighborhood of
each equilibrium point; $r=r_{ei}+\epsilon_i$, so that, up to
terms linear in the perturbations $\epsilon_i$:
$f(r)=(df/dr)_{r_{ei}}\epsilon_i+{\cal O}(\epsilon_i)\Rightarrow
\epsilon_i'=(df/dr)_{r_{ei}}\epsilon_i$. This last equation can be
integrated to yield the evolution of the perturbations in time
$\tau$:

\be
\epsilon_i=\epsilon_{0i}\exp{[(df/dr)_{r_{ei}}\tau]},\label{perts}\ee
where $\epsilon_{0i}$ are arbitrary integration constants. It is
seen from (\ref{perts}) that, only those perturbations for which:

\be (df/dr)_{r_{ei}}<0,\ee decay with time $\tau$, and the
corresponding equilibrium point is stable. The necessary condition
to evade the coincidence problem is then given by the requirement
that the point $\rho_M/\rho_{\phi}=r_{ei}\lesssim 1$ be stable
against small linear perturbations of the kind explained above.

If we take into account the conservation equations
\eqref{backconteq} and \eqref{phconteq}, and the definition of the
interaction term $Q$ given in equation \eqref{interactionterm},
then, the function $f$ can be given by the following expression:

\be
f(r)=r\left[(\ln\bar\chi)'(r+1)+3(\gamma_{\phi}-\gamma_m)\right].\label{function}\ee
Note that, for a model without interaction ($(\ln\bar\chi)'=0$)
and with a constant DE barotropic parameter
$\gamma_{\phi}=\gamma_{\phi,0}$ (consider, for simplicity,
dust-like background fluid so that $\gamma_m=1$);
$f(r)=3(\gamma_{\phi,0}-1)\;r$ and the only (stable) equilibrium
point is the dark energy dominated solution $r=0\Rightarrow
\Omega_\phi=1$. In consequence the coincidence does arise in this
case. \\

\underline{Varying-mass dark matter particles in the framework of
phantom} \underline{cosmologies}\\ An equivalent approach to the
interacting dark energy (for instance, inspired in the action
\eqref{staction}) is to assume that dark energy and dark matter
sectors interact in such a way that the dark matter particles
acquire a varying mass, dependent on the scalar field which
reproduces dark energy \cite{Anderson:1997un}. This consideration
allows for a better theoretical justification, since a
scalar-field-dependent varying-mass can arise from string or
scalar-tensor theories \cite{Damour:1990tw}. Indeed, in such
higher dimensional frameworks one can formulate both the
appearance of the scalar field (which is related to the dilaton
and moduli fields) and its effect on  matter particle masses
(determined by string dynamics, supersymmetry breaking, and the
compactification mechanism) \cite{Casas:1991ky}. In quintessence
scenario, such varying-mass dark matter models have been explored
in cases of linear
\cite{Anderson:1997un,Casas:1991ky,interacting2,quirosvamps},
power-law \cite{Zhang:2005rg} or exponential
\cite{Amendola:1999er} scalar-field dependence. The exponential
case is the most interesting since, apart from solving the
coincidence problem, it allows for stable scaling behavior, that
is for a large class of initial conditions the cosmological
evolution converges to a common solution at late times
\cite{Amendola:1999er}.

Let us construct a cosmological model where dark energy is
attributed to a phantom field, in which the dark matter particles
have a varying mass depending on this field. Throughout the work
we consider a flat Robertson-Walker metric:
\begin{equation}\label{metric}
ds^{2}=dt^{2}-a^{2}(t)d\bf{x}^2,
\end{equation}
with $a$ the scale factor and  $t$ the comoving time.

In the phantom cosmological paradigm the energy density and
pressure of the phantom scalar field $\phi$ are:
\begin{eqnarray}\label{rhophi}
 \rho_{\phi}&=& -\frac{1}{2}\dot{\phi}^{2} + V(\phi)\\
 \label{pphi}
 p_{\phi}&=& - \frac{1}{2}\dot{\phi}^{2} - V(\phi),
\end{eqnarray}
where $V(\phi)$ is the phantom potential and the dot denotes
differentiation with respect to comoving time. In such a scenario,
the dark energy is attributed to the phantom field, and its
equation of state is given by
\begin{equation}
w_{DE}\equiv w_{\phi}=\frac{p_\phi}{\rho_\phi}.
\end{equation}

In varying-mass dark matter models the central assumption is that
the dark-matter particles have a $\phi$-dependent mass
$M_{DM}(\phi)$, while dark matter is considered as dust. Thus, for
the dark matter energy density we have the standard definition
\begin{equation}\label{rhodm}
\rho_{DM}=M_{DM}(\phi)\,n_{DM},
\end{equation}
where $n_{DM}$ is the number density of the dark-matter particles.
As usual, in the case of FRW geometry, it is determined by the
equation
\begin{equation}\label{ndmm}
\dot{n}_{DM}+3Hn_{DM}=0,
\end{equation}
 with $H$ the Hubble parameter. Therefore, differentiating
 (\ref{rhodm}) and using  (\ref{ndmm}) we obtain the evolution
 equation for $\rho_{DM}$, namely:
\begin{equation}\label{rhodmeom}
\dot{\rho}_{DM}+3H\rho_{DM}=\frac{1}{M_{DM}(\phi)}\frac{dM_{DM}(\phi)}{d\phi}\,\dot{\phi}\,\rho_{DM}.
\end{equation}
Obviously, in a case of $\phi$-independent dark-matter particle
mass, we re-obtain the usual evolution equation
$\dot{\rho}_{DM}+3H\rho_{DM}=0$. Therefore, we observe that the
$\phi$-dependent mass reveals the interaction between dark matter
and dark energy (that is the phantom field) sectors that lies
behind it.

Since general covariance leads to total energy conservation, we
deduce that the evolution equation for the phantom energy density
will be:
\begin{equation}\label{rhophieom}
\dot{\rho}_\phi+3H(\rho_\phi+p_\phi)=-\frac{1}{M_{DM}(\phi)}\frac{dM_{DM}(\phi)}{d\phi}\,\dot{\phi}\,\rho_{DM}.
\end{equation}
Thus, $\frac{dM_{DM}(\phi)}{d\phi}\,\dot{\phi}<0$ corresponds to
energy transfer from dark matter to dark energy, while
$\frac{dM_{DM}(\phi)}{d\phi}\,\dot{\phi}>0$ corresponds to dark
energy transformation into dark matter.

Equivalently, using the definitions (\ref{rhophi}) and
(\ref{pphi}), the phantom evolution equation can be written in
field terms as:
\begin{equation}\label{phiddot}
\ddot{\phi}+3H\dot{\phi}-\frac{\partial
V(\phi)}{\partial\phi}=\frac{1}{M_{DM}(\phi)}\frac{dM_{DM}(\phi)}{d\phi}\,\rho_{DM}.
\end{equation}
Finally, the system of equations closes by considering the
Friedmann equations:
\begin{equation}\label{FR1}
H^{2}=\frac{\kappa^{2}}{3}(\rho_{\phi}+\rho_{DM}),
\end{equation}
\begin{equation}\label{FR2}
\dot{H}=-\frac{\kappa^2}{2}\Big(\rho_{\phi}+p_{\phi}+\rho_{DM}\Big),
\end{equation}
where we have set $\kappa^2\equiv 8\pi G$. Although we could
straightforwardly include baryonic matter and radiation in the
model, for simplicity reasons we neglect them.

Alternatively, one could construct the equivalent uncoupled model
described by:
\begin{eqnarray}
\dot{\rho}_{DM}+3H(1+w_{DM,eff})\rho_{DM}=0\\
\dot{\rho}_\phi+3H(1+w_{\phi,eff})\rho_\phi=0,
\end{eqnarray}
where
\begin{eqnarray}
&&w_{DM,eff}=-\frac{1}{M_{DM}(\phi)}\frac{dM_{DM}(\phi)}{d\phi}\,\frac{\dot{\phi}}{3H}\\
&&w_{\phi,eff}=w_\phi
+\frac{1}{M_{DM}(\phi)}\frac{dM_{DM}(\phi)}{d\phi}\,\frac{\dot{\phi}}{3H}\,\frac{\rho_{DM}}{\rho_\phi}.\
\ \ \
\end{eqnarray}
However, it is more convenient to introduce the ``total''  energy
density $\rho_{tot}\equiv\rho_{DM}+\rho_\phi$, obtaining:
\begin{equation}
\label{rhot}
 \dot{\rho}_{tot}+3 H(1+w_{tot})\rho_{tot}=0,
\end{equation}
with
\begin{equation}
w_{tot}=\frac{p_\phi}{\rho_\phi+\rho_{DM}}=w_\phi\Omega_\phi,
\end{equation}
where
$\Omega_\phi\equiv\frac{\rho_\phi}{\rho_{tot}}\equiv\Omega_{DE}$.
Obviously, since $\rho_{tot}=3H^2/\kappa^2$, (\ref{rhot}) leads to
a scale factor evolution of the form $a(t)\propto
t^{2/(3(1+w_{tot}))}$, in the constant $w_{tot}$ case. However, at
the late-time stationary solutions that we are studying in the
present work, $w_{tot}$ has reached to a constant value and thus
the above behavior is valid. Therefore, we conclude that in such
stationary solutions the condition for acceleration is just
$w_{tot}<-1/3$.

In the reference \cite{varyingmass} were investigated varying-mass
dark matter models in scenarios where dark energy is attributed to
a phantom field. By imposing exponential or power-law potentials
and exponential or power-law mass dependence, was proved there
that the coincidence problem cannot be solved or even alleviated
if dark energy is attributed to the phantom paradigm. Apart from
the case of an exponential potential with an
exponentially-dependent dark-matter particle mass, which possesses
a relevant late-time (phantom) attractor, in all the other models
we found that physical, well-motivated solutions have a very small
chance to attract the universe at late times. In addition, in all
the examined cases, solutions having
$\Omega_{{\text{DE}}}/\Omega_{{\text{DM}}}\approx{\cal{O}}(1)$
were not relevant attractors at late times. Therefore, the
coincidence problem cannot be solved or even alleviated in
varying-mass dark matter particles models in the framework of
phantom cosmology, in a radical contrast with the corresponding
quintessence case. However, there exists few interacting phantom
models that solves the coincidence \footnote{Of course with
different potentials and varying-mass functions (or, equivalently,
coupling functions) than the previous mentioned.} but paying the
price of introducing new problems such is the justification of a
non-trivial, almost tuned, sequence of cosmological epochs. This
is the case of \cite{Curbelo:2005dh}, where it is described a
methodology to generate solutions free of the Big-Rip singularity
that solves the coincidence.

\underline{Scalar-tensor theories}\\ Scalar-tensor theories (STT)
of gravity
\cite{Brans:1961sx,Wagoner:1970vr,O'Hanlon:1972my,O'Hanlon:1972hq,Bekenstein:1977rb,Bergmann:1968ve,Nordtvedt:1970uv}
can be supported by fundamental physical theories like superstring
theory \cite{Green:1996bh}.

BDT is the first prototype of STT. In BDT, a scalar field, $\chi,$
acts as the source for the gravitational coupling with a varying
Newtonian 'constant'  $G\sim \chi^{-1}.$  It is worthy to mention
that BDT survive several observational tests including Solar
System tests \cite{Abramovici:1992ah} and Big-Bang nucleosynthesis
constraints \cite{Serna:2002fj,Serna:1995tr}. More general STT
with a non-constant BD parameter $\omega(\chi),$ and non-zero
self-interaction potential $V(\chi),$ have been formulated, and
also survive astrophysical tests
\cite{Will:1993ns,Will:2005va,Barrow:1996kc}.

``Extended'' inflation models
\cite{Barrow:1990nv,Faraoni:2006ik,Liddle:1991am,La:1989za} use
the Brans-Dicke theory (BDT) \cite{Brans:1961sx} as the correct
theory of gravity, and in this case the vacuum energy leads
directly to a powerlaw solution \cite{Mathiazhagan:1984vi} while
the exponential expansion can be obtained if a cosmological
constant is explicitly inserted into the field equations
\cite{Barrow:1990nv,Kolitch:1994kr,Romero:1992ci}.

The action for a general class of STT, written in the so-called
Einstein frame (EF), is given by \cite{Kaloper:1997sh}:
\begin{eqnarray}&& S_{EF}=\int_{M_4} d{ }^4 x \sqrt{|g|}\left\{\frac{1}{2} R-\frac{1}{2}(\nabla\phi)^2-V(\phi)+\right.
\nonumber\\ && \ \ \ \ \ \ \ \ \ \ \ \ \ \ \ \left.
\chi(\phi)^{-2}
\mathcal{L}_{m}(\mu,\nabla\mu,\chi(\phi)^{-1}g_{\alpha\beta})\right\}\label{eq1}
\end{eqnarray}
where $R$ is the curvature scalar, $\phi$ is the a scalar field,
related via conformal transformations with the dilaton field,
$\chi,$ \footnote{For a discussion about the regularity of the
conformal transformation, or the equivalence issue of the two
frames, see for example
\cite{Magnano:1990qu,Cotsakis:1993vm,Teyssandier:1995wr,Schmidt:1995ws,Cotsakis:1995wt,Capozziello:1996xg,Magnano:1993bd,Faraoni:1998qx,Faraoni:2007yn,Faraoni:2006fx}
and references therein.} $(\nabla\phi)^2$ denotes $g^{\mu
\nu}\nabla_\mu\phi\nabla_\nu\phi$, with $\nabla_\alpha$ the
covariant derivative (repeated indexes means sum over them).
$V(\phi)$ is the quintessence self-interaction potential,
$\chi(\phi)^{-2}$ is the coupling function, $\mathcal{L}_{m}$ is
the matter Lagrangian, $\mu$ is a collective name for the matter
degrees of freedom. The energy-momentum tensor of background
matter is defined by
\begin{equation}T_{\alpha
\beta}=-\frac{2}{\sqrt{|g|}}\frac{\delta}{\delta g^{\alpha
\beta}}\left\{\sqrt{|g|}
 \chi^{-2}\mathcal{L}(\mu,\nabla\mu,\chi^{-1}g_{\alpha
 \beta})\right\}.\label{Tab}\end{equation}

By considering the conformal transformation $\bar{g}_{\alpha
\beta}=\chi(\phi)^{-1}g_{\alpha \beta}$ and defining the
Brans-Dicke coupling 'constant' $\omega(\chi)$ in such way that
$d\phi=\pm \sqrt{\omega(\chi)+3/2}\chi^{-1} d\chi$ and recalling
$\bar{V}(\chi)=\chi^2 V(\phi(\chi))$ the action (\ref{eq1}) can be
written in the Jordan frame (JF) as (see \cite{Coley:2003mj}):

\begin{align}& S_{JF}=\int_{M_4} d{ }^4 x \sqrt{|\bar{g}|}\left\{\frac{1}{2}\chi \bar{R}
-\frac{1}{2}\frac{\omega(\chi)}{\chi}(\bar{\nabla}\chi)^2-\bar{V}(\chi)
+\right. \nonumber\\ & \ \ \ \ \ \ \ \ \ \ \ \ \ \ \ \left.
\mathcal{L}_{m}(\mu,\nabla\mu,\bar{g}_{\alpha
\beta})\right\},\label{eq1JF}
\end{align}
where a bar is used to denote geometrical objects defined with
respect to the metric $\bar{g}_{\alpha \beta}.$

In the STT given by (\ref{eq1JF}), the energy-momentum of the
matter fields is separately conserved. That is
$$\bar{\nabla}^\alpha \bar{T}_{\alpha \beta}=0,$$ where $$\bar{T}_{\alpha
\beta}=-\frac{2}{\sqrt{|\bar{g}|}}\frac{\delta}{\delta
\bar{g}^{\alpha \beta}}\left\{\sqrt{|\bar{g}|}
\mathcal{L}(\mu,\nabla\mu,\bar{g}_{\alpha
 \beta})\right\}.$$ However, when written in
de EF (\ref{eq1}), this is no longer the case (although the
overall energy density is conserved). In fact in the EF we find
that
$$Q_\beta\equiv\nabla^\alpha T_{\alpha \beta}=-\frac{1}{2}T\frac{1}{\chi(\phi)}\frac{\mathrm{d}\chi(\phi)}{\mathrm{d}\phi}\nabla_{\beta}\phi,\;
 T=T^\alpha_\alpha.$$

By making use of the above 'formal' conformal equivalence between
the Einstein and Jordan frame we can find, for example, that the
theory formulated in the EF with the coupling function
$\chi(\phi)=\chi_0 \exp((\phi-\phi_0)/\varpi),\; \varpi\equiv
\pm\sqrt{\omega_0+3/2}$ and potential $V(\phi)=\beta
\exp({(\alpha-2){\varpi}/(\phi-\phi_0)})$ corresponds to the
Brans-Dicke theory (BDT) with a powerlaw potential, i.e.,
$\omega(\chi)=\omega_0,\; \bar{V}(\chi)=\beta \chi^\alpha.$ Exact
solutions with exponential couplings and exponential potentials
(in the EF) were investigated in \cite{Gonzalez:2006cj}.

It was found (see \cite{Coley:2003mj} and references therein) that
typically at early times ($t\rightarrow 0$) the BDT solutions are
approximated by the vacuum solutions and at late times
($t\rightarrow\infty$) by matter dominated solutions, in which the
matter is dominated by the BD scalar field (denoted by $\chi$ in
the Jordan frame). Exact perfect fluid solutions in STT of gravity
with a non-constant BD parameter $\omega(\chi)$ have been obtained
by various authors (see \cite{coley2}).

\underline{$f(R)$ Cosmology}\\ In the metric formalism the action
for $f(R)$-gravity is given by
\cite{Sotiriou:2008rp,DeFelice:2010aj}
\begin{equation}
 {\cal S}_{\text{metric}}=\int  {d} x^4
\sqrt{-g}\left[f(R)-2\Lambda+{\cal
L}^{(m)}\right]\label{fRaction},
 \end{equation}
where $f(R)$ is a function of the Ricci scalar $R$, and  ${\cal
L}^{(m)}$ accounts for the matter content of the universe.
Additionally, we use the metric signature $(-1,1,1,1)$, 
Greek indices run from $0$ to $3$, and we impose the standard
units in which $c=8\pi G=1$. Finally, in the following, and
without loss of generality, we set the usual cosmological constant
$\Lambda=0$.

The fourth-order equations obtained by varying the action
(\ref{fRaction}) with respect to the metric write:
\begin{equation}
 G_{\mu \nu}=\frac{T_{\mu \nu}^{(m)}}{f'(R)}+T_{\mu \nu}^R,\label{EFE}
 \end{equation}
 where the prime
denotes differentiation with respect to $R$. In this expression
$T_{\mu \nu}^{(m)}$ denotes the matter energy-momentum tensor,
which is assumed to correspond to a perfect fluid with energy
density $\rho_M$ and pressure $p_m$. Additionally, $T_{\mu \nu}^R$
denotes a correction term describing a ``curvature-fluid''
energy-momentum tensor of geometric origin
\cite{Goheer:2008tn,Goswami:2008fs}: {
\begin{eqnarray}
T_{\mu \nu}^R=\frac{1}{f'(R)}\left[\frac{1}{2} g_{\mu
\nu}\left(f(R)-R f'(R)\right) \ \ \ \ \ \ \ \ \ \right.\nonumber\\
\left. \ \ \ \ \ \ \ \  \ \ \ \ \ \ + \nabla_\mu \nabla_\nu
f'(R)-g_{\mu \nu}\Box f'(R)\right], \label{feqs}
 \end{eqnarray}}
where $\nabla_\mu$ is the covariant derivative associated to the
Levi-Civita connection of the metric and $\Box\equiv
\nabla^\mu\nabla_\mu.$ Note that in the last two terms of the
right hand side there appear fourth-order metric-derivatives,
justifying the name ``fourth order gravity'' used for this class
of theories \cite{DeFelice:2010aj}. By taking the trace of
equation (\ref{EFE}) and re-ordering terms one obtains the ``trace
equation'' (equation (5) in section IIA of
\cite{Capozziello:2009nq})
\begin{equation}
3 \Box f'(R)+R f'(R)-2 f(R)=T\label{Trace},
\end{equation} where $T=T_\mu^\mu$ is the trace of the energy-momentum tensor of ordinary matter.

In the phenomenological fluid description of a general matter
source, the standard decomposition of the energy-momentum tensor
$T_{\mu \nu}$ with respect to a timelike vector field $u^\mu$ is
given by
\begin{equation}
T_{\mu \nu}=\mu u_\mu u_\nu +2 q_{(\mu} u_{\nu)} +P h_{\mu
\nu}+\pi_{\mu \nu},
\end{equation}
 where $\mu$ denotes the energy density scalar, $P$ is the isotropic pressure
 scalar,
$q_{\mu}$ is the energy current density vector ($q_\mu\,u^\mu =
0$) and $\pi_{\mu\nu}$ is the trace-free anisotropic pressure
tensor ($\pi_{\mu \nu}u^\nu=0,\, \pi_\mu^\mu=0, \pi_{\mu
\nu}=\pi_{\nu \mu}$).

The matter fields need to be related through an appropriate
thermodynamical equation of state in order to provide a coherent
picture of the physics underlying the fluid spacetime scenario.
Applying this covariant decomposition to the ``curvature-fluid''
energy-momentum tensor \eqref{feqs} we obtain
\begin{align}
&\mu=-\frac{1}{2}\left[\frac{f(R)-R
f'(R)+6 H\frac{ {d}}{ {d}t}f'(R)}{f'(R)}\right]\nonumber\\
& P=-\frac{1}{2}\left[\frac{-f(R)+R f'(R)-4 H\frac{ {d}}{
{d}t}f'(R)-2\frac{ {d}^2}{
{d}t^2}f'(R)}{f'(R)}\right].\label{matterfields1}
\end{align}
Finally, the anisotropic pressure tensor is given by $\pi^\mu_\nu
= \text{diag}(0,-2\pi_+,\pi_+,\pi_+)$, where
\begin{equation}
\pi_+=-\frac{\frac{ {d}}{
{d}t}f'(R)}{f'(R)}\sp.\label{matterfields2}
\end{equation}

For HOG theories derived from Lagrangians of the form
\begin{equation}
L=\frac{1}{2}F\left(  R\right)
\sqrt{-g}+\mathcal{L}_{matter}(\mu,\nabla\mu,g_{\alpha \beta}),
\label{lagr}%
\end{equation} it is well known that under the conformal transformation, $\widetilde{g}%
_{\mu\nu}=F^{\prime}\left(  R\right)  g_{\mu\nu}$, the field
equations reduce to the Einstein field equations with a scalar
field $\phi$ as an additional matter source, where
\begin{equation}
\phi=\sqrt{\frac{3}{2}}\ln F^{\prime}\left(  R\right)  . \label{scfi}%
\end{equation}

Assuming that (\ref{scfi}) can be solved for $R$ to obtain a
function $R\left(  \phi\right) ,$ the potential of the scalar
field is given by
\begin{equation}
V\left(  R\left(  \phi\right)  \right)  =\frac{1}{2\left(
F^{\prime}\right)
^{2}}\left(  R F^{\prime}-f\right)  , \label{pote}%
\end{equation}
and quadratic gravity, $F(R)=R+\alpha R^2$ with the potential
$$V\left( \phi\right) =\frac{1}{8\alpha}\left(
1-e^{-\sqrt{2/3}\phi}\right) ^{2}$$ is a typical example. The
restrictions on the potential in the papers
\cite{Rendall:2004ic,Rendall:2006cq,Rendall:2005if,Rendall:2005fv}
were used in \cite{Macnay:2008nw} to impose conditions on the
function $f\left( R\right)  $ with corresponding potential
(\ref{pote}). The conformal equivalence can be formally obtained
by conformally transforming the Lagrangian (\ref{lagr}) and the
resulting action becomes \cite{Bean:2006up},
\begin{eqnarray}
&& \widetilde{S}=\int
d^{4}x\sqrt{-\widetilde{g}}\left\{\frac{1}{2} \widetilde{R}-\left(
\nabla\phi\right)  ^{2}-V\left( \phi\right)
 +\right.
\nonumber\\ && \ \ \ \ \ \ \ \ \ \ \ \ \ \ \ \left.
e^{-2\sqrt{2/3}\phi}\mathcal{L}_{m}\left(\mu,\nabla\mu,
e^{-\sqrt{2/3}\phi
}\widetilde{g}\right)  \right\}. \label{action}%
\end{eqnarray}
It is easy to note that the model arising from the action
\eqref{fRaction} can be obtained from \eqref{eq1} with the choice
$\chi(\phi)=e^{\sqrt{2/3}\phi}.$

\underline{Ho\v{r}ava-Lifshitz cosmology}

Recently, a power-counting renormalizable, ultra-violet (UV)
complete theory of gravity was proposed by Ho\v{r}ava in
\cite{hor2,hor1,hor3,hor4}. Although presenting an infrared (IR)
fixed point, namely General Relativity, in the  UV the theory
possesses a fixed point with an anisotropic, Lifshitz scaling
between time and space of the form $\bx\to\ell~\bx$,
$t\to\ell^z~t$, where $\ell$, $z$, $\bx$ and $t$ are the scaling
factor, dynamical critical exponent, spatial coordinates and
temporal coordinate, respectively.

Due to these novel features, there has been a large amount of
effort in examining and extending the properties of the theory
itself
\cite{Volovik:2009av,Cai:2009ar,Cai:2009dx,Orlando:2009en,Nishioka:2009iq,Konoplya:2009ig,Sotiriou:2009bx,Bogdanos:2009uj,Kluson:2009rk,Afshordi:2009tt,Myung:2009ur,Li:2009bg,Visser:2009fg,Chen:2009bu,Chen:2009ka,Shu:2009gc,Charmousis:2009tc}.
Additionally, application of Ho\v{r}ava-Lifshitz gravity as a
cosmological framework gives rise to Ho\v{r}ava-Lifshitz
cosmology, which proves to lead to interesting behavior
\cite{Calcagni:2009ar,Kiritsis:2009sh}. In particular, one can
examine specific solution subclasses
\cite{Lu:2009em,Nastase:2009nk,Colgain:2009fe,Ghodsi:2009rv,Minamitsuji:2009ii,Ghodsi:2009zi},
the perturbation spectrum
\cite{Mukohyama:2009gg,Piao:2009ax,Gao:2009bx,Chen:2009jr,Gao:2009ht,Wang:2009yz,Kobayashi:2009hh},
the gravitational wave production
\cite{Mukohyama:2009zs,Takahashi:2009wc,Koh:2009cy}, the matter
bounce
\cite{Brandenberger:2009yt,Brandenberger:2009ic,Cai:2009in}, the
black hole properties
\cite{Danielsson:2009gi,Cai:2009pe,Myung:2009dc,Kehagias:2009is,Cai:2009qs,Mann:2009yx,Bertoldi:2009vn,Castillo:2009ci,BottaCantcheff:2009mp,Lee:2009rm},
the dark energy phenomenology
\cite{Saridakis:2009bv,Park:2009zr,Wang:2009rw,Appignani:2009dy},
the astrophysical phenomenology \cite{Kim:2009dq,Harko:2009qr}
etc. However, despite this extended research, there are still many
ambiguities if Ho\v{r}ava-Lifshitz gravity is reliable and capable
of a successful description of the gravitational background of our
world, as well as of the cosmological behavior of the universe
\cite{Charmousis:2009tc,Sotiriou:2009bx,Bogdanos:2009uj}.

 We begin with a brief review of Ho\v{r}ava-Lifshitz
gravity. The dynamical variables are the lapse and shift
functions, $N$ and $N_i$ respectively, and the spatial metric
$g_{ij}$ (roman letters indicate spatial indices). In terms of
these fields the full metric is
\begin{eqnarray}
ds^2 = - N^2 dt^2 + g_{ij} (dx^i + N^i dt ) ( dx^j + N^j dt ) ,
\end{eqnarray} 
where indices are raised and lowered using $g_{ij}$. The scaling
transformation of the coordinates reads (z=3):
\begin{eqnarray}
 t \rightarrow l^3 t~~~{\rm and}\ \ x^i \rightarrow l x^i~.
\end{eqnarray}

Decomposing the gravitational action into a kinetic and a
potential part as $S_g = \int dt d^3x \sqrt{g} N ({\cal L}_K+{\cal
L}_V)$, and under the assumption of detailed balance \cite{hor3}
(the extension beyond detail balance will be performed later on),
which apart form reducing the possible terms in the Lagrangian it
allows for a quantum inheritance principle \cite{hor2} (the $D+1$
dimensional theory acquires the renormalization properties of the
D-dimensional one),
 the full action of Ho\v{r}ava-Lifshitz gravity is given by
\begin{eqnarray}
S_g =  \int dt d^3x \sqrt{g} N \left\{ \frac{2}{\kappa^2}
(K_{ij}K^{ij} - \lambda K^2)- \ \ \ \ \ \ \ \ \ \ \ \ \ \ \ \ \  \right. \nonumber \\
\left.
 - \frac{\kappa^2}{2 w^4} C_{ij}C^{ij}
 + \frac{\kappa^2 \mu}{2 w^2}
\frac{\epsilon^{ijk}}{\sqrt{g}} R_{il} \nabla_j R^l_k -
\frac{\kappa^2 \mu^2}{8} R_{ij} R^{ij}+
     \right. \nonumber \\
\left.    + \frac{\kappa^2 \mu^2}{8(1 - 3 \lambda)} \left[ \frac{1
- 4 \lambda}{4} R^2 + \Lambda  R - 3 \Lambda ^2 \right] \right\},
\end{eqnarray}
where
\begin{eqnarray}
K_{ij} = \frac{1}{2N} \left( {\dot{g_{ij}}} - \nabla_i N_j -
\nabla_j N_i \right) \, ,
\end{eqnarray}
is the extrinsic curvature and
\begin{eqnarray} C^{ij} \, = \, \frac{\epsilon^{ijk}}{\sqrt{g}} \nabla_k
\bigl( R^j_i - \frac{1}{4} R \delta^j_i \bigr)
\end{eqnarray}
the Cotton tensor, and the covariant derivatives are defined with
respect to the spatial metric $g_{ij}$. $\epsilon^{ijk}$ is the
totally antisymmetric unit tensor, $\lambda$ is a dimensionless
constant and $\Lambda $ is a negative constant which is related to
the cosmological constant in the IR limit. Finally, the variables
$\kappa$, $w$ and $\mu$ are constants with mass dimensions $-1$,
$0$ and $1$, respectively.

In order to add the dark-matter content in a universe governed by
Ho\v{r}ava gravity, a scalar field is introduced
\cite{Calcagni:2009ar,Kiritsis:2009sh}, with action:
\begin{eqnarray}
S_M\equiv S_\phi = \int dtd^3x \sqrt{g} N \left[
\frac{3\lambda-1}{4}\frac{\dot\phi^2}{N^2}
+m_1m_2\phi\nabla^2\phi-\right.\nonumber\\
\left.-\frac{1}{2}m_2^2\phi\nabla^4\phi +
\frac{1}{2}m_3^2\phi\nabla^6\phi -V(\phi)\right],\ \ \ \ \
\end{eqnarray}
where $V(\phi)$ acts as a potential term and $m_i$ are constants.
Although one could just follow a hydrodynamical approximation and
introduce straightaway the density and pressure of a matter fluid
\cite{Sotiriou:2009bx}, the field approach is more robust,
especially if one desires to perform a phase-space analysis.

Now, in order to focus on cosmological frameworks, we impose the
so called projectability condition \cite{Charmousis:2009tc} and
use an FRW metric,
\begin{eqnarray}
N=1~,~~g_{ij}=a^2(t)\gamma_{ij}~,~~N^i=0~,
\end{eqnarray}
with
\begin{eqnarray}
\gamma_{ij}dx^idx^j=\frac{dr^2}{1-kr^2}+r^2d\Omega_2^2~,
\end{eqnarray}
where $k=-1,0,1$ correspond to open, flat, and closed universe
respectively. In addition, we assume that the scalar field is
homogenous, i.e $\phi\equiv\phi(t)$.

\underline{Ad hoc modifications of the Friedmann equation}\\
Even though the most popular explanation to the late-time
acceleration in the universe is the existence of some kind of dark
energy (perhaps a scalar field), this is not the only possibility.
Recently, Freese and Lewis \cite{Freese:2002sq} proposed the
so-called Cardassian models as an alternative explanation  which
involves only matter and radiation and does not invoke either
vacuum energy or a cosmological constant. In these models the
universe has a flat geometry, as required by measurements of the
cosmic background radiation \cite{Netterfield:2001yq} and it is
filled  only with radiation and matter (baryonic or not). The
Friedman equation is modified with respect to its usual form by
the addition of a term in its right hand side, specifically
\begin{equation}
3H^2=\rho+\sigma\rho^n,
\end{equation} in units such that
${8\pi}/{m_{pl}^2}=1$ and with $\sigma>0$ being an arbitrary
constant.

For $n\!<\!1$ the second term becomes important if
$z\!<\!\mathcal{O}(1)$. From there on it dominates the Friedmann
equation and  yields $a\propto t^{2/3n}$ for ordinary matter,  so
there will be acceleration provided $n<2/3$.  There are  two main
(possibly unrelated) motivations for the  $\rho^n$ modifications:
(1) As shown in \cite{Chung:1999zs}, terms of that form typically
appear in the Friedman equation  when the universe is embedded as
a three-dimensional surface (3-brane) in higher dimensions. (2)
Alternatively, these functions may appear in a purely four
dimensional theory in which the modified right hand side of the
Friedman equation is due to an extra contribution to the total
energy density. One will then regard the right hand side of the
Friedman equation as corresponding to a single fluid, with an
extra contribution to the energy-density tensor in the (ordinary
four dimensional) Einstein equations.

The interpretation of the Cardassian expansion as due to an
interacting dark matter fluid with negative pressure was developed
in \cite{Gondolo:2002fh}. The Cardassian term on the right hand
side of the Friedman equation is interpreted as the interacting
term and gives rise to the effective negative pressure which
drives the cosmological acceleration.

Interestingly, Cardassian models survive several observational
tests,  the most significant being that it allows for a   universe
consisting of just matter and radiation. The energy density giving
a closed universe $\rho_c$ is much smaller that its counterpart
standard cosmology  $\rho_{c \!,\,\rm{old}}$ (specifically
$\rho_c=\rho_{c\!,\,\rm{old}}\left[1+(1+z_eq)^3(1-n)\right]^{-1}$),
 and matter alone is enough to provide a
flat geometry.

To illustrate how the accelerated expansion is possible in such
framework, let us assume for sake that the matter content is dust,
i.e., $\rho\propto a^{-3}\propto (1+z)^{3}$. With this hypothesis,
the correction term dominates for redshifts $z<z_{eq}$ with
$z_{eq}$ defined by \be(1+z_{eq})^{3(1-n)}=\frac{\sigma \rho_0}{3
H^2-\sigma\rho_0},\ee ($\rho_0$ is the current energy density).
Once the correction term dominates, the scale factor ant its
derivatives up to second order are given respectively by:
$$a\propto t^{\frac{2}{3n}},\;\dot a\propto
\frac{2}{3n}\,t^{-1+\frac{2}{3n}},\;\ddot
a\propto\frac{2(2-3n)}{9n^2}\,t^{-2\left(1-\frac{1}{3n}\right)}.$$
Hence, we obtain accelerated expansion provided $n<2/3.$
Therefore, Cardassian models explains very well the current
accelerated expansion only considering ordinary matter and
radiation.

\underline{Crossing the ``phantom divide''}\\
The quintom paradigm is a hybrid construction of a quintessence
component, usually modelled by a real scalar field that is
minimally coupled to gravity, and a phantom field: a real scalar
field --minimally coupled to gravity-- with negative kinetic
energy. Let us define the equation of state parameter of any
cosmological fluid as $w\equiv\text{pressure}/\text{density}$. The
simplest model of dark energy (vacuum energy or cosmological
constant) is assumed to have $w =-1$. A key feature of
quintom-like behavior is the crossing of the so called phantom
divide, in which the equation of state parameter crosses through
the value $w=-1.$

In \cite{Huterer:2004ch}  uncorrelated and nearly model
independent band power estimates (basing on the principal
component analysis \cite{Huterer:2002hy}) of the EoS of DE and its
density as a function of redshift were presented, by fitting to
the SNIa data. Quite unexpectedly, they found marginal ($2\sigma$)
evidence for $w(z)<-1$ at $z < 0.2$, which is consistent with
other results in the literature
\cite{Alam:2004jy,Alam:2003fg,Wang:2004py, Wang:2003gz,
Padmanabhan:2002vv, Zhu:2004cu}.

The aforementioned result implied that the EoS of DE could indeed
vary with time. Therefore, one could use a suitable
parametrization of $w_{\rm DE}$ as a function of the redshift $z$,
in order to satisfactory describe such a behavior. There are two
well-studied parametrizations. The first (ansatz A) is:
\begin{equation}
\label{ansatzA}
 w_{\rm DE}=w_0+w'z~,
\end{equation}
where $w_0$ the DE EoS at present and $w'$ an additional
parameter. However, this parametrization is only valid at low
redshift, since it suffers from severe divergences at high ones,
for example at the last scattering surface $z\sim1100$. Therefore,
a new, divergent-free ansatz (ansatz B) was proposed
\cite{Chevallier:2000qy, Linder:2002et}:
\begin{equation}
\label{ansatzB}
 w_{\rm DE}=w_0+w_1(1-a)=w_0+w_1\frac{z}{1+z}~,
\end{equation}
where $a$ is the scale factor and $w_1=-dw/da$. This
parametrization exhibits a very good behavior at high redshifts.

In \cite{Feng:2004ad} the authors used the ``gold" sample of 157
SNIa, the low limit of cosmic ages and the HST prior, as well as
the uniform weak prior on $\Omega_mh^2$, to constrain the free
parameters of above two DE parameterizations.
\begin{figure}[ht]
\begin{center}
\mbox{\epsfig{figure=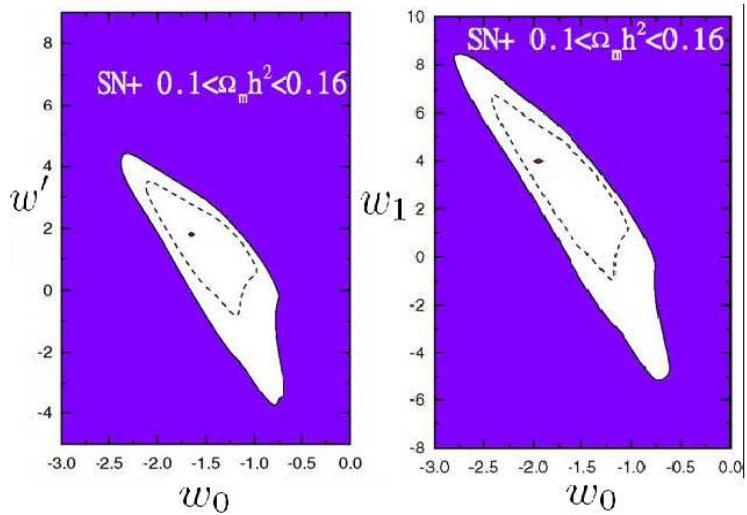,width=9.9cm,angle=0}} \caption{{
Two-dimensional contour-plots of the DE equation-of-state
parameters, in two different parameterizations and using SNIa
data. The left graph corresponds to ansatz A (expression
(\ref{ansatzA})) and the right graph are to ansatz B (expression
(\ref{ansatzB})). From Ref. \cite{Feng:2004ad}.}}
 \label{FFfig4}
\end{center}
\end{figure}
As can be seen in Fig.\ref{FFfig4} they found that the data seem
to favor an evolving DE with the EoS being below $-1$ around the
present epoch, while it was in the range $w>-1$ in the near
cosmological past. This result holds for both parametrizations
(\ref{ansatzA}),(\ref{ansatzB}), and in particular the best fit
value of the EoS at present is $w_0<-1$, while its ``running''
coefficient is larger than $0$.

\begin{figure}[ht]
\begin{center}
\mbox{\epsfig{figure=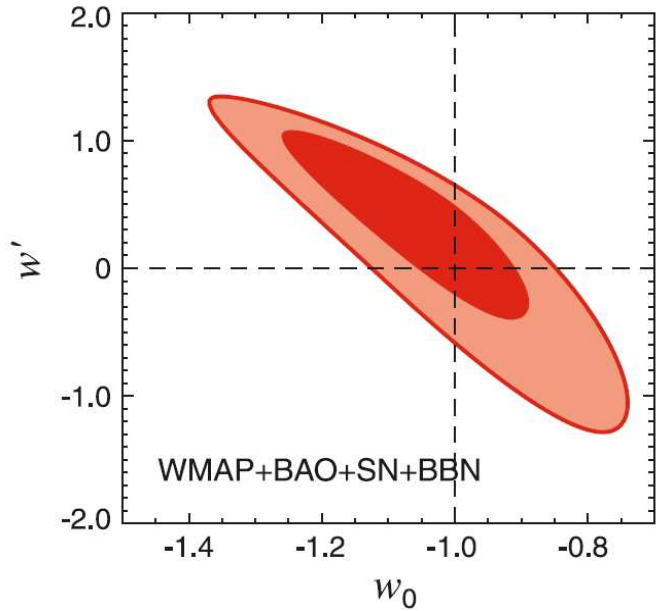,width=8.9cm,angle=0}} \caption{{
Two-dimensional contour-plot of the DE equation-of-state
parameters, in parameterization ansatz B (expression
(\ref{ansatzB})), and using WMAP, BAO, SNIa data. From Ref.
\cite{Komatsu:2008hk}.}}
 \label{fig5}
\end{center}
\end{figure}

Apart from the SNIa data, CMB and LSS data  can be also used to
study the variation of EoS of DE. In \cite{Hannestad:2004cb}, the
authors used the first year WMAP, SDSS and 2dFGRS data to
constrain different DE models. They indeed found that evidently
the data favor a strongly time-dependent $w_{\rm DE}$, and this
result is consistent with similar project of the literature
\cite{Xia:2004rw,Xia:2005ge,Xia:2006cr,Zhao:2006bt,Xia:2006rr,
Xia:2006wd,Zhao:2006qg,Wang:2007mza,Wright:2007vr,Li:2008cj}.
Using the latest 5-year WMAP data, combined with SNIa and BAO
data, the constraints on the DE parameters of ansatz B are:
$w_0=-1.06\pm0.14$ and $w_1=0.36\pm0.62$ \cite{Komatsu:2008hk,
Xia:2008ex, Li:2008vf}, and the corresponding contour plot is
presented in Fig.\ref{fig5}.

In conclusion, as can be observed, the current observational data
mildly favor $w_{\rm DE}$ crossing the phantom divide during the
evolution of universe.

Some interesting general aspects of the problem of the phantom
divide crossing were discussed in \cite{Caldwell:2005ai}, where
the viability requirements on the equation of state and sound
speed were analyzed. Some of realizations of the crossing can have
an extradimensional origin, either in the brane
\cite{Sahni:2002dx,Cai:2005ie,Cai:2005qm,Aref'eva:2005fu}, or the
string gas context \cite{McInnes:2005vp}. Other worth mentioning
alternatives are models in the framework of scalar-tensor theories
\cite{Curbelo:2005dh,Elizalde:2004mq,Perivolaropoulos:2005yv}; a
single field proposal involving high order derivative operators in
the lagrangian \cite{Li:2005fm};  and an interacting Chaplygin gas
\cite{Zhang:2005jj}. The impossibility of the occurrence of the
transition in traditional single field models (minimally coupled
to matter) \cite{Melchiorri:2002ux,Vikman:2004dc}, has motivated
much activity in the construction of two field models that do the
job. Examples of explicit constructions can be found in
\cite{Wei:2005fq,Wei:2005si,Hu:2004kh,Wei:2004nu}, but perhaps the
class of models which have received most attention are quintom
cosmologies
\cite{quintom,quintom1,quintom2,quintom3,quintom4,Guo:2004fq,Zhang:2005eg,Wei:2005fq,Wei:2005nw,Lazkoz:2006pa,stringinspired,stringinspired1,stringinspired2,Cai:2008gk,Saridakis:2009uu,Lazkoz2007,arbitrary,arbitrary1,arbitrary2}.
Quintom behavior (i.e., the $w=-1$ crossing) has been investigated
in the context of h-essence cosmologies
\cite{Wei:2005fq,Wei:2005nw}; in the context of holographic dark
energy
\cite{holographic,holographic1,holographic2,holographic3,holographic4};
inspired by string theory
\cite{stringinspired,stringinspired1,stringinspired2}; derived
from spinor matter \cite{Cai:2008gk}; for arbitrary potentials
\cite{Lazkoz2007,arbitrary,arbitrary1,arbitrary2}; using
isomorphic models consisting of three coupled oscillators, one of
which carries negative kinetic energy (particularly for
investigating the dynamical behavior of massless
quintom)\cite{setare1}. The crossing of the phantom divide is also
possible in the context of scalar tensor theories
\cite{Elizalde2004,Apostolopoulos:2006si,Bamba:2008xa,Bamba:2008hq,Setare:2008mb}
as well as in modified theories of gravity \cite{Nojiri:2006ri}.

The two-field quintom model has Lagrangian:
\begin{equation}\label{expquintomlag}
    \mathcal{L}=\frac{1}{2}\partial_{\mu}\phi\partial^{\mu}\phi
    -\frac{1}{2}\partial_{\mu}\varphi\partial^{\mu}\varphi-V(\phi,\varphi),
\end{equation}

The cosmological evolution of quintom model with exponential
potential has been examined, from the dynamical systems viewpoint,
in \cite{Guo:2004fq} and \cite{Zhang:2005eg,Lazkoz:2006pa}. The
difference between \cite{Guo:2004fq} and
\cite{Zhang:2005eg,Lazkoz:2006pa} is that in the second case the
potential considers the interaction between the conventional
scalar field and the phantom field. In \cite{Zhang:2005eg} it had
been proven that in the absence of interactions, the solution
dominated by the phantom field should be the attractor of the
system and the interaction does not affect its attractor behavior.
In \cite{Lazkoz:2006pa} the case in which the interaction term
dominates against the mixed terms of the potential, was studied.
It was proven there, that the hypothesis in \cite{Zhang:2005eg} is
correct only in the cases in which the existence of the phantom
phase excludes the existence of scaling attractors (in which the
energy density of the quintom field and the energy density of DM
are proportional). Some of this results were extended in
\cite{Lazkoz2007}, for arbitrary potentials. There it was settled
down under what conditions on the potential it is possible to
obtain scaling regimes. It was proved there, that for arbitrary
potentials having asymptotic exponential behavior, scaling regimes
are associated to the limit where the scalar fields diverge.  Also
it has been proven that the existence of phantom attractors in
this framework is not generic and consequently the corresponding
cosmological solutions lack the big rip singularity.

\chapter{Qualitative theory of dynamical systems}

In this chapter we briefly review some results of the qualitative
theory of dynamical systems, settling the theoretical basis for
the qualitative study of concrete cosmological models.

\section{Introduction}

The works of H. J. Poincar\'e in Celestial Mechanics
\cite{150,151} settled the basis for the local and global analysis
of non-linear differential equations, particularly, for the
stability theory of singular points and periodic orbits, stable
and unstable manifolds, etc.

After H. J. Poincar\'e, and following the studies of J. Hadamard
about geodesic fluids \cite{152}, G. D. Birkhoff studied the
complex structure of orbits arising when a complete integrable
system is perturbed \cite{153,154}. Therefore, the basic question
is how prevalent integrability is? The answer was given by A. N.
Kolmogorov (1954), V. I. Arnold (1963) and J. K. Moser (1973), in
which is now called KAM theorem, the basic theorem on caos for
Hamiltonian systems \cite{155,156}. Some of the more important
theorems of stability were given by A. M. Liapunov \cite{157},
offering a method to determine the stability of the singular
points when the information obtained by linearization is not
conclusive. This theory is a vast area belonging to the theory of
dynamical systems \cite{158,Hirsch,160}. Finally, in the XX's, was
possible to formulate a geometric theory of dynamical systems,
mainly due to the works of V. I. Arnold \cite{161,162}.

Qualitative methods have been proved to be a powerful scheme for
investigating the physical behavior of cosmological models. It has
been used three different approaches: approximation by parts,
Hamiltonian methods, and dynamical systems methods \cite{WE}. In
the third case the Einstein's field equations of Bianchi's
cosmologies and its isotropic subclass (FLRW models), can be
written as an autonomous system of first-order differential
equations whose solution curves partitioned to $\mathbb{R}^ n$ in
orbits, defining a dynamical system in $\mathbb{R}^ n$. In the
general case, the elements of the phase space partition (i.e.,
singular points, invariant sets, etc.) can be listed and
described. This study consists of several steps: determining the
singular points, the linearization in a neighborhood of them, the
search for the eigenvalues of the associated Jacobian matrix,
checking the stability conditions in a neighborhood of the
singular points, the finding of the stability and instability sets
and the determination of the basin of attraction, etc. In some
occasions, in order to do that, it is needed to simplify a
dynamical system. Two approaches are applied to this objective:
one, reduce the dimensionality of the system and two, eliminate
the nonlinearity. Two rigorous mathematical techniques that allow
substantial progress along both lines are center manifold theory
and the method of normal forms. We submit the reader to sections
\ref{sectionCM} and \ref{sectionNF} for a summary about such
techniques.

The most general result to be applied in order to determine the
asymptotic stability of a singular point, $a,$ is Lyapunov's
stability theorem. Lyapunov's stability method provides
information not only about the asymptotic stability of a given
singular point but also about its basin of attraction. This cannot
be obtained by the usual methods found in the literature, such as
linear stability analysis or first-order perturbation techniques.
Moreover, Lyapunov's method is also applicable to non-autonomous
systems. To our knowledge, there are few works that use Lyapunov's
method in cosmology
\cite{Setare:2010zd,Cardoso:2008bp,Lavkin:1990gu,Charters:2001hi,Aref'eva:2009xr}.
In \cite{Charters:2001hi} it is investigated the general
asymptotic behavior of Friedman-Robertson-Walker (FRW) models with
an inflaton field, scalar-tensor FRW cosmological models and
diagonal Bianchi-IX models by means of Lyapunov's method. In
\cite{Aref'eva:2009xr} it is investigated the stability of
isotropic cosmological solutions for two-field models in the
Bianchi I metric. The author proved that the conditions sufficient
for the Lyapunov stability in the FRW metric provide the stability
with respect to anisotropic perturbations in the Bianchi I metric
and with respect to the cold dark matter energy density
fluctuations. Sufficient conditions for the Lyapunov's stability
of the isotropic singular points of the system of the Einstein
equations are also found (these conditions coincided with the
previously obtained in \cite{Lazkoz:2007mx} for the quintom
paradigm without using Liapunov's technique). To apply Lyapunov's
stability method it is required the construction of the so-called
strict Lyapunov's function, i.e., a $C^1$ function
$V:U\subset\mathbb{R}^n\rightarrow \mathbb{R}$ defined in a
neighborhood $U$ of $a$ such that $V(a)=0, V(x)>0, x\neq a$ and
$\dot V(x)\leq 0$ ($<0$) in $U\setminus \{a\}.$ The construction
of such V is laborious, and sometimes impossible. One alternative
way is to follow point (5) in section \ref{Procedure}.

For the investigation of hyperbolic singular points of autonomous
vector fields we can use Hartman-Grobman's theorem (theorem
19.12.6 en \cite{wiggins} p. 350) which allows for analyzing the
stability of an singular point from the linearized system around
it. For isolated nonhyperbolic singular points we can use normal
forms theorem (theorem 2.3.1 in \cite{arrowsmith}), which contains
Hartman-Grobman's theorem as particular case. The aim of the
normal form calculation is to construct a sequence of non-linear
transformations which successively remove the non-resonant terms
in the vector field of order $r$ in the Taylor's expansion,
${\mathbf{X}}_r,$ starting from $r=2$. Normal forms (NF) theory
has been used in the context of cosmological models in order to
get useful information about isolated nonhyperbolic singular
points. In \cite{Leon:2009rc} was investigated Ho\v{r}ava-Lifshitz
cosmology from the dynamical system view point. There was proved
the stability of de Sitter solution (corresponding to a
non-hyperbolic singular point) in the case where the
detailed-balance condition is relaxed, using NF expansions. In
\cite{Miritzis:2007yn} were investigated closed isotropic models
in second order gravity. There the normal form of the dynamical
system has periodic solutions for a large set of initial
conditions. This implies that an initially expanding closed
isotropic universe may exhibit oscillatory behavior.

The more relevant concepts of qualitative theory are the concept
of flow, and the concept of invariant manifold. The invariant
manifold theorem (theorem 3.2.1 en \cite{wiggins}), that claims
for the existence of local stable and unstable manifolds (under
suitable conditions for the vector field), only allows to obtain
partial information about the stability of singular points and
does not gives a method for determining the stable and unstable
manifolds. Sometimes it is required to consider higher order terms
in the Taylor's expansion of the vector field (e.g. normal forms).
For the investigation of the asymptotic states of the system the
appropriated concepts are $\alpha$ and $\omega$-limit sets of
$x\in \mathbb{R}^n$ (definition 8.1.2 en \cite{wiggins} p. 105).
To characterize these invariant sets one can use the LaSalle's
Invariance Principle (\cite{LaSalle}; theorem 8.3.1
\cite{wiggins}, p. 111) or the Monotonicity Principle (\cite{WE},
p. 103; \cite{LeBlanc:1994qm} p. 536). To apply the Monotonicity
Principle it is required the construction of a monotonic function.
In some cases a monotonic function is suggested by the form of the
differential equation (see equation \eqref{Eq1} below) and in some
cases by the Hamiltonian formulation of the field equations. In
\cite{Heinzle:2009zb} is given a prescription of how to find
monotonic functions for Bianchi cosmologies using Hamiltonian
techniques, merely as a convenient tool for an intermediate step;
the final results are described in terms of scale-automorphism
invariant Hubble-normalized reduced state vector, which is
independent of a Hamiltonian formulation. Also, one can use the
Poincar\'e-Bendixson (\cite{Coley:2003mj} p. 22, theorem 2
\cite{Coley:1999uh} p. 6, \cite{Hirsch}) theorem and its collorary
to distinguishing among all possible $\omega$-limit sets in the
plane. From the Poincar\'e-Bendixon Theorem follows that any
compact asymptotic set is one of the following; 1.) a singular
point, 2.) a periodic orbit, 3.) the union of singular points and
heteroclinic or homoclinic orbits. As a consequence if the
existence of a closed (i.e., periodic, heteroclinic or homoclinic)
orbit can be ruled out it follows that all asymptotic behavior is
located at a singular point. To ruled out a closed orbits for
two-dimensional systems we can use the Dulac's criterion (theorem
3 \cite{Coley:1999uh} p. 6, se also \cite{WE}, p. 94, and
\cite{Coley:2003mj}). It requires the construction of a Dulac's
function. A Dulac's function, $B,$ is a $C^1$  function defined in
a simply connected open subset $D \subseteq R^2$ such that
$\nabla(Bf) = \frac{\partial}{\partial \sigma_2}
    (Bf_1)+\frac{\partial}{\partial \sigma_3}(Bf_2) > 0, $ or $(<0)$
for all $x \in D$ (see \cite{Coley:1999uh}).

\section{Qualitative theory of dynamical systems}

In this section we discuss some of tools of the qualitative theory
of dynamical systems which are applicable, for instance, to the
study of cosmological systems.

\subsection{Definitions and basic results}

In this book we consider vector fields of the form

\begin{equation}
\mathbf{x}'={\bf X}(\mathbf{x},\tau;\mu)\label{T1.1}
\end{equation}
where $\mathbf{x}\in U\subset\mathbb{R}^n,$ denotes a state vector
defined in an open set $U$ of $\mathbb{R}^n,$ $\tau\in \mathbb{R}$
denotes ``time'', ${\mu}\in V\subset \mathbb{R}^p$ denotes a
vector of parameters defined in an open set $V$ of $\mathbb{R}^p$,
and the comma denotes derivative with respect to $\tau$. It is
assumed that ${\bf X}$ is a function of class $C^r,\;r\geq 1$
(i.e.,  a function with continuous partial derivatives of order r
with respect to their arguments) in order to obtain solutions of
the same differentiable class. In case that ${\bf X}$ do not
depend explicitly of time we say that the vector field is
autonomous. In this case, if the parameters are not relevant for
the discussion, we write
\begin{equation}
\mathbf{x}'={\bf X}(\mathbf{x})\label{T1.2}
\end{equation}
In this case it is assumed that $\mathbf{X}$ is a function of
class $C^r,\;r\geq 1$ defined in an open set
$U\subset\mathbb{R}^n.$

A solution of \eqref{T1.1} is a map, $\mathbf{x}$, defined in an
interval $I\subset\mathbb{R}$ taking values in $\mathbb{R}$ given
by
\begin{equation}\label{T1.3}
\mathbf{x}:I\rightarrow \mathbb{R},\; \tau\rightarrow
\mathbf{x}(\tau),
\end{equation} such that it satisfies equation \eqref{T1.1}, i.e.,
\begin{equation}
{\bf x(\tau)}'={\bf X}({\bf x(\tau)},\tau;\mu)\label{T1.1b}
\end{equation}

The map \eqref{T1.3} is interpreted geometrically as a curve in
$\mathbb{R}^n,$ such that ${\bf X}$ in \eqref{T1.1} represents the
tangent vector at each point on the curve. For that reason ${\bf
X}$ is referred as a vector field. The space of the dependent
variables of \eqref{T1.1} is referred by phase space of
\eqref{T1.1}. In an abstract language, the objective of the
qualitative study of a vector field is to obtain information for
the understanding of the geometry of the solution curves of
\eqref{T1.1} in the phase space, actually without solving equation
\eqref{T1.1} explicitly.

The solution $\mathbf{x}(\tau)$ of \eqref{T1.1}, passing through
the point $\mathbf{x}=\mathbf{x}_0$ at time $\tau=\tau_0,$ is
denoted by $\mathbf{x}(\tau,\tau_0, \mathbf{x}_0; \mu),$ or
$\mathbf{x}(\tau,\tau_0, \mathbf{x}_0),$ if the parameters are not
relevant for the discussion. To $\mathbf{x}(\tau,\tau_0,
\mathbf{x}_0),$ it is referred also as trajectory of phase curve
passing passing through the point $\mathbf{x}=\mathbf{x}_0$ at
time $\tau=\tau_0.$ To the graph
$$\left\{(\mathbf{x},\tau)\in \mathbb{R}^n\times \mathbb{R}| \mathbf{x}=\mathbf{x}(\tau,\tau_0,
\mathbf{x}_0), \tau\in I\right\},$$ it is referred as integral
curve.

Given $\mathbf{x}_0\in U\subset \mathbb{R}^n$ located at the phase
space of \eqref{T1.1}, the orbit passing through the point
$\mathbf{x}=\mathbf{x}_0$ it is denoted and defined as
\begin{equation}\label{T1.4}
O(\mathbf{x}_0)=\left\{\mathbf{x}\in
\mathbb{R}^n|\mathbf{x}=\mathbf{x}(\tau,\tau_0, \mathbf{x}_0),
\tau\in I\right\}.
\end{equation}
It is a fact that for all $T\in I,$ $O(\mathbf{x}(T,\tau_0,
\mathbf{x}_0))=O(\mathbf{x}_0).$

The positive (future) orbit  passing through the point
$\mathbf{x}=\mathbf{x}_0$ it is denoted and defined as
\begin{equation}\label{T1.5}
O^+(\mathbf{x}_0)=\left\{\mathbf{x}\in
\mathbb{R}^n|\mathbf{x}=\mathbf{x}(\tau,\tau_0, \mathbf{x}_0),
\tau\geq \tau_0\right\}.
\end{equation}
Similarly it is defined the negative (past) orbit  passing through
the point $\mathbf{x}=\mathbf{x}_0$, denoted by
$O^-(\mathbf{x}_0),$ changing $\tau\geq \tau_0$ by $\tau\leq
\tau_0$ in \eqref{T1.5}.

In much of the applications the topological structure of the phase
space can be more general than $\mathbb{R}^n;$ common examples are
cylindrical, spherical and toroidal phase spaces. The natural
structure to be considered, whenever phase space topology is
concerned, is that of topological (differentiable) manifold.

\begin{defn}[(Topological) manifold, \cite{163} pp. 3-4]\label{TM}
Let $M$ be a Hausdorff space provided with a numerable basis. Let
$p\in M.$ If there exists a positive number $m$ (possibly
depending on $p$), a neighborhood $V(p)$ of $p$ and an
homeomorphism $h: V(p)\rightarrow \mathbb{R}^m$ such that
$h(V(p))\subset \mathbb{R}^m$ is an open set of $\mathbb{R}^m$,
then, $M$ is a (topological) manifold.
\end{defn}

The positive integer $m$ is unique and it is referred as dimension
of $M.$ To the set $\left\{(V_i, h_i)\right\}$ it is referred as
the collection of local coordinates (local charts) of $M.$

\begin{defn}[Euclidean semi-space, \cite{163} pp. 3-4]\label{SES}
The n-dimensional Euclidean semi-space is denoted and defined by
$\mathbb{H}^n:=\left\{\mathbf{x}\in\mathbb{R}^n: x_n\geq
0\right\}$
\end{defn}

\begin{defn}[(Topological) manifold with boundary, \cite{163} pp. 3-4]\label{MB}
Let $M$ be a Hausdorff space provided with a numerable basis. Let
$p\in M.$ If there exists a positive number $m$ (possibly
depending on $p$), a neighborhood $V(p)$ of $p$ and an
homeomorphism $h: V(p)\rightarrow \mathbb{H}^m$ such that
$h(V(p))\subset \mathbb{H}^m$ is an open set of $\mathbb{H}^m$,
then, $M$ is a (topological) manifold with boundary.
\end{defn}

\begin{defn}[Boundary, \cite{163} pp. 3-4]\label{B}
Let $M$ be a manifold with boundary. The boundary of $M$,
$\partial M,$ is defined by $\partial M:=\left\{p\in M:
h(p)\in\mathbb{R}^{m-1}\times \{0\}\right\}.$
\end{defn}

This means that $\partial M$ is a topological manifold of
dimension $m-1$ consistent of those points $p\in M$ which are
transformed by a chart $(V,h)$ (and so, by all the charts) in a
neighborhood of $p$ in  a point with zero m-th coordinate, i.e.,
$x_m=0.$

\begin{defn}[Interior, \cite{163} pp. 3-4]\label{Int}
Let $M$ be a manifold with boundary. The interior of $M$,
$\text{Int} M,$ is defined by $\text{Int} M:=M\setminus\partial
M.$
\end{defn}

\textbf{Notes.} It is a fact that $\mathbb{R}^{m-1}\subset
\mathbb{H}^m\subset \mathbb{R}^{m}.$ Let $M$ and $N$ topological
manifolds with dimension $n$ and $m$ respectively. Then, $M\times
N$ is a topological manifold (with or without boundary) of
dimension $m+n$ and $\partial(M\times N)=\left(\partial M\times
N\right)\cup\left(M\times\partial N\right).$

\begin{defn}[Differentiable manifold, \cite{163} p. 8]\label{DiffM}
$M$ is a differentiable manifold of class $C^r$ if the local
charts $\left\{(V_i, h_i)\right\}$ satisfy:

\begin{enumerate}
\item ${V_i}$ is a covering of $M,$ i.e., $M\subset\bigcup V_i.$
\item If $(V_1,h_1)$ and $(V_2,h_2)$ are local charts with
$V_1\cap V_2=\emptyset,$ the there exists a local change of charts
$$h_1 \circ \left(h_2\right)^{-1}: h_2(V_1\cap V_2)\rightarrow
\mathbb{R}^m$$ which is of differentiable class $C^r.$ \item The
collection $\left\{(V_i, h_i)\right\}$ is maximal with respect
property 2. This means that, if we include a new (different) local
chart $(V_k,h_k)$ to the collection, then, the local change of
charts $h_j \circ \left(h_k\right)^{-1}$ (the index $j$ is
referred to a local chart in the collection) is of differentiable
class $C^s,\; s<r.$
\end{enumerate}

\end{defn}

In order tho guarantee the existence of solution of \eqref{T1.1}
it is assumed that ${\bf X}(\mathbf{x},\tau)$ is a function of
class $C^r$, ($r\geq 1$) defined in an open set $U\subset
\mathbb{R}^n\times \mathbb{R},$ resulting the

\begin{thm}[Existence and uniqueness, theorem 7.1.1, \cite{wiggins}]\label{existencethm}
Let $(\mathbf{x}_0,\tau_0)\in U.$ Then, there exists a solution of
\eqref{T1.1} passing through the point $\mathbf{x}=\mathbf{x}_0$
at time $\tau=\tau_0,$ denoted by
$\mathbf{x}(\tau,\tau_0,\mathbf{x}_0),$ with
$\mathbf{x}(\tau_0,\tau_0,\mathbf{x}_0)=\mathbf{x}_0,$ for
$|\tau-\tau_0|$ small enough. This solution is unique in the sense
that for any other solution passing through the point
$\mathbf{x}=\mathbf{x}_0$ at time $\tau=\tau_0,$ is the same than
$\mathbf{x}(\tau,\tau_0,\mathbf{x}_0),$ in the common existence
interval. Besides $\mathbf{x}(\tau,\tau_0,\mathbf{x}_0)$ is a
$C^r$, ($r\geq 1$) function of $\tau, \tau_0$ and $\mathbf{x}_0.$
\end{thm}

{\bf Proof}. See \cite{160,Hirsch,165}

Theorem \ref{existencethm} only guarantees existence and
uniqueness for infinitesimal time intervals. This theorem can be
reapplied to extend univocally the time interval of existence as
expressed in the following theorem.

Let $C\subset U\subset \mathbb{R}^n\times \mathbb{R}$ a compact
set containing to $(\mathbf{x}_0,\tau_0).$

\begin{thm}[Extendibility, theorem 7.2.1, \cite{wiggins}]\label{extendingthm}
The solution $\mathbf{x}(\tau,\tau_0,\mathbf{x}_0),$ can be
extended backward and forward in time $\tau$ to the boundary of
$C.$
\end{thm}

{\bf Proof}. See \cite{165}.

The next theorem for autonomous vector fields, shows that if
$\|{\bf X}(\mathbf{x})\|$ does not increase too rapidly as
$\|\mathbf{x}\|\rightarrow +\infty$ then all solutions can be
extended indefinitely.

\begin{thm}[Extendibility, theorem 4.3, \cite{WE}, p. 87]\label{ext}
If ${\bf X}$ is continuous, then there exists a constant $M$ such
that $\|{\bf X}(\mathbf{x})\|\leq M \|\mathbf{x}\|$ for all
$\mathbf{x}\in \mathbb{R}^n,$ then any solution of \ref{T1.2} is
defined for all $\tau\in\mathbb{R}.$
\end{thm}

Theorem \ref{ext} implies that one can modify a given autonomous
vector field \ref{T1.2}, so that the orbits are unchanged, but
such that all solutions are defined for all $\tau\in\mathbb{R}.$
The idea is to re-scale the vector field ${\bf X}$ so as to make
it bounded, $${\bf X}(\mathbf{x})\rightarrow
\lambda(\mathbf{x}){\bf X}(\mathbf{x}),$$ where
$\lambda:\mathbb{R}^n\rightarrow \mathbb{R}$ is a $C^1$ function
which is positive on $\mathbb{R}^n,$ in order to preserve the time
direction of time (e.g., $\lambda(\mathbf{x})=\left(1+\left\|{{\bf
X}(\mathbf{x})}\right\|\right)^{-1}$ will suffice).

\begin{thm}[Corollary, 4.2, \cite{WE}, p. 87]\label{ext2}
If ${\bf X}$ is $C^1$, and $\lambda:\mathbb{R}^n\rightarrow
\mathbb{R}$ is $C^1$ and positive, then \ref{T1.2} and
$\mathbf{x}'=\lambda(\mathbf{x}){\bf X}(\mathbf{x})$ have the same
orbits, and $\lambda$ can be chosen so that all solutions of
\ref{T1.2} are defined for all $\tau\in\mathbb{R}.$
\end{thm}

Given the vector field \eqref{T1.1}, such that ${\bf
X}(\mathbf{x},\tau; \mu)$ is of class $C^r,$ ($r\geq 1$) defined
in an open set $U\subset
\mathbb{R}^n\times\mathbb{R}\times\mathbb{R}^p,$ then it is
verified

\begin{thm}[Differentiability solutions, theorem 7.3.1, \cite{wiggins}]\label{differetiabilitythm}
Let $(\mathbf{x}_0,\tau_0,\mu)\in U.$ Then, the solution
$\mathbf{x}(\tau,\tau_0,\mathbf{x}_0;\mu)$ is a $C^r$ ($r\geq 1$)
function of $\tau, \tau_0, \mathbf{x}_0$ and $\mu.$
\end{thm}

{\bf Proof}. See \cite{160,165}.

\subsection{Desirable stability properties of nonlinear vector fields}

Let be the vector field \eqref{T1.1} such that ${\bf
X}(\mathbf{x},\tau)$ is a function at least continuous in $\tau$
and of class $C^r,$ ($r\geq 2$) with respect to the components of
$\mathbf{x}.$ Let $\mathbf{x}=\bar{\mathbf{x}}(\tau)$ any solution
of \eqref{T1.1}.

Roughly speaking, $\bar{\mathbf{x}}(\tau)$ is stable if the
solutions, close to $\bar{\mathbf{x}}(\tau)$ at some given initial
time, remain close to $\bar{\mathbf{x}}(\tau)$ all future time.
The solution is asymptotically stable if the near solutions not
only remain close, but tend to $\bar{\mathbf{x}}(\tau)$ as time
goes forward.

Let us formalize these ideas. Given the solution
$\mathbf{x}=\bar{\mathbf{x}}(\tau)$ of \eqref{T1.1} defined for
$\tau_0\leq \tau < +\infty,$ then the deviations of the vector
field with respect to $\bar{\mathbf{x}}(\tau)$ is given by the
variable $\mathbf{y}=\mathbf{x}-\bar{\mathbf{x}}(\tau)$ with
vector field
\begin{equation}\label{T1.6}
\mathbf{y}'={\bf Y}(\mathbf{y},\tau),
\end{equation}
where ${\bf Y}(\mathbf{y},\tau)={\bf
X}(\mathbf{y}+\bar{\mathbf{x}}(\tau),\tau)-{\bf
X}(\bar{\mathbf{x}}(\tau),\tau).$ Using this transformation of
variables then $\mathbf{y}=\mathbf{0}$ is the stationary solution
of the vector field \eqref{T1.6}. Hence the stability analysis of
the solution $\bar{\mathbf{x}}(\tau)$ of \eqref{T1.1} is reduce to
the study of the stability of the stationary solution
$\mathbf{y}=0$ of the associated problem \eqref{T1.6}.

For convenience we consider the original notation assuming that
${\bf X}(\mathbf{0},\tau)=\mathbf{0}$ for $\tau_0\leq
\tau<+\infty.$\

Now we enumerate four basic definitions of stability that are
desirable properties of a nonlinear vector field \cite{167}:

\begin{enumerate}
\item The solution $\mathbf{x}=\mathbf{0}$ is said to be
\underline{stable} if given any tolerance $\epsilon>0$ and any
initial time $\tau_0,$ there exists a restriction
$\delta=\delta(\epsilon,\tau_0)>0$ such that
$\|\mathbf{x}_0\|<\delta$ implies that
$\mathbf{x}(\tau,\tau_0,\mathbf{x}_0)$ exists for $\tau_0\leq
\tau<+\infty$ and satisfies
$\|\mathbf{x}(\tau,\tau_0,\mathbf{x}_0)\|<\epsilon$ for all
$\tau\geq\tau_0.$ Therefore, any solution that stars close to
$\mathbf{x}=\mathbf{0}$ will remains close to it all future time.

\item The solution $\mathbf{x}=\mathbf{0}$ is said to be
\underline{asymptotically stable} if it is \underline{stable} and
there exists a restriction $\delta_1=\delta_1(\epsilon,\tau_0)>0$
such that $\|\mathbf{x}_0\|<\delta_1$ implies that
$\mathbf{x}(\tau,\tau_0,\mathbf{x}_0)\rightarrow\mathbf{0}$ when
$\tau\rightarrow +\infty.$ Therefore, $\mathbf{x}=\mathbf{0}$ is
stable and any solution that stars close to the origin tends to it
as time goes forward.

\item The solution $\mathbf{x}=\mathbf{0}$ is said to be
\underline{uniformly asymptotically stable} if it is
\underline{stable} with restriction $\delta$ independently of
$\tau_0$ (i.e., $\delta=\delta(\epsilon)>0$) and for any given
tolerance $\epsilon>0,$ there exists a number $T=T(\epsilon)$ with
the property
$$\lim_{\epsilon \rightarrow 0} T(\epsilon)=+\infty$$ such that
$\tau-\tau_0>T(\epsilon)$ implies
$\|\mathbf{x}(\tau,\tau_0,\mathbf{x}_0)\|<\epsilon.$ Therefore,
$\mathbf{x}$ tends to zero as $\tau-\tau_0\rightarrow +\infty,$
uniformly in $\tau_0$ and in $\mathbf{x}_0.$

\item The solution $\mathbf{x}=\mathbf{0}$ is said to be
\underline{exponentially asymptotically stable} if there exist
positive constants $\delta, K$ and $\alpha$, such that
$\|\mathbf{x}_0\|<\delta$ implies
$$\|\mathbf{x}(\tau,\tau_0,\mathbf{x}_0)\|\leq K
e^{-\alpha(\tau-\tau_0)}\|\mathbf{x}_0\|,$$ for all
$\tau\geq\tau_0.$
\end{enumerate}

The stability definitions 1-4 are used in many contexts. Stability
and asymptotic stability are not robust properties in the sense
that they are not preserved if the system is influenced by small
perturbations of the vector field. However, uniformly asymptotic
stability and exponential asymptotic stability are robust in the
previous sense. For autonomous systems, asymptotic stability and
uniform asymptotic stability are equivalent. A solution which is
exponentially asymptotically stable is also uniformly
asymptotically stable.

The above definitions describe mathematically different types of
stability; however, they do not provide a method for determine
when a given solution is stable or not. Now let us concentrate our
attention to this question

\subsection{Linearization}\label{Linear}

In order to determine the stability of $\bar{\mathbf{x}}(\tau)$ we
need to understand the nature of the solutions nearby to
$\bar{\mathbf{x}}(\tau).$ In order to do so, it is natural to
define the vector $\mathbf{y}$ by
\begin{equation}\label{T1.2.3}
\mathbf{x}=\bar{\mathbf{x}}(\tau)+\mathbf{y}
\end{equation}
Substituting \eqref{T1.2.3} in \eqref{T1.1} and expanding in
Taylor series around $\bar{\mathbf{x}}(\tau)$ we obtain
\begin{equation}\label{T1.2.4}
\mathbf{x}'=\bar{\mathbf{x}}'(\tau)+\mathbf{y}'={\bf
X}(\bar{\mathbf{x}}(\tau))+{\bf
DX}(\bar{\mathbf{x}}(\tau))\mathbf{y}+{\cal O}(\|\mathbf{y}\|^2)
\end{equation}
where ${\bf DX}$ is the derivative of ${\bf X}$ represented in the
canonical basis as the Jacobian matrix ${\bf A}$ having components
$A_{i,j}=\frac{\partial X_i}{\partial x_j};$ $\|\ldots\|$ denotes
a norm in $\mathbb{R}^n.$ To derive \eqref{T1.2.4} we have assumed
that ${\bf X}$ is at least of class $C^2.$ Using the fact that
$\bar{\mathbf{x}}'(\tau)={\bf X}(\bar{\mathbf{x}}(\tau))$ equation
\eqref{T1.2.4} is reduced to
\begin{equation}\label{T1.2.5}
\mathbf{y}'={\bf DX}(\bar{\mathbf{x}}(\tau))\mathbf{y}+{\cal
O}(\|\mathbf{y}\|^2)
\end{equation}

Equation \eqref{T1.2.5} describe the evolution of the orbits near
to $\bar{\mathbf{x}}(\tau).$ For stability questions we will
concentrate ourselves to the behavior of solutions arbitrarily
close to $\bar{\mathbf{x}}(\tau),$ thus it seems reasonably to
think that this question can be resolved studying the associated
linear system
\begin{equation}\label{T1.2.6}
\mathbf{y}'={\bf DX}(\bar{\mathbf{x}}(\tau))\mathbf{y}.
\end{equation}

Hence, the question of the stability of $\bar{\mathbf{x}}(\tau)$
involves the following two steps:
\begin{enumerate}
\item To determine if the trivial solution $\mathbf{y}=\mathbf{0}$
of \eqref{T1.2.6} is stable. \item Prove that the stability
(instability) of the solution $\mathbf{y}=\mathbf{0}$ of
\eqref{T1.2.6} implies the stability (instability) of
$\bar{\mathbf{x}}(\tau).$
\end{enumerate}

The first step is as difficult as to solve the original problem,
since there not exist general analytical methods to find solutions
of linear differential equations with variable coefficients.
However, there is an special type of solutions for which this
problem can be easily resolved: equilibrium solutions.

Let us define an special type of solutions arising for autonomous
vector fields \eqref{T1.2}.

\begin{defn} A {\it singular point} of an autonomous
vector field \ref{T1.2} is a point $\bar{\mathbf{x}} \in
{\mbox{\boldmath R}}^n$ such that  ${\bf
X}(\bar{\mathbf{x}})=\mathbf{0}.$
 \end{defn}

If $\bar{\mathbf{x}}(\tau)$ is an equilibrium solution defined as
$\bar{\mathbf{x}}(\tau)=\bar{\mathbf{x}},$ where
$\bar{\mathbf{x}}$ is a singular point of the vector field
\eqref{T1.2}, then ${\bf DX}(\bar{\mathbf{x}}(\tau))={\bf
DX}(\bar{\mathbf{x}})$ is matrix with constant entries, and the
solution $\mathbf{y}$ of \eqref{T1.2.6} passing through the point
$\mathbf{y}_0\in\mathbb{R}^n$ at $\tau=0$ can be written
immediately as
$$\mathbf{y}(\tau)=e^{{\bf DX}(\bar{\mathbf{x}})\tau}\mathbf{y}_0,$$ where
for any ${\bf A}\in M_n(\mathbb{R})$ (the vector space of real
matrices of order n), we have defined the ``exponential'' matrix
$$e^{{\bf A}\tau}=\textbf{id}+{\bf A}\tau+\frac{1}{2!}{\bf
A}^2\tau^2+\dots+\frac{1}{n!}{\bf A}^n\tau^n+\ldots,$$ with
$\textbf{id}$ denoting the identity matrix $n\times n.$ Then, the
solution $\mathbf{y}=\mathbf{0}$ is asymptotically stable if all
the eigenvalues of ${\bf DX}(\bar{\mathbf{x}})$ have negative real
parts. The answer to step 2 is given in the theorem \ref{Thm1.2.5}
to be discussed next. Its proof require the use of the so-called
Lyapunov functions. Thus, let us formulate the following stability
criteria for autonomous systems due to Lyapunov.

\begin{thm}[Theorem 2.0.1, \cite{wiggins}]\label{Lya} Consider the vector
field \eqref{T1.2}. Let $\bar{\mathbf{x}}$ be a singular point of
\eqref{T1.2} and let $V:U\rightarrow \mathbb{R}$ a $C^1$ function
defined in a neighborhood $U$ of $\bar{\mathbf{x}}$ such that
\begin{itemize}
\item[i)] $V(\bar{\mathbf{x}})=0$ and $V(\mathbf{x})>0$ if
$\mathbf{x}\neq\bar{\mathbf{x}}$ \item[ii)]
$V'(\mathbf{x})\equiv\nabla V\cdot \mathbf{x}'\leq 0$ in
$U\setminus{\bar{\mathbf{x}}}$ then $\bar{\mathbf{x}}$ is stable.
\footnote{Here and through the text, the comma in $V'(\mathbf{x})$
means total derivative with respect time. And it is referred as
derivative through the flow-orbits, or Eulerian derivative.}
Besides, if \item[iii)] $V'(\mathbf{x})< 0$ in
$U\setminus{\bar{\mathbf{x}}}$ then $\bar{\mathbf{x}}$ is
asymptotically stable.
\end{itemize}
\end{thm}

The function $V$ of the theorem \ref{Lya} is referred as Lyapunov
function. If in theorem \ref{Lya} the condition iii) holds, then
$V$ is referred as strict Lyapunov function. If $U=\mathbb{R}^n,$
then $\bar{\mathbf{x}}$ is said to be globally asymptotically
stable if i) and iii) in theorem \ref{Lya} holds.

\subsubsection{Linear stability theorem}

The Taylor expansion of ${\bf X}(\mathbf{x},\tau)$ in a
neighborhood of $\mathbf{x}=\mathbf{0}$ can be used to obtain a
linear problem for small deviations of $\mathbf{x}=\mathbf{0}.$

Since ${\bf X}(\mathbf{x},\tau)=\mathbf{0},$ when expanding in
Taylor series in a in a neighborhood of the origin we obtain ${\bf
X}(\mathbf{x},\tau)={\bf A}(\tau)\mathbf{x}+{\bf
G}(\mathbf{x},\tau)$ where ${\bf A}(\tau)={{\bf
DX}(\mathbf{0},\tau)}$ is the jacobian matrix having components
$A_{i,j}=\frac{\partial X_i}{\partial x_j}(\mathbf{0},\tau),\;
i,j=1\ldots n,$ and ${\bf G}$ is the rest in the Taylor formula,
such that for each $\tau$ exists a constant $K_1$ such that
$\|{\bf G}(\mathbf{x},\tau)\|\leq K_1\|\mathbf{x}\|^2.$ That is
${\bf G}(\mathbf{x},\tau)={\mathcal O}(\|\mathbf{x}\|^2)$ as
$\mathbf{x}\rightarrow 0$ for each $\tau.$ If the vector field
${\bf X}$ is linear, then ${\bf G}\equiv \mathbf{0}.$

In the following we assume that the above estimate holds uniformly
for $\tau_0\leq \tau< +\infty$ (i.e., $K_1$ does not depends on
$\tau$).

If the linear problem is exponentially asymptotically stable, then
the solutions of the unperturbed system can be studied, in a
neighborhood of $\mathbf{x}=\mathbf{0},$ using the linear problem
and ignoring ${\bf G}.$

Let be the linear problem exponentially asymptotically stable and
$\Phi(\tau)$ denoting the fundamental solution of the linear
problem
\begin{equation}\label{fundsol}
\Phi'={\bf A}(\tau)\Phi, \, \Phi(\tau_0)=\textbf{id},
\end{equation}
with $\textbf{id}$ denoting the identity matrix $n\times n.$

The connection between the linear problem and the nonlinear one
can be established using the variation of constant formula and
considering ${\bf G}$ as a known matrix. Therefore, the initial
value problem
\begin{equation}\label{fundsoleq}
\mathbf{x}'={\bf A}(\tau)\mathbf{x}+{\bf G}(\mathbf{x},\tau), \,
\mathbf{x}(\tau_0)=\mathbf{x}_0,
\end{equation}
is equivalent to the integral equation
\begin{equation}
\mathbf{x}(\tau)=\Phi(\tau)\mathbf{x}_0+\int_{\tau_0}^\tau
\Phi(\tau)\Phi(s)^{-1}{\bf G}(\mathbf{x}(s),s)\mathrm{d}s.
\end{equation}

The matrix $\Phi$ encodes information about the behavior of the
solutions of the linear problem and ${\bf G}$ encodes information
about the nonlinearity of the problem. As commented before the
assumption of the exponential asymptotically stability of the
linear problem will be crucial for the application of the
linearization technique.

{\bf Hypothesis H}: There exist positive constants $K$ and
$\alpha,$ such that
$$\left\|\Phi(\tau)\Phi(s)^{-1}\right\|\leq K
e^{-\alpha(\tau-s)},\; \forall\tau_0\leq s\leq \tau<+\infty.$$

\begin{thm}[Linear stability]\label{linearstthm}
If {\bf Hypothesis H} holds, and  ${\bf
G}(\mathbf{x},\tau)={\mathcal O}(\|\mathbf{x}\|^2)$ as
$\mathbf{x}\rightarrow 0$ uniformly for $\tau_0\leq \tau<+\infty.$
Then, there exists $\delta_0>0$ such that, if
$\|\mathbf{x}(\tau_0)\|<\delta_0,$ then, exists a positive
constant $\alpha_0$ such that
$$\|\mathbf{x}(\tau)\|<K\|\mathbf{x}(\tau_0)\|
e^{-\alpha_0(\tau-\tau_0)}$$ for all $\tau\geq \tau_0.$ In this
way the solution emerging from this initial state tends to
$\mathbf{x}=\mathbf{0}$ as $\tau\rightarrow +\infty.$
\end{thm}

{\bf Proof}. See \cite{167}, pp. 93-94.

Therefore, if the small perturbations problem is exponentially
asymptotically stable, the nonlinear problem also is.

For autonomous systems the matrix ${\bf A}$ have constant entries
$A_{i,j}=\frac{\partial X_i}{\partial x_j}(\mathbf{0}).$ In such
case a sufficient condition for {\bf Hypothesis H} (only valid for
autonomous systems) is that

\begin{thm}[Sufficient condition for {\bf Hypothesis H}]\label{Hcondition}
Let ${\bf A}\in M_n(\mathbb{R})$ (the vector space of real
matrices of order n) a constant matrix. If there exists a positive
constant $\alpha,$ such that the eigenvalues $\lambda$ of ${\bf
A}$ satisfy $\Re(\lambda)<-\alpha,$ then {\bf Hypothesis H} holds.
\end{thm}

{\bf Proof}. Let ${\bf A}\in M_n(\mathbb{R})$ a constant matrix.
Then each entry of the matrix $e^{{\bf A}\tau}$ is a linear
combination of the functions $\tau^k e^{l\tau}\cos m\tau, \tau^k
e^{l\tau}\cos m\tau,$ where $l+ m \imath$ denotes the eigenvalues
of ${\bf A}$ with $m\geq 0$ ($m=0$ corresponds to real
eigenvalues) and $k$ is an integer taking values $0,1,2,\ldots,
n-1$ less than the multiplicity of the corresponding eigenvalue
(\cite{Hirsch} p. 135). The fundamental solution of the linear
part of the problem satisfies $\Phi(\tau)\Phi(s)^{-1}=e^{{\bf
A}(\tau-s)},$ for all $\tau_0\leq s\leq \tau<+\infty.$ From the
previous result and given the fact that for any $\epsilon>0$ and
$n>0,$ there exists a constant $C$ such that $\tau^n<C
e^{\epsilon\tau},$ for all $\tau\geq \tau_0$ follows {\bf
Hypothesis H}.

Now let us formulate a version of the linear stability theorem
\ref{linearstthm} which is adequate to investigate the stability
of singular points of autonomous systems.

\begin{thm}[Theorem 1.2.5 \cite{wiggins}]\label{Thm1.2.5}
Suppose that the eigenvalues of ${\bf DX}(\bar{\mathbf{x}})$ have
negative real parts. Then, the equilibrium solution
$\mathbf{x}=\bar{\mathbf{x}}$ of the nonlinear vector field
\eqref{T1.2} is asymptotically stable.
\end{thm}

Sketch of the Proof. Express the the nonlinear vector field
\eqref{T1.2} in the form
\begin{equation}\label{2.0.7}
\mathbf{y}'={\bf DX}(\bar{\mathbf{x}}) \mathbf{y}+{\bf
R}(\mathbf{y})
\end{equation}
where $\mathbf{y}=\mathbf{x}-\bar{\mathbf{x}}(\tau)$ and ${\bf
R}(\mathbf{y})={\cal O}(\|\mathbf{y}\|^2).$

Introduce the coordinate re-scaling
\begin{equation}\label{2.0.8}
\mathbf{y}=\epsilon {\bf u},\; 0<\epsilon<1.
\end{equation}
Then, taking an small $\epsilon$ implies taking an small
$\mathbf{y}.$

Under \eqref{2.0.8}, the system \eqref{2.0.7} becomes
$${\bf u}'={\bf DX}(\bar{\mathbf{x}}) {\bf u}+\bar{\bf R}({\bf u},\epsilon)$$
where $\bar{\bf R}({\bf u},\epsilon)={\bf R}(\epsilon {\bf
u})/\epsilon.$ It is clear that $\bar{\bf R}({\bf u},0)=0$ since
${\bf R}(\mathbf{y})={\cal O}(\|\mathbf{y}\|^2).$

Choose as  Lyapunov function the function
$$V({\bf u})=\frac{1}{2}\|{\bf u}\|^2.$$ Then
\begin{equation}\label{2.0.10}
V'({\bf u})\equiv \nabla V({\bf u}){\bf u}'=\left({\bf u}\cdot
{\bf DX}(\bar{\mathbf{x}}) {\bf u}\right) +\left({\bf u}\cdot
\bar{\bf R}({\bf u},\epsilon)\right).
\end{equation}

From linear algebra we have that if all the eigenvalues of ${\bf
DX}(\bar{\mathbf{x}})$ have negative real parts, then there exists
a basis such that

\begin{equation}\label{2.0.11}
\left({\bf u}\cdot {\bf DX}(\bar{\mathbf{x}}) {\bf u}\right)<K
\|{\bf u}\|^2
\end{equation} for some real number $K$ and all ${\bf u}$
(see \cite{Hirsch}). Then, choosing $\epsilon$ small enough,
\eqref{2.0.10} will be strictly negative, which implies, using
theorem \ref{Lya}, that the singular point $\bar{\mathbf{x}}$ is
asymptotically stable. To finish the proof one need to show that
this result does not depends on the particular basis for which
\eqref{2.0.11} holds.

\subsection{Flow for autonomous vector fields}

Qualitative analysis of a system begins with the location of
singular points. Once the singular points of the vector field are
obtained, it is of interest to consider the dynamics in a local
neighborhood of each of the points.  Assuming that the vector
field ${\bf X}(\mathbf{x})$ is of class $C^1$ the process of
determining the local behavior is based on the linear
approximation of the vector field in the local neighborhood of the
singular point $\bar{x}$. In this neighborhood \be {{\bf
X}(\mathbf{x}) \approx {\bf
DX}(\bar{\mathbf{x}})(\mathbf{x}-\bar{\mathbf{x}})}\label{linapp}
\ee where ${\bf DX}(\bar{\mathbf{x}})$ is the Jacobian of the
vector field at the singular point $\bar{\mathbf{x}}$. The system
(\ref{linapp}) is referred to as the {\it linearization of the DE
at the singular point}. Each of the singular points can then be
classified according to the eigenvalues of the Jacobian of the
linearized vector field at the point.

\begin{defn} Let $\bar{x}$ be a singular point of the vector field \ref{T1.2}.  The
point $\bar{x}$ is called a {\it hyperbolic} singular point if
$\Re(\lambda_i) \neq 0$ for all eigenvalues, $\lambda_i$, of the
Jacobian of the vector field ${\bf X}(\mathbf{x})$ evaluated at
$\bar{\mathbf{x}}$. Otherwise the point is called {\it
non-hyperbolic}.
\end{defn}

The notion of ``hyperbolicity of a singular point'' is defined in
terms of the linearization around the singular point. This notion
is extended to more general trajectories and to invariant sets
(manifolds). In all these cases the hyperbolicity is defined also
in terms of the linearization around the trajectories or to the
invariant sets (manifolds). Hyperbolicity persists under small
perturbations of the vector field \cite{wiggins}.

\begin{defn} A set of non-isolated singular points is said to be normally
hyperbolic if the only eigenvalues with zero real parts are those
whose corresponding eigenvectors are tangent to the set.
\end{defn}

Since by definition any point on a set of non-isolated singular
points will have at least one eigenvalue which is zero, all points
in the set are {\it non-hyperbolic}.  A set which is normally
hyperbolic can, however, be completely classified as per it's
stability by considering the signs of the eigenvalues in the
remaining directions  (i.e., for a curve, in the remaining $n-1$
directions) \cite{Aulbach1984a}.

The classification then follows from the fact that if the singular
point is hyperbolic in nature the flows of the non-linear system
and it's linear approximation are {\it topologically equivalent}
in a neighborhood of the singular point.  This result is given in
the form of the following theorem:

\begin{thm}[Hartman-Grobman Theorem]\label{Hartman-Grobbman}
Consider the vector field \ref{T1.2}, where  ${\bf X}$ is of class
$C^1$. If $\bar{x}$ is a hyperbolic singular point of the
\ref{T1.2} then there exists a neighborhood, $\bar{U},$ of
$\bar{x}$ on which the flow is topologically equivalent to the
flow of the linearization of the DE at $\bar{x}$. That is, there
exists an homeomorphism $h:U\rightarrow \bar{U}$ defined in a
neighborhood $U$ of the origin such that $h\left(e^{{\bf
DX}(\bar{x})\tau}y\right)=\mathbf{x}\left(\tau,\tau_0,h(y)\right)$
for all $y\in U, t\in\mathbb{R}$ (theorem 19.12.6 in
\cite{wiggins} p. 350).
\end{thm}

{\bf Proof}. See \cite{160}.

Given the autonomous vector field \ref{T1.2}, without loss of
generality we can assume that the solutions exists for all
$\tau\in \mathbb{R}$ (if not we can apply theorem \ref{ext} or \ref{ext2} in
order to do so). Thus we can define the concept of flow:

\begin{defn}[Flow for autonomous vector fields]\label{flow}
Given the vector field \ref{T1.2}, such that ${\bf X}$ is of class
$C^1,$ and whose orbits are defined for all $\tau\in\mathbb{R}.$
Let $\mathbf{x}(\tau,\mathbf{x}_0)$ the unique maximal solution
that satisfies $\mathbf{x}(0,\mathbf{x}_0)=\mathbf{x}_0.$ The flow
is defined as the one-parametric family of mappings
$\left\{\mathbf{g}^\tau\right\}_{\tau\in\mathbb{R}}$ such that
$\mathbf{g}^\tau:\mathbb{R}^n\rightarrow\mathbb{R}^n,
\mathbf{g}^\tau(\mathbf{y})=\mathbf{x}(\tau,\mathbf{y})$ for all
$\mathbf{y}\in\mathbb{R}^n$ (definition 4.1, \cite{WE}, p. 88).
\end{defn}

If the solutions of \eqref{T1.2} are extensible as
$\tau\rightarrow +\infty,$ but not as $\tau\rightarrow -\infty,$
we can define the positive semi-flow, ${\mathbf{g}^\tau}_+,$ of
the vector field replacing $\tau \in\mathbb{R}$ by $\tau
\in\mathbb{R}^+$ in definition \eqref{flow}. Similarly, if the
solutions of \eqref{T1.2} are extensible as $\tau\rightarrow
-\infty,$ but not as $\tau\rightarrow +\infty,$ we can define the
negative semi-flow ${\mathbf{g}^\tau}_-,$ of the vector field
replacing $\tau \in\mathbb{R}$ by $\tau \in\mathbb{R}^-$ in
definition \eqref{flow}.

The conceptual difference between a solution
$\mathbf{x}(\cdot,\mathbf{y})$ and a flow $\mathbf{g}^\tau(\cdot)$
is in that

\begin{itemize}
\item For a fixed $\mathbf{y}\in \mathbb{R}^n,$ the map
$\mathbf{x}(\cdot,\mathbf{y}):\mathbb{R}\rightarrow\mathbb{R}^n$
represents the state, $\mathbf{x}(\tau,\mathbf{y}),$ of the system
for all $\tau\in\mathbb{R},$ such that
$\mathbf{x}(0,\mathbf{y})=\mathbf{y}$ initially. \item For a fixed
$\tau\in \mathbb{R},$ the map
$\mathbf{g}^\tau(\cdot):\mathbb{R}^n\rightarrow\mathbb{R}^n$
represents the state, $\mathbf{g}^\tau(\mathbf{y}),$ of the system
at time $\tau$ for all initial states $\mathbf{y}.$
\end{itemize}

\begin{thm}[Smoothness of the flow]\label{Smoothness}
Given the vector field \eqref{T1.2} such that ${\bf X}$ is of
class $C^1.$ Then the flow of the vector field \eqref{T1.2}
consists of $C^1$ mappings.
\end{thm}

From this result follows that the solutions of \eqref{T1.2} are in
a smooth dependence with respect to the initial conditions.

Some of the basic properties of a flow are the following

\begin{prop}[Properties of the flow]\label{7.4.3}

\begin{enumerate}
\item $\mathbf{g}^\tau(\mathbf{y})$ is $C^r$ (see theorem
\ref{Smoothness}). \\ \item
$\mathbf{g}^0(\mathbf{y})=\mathbf{y}.$\\
\item
$\mathbf{g}^{\tau+s}(\mathbf{y})=\mathbf{g}^\tau\left(\mathbf{g}^s(\mathbf{y})\right)
=\mathbf{g}^s\left(\mathbf{g}^\tau(\mathbf{y})\right).$
\end{enumerate}
\end{prop}

\subsection{Invariant sets}

Now let us introduce one of the fundamental concepts in the
analysis of the properties of the flow associated to a vector
field: the concept of invariant set.

\begin{defn}[Invariant set]
Let $S\subset\mathbb{R}^n$ be a set. $S$ is called an invariant
set under the vector field \eqref{T1.2} if $\mathbf{y}\in S$
implies $\mathbf{x}(\tau, \mathbf{y})\in S$ (where $\mathbf{x}(0,
\mathbf{y})=\mathbf{y}$) for all $\tau\in\mathbb{R}.$ If we
consider the property valid for $\tau\geq 0$ we say that $S$ is
positively invariant. On the other hand, if the property is valid
for $\tau\leq 0$ we say that $S$ is negatively invariant.
\end{defn}

That is,  $S \subseteq \mathbb{R}^n$ is called an invariant set
for the flow of the vector field \eqref{T1.2} if for any point
$\mathbf{y} \in S$ the orbit through $\mathbf{y}$ lies entirely in
$S$, that is $O(\mathbf{y}) \subseteq S$. Succinctly, the
invariant sets have the property that all trajectories initially
at the invariant set, remain in the invariant set all past and
future evolution. From the invariance of $S$ under the flow of the
vector field \eqref{T1.2} follows that it acts as a dynamically
independent object. Thus, when studying the dynamical properties
of the flow of a vector field, we can investigate all possible
invariant sets, and then, investigate the properties of the flow
restricted to all of them.

We have a practical tool to determine some but not all the
invariant sets of a vector field:

\begin{prop}[Proposition 4.1, \cite{WE} p. 92]\label{Proposition 4.1} Let us consider the
autonomous vector field \eqref{T1.2} with flow $\mathbf{g}^\tau.$
Let be defined a $C^1$ function $Z:\mathbb{R}^n\rightarrow
\mathbb{R}$ which satisfies $Z'\equiv \nabla Z\cdot {\bf
X}(\mathbf{x})=\alpha Z$ where
$\alpha:\mathbb{R}^n\rightarrow\mathbb{R}$ is a continuous
function. Then, the subsets of $\mathbb{R}^n$ defined by
$\left\{\mathbf{x}\in\mathbb{R}^n| Z(\mathbf{x})\lesseqqgtr
0\right\}$ are invariant sets for the flow $\mathbf{g}^\tau.$
\end{prop}

\begin{defn}[Invariant Manifold] An invariant set $S$ for the flow of the vector field
\eqref{T1.2} is called  a $C^r$ ($r\geq 1$) invariant manifold if
$S$ has the structure of a differentiable $C^r$ manifold. In that
case, if the set $S$ is positively (negatively) invariant, then it
is called a positively (negatively) invariant manifold.
\end{defn}

The general concept of topological (resp., differentiable)
manifold is given in definition \ref{TM} (resp., \ref{DiffM}).
Speaking without mathematical rigor, a manifold is a set that
locally have the structure of an Euclidean space. In applications
the manifolds are given more often as $m$-dimensional
hypersurfaces immersed in $\mathbb{R}^n.$ If the surface has not
singular points, i.e., the derivative of the function representing
the surface has maximal rank, then from the implicit function
theorem follows that it can be represented locally as a graph. The
surface is a manifold if the associated graph is $C^r.$

Another basic example of manifold is the following. Let
$\left\{s_1,\ldots, s_n\right\}$ denoting the standard basis in
$\mathbb{R}^n.$ Let $\left\{s_{{i}_1},\ldots, s_{{i}_j}\right\},\,
j<n$ denoting any $j$-basis of vectors from this set; the set
spanned by $\left\{s_{{i}_1},\ldots, s_{{i}_j}\right\}$ forms a
$j$-dimensional subspace of $\mathbb{R}^n$ which is trivially a
$C^\infty$ $j$-dimensional manifold. The main reason tho choose
these examples in in that, in the major part of our discussion
whenever we use the term manifold, it will be sufficient to think
in one of the following situations:

\begin{enumerate}
\item {\bf Linear formulation}: a vector subspace of
$\mathbb{R}^n$ \item {\bf Nonlinear formulation}: a surface
immersed in $\mathbb{R}^n$ which can be represented locally as a
graph (this can be justified by means of the implicit function
theorem).
\end{enumerate}

\subsubsection{Stable, Unstable and Center subspaces of singular points of linear autonomous vector fields}

Now let us return to our study of the structure of the orbits near
a singular point $\mathbf{x}=\bar{\mathbf{x}}$ of \eqref{T1.2} in
order to describe some important invariant manifolds that arise in
such investigation.

Let $\bar{\mathbf{x}}$ a singular point of the vector field
\eqref{T1.2} defined in $\mathbb{R}^n.$ Following the discussion
in section \ref{Linear}, it is natural to investigate the
associated linear system \be{\mathbf{y}'={\bf
A}\mathbf{y}}\label{linsys}\ee where ${\bf A}$ is a matrix with
constant coefficients ${\bf DX}(\bar{\mathbf{x}}).$

It is a straightforward matter to show that if the eigenvalues of
the matrix ${\bf A}$ are all positive the solutions in the
neighborhood of $\bar{\mathbf{y}}=\mathbf{0}$ all diverge from
that point. This point is then referred to as a source. Similarly,
if the eigenvalues all have negative real parts all solutions
converge to the singular point $\bar{\mathbf{y}}=0$, and the point
is referred to as a sink. Therefore, it follows from topological
equivalence that if all eigenvalues of the Jacobian of the vector
field for a non-linear system of ODEs have positive real parts the
point is classified as a source (and all orbits diverge from the
singular point), and if the eigenvalues all have negative real
parts the point is classified as a sink.

In most cases the eigenvalues of the linearized system
(\ref{linapp}) will have eigenvalues with both positive, negative
and/or zero real parts.  In these cases it is important to
identify which orbits are attracted to the singular point, and
which are repelled away as the independent variable tends to
infinity.

For a linear system of ODEs, (\ref{linsys}), the phase space
$\mathbb{R}^n$ is spanned by the eigenvectors of $A$.  These
eigenvectors divide the phase space into three distinct subspaces;
namely:
\begin{center}
\begin{tabular}{ll}
The {\it stable subspace}     & $E^s=$ span$(\e_1, \e_2, ... \e_{s})$ \\
The {\it unstable subspace}   & $E^u=$ span$(\e_{s+1}, \e_{s+2}, ... \e_{s+u})$\\
The {\it center subspace}     & $E^c=$ span$(\e_{s+u+1},
\e_{s+u+2}, ... \e_{s+u+c})$
\end{tabular}
\end{center}
where $\{\e_1, \e_2, ... \e_{s}\}$ are the generalized
eigenvectors of ${\bf A}$ having associated eigenvalues with
negative real part; $\{\e_{s+1}, \e_{s+2}, ... \e_{s+u}\}$ are
those whose eigenvalues have positive real part, and
$\{\e_{s+u+1}, \e_{s+u+2}, ... \e_{s+u+c}\}$ are those whose
eigenvalues have zero eigenvalues. These ones are examples of
invariant subspaces (manifolds), since the solutions of
\eqref{linsys} with initial conditions entirely contained in
$E^s,$ $E^u,$ or $E^c$ should be remain forever in this particular
subspace. Besides, flows (or orbits) in the stable subspace
asymptote in the future to the singular point, and those in the
unstable subspace asymptote in the past to the singular point.

\paragraph{Invariance of the stable, unstable and center manifold}

Consider the matrix ${\bf A}$ associated to the linear vector
field \eqref{linapp} as a linear application from $\mathbb{R}^n$
to itself. It is clear that $E^s, E^u$ and $E^c$ are invariant
subspaces under this linear application since each subspace is
generated by a collection of generalized eigenvectors
\cite{wiggins}. We want to prove that they are also invariant
under the linear application $e^{{\bf A}\tau}.$

Suppose that $V\subset\mathbb{R}^n$ is an invariant set under the
linear map ${\bf A}.$ Then,

\begin{itemize}
\item For each $c\in\mathbb{R},$ $V$ is invariant with respect to
$c{\bf A}.$ \item For each integer $n>1,$ $V$ is invariant with
respect to ${\bf A}^n.$ \item Suppose that ${\bf A}_1$ and ${\bf
A}_2$ are linear maps that leave $V$ invariant, then $V$ is
invariant with respect to ${\bf A}_1+{\bf A}_2.$ This results
follows for a finite number of linear applications ${\bf A}_i$
that leaves $V$ invariant.
\end{itemize}

Using all of these points follows that $V$ is invariant under the
linear application
$$L_n(\tau)\equiv\textbf{id}+{\bf
A}\tau+\frac{1}{2!}{\bf A}^2\tau^2+\dots+\frac{1}{n!}{\bf
A}^n\tau^n=\sum_{i=0}^n\frac{1}{i!}{\bf A}^i\tau^i,$$ for each
$n,$ where $\textbf{id}$ is the $n\times n$ identity matrix.

Using the fact that $V$ is closed and that $L_n(\tau)$ converges
uniformly to $e^{{\bf A}\tau},$ we conclude that $V$ is invariant
under $e^{{\bf A}\tau}.$

Now let us discuss several examples.

In the example example 3.1.1 in \cite{wiggins}, the eigenvalues of ${\bf A}$ are reals and different,
denoted by $\lambda_1,\lambda_2<0, \lambda_3>0.$ Then, ${\bf A}$
have three linearly independent (l.i) eigenvectors, $\e_1, \e_2,
\e_3$ corresponding respectively to
$\lambda_1,\lambda_2,\lambda_3.$ Let us define the $3\times 3$
matrix ${\bf T}$ taking as columns the eigenvectors
\begin{equation}\label{3.1.9}
{\bf T}\equiv \left(%
\begin{array}{ccc}
  \vdots & \vdots & \vdots \\
  \e_1 & \e_2 & \e_3 \\
  \vdots & \vdots & \vdots \\
\end{array}%
\right),
\end{equation}
Then, we have ${\bf A}={\bf T}\Lambda {\bf T}^{-1}$ where

\begin{equation}
\Lambda\equiv \left(%
\begin{array}{ccc}
  \lambda_1 & 0 &0 \\
  0 & \lambda_2 & 0 \\
 0 & 0 & \lambda_3 \\
\end{array}%
\right)
\end{equation}

Recall that the solution of \eqref{linsys} passing through
$\mathbf{y}_0\in\mathbb{R}^n$ at $\tau=0$ is given by
$\mathbf{y}(\tau)=e^{{\bf A}\tau} \mathbf{y}_0=e^{{\bf T}\Lambda
{\bf T}^{-1}\tau}\mathbf{y}_0.$

It is easy to see that
\begin{equation}\label{3.1.12}\mathbf{y}(\tau)=\left(\begin{array}{ccc}
  \vdots & \vdots & \vdots \\
  \e_1 e^{\lambda_1\tau} & \e_2 e^{\lambda_2\tau}& \e_3 e^{\lambda_3\tau} \\
  \vdots & \vdots & \vdots \\
\end{array}%
\right) {\bf T}^{-1}\mathbf{y}_0.\end{equation}

According to the previous results we have that $E^s=$ span$(\e_1,
\e_2)$ and $E^u=$ span$(\e_3).$ To see illustrate the invariance
we choose a point $\mathbf{y}_0\in\mathbb{R}^3.$ Then, ${\bf
T}^{-1}$ induce a coordinate transformation that change the
coordinates of $\mathbf{y}_0$ with respect to the standard basis
in $\mathbb{R}^3$ (i.e., $(1,0,0)^T, (0,1,0)^T, (0,0,1)^T$) in
coordinates with respect to the eigenbasis $\e_1, \e_2,\e_3.$

Then, for $\mathbf{y}_0\in E^s,$
\begin{equation}\label{3.1.13}{\bf T}^{-1}\mathbf{y}_0=\left(%
\begin{array}{c}
  z_{01} \\
  z_{02} \\
  0 \\
\end{array}%
\right)\end{equation} and for $\mathbf{y}_0\in E^u,$ \begin{equation}\label{3.1.14}{\bf T}^{-1}\mathbf{y}_0=\left(%
\begin{array}{c}
 0 \\
  0\\
 z_{03} \\
\end{array}%
\right)\end{equation}

Hence, by substituting \eqref{3.1.13} (resp., \eqref{3.1.14}) in
\eqref{3.1.12}, it is easy to see that $\mathbf{y}_0\in E^s$
(resp., $\mathbf{y}_0\in E^u$) implies $e^{{\bf
A}\tau}\mathbf{y}_0\in E^s$ (resp., $e^{{\bf
A}\tau}\mathbf{y}_0\in E^u$). Then, $E^s$ and $E^u$ are invariant
manifolds. Also, for each $\mathbf{y}_0\in E^s$ (resp.,
$\mathbf{y}_0\in E^u$) we have $e^{{\bf
A}\tau}\mathbf{y}_0\rightarrow \mathbf{0}$ as $\tau\rightarrow
+\infty$ (resp., $\tau\rightarrow -\infty$).

In the example 3.1.2 in  \cite{wiggins} the matrix ${\bf A}$  have
two complex conjugated eigenvalues $\varrho\pm \imath \varpi,$
$\varrho<0,\varphi\neq 0$ and a real value $\lambda>0.$ Then,
${\bf A}$ have three real generalized eigenvectors, $\e_1, \e_2,
\e_3$ which can be used as columns of the transformation matrix
${\bf T}$ such that
\begin{equation}
\Lambda\equiv \left(%
\begin{array}{ccc}
  \varrho & \varpi &0 \\
  -\varpi & \varrho & 0 \\
 0 & 0 & \lambda \\
\end{array}%
\right)={\bf T}^{-1}{\bf A}{\bf T}.
\end{equation}

In this example
\begin{equation}\label{3.1.16}
\mathbf{y}(\tau)={\bf T}e^{{\bf \Lambda}\tau}{\bf T}^{-1}\mathbf{y}_0={\bf T}\left(%
\begin{array}{ccc}
  e^{\varrho\tau} \cos \varpi\tau &  e^{\varrho\tau} \sin \varpi\tau  &0 \\
  - e^{\varrho\tau} \sin \varpi\tau  &  e^{\varrho\tau} \cos \varpi\tau  & 0 \\
 0 & 0 & e^{\lambda\tau} \\
\end{array}%
\right){\bf T}^{-1}\mathbf{y}_0
\end{equation}

Using the arguments given in the example 3.1.2, it is clear
that  $E^s=$ span$(\e_1, \e_2)$ is an invariant manifold of
solutions exponentially decaying to zero as $\tau\rightarrow
+\infty$ and $E^u=$ span$(\e_3)$ is an invariant manifold of
solutions exponentially decaying to zero as $\tau\rightarrow
-\infty.$

In the example 3.1.3 in \cite{wiggins} the matrix ${\bf A}$  have
two real repeated eigenvalues, $\lambda<0,$ and a third distinct
eigenvalue $\gamma>0$ such that there exist three real generalized
eigenvectors, $\e_1, \e_2, \e_3$ which can be used as columns of
the transformation matrix ${\bf T}$ such that ${\bf A}$ is
transformed according to
\begin{equation}
\Lambda\equiv \left(%
\begin{array}{ccc}
 \lambda & 1 &0 \\
0 & \lambda & 0 \\
 0 & 0 & \gamma \\
\end{array}%
\right)={\bf T}^{-1}{\bf A}{\bf T}.
\end{equation}

Following the ideas of the previous examples, in this
case, the solution passing through the point $\mathbf{y}_0$ at
time $\tau=0$ is given by
\begin{equation}\label{3.1.18}
\mathbf{y}(\tau)={\bf T}e^{{\bf \Lambda}\tau}{\bf T}^{-1}\mathbf{y}_0={\bf T}\left(%
\begin{array}{ccc}
  e^{\lambda\tau} &  \tau e^{\lambda\tau} \sin \varpi\tau  &0 \\
  0  &  e^{\lambda\tau}  & 0 \\
 0 & 0 & e^{\gamma\tau} \\
\end{array}%
\right){\bf T}^{-1}\mathbf{y}_0
\end{equation}

Using the same arguments as in example 3.1.1, it is clear
that  $E^s=$ span$(\e_1, \e_2)$ is an invariant manifold of
solutions decaying to zero as $\tau\rightarrow +\infty$ and $E^u=$
span$(\e_3)$ is an invariant manifold of solutions decaying to
zero as $\tau\rightarrow -\infty.$

In the non-linear case, the topological equivalence of flows
allows for a similar classification of the singular points. The
equivalence only applies in directions where the eigenvalue has
non-zero real parts. In these directions, since the flows are
topologically equivalent, there is a flow {\it tangent} to the
eigenvectors.

Unlike a linear system of ODEs, a non-linear system allows for
singular structures which are more complicated than that of the
singular points, fixed lines or periodic orbits. These structures
include, though are not limited to, such things as heteroclinic
and/or homoclinic orbits and non-linear invariant sub-manifolds
(for definitions see \cite{wiggins}).

\subsubsection{Stable, Unstable and Center manifolds of singular points of
nonlinear autonomous vector fields}

It is well-known that a nonlinear autonomous vector field can be
expressed locally in a neighborhood of a singular point,
$\bar{\mathbf{x}},$ as

\begin{equation}\label{T1.10}
\mathbf{y}'={\bf A} \mathbf{y}+{\bf R}(\mathbf{y}),\;
\mathbf{y}\in\mathbb{R}^n,
\end{equation}
where ${\bf A}={\bf DX}(\bar{\mathbf{x}}),$ and ${\bf
R}(\mathbf{y})={\cal O}(\|\mathbf{y}\|^2).$

Using elementary algebra \cite{Hirsch} follows that there exists a
lineal transformation, ${\bf T},$ such that the linear part in
\eqref{T1.10}, $\mathbf{y}'={\bf A} \mathbf{y},$ can be expressed
in the real Jordan form
\begin{eqnarray}\label{eq1.10'}
{\bf u}'={\bf A}_s {\bf u},\nonumber\\
{\bf v}'={\bf A}_u {\bf v},\nonumber\\
{\bf w}'={\bf A}_c {\bf w},
\end{eqnarray}
where $${\bf T}^{-1}(\mathbf{y}_1,\mathbf{y}_2,\mathbf{y}_3)\equiv
({\bf u},{\bf v},{\bf w})\in \mathbb{R}^s\times \mathbb{R}^u\times
\mathbb{R}^c,\, s+u+c=n;$$ ${\bf A}_s$ is the $s\times s$ matrix
having eigenvalues with negative real parts; ${\bf A}_u$ is the
$u\times u$ matrix having eigenvalues with positive real parts;
and ${\bf A}_c$ is the $c\times c$ matrix having eigenvalues with
zero real parts. By the change of coordinates induced by ${\bf T}$
the nonlinear vector field \eqref{T1.2} can be expressed as
\begin{eqnarray}\label{eq1.10''}
{\bf u}'={\bf A}_s {\bf u}+{\bf R}_s ({\bf u},{\bf v},{\bf w}),\nonumber\\
{\bf v}'={\bf A}_u {\bf v}+{\bf R}_u ({\bf u},{\bf v},{\bf w}),\nonumber\\
{\bf w}'={\bf A}_c {\bf w}+{\bf R}_c ({\bf u},{\bf v},{\bf w}),
\end{eqnarray}
where ${\bf R}_s ({\bf u},{\bf v},{\bf w}), {\bf R}_u ({\bf
u},{\bf v},{\bf w}), {\bf R}_c ({\bf u},{\bf v},{\bf w}),$ are,
respectively the first $s, u$ and $c$ components of the vector
field ${\bf T}^{-1}{\bf R}({\bf T y}).$

Let us consider the the linear vector field \eqref{eq1.10'}.
Following the previous discussion, the origin of  \eqref{eq1.10'}
have a $s$-dimensional stable invariant manifold; a
$u$-dimensional unstable invariant manifold; and a $c$-dimensional
center invariant manifold, all of them intersecting the origin.
The following theorem shows how the structure of the invariant
subspaces of the origin change when passing from the study of the
linear system \eqref{eq1.10'} to nonlinear one \eqref{eq1.10''}.

\begin{thm}[Local stable, unstable, and center manifolds at the origin]\label{InvMthm}\ If \eqref{eq1.10''}
 is of class $C^r,\; r\geq 2,$ then the singular point $({\bf u},{\bf v},{\bf w})=\mathbf{0}$
 of \eqref{eq1.10''} have a local invariant stable manifold of dimension $s,$
 $W_{\text{loc}}^s(\mathbf{0});$ a local invariant unstable manifold of dimension
 $u,$ $W_{\text{loc}}^u(\mathbf{0});$ and a local invariant center manifold of dimension
 $c,$ $W_{\text{loc}}^c(\mathbf{0}),$  all of them intersecting at the origin. These manifolds
 are tangent at the origin to the respective invariant subspaces of the linear vector
 field \eqref{eq1.10'}. Then they can expressed locally as the graphs
\begin{eqnarray}
&& W_{\text{loc}}^s(\mathbf{0})=\left\{({\bf u},{\bf v},{\bf
w})\in \mathbb{R}^s\times \mathbb{R}^u\times \mathbb{R}^c|{\bf
v}={\bf h}_{\bf v}^s({\bf u}), {\bf w}={\bf h}_{\bf w}^s({\bf u}),
\|{\bf u}\|<\delta, \right. \nonumber\\ && \left. {\bf h}_{\bf
v}^s(\mathbf{0})=\mathbf{0},{\bf h}_{\bf
w}^s(\mathbf{0})=\mathbf{0},{\bf Dh}_{\bf
v}^s(\mathbf{0})=\mathbf{0},{\bf Dh}_{\bf
w}^s(\mathbf{0})=\mathbf{0}
\right\};\nonumber\\
&& W_{\text{loc}}^u(\mathbf{0})=\left\{({\bf u},{\bf v},{\bf
w})\in \mathbb{R}^s\times \mathbb{R}^u\times \mathbb{R}^c|{\bf
u}={\bf h}_{\bf u}^u({\bf v}), {\bf w}={\bf h}_{\bf w}^u({\bf v}),
\|{\bf v}\|<\delta, \right. \nonumber\\ && \left. {\bf h}_{\bf
u}^u(\mathbf{0})=\mathbf{0},{\bf h}_{\bf
w}^u(\mathbf{0})=\mathbf{0},{\bf Dh}_{\bf
u}^u(\mathbf{0})=\mathbf{0},{\bf Dh}_{\bf
w}^u(\mathbf{0})=\mathbf{0}
\right\};\nonumber\\
&& W_{\text{loc}}^c(\mathbf{0})=\left\{({\bf u},{\bf v},{\bf
w})\in \mathbb{R}^s\times \mathbb{R}^u\times \mathbb{R}^c|{\bf
u}={\bf h}_{\bf u}^c({\bf w}), {\bf v}={\bf h}_{\bf v}^c({\bf w}),
\|{\bf w}\|<\delta, \right. \nonumber\\ && \left. {\bf h}_{\bf
u}^c(\mathbf{0})=\mathbf{0},{\bf h}_{\bf
v}^c(\mathbf{0})=\mathbf{0},{\bf Dh}_{\bf
u}^c(\mathbf{0})=\mathbf{0},{\bf Dh}_{\bf
v}^c(\mathbf{0})=\mathbf{0} \right\},
\end{eqnarray}
where the functions ${\bf h}_{\bf v}^s,{\bf h}_{\bf w}^s,{\bf
h}_{\bf u}^u,{\bf h}_{\bf w}^u,{\bf h}_{\bf v}^c,$ and ${\bf
h}_{\bf v}^c$ are $C^r$-functions and $\delta$ a positive small
enough number. The orbits at $W_{\text{loc}}^s(\mathbf{0})$ and at
$W_{\text{loc}}^s(\mathbf{0})$ have the same asymptotic properties
as the orbits in the invariant subsets $E^s$ and $E^u$
respectively. That is, the orbits of \eqref{eq1.10''} with initial
conditions at $W_{\text{loc}}^s(\mathbf{0})$ (resp.,
$W_{\text{loc}}^u(\mathbf{0})$) tends asymptotically to the origin
at an exponential rate as $\tau\rightarrow +\infty$ (resp.,
$\tau\rightarrow -\infty$) [theorem 3.2.1 in \cite{wiggins}].
 \end{thm}

{\bf Proof}. See \cite{166,168,169}

Stable, unstable or center manifold should be referred to
something (e.g., singular point, set, etc.) in order to be
meaningful.

The conditions ${\bf Dh}_{\bf v}^s(\mathbf{0})=\mathbf{0},{\bf
Dh}_{\bf w}^s(\mathbf{0})=\mathbf{0}, \ldots$ reflect the fact
that the nonlinear manifolds are tangent to the associated
invariant linear subspaces at the origin.

In the formulation of theorem \ref{InvMthm}, in expressions like
``local invariant stable manifold $\ldots$'', the term ``local''
is referred to the fact that the manifolds are defined as a graph
only in an small neighborhood of the singular point. Consequently,
all these invariant manifolds have a boundary. Hence, they are
only locally invariant in the sense that the orbits initially on
them can abandon the local manifold, but only crossing the
boundary. The invariance maintains because the vector field is
tangent to the manifolds.

In case that the singular point is hyperbolic (i.e.,
$E^c=\emptyset$), the interpretation of theorem \ref{InvMthm} is
that the trajectories of the nonlinear vector field have
qualitatively the same behavior of the orbits of the linear
associated problem in a neighborhood of the singular point. This
fact is explicitly expressed in theorem \ref{Hartman-Grobbman}.

The stable and unstable manifolds are unique. This can be proved
using a contraction mapping argument. For the center manifold, due
to the non-hyperbolicity, the analysis is more difficult, and in
general the center manifold is not unique. However, the center
manifold is unique in all the orders in its Taylor expansion. That
is, the all possible invariant manifolds differ only on small
exponential perturbations depending on the distance from the
origin to the singular point (see \cite{wiggins}).

It is important to note, however, that unlike the case of a linear
system, the center manifold, $W_{\text{loc}}^c(\mathbf{0})$ will
contain all those dynamics not classified by linearization (i.e.,
the non-hyperbolic directions). In particular, this manifold may
contain regions which are stable, unstable or neutral.  The
classification of the dynamics in this manifold can only be
determined by utilizing more sophisticated methods, such as center
manifold theorems or the theory of normal forms (see
\cite{wiggins}).

\subsubsection{Center Manifold Theory}\label{sectionCM}

In this section we offer the main techniques for the construction
of center manifolds for vector fields in $\mathbb{R}^n$. We follow
the approach in \cite{wiggins} chapter 18.

The setup is as follows. We consider vector fields in the form

\begin{align}&\mathbf{x}'=\mathbf{A x}+\mathbf{f}(\mathbf{x},\mathbf{y}),\nonumber\\
&\mathbf{y}'=\mathbf{B x}+\mathbf{g}(\mathbf{x},\mathbf{y}),  \;
(\mathbf{x},\mathbf{y})\in
\mathbb{R}^c\times\mathbb{R}^s,\label{basiceqs}\end{align} where
\begin{align}& \mathbf{f(0,0)}=\mathbf{0},
\mathbf{Df(0,0)}=\mathbf{0},\nonumber\\&\mathbf{g(0,0)}=\mathbf{0},
\mathbf{Dg(0,0)}=\mathbf{0}.\end{align} In the above, $\mathbf{A}$
is a $c\times c$ matrix having eigenvalues with zero real parts,
$\mathbf{B}$ is an $s\times s$ matrix having eigenvalues with
negative real parts, and $\mathbf{f}$ and $\mathbf{g}$ are $C^r$
functions ($r\geq 2$).

\begin{defn}[Center Manifold]\label{CMdef}
An invariant manifold will be called a center manifold for
\eqref{basiceqs} if it can locally be represented as follows
\[
W^{c}\left(  \mathbf{0}\right)  =\left\{  \left(
\mathbf{x},\mathbf{y}\right)
\in\mathbb{R}^c\times\mathbb{R}^s:\mathbf{y}=\mathbf{h}\left(
\mathbf{x}\right),\,\left\vert \mathbf{x}\right\vert
<\delta\right\};\,\mathbf{h}\left( \mathbf{0}\right)
=\mathbf{0},\,D\mathbf{h}\left( \mathbf{0}\right)  =\mathbf{0},
\]
for $\delta$ sufficiently small (cf. \cite{wiggins} p. 246,
\cite{Perko},p. 155).
\end{defn}

The conditions $\mathbf{h}\left( \mathbf{0}\right)
=\mathbf{0},\;D\mathbf{h}\left( \mathbf{0}\right)  =\mathbf{0}$
imply that $W^{c}\left(  \mathbf{0}\right)$ is tangent  to $E^c$
at $\left(\mathbf{x},\mathbf{y}\right)=(\mathbf{0},\mathbf{0}),$
where $E^c$ is the generalized eigenspace whose corresponding
eigenvalues have zero real parts. The following three theorems
(see theorems 18.1.2, 18.1.3 and 18.1.4  in \cite{wiggins} p.
245-248) are the main results to the treatment of center
manifolds. The first two are existence and stability theorems of
the center manifold for \eqref{basiceqs} at the origin. The third
theorem allows to compute the center manifold to any desired
degree accuracy by using Taylor series to solve a quasilinear
partial differential equation that $\mathbf{h}\left(
\mathbf{x}\right)$ must satisfy. The proof of those results is
given in \cite{Carr:1981}.

\begin{thm}[Existence]\label{existenceCM}
There exists a $C^r$ center manifold for \eqref{basiceqs}. The
dynamics of \eqref{basiceqs} restricted to the center manifold is,
for $\mathbf{u}$ sufficiently small, given by the following
c-dimensional vector field \be\label{vectorfieldCM}
\mathbf{u}'=\mathbf{A
u}+\mathbf{f}\left(\mathbf{u},\mathbf{h}\left(\mathbf{u}\right)\right),\;
\mathbf{u}\in\mathbb{R}^c. \ee
\end{thm}

The next results implies that the dynamics of
\eqref{vectorfieldCM} near $\mathbf{u}=0$ determine the dynamics
of \eqref{basiceqs} near
$\left(\mathbf{x},\mathbf{y}\right)=(\mathbf{0},\mathbf{0})$ (see
also Theorem 3.2.2 in \cite{Guckenheimer}).

\begin{thm}[Stability]\label{stabilityCM}
i) Suppose the zero solution of \eqref{vectorfieldCM} is stable
(asymptotically stable) (unstable); then the zero solution of
\eqref{basiceqs} is also stable (asymptotically stable)
(unstable). Then if $(\mathbf{x}(\tau),\mathbf{y}(\tau))$ is a
solution of \eqref{basiceqs} with $(\mathbf{x}(0),\mathbf{y}(0))$
sufficiently small, then there is a solution $\mathbf{u}(\tau)$ of
\eqref{vectorfieldCM} such that, as $\tau\rightarrow\infty$
\begin{align*}
& \mathbf{x}(\tau)=\mathbf{u}(\tau)+{\cal O}(e^{-r \tau}),\\
& \mathbf{x}(\tau)=\mathbf{h}\left(\mathbf{u}(\tau)\right)+{\cal
O}(e^{-r \tau}),
\end{align*} where $r>0$ is a constant.
\end{thm}

\paragraph{Dynamics Captured by the center manifold}

Stated in words, this theorem says that for initial conditions of
the \emph{full system} sufficiently close to the origin,
trajectories through them asymptotically approach a trajectory on
the center manifold. In particular, singular points sufficiently
close to the origin, sufficiently small amplitude periodic orbits,
as well as small homoclinic and heteroclinic orbits are contained
in the center manifold.

The obvious question now is how to compute the center manifold so
that we can use the result of theorem \ref{stabilityCM}? To answer
this question we will derive an equation that
$\mathbf{h}(\mathbf{x})$ must satisfy in order to its graph to be
a center manifold for \eqref{basiceqs}.

Suppose we have a center manifold \[ W^{c}\left(
\mathbf{0}\right)  =\left\{  \left( \mathbf{x},\mathbf{y}\right)
\in\mathbb{R}^c\times\mathbb{R}^s:\mathbf{y}=\mathbf{h}\left(
\mathbf{x}\right),\,\left\vert \mathbf{x}\right\vert
<\delta\right\};\,\mathbf{h}\left( \mathbf{0}\right)
=\mathbf{0},\,D\mathbf{h}\left( \mathbf{0}\right)  =\mathbf{0},
\] with $\delta$ sufficiently small. Using the invariance of $W^{c}\left(  \mathbf{0}\right)$ under the dynamics of \eqref{basiceqs}, we derive a quasilinear partial differential equation that $\mathbf{h}\left( \bf{x}\right)$ must satisfy. This is done as follows:

\begin{enumerate}
\item The $(\mathbf{x},\mathbf{y})$ coordinates of any point on
$W^{c}\left(  \mathbf{0}\right) $ must satisfy
\be\mathbf{y}=\mathbf{h}(\mathbf{x})\label{coordCM}\ee \item
Differentiating \eqref{coordCM} with respect to time implies that
the $(\mathbf{x}',\mathbf{y}')$ coordinates of any point on
$W^{c}\left(  \mathbf{0}\right) $ must satisfy
\be\mathbf{y}'=D\mathbf{h}\left(
\mathbf{x}\right)\mathbf{x}'\label{totalderivative}\ee \item Any
point in $W^{c}\left(  \mathbf{0}\right) $ obey the dynamics
generated by \eqref{basiceqs}. Therefore substituting
\begin{align*}
\mathbf{x}'=\mathbf{A x}+\mathbf{f}\left(\mathbf{x},\mathbf{h}(\mathbf{x})\right),\\
\mathbf{y}'=\mathbf{B}\mathbf{h}(\mathbf{x})+\mathbf{g}\left(\mathbf{x},\mathbf{h}(\mathbf{x})\right)
\end{align*} into \eqref{totalderivative} gives
\be{\cal N}\left(\mathbf{h}(\mathbf{x})\right)\equiv
D\mathbf{h}(\mathbf{x})\left[\mathbf{A
x}+\mathbf{f}\left(\mathbf{x},\mathbf{h}(\mathbf{x})\right)\right]
-\mathbf{B}\mathbf{h}(\mathbf{x})-\mathbf{g}\left(\mathbf{x},\mathbf{h}(\mathbf{x})\right)=0\label{MaineqcM}.\ee
\end{enumerate}

Equation \eqref{MaineqcM} is a quasilinear partial differential
that $\mathbf{h}(\mathbf{x})$ must satisfy in order for its graph
to be an invariant center manifold. To find the center manifold,
all we need to do is solve \eqref{MaineqcM}.

Unfortunately, it is probably more difficult to solve
\eqref{MaineqcM} than our original problem; however the following
theorem give us a method for computing an approximated solution of
\eqref{MaineqcM} to any desired degree of accuracy.

\begin{thm}[Approximation]\label{approximationCM}
Let $\mathbf{\Phi}:\mathbb{R}^c\rightarrow\mathbb{R}^s$ be a $C^1$
mapping with $\mathbf{\Phi}(\mathbf{0})=\mathbf{0}$ and
$D\mathbf{\Phi}(\mathbf{0})=\mathbf{0}$ such that ${\cal
N}\left(\mathbf{\Phi}(\mathbf{x})\right)={\cal
O}(\|\mathbf{x}\|^q)$ as $\mathbf{x}\rightarrow \mathbf{0}$ for
some $q>1.$ Then,
$|\mathbf{h}(\mathbf{x})-\mathbf{\Phi}(\mathbf{x})|={\cal
O}(\|\mathbf{x}\|^q)$ as $\mathbf{x}\rightarrow \mathbf{0}$
\end{thm}

This theorem allows us to compute the center manifold to any
desired degree of accuracy by solving \eqref{MaineqcM} to the same
degree of accuracy. For this task power series expansions will
work nicely. Let us consider a concrete example further in section
\ref{stabilityP4}.

\subsubsection{Normal Forms}\label{sectionNF}

In this section we offer the main techniques for the construction
of normal forms for vector fields in $\mathbb{R}^n$. We follow the
approach in \cite{arrowsmith}.

Let ${\bf X}:\mathbb{R}^n\rightarrow \mathbb{R}^n$ be a smooth
vector field satisfying ${\bf X}({\bf 0})={\bf 0}.$ We can
formally construct the Taylor expansion of $\mathbf{x}$ around
${\bf 0},$ namely, ${\bf X}={\bf X}_1+{\bf X}_2+\ldots +{\bf
X}_k+{O}(\|\mathbf{x}\|^{k+1}),$ where ${\bf X}_r\in H^r,$ the
real vector space of vector fields whose components are
homogeneous polynomials of degree $r$. For $r=1$ to $k$ we write

\ben
&{\bf X}_r(\mathbf{x})=\sum_{m_1=1}^{r}\ldots\sum_{m_n=1}^{r}\sum_{j=1}^{n}{\bf X}_{{\bf m},j}{\mathbf{x}}^{{\bf m}}{\bf e}_j,\nonumber\\
& \sum_i m_i=r, \een

Observe that ${\bf X}_1={\bf DX(\mathbf{0})}\mathbf{x}\equiv {\bf
A}\mathbf{x},$ i.e., the Jacobian matrix.

The aim of the normal form calculation is to construct a sequence
of transformations which successively remove the non-linear term
${\bf X}_r,$ starting from $r=2.$

The transformation themselves are of the form

\be \mathbf{x}=\mathbf{y}+{\bf h}_r
(\mathbf{y}),\label{htransform}\ee where ${\bf h}_r\in H^r,\,r\geq
2.$

The effect of (\ref{htransform}) in ${\bf X}_1$ is as follows
\cite{arrowsmith}: Observe that $\mathbf{x}={O}(\|\mathbf{y}\|).$
Then, the inverse of (\ref{htransform}) takes the form \be
\mathbf{y}=\mathbf{x}-{\bf
h}_r(\mathbf{x})+{O}(\|\mathbf{x}\|^{r+1}).\label{inversehtransform}\ee

By applying total derivatives in both sides, and assuming
$\mathbf{x}'={\bf A}\mathbf{x}+{\bf X}_r(\mathbf{x})$, we find

\be \mathbf{y}'={\bf A} \mathbf{y}-{\bf L_A} {\bf h}_r
(\mathbf{y})+{\bf
X}_r(\mathbf{y})+{O}(\|\mathbf{y}\|^{r+1})\label{evoly}\ee where
${\bf L_A}$ is the linear operator that assigns to ${\bf h(y)}\in
H^r$ the Lie bracket of the vector fields ${\bf A y}$ and ${\bf
h(y)}$:

\ben {\bf L_A}: H^r& &\rightarrow H^r\nonumber\\
     {\bf h}  & & \rightarrow  {\bf L_A} {\bf h (y)}={\bf D h(y)} {\bf A y}- {\bf A h(y)}.
\een

Both ${\bf L_A}$ and $\mathbf{X}_r\in H^r,$ so that the deviation
of the right-hand side of (\ref{evoly}) from ${\bf A y}$ has no
terms of order less than $r$ in $\|\mathbf{y}\|.$ This means that
if ${\bf X}$ is such that ${\bf X}_2=\ldots {\bf X}_{r-1}=0,$ they
will remain zero under the transformation (\ref{htransform}). This
makes clear how we may be able to remove ${\bf X}_r$ from a
suitable choice of ${\bf h}_r.$

The proposition 2.3.2 in \cite{arrowsmith} states that if the
inverse of ${\bf L_A}$ exists, the differential equation \be
\mathbf{x}'={\bf A}\mathbf{x}+{\bf
X}_r(\mathbf{x})+{O}(\|\mathbf{x}\|^{r+1})\label{ODE1}\ee with
${\bf X}_r\in H^r,$ can be transformed to \be \mathbf{y}'={\bf A
y}+{O}(\|\mathbf{y}\|^{r+1})\label{ODE2}\ee by the transformation
(\ref{htransform}) where \be {\bf h}_r(\mathbf{y})={\bf L_A}^{-1}
{\bf X}_r(\bf y)\label{Transform2}\ee

The equation \be {\bf L_A}{\bf h}_r(\mathbf{y})={\bf X}_r(\bf
y)\ee is named the homological equation.

If ${\bf A}$ has distinct eigenvalues $\lambda_i,\, i=1,2,3,$ its
eigenvectors form a basis of $\mathbb{R}^n.$ Relative to this
eigenbasis, ${\bf A}$ is diagonal. It can be proved (see proof in
\cite{arrowsmith}) that ${\bf L_A}$ has eigenvalues $\Lambda_{{\bf
m},i}={\bf m}\cdot {\mathbf{\lambda}}-\lambda_i=\sum_j
m_j\lambda_j-\lambda_i$ with associated eigenvectors ${\bf
x^m}{\bf e}_i.$ The operator, ${\bf L_A}^{-1},$ exists if and only
if the $\Lambda_{{\bf m},i}\neq 0,$ for every allowed ${\bf m}$
and $i=1\ldots r.$

If we were able to remove all the nonlinear terms in this way,
then the vector field can be reduced to its linear part
$$\mathbf{x}'={\bf X}(\mathbf{x})\rightarrow \mathbf{y}'={\bf A}\mathbf{y}.$$
Unfortunately, not all the higher order terms vanishes by applying
these transformations. It is the case if resonance occurs.

The n-tuple of eigenvalues ${\bf
\lambda}=(\lambda_1,\ldots,\lambda_n)^T$ is resonant of order $r$
(see definition 2.3.1 in \cite{arrowsmith}) if there exist some
${\bf m}=(m_1,m_2,\ldots m_n)^T$ (a n-tuple of non-negative
integers) with $m_1+m_2+\ldots m_n=r$ and some $i=1\ldots n$ such
that $\lambda_i={\bf m}\cdot \lambda,$ i.e., if $\Lambda_{{\bf
m},i}=0$ for some ${\bf m}$ and some $i.$

If there is no resonant eigenvalues, and provided they are
different, we can use the eigenvectors of ${\bf A}$ as a basis for
$H^r.$ Then, we can write ${\bf h}_r$ as
$${\bf h}_r(\mathbf{x})=\sum_{{\bf m},i,\sum m_j=r}h_{{\bf m},i}\mathbf{x}^{\bf m}{\bf e}_i$$ and any vector
field ${\bf X}\in H^r$ as $${\bf X}(\mathbf{x})=\sum_{{\bf m},i,\;
\sum m_j=r} {\bf X}_{{\bf m},i}\mathbf{x}^m {\bf e}_i$$ where
${\bf m}=(m_1,m_2,\ldots,m_n)^T,$
$\mathbf{x}^m=x_1^{m_1}\,x_2^{m_2}\,\ldots\,x_n^{m_n}$ and ${\bf
e}_i,\,i=1,\ldots n$ stands for the canonical basis in
$\mathbb{R}^n.$ If the eigenvalues of ${\bf A}$ are not resonant
of order $r,$ then $$h_{{\bf m},i}=X_{{\bf m},i}/\Lambda_{{\bf
m},i}.$$ This gives ${\bf h}_r$ explicitly in terms of ${\bf
X}_r.$

In case that resonance occurs, we proceed as follows. If ${\bf A}$
can diagonalized, then the eigenvectors of ${\bf L}_{\bf A}$ form
a basis of $H^r.$ The subset of eigenvectors of ${\bf L}_{\bf A}$
with non-zero eigenvalues then form a basis of the image, $B^r,$
of $H^r$ under ${\bf L}_{\bf A}.$ It follows that the component of
${\bf X}_r$ in $B^r$ can be expanded in terms of these
eigenvectors and ${\bf h}_r$ chosen such that $$h_{{\bf
m},i}=X_{{\bf m},i}/\Lambda_{{\bf m},i}.$$ to ensure the removal
of these terms. The component, ${\bf w}_r,$ of ${\bf X}_r$ lying
in the complementary subspace, $G^r,$ of $B^r$ in $H^r$ will be
unchanged by the transformations $\mathbf{x}=\mathbf{y}+{\bf
h}_r(\mathbf{y})$ obtained from $B^r.$

Since $$ {\bf X}_r(\mathbf{y}+{\bf h}_{r+k}(\mathbf{y}))={\bf
X}_r(\mathbf{y})+{O}(\|\mathbf{y}\|^{r+k+1}),r\geq 2,
k=1,2,\ldots,$$ these terms are not changed by subsequent
transformations to remove non-resonant terms of higher order.

The above facts are expressed in

\begin{thm}[theorem 2.3.1 in \cite{arrowsmith}]\label{NFTheorem}
Given a smooth vector field $\bf X(\mathbf{x})$ on $\mathbb{R}^n$
with ${\bf X(0)=0},$ there is a polynomial transformation to new
coordinates, $\mathbf{y},$ such that the differential equation
$\mathbf{x}'={\bf X}(\mathbf{x})$ takes the form $\mathbf{y}'={\bf
J}\mathbf{y}+\sum_{r=1}^N {\bf
w}_r(\mathbf{y})+{O}(\|\mathbf{y}\|^{N+1}),$ where ${\bf J}$ is
the real Jordan form of ${\bf A}={\bf D X}({\bf 0})$ and ${\bf
w}_r\in G^r,$ a complementary subspace of $H^r$ on $B^r={\bf
L_A}(H^r).$
\end{thm}

\subsubsection{Asymptotic behavior}\label{appendixA}

No we develop the technical apparatus to dealing with the notions
of ``long term'' and ``observable'' behavior of the orbits in the
phase space. We consider vector fields \eqref{T1.2} with ${\bf X}$
of class $C^r$ ($r\geq 1.$) In the following
$\mathbf{g}^\tau(\mathbf{x})$ denotes the flow generated by the
vector field (or differential equation)

\begin{equation}{\mathbf{x}}'(\tau)={\bf X}(\mathbf{x}(\tau)),\,\mathbf{x}(\tau)\in\mathbb{R}^n,\label{vectorfield}\end{equation} where the prime denote derivative with respect to $\tau.$

\paragraph{Limit Sets}

Let us define the concepts of $\alpha$ and $\omega$-sets.

\begin{defn}[definition 8.1.1, \cite{wiggins} p. 104]\label{omegalimitpoint} A point $\mathbf{x}_0\in\mathbb{R}^n$ is called an $\omega$-limit point of
$\mathbf{x}\in\mathbb{R}^n,$ denoted $\omega(\mathbf{x}),$ if
there exists a sequence $\{\tau_i\},\,\tau_i\rightarrow\infty$
such that $\mathbf{g}^{\tau_i}(\mathbf{x})\rightarrow
\mathbf{x}_0.$ $\alpha$-limits are defined similarly by taking a
sequence $\{\tau_i\},\,\tau_i\rightarrow -\infty. $

\end{defn}

\begin{defn}[definition 8.1.2, \cite{wiggins} p. 105]\label{omegalimitset} The set of all $\omega$-limit points of a flow or map is called a
$\omega$-limit set. The $\alpha$-limit is similarly defined.

\end{defn}

The following result describe some of the basic properties of the
$\alpha$ and $\omega$-limit sets of trajectories.

\begin{prop}[proposition 8.1.3 \cite{wiggins}, p. 105]\label{omegalimitsetproperties} Let $\mathbf{g}^\tau(\cdot)$ be a flow generated by a vector field and
let $M$ be a positively invariant compact set for this flow (see
definition 3.0.3 p. 28 \cite{wiggins}). Then for $\mathbf{p}\in M,$ we
have

\begin{enumerate}\label{prop_omega}
\item[i)] $\omega(\mathbf{p})\neq\emptyset$ \item[ii)]
$\omega(\mathbf{p})$ is closed \item[iii)] $\omega(\mathbf{p})$ is
invariant under the flow, i.e., $\omega(\mathbf{p})$ is a union of
orbits. \item[iv)] $\omega(\mathbf{p})$ is connected.
\end{enumerate}

\end{prop}

{\bf Proof}. See \cite{wiggins}

\begin{enumerate}
\item[i)] Choose a sequence a sequence
$\{\tau_i\},\,\tau_i\rightarrow\infty,$ and let $\left\{\mathbf{
p}_i=\mathbf{g}^{\tau_i}({\mathbf{p}})\right\}.$ Since $M$ is
compact, $\left\{\mathbf{ p}_i\right\}$ is a convergent
subsequence whose limit belongs to $\omega(\mathbf{p}).$ Then
$\omega(\mathbf{p})\neq\emptyset.$

\item[ii)] It is sufficient to show that the complement of
$\omega(\mathbf{p})$ is an open set. Let $\mathbf{q}\notin
\omega(\mathbf{p}).$ Then, there exists a neighborhood of
$\mathbf{q},$ $U(\mathbf{q}),$ which is disjoint to the set of
points $\left\{\mathbf{g}^{\tau}(\mathbf{p})| \tau>T\right\}$ for
some $\tau\geq T.$ Then, $\mathbf{q}$ is contained in an open set
that do not contain any point in $\omega(\mathbf{p}).$ Since
$\mathbf{q}$ is arbitrary, we obtain the desired result.

\item[iii)] Let $\mathbf{q}\in \omega(\mathbf{p})$ and
$\tilde{\mathbf{q}}\in \mathbf{g}^{s}(\mathbf{q}).$ Choose a
sequence $\left\{\tau_i\right\},\; \tau_i\rightarrow +\infty$ when
$i\uparrow +\infty.$ Since
$\mathbf{g}^{\tau_i}(\mathbf{p})\rightarrow \mathbf{q},$ then
$\mathbf{g}^{\tau_i+s}(\mathbf{p})=\mathbf{g}^{s}\left(\mathbf{g}^{\tau_i}(\mathbf{p})\right),$
converges to $\tilde{\mathbf{q}}$ as $i\rightarrow +\infty.$ Thus,
$\tilde{\mathbf{q}}\in \omega(\mathbf{p})$ and then,
$\omega(\mathbf{p})$ is invariant.

In this proof we have assumed that $\mathbf{g}^s(\cdot)$ exist for
all $s.$ However, this fact it not so obvious. Let us prove the
above statement for $\mathbf{q}\in\omega(\mathbf{p}),$ that is,
let us prove that $\mathbf{g}^s(\mathbf{q})$ exist for $s\in
(-\infty,+\infty)$ for all $\mathbf{q}\in\omega(\mathbf{p}).$ It
is clear that this is true for $s\in (0,+\infty)$ since $M$ is a
positively invariant compact set. Then, it is sufficient to prove
that it is true for $s\in (-\infty,0].$

Since $\mathbf{q}\in\omega(\mathbf{p}),$ there exists a time
sequence $\left\{\tau_i\right\},\; \tau_i\rightarrow +\infty$ as
$i\uparrow +\infty,$ such that
$\mathbf{g}^{\tau_i}(\mathbf{p})\rightarrow \mathbf{q}$ as
$i\rightarrow +\infty.$ Let us sort the sequence such that
$\tau_1<\tau_2<\ldots< \tau_n<\ldots,$ and consider
$\mathbf{g}^s\left(\mathbf{g}^{\tau_i}(\mathbf{p})\right).$
Following proposition \ref{7.4.3}, the above composition is
well-defined for $s\in[-\tau_i,0].$ Taking the limit as
$i\rightarrow +\infty,$ using the continuity of the flow and the
fact that $\mathbf{g}^{\tau_i}(\mathbf{p})\rightarrow \mathbf{q}$
as $i\rightarrow +\infty,$ we see  that $\mathbf{g}^s(\mathbf{q})$
exist for $s\in (-\infty,0].$

\item The proof is by contradiction. Suppose that
$\omega(\mathbf{p})$ is not connected. The we can choose open sets
$V_1, V_2$ such that $\omega(\mathbf{p})\in V_1\cup V_2,$
$\omega(\mathbf{p})\cap V_1\neq\emptyset,$ $\omega(\mathbf{p})\cap
V_2\neq\emptyset,$ and $\bar{V_1}\cap\bar{V_2}=\emptyset.$

The orbit $O(\mathbf{p})$ accumulates both in points of $V_1$ and
in points of $V_2;$ thus, for a given $T>0,$ exists $\tau>T$ such
that $\mathbf{g}^\tau(\mathbf{p})\in M\setminus\left(V_1\cup
V_2\right)=K,$ a compact set. Then, we can find a sequence
$\tau_n\rightarrow +\infty$ as $n\uparrow +\infty$ with
$\mathbf{g}^{\tau_n}(\mathbf{p})\in K.$ Passing to a subsequence,
if necessary ($K$ is compact), we have
$\mathbf{g}^{\tau_n}(\mathbf{p})\rightarrow\mathbf{q},\; q\in K.$
This implies that $\mathbf{q}\in V_1\cup V_2.$ However, our
construction indicates that $\mathbf{q}\in \omega(\mathbf{p}).$
This is a contradiction with the hypothesis
$\bar{V_1}\cap\bar{V_2}=\emptyset.$
\end{enumerate}

It can be proved an analogous result for $\alpha$-limit sets if
the hypothesis of proposition \ref{prop_omega} holds for a flow in
reversed time.

\paragraph{Attracting sets, attractors and basin of attraction}

Now we want to develop the idea of an attractor.

\begin{defn}[Attracting set for flows, definition 8.2.1,
\cite{wiggins}]\label{attract} A closed invariant set
$A\subset\mathbb{R}^n,$ is called an attracting set, if there
exists a neighborhood $U$ of $A$ such that $\forall \tau\geq 0,
\mathbf{g}^\tau(U)\subset U$ and
$\cap_{\tau>0}\mathbf{g}^\tau(U)=A.$
\end{defn}

\begin{defn}[Trapping region, definition 8.2.2, \cite{wiggins}]
The open set $U$ id the definition \ref{attract} is often referred
to as a trapping region.
\end{defn}

A similar definition is given in

\begin{defn} Given the vector field \eqref{T1.2} in $\mathbb{R}^n$, with flow $\{\mathbf{g}^\tau\}$, a
subset $S \subseteq \mathbb{R}^n$ is said to be a trapping set of
the DE if it satisfies:
\begin{enumerate} \item $S$ is a closed and bounded set, \item $\mathbf{y} \in S$ implies that
$\mathbf{g}^\tau (\mathbf{y}) \in S$ for all $\tau \geq 0$.
\end{enumerate}
\end{defn}

To find a Lyapunov function is equivalent to finding a trapping
region. By theorem \ref{existencethm} it follows that all solution
starting in a trapping region exists for all positive times. This
is useful in noncompact phase spaces such as $\mathbb{R}^2$ for
proving existence on semi-infinite time intervals. In the
continuous time case, one ``test'' whether or not a region is a
candidate to be a trapping region by evaluating the vector field
on the boundary of the region in question. If, on the boundary of
the region, the vector field is pointing toward the interior of
the region, or it is tangent to the boundary, then, the given
section is a trapping region. However, in order to the test be
carried out, one needs that the boundary of the region must be, at
least, $C^1.$

Another idea related with trapping regions is that of absorbing
set.

\begin{defn}[Absorbing set for flows, definition 8.2.3, \cite{wiggins}]\label{absorbing}
A positive invariant compact subset $B\subset \mathbb{R}^n$ is
called an absorbing set if there exists a bounded subset of
$\mathbb{R}^n,$ $U,$ with $U\supset B,$ and $\tau_U$ such that
$\mathbf{g}^\tau(U)\subset B,\, \forall \tau\geq \tau_U.$
\end{defn}

If we have an attracting set it is natural to ask which points in
phase space approach the attracting set asymptotically.

\begin{defn}[Basin of attraction for flows, definition 8.2.4, \cite{wiggins}]\label{basin}
The domain or basin of attraction of an attracting set $A$ is
given by $$\bigcup_{\tau\leq 0}\mathbf{g}^\tau(U),$$ where $U$ is
any open set satisfying definition \ref{attract}.
\end{defn}

{\bf Note}. The basin of attraction is independent of the choice
of the open set $U,$ provided that $U$ satisfies definition
\ref{attract}.

An  attracting set can contain several sinks (``attractors'') and
almost all the points in the phase space will eventually end up
near one of these sinks (see example 8.2.2, \cite{wiggins}).
Therefore, if we are interested in describing where most points in
the phase space ultimately go, the idea is an attracting set is
not quite precise enough. It is necessary to incorporate into the
definition the notion that it is not only a collection of distinct
attractors, but rater, that all points in the attracting set
eventually come arbitrarily close to every other point in the
attracting set under the evolution of the flow.

This idea can be rigourously stated as

\begin{defn}[Topological transitivity, definition 8.2.5, \cite{wiggins}]
A closed invariant set $A$ is said to be topologically transitive
if, for any two open sets $U, V\subset A$ exists $\tau\in
\mathbb{R}$ such that $\mathbf{g}^\tau(U)\cap V\neq \emptyset.$
\end{defn}

\begin{defn}[Attractor, definition 8.2.6, \cite{wiggins}]
An attractor is a  topologically transitive attracting set.
\end{defn}

\paragraph{LaSalle Invariance Principle, Monotone Functions and Monotonicity
Principle.} In this section we describe an application of the
invariance of $\omega$-limit sets of a trajectory that is very
useful for the stability analysis. It is referred as the LaSalle
Invariance Principle (see \cite{LaSalle}; theorem 8.3.1 in
\cite{wiggins}, p. 111).

Given the autonomous vector field \eqref{T1.2} where $\mathbf{X}$
is of class $C^r$ ($r\geq 1$). Let ${\cal M}\subset\mathbb{R}^n$
be a positively invariant compact set under the flow,
$\mathbf{g}^\tau(\cdot),$ generated by this vector field, which is
the closure of some open set (so that it has a nonempty interior)
and whose boundary is (at least) $C^1.$ Therefore, ${\cal M}$ is a
trapping region. Let $V(\mathbf{x})$ a Lyapunov function on ${\cal
M}.$ By this we mean that $V'(\mathbf{x})\equiv\nabla
V(\mathbf{x}) \cdot \mathbf{X}(\mathbf{x})\leq 0$ on ${\cal M}$
(here we use the notion of Lyapunov function not as in the context
of theorem \ref{Lya}, as a local notion in a neighborhood of a
fixed point, but in a broader global sense). Consider the two sets
$$E\equiv \left\{\mathbf{x}\in{\cal M}| V'(\mathbf{x})=0\right\}$$
and $$M\equiv\left\{%
\begin{array}{c}
  \text{The union of all the trajectories that start in}\, E  \\
  \text{an remain in}\, E\,  \text{for all}\, \tau>0\\
\end{array}%
\right\}.$$

$M$ is a ``positively invariant part'' of $E.$ Now we can state
the LaSalle Invariance Principle.

\begin{thm}[LaSalle Invariance Principle] For all $\mathbf{x}\in{\cal M},$ $\mathbf{g}^\tau
(\mathbf{x})\rightarrow M$ (\cite{LaSalle}; theorem 8.3.1 in
\cite{wiggins}, p. 111).
\end{thm}

{\bf Proof}. (See \cite{wiggins}, p. 111) Fist, let us prove that
$V$ is a constant $\chi$ on $\omega(\mathbf{x}).$ Suppose that
$\bar{\mathbf{x}}\in\omega(\mathbf{x})$ and let
$\chi=V(\bar{\mathbf{x}}),$ then, $\chi$ is greatest lower bound
of the set $\left\{V(\mathbf{g}^\tau(\mathbf{x}))|\tau\geq
0\right\}.$ This follows from the fact that $V$ decreases along
trajectories (hence
$V\left(\mathbf{g}^{\tau_i}(\mathbf{x})\right)\geq
V\left(\mathbf{g}^{\tau}(\mathbf{x})\right)\geq
V\left(\mathbf{g}^{\tau_{i+1}}(\mathbf{x})\right)$ for $\tau_i\leq
\tau\leq \tau_{i+1}$) and by the continuity of $V.$ From
proposition \ref{omegalimitsetproperties}, the omega limit set of
a trajectory is invariant, hence
$\mathbf{g}^\tau(\bar{\mathbf{x}})$ is also an omega limit point
of $\mathbf{g}^\tau(\mathbf{x}).$ Then, since $\chi$ is the
greatest lower bound of the set
$\left\{V(\mathbf{g}^\tau(\mathbf{x}))|\tau\geq 0\right\},$
$V\left(\mathbf{g}^\tau(\bar{\mathbf{x}})\right)=\chi.$ From this
it follows that $V'=0$ on $\omega(\mathbf{x}).$ Then, by the
definition of $E,$ $\omega(\mathbf{x})\subset E.$ Since
$\omega(\mathbf{x})$ is invariant (proposition
\ref{omegalimitsetproperties}), it follows by the definition of
$M$ that $\omega(\mathbf{x})\subset M.$ Therefore $\mathbf{g}^\tau
(\mathbf{x})\rightarrow M$ as $\tau\rightarrow +\infty.$

LaSalle Invariance Principle can be generalized considerably, for
instance to theorem \ref{theorem 4.12}. This extension requires
the introduction of the concept of monotonic function for the
flow.

\begin{defn}[definition 4.8 \cite{WE}, p. 93]\label{Definition 4.8} Let $\mathbf{g}^\tau(\mathbf{x})$ be a flow on $\mathbb{R}^n,$ let $S$ be an invariant set of $\mathbf{g}^\tau(\mathbf{x})$ and let $Z:S\rightarrow\mathbb{R}$ be a continuous function. $Z$
is monotonic decreasing (increasing) function for the flow
$\mathbf{g}^\tau(\mathbf{x})$ means that for all $\mathbf{x}\in
S,$ $Z(\mathbf{g}^\tau(\mathbf{x}))$ is a monotonic decreasing
(increasing) function of $\tau.$

\end{defn}

\begin{thm}[Monotonicity Principle]\label{theorem 4.12}

Let $\mathbf{g}^\tau(\mathbf{x})$ be a flow on $\mathbb{R}^n$ with
$S$ an invariant set. Let $Z: S\rightarrow\mathbb{R}$ be a
$C^1\left(\mathbb{R}^n\right)$ function whose range is the
interval $(a,\;b)$ where $a\in \mathbb{R} \cup \{-\infty\},$ $b\in
\mathbb{R} \cup \{+\infty\},$ and $a<b.$ If $Z$ is decreasing on
orbits in $S,$ then for all $\mathbf{x}\in S$,
$\omega(\mathbf{x})\subset\{\mathbf{s}\in \bar S\setminus
S|lim_{\mathbf{y}\rightarrow \mathbf{s}} Z(\mathbf{y})\neq b \}$
and $\alpha(\mathbf{x})\subset\{\mathbf{s}\in \bar S\setminus
S|lim_{\mathbf{y}\rightarrow \mathbf{s}} Z(\mathbf{y})\neq a \}.$
\end{thm}

\paragraph{Results for Planar Systems}

The Poincar\'e-Bendixon theorem gives us a complete determination
of the asymptotic behavior of a large class of flows on the plane,
cylinder, and two-sphere. It is  remarkable in that it assumes not
detailed information about the vector field, only uniqueness of
the solutions, properties of $\omega$-limit sets, and some
properties of the geometry of the underlying phase plane.

\begin{thm}[Poincar\'e-Bendixon Theorem]\label{PBT} Let ${\cal M}$
be a positively invariant set for the vector field
\eqref{vectorfield} on $\mathbb{R}^2$ (with $\mathbf{X} \in C^2$),
containing at most a finite number of singular points (i.e., no
non-isolated singular points). Let $\mathbf{p}\in {\cal M},$ and
consider $\omega(\mathbf{p}).$ Then one of the following
possibilities holds:
\begin{enumerate}
\item $\omega(\mathbf{p})$ is a singular point. \item
$\omega(\mathbf{p})$ is a closed (periodic) orbit. \item
$\omega(\mathbf{p})$ is the union of singular points and
heteroclinic or homoclinic orbits. That is $\omega(\mathbf{p})$
consists of a finite number of singular points  $\mathbf{p}_1,
\ldots, \mathbf{p}_n$ and orbits $\gamma$ with
$\alpha(\gamma)=\mathbf{p}_i$ and $\omega(\gamma)=\mathbf{p}_j,$
$i,j\in \{1\ldots n\}.$
\end{enumerate}
\end{thm}

\begin{thm}[Corollary of Poincar\'e-Bendixon Theorem \ref{PBT}, \cite{Coley:2003mj}, p. 22]\label{PBT2}
Let be $K$ a positively invariant set for the vector field
\eqref{vectorfield} on $\mathbb{R}^2$ (with $\mathbf{X} \in C^1$).
If $K$ is a bounded and closed set, the $K$ contains either a
closed (periodic) orbit, or a singular point.

\end{thm}

\begin{thm}[Dulac's Criterion]\label{DC}
If $D \subseteq \mathbb{R}^2$ is a simply connected open set and
$B$ is a Dulac's function on $D$, then the differential equation
\eqref{vectorfield} on $\mathbb{R}^2$, with $\mathbf{X} \in C^1$
has no periodic (or closed) orbit which is contained in $D$.
\end{thm}

\section{Procedure for analyzing cosmological dynamical systems}\label{Procedure}

Given a cosmological dynamical system determined by the differential equation

\begin{eqnarray}
\frac{\mathrm{d}y}{\mathrm{d}\tau}&=&f(y), y\in \mathbb{R}^n\label{Eq1},\\
g(y)&=&0\label{eq2},
\end{eqnarray}
the standard procedure to analyze the properties of the flow
generated by \eqref{Eq1} subject to the constraint(s) \eqref{eq2}
(see, for example, the reference \cite{WE}) is the following:

\begin{enumerate}
\item Determine whether the state space, as defined by
\eqref{eq2}, is compact. \item Identify the lower-dimensional
invariant sets, which contains the orbits of more special classes
of models with additional symmetries. \item Find all the singular
points and analyze their local stability. Where possible identify
the stable and unstable manifolds of the singular points, which
may coincide with some of the invariant sets in point (2). \item
Find Dulac's functions or monotone functions in various invariant
sets where possible. \item Investigate any bifurcation that occur
as the equation of state parameter $\gamma$ (or any other
parameters) varies. The bifurcations are associated with changes
in the local stability of the singular points. \item Having all
the information in the points (1)-(5) one can hope to formulate
precise conjectures about the asymptotic evolution, by identifying
the past and the future attractors. The past attractor will
describe the evolution of a typical universe near the initial
singularity while the future attractor will play the same role at
late times. The monotone functions in point (4) above, in
conjunction with theorems of dynamical systems theory, may enable
some of the conjectures to be proved. \item Knowing the stable and
unstable manifolds of the singular points it is possible to
construct all possible heteroclinic sequences that join the past
attractor, thereby gaining insight into the intermediate evolution
of cosmological models.
\end{enumerate}

\chapter{Non-minimally Coupled Dark Energy Models}\label{SFNMC}

In this chapter we investigate, from the dynamical systems
viewpoint flat FRW models in the conformal (Einstein) frame of
scalar-tensor gravity theories including f(R) theories through
conformal transformation. Particularly we are interested in
investigating the stability of the de Sitter solution in this
framework that give an answer to the accelerating expansion. Also
que investigate the stability of scaling solutions. Scaling
late-time attractor solutions provide a hint for solving or
alleviating the Coincidence Problem.

\section{Introduction}

Current astrophysical observations suggest that the universe is
permeated by an exotic form of matter called Dark Energy that is
driving the current accelerated expansion
\cite{Komatsu:2010fb,obs1,obs2,obs3,obs4,Perlmutter:1998np,Bennett:2003bz,Tegmark:2003ud,Allen:2004cd,Tegmark:2003uf,Komatsu:2003fd,Hinshaw:2003ex,Spergel:2003cb,Peiris:2003ff,Riess:1998cb,Riess:2004nr,Jassal:2004ej,Freedman:2000cf,Mould:1999ap,Choudhury:2003tj}
which can be modelled as a self-interacting scalar field.

Scalar fields, and theories including it such as Scalar-tensor
theories (STT) of gravity
\cite{Brans:1961sx,Wagoner:1970vr,O'Hanlon:1972my,O'Hanlon:1972hq,Bekenstein:1977rb,Bergmann:1968ve,Nordtvedt:1970uv}
can be supported by fundamental physical theories like superstring
theory \cite{Green:1996bh}. Quintessential DE models
\cite{Sahni:2002kh,Kolda:1998wq,Padmanabhan:2002ji}, for instance,
are described by an ordinary scalar field minimally coupled to
gravity. A particular choice of the  scalar field self-interacting
potentials can drive the past and current accelerated expansion.
The natural generalizations to quintessence models evolving
independently from the background matter are models that exhibit
non-minimal coupling between both components. The effective
interaction dark energy-dark matter appears when we apply
conformal transformations \footnote{See the reference
\cite{Faraoni:1998qx} for applications of conformal
transformations in both relativity and cosmology.} to the STTs.
Several physical theories which predict the presence of a scalar
field coupled to matter. For example, in string theory the dilaton
field is generally coupled to matter \cite{Gasperini:2007zz}.
Nonminimally coupling occurs also in STT of gravity
\cite{Fujii:2003pa,Faraoni:2004pi}, in HOG theories
\cite{Capozziello:2007ec} and in models of chameleon gravity
\cite{Waterhouse:2006wv}. Coupled quintessence was investigated
also in
\cite{Amendola:1999er,TocchiniValentini:2001ty,Billyard:2000bh} by
using dynamical systems techniques.

The cosmological dynamics of scalar-tensor gravity have been
investigated in \cite{Carloni:2007eu,Tsujikawa:2008uc}.
Phenomenological coupling functions were studied for instance in
\cite{Boehmer:2008av} which can describe either the decay of dark
matter into radiation, the decay of the curvaton field into
radiation or the decay of dark matter into dark energy (see
section III of \cite{Boehmer:2008av} for more information and for
useful references).  In the reference \cite{Tsujikawa:2008uc}, the
authors construct a family of viable scalar-tensor models of dark
energy (which includes pure $F(R)$ theories and quintessence). By
investigating a phase space the authors obtain that the model
posses a phase of late-time acceleration preceded by a standard
matter era, while at the same time satisfying the local gravity
constraints (LGC). In \cite{Gonzalez:2004dh} it is studied a
scalar field responsible for both the early and the late time
inflationary expansion in the context of brane cosmology.

In the inflationary universe scenarios (mainly based on GRT)
matter is modelled, usually, as a scalar field, $\phi,$ with
potential $V(\phi)$, which must meet the requirements necessary to
lead to the early-time accelerating expansion
\cite{Billyard:1999a,Copeland:2004b,Copeland:1993,Kolb:1995,Lidsey:1997}.
If the potential is constant, i.e., if $V (\phi) = V_0$,
space-time is de Sitter and expansion is exponential. If the
potential is exponential, i.e., $V (\phi) = V_0
\exp[-\lambda\phi]$, we get an inflationary powerlaw solution
\cite{Lucchin:1984yf,Burd:1988}. Several gravity theories consider
multiple scalar fields with exponential potential, particularly
assisted inflation scenarios
\cite{Chimento:1998,Guo:2003b,Coley:2000,Copeland:1999,Malik:1999},
quintom dark energy paradigm
\cite{Guo:2004fq,Zhang:2005eg,Lazkoz:2006pa,Lazkoz:2007mx} and
others. Also, have been considered positive and negative
exponential potentials \cite{Heard:2002}, single exponential and
double exponential
\cite{Huang:2006ku,Arias:2003,Gonzalez:2007,Gonzalez:2006}, etc.
Other generalizations with multiple scalar fields are available
\cite{vandenHoogen:2000cf,Damour:1992d}.

The dynamical behavior of space-times based on GRT is so far known
for a large variety of models with  scalar fields with
non-negative potential
\cite{Rendall:2004ic,Rendall:2006cq,Foster:1998sk,Miritzis:2003ym,Miritzis:2005hg,Hertog:2006rr}.
In reference \cite{Hertog:2006rr}, have been extended many of the
results obtained in \cite{Miritzis:2003ym} considering arbitrary
potentials. In \cite{Foster:1998sk} it has been shown that for a
large class of FRW cosmologies with scalar fields with arbitrary
potential, the past attractor is a family of solutions in
one-to-one correspondence with exactly integrable cosmologies with
a massless scalar field. This result has been extended somewhat in
\cite{Leon:2008de} to FRW cosmologies based on STTs. In this
reference was investigated a general model of coupled dark energy
with arbitrary potential $V$ and coupling function $\chi$.  It was
proved there, by using dynamical systems techniques that if the
potential and the coupling function are sufficiently smooth
functions; the scalar field almost always diverges into the past.
Under some regularity conditions for the potential and for the
coupling function in that limit, it was constructed a dynamical
system well suited to investigate the dynamics near the initial
singularity. The singular points therein were investigated and the
cosmological solutions associated to them were characterized.
There was presented asymptotic expansions for the cosmological
solutions near the initial space-time singularity, which extend
previous results of \cite{Foster:1998sk}. On the other hand, in
\cite{Giambo':2009cc} it was investigated flat and negatively
curved Friedmann-Robertson-Walker (FRW) models with a perfect
fluid matter source and a scalar field arising in the conformal
frame of $F(R)$ theories nonminimally coupled to matter. It was
proved there that, for a general class of potentials $V,$ the
equilibrium corresponding to non-negative local minima for $V$ are
asymptotically stable, as well as horizontal asymptotes approached
from above by $V$. For a nondegenerated minimum of the potential
with zero critical value they prove in detail that if $\gamma>1$,
then there is a transfer of energy from the fluid to the scalar
field and the later eventually dominates in a generic way. As we
will see in next sections the results in \cite{Giambo':2009cc} and
in \cite{Leon:2008de} can be obtained by investigating a general
class of models containing both STTs and $F(R)$ gravity.

\section{The Field Equations}

In this section we consider a phenomenological model inspired in
the action \eqref{eq1} for FRW space-times with flat spatial
slices, modelled by the metric:
\begin{align}d s^2 = -d t^2 +a(t)^2\left({d r^2}+r^2\left(d \theta^2+\sin^2\theta d \varphi^2\right)\right).
\label{flatFRW}
\end{align} We use a system of units in which $8\pi G=c=\hbar=1.$
We assume that the energy-momentum tensor \eqref{Tab} is in the
form of a perfect fluid
$$T^\alpha_\beta=\text{diag} \left(-\rho,p,p,p\right),$$ where $\rho$ and
$p$ are respectively the isotropic energy density and the
isotropic pressure (consistently with FRW metric, pressure is
necessarily isotropic \cite{Trodden:2004st}). For simplicity we
will assume a barotropic equation of state $p=(\gamma-1)\rho.$
Also we consider a quintessence scalar field, $\phi,$ interacting
in the action with the background pressureless dark matter fluid.
As in \cite{Gonzalez:2006cj}, here the baryons (a subdominant
component at present, but important in the past of the cosmic
evolution) are included in the background of dark matter. In fact,
there is the possibility of a universal coupling of dark energy to
all sorts of matter, including baryons (but excluding radiation)
\cite{Chimento:2003iea}. We include radiation in the cosmic budget
since it is an important matter source in the early universe.
However, in some cases of interest we set $\rho_r=0.$

The cosmological equations for flat FRW models with a scalar field
coupled to matter an including also radiation are given by

\ben \dot H &=& -\frac{1}{2}\left(\gamma
\rho
+\frac{4}{3}\rho_r+\dot\phi^2\right),\label{Raych}\\
\dot \rho&=& -3\gamma H \rho-\frac{1}{2}(4-3\gamma)\rho\dot\phi
\frac{\mathrm{d}\ln
\chi(\phi)}{\mathrm{d}\phi},\label{consm}\\
\dot \rho_r &=& -4 H \rho_r,\label{consr}\\
\ddot \phi&=& -3 H \dot \phi-\frac{\mathrm{d}
V(\phi)}{\mathrm{d}\phi}+\frac{1}{2}(4-3\gamma)\rho
\frac{\mathrm{d}\ln
\chi(\phi)}{\mathrm{d}\phi}\label{KG},\\
3H^2 &=& \frac{1}{2}\dot\phi^2+V(\phi)+\rho+\rho_r. \label{Fried}
\een where $a$ is the scale factor, $H=\dot a/a$ is the Hubble
parameter, $\rho$ denotes the energy density of barotropic matter,
$\rho_r$ denotes de energy density of radiation, $\phi$ denotes
the scalar field and $V(\phi)$ and $\chi(\phi)$ are, respectively,
the potential and coupling functions.

To maintain the analysis as general as possible, we will not
specify the functional forms of the potential and the coupling
function from the beginning. Instead we consider the general
hypothesis $V(\phi)\in C^3, V(\phi)>0, \chi(\phi)\in C^3$ and
$\chi(\phi)>0.$ We impose they in order to obtain dynamical
systems of class $C^2.$ However, to derive some of our results we
will relax some of this hypothesis, or consider further
assumptions (they will be clearly stated when applicable).  We
consider also $\rho\geq 0$ and $0<\gamma<2, \gamma\neq
\frac{4}{3}.$ These hypotheses for background matter are the
usual. We assume $\gamma\neq \frac{4}{3}$ to exclude the
possibility that the background matter behaves as radiation. The
energy momentum tensor for radiation ($\gamma=4/3$) is traceless,
so it is automatically decoupled from a scalar field non-minimally
coupled to dark matter in the Einstein frame. Radiation source is
added by hand in order to model also the cosmological epoch when
barotropic matter and radiation coexisted since we want to
investigate the possible scaling solutions in the radiation
regime. We neglect ordinary (uncoupled) barotropic matter.

The strength of the coupling between the perfect fluid and the
scalar field is defined by
$$\delta\equiv-\frac{1}{2}(4-3\gamma)\rho\dot\phi
\frac{\mathrm{d}\ln \chi(\phi)}{\mathrm{d}\phi}.$$ One of the
first papers to take seriously the possibility of interaction in
scalar field cosmologies, from the dynamical system perspective,
was \cite{Billyard:2000bh}. In this paper we can find a review on
the subject. There it was investigated the interaction terms (in
the flat FRW geometry) $\delta=-\alpha\dot\phi\rho$ and
$\delta=\alpha\rho H,$ where $\alpha$ is a constant, $\phi$ is the
scalar field, $\rho$ is the energy density of background matter
and $H$ stands for the Hubble parameter. The first choice
corresponds to an exponential coupling function $\chi(\phi)=\chi_0
\exp\left(2 \alpha \phi/(4-3\gamma)\right).$ The second case
corresponds to the choice $\chi=\chi_0 a^{-2\alpha/(4-3\gamma)}$
(and then, $\rho\propto a^{\alpha-3\gamma}$), where $a$ denotes
the scale factor of the Universe (recall that the former
derivations are only valid in a flat FRW model). Other
phenomenological coupling functions were studied elsewhere. We
want to draw the attention of the reader to a physically well
motivated approach to the coupling function in
\cite{Boehmer:2008av}. In that paper it was investigated a
coupling term of the form $\delta=-\alpha \rho,$ where $\alpha$ is
a constant ($\Gamma$ in their notation). As commented in that
reference, if $\alpha>0$, the model can describe either the decay
of dark matter into radiation, the decay of the curvaton field
into radiation or the decay of dark matter into dark energy (see
section III of \cite{Boehmer:2008av} for more information and for
useful references). In the reference \cite{Tsujikawa:2008uc}, the
authors construct a family of viable scalar-tensor models of dark
energy (which includes pure $f(R)$ theories and quintessence).
They consider a  coupling between the scalar field and the
non-relativistic matter in the Einstein frame of the type -in our
notation- $\chi(\phi)=e^{-2 Q \phi}$, with $Q$ constant. By
investigating a phase space the authors obtain that the model
posses a phase of late-time acceleration preceded by a standard
matter era, while at the same time satisfying the local gravity
constraints (LGC).  In fact, by studying the evolution of matter
density perturbations and employing them, the authors place bounds
on the coupling of the order $|Q|<2.5 \times 10^{-3}$ (for the
massless case). By a chameleon mechanism the authors show that
these models can be made compatible with LGC even when $|Q|$ is of
the order of unity if the scalar-field potential is chosen to have
a sufficiently large mass in the high-curvature regions.

In order to classify the global behavior of the solutions of
(\ref{eq1}) it is required a detailed knowledge of the global form
of the scalar field potential (and of the coupling function
$\chi$). However, up to the present, there exist no consensus
about the specific functional form of $V(\phi)$ (and of
$\chi(\phi)$) for make calculations. As a consequence it would be
of interest classify the dynamical behavior of solutions without
specifying the functional form of the potential function (and of
the coupling function). In the literature of General Relativity
(GR) relevant attempts have been made in this more general
direction
\cite{Leon:2008de,Rendall:2004ic,Rendall:2005if,Rendall:2005fv,Rendall:2006cq}.

In this investigation we want to study, from the dynamical systems
point of view, a phenomenological model inspired in a STT with
action (\ref{eq1}) where  the matter and the  (quintessence)
scalar field are coupled in the action (\ref{eq1}) through the
scalar tensor metric  $\chi(\phi)^{-1}g_{\alpha \beta}$
\cite{Kaloper:1997sh}. We consider arbitrary functional form for
self-interaction potential and the coupling function for the
scalar field, $\phi.$ When we take the conformal transformation
allowing writing the action in the JF as in (\ref{eq1JF}) the
coupling function $\chi$ should be interpreted as the dilation
(BD) field and the corresponding $\omega(\chi)$ as the varying BD
parameter.

The aim of the chapter is to extent several results in
\cite{Leon:2008de,Foster:1998sk,Miritzis:2003ym,Giambo':2009cc} to
the more realistic situation when radiation is included in the
cosmic budget (particularly for investigating the early-time
dynamics). We will focus mainly in a particular era of the
universe where matter and radiation coexisted. Otherwise the
inclusion of radiation complicates the study in an unnecessary
manner, since, assuming a perfect barotropic fluid with an
arbitrary barotropic index $\gamma$, for $\gamma= 4/3$, this
matter source corresponds to radiation. Thus we will consider both
ordinary matter described by a perfect fluid with equation of
state $p=(\gamma-1)\rho$ (coupled to a scalar field) and radiation
with energy density $\rho_r$. We are interested in investigate all
possible scaling solutions in this regime. Although we are mainly
interest in describing the early time dynamics of our model, for
completeness we will focus also in the late-time dynamics. As in
\cite{Giambo':2009cc} we obtain for flat FRW models sufficient
conditions under the potential, to establish the asymptotic
stability of the non-negative local minima for $V(\phi).$ Center
manifold theory is employed to analyze the stability solutions
associated to the local degenerated minimum and the inflection
points of the potential. We prove for arbitrary potentials and
arbitrary coupling functions $\chi(\phi),$ of appropriate
differentiable class, that the scalar field almost always diverges
into the past generalizing the results in
\cite{Foster:1998sk,Leon:2008de}). It is de signed a dynamical
system adequate to studying the stability of the singular points
in the limit $|\phi|\rightarrow\infty.$ We obtain there:
radiation-dominated cosmological solutions; power-law scalar-field
dominated inflationary cosmological solutions;
matter-kinetic-potential scaling solutions and
radiation-kinetic-potential scaling solutions. It is discussed, by
means of several worked examples, the link between our results and
the results obtained for specific $F(R)$ frameworks by using
appropriated conformal transformations. We illustrated both
analytically and numerically our principal results. Particularly,
we investigate the important examples of higher order gravity
theories $F(R) = R + \alpha R^2$ (quadratic gravity) and $F(R)
=R^n.$  In the case of quadratic gravity we prove, by an explicit
computation of the center manifold, that the singular point
corresponding to \emph{de Sitter} solution is locally unstable
(saddle point). This result complements the result of the
proposition discussed in \cite{Miritzis:2005hg} p. 5, where it was
proved the local asymptotic instability of the de Sitter universe
for positively curved FRW models with a perfect fluid matter
source and a scalar field which arises in the conformal frame of
the $R+\alpha R^2$ theory. Finally, we investigate a general class
of potentials containing the cases investigated in
\cite{Copeland:1997et,vandenHoogen:1999qq}. In order to provide a
numerical elaboration for our analytical results for this class of
models, we re-examine the model with power-law coupling and
Albrecht-Skordis potential $V(\phi )=e^{-\mu \phi }{\left( A+(\phi
-B)^2\right)}$ investigated in \cite{Leon:2008de} in presence of
radiation. Also, we investigate in detail the invariant set
$\rho_r=0$ obtaining several results for the early- and late-time
universe. In particular we formalize and prove two singularity
theorems.

\section{Late Time Behavior}\label{Futurequalitative}

In the following, we study the late time behavior of solutions of
\eqref{Raych}, \eqref{consm}, \eqref{consr}, \eqref{KG}, which are
expanding at some initial time, i. e., $H(0)>0.$ The state vector
of the system is $\left(\phi, \dot\phi, \rho, \rho_r, H\right).$
Defining $y:=\dot\phi,$ we rewrite the autonomous system as \ben
\dot\phi&=&y,\label{dotphi}\\
\dot y&=& -3 H y-\frac{\mathrm{d}
V(\phi)}{\mathrm{d}\phi}+\frac{1}{2}(4-3\gamma)\rho
\frac{\mathrm{d}\ln
\chi(\phi)}{\mathrm{d}\phi}\label{newKG},\\
\dot \rho&=& -3\gamma H \rho-\frac{1}{2}(4-3\gamma)\rho y
\frac{\mathrm{d}\ln
\chi(\phi)}{\mathrm{d}\phi},\label{newconsm}\\
\dot \rho_r &=& -4 H \rho_r,\label{newconsr}\\
\dot H &=& -\frac{1}{2}\left(\gamma \rho
+\frac{4}{3}\rho_r+y^2\right),\label{newRaych} \een subject to the
constraint \be 3H^2 = \frac{1}{2}y^2+V(\phi)+\rho+\rho_r.
\label{newFried}\ee

\begin{rem}\label{decreasing}
Using standard arguments of ordinary differential equations
theory, follows from equations \eqref{newconsm} and \eqref{newconsr}
that the signs of $\rho$ and $\rho_r,$
respectively, are invariant. This means that if $\rho>0$ and $\rho_r>0$
for some initial time $t_0,$ then $\rho(t)>0,$ and $\rho_r(t)>0$ throughout the solution.
From \eqref{newRaych} and \eqref{newFried} and only if additional conditions are
assumed, for example $V(\phi)\geq 0$ and $V(\phi_*)= 0$ for some $\phi_*$, follows that the sign of $H$ is invariant.
From \eqref{newconsr} and \eqref{newRaych} follows that $\rho_r$ and
$H$ decreases. Also, defining $\epsilon=\frac{1}{2} y^2+V(\phi),$
follows from \eqref{newKG}-\eqref{newconsm} that
\be\dot\epsilon+\dot\rho=-3 H(y^2+\gamma \rho)\label{decr}.\ee
Thus, the total energy density contained in the dark sector is
decreasing.
\end{rem}

First we study the simpler case were $\rho_r=0,$ then, we
investigate the general case.

For $\rho_r=0,$ equations \eqref{dotphi}-\eqref{newRaych} becomes

\ben \dot H &=& -\frac{1}{2}\left(\gamma \rho
+y^2\right),\label{eq3}\\
\dot \rho&=& -3\gamma H \rho-\frac{1}{2}(4-3\gamma)\rho y
\frac{\mathrm{d}\ln
\chi(\phi)}{\mathrm{d}\phi},\label{eq4}\\
\dot y&=& -3 H y-\frac{\mathrm{d}
V(\phi)}{\mathrm{d}\phi}+\frac{1}{2}(4-3\gamma)\rho
\frac{\mathrm{d}\ln
\chi(\phi)}{\mathrm{d}\phi}\label{eq5},\\
\dot\phi &=& y, \label{eq6} \een defining a dynamical system in
the phase space

\begin{equation}\Omega=\{\left(H,\rho,y,\phi\right)\in\mathbb{R}^4|3H^2 =
\frac{1}{2}y^2+V(\phi)+\rho\}\label{eq44}.\end{equation}

First, let us consider a potential function with a local minimum
$V(0)=0.$ With this hypothesis the point $(0,0,0,0)$ is a singular
point of \eqref{eq3}-\eqref{eq6}. This fact can be used to show
that an initially expanding universe ($H>0$) should expand
forever. Indeed, the set  $\left\{(H, \rho, y, \phi)\in\Omega| H=0
\right\}$ is invariant under the flow of \eqref{eq3}-\eqref{eq6} .
Besides, the sign of $H$ is invariant. If the sign of $H$ changes,
a trajectory with $H(0)>0$ can passing through $(0,0,0,0),$
violating the existence an uniqueness theorem \ref{existencethm}.

The proposition 2 of \cite{Miritzis:2003ym} can be generalized to
this context as follows.

\begin{prop}\label{Proposition I} Suppose that $V\geq 0$ and
$V(\phi)=0\Leftrightarrow \phi=0.$ Let $A$ such that $V$ bounded
in $A$ implies $ V'(\phi)$ is bounded in $A.$ If there exists a
constant $K,$ $K\neq 0$ such that
$$\chi'(\phi)/\chi(\phi)\leq 2 K/(2-\gamma)(4-3\gamma).$$
Then, $$\lim_{t\rightarrow\infty}
\rho=0=\lim_{t\rightarrow\infty}y.$$
\end{prop}

{\bf Proof}. Consider the trajectory passing through an arbitrary
point $(H, \rho, y, \phi)\in\Omega $ with $H>0$ at $t=t_0.$ Since
$H$ is positive and decreasing we have that $\lim_{t\rightarrow
\infty} H(t)$ exists and it is a nonnegative number $\eta$;
besides, $H(t)\leq H(t_0)$ for all $t\geq t_0.$ The, from the
restriction \eqref{eq44} follows that each term $\rho,$ $1/2 y^2,$
and $V(\phi)$ is bounded by $3H(t_0)^2$ for all $t\geq t_0.$

Let $A=\left\{\phi: V(\phi)\leq 3H(t_0)^2\right\}.$ Then, the
trajectory is such that $\phi$ remains in the interior of $A.$

From equation \eqref{eq3} we have that
$$-\int_{t_0}^{t}\left(\frac{1}{2}y^2+\frac{\gamma}{2}\rho\right)\mathrm{d}
t = H(t)-H(t_0)$$ and taking the limit as $t\rightarrow\infty,$ we
obtain
$$\frac{1}{2}\int_{t_0}^{\infty}\left(y^2+\gamma\rho\right) d t = H(t_0)- \eta$$ besides,
\begin{equation}
\int_{t_0}^{\infty}\left(y^2+\gamma\rho\right) \mathrm{d} t
<\infty.\label{eq10}
\end{equation}

Taking the time derivative of $f(t)= y^2+\gamma\rho$ and making
use of the hypothesis for $\chi(\phi)$ we obtain
\begin{align}&\frac{d}{d t}\left(y^2+\gamma\rho\right)\leq
y\left(-2 V'(\phi) +K\rho\right).\nonumber\end{align} As we have
seen, $y$ and $\rho$ are bounded, and by the hypothesis for
$V(\phi)$, $ V'(\phi)$ is bounded. From this facts follow that the
time derivative of $f$ is bounded. Since $f$ is a nonnegative
function, the convergence of $\int_{t_0}^{\infty} f(t) \mathrm{d}
t$ implies $\lim_{t\rightarrow\infty}f(t)=0.$ Hence, we have that
$$\lim_{t\rightarrow\infty} \rho=0=
\lim_{t\rightarrow\infty}y.$$ $\blacksquare$

The hypotheses in \ref{Proposition I} are satisfied by a large
class of potentials as commented in \cite{Miritzis:2003ym} (this
result is extensible to the case of non-minimal coupling), an by a
large class of coupling functions including functions dominated by
exponential ones.

Under the same hypothesis of proposition \ref{Proposition I}, we
can generalize the proposition 3 in \cite{Miritzis:2003ym}.
\begin{prop}\label{Proposition II} Suppose that $V'(\phi)>0$ for
$\phi>0$ and $V'(\phi)<0$ for $\phi<0.$ Then, under the same
hypotheses as in proposition \ref{Proposition I},
$\lim_{t\rightarrow\infty}\phi$ exists and is equal to $+\infty$,
$0$ or $-\infty.$
\end{prop}

{\bf Proof}. Using the same argument as in proposition
\ref{Proposition I}, $\exists \lim_{t\rightarrow\infty}
H(t)=\eta.$ If $\eta=0,$ then by the restriction \eqref{eq44} we
obtain $\lim_{t\rightarrow\infty} V(\phi(t))=0.$ Since $V$ is
continuous and $V(\phi)=0\Leftrightarrow \phi=0$ this implies that
$\lim_{t\rightarrow\infty} \phi(t)=0.$

Suppose that $\eta>0.$ From \eqref{eq44} we obtain that
$\lim_{t\rightarrow\infty} V(\phi(t))=3\eta^2.$ Therefore, exists
$t'$ such that $V(\phi)>3\eta^2/2$ for all $t>t'.$ Form this fact
follows that $\phi$ cannot be zero for some $t>t'$ because $\phi=0
\Leftrightarrow V(\phi)=0.$ Then, the sign of $\phi$ is invariant
for all $t>t'.$

Suppose that $\phi$ is positive for all $t>t'.$ Since $V$ is an
increasing function of $\phi$ in $(0,+\infty),$ we have that
$\lim_{t\rightarrow\infty} V(\phi(t))=3\eta^2\leq
\lim_{\phi\rightarrow\infty} V(\phi).$  By the continuity and
monotony of $V$ it is obvious that the equality holds if, and only
if, $\lim_{t\rightarrow\infty} \phi(t)=+\infty.$

If $\lim_{t\rightarrow\infty} V(\phi(t))<
\lim_{\phi\rightarrow\infty} V(\phi),$ then there exists
$\bar{\phi}\geq 0$ such that $$\lim_{t\rightarrow\infty}
V(\phi(t))=V(\bar{\phi}).$$ Since $V$ is continuous and strictly
increasing we have that
$$\lim_{t\rightarrow\infty} \phi=\bar{\phi}.$$

By proposition \ref{Proposition I}, $\lim_{t\rightarrow\infty}
\rho(t)=0=\lim_{t\rightarrow\infty}y(t)$. Besides, $H$ and
$\chi'(\phi)/\chi(\phi)$ are bounded. Therefore, taking the limit
as $t\rightarrow\infty$ in \eqref{eq5} we find that
$$\lim_{t\rightarrow\infty}\frac{d}{d
t}y=-V'(\bar{\phi})<0.$$ Hence, there exists $t''> t'$ such that
$\frac{d}{d t}y<-V'(\bar{\phi})/2$ for all $t\geq t''.$ This
implies $$y(t)- y(t'')=\int_{t''}^{t}\left(\frac{d}{d t}y\right)
dt<-\frac{ V'(\bar{\phi})}{2}(t- t''),$$ that is, $y(t)$ takes
negative values with arbitrary large modulus as $t$ increases,
which is not possible since $\lim_{t\rightarrow\infty}y(t)=0.$

Hence, if $\phi>0$ for all $t>t',$ we have that
$\lim_{t\rightarrow\infty}\phi=+\infty.$ Similarly, when $\phi<0$
for all $t>t',$ we have $\lim_{t\rightarrow\infty}\phi=-\infty.$
$\blacksquare$

From this we conclude that, if initially
$3H(t_0)^2<\min\left\{\lim_{\phi\rightarrow\infty}V(\phi),
\lim_{\phi\rightarrow -\infty}V(\phi)\right\},$  then,
$\lim_{t\rightarrow\infty}H(t)=0.$ Indeed, we have that
$\lim_{t\rightarrow\infty}\phi$ is equal to $+\infty$, $0$ or
$-\infty.$ If $\lim_{t\rightarrow\infty}\phi=+\infty$, then from
the restriction \eqref{eq44}, follows
$$3 \eta^2=\lim_{t\rightarrow\infty}V(\phi(t))=
\lim_{\phi\rightarrow\infty} V(\phi)>3 H(t_0)^2.$$ This is
impossible since $H(t)$ is a decreasing function  and $H(t_0)\geq
\eta.$ In the same way, $\lim_{t\rightarrow\infty}\phi=-\infty$
leads to a contradiction. Then, $\lim_{t\rightarrow\infty}\phi=0$
and this implies $\lim_{t\rightarrow\infty}V(\phi(t))=0,$  and
again by \eqref{eq44}, $\lim_{t\rightarrow\infty}H(t)=0.$

The interpretation of these results is clear.

If the potential has a local minimum at zero, if the derivative of
the potential is bounded in the same set where the potential
itself is, and provided the derivative of the logarithm of the
coupling function is bounded by above, then, de energy density of
DM and the kinetic energy density of DE tends to zero as the time
goes forward. In this case the energy density of the Unverse will
be dominated by the potential energy of DE. Hence, the Universe
would be expand forever in a de Sitter phase.

With the above hypotheses and with the additional assumption of
$V(\phi)$ being strictly decreasing (increasing) if $\phi<0$
($\phi>0$) it is proved (in a similar way as in Proposition 3 in
\cite{Miritzis:2003ym}) that the scalar field can be either zero
or divergent into the future (the former case holds if the Hubble
scalar vanishes asymptotically).

In order to complement the former ideas, we must remark that if
the potential is non negative (and with no necessarily a local
minimum at $(0,0)$) having continuous derivative (bounded in the
same set where the potential itself is); and assuming that the
potential $V(\phi(t)),$ strictly decreasing as a function of $t$,
unbounded when $t\rightarrow\infty.$ Then, the cosmological model
enters a de Sitter expansion, characterized by divergences of the
scalar field into the future. If additionally the potential as a
function of $\phi$ vanishes asymptotically into the future, the
Hubble scalar vanishes too. This fact is true for exponential
potentials.

\begin{prop}\label{Proposition III}
Suppose that there exists a nonzero constant $K,$ such that
$\chi'(\phi)/\chi(\phi)\leq 2 K/(2-\gamma)(4-3\gamma).$ Let  $V$
be a potential function with the properties:
 \begin{enumerate}
    \item $V\geq 0$ and $\lim_{t\rightarrow\infty} V(\phi)=+\infty.$
    \item $V'$ is continuous and $V'(\phi)<0.$
    \item If  $A\subset\mathbb{R}$ is such that $V$ is bounded in $A,$ Then, $ V'(\phi)$ is bounded in $A.$
 \end{enumerate}
Then, $\lim_{t\rightarrow\infty} \rho=0=
\lim_{t\rightarrow\infty}y,$ and
$\lim_{t\rightarrow\infty}\phi=+\infty.$
\end{prop}

{\bf Proof}. From equation \eqref{eq4}, the set $\rho>0$ is
invariant under the flow of \eqref{eq3}-\eqref{eq6} with
restriction \eqref{eq44}; besides $\rho$ is different from zero if
$\rho(t_0)$ is different from zero at the initial time. From this
fact we have that $H$ is never zero (do not changes of sign) since
by \eqref{eq44}, $3 H(t)^2\geq \rho(t)>0$ for all $t>t_0,$ then,
$H$ is always nonnegative if initially is nonnegative. Besides, by
equation \eqref{eq3}, follows that $H$ is decreasing, then
$\exists \lim_{t\rightarrow\infty} H(t)=\eta\geq 0$ and
$$\frac{1}{2}\int_{t_0}^{\infty}\left(y^2+\gamma\rho\right)
d t = H(t_0)-\eta<+\infty.$$ As in proposition \ref{Proposition
I}, the total time derivative of $y^2+\gamma\rho $ is bounded.
Hence $\lim_{t\rightarrow\infty}
\rho=0=\lim_{t\rightarrow\infty}y.$

It can be proved that $\lim_{t\rightarrow\infty}\phi=+\infty$ in
the same way as proved in \ref{Proposition II}.

From equation \eqref{eq44} we have that $\lim_{t\rightarrow\infty}
V(\phi)=3\eta^2.$ Since $V$ is strictly decreasing with respect to
$\phi;$ then $V(\phi)>\lim_{\phi\rightarrow\infty} V(\phi)$ for
all $\phi,$ therefore $\lim_{t\rightarrow\infty} V(\phi(t))\geq
\lim_{\phi\rightarrow\infty} V(\phi).$ We will consider two cases:
 \begin{enumerate}
   \item If $\lim_{t\rightarrow\infty} V(\phi(t))= \lim_{\phi\rightarrow\infty} V(\phi),$
   by the continuity of  $V$ is obvious that
$\lim_{t\rightarrow\infty}\phi=+\infty;$
   \item If  $ \lim_{t\rightarrow\infty} V(\phi(t))> \lim_{\phi\rightarrow\infty} V(\phi),$
   then, there exists a unique $\bar{\phi}$ such that
   $$\lim_{t\rightarrow\infty} V(\phi(t))=V(\bar{\phi}).$$ Since  $V$ is continuous and
   strictly decreasing follows that $$\lim_{t\rightarrow\infty} \phi=\bar{\phi}.$$
From equation \eqref{eq5} follows that
$$\lim_{t\rightarrow\infty}\frac{d}{d
t}y=-V'(\bar{\phi})>0,$$ therefore, exists $t'$ such that
$\frac{d}{d t}y>-V'(\bar{\phi})/2$ for all $ t\geq t'.$ Form this
fact we conclude that
$$y(t)- y(t')
>-\frac{V'(\bar{\phi})}{2}(t- t'),$$ which is impossible since
$\lim_{t\rightarrow\infty}y(t)=0.$ Finally
$\lim_{t\rightarrow\infty}\phi=+\infty.$
 \end{enumerate} $\blacksquare$

If additionally, the potential is such that
$\lim_{\phi\rightarrow\infty} V(\phi)=0,$ then we conclude that
$H\rightarrow 0$ as $t \rightarrow\infty.$

Now, let us consider the general case by including radiation. Our
purpose is to formulated a proposition that extent in some way (we
are considering only flat FRW models) the proposition 1 of
\cite{Giambo':2009cc}, which gives a characterization of the
future attractor of the system \eqref{Raych}, \eqref{consm},
\eqref{consr}, \eqref{KG}.

Let us formalize notion of degenerate local minimum introduced in
\cite{Giambo':2009cc}:

\begin{defn}
The function $V(\phi)$ is said to have a degenerate local minimum
at $\phi_\star$ if  $$V'(\phi),V''(\phi), \dots V^{(2n-1)}$$
vanish at $\phi_*,$ and $V^{(2n)}(\phi_*)>0,$ for some integer
$n.$
\end{defn}

\begin{prop}\label{thmIII} Suppose that $V(\phi)\in C^2(\mathbb{R})$ satisfies the following conditions \footnote{See assumptions 1 in
\cite{Giambo':2009cc}.}
\begin{itemize}
\item[(i)] The (possibly empty) set $\{\phi: V(\phi)<0\}$ is
bounded; \item[(ii)] The (possibly empty) set of singular points
of $V(\phi)$ is finite.
\end{itemize}
Let $\phi_*$ a strict local minimum for $V(\phi),$ possibly
degenerate, with non-negative critical value. Then ${\bf
p}_*:=\left(\phi_*,y_*=0,\rho_*=0,\rho_r=0,
H=\sqrt{\frac{V(\phi_*)}{3}}\right)$ is an asymptotically stable
singular point for the flow of \eqref{dotphi}-\eqref{newRaych}.
\end{prop}

{\bf Proof.}

We adapt the demonstration in \cite{Giambo':2009cc} (for flat FRW
cosmologies) to the case where radiation is considered.
\footnote{From physical considerations we can neglect radiation
for the analysis of the future attractor, but we prefer to offer
the complete {\bf Proof}.}

First let us consider the case $V(\phi_*)>0.$ Let
$\tilde{V}>V(\phi_*)$ be a regular value for $V$ such that the
connected component of $V^{-1}\left((-\infty,\tilde{V}]\right)$
that contains $\phi_*$ is a compact set in $\mathbb{R}.$ Let us
denote this set by $A$ and define $\Psi$ as
$$\Psi=\left\{(\phi,y,\rho,\rho_r,H): \phi\in A,
\epsilon+\rho\leq \tilde{V}, \rho\geq 0, \rho_r\in
\left[0,\tilde{W}\right]\right\},$$ where $\tilde{W}$ is a
positive constant. We can show that $\Psi$ is a compact set as
follows.

\begin{itemize}
\item[(i)] $\Psi$ is a closed set in $\mathbb{R}^5;$
\item[(ii)] $V(\phi_*)\leq V(\phi)\leq \tilde{V},\, \forall \phi\in A;$
\item[(iii)] $\frac{1}{2}y^2+V(\phi_*)\leq \frac{1}{2}y^2+V(\phi)+\rho\leq \tilde{V}$, and therefore $y$ is bounded;
\item[(iv)] $\rho\leq \tilde{V}-\frac{1}{2}y^2-V(\phi)\leq \tilde{V}-V(\phi_*)$ and then $\rho$ is bounded;
\item[(v)] Finally, from \eqref{newFried}, $\frac{V(\phi_*)}{3}\leq H^2\leq \tilde{V}+\tilde{W}.$
\end{itemize}

Let $\Psi_+\subseteq \Psi$ be the connected component of $\Psi$
containing ${\bf p}_*.$ Following similar arguments as in
\cite{Giambo':2009cc} it can be proved that $\Psi_+$ is positively
invariant with respect to  \eqref{dotphi}-\eqref{newRaych}, i.e,
all the solutions with initial data at $\Psi_+$ remains at
$\Psi_+$ for all $t>0.$ Indeed, let $\mathbf{x}(t)$ be such a
solution and $$\bar{t}=\sup\left\{t>0:
H(t)>0\right\}\in\mathbb{R}\cup\{+\infty\}.$$ When $t<\bar{t},$
equation \eqref{decr} imply that $\epsilon+\rho$ decreases
monotonically (cf remark \ref{decreasing}). Moreover, it can be
proved by contradiction that \be\phi(t)\in A\, \forall
t<\bar{t}\label{phiinA},\ee otherwise there would exists some
$t<\bar{t}$ such that $V(\phi(t))>\tilde{V},$ but then
$$\tilde{V}<V(\phi(t))\leq \frac{1}{2}y(t)^2+V(\phi(t))+\rho(t)\leq \tilde{V},$$
a contradiction. Thus, \eqref{phiinA} holds. But since $\rho_r\geq
0$ along the flow (cf remark \ref{decreasing}), it follows that
$$H(t)^2\geq \frac{1}{3}\left(\frac{1}{2}y(t)^2+V(\phi(t))+\rho(t)\right)\geq \frac{V(\phi(t))}{3}\geq \frac{V(\phi_*)}{3}>0.$$

We have proved that as long as $H$ remains positive, it is
strictly bounded away from zero; thus $\bar{t}=+\infty,$ and from
this can be deduced that $\mathbf{x}(t)$ remains in $\Psi_+$ for
all $t>0.$

From all the above $\Psi_+$ satisfies the hypothesis of LaSalle's
invariance theorem (see \cite{LaSalle}; theorem 8.3.1 in
\cite{wiggins}, p. 111). If we consider the monotonic decreasing
functions $\epsilon+\rho$ and $\rho_r$ defined in $\Psi_+$ then
follows that, every solution with initial data at $\Psi_+$ must be
such that $H\left(y^2+\gamma\rho\right)\rightarrow 0$ and
$H\rho_r\rightarrow 0$ as $t\rightarrow +\infty.$ Since $H$ is
strictly bounded away from zero in $\Psi_+$ follows that
$(y,\rho,\rho_r)\rightarrow (0,0,0)$ and
$H^2-\frac{V(\phi)}{3}\rightarrow 0$ as $t\rightarrow +\infty.$

Since $H$ is monotone decreasing (cf remark \ref{decreasing}) and
it is bounded away from zero it must have a limit. This means that
$V(\phi)$ also admits a limit. This limit has to be $V(\phi_*);$
otherwise $V'(\phi)$ would tend to a positive value and so would
the righthand side of \eqref{newKG}, a contradiction. Therefore
the solution approaches the singular point ${\bf p}_*$.

If $V(\phi_*)=0$, the above argument can be easily adapted. In
this case the set $\Omega$ is connected and we choose $\Omega_{+}$
to be its subset characterized by the property $H\geq0$. The only
point in $\Omega_{+}$ with $H=0$ is exactly the singular point
${\bf p}_*$, and so if $H(t)\rightarrow 0$ the solution is forced
to approach the equilibrium since $H$ is monotone; if by
contradiction $H(t)$ had a strictly positive limit, we could argue
as before to find $y\rightarrow0,\,\rho\rightarrow0$ and
$\rho_r\rightarrow0$ and so $H$ must necessarily converge to zero.
$\blacksquare$

\section{Early time behavior in the invariant set $\rho_r=0$}

In order to analyze the initial singularity (and also, the late
time behavior) it is convenient to normalize the variables, since
in the vicinity of an hypothetical initial singularity, the
physical variables would typically diverge, whereas at late times
they commonly vanish \cite{Wainwright:2004cd}. In this section we
rewrite equations \eqref{Raych}, \eqref{consm}, \eqref{consr},
\eqref{KG} as an autonomous system defined on a state space by
introducing Hubble-normalized variables. These variables satisfy
an inequality arising from the Friedmann equation \eqref{Fried}.
We analyze the cosmological model by investigating the flow of the
autonomous system in a phase space by using dynamical systems
tools.

Let us introduce the following normalized variables

\be
x_1=\frac{\dot\phi}{\sqrt{6}H},\,x_2=\frac{\sqrt{\rho}}{\sqrt{3}
H},\,x_3=\frac{1}{H}\label{0vars}\ee and the time coordinate \be
d\tau=3 H dt.\ee
Using the field equations \eqref{eq3}-\eqref{eq6} we
get that the variables (\ref{0vars}) and the scalar field $\phi$
evolve with respect $\tau$ as

\begin{align}
& \phi'=\sqrt{\frac{2}{3}}x_1 \label{T2.9},\\
      &x_1'=x_1^3+\frac{1}{2} \left(x_2^2 \gamma -2\right) x_1-\frac{x_3^2}{3 \sqrt{6}}
      \frac{\mathrm{d}V(\phi)}{\mathrm{d}\phi}+
      \frac{(4-3\gamma)x_2^2 }{2\sqrt{6}}\frac{\mathrm{d}\ln \chi (\phi )}{\mathrm{d}\phi},\label{T2.10}\\
      &x_2'=\frac{1}{2} x_2 \left(2 x_1^2+\left(x_2^2-1\right) \gamma \right)
      -\frac{(4-3\gamma)x_1 x_2}{2\sqrt{6}}\frac{\mathrm{d}\ln \chi (\phi )}{\mathrm{d}\phi},\label{T2.11}\\
  &x_3'=\frac{1}{2} x_3 \left(2 x_1^2+x_2^2 \gamma \right),\label{T2.12}
\end{align}
where the prime denotes derivative with respect $\tau.$ This is an
autonomous system where de variables are subject to the constraint
\begin{eqnarray} x_1^2+x_2^2+\frac{1}{3} x_3^2
V(\phi)=1.\label{eq22}\end{eqnarray}

From the hypotheses $V\in C^3, V(\phi)>0, \chi\in C^3,
\chi(\phi)>0, 0<\gamma<2, \gamma\neq \frac{4}{3}$ follows that
\eqref{T2.9}-\eqref{T2.12} defines a dynamical system of class
$C^2$ in $\mathbb{R}^4.$

\begin{prop}\label{prop2.1}
The sets defined by
\begin{equation}
\Sigma_T:=\left\{\mathbf{p}=(\phi,x_1,x_2,x_3)\in\mathbb{R}^4|
x_1^2+x_2^2+\frac{1}{3} x_3^2 V(\phi)=1 \right\},
\end{equation}
\begin{equation}\label{Vjk}
V_{j k}:=\left\{\mathbf{p}\in\mathbb{R}^4| (-1)^k x_j>0
\right\}\cap \Sigma_T,
\end{equation} and
\begin{equation}\label{Uj}
U_{j}:=\left\{\mathbf{p}\in\mathbb{R}^4| x_j=0 \right\}\cap
\Sigma_T,
\end{equation} with $j=2,3; k=1,2;$ are invariant sets for the
flow of \eqref{T2.9}-\eqref{T2.12} defined in $\mathbb{R}^4.$
\end{prop}

{\bf Proof}. To prove these results we use the proposition
\ref{Proposition 4.1}.

\begin{enumerate}
\item To prove the invariance of $\Sigma_T$ we define \be Z:
\mathbb{R}^4\rightarrow \mathbb{R},\, \mathbf{p}\rightarrow
x_1^2+x_2^2+\frac{1}{3} x_3^2 V(\phi)-1\label{3.29}\ee and \be
\alpha: \mathbb{R}^4\rightarrow \mathbb{R},\,
\mathbf{p}\rightarrow 2 x_1^2+x_2^2\gamma.\ee Observe that $Z$ and
$\alpha$ satisfy the hypothesis of proposition \ref{Proposition
4.1}. Thus $\Sigma_T=\left\{\mathbf{p}\in\mathbb{R}^4|
Z(\mathbf{p})=0\right\}$ is invariant.

\item To prove the invariance of $V_{2 k},\, k=1,2$ and $U_2$ we
define \be Z: \Sigma_T\rightarrow \mathbb{R},\,
\mathbf{p}\rightarrow x_2\ee and \be \alpha: \Sigma_T\rightarrow
\mathbb{R},\, \mathbf{p}\rightarrow \frac{1}{2} \left(2
x_1^2+\left(x_2^2-1\right) \gamma \right)
      -\frac{(4-3\gamma)x_1 }{2\sqrt{6}}\frac{\mathrm{d}\ln \chi (\phi )}{\mathrm{d}\phi}.\ee Observe that $Z$ and $\alpha$ satisfy the
hypothesis of proposition \ref{Proposition 4.1}. Thus $V_{2
1}=\left\{\mathbf{p}\in\Sigma_T| Z(\mathbf{p})<0\right\},$ $V_{2
2}=\left\{\mathbf{p}\in\Sigma_T| Z(\mathbf{p})>0\right\}$ and
$U_2=\left\{\mathbf{p}\in\Sigma_T| Z(\mathbf{p})=0\right\}$ are
invariant.

\item To prove the invariance of $V_{3 k},\, k=1,2$ and $U_3$ we
define \be Z: \Sigma_T\rightarrow \mathbb{R},\,
\mathbf{p}\rightarrow x_3\ee and \be \alpha: \Sigma_T\rightarrow
\mathbb{R},\, \mathbf{p}\rightarrow \frac{1}{2} \left(2
x_1^2+x_2^2\gamma\right).\ee Observe that $Z$ and $\alpha$ satisfy
the hypothesis of proposition \ref{Proposition 4.1}. Thus $V_{3
1}=\left\{\mathbf{p}\in\Sigma_T| Z(\mathbf{p})<0\right\},$ $V_{3
2}=\left\{\mathbf{p}\in\Sigma_T| Z(\mathbf{p})>0\right\}$ and
$U_3=\left\{\mathbf{p}\in\Sigma_T| Z(\mathbf{p})=0\right\}$ are
invariant.
\end{enumerate} $\blacksquare$

According to proposition \ref{prop2.1}, the invariant set
$\Sigma_T$ acts as an independent dynamical object and thus will
be suffice to consider the autonomous system
\eqref{T2.9}-\eqref{T2.12} defined in $\Sigma_T.$ Observe that the
sets $V_{j k},\, U_j,\, j=2,3; k=1,2;$ are invariant sets for the
flow of \eqref{T2.9}-\eqref{T2.12} defined in $\Sigma_T.$

In the following we consider the {\bf notations:} $\|.\|$ denotes
the Euclidean vector norm;
$\mathbb{D}^n:=\left\{\mathbf{x}\in\mathbb{R}^n:
\|\mathbf{x}\|<1\right\}$ denotes the n-dimensional unitary disc.

\subsection{The Topological Properties of the Phase Space}

In the following we discuss the topological properties of
$\Sigma_T$ resulting the

\begin{prop}\label{Prop8}
$\Sigma_T$ is a topological manifold (without boundary).
\end{prop}

{\bf Proof}. First, let us prove that $\Sigma_T$ has dimension
less than $4.$ For this purpose it is sufficient to prove that
$\Sigma_T$ is not an open set with respect to the usual topology
of $\mathbb{R}^4.$ The mapping \eqref{3.29} is continuous for all
$\mathbf{p}\in\mathbb{R}^4.$ Let $\mathbf{p}_0$ an arbitrary point
of $\mathbb{R}^4\setminus \Sigma_T.$ Then, $Z(\mathbf{p}_0)=c\neq
0$ where $c$ is a constant. Since $Z$ is continuous and it is
defined for all $\mathbf{p},$ follows that there exists a real
number $\delta>0$ such that for all $\mathbf{p}$ of
$S_\delta(\mathbf{p}_0)=\left\{\mathbf{p}\in\mathbb{R}^4:
\|\mathbf{p}-\mathbf{p}_0\|<\delta\right\},$ we have
$\|Z(\mathbf{p})-Z(\mathbf{p}_0)\|<\frac{1}{2}|c|.$ Then,
$\|Z(\mathbf{p})-c\|<\frac{1}{2}|c|.$ From this follows that
$Z(\mathbf{p})\neq 0$ for all $\mathbf{p}\in
S_\delta(\mathbf{p}_0),$ where $S_\delta(\mathbf{p}_0)$ is an open
set of $\mathbb{R}^4$ contained in $\mathbb{R}^4\setminus
\Sigma_T.$ We conclude that $\mathbb{R}^4\setminus \Sigma_T$ is an
open set of $\mathbb{R}^4;$ thus, $\Sigma_T$ is a closed subset of
$\mathbb{R}^4.$ Second, since $\Sigma_T$ is a closed set with
respect the usual topology of $\mathbb{R}^4$; it is a Hausdorff
space equipped with a numerable basis. The rest of the proof
requires the construction of a set of local charts.

The sets $V_{j k}, j=1,2,3$ defined by the formula \eqref{Vjk} are
open sets of $\Sigma_T$ with respect to the induced topology in
$\Sigma_T$ by the usual topology of $\mathbb{R}^4,$ which cover
$\Sigma_T.$

Let us define the maps: \begin{align} & h_{1 k}: V_{1
k}\rightarrow\mathbb{R}^3, \nonumber \\& \mathbf{p}\rightarrow
h_{1 k}(\mathbf{p})=\left(\phi, x_2, \sqrt{\frac{V(\phi)}{3}}
x_3\right)=\left(\xi_1, \xi_2,\xi_3\right),\;k=1,2, \end{align}
which satisfy $h_{1 k}(V_{1 k})=\mathbb{R}\times \mathbb{D}^2$
with inverse given by
\begin{align} &  h_{1 k}^{-1}:
\mathbb{R}\times \mathbb{D}^2 \rightarrow V_{1 k}, \nonumber\\ &
\xi\rightarrow h_{1
k}^{-1}(\xi)=\left(\xi_1,(-1)^k\sqrt{1-\xi_2^2-\xi_3^2}, \xi_2,
\sqrt{\frac{3}{V(\xi_1)}} \xi_3\right),\;k=1,2;\end{align}
\begin{align} h_{2 k}: V_{2 k}\rightarrow\mathbb{R}^3,\nonumber\\
&\mathbf{p}\rightarrow h_{1 k}(\mathbf{p})=\left(\phi, x_1,
\sqrt{\frac{V(\phi)}{3}} x_3\right)=\left(\xi_1,
\xi_2,\xi_3\right),\;k=1,2;\end{align} which satisfy $h_{1 k}(V_{1
k})=\mathbb{R}\times \mathbb{D}^2,$ with inverse given by
\begin{align} & h_{2 k}^{-1}: \mathbb{R}\times \mathbb{D}^2 \rightarrow V_{2
k},\nonumber \\ & \xi\rightarrow h_{2
k}^{-1}(\xi)=\left(\xi_1,\xi_2,(-1)^k\sqrt{1-\xi_2^2-\xi_3^2},
\sqrt{\frac{3}{V(\xi_1)}} \xi_3\right),\; k=1,2;\end{align} and
\begin{align} & h_{3 k}: V_{3
k}\rightarrow\mathbb{R}^3,\nonumber\\ & \mathbf{p}\rightarrow h_{3
k}(\mathbf{p})=\left(\phi, x_1, x_2\right)=\left(\xi_1,
\xi_2,\xi_3\right),\; k=1,2\end{align} which satisfy $h_{1 k}(V_{1
k})=\mathbb{R}\times \mathbb{D}^2,$ with inverse given by
\begin{align} & h_{3 k}^{-1}: \mathbb{R}\times \mathbb{D}^2 \rightarrow V_{3
k},\nonumber\\ & \xi\rightarrow h_{3
k}^{-1}(\xi)=\left(\xi_1,\xi_2,\xi_3,(-1)^k\sqrt{\frac{3\left(1-\xi_2^2-\xi_3^2\right)}{V(\xi_1)}}\right),\;
k=1,2.\end{align}

Since $V(\phi)$ is positive and differentiable of class $C^3,$
follows that the above functions and they inverses are of class
$C^3$ (and then continuous). Hence, all the above applications are
homeomorphism (in fact they are diffeomorphism of class $C^3$).

From the previous characterization of $\Sigma_T$ follows that it
is a topological manifold (without boundary) of dimension $4$
immersed in $\mathbb{R}^4,$ since $id:
\Sigma_T\rightarrow\mathbb{R}^4,\mathbf{p}\rightarrow
id(\mathbf{p})=\mathbf{p}$ is a diffeomorphic immersion.
$\blacksquare$

Observation. The system \eqref{T2.9}-\eqref{T2.12} is invariant
under the coordinate transformations $x_2\rightarrow -x_2$ and
$x_3\rightarrow -x_3.$ Thus the analysis can be simplified
considerably if we restrict the flow to the invariant set
\begin{equation} \Sigma:=\left\{
\left(\phi,\,x_1,\,x_2,\,x_3\right)\in\mathbb{R}^2\times\left(\mathbb{R}^0_+\right)^2\right\}\cap\Sigma_T=V_{22}\cup
V_{3 2}\cup U_2\cup U_3.\end{equation} where we have used the
notation $\mathbb{R}^0_+:=\left\{x\in \mathbb{R}| x\geq
0\right\},$ and $\mathbb{R}_+:=\left\{x\in \mathbb{R}| x>
0\right\}.$

\begin{prop}\label{Prop9}
$\Sigma$ is topological manifold with boundary
\end{prop}

{\bf Proof}. Using the same arguments as in the proof of
proposition \ref{Prop8}, we can prove that $\Sigma$ is not an open
set with respect to the induced topology as subset of
$\mathbb{R}^2\times \mathbb{R}_+^2;$ thus it dimension should be
less than $4.$ Let us construct a set of local charts as follows.

Let be defined the sets $W_j=\{\mathbf{p}\in\Sigma, x_j>0\}, \,
j=2,3.$ These sets are open with respect to the topology induced
in $\Sigma$ by the usual topology of $\mathbb{R}^2\times
\mathbb{R}_+^2.$

Let be defined the maps
\begin{equation} h_2:
W_2\rightarrow \mathbb{H}^3, \, \mathbf{p}\rightarrow
h_2(\mathbf{p})=\left(\phi,
x_1,\sqrt{\frac{V(\phi)}{3}}x_3\right)=(\xi_1,\xi_2,\xi_3)\label{h2}\end{equation}
and
\begin{equation} h_3: W_3\rightarrow \mathbb{H}^3, \,
\mathbf{p}\rightarrow h_3(\mathbf{p})=\left(\phi,
x_1,x_2\right)=\left(\xi_1,\xi_2,\xi_3\right)\label{h3}\end{equation}
Observe that  $h_j(\mathbf{p})=\mathbb{R}\times
\left(\mathbb{H}^2\cap \mathbb{D}^2\right)$ which are open sets of
$\mathbb{H}^3$ with respect to the induced topology. The inverse
functions of $h_1$ and $h_2$ are given by
\begin{align} & h_2^{-1}: \mathbb{R}\times \left(\mathbb{H}^2\cap
\mathbb{D}^2\right)\rightarrow W_2, \nonumber\\ & \xi\rightarrow
h_2^{-1}(\xi)=\left(\xi_1,\xi_2\sqrt{1-\xi_2^2-\xi_3^2},\sqrt{\frac{3}{V(\xi_1)}}\xi_3\right)\label{h2inv}\end{align}
and
\begin{align} & h_3^{-1}: \mathbb{R}\times \left(\mathbb{H}^2\cap
\mathbb{D}^2\right)\rightarrow W_3, \nonumber\\ &\xi\rightarrow
h_3^{-1}(\xi)=\left(\xi_1,\xi_2,\xi_3,\sqrt{\frac{3\left(1-\xi_2^2-\xi_3^2\right)}{V(\xi_1)}}\right)
\label{h3inv}\end{align} From the hypotheses on $V$ follows that
all these functions are differentiable of class $C^3.$ Hence,
$\left(W_2,h_2\right), \left(W_3,h_3\right)$ are local charts.
Since the local charts $\left(W_2,h_2\right),
\left(W_3,h_3\right)$ do not cover the sets with $x_2=x_3=0$ we
define the sets $W_1^\pm=\{\mathbf{p}\in\Sigma: x_1=\pm 1,
x_2=x_3=0\},$ $W_1=W_1^+ \cup W_1^-.$ From the equalities
$W_1^+=V_{1 2}\cap U_2 \cap U_3$ and $W_1^-=V_{1 1}\cap U_2 \cap
U_3$ where $V_{1 k}=\left\{\mathbf{p}\in\mathbb{R}^4| (-1)^k x_1
>0\right\}\cap \Sigma$ follow that $W_1^\pm$ are open sets with
respect to the topology induced in $\Sigma$ by the usual topology
of $\mathbb{R}^2\times \mathbb{R}_+^2$ which are disjoint copies
of $\mathbb{R}.$ $\blacksquare$

We can prove that $W_2$ and $W_3$ are topological manifolds with
boundary. In order to do so we observe that, since $V$ is
positive, the points $\mathbf{p}\in W_2$ transforms by $h_2$ is
points with $\xi_3:=\sqrt{\frac{V(\phi)}{3}}x_3=0,$ if, and  only
if, $x_3=0.$ In an analogous way, it is easy to see that the
points $\mathbf{p}\in W_3$ transforms by $h_3$ is points with
$\xi_3:=x_2=0,$ if, and  only if, $x_2=0.$ Thus, the boundaries of
$W_2$ and $W_3$ are given respectively by
\begin{eqnarray}
\partial W_2&=&\left\{\mathbf{p}=(\phi,x_1,x_2,x_3)\in W_2: h_{2}(\mathbf{p})\in \mathbb{R}^{2}\times
\{0\}\right\}\nonumber\\&=&\left\{\mathbf{p}\in W_2: x_3=0\right\}
\\
\partial W_3&=&\left\{\mathbf{p}=(\phi,x_1,x_2,x_3)\in W_3: h_{3}(\mathbf{p})\in \mathbb{R}^{2}\times
\{0\}\right\}\nonumber\\&=&\left\{\mathbf{p}\in W_3:
x_2=0\right\}.
\end{eqnarray}

Let us define the sets \begin{align}(\partial \Sigma)_1 &=\partial
W_2 \cup W_1=\left\{\mathbf{p}\in \Sigma: x_3=0\right\}\nonumber\\
&=\left\{(\phi,x_1,x_2)\in \mathbb{H}^3:
x_1^2+x_2^2=1\right\}\end{align} and \begin{align}(\partial
\Sigma)_2&=\partial W_3 \cup W_1=\left\{\mathbf{p}\in \Sigma:
x_2=0\right\}\nonumber\\&=\left\{(\phi,x_1,x_3)\in \mathbb{H}^3:
x_1^2+\frac{V(\phi)}{3}x_3^2=1\right\}.\end{align}

Defining the maps
\begin{eqnarray} & g: (\partial
\Sigma)_1\rightarrow \mathbb{R}\times [-1,1],
(\phi, x_1,x_2)\rightarrow (\phi, x_1)=(\xi_1,\xi_2), \\
& g^{-1}: \mathbb{R}\times [-1,1]\rightarrow (\partial \Sigma)_1,
(\xi_1,\xi_2)\rightarrow \left(\xi_1,
\xi_2,\sqrt{1-\xi_2^2}\right);
\end{eqnarray}
\begin{eqnarray}
& h: (\partial \Sigma)_2\rightarrow \mathbb{R}\times [-1,1],
(\phi, x_1,x_3)\rightarrow (\phi, x_1)=(\xi_1,\xi_2), \\
& h^{-1}: \mathbb{R}\times [-1,1]\rightarrow (\partial \Sigma)_2,
(\xi_1,\xi_2)\rightarrow \left(\xi_1,
\xi_2,\sqrt{\frac{3(1-\xi_2^2)}{V(\xi_1)}}\right);
\end{eqnarray} and

we can show that $(\partial \Sigma)_1$ and $(\partial \Sigma)_1$
are 2-dimensional manifolds with boundaries given by $W_1.$

It is easy to show the identities
\begin{eqnarray} \Sigma &=& Int (\Sigma\setminus
W_1)
\cup (\partial \Sigma)_1 \cup (\partial \Sigma)_2 \label{decompositionSigma}\\
Int \Sigma &=& Int (\Sigma\setminus W_1)=\{\mathbf{p}\in \Sigma:
x_2>0, x_3>0\} \label{interiorSigma}
\end{eqnarray}

From the above arguments and expression \eqref{decompositionSigma}
we have the following.

\begin{rem}
\begin{itemize}
\item The interior of $\Sigma$ is given by \eqref{interiorSigma}.
It is a 3- dimensional manifold without boundary.

\item The boundary of $\Sigma$, $\partial\Sigma,$ is the union of
two 2-dimensional topological manifolds with boundary given by
$(\partial \Sigma)_1=\left\{\mathbf{p}\in \Sigma: x_3=0\right\}$
and $(\partial \Sigma)_2=\left\{\mathbf{p}\in \Sigma:
x_2=0\right\},$ contained in $\mathbb{R}\times\mathbb{R}^0_+.$

\item  $(\partial \Sigma)_1$ and $(\partial \Sigma)_2$ share the
same boundary (which is a union of two disjoint a copies of
$\mathbb{R}$) given by $W_1=\left\{\mathbf{p}\in\Sigma|
x_1=1\right\}\cup \left\{\mathbf{p}\in\Sigma| x_1=-1\right\}.$

\end{itemize}

\end{rem}

\begin{lem}\label{Theorem I} Suppose that $0<\chi(\phi)<+\infty$
for any compact set. Then, for all $p\in Int\Sigma$ the  $\alpha$-
and $\omega$-limit sets of $p$ are such that $\alpha(p)\subset
\left(\partial\Sigma\right)_1$ and $\omega(p)\subset
\left(\partial\Sigma\right)_2.$
\end{lem}

{\bf Proof}. By the proposition \ref{Proposition 4.1} we have that
$Int\Sigma=\{(\phi,x_1,x_2,x_3)\in\Sigma: x_2>0, x_3>0\}$ is an
invariant set of the flow  of \eqref{T2.9}-\eqref{T2.12}. Let be
defined on $Int\Sigma$ the function
$$Z(\phi,x_1,x_2,x_3)=\left(\frac{x_2}{x_3}\right)^2
\chi(\phi)^{2-\frac{3\gamma}{2}}.$$  The function $Z$ is a
monotone decreasing function (in the direction of the flow) in
$Int\Sigma$, since its directional derivative through the flow is
$Z'=-\gamma Z.$  The rank of $Z$ is $(0,\infty).$ Let be $s\in
\partial\Sigma.$ From the hypothesis about $\chi$ follows that it
cannot not be zero or infinite unless $|\phi|\rightarrow\infty.$
Then, it is verified that $Z(s)\rightarrow 0$ as $s\rightarrow
\left(\partial\Sigma\right)_2$ and $Z(s)\rightarrow \infty$ as
$s\rightarrow \left(\partial\Sigma\right)_1.$ Hence, applying the
Monotonicity Principle (theorem \ref{theorem 4.12}) follows the
required result.

\subsection{The flow on the invariant set
$\left(\partial\Sigma\right)_1$}

The flow on $\left(\partial\Sigma\right)_1$ is governed by the
differential equations:

\begin{eqnarray}
\phi'&=&\sqrt{\frac{2}{3}} x_1, \label{eq1phi}\\
x_1'&=&\frac{1}{2} \left(1-x_1^2\right)
\left(x_1(\gamma-2)+\frac{(4-3\gamma)}{\sqrt{6}}\frac{{\chi}'\left(\phi\right)}{{\chi}
\left(\phi\right)}\right),\label{eq1y}
\end{eqnarray}

If $\mathbf{p}\in\left(\partial\Sigma\right)_1$ then
$x_1^2+x_2^2=1,$ which implies $|x_1|\leq 1.$ The equality holds
if and only if $x_2=0.$ It is easy to prove that the set
$$W_1=\left\{\mathbf{p}=\left(\phi,\,x_1,\,x_2,\,x_3\right)\in\Sigma: |x_1|=1\right\}$$ is an invariant set of the flow of
\eqref{T2.9}-\eqref{T2.12} in $\Sigma.$ Observe that $\phi$ is
unbounded on $W_1$ since from $|x_1|=1$ and by the equation
\eqref{T2.9}, $$ \phi'=\pm\sqrt{\frac{2}{3}}\implies
\phi=\phi_0\pm\sqrt{\frac{2}{3}}\tau$$ (which is an unbounded
function of $\tau$).

In the set $\left(\partial\Sigma\right)_1$ there exist a (possibly
empty) family of singular points with  $\phi$ bounded
$Q:=\{\mathbf{q}:=(\phi,x_1)=(q_1,0)\in\bar{S}: \chi'(q_1)=0\}.$
\footnote{For definiteness we are assuming that $\chi(\phi_1)\neq
0.$ Also, we are assuming that $\chi$ admits only isolated
singular points.} The eigenvalues of the matrix of derivatives
evaluated at $\mathbf{q}\in Q$ are
\begin{eqnarray}\Delta_1\pm\sqrt{\Delta_1^2
+\Delta_2\frac{\chi''(q_1)}{\chi(q_1)}}\end{eqnarray} with
$\Delta_1=(-2+\gamma)/{4}< 0,$ and
$\Delta_2=\left(4-3\gamma\right)/6.$ The local dynamical character
of the singular point $\mathbf{q}\in Q$ on the invariant set
$\left(\partial\Sigma\right)_1$ is as follows \footnote{Remember
we are assuming that the barotropic index $\gamma$ satisfies
$0<\gamma<2.$}:
\begin{enumerate}
\item $\mathbf{q}$ is an stable focus if  $0<\gamma<4/3$ and
$\chi''(q_1)<-\frac{\Delta_1^2\,\chi(q_1)}{\Delta_2} $ or
$4/3<\gamma<2$ and
$\chi''(q_1)>-\frac{\Delta_1^2\,\chi(q_1)}{\Delta_2}.$

\item $\mathbf{q}$ is an stable node if $0<\gamma<4/3$ and
$-\frac{\Delta_1^2\,\chi(q_1)}{\Delta_2}\leq \chi''(q_1)<0$ or
$4/3<\gamma<2$ and $0<\chi''(q_1)\leq
-\frac{\Delta_1^2\,\chi(q_1)}{\Delta_2}.$ \item $\mathbf{q}$ is a
saddle point if  $0<\gamma<4/3$ and $\chi''(q_1)> 0$ or
$4/3<\gamma<2$ and $\chi''(q_1)< 0.$

\item $\mathbf{q}$ is nonhyperbolic, if $\chi''(q_1)= 0,$ in which
case, there exists a 1-dimensional stable manifold which is
tangent to the axis $x_1$ at $\mathbf{q}.$ There exist also a
1-dimensional center manifold tangent to the line
$(1-\gamma/2)x_1-\sqrt{2/3}\phi=0$ at $\mathbf{q}.$
\end{enumerate}


\begin{figure}[ht]
\begin{minipage}{9pc}
\hspace{0.4cm}
\begin{center}
\includegraphics[width=5cm,height=5cm]{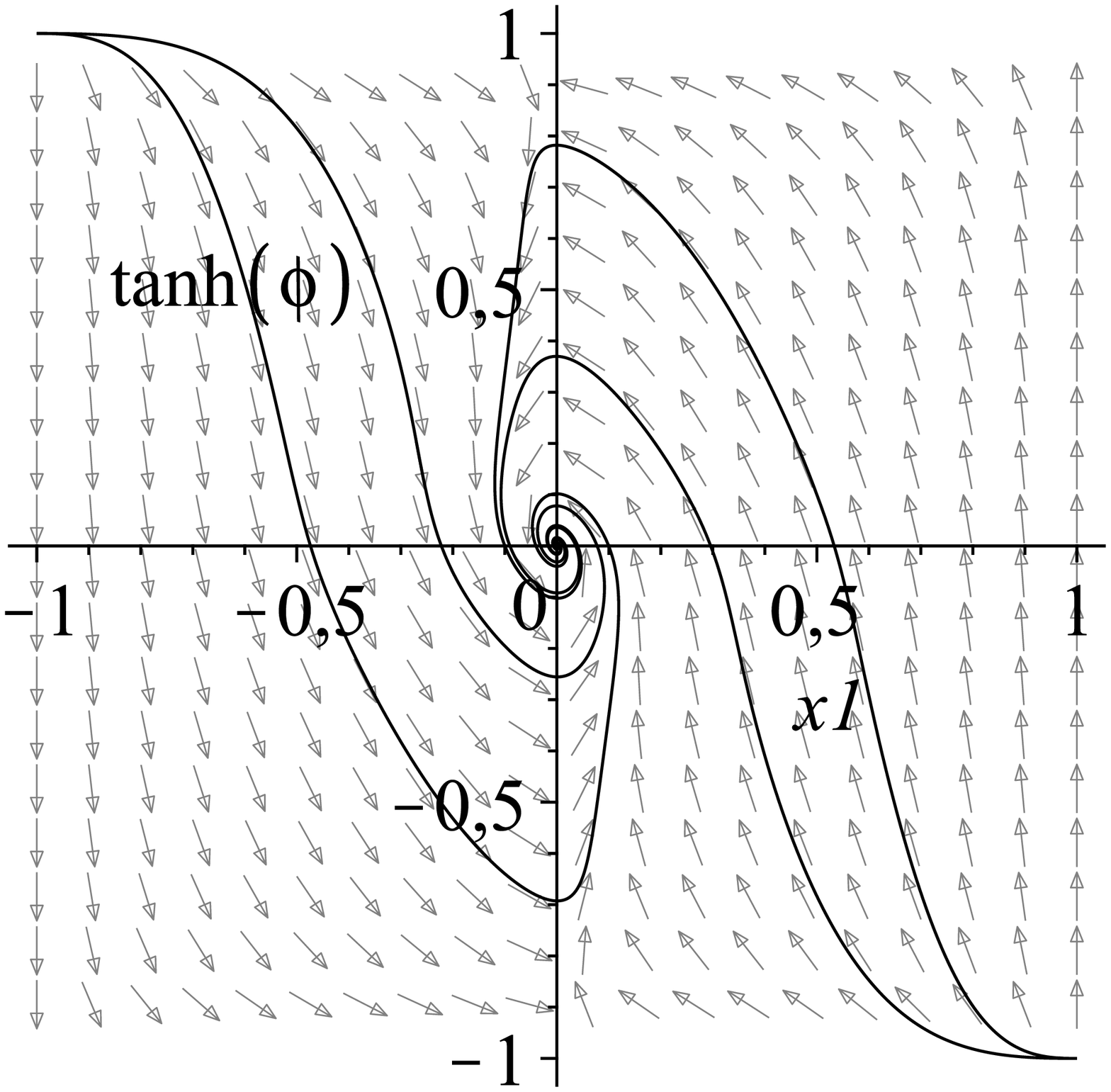}
\begin{center} (a)\end{center}
\end{center}\end{minipage}\hspace{2pc}%
\begin{minipage}{9pc}
\hspace{0.4cm}
\begin{center}
\includegraphics[width=5cm,height=5cm]{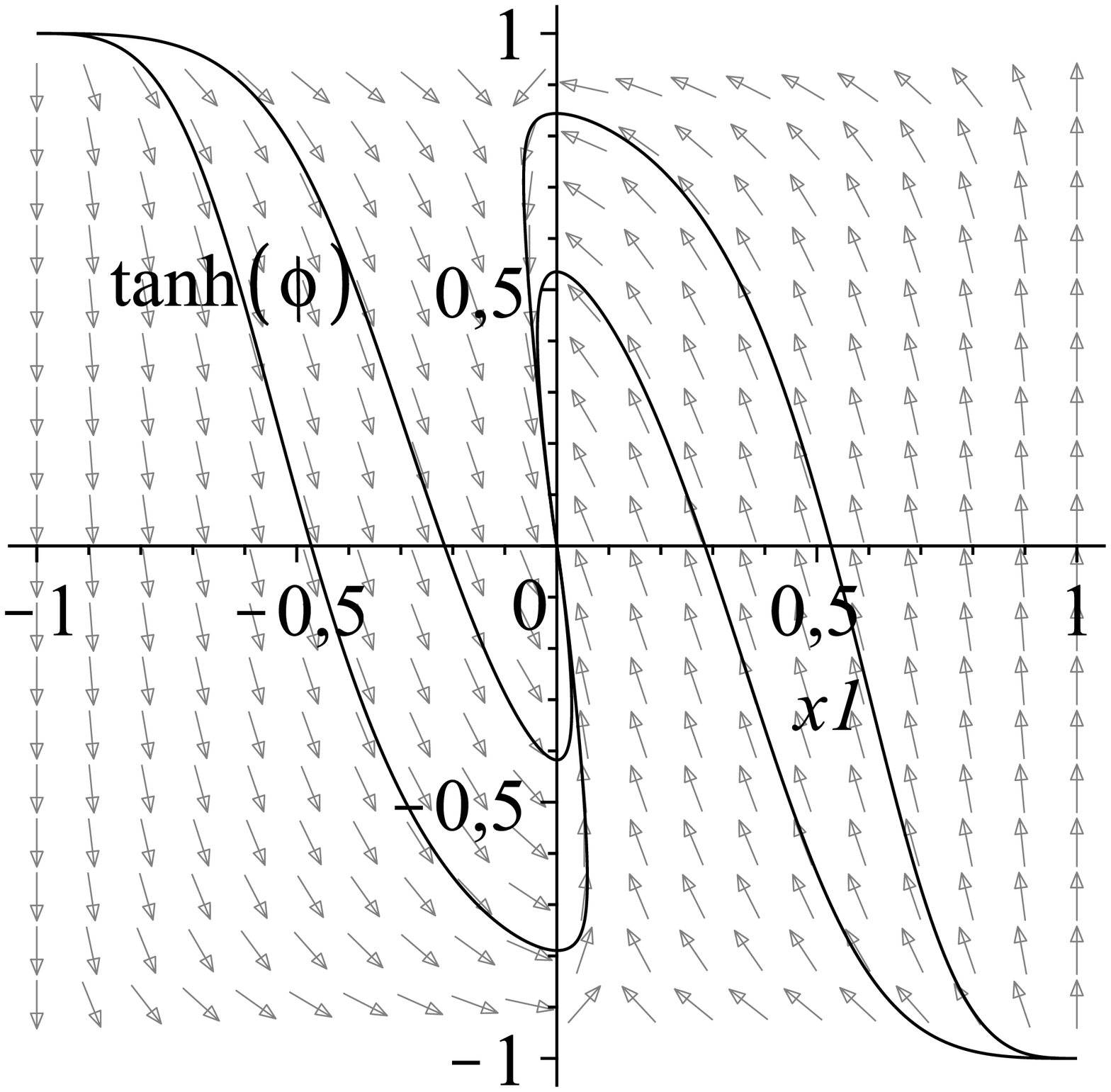}
\begin{center} (b)\end{center}
\end{center}\end{minipage}\hspace{2pc}%
\begin{minipage}{9pc}
\hspace{0.4cm}
\begin{center}
\includegraphics[width=5cm,height=5cm]{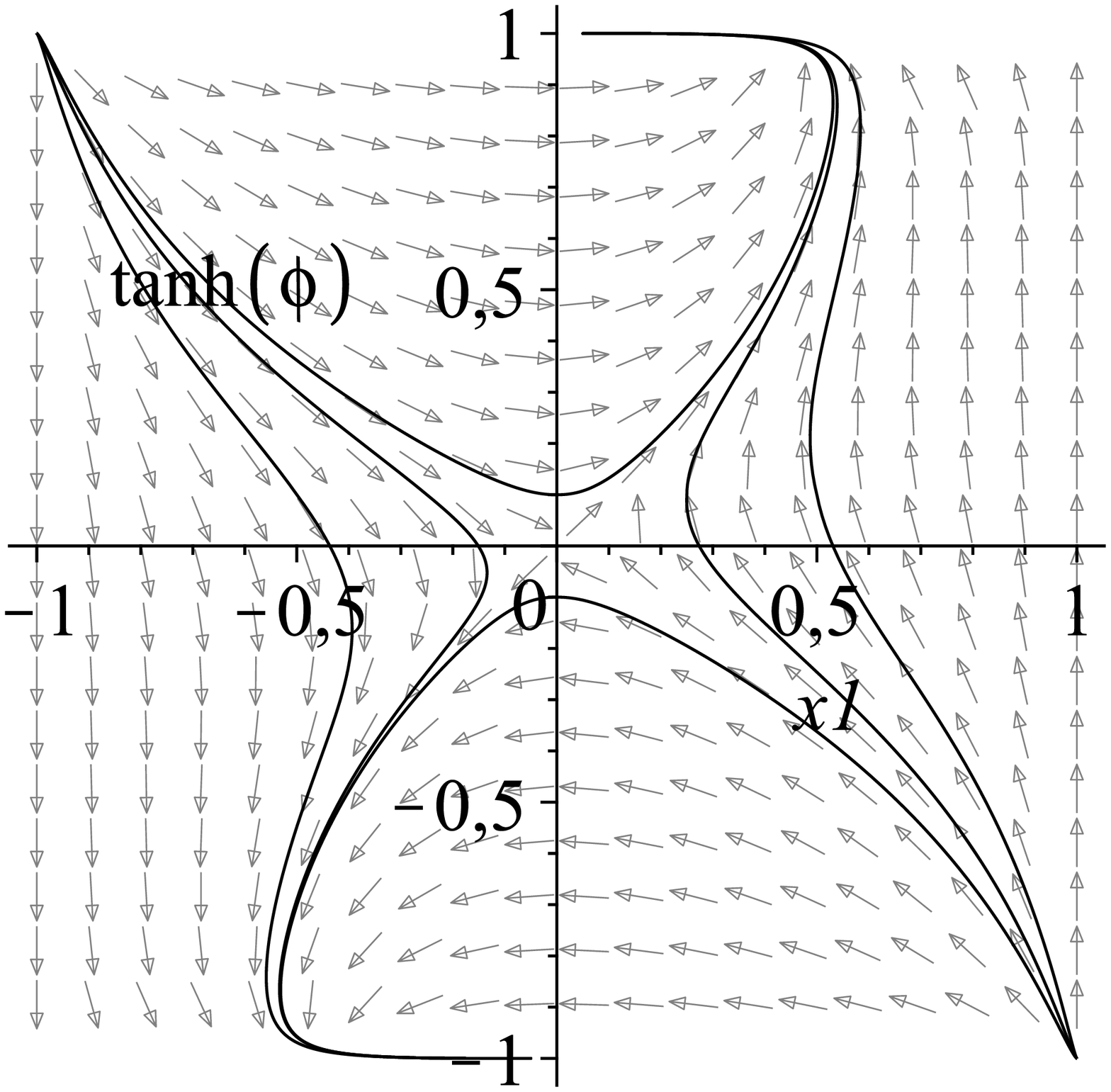}
\begin{center} (c)\end{center}
\end{center}
\end{minipage}
\hspace{2pc}%
\caption{Projection of orbits on the phase plane $(x_1,\varphi)$
for the coupling function  (\ref{couplingexamplea}): (a) for
$n=2,$ $\gamma=1.35$ and $\chi_0=0.05$ the origin is an stable
focus. (b) for $n=2,$ $\gamma=1.4$  and $\chi_0=3$  the origin is
an stable node. (c) for $n=2,$ $\gamma=1$  and $\chi_0=0.3$  the
origin is saddle point. Notice that the singular points
$(x_1,\vphi)=(0,\pm 1)$ seems to be saddle points for the cases
(a) and (b) whereas in the case (c) they are local sinks. Observe
that the singular points $(x_1,\vphi)=(1,-1)$ and
$(y,\vphi)=(-1,1)$ are in all the cases local sources (which is a
suggestive argument in favor for the unboundedness of the scalar
field into the past).} \label{powerlaw_coupling.eps}
\end{figure}


\subsubsection{Powerlaw-coupling function}

Let us consider the coupling function \be
\chi(\phi)=\frac{\lambda}{n}\left(\phi^n+\chi_0\right),\;
\chi_0>0,\,\lambda>0,\,n>1\label{couplingexamplea},\ee where $n$ is
an integer number.

In this case the equations \eqref{eq1phi}-\eqref{eq1y} reduces to

\ben \phi'&=&\sqrt{\frac{2}{3}} x_1,\label{eq1phinew}\\
x_1'&=&\frac{1}{2} \left(1-x_1^2\right) \left(\frac{n (4-3 \gamma
) \phi ^{n-1}}{\sqrt{6}
   \left(\phi ^n+\chi_0\right)}+x_1 (\gamma
-2)\right)\label{eq1ynew}.\een

Observe that
$$\chi(0)=\frac{\lambda\chi_0}{n}> 0,$$ and
\be\frac{d^m}{d\phi^m} \chi(\phi)=\left\{\begin{array}{rcr} {\lambda}{{\phi^{n-m}}}\prod_{j=1}^{m-1}{(n-j)}&,&m< n\\
                                     (n-1)!\,\lambda&,&m=n\\
                                     0&,& m>n \end{array}\right.\label{chi}\ee
which implies, $\frac{d^m \chi}{d\phi^m}(0)=0$  for all $m\neq n$
and $\frac{d^n \chi}{d\phi^n}(0)=(n-1)!\,\lambda.$

Hence, if $n=2$ then
$\chi(0)=\lambda\chi_0/2,\;\chi'(0)=0,\;\chi''(0)=\lambda>0$
(i.e., the coupling function has a local minimum
$\chi(0)=\lambda\chi_0/2>0$), and by the former analysis, the
origin can be either an stable focus if $4/3<\gamma<2$ and
$\chi_0<-2\Delta_2/\Delta_1^2,$ or an stable node if
$4/3<\gamma<2$ and $\chi_0\geq -2\Delta_2/\Delta_1^2,$ or a saddle
point if $0<\gamma<4/3.$ Observe that if the background is a
pressureless dark matter fluid (dust) the origin is a saddle
point. If $n> 2$ then $\chi'(0)=\chi''(0)=\ldots
=\chi^{(m)}(0)=0,$ for $m<n,$ and $\mathbf{q}=\left(0,0\right)$ is
nonhyperbolic. Thus, there exists a 1-dimensional stable manifold
which is tangent to the axis $x_1$ at the origin. There exist also
a 1-dimensional center manifold tangent to the line
$(1-\gamma/2)x_1-\sqrt{2/3}\phi=0$ at the origin.

We have introduced a new scalar field $\varphi=\tanh (\phi)$
(which takes infinity to a finite value) to make the numerics.
Notice the existence of the singular points $(x_1,\vphi)=(1,-1)$
and $(x_1,\vphi)=(-1,1)$ which are in all the cases local sources
(which is a suggestive argument in favor for the unboundedness of
the scalar field into the past; see figures
\ref{powerlaw_coupling.eps}, (a)-(c)). Also, if the origin is a
saddle point, then the singular points $(x_1,\vphi)=(0,\pm 1)$
seems to be local sinks (\ref{powerlaw_coupling.eps}-(c)).
However, if the origin is either an stable focus (figure
\ref{powerlaw_coupling.eps}-(a)) or an stable node (figure
\ref{powerlaw_coupling.eps}-(b)), the singular points
$(x_1,\vphi)=(0,\pm 1)$ are saddle points. In such a case the
orbits spent an infinite amount of time close to the matter
dominated universe ($x_1=0\implies x_2=1$).

\subsection{The flow on the invariant set
$\left(\partial\Sigma\right)_2$}

The dynamics in the invariant set $\left(\partial\Sigma\right)_2$
is governed by the differential equations:

\begin{eqnarray}
x_1'&=&\left(x_1^2-1\right) \left(x_1 +\frac{\sqrt{6}}{6}\frac{
\partial_{\phi} V(\phi
   )}{V(\phi )}\right),\label{eq2y}
\end{eqnarray}  and (\ref{eq1phi})

In the invariant set $\left(\partial\Sigma\right)_2$ there are two
invariant subsets: the set
$$W_1=\left\{p=\left(\phi,\,x_1,\,x_2,\,x_3\right)\in\Sigma:
|x_1|=1\right\},$$ and the (possibly empty) class, $P,$ of
singular points $\mathbf{p}$ with coordinates $\phi=\phi_2$ with
$\chi(\phi_2)\neq 0,\,V'(\phi_2)=0,$ and $x_1=0,$ hence
$x_3=\sqrt{\frac{3}{V(\phi_2)}}.$ We assume that $V$ has only
isolated singular points. Each singular point $\mathbf{p}\in P$ is
\begin{enumerate}

\item a saddle if $V''(\phi_2)<0,$

\item a stable node if $0< V''(\phi_2)\leq\frac{3}{4} V(\phi_2),$
and

\item a stable focus if $V''(\phi_2)>\frac{3}{4} V(\phi_2).$

\end{enumerate}


\begin{figure}[ht]
\begin{center}
\begin{minipage}{12pc}
\hspace{0.4cm}
\begin{center}
\includegraphics[width=6cm,height=6cm]{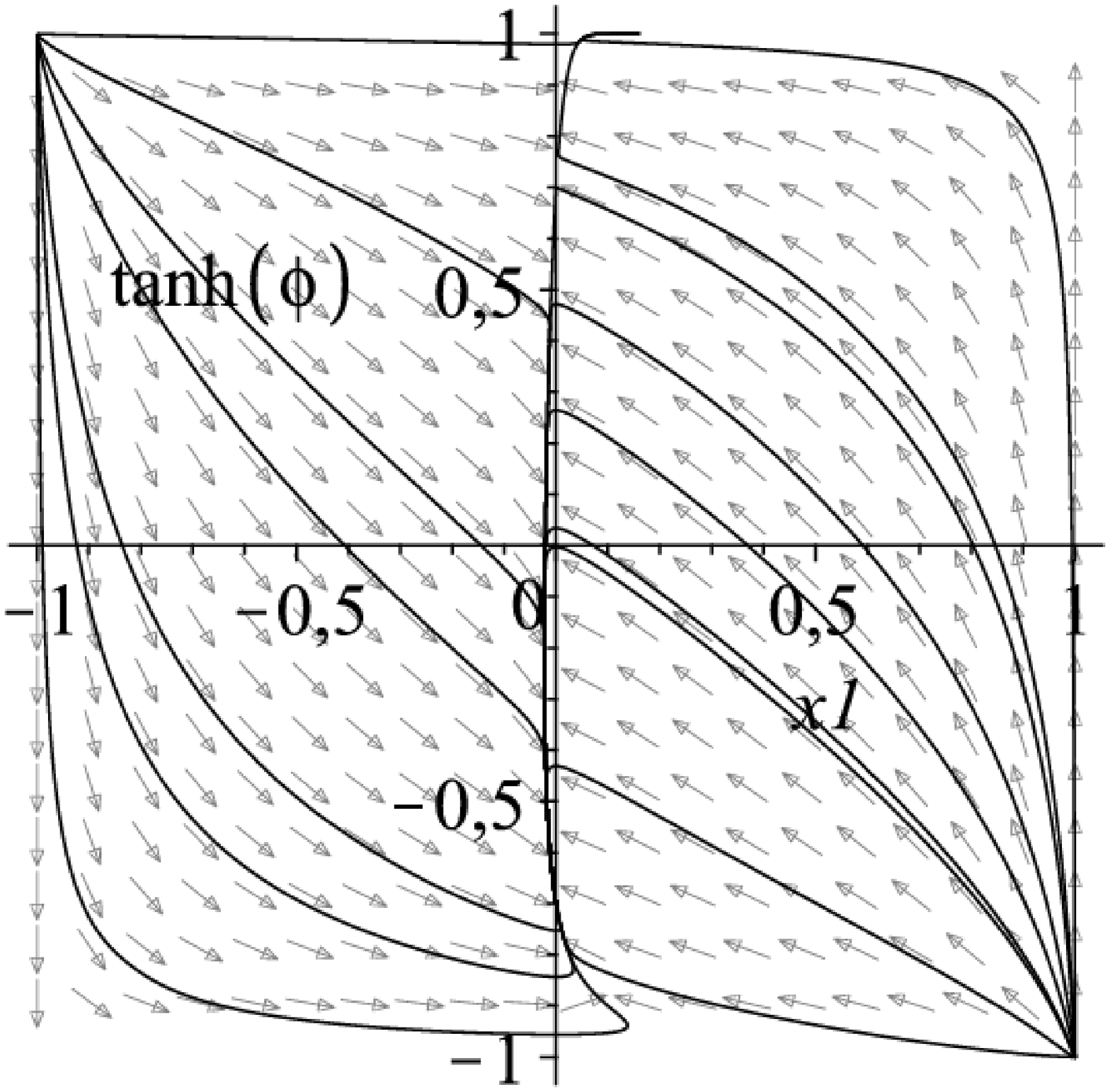}
\begin{center} (a)\end{center}
\end{center}\end{minipage}\hspace{2pc}%
\begin{minipage}{12pc}
\hspace{0.4cm}
\begin{center}
\includegraphics[width=6cm,height=6cm]{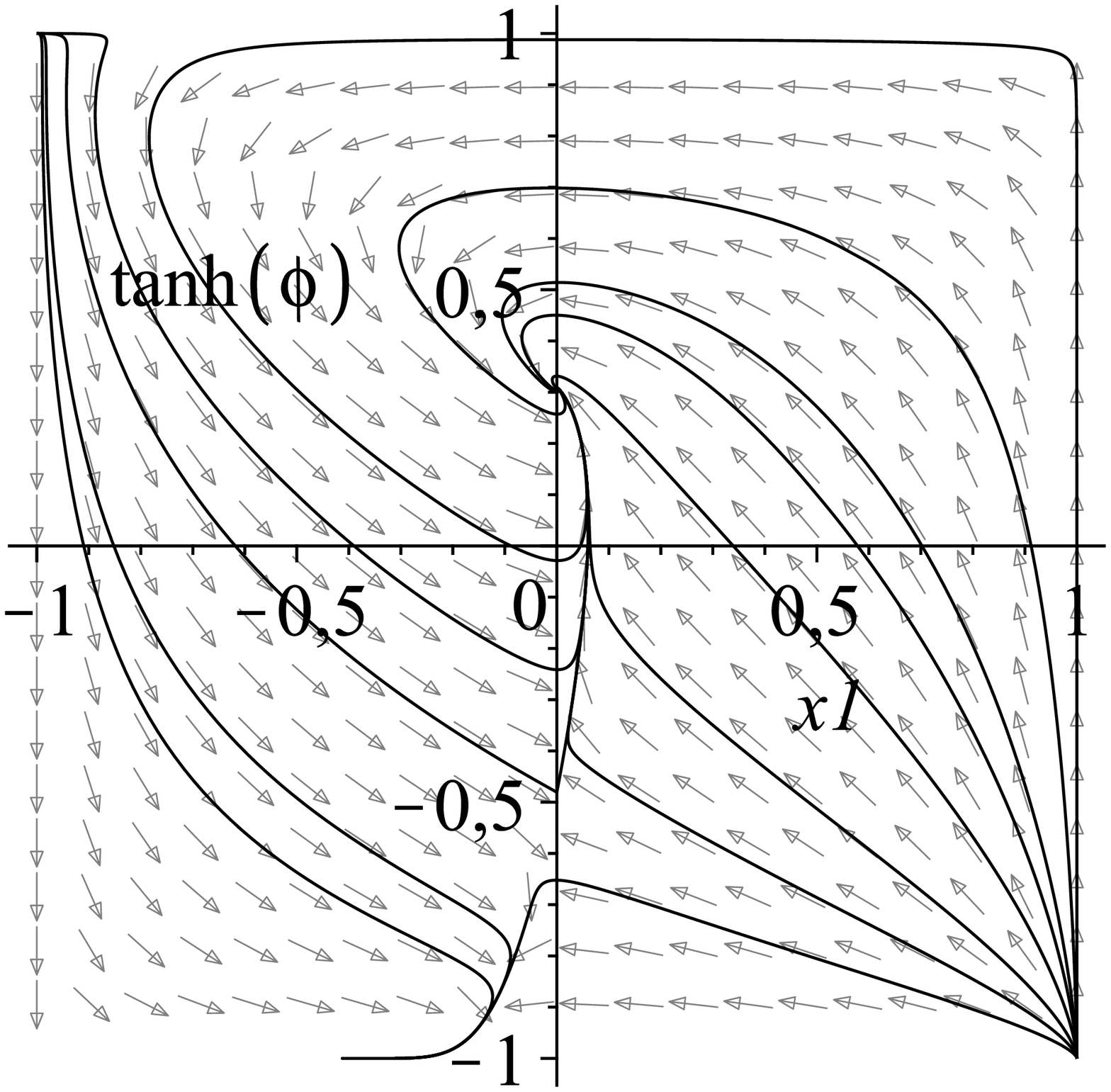}
\begin{center} (b)\end{center}
\end{center}\end{minipage}\hspace{2pc}%
\caption{Phase plane $(x_1,\varphi)$ for the model with potential
(\ref{Albrecht-Skordis}) (a) for $A=3.25,$ $B=-2,$ and $\mu=0.5$
the singular points of the system are a saddle located at
$(x_1,\varphi)=(0,0.6993)$ and a node at $(x_1,\varphi)=(0,
-0.6993)$. (b) for $A=0.62,$ $B=0.79,$ and $\mu=-1.12$ the
singular points of the system are a saddle located at
$(x_1,\varphi)=(0,-0.4806)$ and an stable spiral at
$(x_1,\varphi)=(0, 0.3078)$. } \label{figuraAlbrecht-Skordis}
\end{center}
\end{figure}


\subsubsection{The Albrecht-Skordis potential}

Albrecht and Skordis \cite{Albrecht:1999rm} have proposed a
particularly attractive model of quintessence. It is driven by a
potential which introduces a small minimum to the exponential
potential:
\begin{equation}
V(\phi )=e^{-\mu \phi }{\left( A+(\phi -B)^2\right).}
\label{Albrecht-Skordis}
\end{equation}
Unlike previous quintessence models, late-time acceleration is
achieved without fine tuning of the initial conditions. The
authors argue that such potentials arise naturally in the
low-energy limit of $M$-theory. The constant parameters, $A$ and
$B$, in the potential take values of order $1$ in Planck units, so
there is also no fine tuning of the potential (we suppose also
that $\mu\neq 0$). They show that, regardless of the initial
conditions, $\rho _\phi $ scales, with $\rho \propto \rho _\phi
\propto t^{-2}$ during the radiation and matter eras, but leads to
permanent vacuum domination and accelerated expansion after a time
which can be close to the present.

The extremes of of the potential (\ref{Albrecht-Skordis}) are
located at $\phi^{\pm}=\frac{1+B\mu\pm\sqrt{1-A\mu^2}}{\mu}.$ They
are real if $1\geq \mu ^2A.$ The local minimum (respectively,
local maximum)  is located at $\phi^-$ (respectively $\phi^+$)
since
$$\pm V''(\phi^{\pm})=- 2 V_0\sqrt{1-A\mu^2}e^{-\left(1+B\mu\pm\sqrt{1-A\mu^2}\right)}<0.$$

By using the formalism developed here we find that the singular
point associated to $\phi^+$ is always a saddle point of the
corresponding phase portrait (see figures
\ref{figuraAlbrecht-Skordis} (a)-{b}). The singular point
associated to $\phi^-$ could be either an stable node (see
\ref{figuraAlbrecht-Skordis} (a)) or an stable spiral (see
\ref{figuraAlbrecht-Skordis} (b)) if
$$\frac{8(3+2\mu^2)}{\left(3+4\mu^2\right)^2}<A<
\frac{1}{\mu^2}$$ or
$$A<\frac{8(3+2\mu^2)}{\left(3+4\mu^2\right)^2}.$$

As before we have introduced a new scalar field $\varphi=\tanh
(\phi)$ in order to make the numerics. Observe that, almost all
the initial points in the phase plane past asymptote (in figures
\ref{figuraAlbrecht-Skordis} (a)-(b)) to the points $p_{1,2}$ with
coordinates $(y,\varphi)=(\mp 1,\pm 1)$ (they are associated
respectively with the infinite boundaries $\phi=\pm \infty$). In
the case (a) there are some orbits past asymptotic to $P_{1,2}$
converging to the point $P_4$ with coordinates
$(x_1,\varphi)=(\mu/\sqrt{6},+1)$ which is a singular point
located at the region $\phi=\infty$, in this example, its center
manifold acts as an exponential attractor (for more details see
the next section) whereas the singular point with coordinates
$(x_1,\varphi)=(\mu/\sqrt{6},-1)$ acts as a saddle. In the case
(b) we have a similar situation. Actually, there exist two
singular points with coordinates $x_1=\mu/\sqrt{6}$ each one
contained in the invariant manifolds $\varphi=\pm 1$ (i.e.,
$\phi=\pm\infty$) respectively. Each one has a center manifold
which acts as an exponential attractor for nearby orbits.

These examples suggest the divergence of $\phi$ towards the past.
We proceed to prove the following

\begin{thm}\label{Theorem_2.4}
Let $\chi (\phi)$ and $V(\phi)$ be positive functions of class
$C^3,$ such that $\chi $ satisfies that $0<\chi(\phi)<+\infty$ for
any compact set, and has, at most, a finite number of isolated
singular points. Let  $\gamma \in \left(0,\frac{4}{3}\right)\cup
\left(\frac{4}{3},2\right).$ Let $\mathbf{p}$ a point in   $\Sigma
,$ and $O^-(p)$ the past orbit of $\mathbf{p}$ under the flow
$\mathbf{g}^{\tau }$ of \eqref{T2.9}-\eqref{T2.12} in $\Sigma.$
Then, $\phi$ is unbounded in $O^-(\mathbf{p})$ for almost all
$\mathbf{p}.$

\end{thm}

{\bf Proof}. For the demonstration we will consider only interior
points of $\Sigma,$ since its boundary
$\partial\Sigma=\left(\partial\Sigma\right)_1\cup
\left(\partial\Sigma\right)_2$ is of dimension $2<dim \Sigma.$
Form this fact follows that this set is of zero Lebesgue measure.
\footnote{However, in the previous sections we have presented
several numerical examples concerning the dynamics on
$\left(\partial\Sigma\right)_1$ and on
$\left(\partial\Sigma\right)_2,$ which supported the unboundedness
of the scalar field towards the past.}

Let us consider the time reversal transformation $\left(\tau,
\phi,x_1,x_2,x_3\right)\rightarrow \left(-\tau,
\phi,x_1,x_2,x_3\right).$ Using this transformation the system
becomes

\begin{align}
& \phi'=-\sqrt{\frac{2}{3}}x_1 \label{T2.9b},\\
      &x_1'=-x_1^3-\frac{1}{2} \left(x_2^2 \gamma -2\right) x_1+\frac{x_3^2}{3 \sqrt{6}}
      \frac{\mathrm{d}V(\phi)}{\mathrm{d}\phi}-\frac{(4-3\gamma)x_2^2 }{2\sqrt{6}}\frac{\mathrm{d}\ln \chi (\phi )}{\mathrm{d}\phi},\label{T2.10b}\\
      &x_2'=-\frac{1}{2} x_2 \left(2 x_1^2+\left(x_2^2-1\right) \gamma \right)
      +\frac{(4-3\gamma)x_1 x_2}{2\sqrt{6}}\frac{\mathrm{d}\ln \chi (\phi )}{\mathrm{d}\phi},\label{T2.11b}\\
  &x_3'=-\frac{1}{2} x_3 \left(2 x_1^2+x_2^2 \gamma \right),\label{T2.12b}
\end{align} where new the comma denotes derivative with respect to
$-\tau.$

Let $\mathbf{p}_0=(\phi_{0}, x_{10}, x_{20},x_{30})\in Int \Sigma$
such that exists $K, |\phi|<K$, for all $\mathbf{p}=(\phi, x_{1},
x_{2},x_{3})\in O^+(\mathbf{p}_0),$ where $O^+(\mathbf{p}_0)$
denotes the future orbit of \eqref{T2.9b}-\eqref{T2.12b} passing
through $\mathbf{p}_0.$ Then, the following inequalities hold
\begin{equation}-1\leq
x_1\leq 1,\, 0\leq x_2\leq 1, 0\leq x_3\leq
\sqrt{\frac{3}{\inf_{\phi\in [-K, K]} V(\phi)}}, \forall
\mathbf{p}\in O^+(\mathbf{p}_0).\end{equation} Since
 $V(\phi)>0$ follows that
$O^+(\mathbf{p}_0)$ is contained in a compact subset of $\Sigma$.
This implies that there exists a non empty, closed, connected
invariant manifold $\omega(\mathbf{p}), \forall \mathbf{p}\in
O^+(\mathbf{p}_0).$

Since $\gamma>0,$ follows that the function

\begin{equation} Z: Int \Sigma \rightarrow (0,\infty),
p=(\phi, x_1,x_2,x_3)\rightarrow
Z(p)=\left(\frac{x_3}{x_2}\right)^2\chi(\phi)^{-2+\frac{3
\gamma}{2}}\end{equation} satisfies $Z'=\nabla Z \cdot (\phi',
x_1', x_2', x_3')=-\gamma Z.$ This means that $Z$ is a monotonic
function for the flow of \eqref{T2.9b}-\eqref{T2.12b},  defined in
the invariant set $Int\Sigma$. Since $\chi$ is a positive function
of class $C^3$ follows that $Z$ is $C^3$ in $Int\Sigma$.

The function $Z$ takes values in the interval $(0,+\infty).$ From
the hypothesis about $\chi$ follows that it cannot not be zero or
infinite unless $|\phi|\rightarrow\infty.$ Thus, by construction
follows that $Z(\mathbf{p})\rightarrow 0$ if and only if
$\mathbf{p}\rightarrow\mathbf{q}$ with $\mathbf{q}$ such that
$x_3=0$ and $Z(\mathbf{p})\rightarrow +\infty$ if and only if
$\mathbf{p}\rightarrow\mathbf{q}$ with $\mathbf{q}$ such that
$x_2=0.$

If $x_2\rightarrow 0$ and $x_3\rightarrow 0$ simultaneously, then
by the definition of $\Sigma$ follows that $x_1\rightarrow\pm 1,$
and from \eqref{T2.9b} follows that $\phi\rightarrow \mp \infty,$
which contradicts that $\phi$ is bounded at $O^+(\mathbf{p}_0).$
This implies that $x_2$ and $x_3$ cannot simultaneously tend to
zero in $\omega(\mathbf{p})$ for all $\mathbf{p}\in
O^+(\mathbf{p}_0).$

Applying the Monotonicity Principle (theorem \ref{theorem 4.12})
follows that $\omega(\mathbf{p})\subset \{\mathbf{p}\in\partial
\Sigma: x_2\neq 0\} \cap \{\mathbf{p}\in\Sigma: |\phi|<K\}=
\{\mathbf{p}\in \Sigma: x_2>0, x_3=0\}\cap \{\mathbf{p}\in\Sigma:
|\phi|<K\}\subset (\partial \Sigma)_1\setminus W_1.$

Let us define  $V:=\{\mathbf{p}\in \Sigma: x_2>0, x_3=0\}\cap
\{\mathbf{p}\in\Sigma: |\phi|<K\}={\left\{(\phi,x_1,x_2)\in
\mathbb{H}^3: |\phi|<K, x_1^2+x_2^2=1\right\}}.$

Let $\mathbf{q}_0\in \omega (\mathbf{p}_0).$ From the invariance
of $\omega(\mathbf{p}_0)$ follows that the solution
$\mathbf{x}(\tau,\mathbf{q}_0)\in \omega(\mathbf{p}_0).$

Taking the local chart $(V,g|_V)$ with $g$ defined by $g:
(\partial \Sigma)_1\setminus W_1\rightarrow \mathbb{R}\times
(-1,1), (\phi, x_1,x_2)\rightarrow (\phi, x_1)=(\xi_1,\xi_2), $
follows that flow of  \eqref{T2.9b}-\eqref{T2.12b} is
topologically equivalent in a neighborhood of  $\mathbf{q}_0$ to
the flow of

\begin{eqnarray}
\xi_1'&=& f_1(\xi_1,\xi_2)=-\sqrt{\frac{2}{3}} \xi_2, \label{2d1}\\
\xi_2'&=&
f_2(\xi_1,\xi_2)=\frac{1}{2}\left(1-\xi_2^2\right)\left((2-\gamma)\xi_2-\frac{4-3\gamma}{\sqrt{6}}\frac{\mathrm{d}\ln
\chi(\xi_1)}{\mathrm{d} \xi_1}\right), \label{2d2}
\end{eqnarray}
defined in a neighborhood of  $\mathbf{\xi}_0=h(\mathbf{q}_0).$

From the topological equivalence between flows follows that  there
exists a non-empty  $\omega$-limit set of $\mathbf{\xi}_0.$

Let $S=h(V):=\mathbb{R}\times (-1,1).$  $S$ is an open simply
connected set of $\mathbb{R}^2$. The question is: What invariant
closed subsets of $\bar{S}=\mathbb{R}\times [-1,1]$ can be
candidates to $\omega(\xi_0)$?

The subsets of $\bar S$ with $\xi_2=\pm 1$ are discarded since on
them $\phi$ is unbounded.

Let $L\subset \bar{S}$ be a closed positively invariant set of
\eqref{2d1}-\eqref{2d2}.  From the Dulac's criterion (theorem
\ref{DC}) follows that do not exists periodic orbits in $\bar{S},$
since the function
\begin{equation}B:\mathbb{R}\times (-1,1)\rightarrow \mathbb{R},\xi\rightarrow
B(\xi)=(1-\xi_2^2)^{-1}.\end{equation}  satisfies  $B\in
C^1(\mathbb{R}\times (-1,1))$ and $\nabla\cdot \left(B
f\right)=\frac{\partial}{\partial_{\xi_1}} B
f_1+\frac{\partial}{\partial_{\xi_2}} B f_2=1-\frac{\gamma}{2}>0.$
Thus, it is a Dulac's function in $S.$ Besides, $\partial
S=\left([-K,K]\times\{-1,1\}\right)\cup\left(\{-K,K\}\times
(-1,1)\right)$ is not a closed orbit. Hence $L\subset \bar{S}$ do
not contain periodic orbits.

From the corollary of Poincar\'e-Bendixon Theorem \ref{PBT2},
follows that the only possible invariant sets are the singular
points with $\xi_1$ bounded, or heteroclinic sequences joining
them.

The system \eqref{2d1}-\eqref{2d2} admits a (possibly empty)
family of singular points with  $\xi_1$ bounded
$Q:=\{\mathbf{q}=(q_1,0)\in\bar{S}: \chi'(q_1)=0\}.$

Suppose that $Q=\emptyset$, i.e., $\chi'(q_1)\neq 0, \forall q_1,
|q_1|<K.$ In this case the orbit $O^+(\xi_0)$ tends to a point
with $\xi_1=\pm 1,$ following the unboundedness of $\phi.$ We
conclude that there are not interior points of $\Sigma$ leading to
a bounded past orbit.

Suppose that $Q\neq \emptyset.$ Let $\mathbf{q}\in Q.$ The
eigenvalues of the Jacobian matrix, $\frac{\partial f^i}{\partial
\xi_j}(\mathbf{q}), i,j=1,2,$ are $\mu^\pm=\frac{2-\gamma}{4}\pm
\sqrt{\left(\frac{2-\gamma}{4}\right)^2+\frac{4-3\gamma}{6}
\frac{\chi''(q_1)}{\chi(q_1)}}.$

Denoting the sets $Q^\pm =\{\mathbf{q}\in Q:\pm \chi''(q_1)>0\}$ y
$Q^0=\{\mathbf{q}\in Q: \chi(q_1)=0\}.$ At least one of these sets
is nonempty. Let us define $R=\{\mathbf{p}\in [-K, K]\times
[-1,1]:
\lim_{\tau\rightarrow\infty}\mathbf{g}^\tau(\mathbf{p})=\mathbf{q}\}.$
Are distinguished five cases   \footnote{We denote the stable,
unstable and center manifolds of $q$ by ${\cal E}^s(\mathbf{q}),
{\cal E}^u(\mathbf{q})$ and ${\cal E}^c(\mathbf{q})$ respectively.
By $leb(A)$ we denote de Lebesgue measure of
$A\subset\mathbb{R}^2.$}:
\begin{itemize}
\item $\mathbf{q}\in Q^-, \, 0<\gamma<\frac{4}{3},$ or
$\mathbf{q}\in Q^+, \, \frac{4}{3}<\gamma<2,$ then ${\cal
E}^u(\mathbf{q})$ is 2D, hence $R=\emptyset.$ \item $\mathbf{q}\in
Q^-, \, \frac{4}{3}<\gamma<2,$ or $\mathbf{q}\in Q^+, \,
0<\gamma<\frac{4}{3},$ then ${\cal E}^u(\mathbf{q})$ is 1D and
${\cal E}^s(\mathbf{q})$ is 1D is such way that $R\subset N,$ $leb
(N)=0.$ \item $\mathbf{q}\in Q^0,$ then ${\cal E}^c(\mathbf{q})$
is 1D and ${\cal E}^u(\mathbf{q})$ is 1D, hence, $R\subset {\cal
E}^c(\mathbf{q}),$ $leb ({\cal E}^c(\mathbf{q}))=0.$
\end{itemize}

Therefore, all solutions future asymptotic to $\mathbf{q}$ (and
then with $\phi$ bounded towards the future) must lie on an stable
manifold or center manifold of dimension $r<2,$ and then contained
in a subset of $[-K, K]\times [-1,1]$ with zero Lebesgue measure.
Since there are at most a finite number of such $q$ the result of
the theorem follows. $\blacksquare$

\subsection{The flow in the invariant set $\rho_r=0$ as $\phi\rightarrow +\infty$}

As we commented before, in order to investigate the generic asymptotic behavior of the system \eqref{T2.9}-\eqref{T2.12} restricted to $\Sigma$  it is sufficient to study the region where $\phi=\pm\infty.$ Let us remark, however, the region
$\phi=\pm\infty$ is not exclusively associated to the asymptotic
behavior to the past. As has been investigated in section
\ref{Futurequalitative} (see also the degree thesis
\cite{Daniaetal}):
\begin{enumerate}
 \item[i)] If $V\geq 0$ and $V(\phi)=0\Leftrightarrow \phi=0,$ where $\phi=0$ is a local minimum of the potential;
 \item[ii)] If there exists $A$ such that $V$ bounded in $A$ implies $ V'(\phi)$ is bounded in $A;$
 \item[iii)] If there exists some constant $K$ (either positive or negative) such that  $$\chi'(\phi)/\chi(\phi)\leq 2 K/(2-\gamma)(4-3\gamma);$$
\item[iv)] If  $V'(\phi)>0$ for $\phi>0$ and $V'(\phi)<0$ for
$\phi<0;$ then $$\lim_{t\rightarrow\infty} \rho=0=
\lim_{t\rightarrow\infty}\dot\phi$$ and
$\lim_{t\rightarrow\infty}\phi$ exists and it is equal to
$+\infty$, $0$ or $-\infty.$
\end{enumerate}

The case $\lim_{t\rightarrow\infty}\phi=0$ holds only if
$\lim_{t\rightarrow\infty} H(t)=0,$ which can be achieved, for
instance, if $3
H(t_0)^2<\min\left\{\lim_{\phi\rightarrow\infty}V(\phi),
\lim_{\phi\rightarrow -\infty}V(\phi)\right\}.$

As discussed in section \ref{Futurequalitative} these results are
extensions to the non-minimal coupling context of the results
proved in \cite{Miritzis:2003ym} (see Propositions 2 and 3 of that
reference).

In this section we follow the nomenclature and formalism
introduced in \cite{Foster:1998sk}.

\begin{defn}[Function well-behaved at infinity \cite{Foster:1998sk}]\label{WBI}
Let $V:\mathbb{R}\rightarrow \mathbb{R}$ be a $C^2$ non-negative
function. Let there exist some $\phi_0>0$ for which $V(\phi)> 0$
 for all $\phi>\phi_0$ and some number $N$ such that the function
$W_V:[\phi_0,\infty)\rightarrow \mathbb{R}$,
$$ W_V(\phi)=\frac{V'(\phi)}{V(\phi)} - N $$
 satisfies
\begin{equation}
\lim_{\phi\rightarrow\infty}W_V(\phi)=0.\label{Lim}
\end{equation}
Then we say that $V$ is Well Behaved at Infinity (WBI) of
exponential order $N$.
\end{defn}

It is important to point out that $N$ may be 0, or even negative.
Indeed the class of WBI functions of order 0 is of particular
interest, containing all non-negative polynomials as remarked in
\cite{Foster:1998sk}.

\begin{thm}[Theorem 2, \cite{Foster:1998sk}]\label{thm2.6}
Let $V$ be a WBI function of exponential order $N,$ then, for all $\lambda>N,$
$$\lim_{\phi\rightarrow +\infty}e^{-\lambda \phi}V(\phi)=0.$$
\end{thm}

In order to classify the smoothness of WBI functions at infinity
it is introduced the definition

\begin{defn}\label{bar}
Let be some coordinate transformation $\varphi=f(\phi)$ mapping a
neighborhood of infinity to a neighborhood of the origin. If $g$
is a function of $\phi$,
 $\overline{g}$ is the function of $\varphi$ whose domain is the range of
$f$ plus the origin, which takes the values;

$$
\overline{g}(\varphi)=\left\{\begin{array}{rcr} g(f^{-1}(\varphi))&,&\varphi>0\\
                                     \lim_{\phi\rightarrow\infty} g(\phi)&,&\varphi=0 \end{array}\right.
$$
\end{defn}
\begin{defn}[Class k WBI functions \cite{Foster:1998sk}]\label{CkWBI}
A $C^k$ function $V$ is class k WBI if it is WBI and if there
exists $\phi_0>0$ and a coordinate transformation
$\varphi=f(\phi)$ which maps the interval $[\phi_0,\infty)$ onto
$(0, \epsilon]$, where $\epsilon=f(\phi_0)$ and
$\lim_{\phi\rightarrow\infty} f=0$, with  the following additional
properties:
\begin{tabbing}
i)\hspace{0.4cm}\=  $f$ is $C^{k+1}$ and strictly decreasing.\\
ii)            \>the functions $\overline{W_V}(\varphi)$ and
$\overline{f'}(\varphi)$ are $C^k$ on
the  closed interval $[0,\epsilon]$.\\
iii)           \> ${\displaystyle
\frac{\mathrm{d}\overline{W_V}}{\mathrm{d}\varphi}(0)=\frac{\mathrm{d}\overline{f'}}{\mathrm{d}\varphi}(0)=0.}$
\end{tabbing}
\end{defn}

We designate the set of all class k WBI functions ${\cal E}^k_+.$
In table \ref{WBItransf} are displayed simple examples of WBI
behavior at large $\phi.$

\begin{table}
\begin{center}
{\small \caption{\label{WBItransf} Simple examples of WBI behavior
at large $\phi$. $n$ and $\lambda$ are arbitrary constants.
Adapted from \cite{Foster:1998sk}.}\bigskip
\begin{tabular}{l|l|l|l|l}
\hline
 $V(\phi)$&$W_V(\phi)$ & $\varphi=f(\phi)$&$\overline{W_V}(\varphi)$ &$\overline{f'}(\varphi)$ \\ \hline
$\left|\frac{\lambda}{n}\right|\phi^{n}$&$n\phi^{-1}$
 &$\phi^{-\frac{1}{2}}$
 &$n \varphi^2$&
$-\frac{1}{2}\varphi^3$\\[5pt]
$e^{\lambda\phi}$&0&$\phi^{-1}$&0&$-\varphi^2$\\[3pt]
$2e^{\lambda\sqrt{\phi}}$ &$\lambda\phi^{-\frac{1}{2}}$
&$\phi^{-\frac{1}{4}}$ &$\lambda \varphi^2$&$-\frac{1}{4}\varphi^5$\\[3pt]
$\left(A+(\phi-B)^2\right)e^{-\mu\phi}$ &
$\frac{2\left(\phi-B\right)}{A+(B-\phi)^2}$&$\phi^{-\frac{1}{2}}$
 &$-\frac{2\varphi^2\left(B\varphi^2-1\right)}{A \varphi^4+\left(B\varphi^2-1\right)^2}$&
$-\frac{1}{2}\varphi^3$\\[3pt]
$\left(1-e^{-\lambda^2\phi}\right)^2$ & $ -\frac{2 \lambda
^2}{1-e^{\lambda ^2 \phi }}$
&$\phi^{-1}$& $-\frac{2 \lambda ^2}{1-e^{\frac{\lambda ^2}{\varphi }}}$&$-\varphi^2$\\[3pt]
 $\ln{\phi}$&$(\phi\ln\phi)^{-1}$&$(\ln\phi)^{-1}$&$\varphi e^{-\frac{1}{\varphi}}$&$-\varphi e^{-\frac{2}{\varphi}}$\\[3pt]

$\phi^2\ln{\phi}$&$2\phi^{-1} +(\phi\ln\phi)^{-1}$&
$(\ln\phi)^{-1}$&$(2+\varphi)e^{-\frac{1}{\varphi}}$&$-\varphi e^{-\frac{2}{\varphi}}$\\[3pt]
\hline
\end{tabular}}
\end{center}
\end{table}

Let be $V,\chi\in {\cal E}^2_+$ with exponential orders $N$ and
$M,$ respectively. Let be the set $$\Sigma_\epsilon=\left\{
\mathbf{p}\in (\phi, x_1,x_2,x_3)\in\Sigma|
\phi>\epsilon^{-1}\right\},$$ where $\epsilon$ is any positive
constant which is chosen sufficiently small so as to avoid any
points where $V$ or $\chi=0,$ thereby ensuring that
$\overline{W_V}(\vphi)$ and $\overline{W}_{\chi}(\vphi)$ are
well-defined.

Let be defined the  coordinate transformation
\begin{equation}\label{T2.25}(\phi, x_1,x_2,x_3) \stackrel{\vphi=f(\phi)}
{\longrightarrow} (\vphi, x_1,x_2,x_3)\end{equation} over
$\Sigma_\epsilon,$ where $f(\phi)$ satisfy the conditions in
definition \ref{CkWBI} for $k=2.$

Taking the coordinate transformation
$$(\phi, x_1,x_2,x_3) \stackrel{\vphi=f(\phi)} {\longrightarrow}
(\vphi, x_1,x_2,x_3) \stackrel{g} {\longrightarrow} (\vphi,
x_1,x_2),$$ where $g$ is a homeomorphism with inverse $$(\vphi,
x_1,x_2) \stackrel{g^{-1}} {\longrightarrow} \left(\vphi,
x_1,x_2,\sqrt{\frac{3(1-x_2^2-x_3^2)}{\bar{V}(\vphi)}}\right),$$
we obtain the 3-dimensional dynamical system:
\begin{align}
&\vphi'=\sqrt{\frac{2}{3}} \overline{f'}(\vphi) x_1. \label{eqinfphi}\\
&x_1'=x_1^3+\frac{1}{2} \left(x_2^2 \gamma -2\right)
x_1-\frac{(1-x_1^2-x_2^2)}{
\sqrt{6}}\left(\overline{W_V}(\vphi)+N\right)+\nonumber\\
&      \ \ \ \ \ \ \ \ \ \ \ \ \ \ \ +\frac{x_2^2 (4-3
   \gamma)}{2 \sqrt{6}}\left(\overline{W}_{\chi}(\vphi)+M\right),\label{eqinfy}\\
&x_2'=\frac{1}{2} x_2 \left(2 x_1^2+\left(x_2^2-1\right) \gamma
\right)+\frac{x_1 x_2 (-4+3 \gamma )}{2
\sqrt{6}}\left(\overline{W}_{\chi}(\vphi)+M\right).\label{eqinfz}
  \end{align}

We may identify $\Sigma_\epsilon$ with its projection into
$\mathbb{R}^3$ so that we have $\Sigma_\epsilon=\left\{0<\varphi<
f(\epsilon^{-1}),\, 0< x_1^2+x_2^2<1\right\}.$ The variable $x_3$
can be treated as a function on $\Sigma_\epsilon$ defined by the
constraint equation which becomes

\be x_1^2+x_2^2+\frac{1}{3}x_3^2
\overline{V}(\vphi)=1\label{eq35}.\ee

Since $\overline{f'},$ $\overline{W_V}$ and $\overline{W}_{\chi}$
are $C^2$ at $\vphi=0$ we may extend
\eqref{eqinfphi}-\eqref{eqinfz} onto the boundary of
$\Sigma_\epsilon$ to obtain a $C^2$ system on the closure of
$\Sigma_\epsilon$, i.e., $\overline{\Sigma_\epsilon}.$ From
definition \ref{CkWBI}, $\overline{f'},$ $\overline{W_V}$ and
$\overline{W}_{\chi}$ vanish at the origin and are each of second
order or higher in $\vphi$ and $\overline{f'}$ is negative on
$\Sigma_\epsilon.$

\subsubsection{Location, existence and stability conditions of
the singular points. Cosmological
parameters}\label{criticalpoints}

The system \eqref{eqinfphi}-\eqref{eqinfz} admits the singular
points at infinity (i.e., with $\phi$ unbounded) labelled by
$P_i$, $i\in\{1,2,3,4,5,6\}.$ In the following we discuss the
existence and the stability conditions for the singular points. In
the table \ref{crit} are displayed the values of some cosmological
magnitudes of interest for the singular points (the deceleration
parameter, the effective EoS parameter for the total matter, etc).
\footnote{For some specific examples there exist singular points
of \eqref{eqinfphi}-\eqref{eqinfz} with $\varphi>0.$ We omit the
description of them because they do not correspond to the limit
$|\phi|\rightarrow \infty.$ However, as we will see later, some of
these points can attract orbits located initially at the
``infinity boundary''.}

\begin{table*}[htbp]
\begin{center}
\caption{\label{crit} The properties of the singular points for
the system \eqref{eqinfphi}-\eqref{eqinfz}. We use the notations
$\alpha=3\left(N(\gamma-2)+M(3\gamma-4)\right),$ $\beta=2\left(2
N-M(3\gamma-4)\right),$
$\delta=\displaystyle\frac{M(3\gamma-4)}{\sqrt{6}(\gamma-2)},$ and
$\Gamma=\frac{\sqrt{2(\gamma-2)(3\gamma-2)}}{4-3\gamma}.$}\bigskip
{\small\begin{tabular}
{|@{\hspace{0.03in}}c@{\hspace{0.03in}}|@{\hspace{0.03in}}c
@{\hspace{0.03in}}|
@{\hspace{0.03in}}c@{\hspace{0.03in}}|@{\hspace{0.03in}}c
@{\hspace{0.03in}}| @{\hspace{0.03in}}c@{\hspace{0.03in}}|
@{\hspace{0.03in}}c@{\hspace{0.03in}}|@{\hspace{0.03in}}c|} \hline
Point& $x_1$ & $x_2$  & $\Omega_{de}$ & $w_{\text{tot}}$ &
Acceleration?
\\
[0.1cm] \hline
\hline &&&&&\\[-0.3cm]
$P_1$ &-1 &0 & 1&
 1 &  no \\[0.2cm]
\hline&&&&&\\[-0.3cm]
$P_2$ &1 &0  &
1 & 1 &  no \\
[0.2cm]
\hline&&&&&\\[-0.3cm]
$P_3$ &$\delta$ &$\displaystyle{\sqrt{1-\delta^2}}$   &
$\delta^2$ & $\displaystyle{\gamma+(\gamma-1)\delta}$  & $\displaystyle{0<\gamma<\frac{2}{3}}$ \\&&&&& and $\displaystyle{\left|M\right|<\Gamma}$\\[0.2cm]
\hline&&&&&\\[-0.3cm]
$P_4$ &$-\frac{N}{\sqrt{6}}$ & 0  &
1 & $-1+\frac{N^2}{3}$ &  $N^2<2$\\
[0.2cm]
\hline&&&&&\\[-0.3cm]
$P_{5,6}$ &$-\frac{6 \sqrt{6} \gamma }{\beta
}$&$\mp\frac{\sqrt{\frac{2 \beta  (2 \alpha +\beta )}{\gamma }-432
\gamma }}{\beta
   }$& $-\frac{2 (2 \alpha +\beta )}{\beta  \gamma }+\frac{432 \gamma }{\beta ^2}+1$ & $\frac{(\gamma +2) \beta ^2+4 \alpha  (\gamma +1) \beta -432 \gamma
   ^2}{\beta ^2 \gamma }$& $\frac{\alpha}{\beta}<-\frac{1}{3}$\\[0.2cm]
\hline
\end{tabular}}
\end{center}
\end{table*}

\begin{enumerate}

\item The singular point $P_1$ with coordinates $\vphi=0,$
$x_1=-1,$ and $x_2=0$ exists for all the values of the free
parameters. The eigenvalues of the linearized system around $P_1$
are $\lambda_{1,1}=2-\sqrt{2/3}
N,\,\lambda_{1,2}=\frac{2-\gamma}{2}-\frac{M(-4+3\gamma)}{2\sqrt{6}}$
and $\lambda_{1,3}=0.$ Hence the singular point is nonhyperbolic,
then, the Hartman-Grobman theorem does not applies. By the
Center-manifold theorem there exist:
\begin{enumerate}
\item an stable invariant subspace of dimension two (tangent to
the $x_1$-$x_2$ plane) if: i) the potential is a WBI function of
exponential order $ N>\sqrt{6}$ and the coupling function is a WBI
function of exponential order $M<-\frac{\sqrt{6} (\gamma
   -2)}{3 \gamma -4}$ (provided $0<\gamma <\frac{4}{3}$), or
ii) the barotropic index satisfies $\frac{4}{3}<\gamma <2$, the
potential is a WBI function of exponential order $N>\sqrt{6}$ and
the coupling function is a WBI function of exponential order
$M>-\frac{\sqrt{6} (\gamma  -2)}{3 \gamma -4};$ \item an unstable
invariant subspace of dimension two (tangent to the $x_1$-$x_2$
plane) provided the potential is a WBI function of exponential
order $N<\sqrt{6}$ and the coupling function is a WBI function of
exponential order $M$ such that $M>-\frac{\sqrt{6} (\gamma
   -2)}{3 \gamma -4}$ (respectively, $M<-\frac{\sqrt{6} (\gamma   -2)}{3 \gamma -4}$)
provided $0<\gamma <\frac{4}{3}$ (respectively, $\frac{4}{3}<\gamma <2$);
\item a 1-dimensional center manifold which is tangent to the
singular point in the direction of the axis $\vphi$. This center
manifold  can be 2-dimensional  or even 3-dimensional (see the
discussion on the point 3).
\end{enumerate}

\item The singular point $P_2$ with coordinates $\vphi=0,$
$x_1=1,$ and $x_2=0$ exists for all the values of the free
parameters. The eigenvalues of the linearized system around $P_2$
are $\lambda_{2,1}=2+\sqrt{2/3} N,$ $\lambda_{2,2}=\lambda_{1,2}$
and $\lambda_{2,3}=0$ (see point 1). Hence the singular point is
nonhyperbolic, then, the Hartman-Grobman theorem does not applies.
By the Center-manifold theorem there exist:
\begin{enumerate}
\item an stable invariant subspace of dimension two (tangent to
the $x_1$-$x_2$ plane) if: i) $ N<-\sqrt{6},$  $M>\frac{\sqrt{6}
(\gamma
   -2)}{3 \gamma -4}$ for $0<\gamma <\frac{4}{3}$, or
ii)  $\frac{4}{3}<\gamma <2,$  $N<-\sqrt{6}$ and  $M<\frac{\sqrt{6} (\gamma  -2)}{3 \gamma -4};$
\item an unstable invariant subspace of dimension two (tangent to
the $x_1$-$x_2$ plane) provided  $N>-\sqrt{6},$ and $M$ such that
$M<\frac{\sqrt{6} (\gamma
   -2)}{3 \gamma -4}$ (respectively  $M>\frac{\sqrt{6} (\gamma   -2)}{3 \gamma -4}$)
provided $0<\gamma <\frac{4}{3}$ (respectively $\frac{4}{3}<\gamma <2$);
\item a 1-dimensional center manifold which is tangent to the
singular point in the direction of the axis $\vphi$. This center
manifold  can be 2-dimensional or even 3-dimensional (see the
discussion on the point 3).
\end{enumerate}

In the following section we shall study the initial spacetime (big
bang) singularity. The singular points $P_{1,2}$ can account for
that singularity. They are in the same phase portrait for the
values $-\sqrt{6}<N<\sqrt{6}$ and $-\frac{\sqrt{6} (\gamma -2)}{3
\gamma -4}<M<\frac{\sqrt{6} (\gamma   -2)}{3 \gamma -4}$ and
$0<\gamma<\frac{4}{3}$ (in which case they have a 2-dimensional
unstable manifold and a 1-dimensional center center respectively).
It is easy to show that the Hubble parameter (and the matter
density) of the cosmological solutions associated to these points
diverges into the past. The scalar field also diverges, it equals
to $+\infty$ (respectively $-\infty$) for $P_1$ (respectively
$P_2$). However, even in this case, the past attractor corresponds
to $P_1$ since $\overline{f'}<0$ and for $x_1>0$ the orbits enter
the phase portrait and $P_2$ acts as a saddle. The last point can
be a past attractor only on a set of measure zero (if $\vphi=0$).

\item The singular point $P_3$  with coordinates $\vphi=0,$  $x_1=
\frac{M (-4+3 \gamma )}{\sqrt{6} (-2+\gamma )},$ and
$x_2=\sqrt{1-\frac{M^2 (4-3 \gamma )^2}{6(-2+\gamma )^2}}$ exists
if $0<\gamma <\frac{4}{3}$ and $-\frac{\sqrt{6} (-2+\gamma )}{-4+3
\gamma }\leq M\leq \frac{\sqrt{6} (-2+\gamma )}{-4+3 \gamma }.$
The eigenvalues of the matrix of derivatives evaluated at the
singular point are $\lambda_{3,1}=\frac{6 (\gamma -2)^2-M^2 (4-3
\gamma )^2}{12 (\gamma -2)},\; \lambda_{3,2}=-\frac{3 \gamma
M^2}{2}+(M+N) M+\frac{2 (N-M) M}{3 (\gamma -2)}+\gamma,$ and
$\lambda_{3,3}=0.$ Hence the singular point is nonhyperbolic,
then, the Hartman-Grobman theorem does not applies.  Under the
above existence conditions we find, by the Center Manifold
Theorem, that there exists an stable manifold  of dimension two
for the values of the parameters:  i) $M<0$ and $N>\frac{M^2 (4-3
\gamma )^2-6 (\gamma -2) \gamma }{2 M (3 \gamma -4)}$ or ii) $M>0$
and $N<\frac{M^2 (4-3 \gamma )^2-6 (\gamma -2) \gamma }{2 M (3
\gamma -4)}$. Otherwise there exist an unstable manifold of
dimension one (in this case the stable subspace is 1-dimensional).
The center manifold  is  in both cases 1-dimensional.  If
$M=\mp\frac{\sqrt{6} (-2+\gamma )}{-4+3 \gamma }$ this singular
point reduces to $P_{1,2}.$ In this case the center subspace is
the 2-dimensional $\vphi$-$x_2$ plane. The center manifold is
tangent to the center subspace at the singular point. If
additionally $\left|N\right|=\sqrt{6},$ the center manifold is
3-dimensional.

\item The singular point $P_4$ with coordinates $\vphi=0,$
$x_1=-\frac{N}{\sqrt{6}},\, x_2=0$ exists if
$\left|N\right|\leq\sqrt{6}.$ Observe that this point reduces to
$P_{1,2}$ if $N^2=6.$ The eigenvalues of the matrix of derivatives
evaluated at the singular point has the eigenvalues
$\lambda_{4,1}=\frac{1}{6} \left(N^2-6\right)\leq 0,$
$\lambda_{4,2}=\frac{1}{6} N (2 M+N)-\frac{1}{4} (M N+2) \gamma$
and $\lambda_{4,3}=0.$ Hence the singular point is nonhyperbolic
and, as before, the Hartman-Grobman theorem does not applies.
However, we can use the Center Manifold Theorem to investigate the
stability of this singular point. The structure of the center
manifold is as follows:
\begin{enumerate}
 \item if $\lambda_{4,1}<0$ and $\lambda_{4,2}\neq 0$ the center
manifold is tangent to the $\vphi$-axis. Then, it is
1-dimensional. Before analyze this case in detail, we will provide
additional information about the structure of the center manifold;
 \item if $M=\frac{2 \left(N^2-3 \gamma \right)}{N (3 \gamma -4)}$ and $N^2<6$,
the center manifold is tangent to the $\vphi$-$x_2$ plane; \item
if $N^2=6$ and $M\neq\mp\frac{\sqrt{6} (-2+\gamma )}{-4+3 \gamma
},$ it the center manifold is tangent to the $\vphi$-$x_1$ plane;
\item if $N^2=6$ and $M=\mp\frac{\sqrt{6} (-2+\gamma )}{-4+3
\gamma },$ the center manifold is 3-dimensional.
\end{enumerate}

The local behavior described in the cases above (excluding the
first case) is in some way special. It requires fine tuning of the
free parameter. However, the typical behavior is the existence of
a one dimensional center manifold $C_N$ through $P_4$, which is
tangent to the $x_2$-axis (if $\lambda_{4,1}<0$ and
$\lambda_{4,2}\neq 0$). In accordance with our previous comments
on the properties of center manifolds it is clear that $C_N$ is an
exponential attractor on a sufficiently small neighborhood of
$P_4$ and it is intuitively obvious from the geometry (for
instance, observe that in the figure \ref{FIG1} there exists a
line passing through $P_4$ which is an exponential attractor) that
any solutions past asymptotic to $P_4$ must lie on the center
manifold.

Let us investigate the case in which $\lambda_{4,1}<0$ and
$\lambda_{4,2}\neq 0.$ Of course, in this case the stable manifold
is at least 1-dimensional (and as we mentioned before, the center
manifold is 1-dimensional).

The structure of the stable subspace is as follows:

\begin{enumerate}
\item if the potential is of exponential order zero ($N=0$), then,
the singular point has coordinates $(0,0,0).$ The eigenvalues of
the linearization are $\left(-1,0,-\frac{\gamma}{2}\right)$ and in
this case, the stable subspace is tangent to the $x_1$-$x_2$
plane; \item if $0<\gamma<\frac{4}{3},$ $-\sqrt{6}<N<0,$ and
$M>\frac{2 \left(N^2-3 \gamma \right)}{N (3 \gamma -4)};$ or \item
if $\frac{4}{3}<\gamma<2,$ $-\sqrt{6}<N<0,$ and $M<\frac{2
\left(N^2-3 \gamma \right)}{N (3 \gamma -4)};$ or \item if
$0<\gamma<\frac{4}{3},$ $0<N<\frac{4}{3},$ and $M<\frac{2
\left(N^2-3 \gamma \right)}{N (3 \gamma -4)};$ or \item if
$\frac{4}{3}<\gamma<2,$ $0<N<\sqrt{6},$ and $M>\frac{2 \left(N^2-3
\gamma \right)}{N (3 \gamma -4)}$ the stable manifold is is
tangent to the $x_1$-$x_2$ plane; \item By interchanging $>$ and
$<$ in the inequalities for $M$ in the cases (b)-(e) we find that
stable manifold is 1-dimensional and is tangent to the singular
point in the direction of the $x_1$-axis (accordingly to that, the
unstable subspace is spanned by $x_2$-axis).
\end{enumerate}

\item The singular points $P_{5,6}$ with coordinates
$y=\frac{\sqrt{6} \gamma }{M (3 \gamma -4)-2 N},\;
z=\mp\frac{\sqrt{4 N (2
   M+N)-6 (M N+2) \gamma }}{2 N+M (4-3 \gamma )}$
(respectively) exists if the following conditions
are simultaneously satisfied: $4 N (2 M+N)-6 (M N+2) \gamma \geq 0,$
$\mp\left({2 N+M (4-3 \gamma )}\right)> 0$ and
$\frac{4 N^2+M (8-6 \gamma ) N+6 (\gamma -2) \gamma }{(2 N+M (4-3 \gamma
   ))^2}\leq 1$ (i.e., the singular points are real-valued,
 and they are inside the cylinder $\overline{\Sigma_\epsilon}$).

The associated eigenvalues are
$\lambda_{5,6}^\pm=\frac{\alpha}{\beta}\pm\frac{\sqrt{8
\left(\beta^2+27 \gamma ^2\right) \alpha^2-2 \beta (\gamma
   -4) \left(\beta^2-216 \gamma ^2\right) \alpha-(\gamma -2) \left(\beta^2-216 \gamma
   ^2\right)^2}}{6 \sqrt{6} \beta \gamma }$
and $\lambda_{5,6}=0,$ where
$\alpha=3\left(N(\gamma-2)+M(3\gamma-4)\right)$ and
$\beta=2\left(2 N-M(3\gamma-4)\right).$ Assuming that the
conditions for existence are satisfied, we can analyze the
stability of the singular points by means of the Center manifold
theorem. We find that the non null eigenvalues can not be either
complex conjugated with positive real parts or real-valued with
different sign,  then, the unstable subspace of $P_{5,6}$ is the
empty set. Then, the stable subspace is 2-dimensional (provided
$\lambda_{5,6}$ is the only null eigenvalue). When the orbits are
restricted to this invariant set, the point $P_{5,6}$ acts as an
stable spiral (if the eigenvalues are complex conjugated) or as a
node (if the eigenvalues are negative reals). The conditions on
the parameters for those cases are very complicated to displayed
them here. If $M=\frac{2 \left(N^2-3 \gamma \right)}{N (3 \gamma
-4)}$ the points $P_{5,6}$ reduces to $P_4$ but in this case, the
center manifold is 2-dimensional and tangent at the singular point
to the $\vphi$-$x_2$ plane.
\end{enumerate}

\subsection{The flow in the invariant set $\rho_r=0$ near $\phi=-\infty$}\label{minusinfinity2}

With the purpose of complementing the global analysis of the
system \eqref{T2.9}-\eqref{T2.12} defined in $\Sigma$ it is
necessary investigate its behavior near $\phi=-\infty.$ It is an
easy task since the system \eqref{T2.9}-\eqref{T2.12} is invariant
under the transformation of coordinates
\begin{equation}\label{T2.38}(\phi,
x_1)\rightarrow (-\phi, -x_1),\; V\rightarrow U,\; \chi\rightarrow
\Xi,\end{equation} where $U(\phi)=V(-\phi)$ and
$\Xi(\phi)=\chi(-\phi).$ Hence, for a particular potential $V,$
and a particular coupling function $\chi$, the behavior of the
solutions of the equations \eqref{T2.9}-\eqref{T2.12} around
$\phi=-\infty$ is equivalent (except for the sign of $\phi$) to
the behavior of the system near $\phi=\infty$ with potential and
coupling functions $U$ and $\Xi,$ respectively.

If $U$ and $\Xi$ are of class $\mathcal{E}^2_+,$ the preceding
analysis in $\bar{\Sigma}_\epsilon$ can be applied (with and
adequate choice of $\epsilon$).

In the following we will denote $\mathcal{E}^k$ to the set of
class $C^k$ functions well behaved in both $+\infty$ and
$-\infty.$ We will use Latin uppercase letters with subscripts
$+\infty$ and $-\infty,$ respectively to indicate the exponential
order of $\mathcal{E}^k$ functions in $+\infty$ and in $-\infty.$

\subsection{The topological structure of the invariant set $\rho_r=0$ at the
past attractor}\label{estructura}

Let be  $x\in\mathbb{R}_+$ fixed and let be defined the set
\begin{equation}\label{T2.39}
\Sigma(x):=\left\{(\phi, x_1,x_2,x_3)\in\Sigma| x_3<x\right\}.
\end{equation}
Since $x_3$ is a monotonic increasing function of $\tau$ follows
that $\Sigma(x)$ is an invariant set and coincides with the union
of its past orbits. Hence, in order to investigate the topological
properties of the past attractor it is suffice to investigate the
topological properties of $\Sigma(x).$

Let $\left(V,h\right)$ a local chart for $\Sigma,$ and let be
$h|_{V\cap \Sigma(x)}$ the restriction of $h$ to $\Sigma(x).$
Since $\Sigma(x)$ is an open set with respect to the induced
topology, then $\left(V\cap\Sigma(x), h|_{V\cap \Sigma(x)}\right)$
is a local chart for $\Sigma(x).$ From this fact follows that

\begin{rem}
$\Sigma(x)$ is a topological manifold with boundary
$$\partial\Sigma(x):=\left\{\mathbf{p}\in \Sigma|
\phi\in\mathbb{R},x_2=0,0\leq x_3<x\right\}\cup
\left\{\phi\in\mathbb{R},x_2\geq 0,x_3=0\right\},$$ for all
$x\in\mathbb{R}_+.$
\end{rem}

In order to describing the global behavior of the system
\eqref{T2.9}-\eqref{T2.12} towards the past, it is required to
make the immersion of $\Sigma(x)$ in a compact differentiable
4-dimensional manifold $\Omega(x),$ such that the vector field
defined by \eqref{T2.9}-\eqref{T2.12} can be smoothly extended
over $\Omega(x).$ To proceed forward, we define the covering of
$\Sigma(x)$ by open sets with respect to the induced topology as
follows. Let be $\epsilon>0,$ a real number, an let be defined the
sets
\begin{eqnarray}
\Sigma(x,\epsilon)^-&=&\left\{\mathbf{p}\in \Sigma(x)|
\phi<-\epsilon^{-1}\right\},\label{T2.40}\\
\Sigma(x,\epsilon)&=&\left\{\mathbf{p}\in \Sigma(x)|
-1-\epsilon^{-1}<\phi<1+\epsilon^{-1}\right\},\label{T2.41}\\
\Sigma(x,\epsilon)^+&=&\left\{\mathbf{p}\in \Sigma(x)|
\phi>\epsilon^{-1}\right\}\label{T2.42}.
\end{eqnarray}
From the inequality
$-1-\epsilon{-1}<-\epsilon^{-1}<1+\epsilon^{-1}$ follows that
\eqref{T2.40}-\eqref{T2.42} is a covering of $\Sigma(x).$ By
construction they are open sets with respect to the induced
topology.

Let be defined the sets:
\begin{eqnarray}
\Omega(x,\epsilon)&=&\left\{\left(\vphi, x_1,x_2,x_3\right)\in
\mathbb{R}^4| -1-\epsilon^{-1}<\varphi<1+\epsilon^{-1},
\right. \nonumber\\
&& \left. -1<x_1<1,0<x_2<1, 0<x_3<x\right\}.\label{T2.40b}
\end{eqnarray}
This set contains a submanifold of dimension $3$ which is
homeomorphic under the identity map to $\Sigma(x,\epsilon).$
\begin{eqnarray}
\Omega(x,\epsilon)^+&=&\left\{\mathbf{p}=\left(\vphi,
x_1,x_2,x_3\right)\in \mathbb{R}^4| 0<\varphi<f(\epsilon^{-1}),
\right. \nonumber\\
&& \left. -1<x_1<1, 0<x_2<1, 0<x_2<x\right\},\label{T2.40bb}
\end{eqnarray}
(where $f$ is the function defined in definition \eqref{CkWBI}).
This set contains a submanifold of dimension $3$ which is
homeomorphic under the map $$(\varphi,
x_1,x_2,x_3)\stackrel{\phi=f^{-1}(\varphi)} {\longrightarrow}
\left(\phi,
x_1,x_2,\sqrt{\frac{3\left(1-x_1^2-x_2^2\right)}{V(\phi)}}\right)$$
to $\Sigma(x,\epsilon)^+.$  Finally, it is constructed a set
$\Omega(x,\epsilon)^-$ in an analogous way as proceeded to
construct $\Omega(x,\epsilon)^+$ using the successive coordinate
transformations \eqref{T2.38} and \eqref{T2.25} with the
identifications $U(\phi)=V(-\phi)$ and $\Xi(\phi)=\chi(-\phi).$
The interior of $\Omega(x)$ it is defined by
$$\Omega(x,\epsilon)^{-}\cup \Omega(x,\epsilon)\cup \Omega(x,\epsilon)^{+}.$$
The construction is completed by attaching a boundary, denoted by
$\partial\Omega(x),$ which is defined taking the union of
$x_3=0,\, x_3=x,\, \varphi=0$ and the circumference
$x_1^2+x_2^2=1$ to each local chart.

By construction, $\Omega(x)$ is compact and it is embedded in
$\mathbb{R}^4.$

Thus, the vector field defined by \eqref{T2.9}-\eqref{T2.12}  can
be smoothly extended over the boundary of $\Omega(x)$ such that
$\Omega(x)$ is the union of its past orbits. It is important to
note that $\Omega(x)$ approaches the non-physical boundary
$\partial \Omega,$ along the intersection of the plane $x_3=0$
with the plane $\varphi=0$ and the circumference $x_1^2+x_2^2=1.$

\subsubsection{The initial space-time singularity.}

In this section we will study the initial space-time (Big-Bang)
singularity. The singular points $P_{1,2}$ can represent such a
singularity. They live at the same phase space for the values of
$M,$ $N$ and $\gamma$ in the intervals $-\sqrt{6}<N<\sqrt{6},$
$-\frac{\sqrt{6} (\gamma   -2)}{3 \gamma -4}<M<\frac{\sqrt{6}
(\gamma   -2)}{3 \gamma -4}$ and $0<\gamma<\frac{4}{3}$ (in this
case, they have an unstable 2-dimensional manifold and a center
1-dimensional manifold). It is easy to show that the Hubble
parameter and the matter energy density of the associated
cosmological solutions diverge towards the past. The scalar field
diverges too, and it is equal to $+\infty$ and to $-\infty$ for
$P_1$ and $P_2$ respectively. However, even in this case, the
possible past attractor corresponds to $P_1$ since
$\overline{f'}<0$ whereas for  $x_1>0$ the orbits enter the phase
space and $P_2$ acts as a saddle. The singular point $P_2$ can act
as a past attractor only in a set of initial conditions of measure
zero (when $\varphi=0$).

\subparagraph{Analysis near $P_1.$}

From the analysis in section \ref{criticalpoints}, it seems
reasonable to think that the initial space-time singularity can be
associated to the singular point $P_1.$ Its unstable manifold is
2-dimensional provided $N<\sqrt{6}.$

The asymptotic behavior of neighboring solutions to $P_1$ can be
approximated, for $\tau$ negative large enough, as \be
x_1(\tau)=-1+O(e^{\lambda_{1,1}\tau}),\,
x_2(\tau)=O(e^{\lambda_{1,2}\tau})\label{yzapprox}.\ee  By
substitution of (\ref{yzapprox}) in \eqref{T2.9}, and integrating
the resulting equation, we obtain
\be\phi(\tau)=\sqrt{\frac{2}{3}}\left(-\tau+\tilde{\phi}\right)+O(e^{\lambda_{1,1}\tau})\label{phiapprox}.\ee
Then, by expanding around $\tau=-\infty$ up to first order, we get
\begin{align*} &\varphi=f\left(\sqrt{\frac{2}{3}}\left(-\tau+\tilde{\phi}\right)+
O(\frac{1}{\tau})^2\right)+O(e^{\lambda_{1,1}\tau})\nonumber\\&=
f\left(\sqrt{\frac{2}{3}}\left(-\tau+\tilde{\phi}\right)\right)+O(e^{\lambda_{1,1}\tau})+h,\end{align*}
where $h$ denotes higher order terms to be discarded.

Then we have a first order solution to (\ref{eqinfphi}-\ref{eqinfz}). Also, by substitution of (\ref{yzapprox}) in
\eqref{T2.12} and solving the resulting differential equation with initial condition $x(0)=x_0$ we get the first order
solution \be x=x_0 e^\tau.\label{xfirstorder}\ee Then, we have $t-t_i=\frac{1}{3}\int x(\tau) d\tau=1/3 x_0 e^\tau.$
For simplicity let us set $t_i=0.$

Neglecting the error terms, we have the following expressions
\begin{eqnarray}
H&=&x^{-1}=\left(x_0 e^\tau\right)^{-1}=\frac{1}{3 t},\;
\phi=\sqrt{\frac{2}{3}}\left(-\tau+\tilde{\phi}\right)=-\sqrt{\frac{2}{3}}\ln\frac{ t}{c},\nonumber\\
\dot\phi&=&-\sqrt{\frac{2}{3}}
t^{-1},\;\rho=0,\label{massless}\end{eqnarray} where
$c=1/3x_0e^{\tilde{\phi}}.$ This asymptotic solution corresponds
to the exact solution of the cosmological equations when $V$ vanishes identically and $\chi$ is a
constant (the minimal coupling case). Hence, there exists a
generic class of massless minimally coupled scalar field
cosmologies in a vicinity of the initial space-time singularity.

The above idea can be stated, more precisely, as the

\begin{thm}[Local singularity theorem]\label{initialsingularity} Let be $V\in
{{\mathcal E}}^2_+$ with exponential order $N$  such that
$N<\sqrt{6}$ and let be $\chi \in {{\mathcal E}}^2_+$ with exponential order
$M$ such that

\begin{enumerate}
\item $0<\gamma <\frac{4}{3}$ y $M>\left(2\ N-\sqrt{6}\gamma
\right)/\left(3\gamma -4\right)$ or

\item $\frac{4}{3}<\gamma <2$  y $M<\left(2\ N-\sqrt{6}\gamma
\right)/\left(3\gamma -4\right)$
\end{enumerate}

Then, there exists a neighborhood ${\mathcal N}(P_1)$ of  $P_1$ such that
for all $p\in {\mathcal N}\left(P_1\right),$ the orbit ${\psi
}_p$ is past asymptotic to $P_1$ and the associated cosmological solution is given by:

\begin{equation} \label{GrindEQ__2_43_}
H=\frac{1}{3t}+O\left({\varepsilon }_V(t)\right),
\end{equation}

\begin{equation} \label{GrindEQ__2_44_}
\phi=-\sqrt{\frac{2}{3}}\ln\frac{t}{c}+O\left(t{\varepsilon
}_V(t)\right),
\end{equation}

\begin{equation} \label{GrindEQ__2_45_}
\dot{\phi}=-\sqrt{\frac{2}{3}}t^{-1}+O\left({\varepsilon
}_V(t)\right),
\end{equation}

\begin{equation} \label{GrindEQ__2_46_}
\rho =\frac{b^2_0}{3}t^{-\gamma }\chi
{\left(-\sqrt{\frac{2}{3}}\ln\frac{t}{c}\right)}^{\frac{3\gamma
}{2}-2}\left(1+O\left(t{\varepsilon }_V(t)\right)\right),
\end{equation}

where ${\varepsilon
}_V(t)=tV\left(-\sqrt{\frac{2}{3}}\ln\frac{t}{c}\right).$
\end{thm}

{\bf Comments}. Since $V\in{\cal E}_+^2$ has exponential order $N,$ then, applying theorem \ref{thm2.6}, we have
$$\lim_{t\rightarrow 0}t^{\alpha}V\left(-\sqrt{\frac{2}{3}}\ln\frac{t}{c}\right)=\lim_{\phi\rightarrow\infty}e^{-\sqrt{\frac{3}{2}}\alpha \phi} V(\phi)=0,\, \forall \alpha>\sqrt{\frac{2}{3}}N.$$ Thus, for $N<\sqrt{6},$ the error terms
$O(\varepsilon_V(t))$ and $O(t\varepsilon_V(t))$ in \eqref{GrindEQ__2_43_}-\eqref{GrindEQ__2_46_} are dominated by the first order terms. If $N<\sqrt{\frac{3}{2}},$ both terms tends to zero as $t\rightarrow 0^+.$

Since $\chi\in {\cal E}_+^2$ has exponential order $M,$ then:

\begin{enumerate}
\item The term
$t^{-\gamma}\chi\left(-\sqrt{\frac{2}{3}}\ln\frac{t}{c}\right)^{\frac{3\gamma}{2}-2}
O(t\varepsilon_V(t))=O(t^{2-\gamma+\frac{M(4-3\gamma)}{\sqrt{6}}-\sqrt{\frac{2}{3}}N})$
tends to zero as $t\rightarrow 0^+,$ in the cases
\begin{enumerate}
\item $0<\gamma<\frac{4}{3}, \, N\leq \sqrt{\frac{3}{2}},\, M>\frac{(2N-\sqrt{6}\gamma)}{3\gamma-4}$ or
\item $0<\gamma<\frac{4}{3}, \, \sqrt{\frac{3}{2}}<N\leq\sqrt{6},\, M>-\frac{(2N-\sqrt{6}(2-\gamma))}{3\gamma-4}$ or
\item $\frac{4}{3}<\gamma<2, \, N\leq \sqrt{\frac{3}{2}},\, M<\frac{(2N-\sqrt{6}\gamma)}{3\gamma-4}$ or
\item $\frac{4}{3}<\gamma<2, \, \sqrt{\frac{3}{2}}<N\leq\sqrt{6},\, M<-\frac{(2N-\sqrt{6}(2-\gamma))}{3\gamma-4}.$
\end{enumerate}
\item The term
$t^{-\gamma}\chi\left(-\sqrt{\frac{2}{3}}\ln\frac{t}{c}\right)^{\frac{3\gamma}{2}-2}$
tends to zero as $t\rightarrow 0^+,$ in the cases
\begin{enumerate}
\item $0<\gamma<\frac{4}{3},\, M>-\frac{\sqrt{6}\gamma}{3\gamma-4}$ or
\item  $\frac{4}{3}<\gamma<2,\, M<-\frac{\sqrt{6}\gamma}{3\gamma-4}.$
\end{enumerate}
\end{enumerate}

First, us prove the

\begin{lem}\label{Lemma2} Let be $V\in {{\mathcal
E}}^k_{{\rm +}}$ a function of exponential order  $N.$ Let
$n>\sqrt{\frac{2}{3}}\ N$ and $\ \lambda
>0.$ Let $\varphi =f(\phi)$ the coordinate transformation referred in the definition
\eqref{CkWBI}. Then, if $\tau \to
-\infty,$ are valid the estimates
\begin{equation}\label{A2i}\int_{-\infty}^\tau{\overline{V}\left(\varphi \right)e^{n\tau }\mathrm{d}\tau =\
\frac{3}{3n-\sqrt{6}N}\ \overline{V}\left(\varphi \right)e^{n\tau
}+h,}\end{equation} and
\begin{equation}\label{A2iii}\int_{-\infty}^\tau{e^{\lambda \tau
}\left(\overline{W_V}\left(\varphi \right)+N\right)\mathrm{d}\tau
=\frac{N}{\lambda}e^{\lambda \tau }+h.}\end{equation}

where $h$ denotes terms of higher order to be discarded.
\end{lem}

{\bf Proof}. We proceed similarly as in the proof of theorem 4 in \cite{Foster:1998sk}.

\begin{enumerate}
\item Let us define the function $T(\phi)=V(\phi)e^{-N \phi}.$
\footnote{If $V$ is of exponential order $N,$ then $T$ is of
exponential order zero since
$W_T(\phi)=T'(\phi)/T(\phi)=W_V(\phi)\rightarrow 0$ as
$\phi\rightarrow +\infty.$ Hence, for all $\lambda>0,$
$\lim_{\tau\rightarrow +\infty}e^{-\lambda\phi}T(\phi)=0.$} Since
$V$ is of exponential order $N$ follows from $\phi=
f^{-1}(\varphi(\tau))\approx-\sqrt{\frac{2}{3}}(\tau-\tilde{\phi})$
that $$\int{\overline{V}\left(\varphi \right)e^{n\tau
}\mathrm{d}\tau} = \int{\overline{T}\left(\varphi
\right)e^{\left(-\sqrt{\frac{2}{3}}N+n\right)\tau
}\mathrm{d}\tau}+h$$  where $h$ denotes terms of higher order to
be discarded. By an explicit computation we have that
\be\label{A6} {\mathrm{d}\ln \overline{T}\left(\varphi \right)
}=\frac{1}{f'\left(f^{-1}(\varphi)\right)}\left(\frac{V'\left(f^{-1}(\varphi)\right)}{V\left(f^{-1}(\varphi)\right)}-N\right){\mathrm{d}\varphi}=\frac{\overline{W_V}(\varphi)}{\overline{f'}(\varphi)}{\mathrm{d}\varphi}.\ee
Using \eqref{eqinfphi} and  \eqref{A6} we obtain \be\label{A7}
{\mathrm{d}\overline{T}\left(\varphi \right)
}=\sqrt{\frac{2}{3}}x_1 \overline{T}(\varphi)
\overline{W_V}(\varphi){\mathrm{d}\tau}.\ee

Integrating by parts gives
\begin{equation}\label{A8}
I(\tau)=\int_{-\infty}^\tau\overline{T}
(\varphi)e^{\left(-\sqrt{\frac{2}{3}}N+n\right)\tau}\mathrm{d}\tau=I_1(\tau)+I_2(\tau)
\end{equation} where
\begin{eqnarray}I_1(\tau)&=&\left.\frac{3}{3n-\sqrt{6}N}\overline{T}
(\varphi)e^{\left(-\sqrt{\frac{2}{3}}N+n\right)\tau}\right|_{-\infty}^\tau\nonumber\\
&=&\frac{3}{3n-\sqrt{6}N}\overline{V}
(\varphi)e^{n\tau}+h,\nonumber\end{eqnarray} and
\begin{eqnarray}I_2(\tau)&=&-\frac{3}{3 n-\sqrt{6}N}\int_{-\infty}^\tau
e^{\left(-\sqrt{\frac{2}{3}}N+n\right)\tau}\mathrm{d}\overline{T}(\varphi)\nonumber\\&=&
-\frac{2}{\sqrt{6}n-2N}\int_{-\infty}^\tau
e^{\left(-\sqrt{\frac{2}{3}}N+n\right)\tau}x_1\overline{T}(\varphi)\overline{W_V}(\varphi)\mathrm{d}\tau.
\nonumber\end{eqnarray}

Consider the second term $I_2(\tau)$. Let $\delta(\tau)=\sup_{\tau'<\tau}\frac{2}{\sqrt{6}n-2N}\left|\overline{W_V}\left(\varphi(\tau')\right)\right|$, then, recalling
that $x_1^2 < 1$;
\begin{eqnarray}\left|I_2(\tau)\right|&=&\frac{2}{\sqrt{6}n-2N}\left|\int_{-\infty}^\tau e^{\left(-\sqrt{\frac{2}{3}}N+n\right)\tau}x_1\overline{T}(\varphi)\overline{W_V}(\varphi)\mathrm{d}\tau\right|\nonumber\\
&<& \delta(\tau)\left|I(\tau)\right|\nonumber\\
&\leq& \delta(\tau)\left(\left|I_1(\tau)\right|+\left|I_2(\tau)\right|\right)\nonumber\\
&<& \frac{\delta(\tau)}{1-\delta(\tau)}\left|I_1(\tau)\right|\nonumber.
\nonumber\end{eqnarray}
for $\tau$ sufficiently small. Letting $\tau$ go to $-\infty$ demonstrates \eqref{A2i}, as required.

\item Let $$I(\tau)=\int_{-\infty}^\tau
e^{\lambda\tau}\frac{V'\left(f^{-1}(\varphi)\right)}{V\left(f^{-1}(\varphi)\right)}\mathrm{d}\tau.$$
Integrating by parts we have that
\begin{eqnarray}I(\tau)&=&\frac{1}{\lambda}
\left.
e^{\lambda\tau}\frac{V'\left(f^{-1}(\varphi)\right)}{V\left(f^{-1}(\varphi)\right)}\right|_{-\infty}^\tau
-\frac{1}{\lambda}\int_{-\infty}^\tau
e^{\lambda\tau}\mathrm{d}\left[\frac{V'\left(f^{-1}(\varphi)\right)}{V\left(f^{-1}(\varphi)\right)}\right]\nonumber\\&=&
\left.\frac{1}{\lambda}e^{\lambda \tau
}\left(\overline{W_V}\left(\varphi
\right)+N\right)\right|_{-\infty}^\tau-\frac{1}{\lambda}\int_{-\infty}^\tau
e^{\lambda\tau}\mathrm{d}\left[\frac{V'\left(f^{-1}(\varphi)\right)}{V\left(f^{-1}(\varphi)\right)}\right]
\nonumber\\&=&I_1(\tau)+I_2(\tau)\nonumber,\end{eqnarray} where
$I_1(\tau)=\left.\frac{1}{\lambda}e^{\lambda \tau
}\left(\overline{W_V}\left(\varphi
\right)+N\right)\right|_{-\infty}^\tau=\frac{1}{\lambda}e^{\lambda
\tau }\left(\overline{W_V}\left(\varphi
\right)+N\right)=\frac{N}{\lambda}e^{\lambda \tau }+h,$ since
$\overline{W_V}\left(\varphi \right)$ goes to zero, as
$\tau\rightarrow-\infty;$  and

\begin{eqnarray}
&&I_2(\tau)=-\frac{1}{\lambda}\int_{-\infty}^\tau e^{\lambda\tau}\mathrm{d}
\left[\frac{V'\left(f^{-1}(\varphi)\right)}{V\left(f^{-1}(\varphi)\right)}\right]\nonumber\\
&&= -\frac{1}{\lambda}\int_{-\infty}^\tau e^{\lambda \tau
}g(\varphi)\left(\frac{\mathrm{d}
\left(\ln\left(g(\varphi)\right)\right)}
{\mathrm{d}\varphi}\frac{\mathrm{d}\varphi}{\mathrm{d}\tau}\right)\mathrm{d}\tau\nonumber\\
&&=-\frac{1}{\lambda}\sqrt{\frac{2}{3}}\int_{-\infty}^\tau
e^{\lambda \tau
}g(\varphi)\left(\frac{\mathrm{d}\left(\ln\left(g(\varphi)\right)\right)}{\mathrm{d}\varphi}\overline{f'}(\varphi)x_1\right)\mathrm{d}\tau.\nonumber
\end{eqnarray}
where $$g(\varphi)=\overline{W_V}\left(\varphi \right)+N.$$ Let
$$\delta_1(\tau)=\sup_{\tau'<\tau}\frac{1}{\lambda}\sqrt{\frac{2}{3}}\left|\overline{f'}(\varphi)\frac{\mathrm{d}}{\mathrm{d}\varphi}\left(\ln\left(\overline{W_V}\left(\varphi(\tau')
\right)+N\right)\right)\right|,$$ then, recalling that $x_1^2 <
1$;
\begin{eqnarray}\left|I_2(\tau)\right|
&<& \delta(\tau)\left|I(\tau)\right|\nonumber\\
&\leq& \delta(\tau)\left(\left|I_1(\tau)\right|+\left|I_2(\tau)\right|\right)\nonumber\\
&<& \frac{\delta(\tau)}{1-\delta(\tau)}\left|I_1(\tau)\right|\nonumber.
\nonumber\end{eqnarray}
for $\tau$ sufficiently small. Letting $\tau$ go to $-\infty$ demonstrates \eqref{A2iii}, as required.

\end{enumerate}
$\blacksquare$

{\bf Proof of theorem \ref{initialsingularity}} Let us consider
the system given by the differential equations
\eqref{eqinfphi}-\eqref{eqinfz}  and \eqref{T2.12}. Using the
restriction \eqref{eq35} and making the coordinate transformation
$x_1\rightarrow -1+u$ the system  \eqref{eqinfphi}-\eqref{eqinfz},
\eqref{T2.12} can be expressed in integral form as

\begin{align}
&\varphi(\tau)=-\sqrt{\frac{2}{3}}\int_{-\infty}^\tau
\overline{f'}(\vphi(s))\left(1-u(s)\right)\mathrm{d} s\label{3.90}\\
&u(\tau)=u_{0} e^{\lambda_{1,1}\tau}+\int_{-\infty}^\tau
e^{\lambda_{1,1}(\tau-s)} h_1\left(\varphi(s),u(s),x_2(s),x_3(s)\right)\mathrm{d} s,\label{3.91}\\
&x_2(\tau)=x_{2 0} e^{\lambda_{1,2}\tau} +\int_{-\infty}^\tau
e^{\lambda_{1,2}(\tau-s)} h_2\left(\varphi(s),u(s),x_2(s),x_3(s)\right)\mathrm{d} s,\label{3.92}\\
&x_3(\tau)=x_{3 0} e^{\tau}+\int_{-\infty}^\tau e^{(\tau-s)}
h_3\left(\varphi(s),u(s),x_2(s),x_3(s)\right)\mathrm{d} s,
\label{3.93}
\end{align}
for $\tau\rightarrow -\infty.$ Where
\begin{align}
&h_1(\varphi, u,x_2,x_3)=u^3-3
   u^2+\frac{\gamma  x_2^2}{2}\left(u-1\right)
 -\sqrt{\frac{2}{3}}
   u \overline{W_V}(\varphi
   ) +\nonumber\\&+\left(\frac{u^2}{\sqrt{6}}+\frac{x_2^2}{\sqrt{6}}\right)(N+\overline{W_V}(\varphi
   ))+\frac{x_2^2 (4-3 \gamma ) (M+\overline{W_\chi}(\varphi ))}{2
   \sqrt{6}}
 ,\\
&h_2(\varphi, u,x_2,x_3)=x_2 u^2-2 x_2 u+\frac{1}{2}
   \gamma  x_2^3-\frac{
   x_2 (3 \gamma -4) \overline{W_\chi}(\varphi )}{2
   \sqrt{6}}+\nonumber\\&+\frac{u
   x_2 (3 \gamma -4) (M+\overline{W_\chi}(\varphi ))}{2
   \sqrt{6}},\\
&h_3(\varphi,
u,x_2,x_3)=-\frac{1}{3}x_3^3\overline{V}(\varphi)-\frac{1}{2}(2-\gamma)x_2^2
x_3.\end{align} To derive \eqref{3.90}-\eqref{3.93}  we have used
the fact that
$$\lim_{\tau\rightarrow -\infty} \left(\varphi(\tau),e^{-\lambda_{1,1}\tau}u(\tau),
e^{-\lambda_{1,2}\tau}x_2(\tau),e^{-\tau}x_3(\tau)\right)=(0,x_{10},x_{20},x_{30}),$$
where $u_{0}>0, x_{20}>0,x_{30}\neq 0$ are sufficiently small real
constants. This fact can be proved using the first order
approximations, as $\tau\rightarrow -\infty$:
$f^{-1}(\varphi(\tau))\approx-\sqrt{\frac{2}{3}}(\tau-\tilde{\phi}),
u(\tau)=u_{0}e^{\lambda_{1,1}\tau},x_2(\tau)=x_{20}e^{\lambda_{1,2}\tau},
x_3(\tau)=x_{30}e^{\tau}.$

To obtain the required expansions, we substitute in the equations
\eqref{3.90}-\eqref{3.93} the above first order approximations;
then, will be suffice to estimate integrals of types
\begin{equation}\int_{-\infty}^\tau{\overline{V}\left(\varphi(s) \right)e^{n\tau }d\tau},\, \int_{-\infty}^\tau \overline{W_V}
(\varphi(s)) \mathrm{d} s, \, \int_{-\infty}^\tau e^{\lambda
s}\left(N+\overline{W_V} (\varphi(s))\right) \mathrm{d} s,
\label{3.106}\end{equation} as $\tau\rightarrow -\infty$ where
$n=2$ and
$\lambda=\left\{\lambda_{1,1},2\lambda_{1,2}-\lambda_{1,1}\right\}$
where we are assuming that $V$ is a generic WBI function of
exponential order $N$ and $\lambda>0.$

Since $\overline{W_V}\left(\varphi \right)$ goes to zero, as
$\tau\rightarrow-\infty;$ the second integral in \eqref{3.106} is
an infinitesimal of higher order as $\tau\rightarrow-\infty.$ In
fact, $\int_{-\infty}^\tau
\overline{W_V}(\varphi)\mathrm{d}\tau=\int_{-\infty}^\tau
\overline{W_T}(\varphi)\mathrm{d}\tau +h=\int_{-\infty}^\tau
\frac{{T'}(f^{-1}(\varphi))}{{T}(f^{-1}(\varphi))}\mathrm{d}\tau
+h.$ Using the first order solution $\phi=
f^{-1}(\varphi(\tau))\approx-\sqrt{\frac{2}{3}}(\tau-\tilde{\phi}),$
the variation of constants formulae and the smoothness of
$T(\phi)$ as $\phi\rightarrow +\infty$ we have that
$\int_{-\infty}^\tau
\overline{W_V}(\varphi)\mathrm{d}\tau=-\sqrt{\frac{2}{3}}\ln
\left[\frac{T(-\sqrt{\frac{2}{3}}\left(\tau-\tilde{\phi}\right))}{\lim_{\phi\rightarrow
+\infty} T(\phi)}\right]\rightarrow 0$ as $\tau\rightarrow
-\infty$.

Using the above procedure and the results of lemma \ref{Lemma2}  we obtain
\begin{align}&x_3(\tau)=x_{30}e^{\tau}-\frac{x_{3 0}^3}{3} e^{\tau}  \int_{-\infty}^\tau
e^{2 s}\overline{V}(\varphi(s))\mathrm{d}
s+\nonumber\\ & \ \ \ \ \ \ \ \ \ \ \ \ \ \ \ \ \ \ -\frac{1}{2}(2-\gamma)x_{20}^2 x_{3 0}e^\tau\int_{-\infty}^\tau e^{2\lambda_{1,2}s}\mathrm{d}s\nonumber\\
&=x_{30}e^{\tau}-\frac{x_{3 0}^3}{3\lambda_{1,1}} e^{\tau} e^{2
\tau}\overline{V}(\varphi(\tau))-\frac{(2-\gamma)x_{20}^2 x_{3
0}}{4\lambda_{1,2}}e^\tau e^{2\lambda_{1,2}\tau}+
h\label{3.104}.\end{align} Since $V$ is of exponential order $N$
follows (at first order) that
$e^{2\tau}\overline{V}(\varphi)=O(e^{\lambda_{1,1}\tau}).$ Then,
from the hypothesis $2\lambda_{1,2}>\lambda_{1,1}$ follows that
the third term in the second line of equation \eqref{3.104} is an
infinitesimal or higher order than the second one as
$\tau\rightarrow -\infty.$ Thus,
\begin{eqnarray}x_3(\tau)&=& x_{30}e^{\tau}-\frac{\left(
x_{30}e^{\tau}\right)^3}{3\lambda_{1,1}}
\overline{V}(\varphi(\tau))+h.\label{3.105}\end{eqnarray}

Integrating again to obtain a second order expression for $t$ and
using the fact that $t\rightarrow 0^+$ as $\tau\rightarrow
-\infty,$ we get
\begin{align}
&3t=\int_{-\infty}^\tau x_3(s)\mathrm{d}s={
x_{30}}e^\tau-\frac{x_{30}^3}{3\lambda_{1,1}}\int_{-\infty}^\tau
e^{3 s}\overline{V}(\varphi(s))\mathrm{d} s+h\nonumber\\
     &= {
     x_{30}}e^\tau-\frac{x_{30}^3}{\lambda_{1,1}(9-\sqrt{6}N)}
e^{3
\tau}\overline{V}(\varphi(\tau))+h\nonumber\\
    &= \left({
     x_{30}}e^\tau\right)-\frac{\left({
     x_{30}}e^\tau\right)^3}{\lambda_{1,1}(9-\sqrt{6}N)}\overline{V}(\varphi(\tau))+h
\end{align}
This expression may be inverted, to second order, to give:
\begin{equation}{
     x_{30}}e^\tau= 3 t+\frac{27 t^3}{\lambda_{1,1}(9-\sqrt{6}N)}{V}(\phi(
     t))+h,\label{tau-t}
\end{equation}
Applying the inversion formula to $x_3$ as given by formula \eqref{3.105} follows:
\begin{eqnarray}\label{3.110}
x_3(t) &=&3 t-\frac{27}{(9-\sqrt{6}N)}{ t^3}{V}(\phi( t)) +h.
\end{eqnarray}
Hence,
\begin{equation}\label{3.111}
H(t)=x_3(t)^{-1}=\frac{1}{3 t}+\frac{3 V(\phi(t))
t}{(9-\sqrt{6}N)}+h.
\end{equation}

Using the formula \eqref{tau-t} we can obtain a second order
approximation for $\rho$ as follows.

Integrating equation \eqref{eq4} for $\rho$ we obtain the
expression
$$\rho=\rho_0 a^{-3 \gamma}\chi^{-3+\frac{3\gamma}{2}},$$ where
$\rho_0 $is an integration constant. Recalling $\tau=\ln a^3$
follows that $$\rho=\frac{\rho_0 x_{3 0}^\gamma
\chi(\phi)^{-3+\frac{3\gamma}{2}}}{\left(x_{3
0}e^\tau\right)^\gamma}.$$ Substituting back the formula
\eqref{tau-t} we obtain
\begin{equation}\label{3.112}
\rho(t)=\frac{1}{3} b_0^2 t^{-\gamma } \left(\frac{9 \gamma
   V\left(\phi(t)\right)
   t^2}{\left(\sqrt{6} N-9\right) \lambda_{1,1}}+1\right) \chi
   \left(\phi(t)\right)^{\frac{3 \gamma }{2}-2},
\end{equation} where $b_0=3^{\frac{1-\gamma }{2}} x_{3 0}^{\gamma /2}
   \sqrt{\rho_0}.$

In an analogous way as we have obtained a second order estimate
for $x_3(\tau)$  we proceed for $u(\tau).$

\begin{align} & u(\tau)=\frac{e^{3 {\lambda_{1,1}} \tau } u_0^3}{2 {\lambda_{1,1}}}
+\frac{e^{2 {\lambda_{1,1}} \tau } \left(\sqrt{6}
   N-18\right) u_0^2}{6 {\lambda_{1,1}}}+e^{{\lambda_{1,1}} \tau } u_0+\nonumber\\+&{x_{2 0}}^2
   \left(\frac{e^{({\lambda_{1,1}}+2 {\lambda_{1,2}}) \tau } u_0 \gamma }{4 {\lambda_{1,2}}}+\frac{1}{2} e^{2
   {\lambda_{1,2}} \tau }\right)
\label{3.114}\end{align}

We have the following three cases:

\begin{enumerate}
\item  $$\lambda_{1,2}<\lambda_{1,1}<2\lambda_{1,2}$$ implies
$$\lambda_{1,1}<2\lambda_{1,2}<2\lambda_{1,1}<\lambda_{1,1}+2\lambda_{1,2}<3\lambda_{1,1};$$
thus
\begin{equation} u(\tau)=e^{{\lambda_{1,1}} \tau } u_0+\frac{{x_{2
0}}^2}{2}
  e^{2
   {\lambda_{1,2}} \tau }+h
\label{3.114a}.\end{equation} Then, from the hypothesis
$2\lambda_{1,2}>\lambda_{1,1}$ follows that the second term in
equation \eqref{3.114a} is an infinitesimal or higher order than
the first one as $\tau\rightarrow -\infty.$ Thus
\begin{equation} u(\tau)=e^{{\lambda_{1,1}} \tau } u_0+h
\label{3.114a1}.\end{equation}

\item  $$\frac{2\lambda_{1,2}}{3}<\lambda_{1,1}\leq\lambda_{1,2}$$
implies $$\lambda_{1,1}<2\lambda_{1,1}\leq
2\lambda_{1,2}<3\lambda_{1,1}<\lambda_{1,1}+2\lambda_{1,2};$$ thus
\begin{equation} u(\tau)=e^{{\lambda_{1,1}} \tau } u_0+\frac{e^{2 {\lambda_{1,1}} \tau } \left(\sqrt{6}
   N-18\right) u_0^2}{6 {\lambda_{1,1}}}+\frac{{x_{2
0}}^2}{2}
  e^{2
   {\lambda_{1,2}} \tau }+h
\label{3.114b}\end{equation} Then, from the hypothesis
$2\lambda_{1,2}>\lambda_{1,1}$ follows that the third term in
equation \eqref{3.114b} is an infinitesimal or higher order than
the second one as $\tau\rightarrow -\infty.$ Thus
\begin{equation}
u(\tau)=e^{{\lambda_{1,1}} \tau } u_0+\frac{e^{2 {\lambda_{1,1}}
\tau } \left(\sqrt{6}
   N-18\right) u_0^2}{6 {\lambda_{1,1}}}+h
\label{3.114b1}\end{equation}

\item $$0<\lambda_{1,1}\leq \frac{2\lambda_{1,2}}{3}$$ implies
$$\lambda_{1,1}<2\lambda_{1,1}<3\lambda_{1,1}\leq
2\lambda_{1,2}<\lambda_{1,1}+2\lambda_{1,2};$$ thus
\begin{equation} u(\tau)=e^{{\lambda_{1,1}} \tau } u_0+\frac{e^{2 {\lambda_{1,1}} \tau } \left(\sqrt{6}
   N-18\right) u_0^2}{6 {\lambda_{1,1}}}+\frac{e^{3 {\lambda_{1,1}} \tau } u_0^3}{2 {\lambda_{1,1}}}+\frac{{x_{2
0}}^2}{2}
  e^{2
   {\lambda_{1,2}} \tau }+h
\label{3.114c}\end{equation} Then, from the hypothesis
$2\lambda_{1,2}>\lambda_{1,1}$ follows that the fourth term in
equation \eqref{3.114c} is an infinitesimal or higher order than
the third one as $\tau\rightarrow -\infty.$ Thus
\begin{equation}
u(\tau)=e^{{\lambda_{1,1}} \tau } u_0+\frac{e^{2 {\lambda_{1,1}}
\tau } \left(\sqrt{6}
   N-18\right) u_0^2}{6 {\lambda_{1,1}}}+\frac{e^{3 {\lambda_{1,1}} \tau } u_0^3}{2 {\lambda_{1,1}}}+h
\label{3.114c1}\end{equation}
\end{enumerate}

In the previous three cases  $$u(\tau)=P(e^{{\lambda_{1,1}} \tau }
u_0)+h$$ with $P(x)=x+A x^2+B x^3,$ where $A=B=0$ in case 1;
$A=\frac{\sqrt{6}N-18}{6\lambda_{1,1}}$ and $B=0$ in case 2; and
in the case 3, $A=\frac{\sqrt{6}N-18}{6\lambda_{1,1}},
B=\frac{1}{2\lambda_{1,1}}.$ Then,
\begin{equation}\label{3.120}u(t)=P(a_0 t^{\lambda_{1,1}})-\frac{9 a_0 t^{\lambda_{1,1}+2} \left(2 A a_0 t^{\lambda_{1,1}}+3 a_0^2 B t^{2
   \lambda_{1,1}}+1\right) V(\phi (t))}{\sqrt{6}
n-9},\end{equation} where $a_0=3^{\lambda_{1,1}} u_0
\left(\frac{1}{x_{3 0}}\right)^{\lambda_{1,1}}.$

Substituting the expansions \eqref{3.120} and \eqref{3.110} in the
expression $$\dot\phi=\frac{\sqrt{6}(u-1)}{x_3}$$ we obtain
\begin{eqnarray}\label{3.121}
&&\dot\phi=-\frac{\sqrt{\frac{2}{3}}}{t}+\frac{3 \sqrt{6} V(\phi
(t)) t}{\sqrt{6}
   N-9}+\left(\frac{\sqrt{\frac{2}{3}}
a_0}{t}-\frac{6 \sqrt{6} a_0 t V(\phi (t))}{\sqrt{6} N-9}\right)
t^{\lambda_{1,1}}+\nonumber\\ &&+\left(\frac{\sqrt{\frac{2}{3}} A
a_0^2}{t}-\frac{9 \sqrt{6} A a_0^2 t V(\phi (t))}{\sqrt{6}
N-9}\right)
   t^{2 \lambda_{1,1}}+\nonumber\\ &&+\left(\frac{\sqrt{\frac{2}{3}} a_0^3 B}{t}-\frac{12 \sqrt{6} a_0^3 B t V(\phi
   (t))}{\sqrt{6} N-9}\right) t^{3 \lambda_{1,1}}+h
\end{eqnarray}

Finally, from  $$\phi'(\tau)\equiv
\sqrt{\frac{2}{3}}(u(\tau)-1),$$ and \eqref{3.114c1} follows, by
integration,
\begin{eqnarray}
&&\phi(\tau)=
-\sqrt{\frac{2}{3}}(\tau-\tilde{\phi})+\sqrt{\frac{2}{3}}\int_{-\infty}^\tau
u(\tau)\mathrm{d}\tau\nonumber\\
&&= -\sqrt{\frac{2}{3}}(\tau-\tilde{\phi})+\frac{e^{3
\lambda_{1,1} \tau } u_0^3}{3 \sqrt{6} \lambda_{1,1}^2}+\frac{e^{2
\lambda_{1,1} \tau }
   \left(N-3 \sqrt{6}\right) u_0^2}{6 \lambda_{1,1}^2}+\nonumber\\
&&+\frac{\sqrt{\frac{2}{3}} e^{\lambda_{1,1} \tau }
   u_0}{\lambda_{1,1}}.
\end{eqnarray}
Using the inversion formula \eqref{tau-t} we obtain
\begin{eqnarray}\label{3.123}
&&\phi(t)=-\sqrt{\frac{2}{3}} \ln
   \left(\frac{t}{c}\right)+ \left(\frac{\sqrt{\frac{2}{3}}
a_0}{\lambda_{1,1}}-\frac{3 \sqrt{6} a_0 t^2
V\left(\phi(t)\right)}{\left(\sqrt{6} N-9\right)
\lambda_{1,1}}\right)
t^{\lambda_{1,1}}+\nonumber\\&&+\left(\frac{a_0^2 \left(N-3
   \sqrt{6}\right)}{6 \lambda_{1,1}^2}-\frac{3 a_0^2 \left(N-3 \sqrt{6}\right) t^2 V\left(\phi(t)\right)}{\left(\sqrt{6} N-9\right) \lambda_{1,1}^2}\right) t^{2 \lambda_{1,1}}+\nonumber\\&&+\left(\frac{a_0^3}{3
   \sqrt{6} \lambda_{1,1}^2}-\frac{3 \sqrt{\frac{3}{2}} a_0^3 t^2 V\left(\phi(t)\right)}{\left(\sqrt{6} N-9\right) \lambda_{1,1}^2}\right) t^{3 \lambda_{1,1}}
\end{eqnarray}
where $c=1/3x_0e^{\tilde{\phi}}.$

Now, Taylor-expanding $V$ and $\chi$ around
$\phi^\star=-\sqrt{\frac{2}{3}}\ln\left(\frac{t}{c}\right)$ and
substituting the results in the left hand side of equations
\eqref{3.111}, \eqref{3.123}, \eqref{3.121}, and \eqref{3.112}
completes the proof. $\blacksquare$

\subsubsection{A global singularity theorem}

Finally, we will state (without a rigorous proof) a global singularity theorem which is in some way an extension of
Theorem 6 in \cite{Foster:1998sk} (page 3501). It is not totally an extension of this theorem, since in our framework it
is very difficult to prove that the correspondence with the massless minimally coupled scalar field cosmologies is
one-to-one.

The theorem states the following:

\begin{thm}[Global singularity theorem]\label{globalinitialsingularity} Let be $V\in
{{\mathcal E}}^2$ with exponential orders $N_{\pm\infty}$ as $\phi\rightarrow\pm\infty$  such that
$N_{\pm\infty}<\sqrt{6}$ and let be $\chi \in {{\mathcal E}}^2$ with exponential orders
$M_{\pm\infty}$ as $\phi\rightarrow\pm\infty$  such that

\begin{enumerate}
\item $0<\gamma <\frac{4}{3}$ y $M_{\pm\infty}>\left(2\ N_{\pm\infty}-\sqrt{6}\gamma
\right)/\left(3\gamma -4\right)$ or

\item $\frac{4}{3}<\gamma <2$  y $M_{\pm\infty}<\left(2\ N_{\pm\infty}-\sqrt{6}\gamma
\right)/\left(3\gamma -4\right)$
\end{enumerate}

Then, it is verified asymptotically that:

\begin{equation} \label{globalGrindEQ__2_43_}
H=\frac{1}{3t}+O\left({\varepsilon }_V^\pm(t)\right),
\end{equation}

\begin{equation} \label{globalGrindEQ__2_44_}
\phi=\pm\sqrt{\frac{2}{3}}ln\frac{t}{c}+O\left(t{\varepsilon
}_V^\pm(t)\right),
\end{equation}

\begin{equation} \label{globalGrindEQ__2_45_}
\dot{\phi}=\pm\sqrt{\frac{2}{3}}t^{-1}+O\left({\varepsilon
}_V^\pm(t)\right),
\end{equation}

\begin{equation} \label{globalGrindEQ__2_46_}
\rho =\frac{b^2_0}{3}t^{-\gamma }\chi
{\left(-\sqrt{\frac{2}{3}}ln\frac{t}{c}\right)}^{\frac{3\gamma
}{2}-2}\left(1+O\left(t{\varepsilon }_V^\pm(t)\right)\right),
\end{equation}

where ${\varepsilon
}_V^\pm(t)=tV\left(\pm\sqrt{\frac{2}{3}}ln\frac{t}{c}\right).$
\end{thm}

{\bf Sketch of the proof of theorem
\ref{globalinitialsingularity}}

Following the same reasoning as in \cite{Foster:1998sk}, it is
suffice to prove that all the solution, but, perhaps, a set of
Lebesgue measure zero, are past asymptotic to the singular point
$P_1$ (for $\phi\rightarrow +\infty$ or $\phi\rightarrow
-\infty$). Since $x_3$ is monotonic increasing, it is enough to
consider orbits in $\Omega(x)$ for $x$ arbitrary. Since
$\Omega(x)$ is compact and contains its past orbits; then, all the
points $\mathbf{p}\in\Omega(x)$ should have a nonempty
$\alpha$-limit set, $\alpha(\mathbf{p})$.  Particularly, for all
the points in the physical state space $\Omega(x),$ theorem
\ref{Theorem_2.4} implies that $\alpha(\mathbf{p})$ must contain
almost always a singular point with $\varphi=0$
($\phi=\pm\infty$). By the discussion in the section
\ref{estructura}, each point with $\varphi=0$ being a limit point
of the physical trajectory, must be part of the non-physical
boundary $\partial\Omega(x)$ and then must have $x_3=0.$ Since
$x_3$ is monotonically increasing, the set $\alpha(p)$ must be
contained completely in the plane $x_3=0,$ or namely in
$\partial\Omega(x).$ It can be proved that the only conceivable
generic past attractor are the singular points $P_1$ in
$\pm\infty$ (the other singular points cannot be generic sources
by our previous analysis in section \ref{criticalpoints}).

\section{Early-time behavior for the model including radiation}

In this section we investigate the early-time dynamics of the
general model by including radiation.

\subsection{Normalized Variables and Dynamical System}

As a difference with the previous study we introduce the new
dimensionless variables

\be \sigma_1=\phi,
\sigma_2=\frac{\dot\phi}{\sqrt{6}H},\,\sigma_3=\frac{\sqrt{\rho}}{\sqrt{3}H},\,\sigma_4=\frac{\sqrt{V}}{\sqrt{3}H},
\,\sigma_5=\frac{\sqrt{\rho_r}}{\sqrt{3}H}\label{vars}\ee and the time coordinate \be \mathrm{d}\tau=3 H \mathrm{d}t.\ee We considered the scalar field itself as a dynamical variable.

Using these coordinates the equations \eqref{Raych}-\eqref{KG}
recast as an autonomous system satisfying an inequality arising
from the Friedmann equation \eqref{Fried}. This system is given by

\begin{align}
&\sigma_1'=\sqrt{\frac{2}{3}} \sigma_2 \label{eq0phi}\\
&\sigma_2'=\sigma_2^3+\frac{1}{6}\left(3\gamma \sigma_3^2+4 \sigma_5^2-6\right)\sigma_2
-\frac{\sigma_4^2}{\sqrt{6}}\frac{\mathrm{d}\ln V(\sigma_1)}{\mathrm{d}\sigma_1}
+\nonumber\\& \ \ \ \ \ \ \ \ \ \ \ \ \ \ \ \ \ +\frac{\left(4-3\gamma\right)\sigma_3^2}{2\sqrt{6}}\frac{\mathrm{d}\ln \chi(\sigma_1)}{\mathrm{d}\sigma_1},\label{eq0x1}\\
&\sigma_3'=\frac{1}{6}\sigma_3\left(6\sigma_2^2+3\gamma\left(\sigma_3^2-1\right)+4 \sigma_5^2\right)
+\nonumber\\& \ \ \ \ \ \ \ \ \ \ \ \ \ \ \ \ \ -\frac{\left(4-3\gamma\right)\sigma_2 \sigma_3}{2\sqrt{6}}\frac{\mathrm{d}\ln \chi(\sigma_1)}{\mathrm{d}\sigma_1},\label{eq0x2}\\
&\sigma_4'=\frac{1}{6}\sigma_4\left(6\sigma_2^2+3\gamma
\sigma_3^2+4 \sigma_5^2\right)
+\frac{\sqrt{6}}{6}\sigma_2 \sigma_4 \frac{\mathrm{d}\ln V(\sigma_1)}{\mathrm{d}\sigma_1},\label{eq0x3}\\
&\sigma_5'=\frac{1}{6}\sigma_5\left(6\sigma_2^2+3\gamma
\sigma_3^2+4 \sigma_5^2-4\right).\label{eq0x4}
\end{align}
The system \eqref{eq0phi}-\eqref{eq0x4} defines a flow in the phase space

\be \Sigma :=\left\{\sigma\in\mathbb{R}^5: \sum_{j=2}^5
\sigma_j^2=1, \sigma_j\geq 0, j=3, 4, 5\right\}\label{Sigma}.\ee

Now, let us proceed to investigate the topological properties of
the phase space. Knowing the topological structure of the phase
space allows to a better understanding of the dynamics and
provides the geometrical basis for the proof of our main results.

\subsection{The Topological Properties of the Phase Space}

Let us define the sets $\Sigma_0:=\{\sigma\in\Sigma: \sigma_5=0\}$
and $\Sigma_+=\{\sigma\in\Sigma: \sigma_5>0\}.$ By construction
these sets are a partition of $\Sigma.$

\begin{prop}\label{thm1}
$\Sigma_0$ is a manifold with boundary.
\end{prop}

{\bf Proof}. Since $\Sigma_0\subset\mathbb{R}^4$ is a closed set
with respect the usual topology of $\mathbb{R}^4,$ it is a
Hausdorff space equipped with a numerable basis. The rest of the
proof requires the construction of a set of local charts.

Let us define the sets $V_j:=\left\{\sigma\in\mathbb{R}^4:
\sigma_j>0\right\}\cap\Sigma_0,\, j=3,4.$ These sets are open with
respect to the induced topology in $\Sigma_0.$

Let be defined the projection maps
$$h_3:V_3\rightarrow\mathbb{H}^3,\, \sigma\rightarrow
h_3(\sigma)=(\sigma_1,\sigma_2,\sigma_4),$$ and
$$h_4:V_4\rightarrow\mathbb{H}^3,\, \sigma\rightarrow
h_4(\sigma)=(\sigma_1,\sigma_2,\sigma_3).$$ These maps satisfy
$h_j(V_j)=\mathbb{R}\times (\mathbb{H}^2\cap \mathbb{D}^2),
j=3,4,$ which are open sets of $\mathbb{H}^3.$ Their inverses are
given by
$$h_3^{-1}:\mathbb{R}\times (\mathbb{H}^2\cap \mathbb{D}^2)\rightarrow V_3,
\\ (\sigma_1,\sigma_2,\sigma_4)\rightarrow \left(\sigma_1,\sigma_2,\sqrt{1-\sigma_2^2-\sigma_4^2},\sigma_4\right),$$
and $$h_4^{-1}:\mathbb{R}\times (\mathbb{H}^2\cap \mathbb{D}^2)\rightarrow V_4,\, (\sigma_1,\sigma_2,\sigma_3)\rightarrow\left(\sigma_1,\sigma_2,\sigma_3,\sqrt{1-\sigma_2^2-\sigma_3^2}\right).$$ It is clear that they are homoemorphisms.

Observe that $\left\{(V_3,h_3), (V_4,h_4)\right\}$ does not cover
the sets with $\sigma_3=\sigma_4=0.$ The construction is completed by defining the sets $$V_1^\pm:=\left\{\sigma\in\Sigma_0:
\sigma_2=\pm 1, \sigma_3=\sigma_4=0\right\}$$ which are disjoint copies of $\mathbb{R},$ and thus they are 1-dimensional manifolds.
$\blacksquare$

By definition $(V_3,h_3), (V_4,h_4)$ are topological manifolds. It
can be proved that, in fact, each one is a topological manifold
with boundary. The boundaries are $$\partial
V_3=\left\{\sigma\in V_3: \sigma_4=0\right\},$$ and
$$\partial V_4=\left\{\sigma\in V_4: \sigma_3=0\right\}.$$
Both are homeomorphic to $\mathbb{R}\times (-1,1).$

Let be defined $V_1:=V_1^- \cup V_1^+.$ Observe that
\ben\Sigma_0=\left(\Sigma_0\setminus V_1\right)\cup
V_1=\text{Int}\left(\Sigma_0\setminus V_1\right)\cup(\partial
V_3\cup\partial V_4)\cup
V_1\nonumber\\=\text{Int}\left(\Sigma_0\setminus
V_1\right)\cup\left(\partial\Sigma_0\right)_1\cup\left(\partial\Sigma_0\right)_2,\label{relation}\een
where we have defined $\left(\partial\Sigma_0\right)_1=\partial
V_4\cup V_1=\left\{\sigma\in\Sigma: \sigma_3=0\right\}$ and
$\left(\partial\Sigma_0\right)_2=\partial V_3\cup
V_1=\left\{\sigma\in\Sigma: \sigma_4=0\right\}.$

From the above arguments and expression \eqref{relation} we have:

\begin{rem}

\begin{itemize}
\item The interior of $\Sigma_0$ is given by
$\text{Int}\Sigma_0=\text{Int}\left(\Sigma_0\setminus V_1\right)$
which is a 3-dimensional manifold (without boundary). \item The
boundary of $\Sigma_0$ is the union of two 2-dimensional
topological manifolds with boundary given by
$\left(\partial\Sigma_0\right)_1$ and
$\left(\partial\Sigma_0\right)_2.$ \item
$\left(\partial\Sigma_0\right)_1$ and
$\left(\partial\Sigma_0\right)_2$  share the same boundary $V_1$
which is the union of two disjoint copies of $\mathbb{R}.$
\end{itemize}

\end{rem}

\begin{prop}\label{thm2}
$\Sigma_+$ is a topological manifold with boundary.
\end{prop}

{\bf Proof}. It is easy to check that
$\Sigma_+\subset\mathbb{R}^5$ is a Hausdorff space equipped with a
numerable basis. The rest of the proof requires the construction
of a set of local charts.

Let us define the sets $W_j:=\left\{\sigma\in\mathbb{R}^5:
\sigma_j>0\right\}\cap\Sigma_+,\, j=3,4.$ These sets are open with
respect to the induced topology in $\Sigma_+.$

Let us define the maps
$$g_3:W_3\rightarrow\mathbb{H}^4,\, \sigma\rightarrow g_3(\sigma)
=\left(\sigma_1,\sigma_2,\sigma_5,\sigma_4\right),$$ and
$$g_4:W_4\rightarrow\mathbb{H}^4,\, \sigma\rightarrow g_4(\sigma)
=\left(\sigma_1,\sigma_2,\sigma_5,\sigma_3\right).$$ These maps
satisfy $g_j(W_j)=\mathbb{R}\times (\mathbb{H}^3\cap
\mathbb{D}^3),\, j=3,4$ which are open set of $\mathbb{H}^4.$
Their inverses are given by
\begin{align*} &g_3^{-1}:\mathbb{R}\times (\mathbb{H}^3\cap
\mathbb{D}^3)\rightarrow W_3, \nonumber\\
&\left(\sigma_1,\sigma_2,\sigma_5,\sigma_4\right) \rightarrow
\left(\sigma_1,\sigma_2,\sqrt{1-\sigma_2^2-\sigma_4^2-\sigma_5^2},\sigma_4,\sigma_5\right),\end{align*}
and \begin{align*} & g_4^{-1}:\mathbb{R}\times (\mathbb{H}^3\cap
\mathbb{D}^3)\rightarrow W_4,\nonumber\\
&\left(\sigma_1,\sigma_2,\sigma_5,\sigma_3\right)\rightarrow
\left(\sigma_1,\sigma_2,\sigma_3,\sqrt{1-\sigma_2^2-\sigma_3^2-\sigma_5^2},\sigma_5\right).\end{align*}
It is clear that they are homeomorphism.

Observe that $\left\{(W_3,g_3), (W_4,g_4)\right\}$ do not cover
the sets with $\sigma_3=\sigma_4=0.$ The construction is
completed by defining the set $W_1:=\left\{\sigma\in\Sigma_+:
\sigma_1^2+\sigma_5^2=1\right\}.$ Using the projection map
$(\sigma_1,\sigma_2,\sigma_5)\stackrel{g_1}{\rightarrow}
(\sigma_1,\sigma_2)$ it is easily proved that $W_1$ is
homeomorphic to the open set $\mathbb{R}\times (-1,1).$
$\blacksquare$

We have that $W_3$ is a manifold with boundary. Its boundary is
the set $\partial W_3:=\{\sigma\in\Sigma: \sigma_3>0, \sigma_4=0,
\sigma_5>0\}$ and its interior is the set $\{\sigma\in\Sigma:
\sigma_3>0, \sigma_4>0, \sigma_5>0\}.$ Also, $W_4$ is a manifold
with boundary. Its boundary is the set $\partial
W_4:=\{\sigma\in\Sigma: \sigma_3=0, \sigma_4>0, \sigma_5>0\}$ with
same interior as $W_3$ and $W_1$ is a manifold without boundary
which is homeomorphic to $\mathbb{R}\times (-1,1).$

Let us define the sets $$\left(\partial
\Sigma_+\right)_1:=\partial W_4 \cup W_1$$ and
$$\left(\partial \Sigma_+\right)_2:=\partial W_3 \cup W_1.$$
Following the same arguments yielding to \eqref{relation} we get
the formula \ben\Sigma_+=\text{Int}\left(\Sigma_+\setminus
W_1\right)\cup\left(\partial\Sigma_+\right)_1\cup\left(\partial\Sigma_+\right)_2\label{relation2}.\een
Using the above arguments and relation \eqref{relation2} we have
the following

\begin{rem}

\begin{itemize}
\item The interior of $\Sigma_+$ is given by
$\text{Int}\Sigma_+=\text{Int}\left(\Sigma_+\setminus W_1\right)$
which is a 4-dimensional manifold (without boundary). \item The
boundary of $\Sigma_+$ is the union of two 3-dimensional
topological manifolds with boundary given by
$\left(\partial\Sigma_+\right)_1$ and
$\left(\partial\Sigma_+\right)_2.$ \item
$\left(\partial\Sigma_+\right)_1$ and
$\left(\partial\Sigma_+\right)_2$ share the same boundary $W_1$
which is a 2-dimensional manifold without boundary homeomorphic to
$\mathbb{R}\times (-1,1).$
\end{itemize}

\end{rem}

\subsection{Monotonic Functions}

The construction of monotonic functions in the state space is an
important tool in any phase space analysis. The existence of such
functions can rule out periodic orbits, homoclinic orbits, and
other complex behavior in invariant sets. If so, the dynamics is
dominated by singular points (and possibly, heteroclinic orbits
joining it). Additionally, some global results can be obtained. In
table \ref{monotonic} are displayed several monotonic functions
for the flow of \eqref{eq0phi}-\eqref{eq0x4} in $\Sigma$.

\begin{table}[!htb]
\caption{\label{monotonic} Monotonic functions for the flow of
\eqref{eq0phi}-\eqref{eq0x4} in \eqref{Sigma}.}
{\small\begin{tabular}[t]{|l|c|c|c|c|} \hline
$Z$&$\mathrm{d}Z/\mathrm{d}\tau$&Invariant set&Restrictions$^{\rm a}$ \\[1ex]
\hline
\hline &&& \\[-2ex]
$Z_1=\left(\frac{\sigma_3}{\sigma_4}\right)^2 V(\sigma_1) \chi(\sigma_1)^{2-\frac{3\gamma}{2}}$&$-\gamma Z_1$
&$\sigma_3>0, \sigma_4>0$ & $\gamma\in \left(0,\frac{4}{3}\right)\cup\left(\frac{4}{3},2\right)$\\[1ex]
\hline
\hline &&& \\[-2ex]
$Z_2=\left(\frac{\sigma_5}{\sigma_3}\right)^2 \chi(\sigma_1)^{-2+\frac{3\gamma}{2}}$
&$-\left(\frac{4}{3}-\gamma\right) Z_2$&$\sigma_3>0, \sigma_5>0$ & $\gamma\in \left(0,\frac{4}{3}\right)\cup\left(\frac{4}{3},2\right)$\\[1ex]
\hline
\hline &&& \\[-2ex]
$Z_3=\left(\frac{\sigma_5}{\sigma_4}\right)^2 V(\sigma_1)$&$-\frac{4}{3} Z_3$&$\sigma_4>0, \sigma_5>0$ & none\\[1ex]
\hline
\hline &&& \\[-2ex]
$Z_4=\frac{\sigma_2^2}{1-\sigma_2^2}$&$-\frac{2}{3}
Z_4$&$\begin{array}{c}\sigma_2\neq 0, \sigma_3=0\\ \sigma_4=0,
\sigma_5>0
\end{array}$ & none \\[1ex]

\hline
\end{tabular}}
\begin{flushleft}
$^{\rm a}$ We assume the general conditions $\chi, V\in
C^3,\,\chi(\sigma_1)>0,V(\sigma_1)>0$.
\end{flushleft}
\end{table}

\begin{rem}
From the definition of $Z_1$ it follows that it is a monotonic
decreasing function for the flow of \eqref{eq0phi}-\eqref{eq0x4}
restricted to the invariant set $\sigma_3>0, \sigma_4>0.$ Applying
the Monotonicity Principle (theorem \ref{theorem 4.12}) it follows
that the past attractor the flow of \eqref{eq0phi}-\eqref{eq0x4}
restricted to the invariant set $\sigma_3>0, \sigma_4>0$ is
contained in the set where $\sigma_4=0$ and the future attractor
in contained in the set where $\sigma_3=0.$
\end{rem}

\begin{rem}
Using the same argument follows from the definition of $Z_2$ that
the past attractor of the flow of \eqref{eq0phi}-\eqref{eq0x4}
restricted to the invariant set $\sigma_3>0, \sigma_5>0$ is
contained in the invariant set where $\sigma_3=0$ and the future
asymptotic attractor is contained in the invariant set
$\sigma_5=0$ provided $\gamma<\frac{4}{3}.$ If
$\gamma>\frac{4}{3}$ the asymptotic behavior is the reverse of the
previously described.
\end{rem}

\begin{rem}
From the definition of $Z_3$ follows that the past attractor of
the flow of \eqref{eq0phi}-\eqref{eq0x4} restricted to the
invariant set $\sigma_4>0, \sigma_5>0$ is contained in the
invariant set where $\sigma_4=0$ and the future asymptotic
attractor is contained in the invariant set $\sigma_5=0.$
\end{rem}

\begin{rem}
From the definition of $Z_4$ follows that the past attractor of
the flow of \eqref{eq0phi}-\eqref{eq0x4} restricted to the
invariant set $\sigma_2\neq 0, \sigma_3=\sigma_4=0, \sigma_5>0$ is
contained in the invariant set where $\sigma_2=0$ (i.e., where
$\sigma_5=1$) and the future asymptotic attractor is contained in
the invariant set $\sigma_2=\pm 1.$
\end{rem}

\subsection{Singular points with $\phi$
Bounded}\label{Qualitative}

Let us make a preliminary analysis of the linear stability of the
singular points of the flow of \eqref{eq0phi}-\eqref{eq0x4}
defined in $\Sigma.$ It is a classic result that the linear
stability of the singular points does not change under
homeomorphisms.

Since $\Sigma$ is a 4-dimensional manifold (with boundary) we will consider the projection of $\Sigma$ in a real 4-dimensional manifold (with boundary).

Let be defined the projection map

\ben && \pi: \Sigma \rightarrow \Omega \nonumber\\
&&(\sigma_1, \sigma_2,\sigma_3,\sigma_4,\sigma_5)\rightarrow
(\sigma_1, \sigma_2,\sigma_3,\sigma_5)\een where
\be\Omega:=\left\{\sigma\in\mathbb{R}^4:
\sigma_2^2+\sigma_3^2+\sigma_5^2\leq 1, \sigma_j\geq 0, j= 3,
5\right\}.\ee

The flow of \eqref{eq0phi}-\eqref{eq0x4} defined on $\Sigma$ is topologically equivalent (under $\pi$) to the flow of

\begin{align}
&\sigma_1'=\sqrt{\frac{2}{3}} \sigma_2 \label{eqsigma1}\\
&\sigma_2'=\sigma_2^3+\frac{1}{6}\left(3\gamma \sigma_3^2+4 \sigma_5^2-6\right)\sigma_2-\frac{(1-\sigma_2^2-\sigma_3^2-\sigma_5^2)}{\sqrt{6}}\frac{\mathrm{d}\ln V(\sigma_1)}{\mathrm{d}\sigma_1}+\nonumber\\
& \ \ \ \ \ \ \ \ \ \ \ \ \ \ \ \ +\frac{\left(4-3\gamma\right)\sigma_3^2}{2\sqrt{6}}\frac{\mathrm{d}\ln \chi(\sigma_1)}{\mathrm{d}\sigma_1},\label{eqsigma2}\\
&\sigma_3'=\frac{1}{6}\sigma_3\left(6\sigma_2^2+3\gamma\left(\sigma_3^2-1\right)+4 \sigma_5^2\right)-\frac{\left(4-3\gamma\right)\sigma_2 \sigma_3}{2\sqrt{6}}\frac{\mathrm{d}\ln \chi(\sigma_1)}{\mathrm{d}\sigma_1},\label{eqsigma3}\\
&\sigma_5'=\frac{1}{6}\sigma_5\left(6\sigma_2^2+3\gamma \sigma_3^2+4 \sigma_5^2-4\right),\label{eqsigma5}
\end{align}
defined in $\Omega.$

\begin{table}[!htb]
\begin{center}
{\caption{\label{crit0A} Location of the singular points of the
flow of \eqref{eqsigma1}-\eqref{eqsigma5} defined in $\Omega$.}
\begin{tabular}[t]{|l|c|c|c|c|}
\hline
Label&$\sigma_1$$^{\rm a}$ &$\sigma_2$&$\sigma_3$&$\sigma_5$\\[1ex]
\hline
\hline &&& \\[-2ex]
$Q_1$&$\sigma_{1c}: \chi'(\sigma_{1c})=0$&$0$&$1$&$0$\\[1ex]
\hline
\hline &&& \\[-2ex]
$Q_2$&$\sigma_{1c}: V'(\sigma_{1c})=0$&$0$&$0$&$0$\\[1ex]
\hline
\hline &&& \\[-2ex]
$Q_3$&$\sigma_{1c}\in\mathbb{R}$&$0$&$0$&$1$\\[1ex]
\hline
\end{tabular}}
\end{center}
\begin{center}
$^{\rm a}$ We are assuming $V(\sigma_{1c})\neq
0,\,\chi(\sigma_{1c})\neq 0$ and $\gamma\in
\left(0,\frac{4}{3}\right)\cup\left(\frac{4}{3},2\right).$
\end{center}
\end{table}

The system \eqref{eqsigma1}-\eqref{eqsigma5} defined in $\Omega$
admits three classes of singular points located at  $\Omega,$
denoted by $Q_j,\, j=1,2,3.$  In table \ref{crit0A} are displayed
the coordinates of such points. The dynamics near the singular
points (and its stability properties) is dictated by the signs of
the real parts of the eigenvalues of the Jacobian matrix evaluated
at each singular point as follows:

\begin{enumerate}

\item The eigenvalues of the linearization around $Q_1$ are
$-\Delta_2,\,\gamma,\,\Delta_1\pm\sqrt{\Delta_1^2
+\Delta_2\frac{\chi''(\phi_1)}{\chi(\phi_1)}},$ where
$\Delta_1=(-2+\gamma)/{4}< 0,$ and
$\Delta_2=\left(4-3\gamma\right)/6.$ Then, the local stability of
singular point $P_1$ is as follows:

\begin{enumerate}
\item If $-\frac{\Delta_1^2 \chi (\sigma_{1c} )}{\Delta_2}\leq \chi
   ''(\sigma_{1c} )<0$ and $0< \gamma< \frac{4}{3}$ there exist a 3-dimensional
   stable manifold and a 1-dimensional unstable manifold of $Q_1.$
\item If $0<\gamma<\frac{4}{3}$ and $\chi
   ''(\sigma_{1c} )>0$ or $\frac{4}{3}<\gamma<2$ and $0<\chi
   ''(\sigma_{1c} )<-\frac{\Delta_1^2 \chi (\sigma_{1c} )}{\Delta_2},$
   there exist a 2-dimensional stable manifold and a 2-dimensional unstable manifold of $Q_1.$
\item If $0<\gamma<\frac{4}{3}$ and $\chi
   ''(\sigma_{1c} )=0$ there exist a 2-dimensional stable manifold,
   a 1-dimensional unstable manifold and a 1-dimensional center manifold of $Q_1.$
\item If $\frac{4}{3}<\gamma<2$ and $\chi
   ''(\sigma_{1c} )=0$ there exist a 1-dimensional stable manifold, a 2-dimensional
   unstable manifold and a 1-dimensional center manifold of $Q_1.$
\end{enumerate}

\item The eigenvalues of the linearization around $Q_2$ are:
$-\frac{2}{3}, -\frac{\gamma}{2}, -\frac{1}{2}\pm
\frac{1}{2}\sqrt{1-\frac{4}{3}\frac{V''(\sigma_{1c})}{V(\sigma_{1c})}}.$
Then, the local stability of singular point $P_2$ is as follows:
\begin{enumerate}
\item If $V''(\sigma_{1c})< 0$ there exists a 3-dimensional stable
manifold and a 1-dimensional unstable manifold of $Q_2.$ \item If
$V''(\sigma_{1c})= 0$ the eigenvalues are
$-1,-\frac{2}{3},0,-\frac{\gamma }{2}.$ Then there exists a
1-dimensional center manifold tangent to the $\sigma_1$-axis and a
3-dimensional stable manifold tangent to the 3-dimensional surface
$$\left\{(\sigma_1,\sigma_2,\sigma_3,\sigma_5)\in\mathbb{R}^4:
\sigma_1=-\sqrt{\frac{2}{3}}\sigma_2\right\}.$$ \item If
$0<V''(\sigma_{1c})\leq \frac{3}{4}V(\sigma_{1c})$ the stable
manifold of $Q_2$ is 4-dimensional (the singular point is
asymptotically stable and it is an stable node). If
$V''(\sigma_{1c})> \frac{3}{4}V(\sigma_{1c})$ the stable manifold
of $Q_2$ is 4-dimensional (the singular point is asymptotically
stable and it is an stable focus).
\end{enumerate}

\item The eigenvalues of the linearization around $Q_3$ are
$\frac{4}{3},-\frac{1}{3},0,\frac{1}{6} (4-3 \gamma ).$ Then, the
local stability of singular point $Q_3$ is as follows:
\begin{enumerate}
\item There exists a 1-dimensional center manifold tangent to the
$\sigma_1$-axis \item If $\gamma<\frac{4}{3},$ $Q_3$ has a
2-dimensional unstable manifold and a 1-dimensional stable
manifold tangent to the line
$$\left\{(\sigma_1,\sigma_2,\sigma_3,\sigma_5)\in\mathbb{R}^4:
\sigma_1=-\sqrt{\frac{2}{3}}\sigma_2,\,\sigma_3=\sigma_5=0\right\}.$$
 \item If $\gamma>\frac{4}{3},$ $Q_3$ has a 1-dimensional unstable manifold and a 2-dimensional stable
manifold tangent to the 2-dimensional
subspace $$\left\{(\sigma_1,\sigma_2,\sigma_3,\sigma_5)\in\mathbb{R}^4: \sigma_1=-\sqrt{\frac{2}{3}}\sigma_2,\,\sigma_5=0\right\}.$$

\end{enumerate}

\end{enumerate}

Let us comment on the physical interpretation of the singular
points listed above. The singular points $Q_1$ represent
matter-dominated cosmological solutions with infinite curvature
($H\rightarrow +\infty$). Since $\chi'(\sigma_{1c})=0$ (i.e.,
$\sigma_{1c}$ is an stationary point of the coupling function) and
$\chi'(\sigma_{1c})\neq 0$ they are solutions with minimally
coupled scalar field and negligible kinetic energy. The potential
function does not influence its dynamical character.

The singular points $Q_2$ represent a \emph{de Sitter}
cosmological solution. When one of these singular points is
approached, the energy density of dark matter and the kinetic
energy of the scalar field go to zero. In this case the potential
energy of the scalar field becomes dominant. Hence, the universe
would be expanding forever in a \emph{de Sitter} phase.

The singular points $Q_3$ represent radiation-dominated
cosmological solutions. These solutions are very important during
the radiation era.

\subsection{Center manifold of $Q_2.$}

In this section we apply the center manifold theorem to examine the stability of \eqref{eqsigma1}-\eqref{eqsigma5} around $Q_2.$ We exclude from the analysis the case where  $V'(\sigma_{1c})=0$ and $V''(\sigma_{1c})>0$ (i.e., $V(\phi)$ has a local minimum at $\sigma_{1c}$) since from the linear analysis (see point 2c in subsection \ref{Qualitative}) follows the asymptotic stability of $Q_2.$

Let $V(\phi)$ and $\chi(\phi)$ be smooth ($C^\infty$) functions. The barotropic index of the fluid satisfies $0<\gamma<2, \gamma\neq \frac{4}{3}.$

\begin{prop}\label{Q2inflection} Let $\sigma_{1c}$ such that
$V'(\sigma_{1c})=V''(\sigma_{1c})=0$ and $V^{(3)}(\sigma_{1c})\neq
0,$ i.e., $\sigma_{1c}$ is an inflection point  of $V(\phi).$
Then, the singular point $Q_2$ of the system
\eqref{eqsigma1}-\eqref{eqsigma5} is locally unstable.
\end{prop}

{\bf Proof}. Let us consider the coordinate transformation $(\sigma_1,\sigma_2,\sigma_3,\sigma_5)\rightarrow
(\sigma_1+\sigma_{1c},\sigma_2,\sigma_3,\sigma_5).$ The Taylor expansion up to third order of the system arising from \eqref{eqsigma1}-\eqref{eqsigma5} under such coordinate transformation around the origin reads

\begin{align} &\sigma_1'=\sqrt{\frac{2}{3}}\sigma_2 +{\cal O}(3),\nonumber\\
&\sigma_2'=-\sigma_2-\frac{\sigma_1^2}{2\sqrt{6}}\frac{V^{(3)}(\sigma_{1c})}{V(\sigma_{1c})}
+\frac{(4-3\gamma)\sigma_3^2}{2\sqrt{6}}
\frac{\chi'(\sigma_{1c})}{\chi(\sigma_{1c})}+{\cal
O}(3),\nonumber\\
&\sigma_3'=-\frac{\gamma}{2}\sigma_3-\frac{(4-3\gamma)\sigma_2\sigma_3}{2\sqrt{6}}
\frac{\chi'(\sigma_{1c})}{\chi(\sigma_{1c})}+{\cal
O}(3),\nonumber\\
&\sigma_5'=-\frac{2}{3}\sigma_5+{\cal O}(3).
\label{thirdorder}\end{align}

The change of coordinates $$\sigma_1\to x-\sqrt{\frac{2}{3}} y_2,\sigma_2\to y_2,\sigma_3\to y_3,\sigma_5\to
   y_1$$ allows to reduce the linear part of \eqref{thirdorder} to its real Jordan form
 \begin{align} &x'=-\frac{V^{(3)}(\sigma_{1c}) x^2}{6 V(\sigma_{1c})}+\frac{\sqrt{\frac{2}{3}} y_2 V^{(3)}(\sigma_{1c}) x}{3
   V(\sigma_{1c})}+\frac{2 y_3^2 \chi '(\sigma_{1c})}{3
   \chi (\sigma_{1c})}-\frac{y_3^2 \gamma  \chi '(\text{$\sigma
   $1c})}{2 \chi (\sigma_{1c})}+\nonumber \\ &-\frac{y_2^2
   V^{(3)}(\sigma_{1c})}{9 V(\sigma_{1c})}+{\cal O}(3),\nonumber\\
&y_1'=-\frac{2
   y_1}{3}+{\cal
O}(3),\nonumber\\
&y_2'=-\frac{V^{(3)}(\sigma_{1c}) x^2}{2 \sqrt{6}
   V(\sigma_{1c})}+\frac{y_2 V^{(3)}(\sigma_{1c}) x}{3
   V(\sigma_{1c})}-y_2+\frac{\sqrt{\frac{2}{3}} y_3^2 \chi
   '(\sigma_{1c})}{\chi (\sigma_{1c})}+\nonumber\\ &-\frac{\sqrt{\frac{3}{2}}
   y_3^2 \gamma  \chi '(\sigma_{1c})}{2 \chi (\text{$\sigma
   $1c})}-\frac{y_2^2 V^{(3)}(\sigma_{1c})}{3 \sqrt{6}
   V(\sigma_{1c})}+{\cal
O}(3),\nonumber\\
&y_3'=-\frac{y_3 \gamma
   }{3}+\frac{\sqrt{\frac{3}{2}} y_2 y_3 \chi '(\sigma_{1c}) \gamma }{2 \chi (\sigma_{1c})}-\frac{\sqrt{\frac{2}{3}}
   y_2 y_3 \chi '(\sigma_{1c})}{\chi (\sigma_{1c})}+{\cal O}(3).
\label{thirdordercenter}\end{align}
Hence, the system \eqref{thirdordercenter} is written in diagonal form
\begin{align}
x'  &  =Cx+f\left(  x,\mathbf{y}\right) \nonumber\\
\mathbf{y}'  &  =P\mathbf{y}+\mathbf{g}\left(  x,\mathbf{y}\right)  ,
\label{thirdordercenter3}
\end{align}
where $\left(  x,\mathbf{y}\right)  \in\mathbb{R}\times\mathbb{R}^{3},$ $C$ is
the zero $1\times1$ matrix, $P$ is a $3\times3$ matrix with negative eigenvalues and $f,\mathbf{g}$ vanish at $\mathbf{0}$ and have vanishing derivatives at $\mathbf{0.}$ The center manifold theorem \ref{existenceCM} asserts that there exists a 1-dimensional invariant local center manifold $W^{c}\left(
\mathbf{0}\right) $ of \eqref{thirdordercenter3} tangent to the center subspace (the
$\mathbf{y}=\mathbf{0}$ space) at $\mathbf{0}.$ Moreover, $W^{c}\left(
\mathbf{0}\right)  $ can be represented as
\[
W^{c}\left(  \mathbf{0}\right)  =\left\{  \left(
x,\mathbf{y}\right)
\in\mathbb{R}\times\mathbb{R}^{3}:\mathbf{y}=\mathbf{h}\left(
x\right), \left\vert x\right\vert <\delta\right\};
\mathbf{h}\left(  0\right) =\mathbf{0}, D\mathbf{h}\left( 0\right)
=\mathbf{0},
\]
for $\delta$ sufficiently small (see definition \ref{CMdef}). The restriction of
(\ref{thirdordercenter3}) to the center manifold is (see definition \ref{vectorfieldCM})
\begin{equation}
x'=f\left( x,\mathbf{h}\left(  x\right)  \right)  . \label{thirdorderrest}
\end{equation}
According to Theorem \ref{stabilityCM}, if the origin $x=0$ of \eqref{thirdorderrest}
is stable (asymptotically stable) (unstable) then the origin of \eqref{thirdordercenter3} is also stable (asymptotically stable) (unstable). Therefore, we have to find the local center manifold, i.e.,
the problem reduces to the computation of $\mathbf{h}\left(  x\right).$

Substituting $\mathbf{y}=\mathbf{h}\left(  x\right)  $ in the second
component of \eqref{center3} and using the chain rule, $\mathbf{y
}'=D\mathbf{h}\left(  x\right)  x'$, one can show that the function
$\mathbf{h}\left(  x\right)  $ that defines the local center manifold
satisfies%
\begin{equation}
D\mathbf{h}\left(  x\right)  \left[  f\left(  x,\mathbf{h}\left(  x\right)
\right)  \right]  -P\mathbf{h}\left(  x\right)  -\mathbf{g}\left(
x,\mathbf{h}\left(  x\right)  \right)  =0. \label{h0}
\end{equation}
According to Theorem \ref{approximationCM}, equation \eqref{h0} can be solved approximately by using an approximation of $\mathbf{h}\left(  x\right)  $
by a Taylor series at $x=0.$ Since $\mathbf{h}\left(  0\right)  =\mathbf{0\ }
$and $D\mathbf{h}\left(  0\right)  =\mathbf{0},$ it is obvious that
$\mathbf{h}\left(  x\right)  $ commences with quadratic terms. We substitute
\[
\mathbf{h}\left(  x\right)  =:\left[
\begin{array}
[c]{c}%
h_{1}\left(  x\right) \\
h_{2}\left(  x\right) \\
h_{3}\left(  x\right)
\end{array}
\right]  =\left[
\begin{array}
[c]{c}%
a_{1}x^{2}+a_{2}x^{3}+O\left(  x^{4}\right) \\
b_{1}x^{2}+b_{2}x^{3}+O\left(  x^{4}\right) \\
c_{1}x^{2}+c_{2}x^{3}+O\left(  x^{4}\right)
\end{array}
\right]
\]
into (\ref{h0}) and set the coefficients of like powers of $x$
equal to zero to find the unknowns $a_{1},b_{1},c_{1},...$. It is
straightforward to find that
$$(a_1,a_2,b_1,b_2,c_1,c_2)=\left(0,0,-\frac{V^{(3)}(\sigma_{1c})}{2
\sqrt{6} V(\sigma_{1c})},-\frac{V^{(3)}(\sigma_{1c})^2}{3 \sqrt{6}
V(\sigma_{1c})^2},0,0\right).$$ Thus, the restriction of the
vector field to the center manifold of the origin is given by
\be\label{restrictcenter}
   x'=-\frac{V^{(3)}(\sigma_{1c}) x^2}{6 V(\sigma_{1c})}-\frac{V^{(3)}(\sigma_{1c})^2 x^3}{18 V(\sigma_{1c})^2}-\frac{V^{(3)}(\sigma_{1c})^3 x^4}{24 V(\sigma_{1c})^3}+{\cal O}\left(x^5\right).
\ee

This is a gradient like vector field whose potential function has
an inflection point at the origin irrespective the sign of the
ratio  $\frac{V^{(3)}(\sigma_{1c})}{ V(\sigma_{1c})}.$ From this
follows that, under the conditions discussed here, the origin of
coordinates is locally unstable (of saddle type) for the flow of
the original system \eqref{thirdorder}.

$\blacksquare$

\begin{prop}\label{Q2extremum} Let the function $V(\phi)$ to have a degenerate
local minimum (maximum) at $\sigma_{1c}$ of order $n=2$. Then, the singular point $Q_2$ of the system \eqref{eqsigma1}-\eqref{eqsigma5} is locally asymptotically stable (unstable).
\end{prop}

{\bf Proof}. By considering the coordinate transformation $(\sigma_1,\sigma_2,\sigma_3,\sigma_5)\rightarrow
(\sigma_1+\sigma_{1c},\sigma_2,\sigma_3,\sigma_5).$ By Taylor expanding up to fourth order the system \eqref{eqsigma1}-\eqref{eqsigma5} under such coordinate transformation around the origin reads

\begin{align}
&\sigma_1'=\sqrt{\frac{2}{3}}\sigma_2 +{\cal O}(4),\nonumber \\
&\sigma_2'=-\frac{V^{(4)}(\sigma_{1c}) \sigma_1^3}{6 \sqrt{6} V(\sigma_{1c})}+\frac{(3
   \gamma -4) \sigma_3^2 \chi '(\sigma_{1c})^2
   \sigma_1}{2 \sqrt{6} \chi (\sigma_{1c})^2}+\frac{(4-3
   \gamma ) \sigma_3^2 \chi ''(\sigma_{1c})
   \sigma_1}{2 \sqrt{6} \chi (\sigma_{1c})}+\nonumber\\&+\sigma_2^3-\sigma_2+\frac{1}{6}
   \sigma_2 \left(3 \gamma  \sigma_3^2+4 \sigma_5^2\right)+\frac{(4-3 \gamma ) \sigma_3^2 \chi
   '(\sigma_{1c})}{2 \sqrt{6} \chi (\sigma_{1c})}+{\cal O}(4),\nonumber\\
   &\sigma_3'=\frac{(4-3 \gamma ) \sigma_1 \sigma_2
   \sigma_3 \chi '(\sigma_{1c})^2}{2 \sqrt{6} \chi
   (\sigma_{1c})^2}+\frac{(3 \gamma -4) \sigma_2
   \sigma_3 \chi '(\sigma_{1c})}{2 \sqrt{6} \chi
   (\sigma_{1c})}-\frac{\gamma  \sigma_3}{2}+\nonumber\\&+\frac{1}{6}
   \sigma_3 \left(6 \sigma_2^2+3 \gamma  \sigma_3^2+4 \sigma_5^2\right)+\frac{(3 \gamma -4) \sigma_1 \sigma_2 \sigma_3 \chi ''(\sigma_{1c})}{2 \sqrt{6} \chi (\sigma_{1c})}+{\cal O}(4),\nonumber\\
   & \sigma_5'=\frac{1}{6}
   \sigma_5 \left(6 \sigma_2^2+3 \gamma  \sigma_3^2+4 \sigma_5^2\right)-\frac{2 \sigma_5}{3}+ {\cal O}(4).\label{fourthorder}
\end{align}

The change of coordinates $$\sigma_1\to x-\sqrt{\frac{2}{3}} y_2,\sigma_2\to y_2,\sigma_3\to y_3,\sigma_5\to
   y_1$$ allows to reduce the linear part of \eqref{fourthorder} to its real Jordan form

 \begin{align} &x'=-\frac{V^{(4)}(\sigma_{1c}) x^3}{18 V(\sigma_{1c})}+\frac{y_2 V^{(4)}(\sigma_{1c}) x^2}{3 \sqrt{6}
   V(\sigma_{1c})}-\frac{2 y_3^2 \chi '(\sigma_{1c})^2
   x}{3 \chi (\sigma_{1c})^2}+\frac{y_3^2 \gamma  \chi
   '(\sigma_{1c})^2 x}{2 \chi (\sigma_{1c})^2}\nonumber\\&+\frac{2
   y_3^2 \chi ''(\sigma_{1c}) x}{3 \chi (\sigma_{1c})}-\frac{y_3^2 \gamma  \chi ''(\sigma_{1c}) x}{2 \chi
   (\sigma_{1c})}-\frac{y_2^2 V^{(4)}(\sigma_{1c}) x}{9
   V(\sigma_{1c})}+\sqrt{\frac{2}{3}} y_2^3\nonumber\\&+\frac{2
   \sqrt{\frac{2}{3}} y_2 y_3^2 \chi '(\sigma_{1c})^2}{3
   \chi (\sigma_{1c})^2}-\frac{y_2 y_3^2 \gamma  \chi
   '(\sigma_{1c})^2}{\sqrt{6} \chi (\sigma_{1c})^2}+\frac{2}{3}
   \sqrt{\frac{2}{3}} y_1^2 y_2+\frac{y_2^3 \gamma
   }{\sqrt{6}}+\frac{2 y_3^2 \chi '(\sigma_{1c})}{3 \chi
   (\sigma_{1c})}\nonumber\\&-\frac{y_3^2 \gamma  \chi '(\sigma_{1c})}{2 \chi (\sigma_{1c})}-\frac{2 \sqrt{\frac{2}{3}} y_2
   y_3^2 \chi ''(\sigma_{1c})}{3 \chi (\sigma_{1c})}+\frac{y_2 y_3^2 \gamma  \chi ''(\sigma_{1c})}{\sqrt{6} \chi (\sigma_{1c})}+\frac{\sqrt{\frac{2}{3}}
   y_2^3 V^{(4)}(\sigma_{1c})}{27 V(\sigma_{1c})}+{\cal O}(4),\nonumber\end{align}
\begin{align}
&y_1'=\frac{2
   y_1^3}{3}+y_2^2 y_1+\frac{1}{2} y_3^2 \gamma
   y_1-\frac{2 y_1}{3}+{\cal
O}(4),\nonumber\end{align}
\begin{align}
&y_2'=-\frac{V^{(4)}(\sigma_{1c}) x^3}{6
   \sqrt{6} V(\sigma_{1c})}+\frac{y_2 V^{(4)}(\sigma_{1c}) x^2}{6 V(\sigma_{1c})}-\frac{\sqrt{\frac{2}{3}} y_3^2
   \chi '(\sigma_{1c})^2 x}{\chi (\sigma_{1c})^2}+\nonumber\\&+\frac{\sqrt{\frac{3}{2}} y_3^2 \gamma  \chi '(\sigma_{1c})^2 x}{2 \chi (\sigma_{1c})^2}+\frac{\sqrt{\frac{2}{3}}
   y_3^2 \chi ''(\sigma_{1c}) x}{\chi (\sigma_{1c})}-\frac{\sqrt{\frac{3}{2}} y_3^2 \gamma  \chi ''(\sigma_{1c}) x}{2 \chi (\sigma_{1c})}-\frac{y_2^2
   V^{(4)}(\sigma_{1c}) x}{3 \sqrt{6} V(\sigma_{1c})}+\nonumber\\&+y_2^3+\frac{2 y_2 y_3^2 \chi '(\sigma_{1c})^2}{3 \chi (\sigma_{1c})^2}-\frac{y_2 y_3^2 \gamma
    \chi '(\sigma_{1c})^2}{2 \chi (\sigma_{1c})^2}+\frac{2
   y_1^2 y_2}{3}-y_2+\frac{y_2^3 \gamma
   }{2}+\frac{\sqrt{\frac{2}{3}} y_3^2 \chi '(\sigma_{1c})}{\chi
   (\sigma_{1c})}\nonumber\\&-\frac{\sqrt{\frac{3}{2}} y_3^2 \gamma  \chi
   '(\sigma_{1c})}{2 \chi (\sigma_{1c})}-\frac{2 y_2
   y_3^2 \chi ''(\sigma_{1c})}{3 \chi (\sigma_{1c})}+\frac{y_2 y_3^2 \gamma  \chi ''(\sigma_{1c})}{2
   \chi (\sigma_{1c})}+\frac{y_2^3 V^{(4)}(\sigma_{1c})}{27 V(\sigma_{1c})}+{\cal
O}(4),\nonumber\end{align}
\begin{align}
&y_3'=\frac{\gamma  y_3^3}{2}+\frac{2
   y_1^2 y_3}{3}+y_2^2 y_3-\frac{2 y_2^2 \chi
   '(\sigma_{1c})^2 y_3}{3 \chi (\sigma_{1c})^2}+\frac{\sqrt{\frac{2}{3}} x y_2 \chi '(\sigma_{1c})^2
   y_3}{\chi (\sigma_{1c})^2}\nonumber\\&+\frac{y_2^2 \gamma  \chi
   '(\sigma_{1c})^2 y_3}{2 \chi (\sigma_{1c})^2}-\frac{\sqrt{\frac{3}{2}} x y_2 \gamma  \chi '(\sigma_{1c})^2 y_3}{2 \chi (\sigma_{1c})^2}-\frac{\gamma
   y_3}{2}-\frac{\sqrt{\frac{2}{3}} y_2 \chi '(\sigma_{1c}) y_3}{\chi (\sigma_{1c})}\nonumber\\&+\frac{\sqrt{\frac{3}{2}}
   y_2 \gamma  \chi '(\sigma_{1c}) y_3}{2 \chi
   (\sigma_{1c})}+\frac{2 y_2^2 \chi ''(\sigma_{1c})
   y_3}{3 \chi (\sigma_{1c})}-\frac{\sqrt{\frac{2}{3}} x
   y_2 \chi ''(\sigma_{1c}) y_3}{\chi (\sigma_{1c})}\nonumber\\&-\frac{y_2^2 \gamma  \chi ''(\sigma_{1c}) y_3}{2
   \chi (\sigma_{1c})}+\frac{\sqrt{\frac{3}{2}} x y_2 \gamma
   \chi ''(\sigma_{1c}) y_3}{2 \chi (\sigma_{1c})}+{\cal O}(4).
\end{align}

Then, we proceed to the calculation of the center manifold. The
procedure is fairly systematic and since we present it completely
in the previous analysis we consider do not repeat it here.
Instead, we present the relevant calculations. We obtain
$a_1=0,a_2=0,b_1=0,b_2=-\frac{V^{(4)}(\sigma_{1c})}{6 \sqrt{6}
V(\sigma_{1c})},c_1=0,c_2=0$ for the Taylor expansion coefficients
of \[ \mathbf{h}\left(  x\right)  =:\left[
\begin{array}
[c]{c}%
h_{1}\left(  x\right) \\
h_{2}\left(  x\right) \\
h_{3}\left(  x\right)
\end{array}
\right]  =\left[
\begin{array}
[c]{c}%
a_{1}x^{2}+a_{2}x^{3}+O\left(  x^{4}\right) \\
b_{1}x^{2}+b_{2}x^{3}+O\left(  x^{4}\right) \\
c_{1}x^{2}+c_{2}x^{3}+O\left(  x^{4}\right)
\end{array}
\right].
\] By substituting this values of the unknowns  $a_{1},b_{1},c_{1},...$ we obtain that the dynamics of the center manifold in given by equation
\be\label{restrictcentercase2}
   x'=-\frac{V^{(4)}(\sigma_{1c}) x^3}{18 V(\sigma_{1c})}-\frac{V^{(4)}(\sigma_{1c})^2 x^5}{108 V(\sigma_{1c})^2}+{\cal O}\left(x^6\right).
\ee This is a gradient-like vector field $x'=-\nabla U(x)$ whose
potential is given by $$U(x)=\frac{V^{(4)}(\sigma_{1c})^2 x^6}{648
V(\sigma_{1c})^2}+\frac{V^{(4)}(\sigma_{1c}) x^4}{72
V(\sigma_{1c})}.$$ Since $V(\sigma_{1c})>0,$ $U(x)$ has a
degenerate local minimum (maximum) at the origin for
$V^{(4)}(\sigma_{1c})<0$ ($V^{(4)}(\sigma_{1c})<0$). Thus, follows
the desired result. $\blacksquare$
{\bf Comment}. This analysis is in agreement with the result of
proposition \ref{thmIII} which states that if $V(\sigma_1)$ have
a strict degenerate local minimum at $\sigma_{1c}$ with
$V(\sigma_{1c})>0,$ then $Q_2:=(\sigma_{1c},0,0,0)$ is
asymptotically stable.

Now, let us formulate  generalization of theorem \ref{Theorem_2.4} (see also theorem 3.2 of \cite{Leon:2008de}
p. 8.)

\begin{thm}\label{thm4}
Assume that $\chi(\phi)$ and $V(\phi)$ are positive functions of
class $C^3,$ such that $\chi$ has at most a finite number of
stationary points and does not tend to zero in any compact set of $\mathbb{R}$. Let $\gamma\in \left(0,\frac{4}{3}\right)\cup\left(\frac{4}{3},2\right)$ and let $p$ be a point in
$\Sigma_+.$ Let $O^{-}(p)$ be the past orbit of $p$ under the flow
of \eqref{eq0phi}-\eqref{eq0x4} restricted to $\Sigma_+$. Then,
$\phi$ is always unbounded on $O^{-}(p)$ for almost all $p$.
\end{thm}

{\bf Proof of theorem \ref{thm4}}

In order to prove the theorem it
is sufficient to consider interior points of $\Sigma_+.$ Also, in
order to apply results concerning future attractors and
$\omega$-limit sets we perform the time reversal $\tau\rightarrow
-\tau.$ Thus we get the system
\begin{align}
&\sigma_1'=-\sqrt{\frac{2}{3}} \sigma_2 \label{eqrevphi}\\
&\sigma_2'=-\sigma_2^3-\frac{1}{6}\left(3\gamma \sigma_3^2+4
\sigma_5^2-6\right)\sigma_2
+\frac{\sigma_4^2}{\sqrt{6}}\frac{\mathrm{d}\ln
V(\sigma_1)}{\mathrm{d}\sigma_1}\nonumber\\ & \ \ \ \ \ \ \ \ \ \
\ \ \ \ \ \ \ \
-\frac{\left(4-3\gamma\right)\sigma_3^2}{2\sqrt{6}}\frac{\mathrm{d}\ln \chi(\sigma_1)}{\mathrm{d}\sigma_1},\label{eqrevx1}
\end{align}
\begin{align}
&\sigma_3'=-\frac{1}{6}\sigma_3\left(6\sigma_2^2+3\gamma\left(\sigma_3^2-1\right)+4
\sigma_5^2\right) \nonumber\\ & \ \ \ \ \ \ \ \ \ \ \ \ \ \ \ \ \
\ +\frac{\left(4-3\gamma\right)\sigma_2
\sigma_3}{2\sqrt{6}}\frac{\mathrm{d}\ln
\chi(\sigma_1)}{\mathrm{d}\sigma_1},\label{eqrevx2}\\
&\sigma_4'=-\frac{1}{6}\sigma_4\left(6\sigma_2^2+3\gamma
\sigma_3^2+4 \sigma_5^2\right)
-\frac{\sqrt{6}}{6}\sigma_2 \sigma_4 \frac{\mathrm{d}\ln V(\sigma_1)}{\mathrm{d}\sigma_1},\label{eqrevx3}\\
&\sigma_5'=-\frac{1}{6}\sigma_5\left(6\sigma_2^2+3\gamma
\sigma_3^2+4 \sigma_5^2-4\right).\label{eqrevx4}
\end{align}
where the prime denotes now derivative with respect to $-\tau.$

Let
$p_0:=(\sigma_{10},\sigma_{20},\sigma_{30},\sigma_{40},\sigma_{50})\in
\text{Int} \Sigma_+$ such that there exist a real positive number
$K$ with $|\sigma_1|<K$ for all
$p:=(\sigma_1,\sigma_2,\sigma_3,\sigma_4,\sigma_5)\in O^+(p_0),$
where $O^+(p_0)$ denotes the positive (future) orbit for the flow
of \eqref{eqrevphi}-\eqref{eqrevx4}.  Then, for all $p\in
O^+(p_0)$ we have $$-1\leq \sigma_2\leq 1,\, 0\leq \sigma_3\leq
1,\,0\leq \sigma_4\leq 1,\, 0\leq \sigma_5\leq 1.$$ Hence
$O^+(p_0)$ is contained in a compact set of (the closure of)
$\Sigma_+.$

Since $O^+(p_0)$ is a positive invariant set, then using proposition \ref{omegalimitsetproperties} we ensure the existence of a non empty, closed, connected and invariant $\omega$-limit of $p_0$ denoted by $\omega(p_0).$

First we demonstrate by contradiction that that $\sigma_3$ and $\sigma_4$ cannot be simultaneously zero at $\omega(p_0).$ Suppose that $\omega(p_0)$ is contained in the set where
$\sigma_3=\sigma_4=0.$ Let us define the function $M_1=Z_4^{-1}$
(see table \ref{monotonic} for the definition of $Z_4$) defined in
the invariant set $\sigma_3=\sigma_4=0, 0<\sigma_5<1.$ From the definition of $M_1$ and applying the Monotonicity principle (theorem \ref{theorem 4.12}) follows that the future asymptotic attractor of the flow of \eqref{eqrevphi}-\eqref{eqrevx4} restricted to the invariant set $\sigma_3=\sigma_4=0$ is contained in the invariant set $\sigma_2=\pm 1.$ Thus, from \eqref{eqrevphi} follows that $\phi\rightarrow \mp \infty$ as $\omega(p_0)$ is approached, a contradiction.

Second, let be defined in $Int \Sigma_+$ the function $M_2=Z_1^{-1}$ (see table \ref{monotonic} for the definition of $Z_1$). The derivative of $M_2$ along any orbit of \eqref{eqrevphi}-\eqref{eqrevx4} is given by $M_2'=-\gamma M_2.$
Then $M_2$ is a $C^3$ monotonic decreasing function for the flow taking values in the interval $(0,+\infty).$ Since $\sigma_3$ and
$\sigma_4$ cannot tend to zero simultaneously in $\omega(p)$ for $p\in O^+(p_0),$ then the function $M_2$ tends asymptotically to a well defined limit. By construction $M_2(p)\rightarrow 0$ if and
only if $p\rightarrow q$ with $q$ satisfying $\sigma_4=0$ (we are using here the condition that $\chi(\phi)$ does not tend to zero in any compact of $\mathbb{R}$) and $M_2(p)\rightarrow +\infty$ if and only if $p\rightarrow q$ with $q$ satisfying $\sigma_3=0.$
Thus, applying the Monotonicity principle (theorem \ref{theorem
4.12}) follows that
$$\omega(p_0)\subset \left\{p\in\Sigma_+: |\sigma_1|<K, \sigma_3>0, \sigma_4=0\right\}=S_1.$$
Let $q_0\in\omega(p_0).$ By the invariance of the $\omega$-limit set follows that $\omega(q_0)=\omega(p_0).$

Observe that $g_3(S_1)=\left\{\sigma\in W_3: |\sigma_1|<K,
\sigma_4=0\right\}$ where $g_3$ is defined in the proof of proposition \ref{thm2}.

Let us define the projection map $$g:
(\sigma_1,\sigma_2,\sigma_3,\sigma_5)\rightarrow
(\sigma_1,\sigma_2,\sigma_5)$$ and let $\sigma_0=g\circ g_3(q_0)$
then the flow of \eqref{eqrevphi}-\eqref{eqrevx4} in a
neighborhood of $q_0$ contained in $S_1$, is topologically
equivalent to the flow of
\begin{align}
&\sigma_1'=-\sqrt{\frac{2}{3}} \sigma_2,\nonumber\\
&\sigma_2'=\frac{1}{2}\left(1-\sigma_2^2\right)\left((2-\gamma)\sigma_2
-\frac{(4-3\gamma)}{\sqrt{6}}\frac{\mathrm{d}\ln\chi(\sigma_1)}{\mathrm{d}\sigma_1}\right)
+ \nonumber\\ & \ \ \ \ \ \ \ \ \ \
-\frac{\left(4-3\gamma\right)\sigma_5^2}{6}\left(\sigma_2
-\frac{\sqrt{6}}{2}\frac{\mathrm{d}\ln\chi(\sigma_1)}{\mathrm{d}\sigma_1}\right),\nonumber\\
&\sigma_5'=-\frac{1}{6} \sigma_5 \left(3(2-\gamma)\sigma_2^2-(4-3\gamma)(1-\sigma_5^2)\right), \label{systthm4}
\end{align}
in a neighborhood of $\sigma_0$ contained in
$$S=\left\{(\sigma_1,\sigma_2,\sigma_5): -K<\sigma_1<K,
\sigma_2^2+\sigma_5^2<1, \sigma_5>0\right\}.$$ Since the vector field is $C^2$ we can extent the flow of \eqref{systthm4} to the closure of $S$ (denoted by $\bar{S}$).

Let us investigate all possible compact, non empty, and connected invariant sets of \eqref{proxi1}-\eqref{proxi2} located in the closure of $S$ (these ones can be candidates to the $\omega$-limit $\omega(\sigma_0)$).

Let us consider two cases:

\begin{enumerate}
\item[i)] $0<\gamma<\frac{4}{3}.$ Let be defined in $S$, the function \be M_3(\sigma)=\frac{\left(1-\sigma_2^2-\sigma_5^2\right)^2 \chi
   (\sigma_1)^{4-3 \gamma }}{\sigma_5^4}.\ee  The derivative of $M_3$ through an arbitrary orbit of \eqref{systthm4} is given by $$M_3'=-2\left(\frac{4}{3}-\gamma\right)M_3.$$ Then
$M_3$ is a $C^3$ monotonic decreasing function for the flow taking
values in the interval $(0,+\infty).$ By construction $M_3(p)\rightarrow 0$ if and only if
$p\rightarrow q$ with $q$ satisfying $\sigma_2^2+\sigma_5^2=1$ (since $\chi$ does tends to zero in $[-K,K]$) and $M_3(p)\rightarrow
+\infty$ if and only if $p\rightarrow q$ with $q$ satisfying
$\sigma_5=0.$  Thus, applying the Monotonicity principle (theorem \ref{theorem 4.12})  follows that $$\omega(\sigma_0)\subset\left\{\sigma\in \bar{S}\setminus S: \sigma_2^2+\sigma_5^2=1\right\}.$$ Let $q_0\in \omega(\sigma_0).$ By the invariance of the $\omega$-limit follows that $\omega(\sigma_0)=\omega(q_0).$

Let us define the projection map $$g': (\sigma_1,\sigma_2,\sigma_5)\rightarrow (\sigma_1,\sigma_2)$$ and let $\sigma_0'=g'(q_0)$ then the flow of \eqref{systthm4} in a neighborhood of $q_0$ contained in $S$, is topologically equivalent to the flow of
\begin{align}
&\sigma_1'=-\sqrt{\frac{2}{3}}\sigma_2,\nonumber\\
&\sigma_2'=\frac{1}{3}\sigma_2\left(1-\sigma_2\right)\left(1+\sigma_2\right).\label{casei)}
\end{align} in a neighborhood of $\sigma_0'$ (contained in
$S':=(-K,K)\times(-1,1)$).

Let be defined in $S'$ the function
$$M_4(\sigma)=\frac{1-\sigma_2^2}{\sigma_2^2}$$ which satisfies
$M_4'=-\frac{2}{3}M_4$ along an arbitrary orbit of \eqref{casei)}.
Thus $M_4$ is a $C^3$ monotonic decreasing function in $S'$.
Applying the monotonicity principle (\ref{theorem 4.12}) follows
that $$\omega(\sigma_0')\subset\left\{\sigma\in \bar{S'}\setminus
S': \sigma_2^2=1\right\}.$$ Thus $\omega(\sigma_0')$ is contained
in one of the invariant sets of \eqref{proxi1}-\eqref{proxi2}
given by $\sigma_2=\pm 1$ but this would imply the divergence of
$\phi.$ A contradiction.

\item[ii)] $\frac{4}{3}<\gamma<2.$ Let be defined in $S$, the function \be M_5(\sigma)=\frac{\sigma_5^2 \chi
   (\sigma_1)^{3 \gamma -4}}{\left(1-\sigma_2^2-\sigma_5^2\right)^2}.\ee  The derivative of $M_5$ through an arbitrary orbit of \eqref{systthm4} is given by $$M_5'=-2\left(\gamma-\frac{4}{3}\right)M_5.$$ Then
$M_5$ is a $C^3$ monotonic decreasing function for the flow taking values in the interval $(0,+\infty).$ By construction $M_5(p)\rightarrow 0$ if and only if
$p\rightarrow q$ with $q$ satisfying $\sigma_5=0$ (since $\chi$ does not tends to zero in $[-K,K]$) and $M_5(p)\rightarrow
+\infty$ if and only if $p\rightarrow q$ with $q$ satisfying
$\sigma_2^2+\sigma_5^2=1.$ Applying the Monotonicity principle (theorem \ref{theorem 4.12}) follows that $$\omega(\sigma_0)\subset \left\{\sigma\in \bar{S}\setminus S: \sigma_5=0\right\}.$$ Let $q_0\in \omega(\sigma_0).$ By the invariance of the $\omega$-limit follows that $\omega(\sigma_0)=\omega(q_0).$

Let us define the projection
map \be h: (\sigma_1,\sigma_2,\sigma_5)\rightarrow (\sigma_1,\sigma_2)\label{proj}.\ee Let $\sigma_0'=h(q_0).$ Then, then the flow of \eqref{systthm4} in a neighborhood of $q_0$ contained in $S$, is topologically equivalent to the flow of

\begin{align}
&\sigma_1'=-\sqrt{\frac{2}{3}}\sigma_2,\label{proxi1}\\
&\sigma_2'=\frac{1}{2}\left(1-\sigma_2^2\right)\left((2-\gamma)\sigma_2-\frac{(4-3\gamma)}{\sqrt{6}}\frac{\mathrm{d}\ln\chi(\sigma_1)}{\mathrm{d}\sigma_1}\right),\label{proxi2}
\end{align} in a neighborhood of $\sigma_0'$ (contained in
$S'$).

Let us investigate the possible compact invariant sets of
\eqref{proxi1}-\eqref{proxi2} located in the closure of $S'$ which can be candidates to the $\omega$-limit $\omega(\sigma_0').$

First  $\omega(\sigma_0')$ cannot be contained in the invariant
sets of \eqref{proxi1}-\eqref{proxi2} given by $\sigma_2=\pm 1$
because this would imply the divergence of $\phi,$ a
contradiction. Second, combining the results of the
Poincar\'e-Bendixon Theorem (theorem \ref{PBT}) and Dulac's
criterion (theorem \ref{DC}) with $B(\xi)=(1-\sigma_2^2)^{-1}$
follows that the only possible compact invariant sets are the
singular points with $\sigma_1$ bounded (or heteroclinic orbits
joining such singular points).

Let us consider $\chi(\sigma_1)$ other than exponential.
\footnote{As we will see next in section \ref{application1}, the
following analysis applies also to exponential coupling
functions.} In this case the system \eqref{proxi1}-\eqref{proxi2}
admits a (possibly empty) family of singular points
$$Q:=\left\{(q_1,0)\in[-K,K]\times[-1,1]: \chi'(q_1)=0\right\}.$$
If $Q=\emptyset,$ i.e.,  $\chi'(q_1)\neq 0$ for all $|q_1|<K,$
then the future orbit $O^+(\sigma_0)$ tends to a point with
$\sigma_1=\pm 1.$ From this follows that $\phi$ is unbounded (a
contradiction) and the proof is done.

Let us assume that $Q\neq \emptyset.$ Let $q\in Q.$ The
eigenvalues of the Jacobian matrix $\frac{\partial f^i}{\partial
\sigma_j}(q), i,j=1,2$ are $\mu^\pm=\Delta_1\pm
\sqrt{\Delta_1^2+\Delta_2 \frac{\chi''(q)}{\chi(q)}},$ where
$\Delta_1=\frac{2-\gamma}{4}>0, \Delta_2=\frac{4-3\gamma}{6}.$
Hence, at least one of its associated eigenvalues has positive
real part. Let be defined the sets $Q^\pm =\{q\in Q:\pm
\chi''(q)>0\}$ and $Q^0=\{q\in Q: \chi(q)=0\}.$ At least one of
these sets is not empty. Let be define $R=\{p\in [-K, K]\times
[-1,1]: \lim_{\tau\rightarrow\infty}\mathbf{g}^\tau(p)=q\}.$ There
are the following cases
\begin{itemize}
\item  $q\in Q^+, \,
\frac{4}{3}<\gamma<2,$ then ${\cal E}^u(q)$ is 2-dimensional implying
$R=\emptyset.$ \item $q\in Q^-, \, \frac{4}{3}<\gamma<2,$ then ${\cal E}^u(q)$ is 1-dimensional and
${\cal E}^s(q)$ is 1-dimensional. Then $R\subset N,$ $leb (N)=0.$
\item $q\in Q^0,$ then ${\cal E}^c(q)$ is 1-dimensional and ${\cal E}^u(q)$
is 1-dimensional in such way that $R\subset {\cal E}^c(q),$ $leb ({\cal
E}^c(q))=0.$
\end{itemize}

Therefore, all solutions future asymptotic to $q$ (and then with
$\phi$ bounded towards the future) must lie on an stable manifold or center manifold of dimension $r<2,$ and then contained in a subset of $[-K, K]\times [-1,1]$ with zero Lebesgue measure. Since there are at most a finite number of such $q$ the result of the theorem follows. $\blacksquare$
\end{enumerate}

Theorem \ref{thm4} allow us to conclude that in order to investigate the generic asymptotic behavior of the system \eqref{eq0phi}-\eqref{eq0x4} restricted to $\Sigma_+$ it is sufficient to study the region where $\phi=\pm\infty.$

\subsection{Analysis in the Limit $\phi\rightarrow\infty.$}

In this section we will investigate the flow as $\phi\rightarrow
\infty$ following the nomenclature and formalism introduced in
\cite{Foster:1998sk} (see also \cite{Giambo:2008sa}). Analogous
results hold as $\phi\rightarrow-\infty.$

By assuming that $V,\chi\in {\cal E}^3_+,$ with exponential orders
$N$ and $M$ respectively, we can define a dynamical system well
suited to investigate the dynamics near the initial singularity.
We will investigate the singular points therein. Particularly
those representing scaling solutions and those associated with the
initial singularity.

Let $\Sigma_\epsilon=\left\{(\sigma_1,
\sigma_2,\sigma_3,\sigma_4,\sigma_5)\in \Sigma:
\sigma_1>\epsilon^{-1}\right\}$ where $\epsilon$ is any positive constant which is chosen sufficiently small so as to avoid any points where $V$ or $\chi=0,$ thereby ensuring that $\overline{W_V}(\varphi)$ and $\overline{W}_{\chi}(\varphi)$ are well-defined. \footnote{See \ref{bar} for the definition of functions with bar.}

Let be defined the projection map
\ben && \pi_1: \Sigma_\epsilon \rightarrow \Omega_\epsilon \nonumber\\
&&(\sigma_1, \sigma_2,\sigma_3,\sigma_4,\sigma_5)\rightarrow
(\sigma_1, \sigma_2,\sigma_4,\sigma_5)\een where
\be\Omega_\epsilon:=\left\{\sigma\in\mathbb{R}^4:\sigma_1>\epsilon^{-1},
\sigma_2^2+\sigma_4^2+\sigma_5^2\leq 1, \sigma_j\geq 0, j= 4,
5\right\}.\ee Let be defined in $\Omega_\epsilon$ the coordinate
transformation $(\sigma_1, \sigma_2,\sigma_4,\sigma_5)
\stackrel{\varphi=f(\sigma_1)}{\longrightarrow} (\varphi,
\sigma_2,\sigma_4,\sigma_5)$ where $f(\sigma_1)$ tends to zero as
$\sigma_1$ tends to $+\infty$ and has been chosen so that the
conditions i)-iii) of definition \ref{CkWBI} are satisfied with
$k=2.$

The flow of \eqref{eq0phi}-\eqref{eq0x4} defined on
$\Sigma_\epsilon$ is topologically equivalent (under $f\circ
\pi_1$) to the flow of the 4-dimensional dynamical system

\begin{align}
&\varphi'=\sqrt{\frac{2}{3}} \overline{f'} \sigma_2,\label{eqvphi}\\
&\sigma_2'= \sigma_2^3+\left(\frac{2 \sigma_5^2}{3}-1\right)
\sigma_2-\frac{\left(\overline{W_V}+N\right)
\sigma_4^2}{\sqrt{6}}+\nonumber\\&+\left(\frac{\left(\overline{W}_{\chi}+M\right)
(4-3 \gamma )}{2 \sqrt{6}}+\frac{\sigma_2 \gamma }{2}\right)
   \left(1-\sigma_2^2-\sigma_4^2-\sigma_5^2\right)\label{eqxrad}\\
&\sigma_4'=\frac{1}{6} \sigma_4 \left(\sqrt{6} \left(\overline{W_V}+N\right) \sigma_2+3(2-\gamma)\sigma_2^2+3\gamma(1-\sigma_4^2)+(4-3\gamma)\sigma_5^2\right), \label{eqyrad}\\
&\sigma_5'=\frac{1}{6} \sigma_5 \left(3(2-\gamma)\sigma_2^2-3\gamma \sigma_4^2-(4-3\gamma)(1-\sigma_5^2)\right),
\label{eqzrad}
\end{align}
defined in the phase space \footnote{For notational simplicity we will denote the image of $\Omega_\epsilon$ under $f$ by the same symbol.} \be\Omega_\epsilon=\{(\varphi, \sigma_2,
\sigma_4,\sigma_5)\in\mathbb{R}^4: 0\leq\varphi\leq
f(\epsilon^{-1}), \sigma_2^2+\sigma_4^2+\sigma_5^2\leq 1,
\sigma_4\geq 0, \sigma_5\geq 0\}.\label{PhaseSpace}\ee It can be easily proved that \eqref{PhaseSpace} defines a manifold with boundary of dimension 4. Its boundary, $\partial\Psi,$  is the union of the sets $\{p\in\Omega_\epsilon:
\varphi=0\},\,\{p\in\Omega_\epsilon:
\varphi=f(\epsilon^{-1})\},\,\{p\in\Omega_\epsilon:
\sigma_4=0\},\,\{p\in\Omega_\epsilon: \sigma_5=0\}$ with the
unitary 3-sphere.

\subsubsection{Singular points of the flow of \eqref{eqvphi}-\eqref{eqzrad} in the phase space \eqref{PhaseSpace}.}

The system \eqref{eqvphi}-\eqref{eqzrad} admits the following
singular points

\begin{enumerate}
\item The singular point $P_1$ with coordinates
$\varphi=0,\sigma_2=-1,\sigma_4=0,\sigma_5=0$ always exists. The
eigenvalues of the linearization around the singular point are $0,
\frac{1}{3},1-\frac{N}{\sqrt{6}},\frac{M (4-3 \gamma
)}{\sqrt{6}}-\gamma +2.$ This means that the singular point is
non-hyperbolic thus the Hartman-Grobmann theorem does not apply.
However, by applying the Invariant Manifold theorem, we obtain
that:
\begin{enumerate}
\item $P_1$ has a 1-dimensional center manifold tangent to the $\varphi$-axis provided $N\neq\sqrt{6}$ and $M\neq -\frac{\sqrt{6} (\gamma -2)}{3 \gamma -4}$ (otherwise the center manifold would be 2- or 3-dimensional).
\item $P_1$ admits a 3-dimensional unstable manifold and a 1-dimensional center manifold for
\begin{enumerate}\item  $N<\sqrt{6},\, 0<\gamma<\frac{4}{3},\,M>-\frac{\sqrt{6} (\gamma -2)}{3 \gamma -4};$ or
\item $N<\sqrt{6},\, \frac{4}{3}<\gamma<2,\,M<-\frac{\sqrt{6} (\gamma -2)}{3 \gamma -4}.$
\end{enumerate} In this case the center manifold of $P_1$ acts as a local source for an open set of orbits in \eqref{PhaseSpace}.
\item $P_1$ admits a 2-dimensional unstable manifold, a 1-dimensional stable manifold and a 1-dimensional center if
\begin{enumerate}\item  $N>\sqrt{6},\, 0<\gamma<\frac{4}{3},\,M>-\frac{\sqrt{6} (\gamma -2)}{3 \gamma -4};$ or
\item $N>\sqrt{6},\, \frac{4}{3}<\gamma<2,\,M<-\frac{\sqrt{6} (\gamma -2)}{3 \gamma -4};$ or
\item  $N<\sqrt{6},\, 0<\gamma<\frac{4}{3},\,M<-\frac{\sqrt{6} (\gamma -2)}{3 \gamma -4};$ or
\item  $N<\sqrt{6},\, \frac{4}{3}<\gamma<2,\,M>-\frac{\sqrt{6} (\gamma -2)}{3 \gamma -4}.$
\end{enumerate}
\item $P_1$ admits a 1-dimensional unstable manifold, a 2-dimensional stable manifold and a 1-dimensional center manifold for \begin{enumerate}\item  $N>\sqrt{6},\, 0<\gamma<\frac{4}{3},\,M<-\frac{\sqrt{6} (\gamma -2)}{3 \gamma -4};$ or
\item  $N>\sqrt{6},\, \frac{4}{3}<\gamma<2,\,M>-\frac{\sqrt{6} (\gamma -2)}{3 \gamma -4}.$
\end{enumerate}
\end{enumerate}

\item The singular point $P_2$ with coordinates
$\varphi=0,\sigma_2=1,\sigma_4=0,\sigma_5=0$ always exists. The
eigenvalues of the linearization around the singular point are $0,
\frac{1}{3},1+\frac{N}{\sqrt{6}},-\gamma +\frac{M (3 \gamma
-4)}{\sqrt{6}}+2.$ As before, let us determine conditions on the
free parameters for the existence of center, unstable and stable
manifolds for $P_2$.
\begin{enumerate}
\item If $N\neq-\sqrt{6}$ and $M\neq \frac{\sqrt{6} (\gamma -2)}{3 \gamma -4}$ there exists a 1-dimensional center manifold tangent to the $\varphi$-axis, otherwise the center manifold would be 2- or 3-dimensional.
\item $P_2$ has a 3-dimensional unstable manifold a a 1-dimensional center manifold (tangent the $\varphi$-axis) if \begin{enumerate}\item $N>-\sqrt{6},\,0<\gamma<\frac{4}{3},\,M<\frac{\sqrt{6} (\gamma -2)}{3 \gamma -4};$ or \item $N>-\sqrt{6},\,\frac{4}{3}<\gamma<2,\,M>\frac{\sqrt{6} (\gamma -2)}{3 \gamma -4}.$
\end{enumerate} In this case the center manifold of $P_2$ acts as a local source for an open set of orbits in \eqref{PhaseSpace}.
\item $P_2$ has a 2-dimensional unstable manifold a 1-dimensional stable and a 1-dimensional center manifold if
\begin{enumerate}\item $N<-\sqrt{6},\,0<\gamma<\frac{4}{3},\,M<\frac{\sqrt{6} (\gamma -2)}{3 \gamma -4};$ or
\item $N<-\sqrt{6},\,\frac{4}{3}<\gamma<2,\,M>\frac{\sqrt{6} (\gamma -2)}{3 \gamma -4};$ or
\item $N>-\sqrt{6},\,0<\gamma<\frac{4}{3},\,M>\frac{\sqrt{6} (\gamma -2)}{3 \gamma -4};$ or
\item $N>-\sqrt{6},\,\frac{4}{3}<\gamma<2,\,M<\frac{\sqrt{6} (\gamma -2)}{3 \gamma -4}.$
\end{enumerate}
\item $P_2$ has a 1-dimensional unstable manifold a 2-dimensional stable and a 1-dimensional center manifold if
\begin{enumerate}
\item $N<-\sqrt{6},\,0<\gamma<\frac{4}{3},\,M>\frac{\sqrt{6} (\gamma -2)}{3 \gamma -4};$ or
\item $N<-\sqrt{6},\,\frac{4}{3}<\gamma<2,\,M<\frac{\sqrt{6} (\gamma -2)}{3 \gamma -4}.$
\end{enumerate}
\end{enumerate}
\item The singular point $P_3$ with coordinates
$\varphi=0,\sigma_2=\frac{M (3 \gamma -4)}{\sqrt{6} (\gamma
-2)},\sigma_4=0,\sigma_5=0$ exists for
\begin{enumerate}\item $0<\gamma <\frac{4}{3},\,-\frac{\sqrt{6} (\gamma -2)}{3 \gamma -4}\leq M\leq \frac{\sqrt{6} (\gamma -2)}{3 \gamma
   -4};$ or \item $\frac{4}{3}<\gamma <2,\,\frac{\sqrt{6} (\gamma -2)}{3 \gamma -4}\leq M\leq -\frac{\sqrt{6} (\gamma
   -2)}{3 \gamma -4}.$\end{enumerate}
The eigenvalues of the linearization are $0,\,\lambda_1=-\frac{(3
\gamma -4) \left((3 \gamma -4) M^2-2 \gamma +4\right)}{12 (\gamma
-2)},\,\lambda_2=\frac{-M^2 (4-3 \gamma )^2+6 (\gamma
   -2) \gamma +2 M N (3 \gamma -4)}{12 (\gamma -2)},\,\lambda_3=\frac{6 (\gamma -2)^2-M^2 (4-3 \gamma )^2}{12 (\gamma -2)}.$
As before, let us determine conditions on the free parameters for the existence of center, unstable and stable manifolds for $P_3$.
\begin{enumerate}
\item For $\gamma, N$ and $M$ such that $\lambda_{1}\neq 0, \lambda_{2}\neq 0, \lambda_{3}\neq 0$ the center manifold is 1-dimensional and tangent to the $\varphi$-axis. Otherwise the center manifold coud be 2-, or 3-dimensional (it is never 4-dimensional).
\item $P_3$ admits a 1-dimensional center manifold and a 3-dimensional stable manifold for
\begin{enumerate}\item $0<\gamma<\frac{4}{3},\,-\frac{\sqrt{2} \sqrt{\gamma -2}}{\sqrt{3 \gamma -4}}<M<0,\,\ N>\frac{M^2 (4-3 \gamma
   )^2-6 (\gamma -2) \gamma }{2 M (3 \gamma -4)};$ or
\item $0<\gamma<\frac{4}{3},\,0<M<\frac{\sqrt{2} \sqrt{\gamma -2}}{\sqrt{3 \gamma -4}},\,
   N<\frac{M^2 (4-3 \gamma )^2-6 (\gamma -2) \gamma }{2 M (3 \gamma -4)}.$
\end{enumerate}
\item In the cases
\begin{enumerate}
\item $0<\gamma <\frac{4}{3},\,-\frac{\sqrt{6} (\gamma -2)}{3 \gamma -4}<M<-\frac{\sqrt{2} \sqrt{\gamma -2}}{\sqrt{3 \gamma -4}},N<\frac{M^2 (4-3 \gamma
   )^2-6 (\gamma -2) \gamma }{2 M (3 \gamma -4)};$ or
\item $0<\gamma <\frac{4}{3},\,\frac{\sqrt{2} \sqrt{\gamma -2}}{\sqrt{3 \gamma -4}}<M<\frac{\sqrt{6} (\gamma -2)}{3 \gamma -4},N>\frac{M^2 (4-3 \gamma
   )^2-6 (\gamma -2) \gamma }{2 M (3 \gamma -4)};$ or
\item $\frac{4}{3}<\gamma <2,\frac{\sqrt{6} (\gamma -2)}{3 \gamma -4}<M<0,N>\frac{M^2 (4-3 \gamma )^2-6 (\gamma -2) \gamma }{2 M
   (3 \gamma -4)};$ or
\item $\frac{4}{3}<\gamma <2, M=0, N\in\mathbb{R};$ or
\item $\frac{4}{3}<\gamma <2,0<M<-\frac{\sqrt{6} (\gamma -2)}{3 \gamma -4},N<\frac{M^2 (4-3 \gamma )^2-6 (\gamma -2) \gamma }{2 M
   (3 \gamma -4)},$  the unstable manifold is 2-dimensional (hence the stable manifold and the center manifold are both 1-dimensional).
\item Otherwise, $P_3$ has a 1-dimensional unstable manifold. Thus, it is never a local source since its unstable manifold is of dimension less than $3.$

\end{enumerate}

\end{enumerate}
\item The singular point $R_1$ with coordinates
$\varphi=0,\sigma_2=0,\sigma_4=0,\sigma_5=1$ always exists. The
eigenvalues of the linearization are

$0,\frac{2}{3},-\frac{1}{3},\frac{4}{3}-\gamma.$ The center
manifold is 1-dimensional and tangent to the $\varphi$-axis. The
unstable (stable) manifold is 1-dimensional (2-dimensional) if
$\gamma>\frac{4}{3}$ otherwise it is 2-dimensional
(1-dimensional). \item The singular point $R_2$ with coordinates
$\sigma_2=\frac{\sqrt{\frac{2}{3}}}{M},\sigma_4=0,\sigma_5=\frac{\sqrt{\frac{4-2
\gamma }{M^2}+3 \gamma -4}}{\sqrt{3 \gamma -4}}$ exists for
$0<\gamma<\frac{4}{3},\,M^2\geq \frac{2\left(\gamma -2\right)}{3
\gamma -4}.$ The eigenvalues of the linearization are

$0,-\frac{M+\sqrt{3 M^2 (4 \gamma -5)-8 (\gamma -2)}}{6 M},\frac{\sqrt{3 M^2 (4 \gamma -5)-8 (\gamma -2)}-M}{6
   M},\frac{1}{3} \left(\frac{N}{M}+2\right).$
Let us determine conditions on the free parameters for the existence of center, unstable and stable manifolds for $R_2$.
\begin{enumerate}
\item $R_2$ has a 3-dimensional stable manifold and a 1-dimensional center manifold if
\begin{enumerate}
\item $0<\gamma <\frac{5}{4},-2 \sqrt{\frac{2}{3}} \sqrt{\frac{\gamma -2}{4 \gamma -5}}\leq M<-\sqrt{2} \sqrt{\frac{\gamma -2}{3
   \gamma -4}},N>-2 M;$ or
\item $0<\gamma <\frac{5}{4},\sqrt{2} \sqrt{\frac{\gamma -2}{3 \gamma -4}}<M\leq 2 \sqrt{\frac{2}{3}} \sqrt{\frac{\gamma -2}{4
   \gamma -5}},N<-2 M;$ or
\item $\frac{5}{4}\leq \gamma <\frac{4}{3},M<-\sqrt{2} \sqrt{\frac{\gamma -2}{3 \gamma -4}},N>-2 M;$  or
\item $\frac{5}{4}\leq \gamma <\frac{4}{3},M>\sqrt{2} \sqrt{\frac{\gamma -2}{3 \gamma -4}},N<-2 M;$ or
\item $0<\gamma <\frac{5}{4},M<-2 \sqrt{\frac{2}{3}} \sqrt{\frac{\gamma -2}{4 \gamma -5}}, N>-2 M;$ or
\item $0<\gamma <\frac{5}{4},M>2 \sqrt{\frac{2}{3}} \sqrt{\frac{\gamma -2}{4 \gamma -5}}, N<-2 M.$
\end{enumerate}
\item By reversing the sign of the last inequality, i.e., the inequality solved for $N$, in the previous six cases we obtain conditions for $R_2$ having a 2-dimensional stable manifold, a 1-dimensional unstable manifold and a 1-dimensional center manifold.
\end{enumerate}
\item The singular point $P_4$ with coordinates $\varphi=0,
\sigma_2=-\frac{N}{\sqrt{6}},\sigma_4=\sqrt{1-\frac{N^2}{6}},
\sigma_5=0$ exists whenever $N^2<6.$ The eigenvalues of the
linearization are

$0,\frac{1}{6} \left(N^2-6\right),\frac{1}{6} \left(N^2-4\right),\frac{1}{3} N (2 M+N)-\frac{1}{2} (M N+2) \gamma.$ The conditions for the existence of stable, unstable and center manifolds is as follows.
\begin{enumerate}
\item The center manifold is 1-dimensional and the stable manifold is 3-dimensional provided
\begin{enumerate}
\item $N=0,\,M\in\mathbb{R},\gamma\neq \frac{4}{3};$ or
\item $0<\gamma <\frac{4}{3},-2<N<0,M>\frac{2 \left(N^2-3 \gamma \right)}{N (3 \gamma -4)};$ or
\item $0<\gamma <\frac{4}{3},0<N<2,M<\frac{2 \left(N^2-3 \gamma \right)}{N (3 \gamma -4)};$ or
\item $\frac{4}{3}<\gamma <2,-2<N<0,M<\frac{2 \left(N^2-3 \gamma \right)}{N (3 \gamma -4)};$ or
\item $\frac{4}{3}<\gamma <2,0<N<2,M>\frac{2 \left(N^2-3 \gamma \right)}{N (3 \gamma -4)}.$
\end{enumerate}
\item The stable manifold is 2-dimensional, the unstable manifold is 1-dimensional and the center manifold is 1-dimensional provided
\begin{enumerate}
\item $0<\gamma <\frac{4}{3},-\sqrt{6}<N<-2,M>\frac{2 \left(N^2-3 \gamma \right)}{N (3 \gamma -4)};$ or
\item $0<\gamma <\frac{4}{3},-2<N<0,M<\frac{2 \left(N^2-3 \gamma \right)}{N (3 \gamma -4)};$ or
\item $0<\gamma <\frac{4}{3},0<N<2,M>\frac{2 \left(N^2-3 \gamma \right)}{N (3 \gamma -4)};$ or
\item $0<\gamma <\frac{4}{3},2<N<\sqrt{6},M<\frac{2 \left(N^2-3 \gamma \right)}{N (3 \gamma -4)};$ or
\item $\frac{4}{3}<\gamma <2,-\sqrt{6}<N<-2,M<\frac{2 \left(N^2-3 \gamma \right)}{N (3 \gamma -4)};$ or
\item $\frac{4}{3}<\gamma <2,-2<N<0,M>\frac{2 \left(N^2-3 \gamma \right)}{N (3 \gamma -4)};$ or
\item $\frac{4}{3}<\gamma <2,0<N<2,M<\frac{2 \left(N^2-3 \gamma \right)}{N (3 \gamma -4)};$ or
\item $\frac{4}{3}<\gamma <2,2<N<\sqrt{6},M>\frac{2 \left(N^2-3 \gamma \right)}{N (3 \gamma -4)}.$
\end{enumerate}
\item The stable manifold is 1-dimensional, the unstable manifold is 2-dimensional and the center manifold is 1-dimensional provided
\begin{enumerate}
\item $0<\gamma <\frac{4}{3},-\sqrt{6}<N<-2,M<\frac{2 \left(N^2-3 \gamma \right)}{N (3 \gamma -4)};$ or
\item $0<\gamma <\frac{4}{3},2<N<\sqrt{6},M>\frac{2 \left(N^2-3 \gamma \right)}{N (3 \gamma -4)};$ or
\item $\frac{4}{3}<\gamma <2,-\sqrt{6}<N<-2,M>\frac{2 \left(N^2-3 \gamma \right)}{N (3 \gamma -4)};$ or
\item $\frac{4}{3}<\gamma <2,2<N<\sqrt{6},M<\frac{2 \left(N^2-3 \gamma \right)}{N (3 \gamma -4)}.$
\end{enumerate}
\end{enumerate}
\item The singular point $R_3$ with coordinates
$\varphi=0,\sigma_2=-\frac{2
\sqrt{\frac{2}{3}}}{N},\sigma_4=\frac{2}{\sqrt{3}
|N|},\sigma_5=\frac{\sqrt{N^2-4}}{|N|}$ exists for $N^2\geq 4.$
The eigenvalues of the linearization are

$0,\frac{1}{6} \left(-\frac{\sqrt{64 N^2-15 N^4}}{N^2}-1\right),\frac{1}{6} \left(\frac{\sqrt{64 N^2-15
   N^4}}{N^2}-1\right),-\frac{(2 M+N) (3 \gamma -4)}{3 N}.$
The conditions for the existence of stable, unstable and center manifolds are as follows.
\begin{enumerate}
\item The stable manifold is 3-dimensional and the center manifold is 1-dimensional provided
\begin{enumerate}
\item $0<\gamma <\frac{4}{3},N<-\frac{8}{\sqrt{15}}, M>-\frac{N}{2};$ or
\item $0<\gamma <\frac{4}{3},-\frac{8}{\sqrt{15}}\leq N<-2,M>-\frac{N}{2};$ or
\item $0<\gamma <\frac{4}{3},2<N\leq \frac{8}{\sqrt{15}},M<-\frac{N}{2};$ or
\item $0<\gamma <\frac{4}{3},N>\frac{8}{\sqrt{15}},M<-\frac{N}{2};$ or
\item $\frac{4}{3}<\gamma <2,N<-\frac{8}{\sqrt{15}},M<-\frac{N}{2};$ or
\item $\frac{4}{3}<\gamma <2,-\frac{8}{\sqrt{15}}\leq N<-2,M<-\frac{N}{2};$ or
\item $\frac{4}{3}<\gamma <2,2<N\leq \frac{8}{\sqrt{15}},M>-\frac{N}{2};$ or
\item $\frac{4}{3}<\gamma <2,N>\frac{8}{\sqrt{15}},M>-\frac{N}{2}.$
\end{enumerate}
\item By reversing the sign of the last inequality, i.e., the inequality solved for $M$, in the previous eight cases we obtain conditions for $R_3$ having a 2-dimensional stable manifold, a 1-dimensional unstable manifold and a 1-dimensional center manifold.
\end{enumerate}
\item The singular point $P_5$ with coordinates

$\varphi=0,\sigma_2=\frac{\sqrt{6} \gamma }{M (3 \gamma -4)-2 N},$ $\sigma_4=\frac{\sqrt{M^2 (4-3 \gamma )^2+M N (8-6 \gamma )-6 (\gamma -2) \gamma }}{2 N+M (4-3 \gamma )},\sigma_5=0$ exists for $2 (2 M+N)>3 M \gamma,\, M (3 \gamma -4) (M (3 \gamma -4)-2 N)\geq 6 (\gamma -2) \gamma,$
and
$\frac{3 (M N+2) \gamma -2 N (2
   M+N)}{(2 N+M (4-3 \gamma ))^2}\leq 0.$ The eigenvalues of the linearization are
$0,\frac{12 M+6 N-3 (3 M+N) \gamma +\sqrt{3} \sqrt{f(\gamma
,M,N)}}{6 (M (3 \gamma -4)-2 N)},\frac{3 N (\gamma -2)+3 M (3
   \gamma -4)+\sqrt{3} \sqrt{f(\gamma ,M,N)}}{6 (2 N+M (4-3 \gamma ))}$, and $\frac{(2 M+N) (3 \gamma -4)}{6 N+3 M (4-3 \gamma
   )},$
   where $f(\gamma ,M,N)=2 M^3 N (3 \gamma -4)^3+2 M N \left(4 N^2-6 \gamma ^2+3 \gamma -6\right) (3 \gamma -4)-M^2 \left(8 N^2-12 \gamma
   -3\right) (4-3 \gamma )^2+3 (\gamma -2) \left(N^2 (9 \gamma -2)-24 \gamma ^2\right).$ The stability conditions of $P_5$ are very complicated to display them here. Thus we must rely on numerical experimentation. We can obtain, however, some analytic results. For instance, there exists at least a 1-dimensional center manifold. The unstable manifold is always of dimension lower than 3. Thus the singular point is never a local source. If all the eigenvalues, apart form the zero one, have negative reals parts, then the center manifold of $P_5$ acts as a local sink. This means that the orbits in the stable manifold approach the center manifold of $P_5$ when the time goes forward.
\item The singular point $P_6$ with coordinates

$\varphi=0,\sigma_2=\frac{\sqrt{6} \gamma }{M (3 \gamma -4)-2 N},$ $\sigma_4=-\frac{\sqrt{M^2 (4-3 \gamma )^2+M N (8-6 \gamma )-6 (\gamma -2) \gamma }}{2 N+M (4-3 \gamma )},\sigma_5=0$ exists for $M (3 \gamma -4) (M (3 \gamma -4)-2 N)\geq 6 (\gamma -2) \gamma ,\,2 (2 M+N)<3 M \gamma,$
and
$\frac{3 (M N+2) \gamma -2 N (2
   M+N)}{(2 N+M (4-3 \gamma ))^2}\leq 0.$ The eigenvalues of the linearization are the same displayed in the previous point. However the stability conditions are rather different (since the existence conditions are different from those of $P_5$). As before, the stability conditions are very complicated to display them here, but similar conclusions concerning the center and unstable manifold, as for $P_5,$ are obtained. For get further information about its stability we must to resort to numerical experimentation.
\end{enumerate}

\subsubsection{Physical description of the solutions and connection with observables.}

Let us now present the formalism of obtaining the physical
description of a singular point, and also connect with the basic
observables relevant for a physical discussion. These will allow
us to describe the cosmological behavior of each singular point,
in the next section.

Firstly, around a singular point we obtain first-order expansions
for $H,a,\phi,$ and $\rho$ and $\rho_r$ in terms of $t$,
considering equations: \eqref{Raych}; the definition of the scale
factor $a$ in terms of the Hubble factor $H$; the definition of
$\sigma_2;$  the matter conservation equations \eqref{consm} and
\eqref{consr}, respectively, given by
\begin{align}
& 2 \dot H(t)=H(t)^2 \left(3 (\gamma -2) {\sigma_2^\star}^2+3
\gamma  \left({\sigma_4^\star}^2+{\sigma_5^\star}^2-1\right)-4
   {\sigma_5^\star}^2\right),\nonumber\\& \dot a(t)=a(t) H(t),\nonumber\\& \dot\phi(t)=\sqrt{6} {\sigma_2^\star} H(t),\nonumber\\& \dot\rho(t)=-\frac{3}{2} H(t)^3
   \left(\sqrt{6} M (3 \gamma -4) {\sigma_2^\star}-6 \gamma \right) \left({\sigma_2^\star}^2+{\sigma_4^\star}^2+{\sigma_5^\star}^2-1\right),\nonumber\\& \dot\rho_r(t)=-12 {\sigma_5^\star}^2 H(t)^3,\label{APPROX}
\end{align}
where the star-superscript denotes the evaluation at a specific
singular point. The equation \be \ddot \phi(t)= \frac{3}{2} H(t)^2
\left(M (3 \gamma -4)
\left({\sigma_2^\star}^2+{\sigma_4^\star}^2+{\sigma_5^\star}^2-1\right)-2
\left(N {\sigma_4^\star}^2+\sqrt{6}
{\sigma_2^\star}\right)\right),\label{consistency}\ee derived from
the equation of motion for the scalar field \eqref{KG} should be
used as a consistency test for the above procedure.  Solving the
differential equations \eqref{APPROX} and substituting the
resulting expressions in the equation \eqref{consistency} results
in
\begin{align}-6 M (3 \gamma -4) \left({\sigma_2^\star}^2+{\sigma_4^\star}^2+{\sigma_5^\star}^2-1\right)+12 N {\sigma_4^\star}^2+\nonumber\\+2 \sqrt{6}
   {\sigma_2^\star} \left(3 \gamma  \left({\sigma_2^\star}^2+{\sigma_4^\star}^2+{\sigma_5^\star}^2-1\right)-6 {\sigma_2^\star}^2-4 {\sigma_5^\star}^2+6\right)=0.\end{align}

This integrability condition should be (at least asymptotically) fulfilled.

\begin{table*}[!ht]
\caption{\label{tab2b} Observable cosmological quantities, and
physical behavior of the solutions, at the singular points of the
cosmological system. We use the notations
$M_1(\gamma)=\frac{\sqrt{2 \gamma  (3 \gamma -8)+8}}{4-3 \gamma
},$ $M_2(\gamma)=\frac{\sqrt{6} \sqrt{(\gamma -3) \gamma +2}}{4-3
\gamma }$
 }\bigskip   \centering
{\small\begin{tabular}{cccc}
  \hline   \hline
  \ \ Cr.P.  \ \  &\  $q$ \ & $w_{\text{eff}}$   & Solution/description\\
  \hline \hline
$P_1$ & 2 & 1 &  Decelerating. \\
   \hline\\
$P_2$ & 2 & 1 & Decelerating. \\
     \hline\\
$P_3$ & $\frac{-M^2 (4-3 \gamma )^2+2 \gamma  (3 \gamma -8)+8}{4 (\gamma -2)}$ & $-\frac{M^2 (4-3 \gamma
   )^2}{6 (\gamma -2)}+\gamma -1$ &  Accelerating for   \\
   \vspace{0.2cm}
   &  & &   $0<\gamma <\frac{2}{3}$ \\
    \vspace{0.2cm}
   &  & &   $-M_1(\gamma)<M<M_1(\gamma)$ \\ \hline \\
$P_4$ & $\frac{1}{2} \left(N^2-2\right)$& $\frac{1}{3}
   \left(N^2-3\right)$ &  Accelerating for   \\
   \vspace{0.2cm}
   &  & &   $-\sqrt{2}<n<\sqrt{2}$ \\
   \vspace{0.2cm}
   &  & &   powerlaw-inflationary \\ \hline \\
$P_5$ & $\frac{3 (M+N) \gamma -2 (2 M+N)}{2 N+M (4-3 \gamma )}$ & $\frac{M
   (4-3 \gamma )-2 N (\gamma -1)}{M (3 \gamma -4)-2 N}$ &  Accelerating for   \\
   \vspace{0.2cm}
   &  & & $\frac{3 (M+N) \gamma -2 (2 M+N)}{2 N+M (4-3 \gamma )}<0$   \\
   &  & &  matter-kinetic-potential- \\
     &  & &   scaling. \\
     \hline\\
$P_6$ & $\frac{3 (M+N) \gamma -2 (2 M+N)}{2 N+M (4-3 \gamma )}$ &
$\frac{M
   (4-3 \gamma )-2 N (\gamma -1)}{M (3 \gamma -4)-2 N}$ &  Accelerating for   \\
   \vspace{0.2cm}
   &  & & $\frac{3 (M+N) \gamma -2 (2 M+N)}{2 N+M (4-3 \gamma )}<0$   \\
   &  & &  matter-kinetic-potential- \\
     &  & &   scaling. \\
     \hline\\
   $R_1$ & 1 & $\frac{1}{3}$ & Decelerating. \\   &  & &   Radiation-dominated. \\
     \hline\\
$R_2$ & 1 & $\frac{1}{3}$ & Decelerating. \\
   &  & &   radiation-kinetic-potential- \\
     &  & &   scaling.\\
     \hline\\
$R_3$ & 1 & $\frac{1}{3}$ & Decelerating. \\
   &  & &   radiation-kinetic-potential- \\
     &  & &   scaling. \\
     \hline\\
\end{tabular}}

\end{table*}

Instead of apply this procedure to a generic singular point here,
we submit the reader to section \ref{applications} for some worked
examples where this procedure has been applied. However we will
discuss on some cosmological observables.

We can calculate the deceleration parameter $q$ defined as usual as \cite{WE}
\begin{equation}
\label{qq}q=-\frac{a \ddot a}{a^2}.
\end{equation}
Additionally, we can calculate the effective (total) equation-of-state parameter of the universe $w_{\text{eff}}$, defined conventionally as
\begin{equation}
\label{weff}
w_{\text{eff}}\equiv\frac{p_{\text{tot}}}{\rho_{\text{tot}}},
\end{equation}
where $p_{\text{tot}}$ and $\rho_{\text{tot}}$ are respectively the
total isotropic pressure and the total energy density. Therefore, in terms of the auxiliary variables we have
\begin{align}
\label{qq2} & q= -\frac{3}{2} (\gamma -2) \sigma_2^2-\frac{3
\gamma  \sigma_4^2}{2}+\frac{1}{2} (4-3
   \gamma ) \sigma_5^2+\frac{1}{2} (3 \gamma -2)\\
&w_{\text{eff}}=(2-\gamma ) \sigma_2^2-\gamma
\sigma_4^2+\frac{1}{3} (4-3 \gamma ) \sigma_5^2+\gamma -1.
\label{weff2}
\end{align}

First of all, for each singular point described in the last
section we calculate the effective (total) equation-of-state
parameter of the universe $w_{\text{eff}}$ using \eqref{weff2},
and the deceleration parameter $q$ using \eqref{qq2}. The results
are presented in Table \ref{tab2b}. Furthermore, as usual, for an
expanding universe $q<0$ corresponds to accelerating expansion and
$q>0$ to decelerating expansion.

\subsection{The Flow as
$\phi\rightarrow-\infty$}\label{minusinfinity}

With the purpose of complementing the global analysis of the system it is necessary investigate its behavior as $\phi\rightarrow-\infty.$ It is an easy task since the system \eqref{eq0phi}-\eqref{eq0x4} is invariant under the transformation of coordinates

$$(\phi, \sigma_2)\rightarrow -(\phi, \sigma_2),\; V\rightarrow U,\; \chi\rightarrow \Xi,$$ where $U(\phi)=V(-\phi)$ and
$\Xi(\phi)=\chi(-\phi).$ Hence, for a particular potential $V,$
and a particular coupling function $\chi$, the behavior of the
solutions of the equations \eqref{eq0phi}-\eqref{eq0x4} around $\phi=-\infty$ is equivalent
(except for the sign of $\phi$) to the behavior of the system near
$\phi=\infty$ with potential and coupling functions $U$ and $\Xi,$
respectively.

If $U$ and $\Xi$ are of class $\mathcal{E}^2_+,$ the preceding  analysis in $\bar{\Sigma}_\epsilon$ can be applied
(with and adequate choice of $\epsilon$).

The set of
class $C^k$ functions well behaved in both $+\infty$ and
$-\infty$ is denoted by $\mathcal{E}^k.$ Latin uppercase letters with subscripts
$+\infty$ and $-\infty,$ are used respectively to indicate the exponential
order of $\mathcal{E}^k$ functions in $+\infty$ and in $-\infty.$

\section{Examples}\label{applications}

In this section we apply the mathematics discussed in previous sections to several worked examples from both analytical and numerical viewpoint.

\subsection{Numerical Evidence of the Result of Theorem \ref{thm4}}\label{application1}

\begin{figure}[ht]
\begin{center}
\mbox{\epsfig{figure=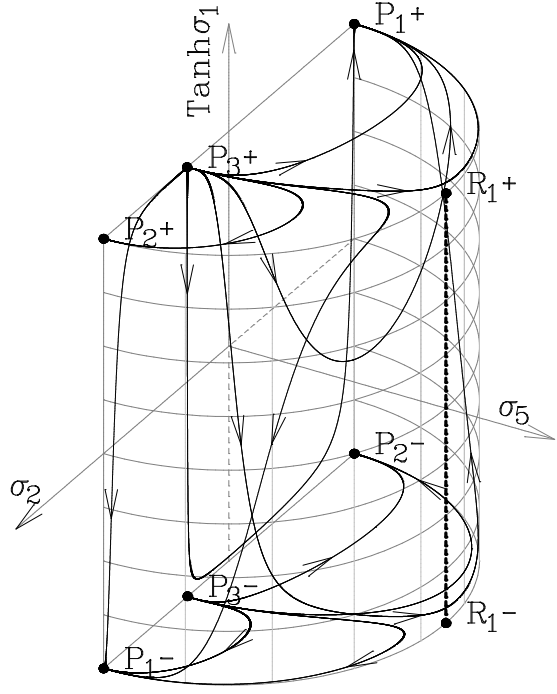,width=9.9cm,angle=0}}
\caption{ \label{fig1}{Qualitative dynamics of the flow of \eqref{systthm4} for the choice $M=\sqrt{2/3}$ and  $\gamma=1.$ Observe that $P_1^+$ or $P_1^-$ are the local attractors (sinks). $P_3^+$ is a local source, $P_3^-$ is a saddle for the full dynamics, but it is a local source in the invariant set $\tanh \sigma_1=-1.$ The thick dashed line is an invariant set which is unstable. In fact all its points including $R_1^+$ and $R_1^-$, act as saddle points. They correspond to cosmological radiation-dominated solutions.}}
\end{center}
\end{figure}

For the particular case $\chi(\sigma_1)=e^{M \sigma_1},$ from equation \eqref{proxi2}, follows that $\sigma_2=\sigma_{2c}:=\frac{M(4-3\gamma)}{\sqrt{6}(2-\gamma)}$ is an invariant set. Given $\frac{4}{3}<\gamma <2,$ the existence conditions lead to $M_1(\gamma)\leq M\leq -\frac{\sqrt{6} (\gamma
   -2)}{3 \gamma -4},$ where $M_1(\gamma)=\frac{\sqrt{2 \gamma  (3 \gamma -8)+8}}{4-3 \gamma }.$ For such values $\frac{\partial f_2}{\partial \sigma_2}|_{\sigma_{2c}}=\frac{M^2 (4-3 \gamma )^2-6 (\gamma -2)^2}{12 (\gamma -2)}\geq 0.$ Thus the asymptotic phase configuration $\sigma_1\rightarrow -\text{sgn}{\sigma_{2c}}\infty, \sigma_2\rightarrow \sigma_{2c}$ is never approached (for an open set of orbits) as $\tau\rightarrow \infty.$ For the original system (i.e., taking the time reversal transformation) this means that this asymptotic phase configuration is never approached towards the past.
In figure \ref{fig1} we show the qualitative dynamics of the flow of \eqref{systthm4} for the choice $M=\sqrt{2/3},$ and $\gamma=1.$ In order to compactify the phase space we have introduce the coordinate transformation $\sigma_1\rightarrow \tanh \sigma_1.$ The mentioned asymptotic configuration is represented in figure \ref{fig1} by $P_3^\pm.$

\begin{figure}[ht]
\begin{center}
\mbox{\epsfig{figure=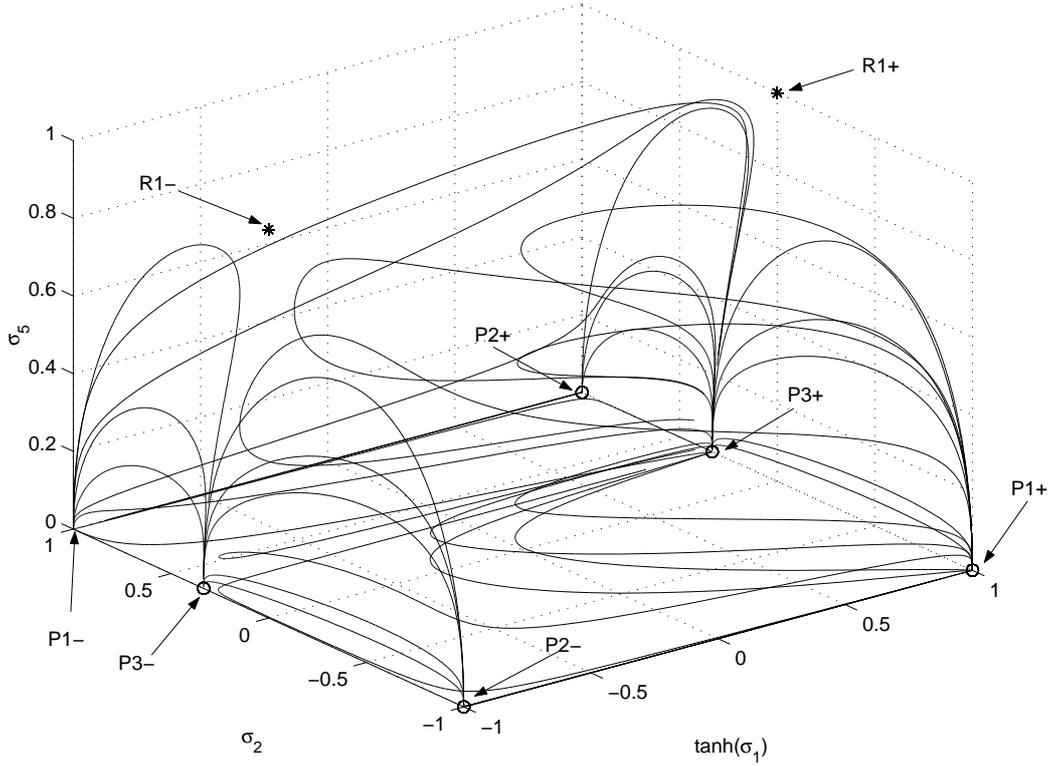,width=14cm,angle=0}}
\caption{ \label{fig1c}{The graphic illustrates the result of theorem \ref{thm4}. We set $M=\sqrt{2/3}$ and  $\gamma=1.$ The point with $\tanh \sigma_1=\pm 1$ are the local sinks (thus for the original system \eqref{eq0phi}-\eqref{eq0x4} the scalar field almost always diverges towards the past).}}
\end{center}
\end{figure}

\subsection{Coupling Functions and Potentials of Exponential Orders $M=0$ and $N=-\mu\neq 0,$ Respectively}

\begin{table}[!htb]
{\caption{\label{critcrit} Location of the singular points of the
flow of \eqref{eqvphi}-\eqref{eqzrad} defined in the invariant set
$\left\{p\in\Omega_\epsilon: \varphi=0\right\}$ for $M=0$ and
$N=-\mu.$}\bigskip {\small \begin{tabular}[t]{|l|c|c|c|c|} \hline
Label&$(\sigma_2,\sigma_4,\sigma_5)$&Existence&Stability$^{\rm a}$\\[1ex]
\hline
\hline &&& \\[-2ex]
$P_1$&$(-1,0,0)$&always & unstable if $\mu>-\sqrt{6}$\\[1ex]
\hline
\hline &&& \\[-2ex]
$P_2$&$(1,0,0)$&always & unstable if $\mu<\sqrt{6}$\\[1ex]
\hline
\hline &&& \\[-2ex]
$P_3$&$(0,0,0)$&always & saddle\\[1ex]
\hline
\hline &&& \\[-2ex]
$R_1$&$(0,0,1)$&always & saddle\\[1ex]
\hline
\hline &&& \\[-2ex]
$P_4$&$\left(\frac{\mu}{\sqrt{6}},\sqrt{1-\frac{\mu^2}{6}},0\right)$&$\mu^2<6$ & stable for  \\
&&& $\left\{\begin{array}{c} 0<\gamma<\frac{4}{3},\,\,\mu^2<3\gamma,\, \text{or}\\
\frac{4}{3}<\gamma<2,\,\, \mu^2<2
\end{array},\right.$\\
&&& saddle otherwise\\[1ex]
\hline
\hline &&& \\[-2ex]
$P_{5,6}$&$\left(\sqrt{\frac{3}{2}}\frac{\gamma}{\mu}, \pm \frac{1}{\mu}\sqrt{\frac{3}{2}(2-\gamma)\gamma},0 \right)$&$\mu^2>3\gamma$ & stable for  \\
&&& $\left\{\begin{array}{c} 0<\gamma<\frac{2}{9},\,\mu^2>3\gamma,\, \text{or}\\
\frac{2}{9}<\gamma<\frac{4}{3},\,\,
3\gamma<\mu^2<\frac{24\gamma^2}{9\gamma-2},\, \text{or}\\
\frac{2}{9}<\gamma <\frac{4}{3},\,\mu ^2>\frac{24 \gamma ^2}{9
\gamma -2}
\end{array},\right.$\\
&&& saddle otherwise\\[1ex]
\hline
\hline &&& \\[-2ex]
$R_3$ & $\left(\frac{2\sqrt{\frac{2}{3}}}{\mu},\frac{2}{\sqrt{3}|\mu|},\frac{\sqrt{\mu^2-4}}{|\mu|}\right)$& $|\mu|>2$ & stable if $\frac{4}{3}<\gamma<2,$ saddle otherwise \\[1ex]
\hline
\end{tabular}}
$^{\rm a}$ The stability is analyzed for the flow restricted to
the invariant set $\varphi=0$.}
\end{table}

As an example let us consider $\chi,V\in {\cal E}^2_+$ of
exponential orders $M=0$ and $N=-\mu,$ respectively. This class of
potentials contains the cases investigated in
\cite{Copeland:1997et,vandenHoogen:1999qq} (there are not
considered coupling to matter, i.e., $\chi(\phi)\equiv 1$, in the
second case, for flat FRW cosmologies), the case investigated in
\cite{Copeland:2009be} (for positive potentials and standard FRW
dynamics), the example examined in \cite{Leon:2008de}, etc. In
table \ref{critcrit} are summarized the location, existence conditions
and stability of the singular points. \footnote{The stability is
analyzed for the flow restricted to the invariant set $\varphi=0$,
i.e., we are not taking into account perturbations in the
$\varphi$-axis.}

Let us discuss some physical properties of the cosmological
solutions associated to the singular points displayed in table
\ref{crit}.
\begin{itemize}
\item  $P_{1,2}$  represent kinetic-dominated cosmological
solutions. They behave as stiff-like matter. The associated
cosmological solution satisfies $H=\frac{1}{3 t-c_1},a=\sqrt[3]{3
t-c_1} c_2,\phi =c_3\pm\sqrt{\frac{2}{3}} \ln \left(3
   t-c_1\right),$ where $c_j,\,j=1,2,3$ are integration constants. These solutions are associated with the local past attractors of the systems for an open set of values of the parameter $\mu.$
\item  $P_3$ represents matter-dominated cosmological solutions
that satisfy $H=\frac{2}{3 t \gamma -2 c_1},a=\left(3 t \gamma -2
c_1\right){}^{\frac{2}{3 \gamma }} c_2,\rho
   =\frac{12}{\left(3 t \gamma -2 c_1\right){}^2}+c_3.$
\item $R_1$ represents a radiation-dominated cosmological
solutions satisfying $H=\frac{1}{2 t-c_1},a=\sqrt{2 t-c_1}
c_2,\rho_r=\frac{3}{\left(2 t-c_1\right){}^2}+c_3.$ \item $P_4$
represents power-law scalar-field dominated inflationary
cosmological solutions. As $t\rightarrow 0^+$ the potential
behaves as $V\sim V_0 \exp[-\mu \phi]$. Thus it is easy to obtain
the asymptotic exact solution: $H=\frac{2}{t \mu ^2-2
c_1},a=\left(t \mu ^2-2 c_1\right){}^{\frac{2}{\mu ^2}} c_2,\phi
\sim\frac{1}{\mu}\ln\left[\frac{V_0(t\mu^2-2c_1)^2}{2(6-\mu^2)}\right].$
\item $P_{5,6}$ represent matter-kinetic-potential scaling
solutions. As before, in the limit $t\rightarrow 0^+$ we obtain
the asymptotic expansions: $H=\frac{2}{3 t \gamma -2
c_1},a=\left(3 t \gamma -2 c_1\right){}^{\frac{2}{3 \gamma }}
c_2,\phi \sim \frac{1}{\mu}\ln\left[\frac{V_0\mu^2(3t\gamma-2
c_1)^2}{18(2-\gamma)\gamma}\right].$ \item $R_3$ represent
radiation-kinetic-potential scaling solutions. As before are
deduced the following asymptotic expansions: $H=\frac{1}{2
t-c_1},a=\sqrt{2 t-c_1} c_2,\phi \sim
\frac{1}{\mu}\ln\left[\frac{v_0\mu^2(2t-c_1)^2}{4}\right].$
\end{itemize}

\subsubsection{Powerlaw coupling and Albrecht-Skordis
potential in the invariant set $\rho_r=0$}\label{toy}

Let us consider the coupling function \be
\chi(\phi)=\left(\frac{3\alpha}{8}\right)^{\frac{1}{\alpha}}\chi_0(\phi-\phi_0)^
{\frac{2}{\alpha}},\; \alpha>0,\text{const.},\,\phi_0\geq
0.\label{couplingexample}\ee

Observe that $$\frac{d\ln
\chi(\phi)}{d\chi}=\frac{2}{\alpha(\phi-\phi_0)}\neq 0$$ for all
finite value of $\phi.$ Since $\ln \chi(\phi)$ has not stationary
points thus the early time dynamics is associated to the limit
where the scalar field diverges.

\begin{figure}[ht]
\begin{center}
\hspace{0.8cm} \put(195,-3){${P_4}$} \put(215,-3){${P_2}$}
\put(0,-3){${P_1}$}\put(107,205){${\varphi}$}\put(235,3){${x_1}$}
\includegraphics[width=8cm, height=7cm]{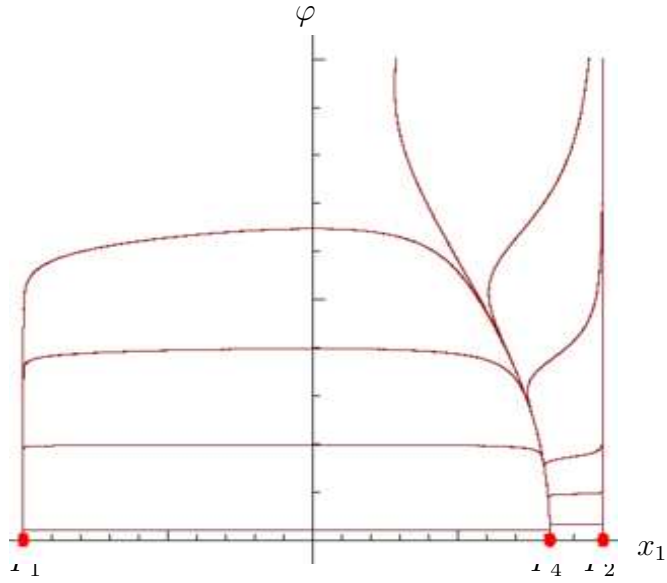}
\caption[Orbits in the invariant set  $\{x_2=0\}\subset
\bar{\Sigma}_\epsilon$ for the coupling function
(\ref{couplingexample}) and potential
(\ref{Albrecht-Skordis}).]{Orbits in the invariant set
$\{x_2=0\}\subset \bar{\Sigma}_\epsilon$ for the model with
coupling function (\ref{couplingexample})  potential
(\ref{Albrecht-Skordis}). We select the values of the parameters:
$\epsilon=1.00,$ $\mu= 2.00, A = 0.50, \alpha = 0.33, B = 0.5,$
 $\phi_0=0,$ and $\gamma=1.$ Observe that i) almost all the orbits are past
asymptotic to $P_1;$ ii) $P_2$ is a saddle, and iii) the center
manifold of $P_4$ attracts all the orbits in the $\{x_2=0\}$.
However, it is no more an attractor in the invariant set
$x_2>0,\,\varphi=0$ (see figure \ref{FIG2})} \label{FIG1}
\end{center}
\end{figure}

\begin{figure}[ht]
\begin{center}
\hspace{0.4cm} \put(195,-3){${P_4}$} \put(215,-3){${P_2}$}
\put(0,-3){${P_1}$}\put(180,105){${P_5}$} \put(100,175){${P_3}$}
\put(107,205){${x_2}$}\put(235,3){${x_1}$}
\includegraphics[width=8cm, height=7cm]{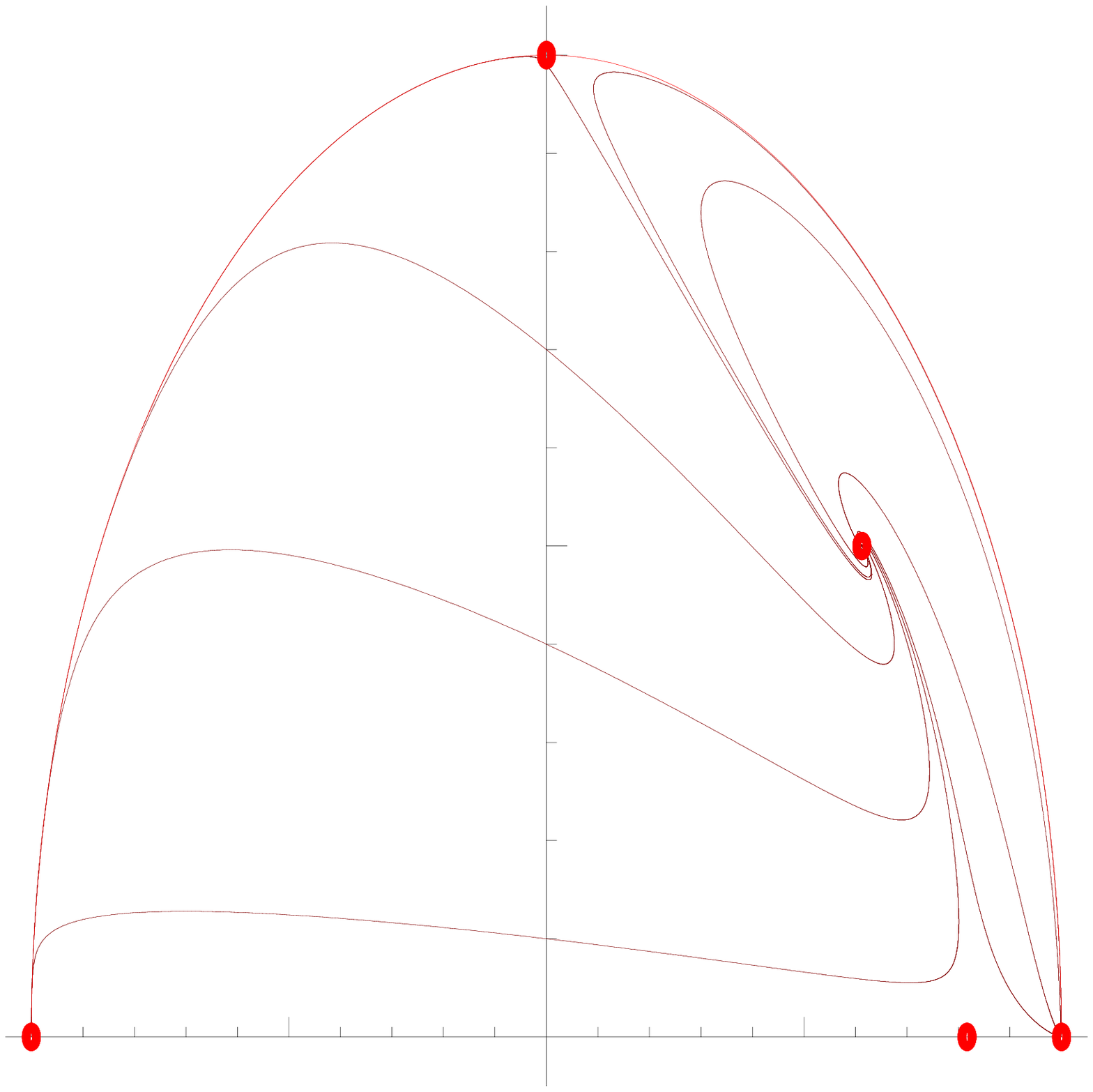}
\caption[Orbits in the invariant set $\{\varphi=0\}\subset
\bar{\Sigma}_\epsilon$ for the coupling function
(\ref{couplingexample}) and potential
(\ref{Albrecht-Skordis}).]{Orbits in the invariant set
$\{\varphi=0\}\subset \bar{\Sigma}_\epsilon$ for the model with
coupling function (\ref{couplingexample}) and potential
(\ref{Albrecht-Skordis}). We select the values of the parameters:
$\epsilon=1.00,$ $\mu= 2.00, A = 0.50, \alpha = 0.33, B = 0.5,$
 $\phi_0=0,$ and $\gamma=1.$ In the figure i) $P_{1,2}$ are local past
attractors, but $P_1$ is the global past attractor; ii) $P_{3,4}$
are saddles, and iii) $P_5$ is a local future
attractor.}\label{FIG2}
\end{center}
\end{figure}

This choice produces a coupling BD parameter given by

$$2\omega(\chi)+3=\frac{4}{3}\alpha\left(\frac{\chi}{\chi_0}\right)^\alpha.$$

This types of power law couplings were investigated in
\cite{DTorres} from the astrophysical viewpoint. For STTs without
potential, the cosmological solutions for the matter domination
era (in a Robertson-Walker metric) are $a(t)\propto (\ln
t)^{(\alpha-1)/3\alpha} t^{\frac{2}{3}},\; \phi(t)\propto (\ln
t)^{\frac{1}{\alpha}}.$ The values of the parameter $\alpha$ in
concordance with the predictions of ${}^4 H$ are $\alpha=1, \,
0.33,\, 3$ (see table 4.2 in \cite{DTorres}). Let us consider also
the Albrecht-Skordis potential given by \eqref{Albrecht-Skordis}.

Observe first that \be W_\chi(\phi)=
\chi'(\phi)/\chi(\phi)=\frac{2}{\alpha  (\phi -\phi_0)}
\Rightarrow \lim_{\phi\rightarrow +\infty} W_\chi(\phi)=0\ee and
\be W_V(\phi)= V'(\phi)/V(\phi)+\mu=\frac{2(\phi-B)}{A+(B-\phi)^2}
\Rightarrow \lim_{\phi\rightarrow +\infty} W_V(\phi)=0.\ee  In
other words, the coupling function (\ref{couplingexample}) and the
potential (\ref{Albrecht-Skordis}) are WBI of exponential orders
$M=0$ and $N=-\mu,$ respectively.

It is easy to prove that Power-law coupling and the
Albrecht-Skordis potential are at least  ${\cal E}^2_+,$  under
the admissible coordinate transformation \footnote{We fix here an
error in formulas B6-B9 in \cite{Leon:2008de}. With the choice
$\vphi=\phi^{-1}$ the resulting barred functions given by B7-B9
there, are not of the desired differentiable class.}

\be \vphi=\phi^{-\frac{1}{2}}=f(\phi)\label{transformAS}.\ee

Using the above coordinate transformation we find

\be\overline{W}_{\chi}(\vphi)=\left\{\begin{array}{rcr} \frac{2 \varphi^2 }{\alpha  (1-\varphi^2  \phi_0)}&,&\vphi>0\\
                                     0&,&\vphi=0 \end{array}\right.\label{WchiAS}\ee

\be\overline{W}_V(\vphi)=\left\{\begin{array}{rcr} -\frac{2\vphi^2(B\vphi^2-1)}{A\vphi^4+(B\vphi^2-1)^2}&,&\vphi>0\\
                                     0 &,&\vphi=0 \end{array}\right.\label{WVAS}\ee
and

\be\overline{f'}(\vphi)=\left\{\begin{array}{rcr} -\frac{1}{2}\vphi^3&,&\vphi>0\\
                                     0&,&\vphi=0 \end{array}\right.\label{fAS}\ee

In this example, the evolution equations for $\vphi,$ $y,$ and $z$
are given by the equations \eqref{eqinfphi}-\eqref{eqinfz} with
$M=0,$ $N=-\mu$ and $\overline{W_V},$  $\overline{f'},$
$\overline{W}_{\chi}$ given respectively by  (\ref{WVAS}),
(\ref{fAS}) and (\ref{WchiAS}). The state space is defined by
$\overline{\Sigma}_\epsilon=\left\{(\vphi, x_1,\,x_2)|
0\leq\varphi\leq \sqrt{\epsilon}, 0\leq x_1^2+x_2^2\leq
1\right\}.$

The singular points of the system (\ref{eqinfy}-\ref{eqinfphi}) in
this example are $P_{1,2}=(0, \mp 1,0)$, $P_3=\left(0,0,1\right),$
$P_4=\left(0, \frac{\mu}{\sqrt{6}},0\right)$, and
$P_{5,6}=\left(0,
\sqrt{\frac{3}{2}}\frac{\gamma}{\mu},\mp\frac{\sqrt{-12\gamma+4\mu^2}}{2\mu}\right).$
The points $P_{1,2,3}$ exist for all the values of the free
parameters. The singular point $P_4$ exists for $\mu^2\leq 6.$ The
singular point $P_5$ exists if $\mu\leq -\sqrt{3\gamma}$ whereas
the singular point $P_6$ exists if $\mu\geq \sqrt{3\gamma}.$ We
will characterize the singular points $P_{5,6}$ in more detail
(for the analysis of the other singular points we submit the
reader to table \ref{crit}). The singular points $P_{5,6}$
corresponds to those studied in the book \cite{Coley:2003mj} (see
equation 4.23 p 49) with the identifications $\Psi=x_1,$
$\Phi^2=\frac{V(\phi)}{3 H^2}={\frac{3\gamma(2-\gamma)}{2\mu^2}}$
and $k=-\mu.$ As stated in that reference the scalar field
'inherits' the equation of state of the fluid, i.e.,
$\gamma_\phi=\gamma$. Then this solutions represents cosmological
kinetic-matter scaling solutions \footnote{See reference
\cite{Copeland:1997et,Holden:1999hm} for a notion of 'scaling'
solutions, particularly, kinetic-matter scaling solutions.} (the
potential energy density is negligible). Because the scalar field
mimics the perfect fluid with exact the same EoS at these points
it seems reasonably to think that if one combine the two 'fluids'
via $p_{\text{tot}}=p_\phi+\rho$ and $\rho_{\text{tot}}$ those
singular points will corresponds to exact perfect fluid models
with total EoS parameter equal to $\gamma-1$ (see
\cite{Coley:2003mj} p 54). This is the case but the effective EoS
parameter of total matter would be  $\omega_{tot}=\gamma
\left(1-\frac{3}{\mu ^2}\right)$ instead $\gamma-1.$ This fact is
due to the existence of the coupling. These singular points
represent accelerating cosmologies for $0<\gamma<\frac{2}{3}.$ The
eigenvalues of the matrix of derivatives evaluated at $P_{5,6}$
are $\left(0,-\frac{2-\gamma}{4\mu}\pm
\frac{1}{4\mu}\sqrt{(2-\gamma)\left(24\gamma^2+\mu^2(2-9\gamma)\right)}\right).$
The orbits initially in the stable subspace of $P_{5,6}$ spiral-in
around $P_{5,6}$ if $\mu^2>24\gamma^2/(-2+9\gamma)$ provided
$\frac{2}{9}<\gamma<2,\gamma\neq \frac{4}{3}.$ Otherwise
$P_{5,6},$ looks like an stable node for the orbits lying in the
stable subspace. The center subspace is tangent to the singular
points in the direction of the $\vphi$ axis.

The system (\ref{eqinfy}-\ref{eqinfphi}) can admit also the
singular points  $$P_{7,8}=\left(\sqrt{\frac{\mu }{(B \mu +1)\pm
\sqrt{1-A \mu ^2}}},0,0\right).$$ $P_7$ exists for $\mu<0,A\leq
\frac{1}{\mu^2}, B>\frac{\mu -\epsilon }{\epsilon \mu
}+\sqrt{\frac{1-A \mu ^2}{\mu ^2}}$ or $\mu>0,A\leq
\frac{1}{\mu^2}, B>\frac{\mu -\epsilon }{\epsilon  \mu
}-\sqrt{\frac{1-A \mu ^2}{\mu ^2}},$ whereas, $P_8$ exists for
$\mu <0,A\leq \frac{1}{\mu ^2},B>\frac{\mu -\epsilon }{\epsilon
\mu }-\sqrt{\frac{1-A \mu ^2}{\mu ^2}},$ or $\mu >0,A<\frac{1}{\mu
^2}, B>\frac{\mu -\epsilon }{\epsilon  \mu }+\sqrt{\frac{1-A \mu
^2}{\mu ^2}},$ or $\mu >0,A=\frac{1}{\mu ^2},B>\frac{\mu -\epsilon
}{\epsilon  \mu }-\sqrt{\frac{1-A \mu ^2}{\mu ^2}}.$

The eigenvalues of the linearization at $P_7$ are $-\frac{\gamma
}{2},-\frac{1}{2}-\frac{\sqrt{9 A^2+12 \left(A \mu ^2+\sqrt{1-A
\mu ^2}-1\right) A}}{6 A}, -\frac{1}{2}+\frac{\sqrt{9 A^2+12
\left(A \mu ^2+\sqrt{1-A \mu ^2}-1\right) A}}{6 A}.$ From the
existence conditions follows that it is a saddle point.

The eigenvalues of the linearization at $P_8$ are $-\frac{\gamma
}{2},-\frac{1}{2}-\frac{\sqrt{9 A^2-12 A \left(-A \mu ^2+\sqrt{1-A
\mu ^2}+1\right)}}{6 A},-\frac{1}{2}+\frac{\sqrt{9 A^2-12 A
\left(-A \mu ^2+\sqrt{1-A \mu ^2}+1\right)}}{6 A}.$ Thus, $P_8$ is
an attractor for
\begin{enumerate}
\item[i)] $0<A<\frac{8 \left(2 \mu ^2+3\right)}{\left(4 \mu
^2+3\right)^2}, B>\frac{1}{\epsilon }-\frac{1}{\mu
}+\sqrt{\frac{1}{\mu
   ^2}-A},\mu >0$ or $0<A<\frac{8 \left(2 \mu ^2+3\right)}{\left(4 \mu
^2+3\right)^2}, B>\frac{1}{\epsilon }-\frac{1}{\mu
}-\sqrt{\frac{1}{\mu
   ^2}-A},\mu <0$ (two complex eigenvalues with negative real part and one negative real eigenvalue) or
\item[ii)] $\frac{8 \left(2 \mu ^2+3\right)}{\left(4 \mu
^2+3\right)^2}\leq A<\frac{1}{\mu ^2}, B>\frac{1}{\epsilon
}-\frac{1}{\mu }+\sqrt{\frac{1}{\mu
   ^2}-A},\mu >0$ or $\frac{8 \left(2 \mu ^2+3\right)}{\left(4 \mu
^2+3\right)^2}\leq A<\frac{1}{\mu ^2}, B>\frac{1}{\epsilon
}-\frac{1}{\mu }+\sqrt{\frac{1}{\mu
   ^2}-A},\mu <0$  (three real negative eigenvalues).
\end{enumerate}

To finish this section let us re-examine the example discussed in
\cite{Leon:2008de} section B.1 in presence of radiation.

\subsubsection{Powerlaw coupling and Albrecht-Skordis
potential for the general model including radiation}

As we investigated in section \ref{Qualitative}, the late time
dynamics of the flow of \eqref{eqsigma1}-\eqref{eqsigma5} is
associated with the extremes of the potential (the singular point
$P_2=(0,0,0)$). When we restrict ourselves to this invariant set,
we find that the singular point associated to $\phi^+$ is always a
saddle point of the corresponding phase portrait. The singular
point associated to $\phi^-$ could be either a stable node or a
stable spiral if
$$\frac{8(3+2\mu^2)}{\left(3+4\mu^2\right)^2}<A\leq
\frac{1}{\mu^2}$$ or
$$A<\frac{8(3+2\mu^2)}{\left(3+4\mu^2\right)^2}.$$
The early time dynamics of the flow of
\eqref{eqsigma1}-\eqref{eqsigma5} corresponds to the limit
$\phi=+\infty.$

In this example, the evolution equations for $\varphi,$
$\sigma_2,$ $\sigma_4,$ and $\sigma_5$ are given by the equations
\eqref{eqvphi}-\eqref{eqzrad} with $M=0,\,N=-\mu,$ and
$\overline{W}_{\chi}(\varphi),\,\overline{W}_{V}(\varphi)=0,$ and
$\overline{f'},$ given by \eqref{WchiAS}, \eqref{WVAS} and
\eqref{fAS} respectively. The state space is defined by
$$\Omega_\epsilon=\{(\varphi, \sigma_2,
\sigma_4,\sigma_5)\in\mathbb{R}^4: 0\leq\varphi\leq
 \sqrt{\epsilon}, \sigma_2^2+\sigma_4^2+\sigma_5^2\leq 1,
\sigma_4\geq 0, \sigma_5\geq 0\}.$$

\begin{figure}[ht]
\begin{center}
\mbox{\epsfig{figure=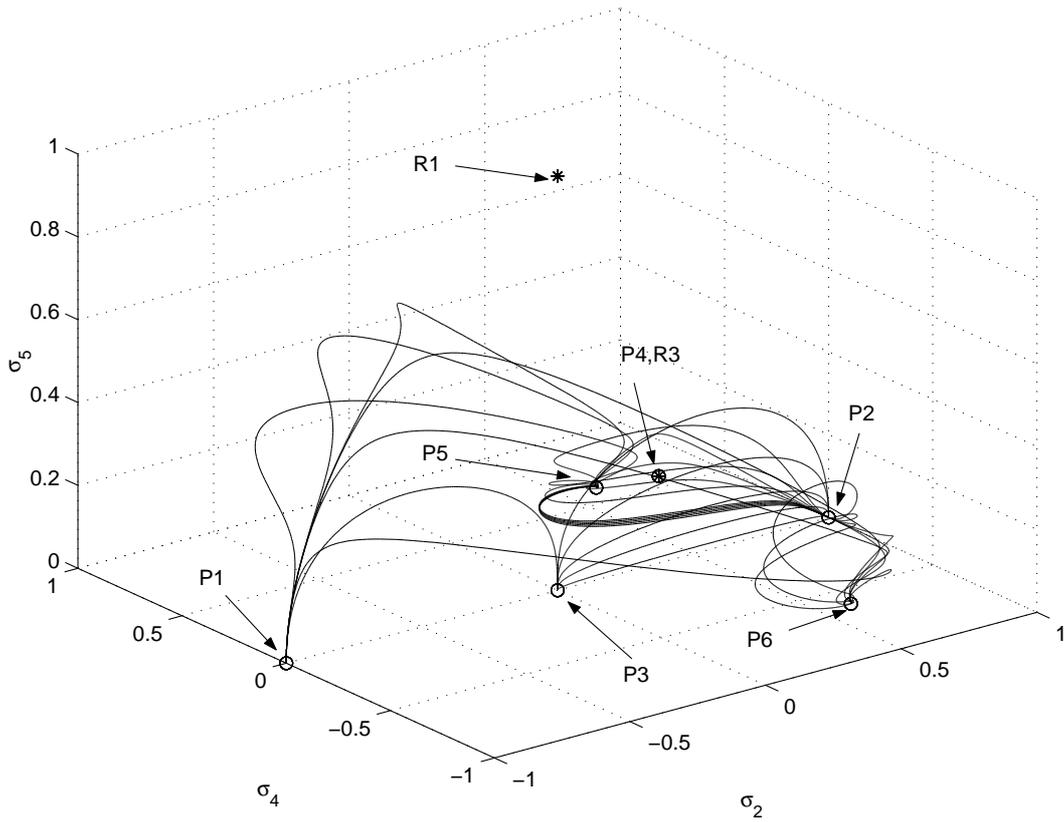,scale=0.8,angle=0}}
\caption{\label{fig4}{Some orbits in the invariant set
$\sigma_2^2+\sigma_4^2+\sigma_5^2\leq 1$ for the choice of
$\varphi=0$ for the model with coupling function
(\ref{couplingexample})  potential (\ref{Albrecht-Skordis}). We
select the values of the parameters: $\gamma=1,$ $\epsilon=1.00,$
$\mu= 2.00, A = 0.50, \alpha = 0.33, B = 0.5,$ and $\phi_0=0.$}}
\end{center}
\end{figure}

\begin{figure}[ht]
\begin{center}
\mbox{\epsfig{figure=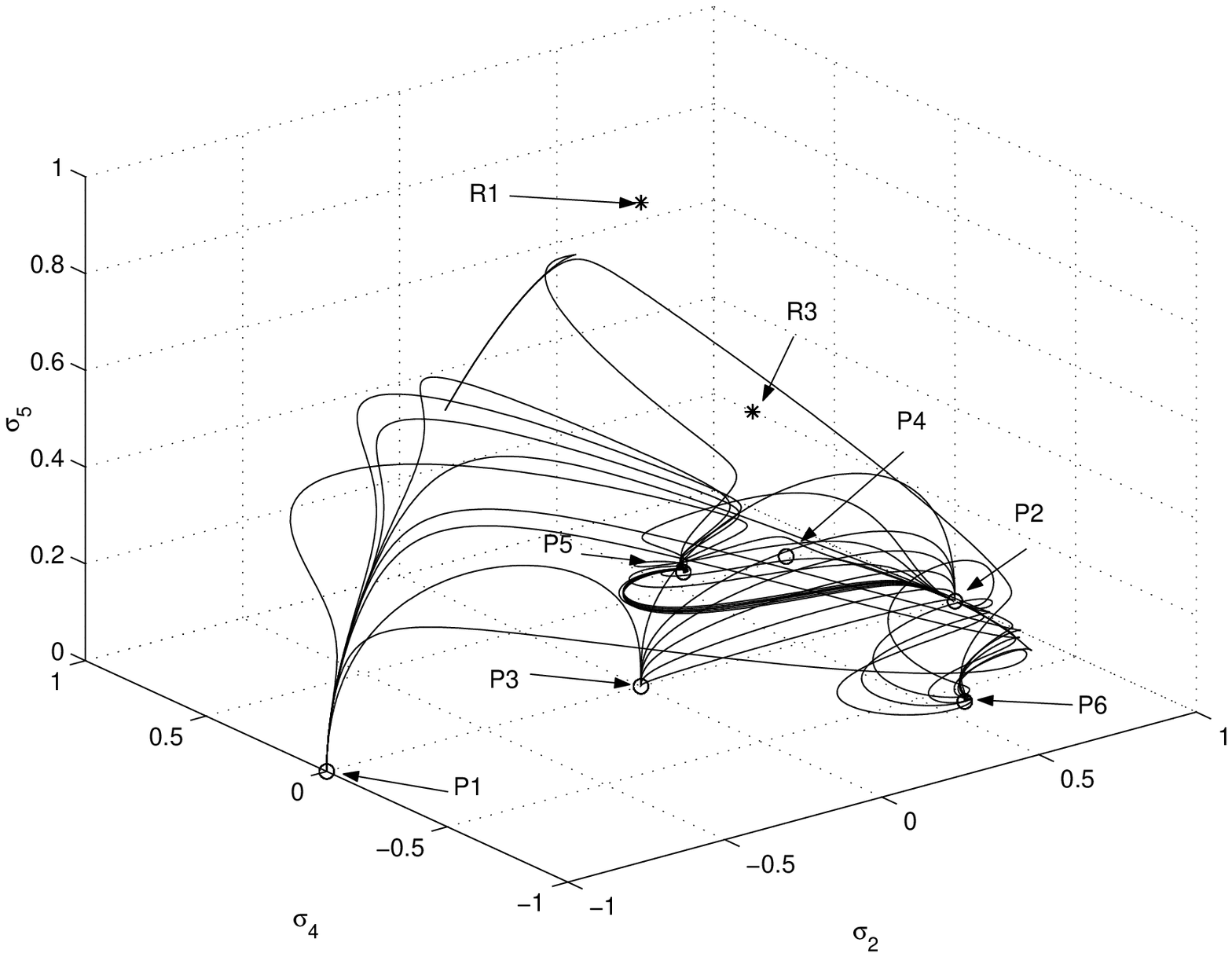,scale=0.8,angle=0}}
\caption{\label{fig4444}{Some orbits in the invariant set
$\sigma_2^2+\sigma_4^2+\sigma_5^2\leq 1$ for the choice of
$\varphi=0$ for the model with coupling function
(\ref{couplingexample})  potential (\ref{Albrecht-Skordis}). We
select the values of the parameters: $\gamma=1,$ $\epsilon=1.00,$
$\mu= 2.10, A = 0.50, \alpha = 0.33, B = 0.5,$ and $\phi_0=0.$}}
\end{center}
\end{figure}

In figure \ref{fig4} we show some orbits in the invariant set
$\sigma_2^2+\sigma_4^2+\sigma_5^2\leq 1$ for $\varphi=0$ for the
model with coupling function (\ref{couplingexample})  potential
(\ref{Albrecht-Skordis}). We select the values of the parameters:
$\epsilon=1.00,$ $\mu= 2.00, A = 0.50, \alpha = 0.33, B = 0.5,$
and $\phi_0=0.$ In this case $P_5$ is the local sink in this
invariant set. For this choice of parameters the points $P_{7,8}$
do not exist. In the figure \ref{fig4444} are displayed some orbits in the invariant set
$\sigma_2^2+\sigma_4^2+\sigma_5^2\leq 1$ for the choice of
$\varphi=0$ for the model with coupling function
(\ref{couplingexample})  potential (\ref{Albrecht-Skordis}). We
select the values of the parameters: $\gamma=1,$ $\epsilon=1.00,$
$\mu= 2.10, A = 0.50, \alpha = 0.33, B = 0.5,$ and $\phi_0=0.$  The dynamics is essentially the same as
in the figure \ref{fig4} with the difference that in this case $R_3$ exists and it is a saddle.

\subsection{Quadratic Gravity: $F(R)=R+\alpha R^2$.}

\begin{table}[!hb]
\begin{center}
{\small \caption{\label{crit0} Location of the singular points of
the flow of \eqref{eqvphi}-\eqref{eqzrad} defined in the invariant
set $\left\{p\in\Omega_\epsilon: \varphi=0\right\}$ for
$M=\sqrt{2/3}$ and $N=0.$}\bigskip
\begin{tabular}[t]{|l|c|c|c|c|}
\hline
Label&$(\sigma_2,\sigma_4,\sigma_5)$&Existence&Stability$^{\rm a}$\\[1ex]
\hline
\hline &&& \\[-2ex]
$P_1$&$(-1,0,0)$&always & unstable for $\left\{\begin{array}{c} 0<\gamma <\frac{4}{3},\, \text{or}\\ \frac{4}{3}<\gamma <\frac{5}{3}\end{array}\right.$ \\[1ex]
     &          &       & saddle, otherwise\\[1ex]
\hline
\hline &&& \\[-2ex]
$P_2$&$(1,0,0)$&always & unstable\\[1ex]
\hline
\hline &&& \\[-2ex]
$P_3$&$\left(1-\frac{2}{3 (2-\gamma)},0,0\right)$& $\left\{\begin{array}{c} 0<\gamma <\frac{4}{3},\, \text{or}\\ \frac{4}{3}<\gamma <\frac{5}{3}\end{array}\right.$ & saddle\\[1ex]
\hline
\hline &&& \\[-2ex]
$R_1$&$(0,0,1)$&always & saddle\\[1ex]
\hline
\hline &&& \\[-2ex]
$P_4$&$\left(0,1,0\right)$& always & stable \\[1ex]
\hline
\end{tabular}}
\end{center}
$^{\rm a}$ The stability is analyzed for the flow restricted to
the invariant set $\varphi=0$.
\end{table}

Quadratic gravity, $F(R)=R+\alpha R^2$, is equivalent to a
non-minimally coupled scalar field with the potential \be V\left(
\phi\right) =\frac{1}{8\alpha}\left( 1-e^{-\sqrt{2/3}\phi}\right)
^{2}\label{pot1}\ee and coupling function
\be\chi(\phi)=e^{\sqrt{\frac{2}{3}}\phi}.\label{coup1}\ee

Observe first that \be W_\chi(\phi)=\chi'(\phi)/\chi(\phi)-\sqrt{2/3}=0 \Rightarrow
\lim_{\phi\rightarrow +\infty} W_\chi(\phi)=0\ee and  \be W_V(\phi)= V'(\phi)/V(\phi)=-\left({\sqrt{{8}/{3}}}\right)/{\left(1-e^{\sqrt{\frac{2}{3}}\phi}\right)} \Rightarrow \lim_{\phi\rightarrow +\infty} W_V(\phi)=0.\ee  In other words, the coupling function \eqref{coup1} and the potential \eqref{pot1} are WBI of exponential orders $M=\sqrt{2/3}$ and $N=0,$ respectively.

It is easy to prove that the coupling function \eqref{coup1} and the potential \eqref{pot1} are at least  ${\cal E}^2_+,$  under the
admissible coordinate transformation \be \varphi=\phi^{-1}=f(\phi)\label{transformQG}.\ee

Using the coordinate transformation \eqref{transformQG} we find

\be\overline{W}_{\chi}(\varphi)=0.\label{WchiQG}\ee

\be\overline{W}_V(\varphi)=\left\{\begin{array}{rcr} -\frac{2 \sqrt{\frac{2}{3}}}{1-e^{{\sqrt{\frac{2}{3}}}/{\varphi }}}&,&\varphi>0\\
                                     0 &,&\varphi=0 \end{array}\right.\label{WVQG}\ee
and

\be\overline{f'}(\varphi)=\left\{\begin{array}{rcr} -\varphi^2&,&\varphi>0\\
                                     0&,&\varphi=0 \end{array}\right.\label{fQG}\ee

In this example, the evolution equations for $\varphi,$ $\sigma_2,$ $\sigma_4,$ and $\sigma_5$ are given by the equations
\eqref{eqvphi}-\eqref{eqzrad} with $M=\sqrt{2/3}$ and $N=0,$ and $\overline{W}_{\chi},$ $\overline{W_V},$ and $\overline{f'},$
given respectively by (\ref{WchiQG}), (\ref{WVQG}) and (\ref{fQG}). The state space is defined by
$$\Omega_\epsilon=\{(\varphi, \sigma_2,
\sigma_4,\sigma_5)\in\mathbb{R}^4: 0\leq\varphi\leq
 \epsilon, \sigma_2^2+\sigma_4^2+\sigma_5^2\leq 1,
\sigma_4\geq 0, \sigma_5\geq 0\}.$$ Let us analyze the local
stability of the singular points of the corresponding system. In
the above analysis we are not taking into account perturbations in
the $\varphi$-axis. It is obvious, from the previous analysis,
that the center manifold of these singular points contains the
$\varphi$-axis as a proper eigenvector. In table \ref{crit0} are
summarized the location, existence conditions and stability
\footnote{The stability is analyzed for the flow restricted to the
invariant set $\varphi=0$.} of the singular points.

\begin{figure}[ht]
\begin{center}
\mbox{\epsfig{figure=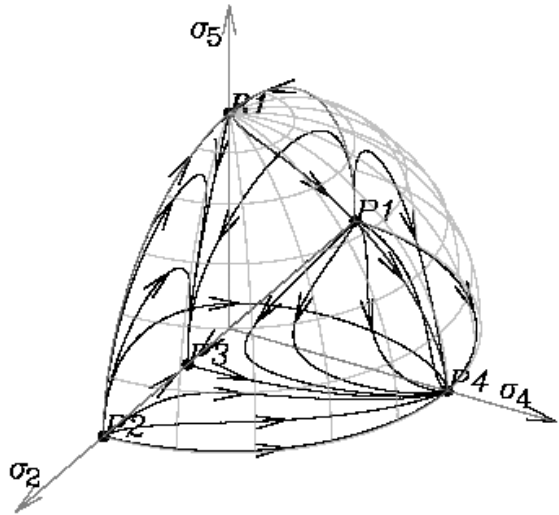,width=9.9cm,angle=0}}
\caption{\label{fig2}{Projection of some orbits of
\eqref{eqvphi}-\eqref{eqzrad} in the invariant set $\varphi=0$ for
the coupling function \eqref{coup1} and the potential \eqref{pot1}
for $\gamma=1.$ Observe that $P_1$ and $P_2$ are local sources,
$R_1,$ and $P_3$ are saddles ($P_3$ is the local attractor in the
invariant set $y=0$) and $P_4$ (the de Sitter solution) is the
local sink in the invariant set $\varphi=0.$}}

\end{center}
\end{figure}

Let us discuss the stability properties of the singular points
displayed in table \ref{crit0}.

The singular point $P_1$ always exists. Its unstable manifold is
$3D$ provided $0<\gamma<\frac{4}{3}$ or
$\frac{4}{3}<\gamma<\frac{5}{3}.$ Otherwise its unstable manifold
is lower dimensional.

The singular point $P_2$ always exists. It has a 3D unstable
manifold. Although non-hyperbolic our numerical experiments
suggest that it is a local source.

The singular point $P_3$ exists for $0<\gamma<\frac{4}{3}$ or
$\frac{4}{3}<\gamma<\frac{5}{3}$ and it is neither a sink nor a
local source.

The singular point $R_1$ always exists and it is neither a sink
nor a local source.

The singular point $P_4$ (corresponding to the \emph{de Sitter}
solution) always exists. Its stable manifold is 3D. Since $P_4$ is
nonhyperbolic the linear stability analysis is not conclusive.
Thus we need to resort to numerical experimentation or
alternatively we can use more sophisticated techniques such as
normal forms expansion or center manifold theorem. Due its
relevance, the full stability analysis of $P_4$ is deserved to
section \ref{stabilityP4}

Let us discuss some physical properties of the cosmological
solutions associated to the singular points displayed in table
\ref{crit0}.
\begin{itemize}
\item  $P_{1,2}$  represent kinetic-dominated cosmological solutions. They behave as stiff-like matter. The associated cosmological solution satisfies $H=\frac{1}{3 t-c_1},a=\sqrt[3]{3 t-c_1} c_2,\phi =c_3\pm\sqrt{\frac{2}{3}} \ln \left(3
   t-c_1\right),$ where $c_j,\,j=1,2,3$ are integration constants. These solutions are associated with the local past attractors of the systems for an open set of values of the parameter $\gamma.$
\item  $P_3$ represents matter-kinetic scaling cosmological solutions such that
$H=\frac{3 (\gamma -2)}{t (3 \gamma -8)-3 (\gamma -2) c_1},$ $a= \left(t (3 \gamma -8)-3 (\gamma -2)
   c_1\right){}^{1+\frac{2}{3 \gamma -8}} c_2,\rho =\frac{60-36 \gamma }{\left(t (3 \gamma -8)-3 (\gamma -2)
   c_1\right){}^2}+c_3,$ and $\phi = c_4+\frac{\sqrt{6} (3 \gamma -4) \ln \left(t (3 \gamma -8)-3 (\gamma -2) c_1\right)}{3
   \gamma -8}$ where $c_j,\,j=1,2,3,4$ are integration constants.
\item $R_1$ represents a radiation-dominated cosmological solutions satisfying $H=\frac{1}{2 t-c_1},a=\sqrt{2 t-c_1} c_2,\rho_r=\frac{3}{\left(2 t-c_1\right){}^2}+c_3.$
\item $P_4$ represents a de Sitter solution with $H=\sqrt{\frac{V_0}{3}}, a=c_1 \exp\left[\sqrt{\frac{V_0}{3}} t\right], V(\phi)=V_0.$
\end{itemize}

In the figure \ref{fig2} are are displayed typical orbits of
\eqref{eqvphi}-\eqref{eqzrad} in the invariant set $\varphi=0.$
The singular points $P_1$ and $P_2$ are local sources, $R_1,$ and
$P_3$ are saddles ($P_3$ is the local attractor in the invariant
set $y=0$) and $P_4$ (the de Sitter solution) is the local
attractor in the invariant set $\varphi=0$. However, concerning
the full dynamics, it is locally unstable as we prove in next
section by explicit calculation of the center manifold at $P_4.$

\subsubsection{Stability Analysis of the \emph{de Sitter} Solution in Quadratic Gravity}\label{stabilityP4}

In order to analyze the stability of \emph{de Sitter} solution we
can use center manifold theorem. Let us proceed as follows. First,
in order to remove the transcendental function in
$\overline{W_V},$ let us introduce the new variable
$$u=\frac{1}{1-\exp\left[{{\sqrt{\frac{2}{3}}}/{\varphi}}\right]},$$
taking values in the range
$$\frac{1}{1-\exp\left[{{\sqrt{\frac{2}{3}}}/{\epsilon}}\right]}\leq
u\leq 0.$$

In this way we obtain the new system of ordinary differential
equations

\begin{align}
&u'=\frac{2 u^2 \sigma_2}{3}-\frac{2 u \sigma_2}{3},\nonumber\\
&\sigma_2'=   \frac{2 u \sigma_4^2}{3}+\left(1-\frac{\gamma }{2}\right)
   \sigma_2^3+\frac{1}{6} (3 \gamma -4) \sigma_2^2 \nonumber\\& +
   \sigma_2 \left(-\frac{\gamma  \sigma_4^2}{2}+\frac{1}{6} (4-3 \gamma ) \sigma_5^2+\frac{\gamma
   -2}{2}\right) \nonumber\\&+\frac{1}{6} (3 \gamma -4) \sigma_4^2+\frac{1}{6} (3
   \gamma -4) \sigma_5^2+\frac{1}{6} (4-3 \gamma ),\nonumber
\\&\sigma_4'= -\frac{2 u
   \sigma_2 \sigma_4}{3}+\left(1-\frac{\gamma }{2}\right)
   \sigma_2^2 \sigma_4-\frac{\gamma  \sigma_4^3}{2}+\sigma_4 \left(\frac{1}{6} (4-3 \gamma ) \sigma_5^2+\frac{\gamma }{2}\right),\nonumber\\
&\sigma_5'= \left(1-\frac{\gamma }{2}\right)
  \sigma_2^2 \sigma_5-\frac{1}{2} \gamma  \sigma_4^2 \sigma_5+\frac{1}{6} (4-3 \gamma ) \sigma_5^3+\frac{1}{6} (3 \gamma -4) \sigma_5\label{quadraticsyst}
\end{align} describing the dynamics of quadratic gravity as $\phi\rightarrow +\infty.$

\begin{prop}\label{centerP4} The singular point $q:=(u,\sigma_2,\sigma_4,\sigma_5)=(0,0,1,0)$ of the system \eqref{quadraticsyst} is locally unstable.
\end{prop}
In order to determine the local center manifold of \eqref{quadraticsyst} at $q$ we have to transform the system into a form suitable for the application of the center manifold theorem (see section \ref{sectionCM} for a summary of the techniques involved in the proof).

{\bf Proof}.

{\bf Case $\gamma\neq 1$}

Let be $\gamma\neq 1.$ The Jacobian of \eqref{quadraticsyst} at
$q=(0,0,1,0)$ has eigenvalues $0,-1,-\frac{2}{3},$ and $-\gamma$
with corresponding eigenvectors
$(0,0,0,1)^T,\left(0,0,\frac{3}{2},0\right)^T,
\left(1,0,1,\frac{1}{3 (\gamma -1)}-1\right)^T,$ and
$(0,1,0,0)^T.$ We shift the singular point to the origin by
setting $\hat{\sigma_4}=\sigma_4-1.$ In order to transform the
linear part of the vector field into Jordan canonical form, we
define new variables $(x,y_1,y_2,y_3)\equiv\mathbf{x}$, by the
equations
\begin{align*}& u=\frac{3 x}{2},\,\sigma_2=x+y_1+y_3 \left(\frac{1}{3 (\gamma
   -1)}-1\right),\nonumber\\&\hat{\sigma}_4=y_3,\,\sigma_5=y_2,\nonumber\end{align*} so that
\be \left(\begin{array}{c}x'\\y_1'\\y_2'\\y_3'
\end{array}\right)=\left(\begin{array}{cccc}0& 0 &0 &0\\0& -1 &0 &0\\0& 0 &-\frac{2}{3}&0\\0& 0 &0 &-\gamma
\end{array}\right)\left(\begin{array}{c}x\\y_1\\y_2\\y_3
\end{array}\right)+\left(\begin{array}{c}f(x,y_1,y_2,y_3)\\g_1(x,y_1,y_2,y_3)\\g_2(x,y_1,y_2,y_3)\\g_3(x,y_1,y_2,y_3)\end{array}\right)\label{center2}
\ee
where

$f(x,y_1,y_2,y_3)=x^3+x^2 y_1+\frac{x^2 y_3}{3 (\gamma -1)}-x^2 y_3-\frac{2 x^2}{3}-\frac{2 x y_1}{3}-\frac{2 x y_3}{9 (\gamma
   -1)}+\frac{2 x y_3}{3},$
\\ $g_1(x,y_1,y_2,y_3)=-\frac{x^3 \gamma }{2}-\frac{3}{2} x^2 y_1 \gamma +2 x^2 y_1+x^2 y_3 \gamma +\frac{x^2 y_3}{3 (\gamma -1)}-\frac{7 x^2
   y_3}{3}+\frac{x^2 \gamma }{6 (\gamma -1)}-\frac{3}{2} x y_1^2 \gamma +3 x y_1^2+2 x y_1 y_3 \gamma +\frac{x
   y_1 y_3}{\gamma -1}-\frac{17 x y_1 y_3}{3}+\frac{2 x y_1}{3}-\frac{1}{2} x y_2^2 \gamma +\frac{2 x
   y_2^2}{3}-\frac{x y_3^2 \gamma ^3}{(\gamma -1)^2}+\frac{16 x y_3^2 \gamma ^2}{3 (\gamma -1)^2}-\frac{157 x y_3^2
   \gamma }{18 (\gamma -1)^2}+\frac{41 x y_3^2}{9 (\gamma -1)^2}-x y_3 \gamma +\frac{2 x y_3}{9 (\gamma -1)}+\frac{4 x
   y_3}{3}-\frac{y_1^3 \gamma }{2}+y_1^3+y_1^2 y_3 \gamma +\frac{y_1^2 y_3}{3 (\gamma -1)}-\frac{7
   y_1^2 y_3}{3}-\frac{3 y_1^2 \gamma }{6-6 \gamma }+\frac{4 y_1^2}{6-6 \gamma }-\frac{1}{2} y_1 y_2^2 \gamma
   +\frac{2 y_1 y_2^2}{3}-\frac{y_1 y_3^2 \gamma ^3}{(\gamma -1)^2}+\frac{10 y_1 y_3^2 \gamma ^2}{3 (\gamma
   -1)^2}-\frac{73 y_1 y_3^2 \gamma }{18 (\gamma -1)^2}+\frac{16 y_1 y_3^2}{9 (\gamma -1)^2}-\frac{y_1 y_3
   \gamma ^3}{(\gamma -1)^2}+\frac{y_1 y_3 \gamma ^2}{(\gamma -1)^2}+\frac{5 y_1 y_3 \gamma }{3 (\gamma -1)^2}-\frac{16
   y_1 y_3}{9 (\gamma -1)^2}+\frac{y_2^2}{18-18 \gamma }+\frac{y_2^2}{6}-\frac{y_3^2 \gamma ^2}{3 (\gamma
   -1)^3}+\frac{5 y_3^2 \gamma }{6 (\gamma -1)^3}-\frac{14 y_3^2}{27 (\gamma -1)^3},$
\\ $g_2(x,y_1,y_2,y_3)=-\frac{1}{2} x^2 y_2 \gamma +x^2 y_2-x y_1 y_2 \gamma +2 x y_1 y_2+x y_2 y_3 \gamma -\frac{x
   y_2 y_3 \gamma }{3 (\gamma -1)}+\frac{2 x y_2 y_3}{3 (\gamma -1)}-2 x y_2 y_3-\frac{1}{2} y_1^2
   y_2 \gamma +y_1^2 y_2+y_1 y_2 y_3 \gamma -\frac{y_1 y_2 y_3 \gamma }{3 (\gamma
   -1)}+\frac{2 y_1 y_2 y_3}{3 (\gamma -1)}-2 y_1 y_2 y_3-\frac{y_2^3 \gamma }{2}+\frac{2
   y_2^3}{3}-y_2 y_3^2 \gamma +\frac{y_2 y_3^2 \gamma }{3 (\gamma -1)}-\frac{y_2 y_3^2 \gamma }{18
   (\gamma -1)^2}-\frac{2 y_2 y_3^2}{3 (\gamma -1)}+\frac{y_2 y_3^2}{9 (\gamma -1)^2}+y_2 y_3^2-y_2
   y_3 \gamma$ and
\\
$g_3(x,y_1,y_2,y_3)=-\frac{1}{2} x^2 y_3 \gamma -\frac{x^2 \gamma }{2}-x y_1 y_3 \gamma +x y_1 y_3-x y_1 \gamma +x y_1+x
   y_3^2 \gamma -\frac{x y_3^2 \gamma }{3 (\gamma -1)}+\frac{x y_3^2}{3 (\gamma -1)}-x y_3^2+x y_3 \gamma -\frac{x
   y_3 \gamma }{3 (\gamma -1)}+\frac{x y_3}{3 (\gamma -1)}-x y_3-\frac{1}{2} y_1^2 y_3 \gamma +y_1^2
   y_3-\frac{y_1^2 \gamma }{2}+y_1^2+y_1 y_3^2 \gamma -\frac{y_1 y_3^2 \gamma }{3 (\gamma -1)}+\frac{2
   y_1 y_3^2}{3 (\gamma -1)}-2 y_1 y_3^2+y_1 y_3 \gamma -\frac{y_1 y_3 \gamma }{3 (\gamma
   -1)}+\frac{2 y_1 y_3}{3 (\gamma -1)}-2 y_1 y_3-\frac{1}{2} y_2^2 y_3 \gamma +\frac{2 y_2^2
   y_3}{3}-\frac{y_2^2 \gamma }{2}+\frac{2 y_2^2}{3}+\frac{y_3^3 \gamma }{3 (\gamma -1)}-\frac{y_3^3 \gamma }{18
   (\gamma -1)^2}-y_3^3 \gamma -\frac{2 y_3^3}{3 (\gamma -1)}+\frac{y_3^3}{9 (\gamma -1)^2}+y_3^3+\frac{y_3^2
   \gamma }{3 (\gamma -1)}-\frac{y_3^2 \gamma }{18 (\gamma -1)^2}-2 y_3^2 \gamma -\frac{2 y_3^2}{3 (\gamma
   -1)}+\frac{y_3^2}{9 (\gamma -1)^2}+y_3^2.$
\\

\begin{figure}[ht]
\begin{center}
\mbox{\epsfig{figure=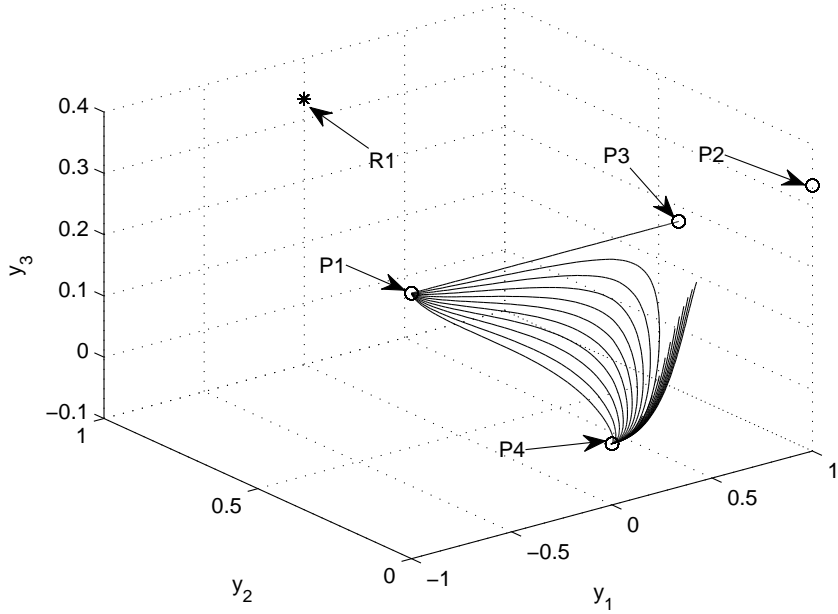,width=12cm,angle=0}} \caption{
\label{fig2a}{Projection of some orbits of \eqref{center2} in the
space $y_1,y_2,y_3$ for the coupling function \eqref{coup1} and
the potential \eqref{pot1} for $\gamma=1.$ The graphic shows the
behavior in the stable manifold of $P_4.$ The bulk of orbits in
front of and at the right hand side of the figure represents a
projection of the center(s) manifold(s).}}

\end{center}
\end{figure}

\begin{figure}[ht]
\begin{center}
\mbox{\epsfig{figure=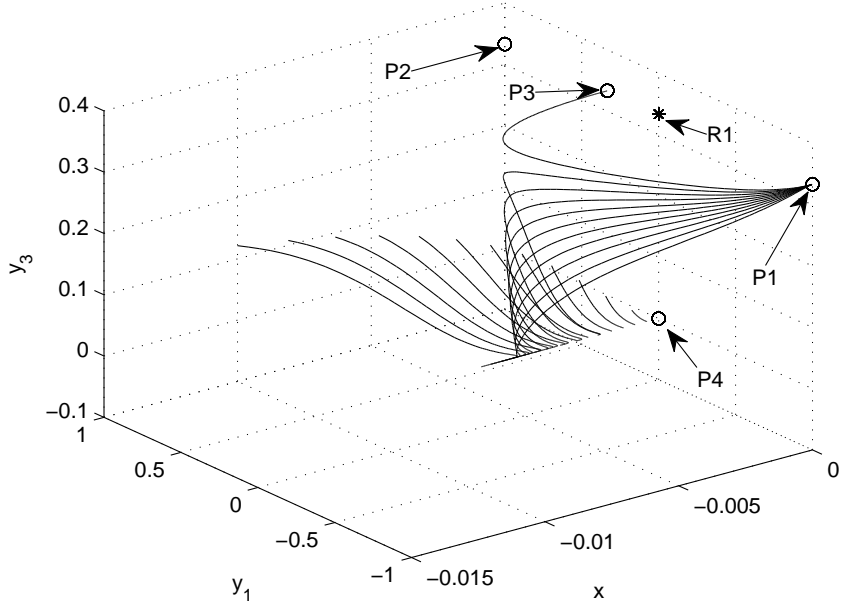,width=12cm,angle=0}} \caption{
\label{fig2b}{Projection of some orbits of \eqref{center2} in the
space $x,y_1,y_3$ for the coupling function \eqref{coup1} and the
potential \eqref{pot1} for $\gamma=1.$ The graphic shows the
unstable character of $P_4$ (trajectories starting at $x<0$ move
away from the origin).}}

\end{center}
\end{figure}

\begin{figure}[ht]
\begin{center}
\mbox{\epsfig{figure=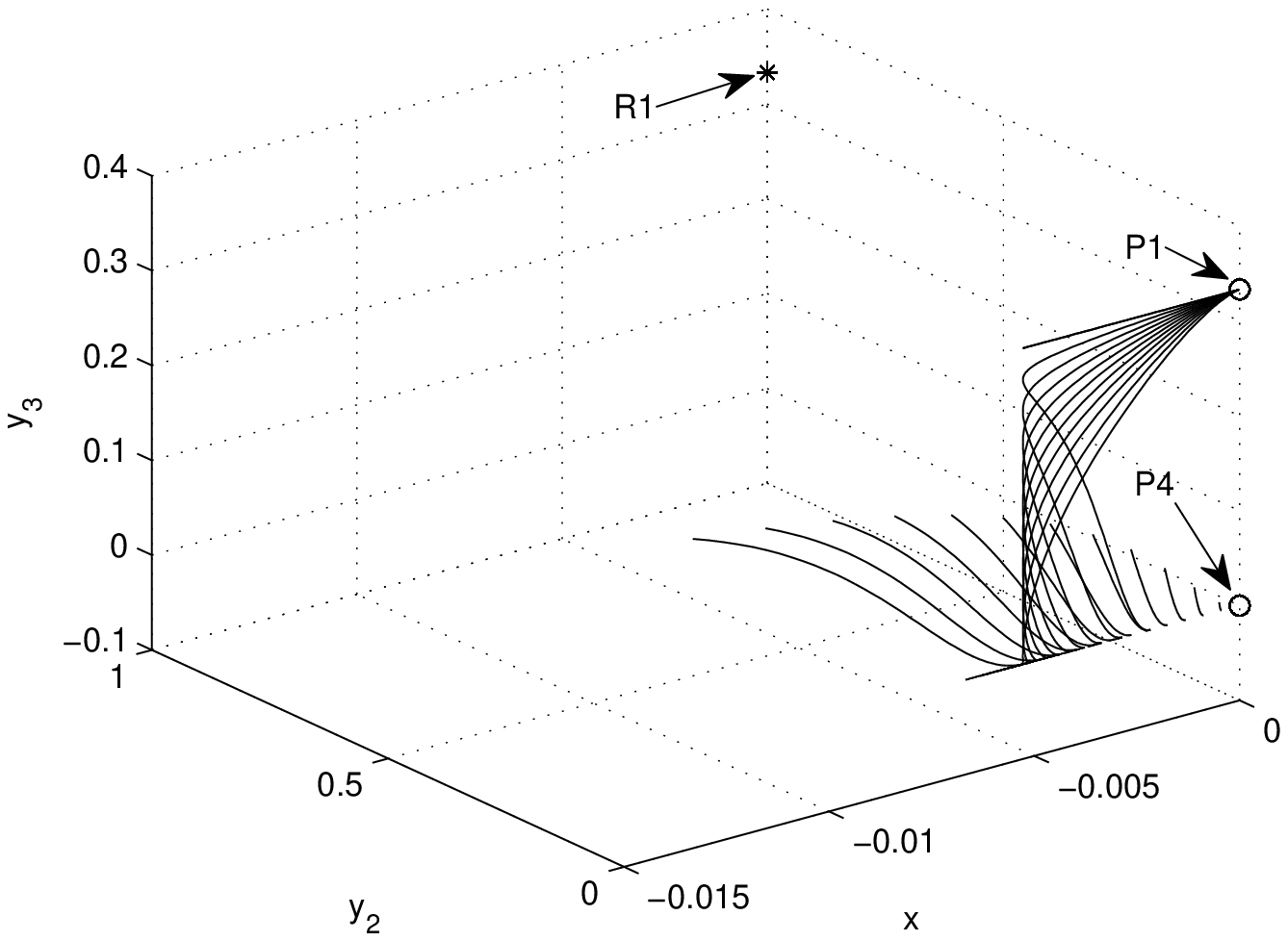,width=12cm,angle=0}} \caption{
\label{fig2c}{Projection of some orbits of \eqref{center2} in the
space $x,y_2,y_3$ for the coupling function \eqref{coup1} and the
potential \eqref{pot1} for $\gamma=1.$ The graphic shows the
unstable character of $P_4$ (the orbits depart from the origin for
$x<0$).}}
\end{center}
\end{figure}

The system \eqref{center2} is written in diagonal form
\begin{align}
x'  &  =Cx+f\left(  x,\mathbf{y}\right) \nonumber\\
\mathbf{y}'  &  =P\mathbf{y}+\mathbf{g}\left(  x,\mathbf{y}\right)  ,
\label{center3}
\end{align}
where $\left(  x,\mathbf{y}\right)
\in\mathbb{R}\times\mathbb{R}^{3},$ $C$ is the zero $1\times1$
matrix, $P$ is a $3\times3$ matrix with negative eigenvalues and
$f,\mathbf{g}$ vanish at $\mathbf{0}$ and have vanishing
derivatives at $\mathbf{0.}$ The center manifold theorem
\ref{existenceCM} asserts that there exists a 1-dimensional
invariant local center manifold $W^{c}\left( \mathbf{0}\right) $
of \eqref{center3} tangent to the center subspace (the
$\mathbf{y}=\mathbf{0}$ space) at $\mathbf{0}.$ Moreover,
$W^{c}\left( \mathbf{0}\right)$ can be represented as
\[
W^{c}\left(  \mathbf{0}\right)  =\left\{  \left(
x,\mathbf{y}\right)
\in\mathbb{R}\times\mathbb{R}^{3}:\mathbf{y}=\mathbf{h}\left(
x\right), \left\vert x\right\vert <\delta\right\} ;
\mathbf{h}\left(  0\right) =\mathbf{0}, D\mathbf{h}\left( 0\right)
=\mathbf{0},
\]
for $\delta$ sufficiently small (see definition \ref{CMdef}). The restriction of
(\ref{center3}) to the center manifold is (see definition \ref{vectorfieldCM})
\begin{equation}
x'=f\left( x,\mathbf{h}\left(  x\right)  \right)  . \label{rest}
\end{equation}
According to Theorem \ref{stabilityCM}, if the origin $x=0$ of \eqref{rest}
is stable (asymptotically stable) (unstable) then the origin of \eqref{center3} is also stable (asymptotically stable) (unstable). Therefore, we have to find the local center manifold, i.e.,
the problem reduces to the computation of $\mathbf{h}\left(  x\right).$

Substituting $\mathbf{y}=\mathbf{h}\left(  x\right)  $ in the second
component of \eqref{center3} and using the chain rule, $\mathbf{y
}'=D\mathbf{h}\left(  x\right)  x'$, one can show that the function
$\mathbf{h}\left(  x\right)  $ that defines the local center manifold
satisfies%
\begin{equation}
D\mathbf{h}\left(  x\right)  \left[  f\left(  x,\mathbf{h}\left(  x\right)
\right)  \right]  -P\mathbf{h}\left(  x\right)  -\mathbf{g}\left(
x,\mathbf{h}\left(  x\right)  \right)  =0. \label{h}
\end{equation}
According to Theorem \ref{approximationCM}, equation \eqref{h} can be solved approximately by using an approximation of $\mathbf{h}\left(  x\right)  $
by a Taylor series at $x=0.$ Since $\mathbf{h}\left(  0\right)  =\mathbf{0\ }
$and $D\mathbf{h}\left(  0\right)  =\mathbf{0},$ it is obvious that
$\mathbf{h}\left(  x\right)  $ commences with quadratic terms. We substitute
\[
\mathbf{h}\left(  x\right)  =:\left[
\begin{array}
[c]{c}%
h_{1}\left(  x\right) \\
h_{2}\left(  x\right) \\
h_{3}\left(  x\right)
\end{array}
\right]  =\left[
\begin{array}
[c]{c}%
a_{1}x^{2}+a_{2}x^{3}+O\left(  x^{4}\right) \\
b_{1}x^{2}+b_{2}x^{3}+O\left(  x^{4}\right) \\
c_{1}x^{2}+c_{2}x^{3}+O\left(  x^{4}\right)
\end{array}
\right]
\]
into (\ref{h}) and set the coefficients of like powers of $x$ equal to zero to find the unknowns $a_{1},b_{1},c_{1},...$.

Since $y_2$ absent from the first of \eqref{center3}, we
give only the result for $h_{1}\left(  x\right)$ and $h_{3}\left(  x\right).$ We find $a_1=\frac{\gamma }{6 (\gamma -1)},a_2=-\frac{3 \gamma -5}{9 (\gamma -1)},c_1=-\frac{1}{2},c_2=-\frac{2}{3}.$ \footnote{We find $b_1=b_2=0.$} Therefore, \eqref{rest} yields
\be x'=-\frac{2 x^2}{3}+\frac{5 x^3}{9}+\frac{4 x^4}{9}+O\left(x^5\right).\label{rest1}\ee

It is obvious that the origin $x=0$ of \eqref{rest1} is locally
unstable (saddle point). Hence, the origin $\mathbf{x}=\mathbf{0}$
of the full four-dimensional system is unstable.

{\bf Case $\gamma=1$}

Let be $\gamma=1.$ The Jacobian of \eqref{quadraticsyst} at
$q=(0,0,1,0)$ has eigenvalues $-1,-1,-\frac{2}{3},$ and $0$ with
corresponding eigenvectors $\left(0,0,0,0,\frac{3}{2}\right)^T,
\left(1,0,0,1\right)^T, (0,-3,0,0)^T$ and $(0,0,1,0)^T.$ As before
we shift the singular point to the origin by setting
$\hat{\sigma_4}=\sigma_4-1$ and define new variables
$(x,y_1,y_2,y_3)\equiv\mathbf{x}$, by the equations \begin{align}&
u=\frac{3
x}{2},\nonumber\\&\sigma_2=x+y_1,\nonumber\\&\hat{\sigma}_4=-3y_3,\nonumber\\&\sigma_5=y_2\nonumber\end{align}
so that

\be
\left(\begin{array}{c}x'\\y_1'\\y_2'\\y_3'
\end{array}\right)=\left(\begin{array}{cccc}0& 0 &0 &0\\0& -1 &0 &1\\0& 0 &-\frac{2}{3}&0\\0& 0 &0 &-1
\end{array}\right)\left(\begin{array}{c}x\\y_1\\y_2\\y_3
\end{array}\right)+\left(\begin{array}{c}f(x,y_1,y_2,y_3)\\g_1(x,y_1,y_2,y_3)\\g_2(x,y_1,y_2,y_3)\\g_3(x,y_1,y_2,y_3)\end{array}\right)\label{center2dust},
\ee
where
$f(x,y_1,y_2,y_3)=x^3+x^2 y_1-\frac{2 x^2}{3}-\frac{2 x y_1}{3},$
\\
$g_1(x,y_1,y_2,y_3)=-\frac{x^3}{2}+\frac{x^2 y_1}{2}+\frac{x^2}{2}+\frac{3 x y_1^2}{2}+\frac{x y_1}{3}+\frac{x y_2^2}{6}+\frac{9 x
   y_3^2}{2}-3 x y_3+\frac{y_1^3}{2}-\frac{y_1^2}{6}+\frac{y_1 y_2^2}{6}-\frac{9 y_1 y_3^2}{2}+3
   y_1 y_3-\frac{y_2^2}{6}-\frac{3 y_3^2}{2},$
\\$g_2(x,y_1,y_2,y_3)=\frac{x^2 y_2}{2}+x y_1 y_2+\frac{y_1^2 y_2}{2}+\frac{y_2^3}{6}-\frac{9 y_2 y_3^2}{2}+3 y_2
   y_3$ and
\\
$g_3(x,y_1,y_2,y_3)=-\frac{x^2 y_3}{2}+\frac{x^2}{6}+\frac{y_1^2 y_3}{2}-\frac{y_1^2}{6}+\frac{y_2^2
   y_3}{6}-\frac{y_2^2}{18}-\frac{9 y_3^3}{2}+\frac{9 y_3^2}{2}.$
\\

Observe that the system \eqref{center2dust} is now in the
canonical form \eqref{center3}. Then, we proceed to the
calculation of the center manifold. The procedure is fairly
systematic and since we present it completely in the previous
analysis we consider do not repeat it here. Instead, we present
the relevant calculations. We obtain
$a_1=\frac{2}{3},a_2=\frac{1}{3},b_1=0,b_2=0,c_1=\frac{1}{6},c_2=\frac{2}{9}$
for the Taylor expansion coefficients of \[ \mathbf{h}\left(
x\right)  =:\left[
\begin{array}
[c]{c}%
h_{1}\left(  x\right) \\
h_{2}\left(  x\right) \\
h_{3}\left(  x\right)
\end{array}
\right]  =\left[
\begin{array}
[c]{c}%
a_{1}x^{2}+a_{2}x^{3}+O\left(  x^{4}\right) \\
b_{1}x^{2}+b_{2}x^{3}+O\left(  x^{4}\right) \\
c_{1}x^{2}+c_{2}x^{3}+O\left(  x^{4}\right)
\end{array}
\right].
\] By substituting this values of the unknowns  $a_{1},b_{1},c_{1},...$ we obtain that, for $\gamma=1,$ the dynamics of the center manifold in given also by equation \eqref{rest1}. The conclusion is straighforward: the origin $x=0$ of \eqref{rest1} is locally unstable (saddle point). Hence, the origin $\mathbf{x}=\mathbf{0}$ of the full four-dimensional system is unstable.

This completes the Proof. $\blacksquare$

The result of proposition \ref{centerP4} complements the result of the proposition discussed in \cite{Miritzis:2005hg} p. 5, where it was proved the local asymptotic instability of the de Sitter universe for positively curved FRW models with a perfect fluid matter source and a scalar field which arises in the conformal frame of the $R+\alpha R^2$ theory.

\subsection{$R^n$-Gravity}

Let us consider the model with $F(R)=R^n$ where we have re-scaled
the usual multiplicative constant. It can be proved that, for
$n>1,$ $R^n$-gravity is conformally equivalent to a non-minimally
coupled scalar field with a positive potential
\begin{equation}V(\phi)=r(n)
e^{\lambda(n)\phi}\label{Rn}\end{equation} where $r(n)=\frac{1}{2}
(n-1) n^{-\frac{n}{n-1}}$ and
$\lambda(n)=-\frac{\sqrt{\frac{2}{3}} (n-2)}{n-1},$ with coupling
function given by \eqref{coup1}.

In this example, the evolution equations for $\varphi,$ $\sigma_2,$ $\sigma_4,$ and $\sigma_5$ are given by the equations
\eqref{eqvphi}-\eqref{eqzrad} with $M=\sqrt{2/3},\,N=-\frac{\sqrt{\frac{2}{3}} (n-2)}{n-1},$  $\overline{W}_{\chi}(\varphi)=\overline{W}_{V}(\varphi)=0,$ and $\overline{f'},$
given by (\ref{fQG}). The state space is defined by
$$\Omega_\epsilon=\{(\varphi, \sigma_2,
\sigma_4,\sigma_5)\in\mathbb{R}^4: 0\leq\varphi\leq
 \epsilon, \sigma_2^2+\sigma_4^2+\sigma_5^2\leq 1,
\sigma_4\geq 0, \sigma_5\geq 0\}.$$

In table \ref{crit01} and \ref{crit01a} are summarized the
location, existence conditions and stability of the singular
points.

\begin{table}[!ht]
\begin{center}
\caption{\label{crit01} Location of the singular points of the
flow of \eqref{eqvphi}-\eqref{eqzrad} defined in the invariant set
$\left\{p\in\Omega_\epsilon: \varphi=0\right\}$ for $M=\sqrt{2/3}$
and $N=-\frac{\sqrt{\frac{2}{3}} (n-2)}{n-1}.$ We use the
notations $n_+=\frac{1}{5}(4+\sqrt{6}),N_+=\frac{2}{27} \left(11+2
\sqrt{10}\right),\Gamma(n)=-\frac{2 n(n-2)}{3-9 n+6 n^2},$ and
$\Gamma_+(\gamma)=\frac{9 \gamma +\sqrt{9 \gamma ^2+48 \gamma
+16}+4}{12 \gamma +4}.$}\bigskip {\small
\begin{tabular}[t]{|l|c|c|c|} \hline
Label&$(\sigma_2,\sigma_4,\sigma_5)$&Existence\\[1ex]
\hline
\hline && \\[-2ex]
$P_1$&$(-1,0,0)$&always  \\[1ex]
\hline
\hline && \\[-2ex]
$P_2$&$(1,0,0)$&always \\[1ex]
\hline
\hline && \\[-2ex]
$P_3$&$\left(\frac{4-3\gamma}{3(2-\gamma)},0,0\right)$& $\left\{\begin{array}{c} 0<\gamma <\frac{4}{3},\, \text{or}\\ \frac{4}{3}<\gamma <\frac{5}{3}\end{array}\right.$ \\[1ex]
\hline
\hline && \\[-2ex]
$R_1$&$(0,0,1)$&always \\[1ex]
\hline
\hline && \\[-2ex]
$P_4$&$\left(\frac{n-2}{3(n-1)},\sqrt{1-\frac{(n-2)^2}{9(n-1)^2}},0\right)$& $n>\frac{5}{4}$ \\[1ex]
    \hline
\hline && \\[-2ex]
$R_3$ & $\left(\frac{2 (n-1)}{n-2},-\frac{\sqrt{2} (n-1)}{n-2}, -\frac{\sqrt{(8-5 n) n-2}}{n-2}\right)$ & $1<n<n_+$ \\[1ex]
\hline
\hline && \\[-2ex]
$P_5$ & $\left(\frac{3 (n-1) \gamma }{3 \gamma  n-2 n-3 \gamma },-\frac{\sqrt{2} \sqrt{n-1} \sqrt{4 n-3 \gamma }}{3 \gamma  n-2 n-3
   \gamma },0\right)$ & $\left\{\begin{array}{c} n=\frac{5}{4}, \, \gamma=\frac{5}{3}\; \text{or}\\
    0<\gamma<\frac{4}{3},\, 1<n\leq \Gamma_+(\gamma),\; \text{or}\\
    \frac{4}{3}<\gamma<\frac{5}{3},\, \frac{3 \gamma }{4}\leq n\leq \Gamma_+(\gamma).
\end{array}\right.$  \\[1ex]
\hline
\hline && \\[-2ex]
$P_6$ & $\left(\frac{3 (n-1) \gamma }{3 \gamma  n-2 n-3 \gamma },\frac{\sqrt{2} \sqrt{n-1} \sqrt{4 n-3 \gamma }}{3 \gamma  n-2 n-3
   \gamma },0\right)$ & $\frac{4}{3}<\gamma \leq \frac{5}{3}, n=\frac{3 \gamma }{4}$ $^{\rm a}$ \\[1ex]
\hline
\end{tabular}}
\end{center}
$^{\rm a}$ In this case $P_6$ and $P_3$ coincides. Thus, the
singular points are nonhyperbolic. There exists a 1-dimensional
stable manifold and a 1-dimensional unstable manifold provided
$\frac{4}{3}<\gamma <\frac{5}{3}.$
\end{table}

\begin{table}[!ht]
\begin{center}
\caption{\label{crit01a} Stability of the singular points of the
flow of \eqref{eqvphi}-\eqref{eqzrad} defined in the invariant set
$\left\{p\in\Omega_\epsilon: \varphi=0\right\}$ for $M=\sqrt{2/3}$
and $N=-\frac{\sqrt{\frac{2}{3}} (n-2)}{n-1}.$ We use the
notations $n_+=\frac{1}{5}(4+\sqrt{6}),N_+=\frac{2}{27} \left(11+2
\sqrt{10}\right),\Gamma(n)=-\frac{2 n(n-2)}{3-9 n+6 n^2},$ and
$\Gamma_+(\gamma)=\frac{9 \gamma +\sqrt{9 \gamma ^2+48 \gamma
+16}+4}{12 \gamma +4}.$ }\bigskip
{\small\begin{tabular}[t]{|l|c|c|} \hline
Label&Stability$^{\rm a}$\\[1ex]
\hline
\hline & \\[-2ex]
$P_1$ & unstable for $0<\gamma<\frac{4}{3}, n>\frac{5}{4};$ \\[1ex]
              & or $\frac{4}{3}<\gamma<\frac{5}{3}, n>\frac{5}{4};$\\[1ex]
                & saddle, otherwise\\[1ex]
\hline
\hline &\\[-2ex]
$P_2$& unstable for $\gamma\neq \frac{4}{3}, n>1$\\[1ex]
                     & saddle, otherwise\\[1ex]
\hline
\hline & \\[-2ex]
$P_3$& saddle\\[1ex]
\hline
\hline & \\[-2ex]
$R_1$& saddle\\[1ex]
\hline
\hline & \\[-2ex]
$P_4$ & stable for\\[1ex]
   &               $\left\{\begin{array}{c}
\gamma\neq\frac{4}{3}, n>2 \; \text{or}\\
n_+<n\leq 2, \Gamma(n)<\gamma<\frac{4}{3} \; \text{or}\\

n_+<n\leq 2, \frac{4}{3}<\gamma<2.
\end{array}\right.$ \\[1ex]
                     & saddle, otherwise\\[1ex]
    \hline
\hline & \\[-2ex]
$R_3$   & stable for\\[1ex]
   &               $\left\{\begin{array}{c}
\frac{4}{3}<\gamma<2, N_+\leq n<n_+ \; \text{or}\\
\frac{4}{3}<\gamma<2, 1< n<N_+ .
\end{array}\right.$ \\[1ex]
    &                  saddle, otherwise\\[1ex]
\hline
\hline & \\[-2ex]
$P_5$ & stable for\\[1ex]
   &                  $\left\{\begin{array}{c} \frac{41}{25}\leq n<2,\, 0<\gamma <\Gamma(n)\;\text{or}\\
   1<n\leq N_+,\, 0<\gamma <\frac{4}{3}\;\text{or}\\N_+<n<\frac{41}{25},\, 0<\gamma <\frac{4 \sqrt{96 n^5-272 n^4+230 n^3-50 n^2}}{3 \left(4
   n^3-24 n^2+29 n-9\right)}-\frac{2 \left(10 n^3-19 n^2+13 n\right)}{3 \left(4 n^3-24 n^2+29 n-9\right)}.\end{array}\right.$ \\[1ex]
    &                  saddle, otherwise\\[1ex]
\hline
\end{tabular}}
\end{center}
$^{\rm a}$ The stability is analyzed for the flow restricted to
the invariant set $\varphi=0$.
\end{table}

\begin{figure}[ht]
\begin{center}
\mbox{\epsfig{figure=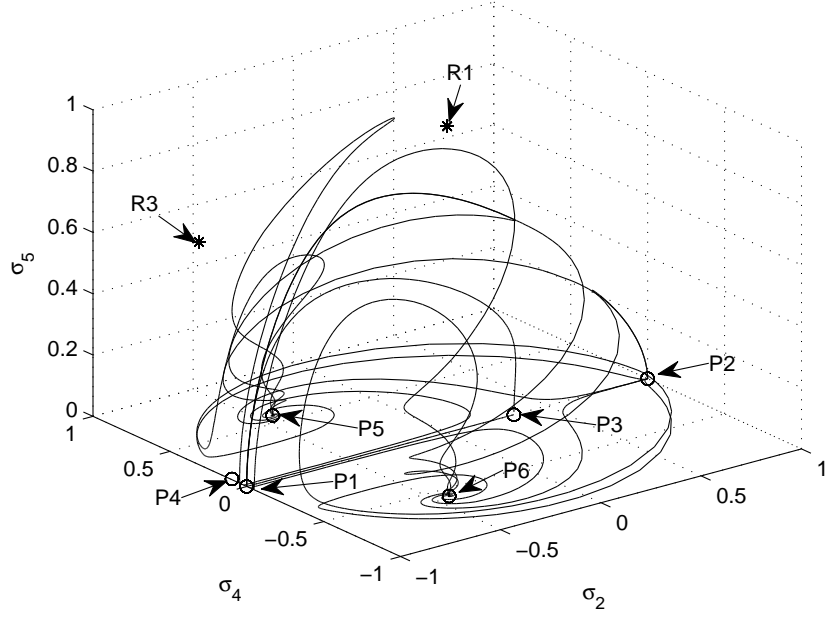,width=12cm,angle=0}}
\caption{\label{fig3a}{Projection in $\varphi=0$ of some orbits of
the flow of \eqref{eqvphi}-\eqref{eqzrad} for
$M=\sqrt{2/3},\,N=-\frac{\sqrt{\frac{2}{3}} (n-2)}{n-1}.$ We set
$n=1.251.$ Observe that $R_1,$ $R_3$ are in the region of physical
interest. These are saddle points. The singular points $P_4$ and
$P_3$ exist and are saddle points. $P_1$ and $P_2$ are local
sources and $P_5$ is a local sink. We display some orbits in the
halfspace $\sigma_4<0$ (corresponding to contracting universes)
for aesthetical purposes. Observe that $P_6$ mirrors the behavior
of $P_5.$}}
\end{center}
\end{figure}

\begin{figure}[ht]
\begin{center}
\mbox{\epsfig{figure=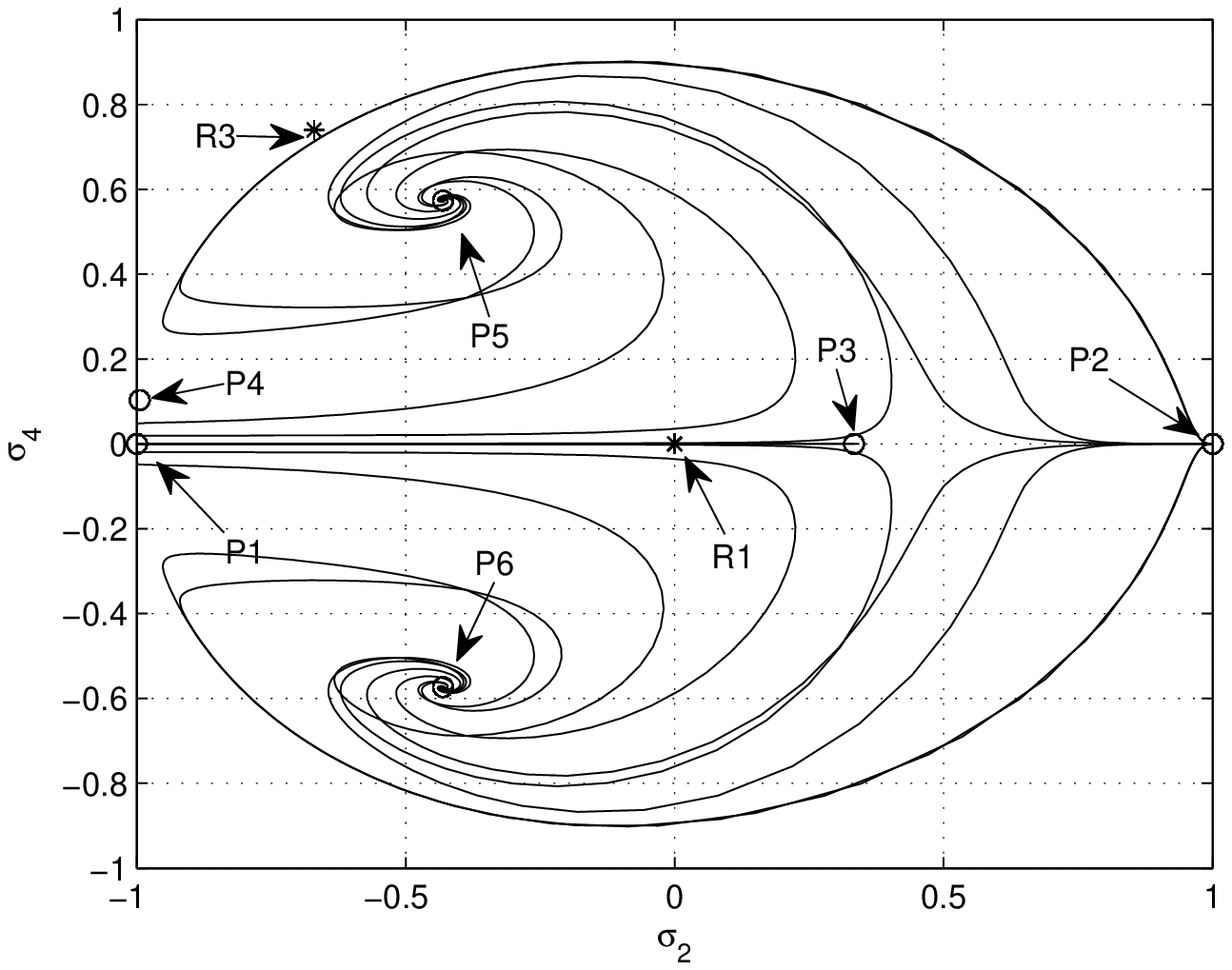,width=9.9cm,angle=0}}
\caption{\label{fig3b}{Projection of the orbits displayed in figure \ref{fig3a} to $\sigma_5=0.$}}
\end{center}
\end{figure}

Let us discuss the stability properties of the singular points
displayed in table \ref{crit01a}.

The singular point $P_1$ always exists. Its unstable manifold is
3D provided $0<\gamma<\frac{4}{3},\, n>\frac{5}{4}$ or
$\frac{4}{3}<\gamma<2, n>\frac{5}{4}.$ Otherwise its unstable
manifold is lower-dimensional.

The singular point $P_3$ always exists and it is a local source.

The singular point $P_3$ exists for $0<\gamma<\frac{4}{3}$ or
$\frac{4}{3}<\gamma<2.$ It is neither a source nor a sink.

The singular point $R_1$ always exists and it is neither a source
nor a sink.

The singular point $P_4$ exists for $n>\frac{5}{4}.$ Its stable
manifold is 3D if $\gamma\neq\frac{4}{3}, n>2$ or $n_+<n\leq 2,
\Gamma(n)<\gamma<\frac{4}{3}$ or $n_+<n\leq 2,
\frac{4}{3}<\gamma<2$. Where we have defined the grouping
constants $n_+=\frac{1}{5}(4+\sqrt{6}),\Gamma(n)=-\frac{2
n(n-2)}{3-9 n+6 n^2}.$ Otherwise its stable manifold is
lower-dimensional.

The singular point $R_3$ exists for $1<n<n_+.$ Thus $P_4$ and
$R_3$ are in the same phase portrait for $\frac{5}{4}<n<n_+.$
$R_3$ admits a 3D stable manifold for $\frac{4}{3}<\gamma<2,
N_+\leq n<n_+$ or $\frac{4}{3}<\gamma<2, 1< n<N_+,$ where we have
defined $N_+=\frac{2}{27} \left(11+2 \sqrt{10}\right).$ Otherwise
its stable manifold is lower-dimensional.

The singular point $P_5$ exists for $n=\frac{5}{4}, \,
\gamma=\frac{5}{3}$ or
    $0<\gamma<\frac{4}{3},\, 1<n\leq \Gamma_+(\gamma),$ or
    $\frac{4}{3}<\gamma<\frac{5}{3},\, \frac{3 \gamma }{4}\leq n\leq \Gamma_+(\gamma),$ where we have defined $\Gamma_+(\gamma)=\frac{9 \gamma +\sqrt{9 \gamma ^2+48 \gamma +16}+4}{12 \gamma +4}.$  $P_5$ admits a 3D stable manifold provided  $\frac{41}{25}\leq n<2,\, 0<\gamma <\Gamma(n),$ or
   $1<n\leq N_+,\, 0<\gamma <\frac{4}{3},$ or $N_+<n<\frac{41}{25},\, 0<\gamma <\frac{4 \sqrt{96 n^5-272 n^4+230 n^3-50 n^2}}{3 \left(4
   n^3-24 n^2+29 n-9\right)}-\frac{2 \left(10 n^3-19 n^2+13 n\right)}{3 \left(4 n^3-24 n^2+29 n-9\right)}.$ Otherwise its stable manifold is lower-dimensional.

The singular point $P_6$ exists for
$\frac{4}{3}<\gamma\leq\frac{5}{3},\, n=\frac{3\gamma}{4}.$ In
this case $P_6$ and $P_3$ coincides; thus, the singular points are
non-hyperbolic with a 2D center manifold, a 1D unstable manifold
and a 1D stable manifold.

Let us discuss some physical properties of the cosmological
solutions associated to the singular points displayed in table
\ref{crit01}.
\begin{itemize}
\item  $P_{1,2}$  represent kinetic-dominated cosmological solutions. They behave as stiff-like matter. The associated cosmological solution satisfies $H=\frac{1}{3 t-c_1},a=\sqrt[3]{3 t-c_1} c_2,\phi =c_3\pm\sqrt{\frac{2}{3}} \ln \left(3
   t-c_1\right),$ where $c_j,\,j=1,2,3$ are integration constants. These solutions are associated with the local past attractors of the systems for an open set of values of the parameter $\mu.$
\item  $P_3$ represents matter-dominated cosmological solutions that satisfy $H=\frac{2}{3 t \gamma -2 c_1},a=\left(3 t \gamma -2 c_1\right){}^{\frac{2}{3 \gamma }} c_2,\rho
   =\frac{12}{\left(3 t \gamma -2 c_1\right){}^2}+c_3.$
\item $R_1$ represents a radiation-dominated cosmological solutions satisfying $H=\frac{1}{2 t-c_1},a=\sqrt{2 t-c_1} c_2,\rho_r=\frac{3}{\left(2 t-c_1\right){}^2}+c_3.$
\item $P_4$ represents power-law scalar-field dominated inflationary cosmological solutions. It is easy to obtain the asymptotic exact solution: $H=\frac{3 (n-1)^2}{(n-2)^2 t-3 (n-1)^2 c_1},a=\left((n-2)^2 t-3 (n-1)^2 c_1\right){}^{\frac{3 (n-1)^2}{(n-2)^2}} c_3,\phi= c_2+\frac{\sqrt{6} (n-1) \ln \left((n-2)^2 t-3 (n-1)^2 c_1\right)}{n-2}.$
\item $P_{5}$ represent matter-kinetic-potential scaling solutions satisfying

$H=\frac{3 (n-1) \gamma -2 n}{3 (n-2) t \gamma +(n (2-3 \gamma )+3 \gamma ) c_1},a= \left(3 (n-2) t
   \gamma +(n (2-3 \gamma )+3 \gamma ) c_1\right){}^{\frac{-3 \gamma  n+2 n+3 \gamma }{6 \gamma -3 n \gamma }} c_2,$ and $\rho =
   c_3-\frac{6 (3 \gamma +n (-9 \gamma +n (6 \gamma +2)-4))}{\left(3 (n-2) t \gamma +(n (2-3 \gamma )+3 \gamma )
   c_1\right){}^2},\phi = c_4+\frac{\sqrt{6} (n-1) \log \left(3 (n-2) t \gamma +(n (2-3 \gamma )+3 \gamma )
   c_1\right)}{n-2}.$
\item $R_3$ represent radiation-kinetic-potential scaling solutions satisfying $H= \frac{1}{2 t-c_1},a= \sqrt{2 t-c_1} c_2,\rho_r= c_3-\frac{3 (n (5 n-8)+2)}{(n-2)^2
   \left(c_1-2 t\right){}^2},\phi =c_4+\frac{\sqrt{6} (n-1) \log \left(2 t-c_1\right)}{n-2}.$
\end{itemize}

To complete the section we present in figures \ref{fig3a}, and
\ref{fig3b}  a numerical elaboration of the model under
consideration.

\section{Conclusion}

In this chapter we have extended several results about flat FRW
models in the conformal (Einstein) frame in scalar-tensor gravity
theories including $f(R)$ theories through conformal
transformation. Particularly we have considered a cosmological
model based on the action \begin{align*}&S_{EF}=\int_{M_4} d{ }^4
x \sqrt{|g|}\left\{\frac{1}{2}
R-\frac{1}{2}(\nabla\phi)^2-V(\phi)\right. \nonumber \\
&\left.+\chi(\phi)^{-2}
\mathcal{L}_{m}(\mu,\nabla\mu,\chi(\phi)^{-1}g_{\alpha\beta})\right\},
\end{align*}
where $R$ is the curvature scalar, $\phi$ is the a scalar field,
and $V$ is the potential of the scalar field and $\chi$ is the
coupling function. We have consider both ordinary matter described
by a perfect fluid with equation of state $p=(\gamma-1)\rho$
(coupled to the scalar field) and radiation $\rho_r$ in order to
describe the dynamics in a cosmological epoch where matter and
radiation coexisted.

We have considered scalar fields with arbitrary (positive)
potentials and arbitrary coupling functions from the beginning.
Then, we have straightforwardly introduced mild assumptions under
such functions (differentiable class, number of singular points,
asymptotes, etc.) in order to clarify the structure of the phase
space of the dynamical system. We have obtained several analytical
results. Also, we have presented several numerical evidences that
confirm some of these results.

Our main results are the following.

\begin{enumerate}

\item We have proved the Proposition \ref{Proposition I}. This
proposition states that if the potential is nonnegative and has a
local zero minimum at $\phi=0;$ its derivative  is bounded in the
same set where the potential is; and provided the derivative of
the logarithm of the coupling function has an upper bound, then
the energy density of the background as well as the kinetic term
tend to zero when the time goes forward. Thus, the total energy
density of the universe will be dominated into the future by the
potential energy of the scalar field and the universe would expand
forever in a de Sitter phase. This result is an extension of the
Proposition 2 in \cite{Miritzis:2003ym} to the non-minimal
coupling context.

\item With the same hypotheses as in \ref{Proposition I} and with
the additional hypothesis of $V(\phi)$ being strictly decreasing
(increasing) for negative (positive) values of the scalar field,
we have proved in \ref{Proposition II} that the scalar field can
be either $-\infty$ or zero or $+\infty.$ This means that under
the above hypothesis the scalar field diverges into the future or
it equals to zero (the last case holds only if the Hubble scalar
vanish towards the future). This proposition is an extension of
proposition 3 in \cite{Miritzis:2003ym}.

\item Assuming that the potential is non-negative (with not
necessarily a local minimum at $(0,0)$) having a continuous
derivative (bounded on a set $A$ provided the potential is bounded
on it). Assuming also that the potential is strictly decreasing
such that $V(\phi(t)),$ viewed as a function of t, diverges when
$t$ approaches infinity. Then the cosmological model enters a de
Sitter phase into the future characterized by the divergence of
the scalar field. If additional the potential vanish
asymptotically to the future as a function of $\phi,$ the Hubble
scalar vanishes too. This fact is true for the exponential
potential. (This result is presented in proposition
\ref{Proposition III}).

\item For the model including radiation we have formulated and
proved proposition \ref{thmIII} generalizing analogous result in
\cite{Giambo':2009cc}. It states that if the potential $V(\phi)$
is such that the (possibly empty set) where it is negative is
bounded and the (possibly empty) set of singular points of
$V(\phi)$ is finite, then, the singular point $${\bf
p}_*:=\left(\phi_*,y_*=0,\rho_*=0,\rho_r=0,
H=\sqrt{\frac{V(\phi_*)}{3}}\right),$$ where $\phi_*$ is a strict
local minimum for $V(\phi),$  is an asymptotically stable singular
point for the flow. From the physical viewpoint this result is
relevant since it provides conditions for the asymptotic stability
of the \emph{de Sitter} solution.

\item After the introduction of modified normalized variables, we
have proved that the phase space of the model has the structure of
a manifold with boundaries (see propositions \ref{Prop8} and
\ref{Prop9} for the invariant set $\rho_r=0$ and propositions
\ref{thm1}, \ref{thm2} for the model including radiation). We have
devised a monotonic function for the flow of the dynamical system
which allow for the identification of some invariant sets (the
more relevant invariant sets are discussed in proposition
\ref{prop2.1}).

\item We have provided approximated center manifolds for the
vector field around the inflection points and the strict
degenerate local minimum of the potential. For inflection and
degenerate local maximum points of order $n=2$ for the potential
the center manifold of $Q_2$ is locally unstable whereas for
degenerate local minimum points of order $n=2$ for the potential
it is locally asymptotically stable (see propositions
\ref{Q2inflection} and \ref{Q2extremum}). The results obtained are
in agreement with the result in proposition \ref{thmIII}.

\item In lemma \ref{Theorem I} it is proved that the orbit passing
through an arbitrary point $p\in\Sigma$ (representing cosmological
solutions  with non-vanishing dimensionless background energy
density and positive finite Hubble parameter) is past asymptotic
to a regime where the Hubble parameter diverges containing an
initial singularity into the past, and is future asymptotic to a
regime where the background density is negligible into the future.
This result is obtained by constructing a monotonic function
defined on an invariant set (subset of the invariant set
$\rho_r=0$) and by applying the LaSalle monotonicity principle
(theorem 4.12, \cite{REZA}).

\item When the scalar field is incorporated  as a dynamical
variable it typically diverges into the past. This fact (proved in
\cite{Foster:1998sk}) is valid also to our general context. The
theorem \ref{Theorem_2.4} is a generalization to the theorem 1 in
the reference \cite{Foster:1998sk}. Theorem \ref{Theorem_2.4}
states that if the potential and the coupling function are
sufficiently smooth functions, then for almost all the points
lying in the Hubble-normalized state space, the scalar field
diverges when the orbit thought $p$ is followed backward in time.
The demonstration of this theorem relies on the topological
properties of the state space and the existence of monotonic
functions.

\item We have proved theorem \ref{thm4} which is a generalization
of the related result in
\cite{Leon:2008de,Foster:1998sk,Miritzis:2003ym} and an extension
of theorem \ref{Theorem_2.4}. This result state that if
$\chi(\phi)$ and $V(\phi)$ are positive functions of class $C^3,$
such that $\chi$ has at most a finite number of stationary points
and does not tend to zero in any compact set of $\mathbb{R}$,
then, given $p,$ an interior point of the phase space manifold,
the scalar field, $\phi,$ is unbounded through the past orbit
$O^{-}(p).$ The relevance from the physical viewpoint of this
result is twofold. First, the inclusion of radiation in the cosmic
budget does not influence radically the early-time behavior of the
scalar field. This result is somewhat expected since for small
scale factor $a,$ the energy densities of radiation and the scalar
field goes respectively as $\rho_r\sim a^{-4},$ and
$\rho_\phi\approx \frac{\dot\phi^2}{2}\sim a^{-6}$ (the last
approximation is supported by theorems 4.1 and 4.2 in
\cite{Leon:2008de}). Second, the result of theorem \ref{thm4}
makes clear that in order to investigate the generic past
asymptotic dynamics of the flow we must scan the region of the
phase space where $|\phi|\rightarrow\infty$.

\item For the analysis of the system as $\phi\rightarrow\infty$ we
have defined a suitable change of variables to bring a
neighborhood of $\phi=\infty$ in a bounded set. This method was
first introduced in \cite{Foster:1998sk} (see also
\cite{Giambo:2008sa}). By assuming some regularity conditions on
the potential and on the coupling function in that regime we have
constructed a dynamical system (well suited to investigate the
dynamics where the scalar field diverges, i.e. near the initial
singularity) defined in the invariant set $\rho_r=0$. The singular
points therein are investigated and the cosmological solutions
associated to them are characterized. We find the existence of
three singular points $P_3,$ $P_5$ and $P_6.$ They are in the
boundary of the phase space $\Sigma_\epsilon.$ They represent
cosmological scaling solutions (where the contribution of the
dimensionless potential energy is negligible). By tuning the free
parameters they can be accelerating. In contrast in the reference
\cite{Foster:1998sk} there exists only one (in our notation,
$P_4$) representing an accelerating cosmology. The solutions
associated to $P_{1,2}$ ($p_\mp$ in the notation in
\cite{Foster:1998sk}) represent stiff and then decelerating
solutions (actually solutions associated to a massless scalar
field). For the general model including radiation we have obtained
in the limit $\phi\rightarrow\infty$: radiation-dominated
cosmological solutions; power-law scalar-field dominated
inflationary cosmological solutions; matter-kinetic-potential
scaling solutions and radiation-kinetic-potential scaling
solutions.

\item For the model without including radiation we have proved a
theorem (theorem \ref{initialsingularity}) which is an extension
of the theorem 4 in \cite{Foster:1998sk} to the STT framework.
Also, we sketch the proof of the global singularity theorem
\ref{globalinitialsingularity}. Theorem
\ref{globalinitialsingularity} indicates that the past asymptotic
structure of non-minimally coupled scalar field theories with FRW
metric, as in the FRW general relativistic case, is independent of
the exact details of the potential and/or the details of the
background matter and the coupling function.

This is a conjecture with solid theoretical and numerical
foundations (see figures 1 and 2 in \ref{toy}). To prove that the
family of solutions which asymptotically approach $P_1$ are
completely characterized by the solution space of the massless
scalar field cosmological model (i.e., $V$ and $\chi$ and then
$\rho$, being dynamically insignificant in the neighborhood of the
singularity $P_1$) it is required to prove that this
correspondence is one-to-one and continuous, which is hard to do
in our scenario.

\item Using the mathematical apparatus developed in the first part
of the chapter, we have investigated, for the general model
including radiation, a general class of potentials containing the
cases investigated in \cite{Copeland:1997et,vandenHoogen:1999qq}.
In order to provide a numerical evidence for our analytical
results for this class of models, we have re-examined the toy
model with power-law coupling and Albrecht-Skordis potential
$V(\phi )=e^{-\mu \phi }{\left( A+(\phi -B)^2\right)}$
investigated in \cite{Leon:2008de} in presence of radiation.

\item Also we have investigated, for the general model including
radiation, the important examples of higher order gravity theories
$F(R) = R + \alpha R^2$ (quadratic gravity) and $F(R) =R^n.$  In
the case of quadratic gravity we have proved in proposition
\ref{centerP4}, by an explicit computation of the center manifold,
that the singular point corresponding to \emph{de Sitter} solution
is locally unstable (saddle point). This result complements the
result of the proposition discussed in \cite{Miritzis:2005hg} p.
5, where it was proved the local asymptotic instability of the
\emph{de Sitter} universe for positively curved FRW models with a
perfect fluid matter source and a scalar field which arises in the
conformal frame of the $R+\alpha R^2$ theory.

\end{enumerate}

\chapter{Phantom dark energy with varying-mass dark matter
particles}

In this chapter we investigate several varying-mass dark-matter
particle models in the framework of phantom cosmology. We examine
whether there exist late-time cosmological solutions,
corresponding to an accelerating universe and possessing dark
energy and dark matter densities of the same order. Imposing
exponential or power-law potentials and exponential or power-law
mass dependence, we conclude that the coincidence problem cannot
be solved or even alleviated. Thus, if dark energy is attributed
to the phantom paradigm, varying-mass dark matter models cannot
fulfill the basic requirement that led to their construction.

\section{Introduction}

The dynamical nature of dark energy introduces a new cosmological
problem, namely why are the densities of vacuum energy and dark
matter nearly equal today although they scale independently during
the expansion history. The elaboration of this ``coincidence''
problem led to the consideration of generalized versions of the
aforementioned scenarios with the inclusion of a coupling between
dark energy and dark matter. Thus, various forms of
``interacting'' dark energy models
\cite{interacting2,interacting,Guo:2004xx,ChenSaridakis} have been
constructed in order to fulfil the observational requirements. In
the case of interacting quintessence one can find accelerated
attractors which moreover give dark matter and dark energy density
parameters of the same order,  thus solving the coincidence
problem
\cite{Curbelo:2005dh,Amendola:1999er,Amendola:1999qq,Bean:2000zm,Chimento:2003ie,Chimento:2000kq,Chimento:2003sb,Zimdahl:2001ar,Gonzalez:2006cj,Wetterich:1994bg,Wetterich:1994bg2},
but paying the price of introducing new problems such is the
justification of a non-trivial, almost tuned, sequence of
cosmological epochs \cite{Amendola:2006qi}. In interacting phantom
models \cite{Guo:2004vg,Guo:2004xx,ChenSaridakis}, the existing
literature remains in some special coupling forms which suggest
that the coincidence problem might be alleviated
\cite{Guo:2004xx,ChenSaridakis}.

An equivalent approach is to assume that dark energy and dark
matter sectors interact in such a way that the dark matter
particles acquire a varying mass, dependent on the scalar field
which reproduces dark energy \cite{Anderson:1997un}. This
consideration allows for a better theoretical justification, since
a scalar-field-dependent varying-mass can arise from string or
scalar-tensor theories \cite{Damour:1990tw}. Indeed, in such
higher dimensional frameworks one can formulate both the
appearance of the scalar field (which is related to the dilaton
and moduli fields) and its effect on  matter particle masses
(determined by string dynamics, supersymmetry breaking, and the
compactification mechanism) \cite{Casas:1991ky}. In quintessence
scenario, such varying-mass dark matter models have been explored
in cases of linear
\cite{Anderson:1997un,interacting2,Casas:1991ky,quirosvamps},
power-law \cite{Zhang:2005rg} or exponential
\cite{Amendola:1999er,Amendola:2001rc,Comelli2003} scalar-field
dependence. The exponential case is the most interesting since,
apart from solving the coincidence problem, it allows for stable
scaling behavior, that is for a large class of initial conditions
the cosmological evolution converges to a common solution at late
times \cite{Amendola:1999er,Comelli2003}.

We are interested in investigating varying-mass dark matter models
in scenarios where dark energy is attributed to a phantom field.
Although such a framework could lead to instabilities at the
quantum level \cite{Cline:2003gs}, there have been serious
attempts in overcoming these difficulties and construct a phantom
theory consistent with the basic requirements of quantum field
theory, with the phantom fields arising as an effective
description \cite{quantumphantom0}. Performing a complete
phase-space analysis using various forms of mass-dependence and
scalar-field potentials, we examine whether there exist stable
late-time accelerating solutions which moreover solve the
coincidence problem. As we will show, the coincidence problem
cannot be solved in any of the investigated models. In the paper
\cite{PLB2010} we examine whether there exist late-time
cosmological solutions, corresponding to an accelerating universe
and possessing dark energy and dark matter densities of the same
order. Imposing exponential or power-law potentials and
exponential or power-law mass dependence, the coincidence problem
cannot be solved or even alleviated. Thus, if dark energy is
attributed to the phantom paradigm, varying-mass dark matter
models cannot fulfill the basic requirement that led to their
construction. In this book we improve the analysis in
\cite{PLB2010} by using the Center Manifold Theory to analyze the
stability of the non-hyperbolic fixed points in the phase space of
dark-matter particle models in the framework of phantom cosmology.
Basically, we use these cosmological models as examples of how to
apply the Center Manifold Theory in cosmology. Also, in this book
we perform a Poincar\'e compactification process allowing to
construct a global phase space containing all the cosmological
information in both finite and infinite regions.

\section{Phase-space analysis} \label{models}

In  section \ref{camposescalares} we constructed a cosmological
scenario where the dark matter particles have a varying mass,
depending on the phantom field. Additionally, we presented the
formalism for its transformation into an autonomous dynamical
system, suitable for a stability analysis (see also the analysis
by one of us in \cite{PLB2010}). In this section we introduce
specific forms for $V(\phi)$ and $M_{DM}(\phi)$, and we perform a
complete phase-space analysis.

For the scalar field potential we consider two well studied cases
of the literature, namely the exponential
\cite{Amendola:1999er,Comelli2003}:
\begin{equation}\label{exppot}
V(\phi)=V_0e^{-\kappa\lambda_1\phi}
\end{equation}
and the power-law one \cite{Zhang:2005rg,Kneller:2003xg}:
\begin{equation}\label{powerpot}
V(\phi)=V_0\phi^{-\lambda_2}.
\end{equation}
For the  dark matter particle mass we consider two possible cases,
namely an exponential dependence
\cite{Amendola:1999er,Amendola:2001rc,Comelli2003}:
\begin{equation}\label{expmass}
M_{DM}(\phi)=M_0e^{-\kappa\mu_1\phi}
\end{equation}
and the power-law one \cite{Zhang:2005rg}:
\begin{equation}\label{powermass}
M_{DM}(\phi)=M_0\phi^{-\mu_2}.
\end{equation}
Therefore, in the following we consider four different models,
arising from the aforementioned combinations.

In order to perform the phase-space and stability analysis of the
phantom model at hand, we have to transform the aforementioned
dynamical system into its autonomous form
\cite{Copeland:1997et,expon,expon1,expon2}. This will be achieved
by introducing the auxiliary variables:
\begin{eqnarray}
&&x=\frac{\kappa\dot{\phi}}{\sqrt{6}H},\nonumber\\
&&y=\frac{\kappa\sqrt{V(\phi)}}{\sqrt{3}H}, \nonumber\\
&&z=\frac{\sqrt{6}}{\kappa\phi} \label{auxilliary}
\end{eqnarray}
together with $M=\ln a$. Thus, it is easy to see that for every
quantity $F$ we acquire $\dot{F}=H\frac{dF}{dM}$.
 Using these
variables we obtain:
\begin{equation}
 \Omega_{\phi}\equiv\frac{\kappa^{2}\rho_{\phi}}{3H^{2}}=-x^2+y^2,
 \label{Omegas}
\end{equation}
\begin{equation}\label{wss}
w_{\phi}=\frac{-x^2-y^2}{-x^2+y^2},
\end{equation}
and
\begin{equation}\label{wtot}
w_{tot}=-x^2-y^2.
\end{equation}
We mention that relations (\ref{wss}) and (\ref{wtot}) are always
valid, that is independently of the specific state of the system
(they are valid in the whole phase-space and not only at the
critical points). Finally, note that in the case of complete dark
energy domination, that is $\rho_{DM}\rightarrow0$ and
$\Omega_\phi\rightarrow1$, we acquire $w_{tot}\approx
w_\phi\leq-1$, as expected to happen in phantom-dominated
cosmology. Finally, the deceleration parameter $q\equiv
-\frac{\ddot a a}{\dot a^2}$ is given by
\begin{equation}
q=\frac{1}{2} \left(1-3 x^2-3 y^2\right). \label{decc}
\end{equation}

\subsection{Model 1: Exponential potential and exponentially-dependent dark-matter particle mass}

Inserting the auxiliary variables (\ref{auxilliary}) into the
equations of motion (\ref{rhodmeom}), (\ref{phiddot}), (\ref{FR1})
and (\ref{FR2}), we result in the following autonomous system:
\begin{eqnarray}
x'=-
3x+\frac{3}{2}x (1-x^2-y^2)-\sqrt{\frac{3}{2}}\la_1\, y^2-\sqrt{\frac{3}{2}}\mu_1(1+x^2-y^2)\nonumber\\
y'=\frac{3}{2}y (1-x^2-y^2)- \sqrt{\frac{3}{2}}\la_1\, x y
\label{autonomous1}.\ \ \ \ \ \ \ \ \ \ \ \ \ \   \ \   \ \ \ \
\end{eqnarray} defined in the phase plane $\{(x,y)| -x^2+y^2\leq 1, y\geq 0\}.$
Note that in this case, the auxiliary variable $z$ is not needed.

\subsubsection{Finite analysis}\label{FiniteVmass1}

The critical points $(x_c,y_c)$ of the autonomous system
(\ref{autonomous1}) are obtained by setting the left hand sides of
the equations to zero. The real and physically meaningful (that is
corresponding to $y>0$ and $0\leq\Omega_\phi\leq1$) of them are:
\begin{eqnarray}
\left(x_{c1}=-\frac{\la_1}{\sqrt{6}},\
y_{c1}=\sqrt{1+\frac{\la_1^2}{6}}\right), \nonumber
\end{eqnarray}
\begin{equation}\left(x_{c2}=\frac{\sqrt{\frac32}}{\la_1-\mu_1},\
y_{c2}=\frac{\sqrt{-\frac32-\mu_1\left(\la_1-\mu_1\right)}}{|\la_1-\mu_1|}\right),
\end{equation}
and in table \ref{stability1} we present the necessary conditions
for their existence and their dynamical character. In table
\ref{densities1} are displayed some basic observables  such that
$\Omega_\s$ \footnote{In this chapter $\Omega_\s$ is referred to
the fractional energy density of the $\s$-field, not to the energy
density of anisotropy.} and $w_{tot}$ and the conditions for
acceleration for the real and physically meaningful critical
points of Model 1.

\begin{table*}[t]
\begin{center}
\caption[The real and physically meaningful critical points of
Model 1 and their behavior.]{\label{stability1} The real and
physically meaningful critical points of Model 1 and their
behavior.}
\begin{tabular}{|c|c|c|c|c|}
\hline
 Cr. P.& $x_c$ & $y_c$ & Existence & Stable for \\
\hline \hline
 A&  $x_{c1}$ & $y_{c1}$ & Always & $\la_1\left(\mu_1-\la_1\right)<3$  \\
B&  $x_{c2}$ & $y_{c2}$ & $\min\{\mu_1^2-3, \la_1^2+3\}\geq
\la_1\mu_1$, & Never \\
 &    &   &  $\mu_1\neq\la_1$ &  \\
\hline
\end{tabular}
\end{center}
\end{table*}

\begin{table*}[t]
\begin{center}
\caption[Basic observables and conditions for acceleration for the
real and physically meaningful critical points of Model 1.
]{\label{densities1} Basic observables and conditions for
acceleration for the real and physically meaningful critical
points of Model 1. }
\begin{tabular}{|c|c|c|c|}
\hline
 Cr. P.&  $\Omega_\s$ &  $w_{tot}$ & Acceleration   \\
\hline \hline
 A & 1 &
 $-\frac13\left(3+\la_1^2\right)$ & Always \\
B & $\frac{\mu_1^2-\la_1 \mu_1-3}{(\la_1-\mu_1)^2}$ &
 $\frac{\mu_1}{\la_1-\mu_1}$ & $\mu_1<0,\, \mu_1<\la_1<-2\mu_1$ \\
  &  &
 & $\mu_1>0,\, -2\mu_1<\la_1<\mu_1$\\
\hline
\end{tabular}
\end{center}
\end{table*}

Therefore, for each critical point of table \ref{stability1}, we
examine the signs of the real parts of the eigenvalues of the
linearization matrix, which determine the type and stability of
this specific critical point. In table \ref{stability1} we present
the results of the stability analysis. In addition, in table
\ref{densities1}, for each critical point we calculate the values
of $w_{tot}$ (given by relation (\ref{wtot})), and of
$\Omega_\phi$ (given by (\ref{Omegas})). Thus, knowing $w_{tot}$
we can express the acceleration condition $w_{tot}<-1/3$ in terms
of the model parameters.

The critical point A  exists always and it is either a saddle
point (the eigenvalues of the linearization matrix have real parts
of different sign) or an attractor (the eigenvalues of the
linearization matrix have negative real parts). The critical point
B, if it exists, it is always a saddle point. The cosmological
model at hand admits another critical point, namely C, which is
unphysical since it leads to $\Omega_\phi<0$. This point has
coordinates $\left(x_{c3}=-\sqrt{\frac{2}{3}}\mu_1,\
y_{c3}=0\right)$ and it is either a saddle point or an attractor.
If $\mu_1(\mu_1-\lambda_1)>3/2$ it is an attractor and in this
case, although unphysical, it can attract an open set of orbits
from the interior of the physical region of the phase space.

\begin{figure}[ht]
\begin{center}
\hspace{0.4cm}
\includegraphics[width=8cm,height=7cm]{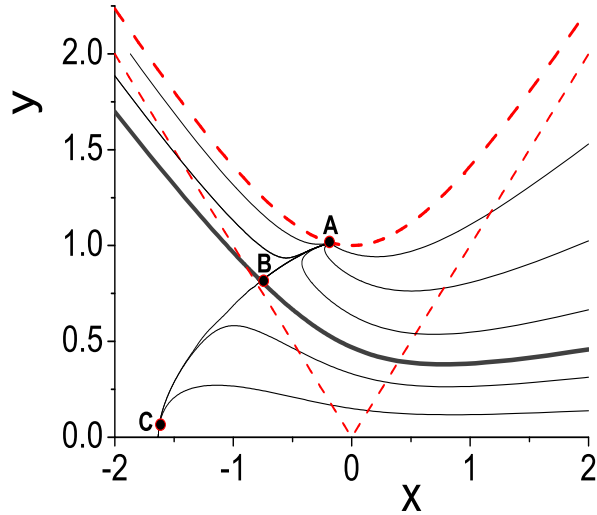}
\caption{{ Phase plane of Model 1 for the parameter values
$\la_1=0.4$ and $\mu_1=2$. The stable manifold of B (thick curve)
divides the physical part of the phase space (region corresponding
to $0\leq \Omega_\phi\leq1$, bounded by the dashed (red) curves)
in two regions. The orbits initially below this curve converge
towards $C.$ The orbits initially above this curve converge to A.
[Taken from \cite{PLB2010}; published with permission of Elsevier
B.V.].}} \label{VmassFig1}
\end{center}
\end{figure}

\begin{figure}[ht]
\begin{center}
\hspace{0.4cm}
\includegraphics[width=8cm,height=7cm]{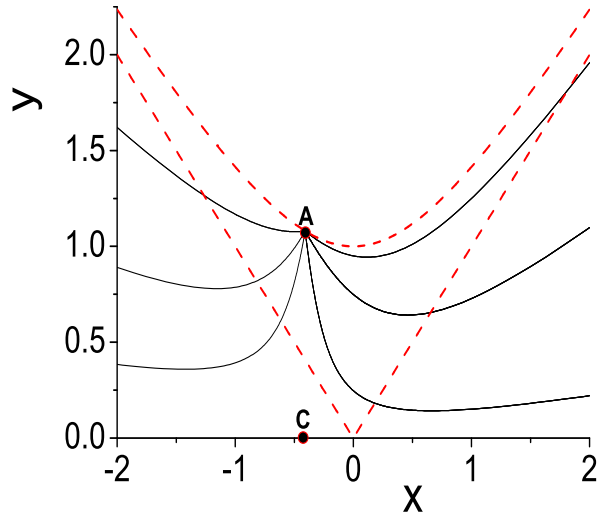}
\caption{{ Phase plane of Model 1 for the parameter values
$\la_1=1$ and $\mu_1=0.5$. In this case the critical point B does
not exist and all orbits initially at the physical region converge
to A. The dashed (red) curves bound the physical part of the phase
space, that is corresponding to
 $0\leq \Omega_\phi\leq1$. [Taken from \cite{PLB2010}; published with permission of Elsevier
B.V.].}} \label{VmassFig2}
\end{center}
\end{figure}

In order to present this behavior more transparently, we
 evolve the autonomous system (\ref{autonomous1}) numerically for
the parameters $\la_1=0.4$ and $\mu_1=2$, and the results are
shown in figure \ref{VmassFig1}.
Depending on which region of the phase-space does the system
initiates, it lies in the basin of attraction of either A or C,
and thus it is attracted by one or the other point. In particular,
the orbits initially below the stable manifold of B-points
converge towards C, while the orbits initially above this curve
converge to A. Interestingly, A is not the global attractor for
points at the physical region (region corresponding to $0\leq
\Omega_\phi\leq1$,  bounded by the dashed (red) curves). However,
if
$\frac{\la_1}{2}-\frac{\sqrt{6+\la_1^2}}{2}<\mu<\frac{\la_1}{2}+\frac{\sqrt{6+\la_1^2}}{2},$
point C is always a saddle one and B does not exist. Thus, in this
case A is the attractor for all the points located at the physical
region. This behavior is presented in figure \ref{VmassFig2}.
Finally, for completeness we mention that in the trivial case
$\mu_1=0$ the origin is also a saddle point. It represents
matter-dominated universe ($\Omega_{DM}\equiv
\frac{\kappa^2\rho_{DM}}{3 H^2}=1$) with $\phi$-independent dark
matter particle mass.

\subsubsection{Analysis at infinity}\label{9.2.1.2}

Owing to the fact that the dynamical system \eqref{autonomous1} is
non-compact, there could be features in the asymptotic regime
which are non-trivial for the global dynamics. Thus, in order to
complete the analysis of the phase space we will now extend our
study using the Poincar\'e central projection method.

For do that we introduce the Poincar\'e variables

\begin{equation}\label{TransfVmass1}
x_r=\rho \cos\theta,\, y_r=\rho \sin\theta
\end{equation} where $\rho=\frac{r}{\sqrt{1+r^2}},\, r=\sqrt{x^2+y^2}.$  Thus, the points at ``infinite''
($r\rightarrow+\infty$) are those having $\rho\rightarrow 1.$
Since $y$ is required to be nonnegative, $\phi$ varies in
$[0,\pi].$ The region of physical interests is given by
$-\frac{\sqrt{2}}{2}\leq x_r\leq\frac{\sqrt{2}}{2},|x_r|\leq
y_r\leq \frac{\sqrt{2}}{2}.$

Performing the transformation \eqref{TransfVmass1}, the system
\eqref{autonomous1} becomes

\begin{align}
&\rho'= -\frac{3 \rho ^3}{2}+\frac{3}{2} \left(\rho ^2-1\right)
\cos (2 \theta ) \rho +\nonumber \\ &+\cos \theta
\left((\mu_1-\lambda_1) \rho ^2 \sqrt{6-6 \rho ^2} \sin
^2\theta-\sqrt{\frac{3}{2}} \mu_1 \sqrt{1-\rho
   ^2}\right),\label{Vmassinfty1a}\\
&   \theta'=3 \cos \theta \sin \theta+\left(\frac{\sqrt{6-6 \rho
^2} \mu_1}{2 \rho } +\frac{\sqrt{\frac{3}{2}}
   (\mu_1-\lambda_1) \rho  \cos (2 \theta )}{\sqrt{1-\rho ^2}}\right) \sin
   \theta\label{Vmassinfty1b}.
\end{align}

In the limit $\rho\rightarrow 1,$ the leading terms in
\eqref{Vmassinfty1a}-\eqref{Vmassinfty1b} are

\begin{eqnarray}
&&\rho'\rightarrow -\frac{3}{2},\label{Vmassinfty1c}\\
&&   \theta'\rightarrow -\frac{\sqrt{\frac{3}{2}}
(\lambda_1-\mu_1) \cos (2 \theta ) \sin \theta}{\sqrt{1-\rho
^2}}\label{Vmassinfty1d}.
\end{eqnarray}

The radial equation does not contain the radial coordinate, so the
singular points  can be obtained using the angular equation only.
Setting $\theta'=0$, we obtain the singular points  which are
listed in table \ref{Vmassinfty1}. The stability of these points
is studied by analyzing first the stability of the angular
coordinate and then deducing, from the sign of equation
\eqref{Vmassinfty1c}, the stability on the radial direction. Since
$\rho'<0,$ the singular points at infinity are either saddles or
sources. For simplicity we assume $\lambda_1\neq\mu_1.$

Performing the above procedure we find that there is no late-time attractors in the infinite region.
Thus, following the discussion in the section \ref{FiniteVmass1} the relevant late-time attractor with
physical sense is the phantom-dominated super-accelerated solution, $A,$ for the choice of parameters in the range
$\la_1\left(\mu_1-\la_1\right)<3.$

\begin{table*}[ht]
\begin{center}\caption{\label{Vmassinfty1} Asymptotic singular points  of the system
\eqref{autonomous1} (case 1) and their stability.}
\begin{tabular}{ccccc}
\hline \hline
 Cr. P & Coordinates: $\theta, x_r, y_r$ & Eigenvalue & $\rho'$ & Stability\\
\hline $Q_1$ & $0, 1, 0$ & $\left\{\begin{array}{cc}
  -\infty & \text{for}\; \mu_1<\lambda_1\\
  +\infty & \text{for}\; \mu_1>\lambda_1\\
\end{array}\right. $  & $-\frac{3}{2}$ &  $\left\{\begin{array}{c} \text{saddle}\\ \text{source} \end{array}\right.$ \vspace{0.2cm}\\
\hline $Q_2$ & $\pi, -1, 0$ & $\left\{\begin{array}{cc}
 +\infty & \text{for}\; \mu_1<\lambda_1\\
  -\infty & \text{for}\; \mu_1>\lambda_1\\
\end{array}\right. $  & $-\frac{3}{2}$ &  $\left\{\begin{array}{c} \text{source}\\ \text{saddle} \end{array}\right.$ \vspace{0.2cm}\\
\hline $Q_3$ & $\frac{\pi}{4}, \frac{\sqrt{2}}{2},
\frac{\sqrt{2}}{2}$ & $\left\{\begin{array}{cc}
  +\infty & \text{for}\; \mu_1<\lambda_1\\
  -\infty & \text{for}\; \mu_1>\lambda_1\\
\end{array}\right. $  & $-\frac{3}{2}$ &  $\left\{\begin{array}{c} \text{source}\\ \text{saddle} \end{array}\right.$ \\
\vspace{-0.35cm}\\
\hline $Q_4$ & $\frac{3\pi}{4}, -\frac{\sqrt{2}}{2},
\frac{\sqrt{2}}{2}$ & $\left\{\begin{array}{cc}
  -\infty & \text{for}\; \mu_1<\lambda_1\\
  +\infty & \text{for}\; \mu_1>\lambda_1\\
\end{array}\right. $  & $-\frac{3}{2}$ &  $\left\{\begin{array}{c} \text{saddle}\\ \text{source} \end{array}\right.$\\
\\\hline \hline
\end{tabular}
\end{center}
\end{table*}

The basic observables \eqref{Omegas}, \eqref{wss}, \eqref{wtot}
and \eqref{decc} are given in terms of the Poincar\'e variables by

\begin{eqnarray}
&&\Omega_\phi\equiv\frac{\left(x_r-y_r\right)
   \left(x_r+y_r\right)}{x_r^2+y_r^2-1}=\frac{\rho ^2 \cos (2 \theta
   )}{\rho ^2-1},\nonumber\\
&& w_\phi\equiv   \frac{x_r^2+y_r^2}{\left(x_r-y_r\right)
   \left(x_r+y_r\right)}=\sec (2 \theta
   ),\nonumber\\
&&w_{\text{tot}}\equiv
\frac{x_r^2+y_r^2}{x_r^2+y_r^2-1}=\frac{\rho
^2}{\rho ^2-1},\nonumber\\
&&q\equiv \frac{4
   x_r^2+4 y_r^2-1}{2 \left(x_r^2+y_r^2-1\right)}=\frac{1-4 \rho ^2}{2-2
   \rho ^2}\label{PoincareObsv}.
\end{eqnarray}
Taking the limit $\rho\rightarrow 1^-$ in the expressions
\eqref{PoincareObsv}, it is easy to see that $q\rightarrow
-\infty$ and $w_{\text{tot}}\rightarrow -\infty.$ That is, the
points at infinity represents supper-accelerating ($q\ll 0$)
phantom solutions ($w_{\text{tot}}\ll -1$), which can be physical
or unphysical depending whether or not $0\leq \Omega_\phi\leq 1,$
or, equivalently,  whether or not $-\frac{\sqrt{2}}{2}\leq
x_r\leq\frac{\sqrt{2}}{2},|x_r|\leq y_r\leq \frac{\sqrt{2}}{2}.$
Thus the solutions $Q_1$ and $Q_2$ are unphysical, with
$\Omega_\phi\rightarrow -\infty.$ At the singular point $Q_3$ we
have that $\Omega_\phi=0, w_\phi=-\infty$ and at the singular
point $Q_4$ we have $\Omega_\phi=0, w_\phi=+\infty.$ Although they
are matter dominated solutions, since $q\rightarrow -\infty$ and
$w_{\text{tot}}\rightarrow -\infty,$ they mimics phantom behavior.
In fact they correspond to big-rip singularities.

The system \eqref{Vmassinfty1a}-\eqref{Vmassinfty1b} has an
apparent singularity in $\rho=0,\sin\theta=0,$ which is due to the
spherical coordinate system. Thus, for numerical examinations it
is more convenient to use the cartesian coordinates $x_r,y_r.$ The
system reads
\begin{eqnarray}
&x_r'=\frac{1}{2} \left(\frac{\sqrt{6} \left(\left(2
x_r^2-1\right) y_r^2 \lambda_1-\left(x_r^2-1\right) \left(2
y_r^2-1\right)
  \mu_1\right)}{\sqrt{1-x_r^2-y_r^2}}-3 x_r \left(2 y_r^2+1\right)\right),\nonumber\\
&y_r'=  -\frac{1}{2} y_r \left(2 y_r^2-1\right)
   \left(\frac{\sqrt{6} x_r
   (\mu_1-\lambda_1)}{\sqrt{1-x_r^2-y_r^2}}+3\right)\label{InfVmassM1}.
\end{eqnarray}

\begin{figure}[ht]
\begin{center}
\hspace{0.4cm}
\includegraphics[width=8cm,height=7cm]{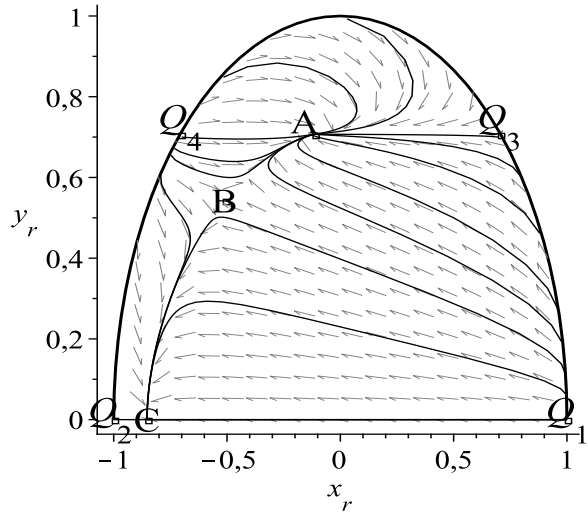}
\caption{{ Poincar\'e (global) phase plane of Model 1 for the
parameter values $\la_1=0.4$ and $\mu_1=2$. The attractors in the
finite region are A which is physical ($0\leq \Omega_\phi\leq1$)
and $C.$ The orbits initially above the stable manifold of B
converge to A (see figure \ref{VmassFig1}). The points at infinity
$Q_1$ and $Q_4$ are sources, whereas  $Q_2$ and $Q_3$ are
saddles.}} \label{InfVmass_la0_4mu_2}
\end{center}
\end{figure}

In figure \ref{InfVmass_la0_4mu_2} we present the Poincar\'e
(global) phase plane of Model 1 for the parameter values
$\la_1=0.4$ and $\mu_1=2$. There are two attractors in the finite
region: a physical one A, and an unphysical state C. The orbits
initially above the stable manifold of B converge to A. The points
at infinity $Q_1$ and $Q_4$ are sources, whereas  $Q_2$ and $Q_3$
are saddles.

\begin{figure}[ht]
\begin{center}
\hspace{0.4cm}
\includegraphics[width=8cm,height=7cm]{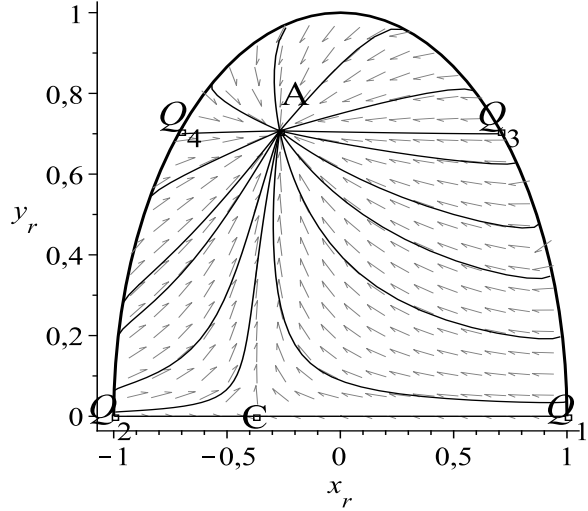}
\caption{{ Poincar\'e (global) phase plane of Model 1 for the
parameter values $\la_1=1.0$ and $\mu_1=0.5$. The attractor in the
finite region is the singular point A which is physical ($0\leq
\Omega_\phi\leq1$) (see figure \ref{VmassFig2}). The unphysical
state $C$ is a saddle. The points at infinity $Q_2$ and $Q_3$ are
sources, whereas $Q_1$ and $Q_4$ are saddles.}}
\label{VmassModel1_la_1_mu_0_5}
\end{center}
\end{figure}

In the figure \ref{VmassModel1_la_1_mu_0_5} we show the Poincar\'e
(global) phase plane of Model 1 for the parameter values
$\la_1=1.0$ and $\mu_1=0.5$. The points at infinity $Q_2$ and
$Q_3$ are sources, whereas $Q_1$ and $Q_4$ are saddles. Thus, the
scalar field dominated solution A is the global attractor for both
finite and infinite regions for the choice of parameters in the
range $\la_1\left(\mu_1-\la_1\right)<3.$

\subsubsection{Cosmological implications and discussion: Model 1}
\label{cosmimpl}

Having performed a complete phase-space analysis we can discuss
the corresponding cosmological behavior. A general remark is that
this behavior is radically different from the corresponding
quintessence scenarios with the same potentials and mass-functions
\cite{Amendola:1999er,Anderson:1997un,Casas:1991ky,interacting2,quirosvamps,Zhang:2005rg,Amendola:2001rc,Comelli2003}.
Additionally, a common feature of almost all the phantom models
previously studied is the existence of attractors with
$w_{\phi}\leq-1$ in the whole phase-space \cite{Guo:2004vg,phant},
and thus, independently of the specific scenario and of the
imposed initial conditions, the universe always lies below the
phantom divide, as it is expected for phantom cosmology. This
global behavior is not always realized  in the case of
exponentially dependent dark-matter mass, and additional
constraints must be imposed.

Apart form acquiring acceleration, in this work we examine whether
the above constructed varying dark-matter-mass models can solve or
alleviate the coincidence problem. Thus, assuming as usual that
the present universe is already at a late-time attractor, we
calculate $\Omega_\phi$ in all stable fixed points, and if
$0<\Omega_\phi<1$ then the coincidence problem is solved since
$\Omega_\phi$ and $\Omega_{DM}$ will be of the same order of
magnitude as suggested by observations. On the contrary,
$\Omega_\phi=1$ corresponds to a universe completely dominated by
dark energy, while $\Omega_\phi=0$ (that is $\Omega_{DM}=1$ ) to
one completely dominated by dark matter, both in contrast with
observations.

Finally, we mention that as long as the interaction responsible
for the varying dark-matter particle mass is not too strong, the
standard cosmology can be always recovered. On the other hand,
since we assume that the universe is currently at an attractor,
its state is independent of the initial conditions. Thus, we can
switch on the interaction and consider as initial conditions the
end of the known epochs of standard Big Bang cosmology, in order
to avoid disastrous interference.

For the model 1, the critical point B is unstable, and therefore
it cannot be a late-time cosmological solution. The only relevant
critical point is A, which is a stable fixed point for
$\la_1\left(\mu_1-\la_1\right)<3$. As can be seen from table
\ref{stability1}, it corresponds to an accelerating universe with
$\Omega_\phi=1$, that is to complete dark-energy domination. Thus,
this specific cosmological solution cannot solve the coincidence
problem. Furthermore, the fact that $w_{tot}$ is not only less
than $-1/3$, as required by the acceleration condition, but it is
always less than $-1$, leads to $\dot{H}>0$ at all times.
Therefore, this solution corresponds to a super-accelerating
universe \cite{Das:2005yj}, that is with a permanently increasing
$H$, resulting to a Big Rip. This behavior is common in phantom
cosmology \cite{Briscese:2006xu,phant}. Another interesting
feature of exponential potential and exponentially-dependent
dark-matter particle mass phantom model is the existence of past
big-rip singularities at infinity (either $Q_{1,4}$ or $Q_{2,3}$).

A remarkable feature of this model, as well as  of Model 3, is
that if there exist scaling solutions, then, for a wide region of
the parameter space, the stable manifold of the corresponding
critical point marks the basin of attraction of either a phantom
attractor or an unphysical attracting state. Thus, there exist an
open set of orbits of the physical region that converge to an
unphysical state instead to a phantom solution. This behavior was
revealed analytically and it was confirmed through numerical
elaboration, and seems to be typical in the case of
exponentially-dependent dark-matter mass in the phantom framework.
To avoid dealing with unphysical states, we can either restrict
the physical portion of the phase-space to the region above the
stable manifold of the scaling solutions, or restrict the
parameter-space itself. In both cases we obtain an additional
constraint, that was not present in previous studies of phantom
cosmology \cite{Guo:2004vg,Guo:2004xx,ChenSaridakis}, which
further weakens the applicability of the model.

In summary, Model 1, that is an exponential potential and an
exponentially-dependent dark-matter particle mass, cannot act as a
candidate for solving the coincidence problem.

\subsection{Model 2: Power-law potential and power-law-dependent dark-matter particle mass}
\label{mod2a}

Inserting the auxiliary variables (\ref{auxilliary}) into the
equations of motion (\ref{rhodmeom}), (\ref{phiddot}), (\ref{FR1})
and (\ref{FR2}), we result in the following autonomous system:
\begin{eqnarray}
x'&=&-
3x+\frac{3}{2}x (1-x^2-y^2)-\frac{\lambda_2y^2 z}{2}-\frac{\mu_2}{2}z(1+x^2-y^2)\nonumber\\
y'&=&\frac{3}{2}y (1-x^2-y^2)-\frac{\lambda_2
xyz}{2}\nonumber\\
z'&=&-xz^2\label{autonomous2}.
\end{eqnarray}

\subsubsection{Finite analysis}

The real and physically meaningful critical points are
\begin{eqnarray}
&&\left(x_{c4}=0,\ y_{c4}=0,\ z_{c4}=0\right), \nonumber\\
&&\left(x_{c5}=0,\ y_{c5}=1,\ z_{c5}=0\right),
\end{eqnarray}
and in table \ref{stability2} we present the necessary conditions
for their existence.
\begin{table*}[t]
\begin{center}\caption[The real and physically
meaningful critical points of Model 2 and their
behavior.]{\label{stability2} The real and physically meaningful
critical points of Model 2 and their behavior.}
{\begin{tabular}{|c|c|c|c|c|c|c|c|c|} \hline
 Cr. P.& $x_c$ & $y_c$ & $z_c$ & Existence & Stable  & $\Omega_\s$ &  $w_{tot}$ & Acceleration   \\
  &  &   &   &   &  manifold &   &    &     \\
\hline \hline
 D&  $x_{c4}$ & $y_{c4}$ &$z_{c4}$ & Always  &1-D  &  0
 &0
    & Never  \\
E&  $x_{c5}$ & $y_{c5}$ &$z_{c5}$ & Always  & 2-D  & 1 & -1
& Always \\
\hline
\end{tabular}}
\end{center}
\end{table*}

In this case, the critical points are non-hyperbolic, that is
there exists always at least a zero eigenvalue. We mention that
for non-hyperbolic critical points the result of linearization
cannot be applied in order to investigate the local stability of
the system (the system can be unstable to small perturbations on
the initial condition or to small perturbations on the parameters)
\cite{arrowsmith,wiggins,REZA}. However, it is possible to get
information about the existence and the dimensionality of the
stable manifold by applying the center manifold theorem
\cite{arrowsmith}. Doing so we deduce that the dimensionality of
the local stable manifold is 1 and 2 for D and E respectively. In
particular, the stable manifold of D is tangent, at the critical
point, to the $x$-axis, while the stable manifold of E is tangent,
at the critical point, to the  $xy$-plane. The existence of an 1D
stable manifold for D, implies that the orbits asymptotic to D as
$t\rightarrow -\infty$ are contained in either an unstable or
center manifold (each one of dimensionality 1, that is a curve).
There are some exceptional orbits converging to D as $t\rightarrow
+\infty$, but these have a zero measure. On the other hand, the
fact that E has a 2D stable manifold implies that there exists a
non-zero-measure set of orbits that converges to E as
$t\rightarrow +\infty$. Finally, there are some exceptional orbits
contained in its center manifold that cannot be classified by
linearization.

\subsubsection{Stability of de Sitter solution for Power-law
potential and power-law-dependent dark-matter particle mass.}

The singular point $E$ represents the de Sitter solution for
Power-law potential and power-law-dependent dark-matter particle
mass. In this section we will analyze the stability of the center
manifold of $E$ for the vector field \eqref{autonomous2}.

\begin{prop}\label{centerE} For $\la_2< 0,$ the singular point $E:\left(x_{c5}=0,\ y_{c5}=1,\ z_{c5}=0\right)$ of the system
\eqref{autonomous2} is locally asymptotically stable. For
$\la_2>0,$ $E$ is locally unstable (saddle type). For $\la_2=0,$
$E$ is stable but not asymptotically stable.
\end{prop}

In order to determine the local center manifold of
\eqref{autonomous2} at the origin we have to transform the system
into a form suitable for the application of the center manifold
theorem (see section \ref{sectionCM} for a summary of the
techniques involved in the proof).

{\bf Proof}.

{\bf Case 1}. Let us assume that $\la_2\neq 0.$

In order to translate $E$ to the origin and transforming the
linear part of the resulting vector field into its Jordan
canonical form, we define new variables
$(u,v_1,v_2)\equiv\mathbf{x}$, by the equations \be
u=-\frac{z\la_2}{6},\,v_1=y-1,\, v_2=x+\frac{z
\la_2}{6}\nonumber\ee so that \be \left(\begin{array}{c}
u'\\v_1'\\v_2'
\end{array}\right)=\left(\begin{array}{ccccc} 0& 0 &0 \\
0& -3 &0
\\  0& 0 & -3
\end{array}\right)\left(\begin{array}{c}u\\v_1\\v_2
\end{array}\right)+\left(\begin{array}{c}f(u,v_1,v_2)\\g_1(u,v_1,v_2)\\g_2(u,v_1,v_2)
\end{array}\right)\label{Ecenter2}
\ee where

$f(u,v_1,v_2)=\frac{6 u^2 (u+v_2)}{\la_2},$
$g_1(u,v_1,v_2)=\frac{3}{2} (v_1+1)
\left(u^2-v_2^2\right)-\frac{3}{2} v_1^2 (v_1+3),$ and
$g_2(u,v_1,v_2)=-\frac{3 (\la_2-2 \mu_2+4) u^3}{2 \la_2}-\frac{3
v_2 (3 \la_2-4 \mu_2+4) u^2}{2
\la_2}+\left(\left(\frac{3}{2}-\frac{3 \mu_2}{\la_2}\right)
   v_1^2+\left(3-\frac{6 \mu_2}{\la_2}\right) v_1+v_2^2 \left(\frac{3 \mu_2}{\la_2}-\frac{9}{2}\right)\right) u-\frac{3 v_2^3}{2}-\frac{3 v_1^2 v_2}{2}-3 v_1
   v_2.$
\\
The system \eqref{Ecenter2} is written in diagonal form
\begin{align}
u'  &  =Cu+f\left(  u,\mathbf{v}\right) \nonumber\\
\mathbf{v}'  &  =P\mathbf{v}+\mathbf{g}\left(  u,\mathbf{v}\right)
, \label{Ecenter3}
\end{align}
where $\left(  u,\mathbf{v}\right)
\in\mathbb{R}\times\mathbb{R}^{2},$ $C$ is the zero $1\times1$
matrix, $P$ is a $2\times 2$ matrix with negative eigenvalues and
$f,\mathbf{g}$ vanish at $\mathbf{0}$ and have vanishing
derivatives at $\mathbf{0.}$ The center manifold theorem
\ref{existenceCM} asserts that there exists a 1-dimensional
invariant local center manifold $W^{c}\left( \mathbf{0}\right) $
of \eqref{Ecenter3} tangent to the center subspace (the
$\mathbf{v}=\mathbf{0}$ space) at $\mathbf{0}.$ Moreover,
$W^{c}\left( \mathbf{0}\right)  $ can be represented as
\[
W^{c}\left(  \mathbf{0}\right)  =\left\{  \left(
u,\mathbf{v}\right)
\in\mathbb{R}\times\mathbb{R}^{2}:\mathbf{v}=\mathbf{h}\left(
u\right), \left\vert u\right\vert <\delta\right\} ;
\mathbf{h}\left(  0\right) =\mathbf{0}, D\mathbf{h}\left( 0\right)
=\mathbf{0}
\]
for $\delta$ sufficiently small (see definition \ref{CMdef}). The
restriction of (\ref{Ecenter3}) to the center manifold is (see
definition \ref{vectorfieldCM})
\begin{equation}
u'=f\left( u,\mathbf{h}\left(  u\right)  \right)  . \label{Erest}
\end{equation}
According to Theorem \ref{stabilityCM}, if the origin $u=0$ of
\eqref{Erest} is stable (asymptotically stable) (unstable) then
the origin of \eqref{Ecenter3} is also stable (asymptotically
stable) (unstable). Therefore, we have to find the local center
manifold, i.e., the problem reduces to the computation of
$\mathbf{h}\left( u\right).$

Substituting $\mathbf{v}=\mathbf{h}\left(  u\right)  $ in the
second component of \eqref{Ecenter3} and using the chain rule,
$\mathbf{v }'=D\mathbf{h}\left(  u\right)  u'$, one can show that
the function $\mathbf{h}\left( u\right)  $ that defines the local
center manifold
satisfies%
\begin{equation}
D\mathbf{h}\left(  u\right)  \left[  f\left(  u,\mathbf{h}\left(
u\right) \right)  \right]  -P\mathbf{h}\left(  u\right)
-\mathbf{g}\left( u,\mathbf{h}\left(  u\right)  \right)  =0.
\label{Eh}
\end{equation}
According to Theorem \ref{approximationCM}, equation \eqref{Eh}
can be solved approximately by using an approximation of
$\mathbf{h}\left(  u\right)  $ by a Taylor series at $u=0.$ Since
$\mathbf{h}\left(  0\right)  =\mathbf{0\ } $and $D\mathbf{h}\left(
0\right)  =\mathbf{0},$ it is obvious that $\mathbf{h}\left(
u\right)  $ commences with quadratic terms. We substitute
\[
\mathbf{h}\left(  u\right)  =:\left[
\begin{array}
[c]{c}%
h_{1}\left(  u\right) \\
h_{2}\left(  u\right)
\end{array}
\right]  =\left[
\begin{array}
[c]{c}%
a_{1}u^{2}+a_{2}u^{3}+O\left(  u^{4}\right) \\
b_{1}u^{2}+b_{2}u^{3}+O\left(  u^{4}\right)
\end{array}
\right]
\]
into (\ref{Eh}) and set the coefficients of like powers of $u$
equal to zero to find the unknowns $a_{1},b_{1},...$.

We find that the non-zero coefficients are $$a_2=\frac{1}{2},\,
b_3= -\frac{2}{\la_2},$$ Therefore, \eqref{Erest} yields \be
u'=\frac{6 u^3}{\la_2}-\frac{12
u^5}{\la_2^2}+O\left(u^6\right).\label{Erest1}\ee Neglecting the
error terms, this  is a gradient-like equation (i.e., $u'=-\nabla
U(u)$) with potential $U(u)=\frac{2 u^6}{\la_2^2}-\frac{3 u^4}{2
\la_2}$ for which the origin is a degenerate minimum provided
$\lambda_2<0$ and a degenerated maximum provided $\lambda_2>0$.
Thus, for $\lambda_2<0,$ the origin $u=0$ of \eqref{Erest1} is
locally asymptotically stable. Hence, the origin
$\mathbf{u}=\mathbf{0}$ of the full three-dimensional system is
asymptotically stable. For $\la_2>0$ the origin is locally
unstable (saddle type)

{\bf Case 2}.
Let us assume that $\la_2=0.$ In this case in order to translate
$E$ to the origin and reducing the linear part of the vector field
to its Jordan canonical form, we define new variables the $u=z,\,
v_1=x,\, v_2=y-1.$
Thus, the system \eqref{autonomous2} reduces to

so that \begin{eqnarray} & \left(\begin{array}{c}
u'\\v_1'\\v_2'
\end{array}\right)=\left(\begin{array}{ccccc} 0& 0 &0 \\
0& -3 &0
\\  0& 0 & -3
\end{array}\right)\left(\begin{array}{c}u\\v_1\\v_2
\end{array}\right)\nonumber \\ & +\left(\begin{array}{c}-u^2 v_1\\\frac{1}{2} \left(-3 v_1^3-u \mu_2 v_1^2-3 v_2 (v_2+2) v_1+u v_2 (v_2+2)
   \mu_2\right)\\-\frac{3}{2} \left((v_2+1) v_1^2+v_2^2 (v_2+3)\right)
\end{array}\right)\label{Ecenter2la20}
\end{eqnarray}

The equations for the center manifold of the origin
\eqref{Eh} reduces to

\begin{align} & u \mu_2 h_2 (h_2+2)=h_1
\left(-2 h_1' u^2+\mu_2 h_1 u+3
   h_1^2+3 h_2 (h_2+2)+6\right),\nonumber\\
   & 3 (h_2+1) \left(h_1^2+h_2
   (h_2+2)\right)=2 u^2 h_1 h_2'.\label{centerEla_0}
\end{align}
We obtain, using a Taylor series at $u=0,$  that the solution of
\eqref{centerEla_0} satisfying $\mathbf{h}(0)=0, D\mathbf{h}\left(
0\right) =\mathbf{0}$ is the trivial solution to arbitrary order.
This means that the center manifold of $E$ is a small segment
contained in the z-axis.

In order to examine the stability of the origin for the flow of \eqref{Ecenter2la20} we proceed as follows.
Using spherical coordinates
\begin{equation}u=r \cos \varphi  \sin \theta,v_1=r \sin \theta \sin \varphi ,v_2=r \cos \theta \label{sphcoord}
\end{equation} and
taking the limit $r\rightarrow 0$ the angular equations
$\theta',\varphi'$ becomes \be \varphi' \to -3 \cos \varphi
\sin \varphi ,\, \theta' \to 3 \cos \theta \cos ^2\varphi
\sin \theta
   .\label{angularla20}\ee

Solving the approximate equations \eqref{angularla20} we obtain
\be\theta(\tau)=\tan ^{-1}\left(e^{2 c_2} \sqrt{e^{6 \tau }+e^{4 c_1}}\right),\, \varphi(\tau)=\tan ^{-1}\left(e^{2 c_1-3 \tau }\right),\label{approxsolsangularla20}\ee where $c_1$ and $c_2$ are integration constants.

By Taylor expanding the radial equation around $r=0$ we obtain the
equation \be r'=-\frac{3}{4} \left(-2 \cos (2 \varphi ) \sin
^2\theta+\cos (2 \theta )+3\right)
r+O\left(r^2\right).\label{approxrla20} \ee

By substituting the first order solution
\eqref{approxsolsangularla20} into the equation
\eqref{approxrla20} and solving the resulting equation we obtain
\be r(\tau)= e^{-3 \tau } \sqrt{1+e^{4 \left(c_1+c_2\right)}+e^{6
\tau +4 c_2}} c_3\label{approxsolrla20},\ee where $c_3$ is an
integration constant. Substituting \eqref{approxsolsangularla20}
and  \eqref{approxsolrla20} in \eqref{sphcoord} and taking the
limit as $\tau\rightarrow +\infty$ we obtain $u\rightarrow u_0,
v_1\rightarrow 0, v_2\rightarrow 0$ where $u_0=u(0).$ Let be
$\epsilon>0$ an arbitrary number. Then there exists a $\delta>0,$
such that $\delta<\epsilon.$ Let us consider the solution with
initial value $u(0)=u_0,v_1(0)=v_{1 0},v_2(0)=v_{2 0},$ with
$u_0^2+v_{1 0}^2+v_{20}^2<\delta^2.$ Since $u\rightarrow u_0,$
satisfying $|u_0|<\delta,$ then the solution, ${\bf x}(\tau,{\bf
x}_0)$ passing through ${\bf x}_0=\left(u_0, v_{1
0},v_{20}\right)$ at $\tau=0,$ satisfies $\|{\bf x}(\tau,{\bf
x}_0)\|<\epsilon,$  for $\tau$ arbitrarily large. In this way we
prove the stability (but not asymptotic stability) of the $E.$
$\blacksquare$

In summary, we indeed find that the center manifold of E attracts
an open set of orbits provided $\la_2\leq 0$. On the other hand, if
$\la_2> 0$ the orbits located near the center manifold of E blow
up in a finite time. This point does not allow for a solution of
the coincidence problem (it always possesses $\Omega_\phi=1$).

\begin{figure}[ht]
\begin{center}
\hspace{0.4cm}
\includegraphics[width=8cm,height=7cm]{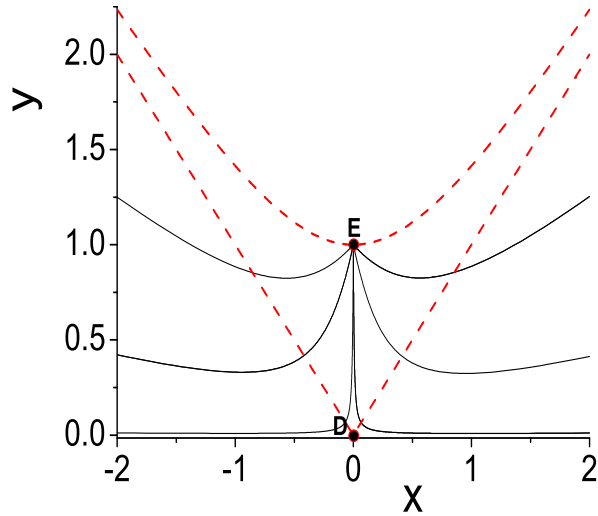}
\caption{{ xy-projection of the phase-space of Model 2, for the
parameter values $\la_2=-0.5$ and $\mu_2=0.5$. The critical point
E (representing de Sitter solutions) is the attractor of the
system. The dashed (red) curves bound the physical part of the
phase space, that is corresponding to
 $0\leq \Omega_\phi\leq1$. [Taken from \cite{PLB2010}; published with permission of Elsevier
B.V.].}} \label{Fig3}
\end{center}
\end{figure}

Numerical investigation reveals the above features.
 In fig.~\ref{Fig3} we depict orbits projected in the xy-plane,
as they arise from numerical evolution in the case of $\la_2=-0.5$
and $\mu_2=0.5$.

\subsubsection{Analysis at infinity}

Introducing the Poincar\'e variables
\begin{equation}\label{TransfVmass2}
x_r=\rho \cos \theta  \sin \psi ,y_r=\rho \sin \theta \sin \psi
,z_r=\rho \cos \psi
\end{equation} where $\rho=\frac{r}{\sqrt{1+r^2}},\, r=\sqrt{x^2+y^2+z^2}.$  Thus, the points at ``infinite''
($r\rightarrow+\infty$) are those having $\rho\rightarrow 1.$
Since $y,z$ are required to be nonnegative, $\theta$ varies in
$[0,\pi],$ and $\psi$ varies in $[0,\frac{\pi}{2}].$

Performing the transformation \eqref{TransfVmass2}, the system
\eqref{autonomous2} becomes

\begin{align}
&\rho'= -\frac{3}{2} \rho ^3 \sin ^4\psi+\frac{1}{4}
(\lambda_2-\mu_2) \rho ^3 \cos (3 \theta ) \cos \psi \sin
   ^3\psi+\nonumber\\&
   +\frac{3}{2} \rho  \left(\rho ^2-1\right) \cos (2 \theta ) \sin ^2\psi+\nonumber\\&
   -\frac{1}{16} \rho
   \left((\lambda_2-3 \mu_2+4) \rho ^2+4 \mu_2\right) \cos \theta \sin (2 \psi )+\nonumber\\&
   +\frac{1}{32} (\lambda_2+\mu_2-4) \rho ^3 \cos \theta \sin (4 \psi ),\label{Vmassinfty2a}\\
&   \theta'=\frac{(\lambda_2-\mu_2) (\sin (3 \theta )-\sin \theta
) \sin (2 \psi ) \rho ^2}{8 \left(\rho
   ^2-1\right)}+\nonumber\\&+\frac{1}{2} (6 \cos \theta+\mu_2 \cot \psi ) \sin \theta\label{Vmassinfty2b},\\
   &   \psi'=-\frac{1}{2} \cos \psi  \left(3 \sin \psi  \left(\frac{\rho ^2 \sin ^2\psi }{1-\rho ^2}
   +\cos (2 \theta )\right)+\right. \nonumber\\ & \left. +\cos
   \theta \cos \psi \left(\frac{\rho ^2 (\lambda_2+(\mu_2-\lambda_2) \cos (2 \theta )-2) \sin
   ^2\psi }{1-\rho ^2}+\mu_2\right)\right)\label{Vmassinfty2c}.
\end{align}

In the limit $\rho\rightarrow 1,$ the leading terms in
\eqref{Vmassinfty2a}-\eqref{Vmassinfty2c} are

\begin{align}
&\rho'\rightarrow -\frac{3}{2} \sin ^4\psi+\frac{1}{4}
(\lambda_2-\mu_2) \cos (3 \theta ) \cos \psi \sin ^3(\psi
   )+\nonumber\\&-\frac{1}{16} (\lambda_2+\mu_2+4) \cos \theta \sin (2 \psi )+
   \frac{1}{32} (\lambda_2+\mu_2-4) \cos \theta \sin (4 \psi ),\label{Vmassinfty2d}\\
&   \theta'\rightarrow
\frac{(\lambda_2-\mu_2) (\sin \theta -\sin (3 \theta )) \sin (2 \psi )}{8 \left(1-\rho ^2\right)}\label{Vmassinfty2e},\\
&   \psi'\rightarrow \frac{\cos \psi  (\cos \theta
(-\lambda_2+(\lambda_2-\mu_2) \cos (2 \theta )+2) \cos \psi -3
   \sin \psi ) \sin ^2\psi }{2 \left(1-\rho ^2\right)}\label{Vmassinfty2f}.
\end{align}

The radial equation does not contain the radial coordinate, so the
singular points  can be obtained using the angular equations only.
Setting $\theta'=0,\psi'=0$, we obtain the singular points which
are listed in table \ref{Vmassinfty2}. The stability of these
points is studied by analyzing first the stability of the angular
coordinates and then deducing, from the sign of equation
\eqref{Vmassinfty2d}, the stability on the radial direction.

\begin{landscape}

\begin{table*}[ht]
\begin{center}\caption{\label{Vmassinfty2} \small Asymptotic singular points  of the system
\eqref{autonomous2} (case 2) and their stability. We use the
notations $\alpha=\frac{3\sqrt{2}}{\sqrt{22-4\la_2+\la_2^2}}$ and
$\beta=\frac{3}{\sqrt{13-4\mu_2+\mu_2^2}},$ $\epsilon=\text{sign}
(\la_2-\mu_2),\delta=\text{sign}(-26+4\la_2+\la_2^2),\eta=\text{sign}(-11-4\mu_2+\mu_2^2),$
and $\mu_-=\frac{1}{3} \left(5-\sqrt[3]{\frac{2}{245-9
\sqrt{741}}}-\sqrt[3]{\frac{1}{2} \left(245-9
\sqrt{741}\right)}\right)\approx -0.47.$ NH stands for
nonhyperbolic.} {\small\begin{tabular}{ccccc} \hline \hline
 Cr. P & Coordinates & Eigenvalues & $\rho'$ & Stability\\
    &  $\theta,\psi, x_r, y_r, z_r$ &  &  & \\
\hline $Q_5$ & $0, 0, 0, 0 , 1$ & $0,0$ & 0 & NH; 3D center manifold \vspace{0.2cm}\\
\hline $Q_6$ & $0, \frac{\pi}{2}, 1, 0, 0$ & $0,+\infty$ & $-\frac{3}{2}$& { NH; 2D unstable manifold} \\
\vspace{-0.35cm}\\
\hline $Q_7$ & $\pi, \frac{\pi}{2}, -1, 0, 0$ & $0,+\infty$ & $-\frac{3}{2}$& {NH; 2D unstable manifold} \\
\vspace{-0.35cm}\\
\hline $Q_8$ & $\frac{\pi}{4}, \frac{\pi}{2}, \frac{\sqrt{2}}{2}, \frac{\sqrt{2}}{2}, 0$ & $0,+\infty$ & $-\frac{3}{2}$& { NH; 2D unstable manifold} \\
\vspace{-0.35cm}\\
\hline $Q_9$ & $\frac{3\pi}{4}, \frac{\pi}{2}, -\frac{\sqrt{2}}{2}, \frac{\sqrt{2}}{2}, 0$ & $0,+\infty$ & $-\frac{3}{2}$& {NH; 2D unstable manifold}\\
\vspace{-0.35cm}\\
\hline $Q_{10}$ & $\frac{\pi}{4}, \cos^{-1}(\alpha),
\frac{\alpha|\la_2-2|}{6}, \frac{\alpha|\la_2-2|}{6},
\alpha$  & $\begin{array}{c}\epsilon\infty, \delta \infty \, \text{if}\, \la_2>2\\
\epsilon\infty, -\infty \,\text{if}\, \la_2<2\end{array}  $ & $<0$
&
${\begin{array}{c} \text{source if}\, \la_2>2+\sqrt{30}\\
\text{and}\, \mu_2<\la_2\\
\text{saddle, otherwise}\end{array}} $ \\
\vspace{-0.35cm}\\
\hline $Q_{11}$ & $\frac{3\pi}{4}, \cos^{-1}(\alpha),
-\frac{\alpha|\la_2-2|}{6}, \frac{\alpha|\la_2-2|}{6},
\alpha$ & $\begin{array}{c}-\epsilon\infty, - \infty \, \text{if}\, \la_2>2\\
-\epsilon\infty, \delta\infty \, \text{if}\, \la_2<2\end{array}  $
&
$\begin{array}{c} <0 \, \text{if}\, \la_2<-1\\
>0 \, \text{if}\, \la_2>-1,\\ \text{and}\, \la_2\neq 2\end{array}$ &
${\begin{array}{c} \text{sink, if}\, \la_2>-1,\\ \text{and}\, \la_2\neq 2, \\
\text{and}\, \mu_2<\la_2\\
\text{saddle, otherwise}\end{array} } $  \\
\vspace{-0.35cm}\\
\hline $Q_{12}$ & $0, \cos^{-1}(\beta), \frac{\beta|\mu_2-2|}{3},
0,
\beta$ &  $\begin{array}{c}-\epsilon\infty, - \infty \, \text{if}\, \mu_2<2\\
-\epsilon\infty, \eta\infty \, \text{if}\, \mu_2>2\end{array}  $ &
$<0$ &
${ \begin{array}{c} \text{source, if}\, \mu_2 >2+\sqrt{15},\\ \text{and}\, \la_2< \mu_2, \\
\text{saddle, otherwise}\end{array} } $   \\
\vspace{-0.35cm}\\
\hline $Q_{13}$ & $\pi, \cos^{-1}(\beta),
-\frac{\beta|\mu_2-2|}{3}, 0,
\beta$ &  $\begin{array}{c}\epsilon\infty, \eta \infty \, \text{if}\, \mu_2<2\\
\epsilon\infty, -\infty \, \text{if}\, \mu_2>2\end{array}  $ &
$\begin{array}{c}<0 \,\text{if}\,\mu_2<\mu_-\\>0
\,\text{if}\,\mu_2>\mu_-, \mu\neq 2 \end{array}$ & ${
\begin{array}{c} \text{source, if}\,
\mu_2<\la_2\leq 2-\sqrt{15}, \\ \text{or}\, \mu_2<2-\sqrt{15}<\la_2\\
\text{sink, if}\,
\la_2<\mu_2, \mu_2>2\\ \text{or}\, \la_2\leq \mu_-<\mu_2<2\\
\text{or}\, \mu_-<\la_2<\mu_2<2;\\
\text{saddle, otherwise}\end{array} } $  \\
\hline \hline
\end{tabular}}
\end{center}
\end{table*}

\end{landscape}

In order to perform the numerical experiments for the system
\eqref{Vmassinfty2a}-\eqref{Vmassinfty2c} it is useful to rewrite
the system in the cartesian coordinates $x_r, y_r, z_r.$ The
system \eqref{Vmassinfty2a}-\eqref{Vmassinfty2c} becomes

\begin{align}
& x_r'=\frac{1}{2} \left(\left(\mu_2-(\mu_2-2) x_r^2\right)
z_r^3+3 x_r \left(-x_r^2+y_r^2+1\right)
   z_r^2+\right. \nonumber\\ & \left.   -\left(\lambda_2 \left(1-2 x_r^2\right) y_r^2+\mu_2 \left(x_r^2-1\right) \left(2 y_r^2-1\right)\right)
   z_r\right. \nonumber\\ & \left. +3 x_r \left(x_r^2+y_r^2-1\right) \left(2 y_r^2+1\right)\right),
\nonumber\\
& y_r'=   \frac{1}{2} y_r \left(-(\mu_2-2) x_r z_r^3-3
   \left(x_r^2-y_r^2+1\right) z_r^2+\right. \nonumber\\ & \left.  +(\lambda_2-\mu_2) x_r \left(2 y_r^2-1\right) z_r+\right. \nonumber\\ & \left. +3
   \left(x_r^2+y_r^2-1\right) \left(2 y_r^2-1\right)\right),
\nonumber\\
& z_r'= \frac{1}{2} z_r \left(\left(6 y_r^2-3 z_r^2+3\right)
x_r^2+\right. \nonumber\\ &  \left.+z_r
   \left(2 (\lambda_2-\mu_2) y_r^2-(\mu_2-2) z_r^2+\mu_2-2\right) x_r+\right. \nonumber\\ &  \left.+3 y_r^2 \left(2
   y_r^2+z_r^2-1\right)\right)\label{Model2cartesian}
\end{align}
where we have used the time re-scaling $$\mathrm{d}\tau\rightarrow
\frac{\mathrm{d}\tau}{1-\rho^2}$$ which leave invariant the orbits
of the phase-space and the time direction (see theorem
\ref{ext2}).

By an explicit calculation we find that that center manifolds of
$Q_6$ and $Q_8$ for the flow of \eqref{Model2cartesian} is the arc
$x_r=\sqrt{1-y_r^2}$ and that the center manifolds of $Q_7$ and
$Q_9$ is the arc $x_r=-\sqrt{1-y_r^2}.$ These center manifolds are
unstable. Thus, the singular points $Q_{6,7,8,9}$ are local
sources.

To examine the stability of $Q_5$ we use Normal forms
calculations. Let us assume $\mu_2\neq 0.$

First, by introducing the linear change of coordinates
$$u_1=\frac{x_r}{\mu_2}, u_2=z_r-1, u_3=y_r$$ the system around $Q_5$
becomes

\begin{eqnarray}
&&u_1'=\mu_2 u_1^2+u_2+\frac{3}{2} u_2 (2 u_1+u_2)+u_3^2
   \left(1-\frac{\lambda_2}{2 \mu_2}\right)+{\cal O}(3),\nonumber\\
&&u_2'=-u_1 u_2 (\mu_2-2) \mu_2+{\cal O}(3),\nonumber\\
   && u_3'=-\frac{1}{2}
   u_3 (6 u_2+u_1 (\lambda_2-2) \mu_2)+{\cal O}(3)\label{Q5secondorder}
\end{eqnarray}

Observe that the phase space is compact since
$$u_1\in\left[-\frac{1}{|\mu_2|},\frac{1}{|\mu_2|}\right],\, u_2\in[-1,1], u_3\in[0,1].$$

The linear part of the vector field \eqref{Q5secondorder} is given
by $$J=\left(
\begin{array}{ccc}
 0 & 1 & 0 \\
 0 & 0 & 0 \\
 0 & 0 & 0
\end{array}
\right).$$

Let us consider the linear operator ${\bf L_J^{(2)}}$ associated
to $J$ that assigns to ${\bf h(u)}\in H^2$ the Lie bracket of the
vector fields ${\bf A u}$ and ${\bf h(u)}$:
\ben {\bf L_J^{(2)}}: H^2& &\rightarrow H^2\nonumber\\
     {\bf h}  & & \rightarrow  {\bf L_J} {\bf h (u)}={\bf D h(u)} {\bf J u}- {\bf J h(u)}.
\een where $H^2$ the real vector space of vector fields whose
components are homogeneous polynomials of degree 2.  The canonical
basis for the real vector space of 3-dimensional vector fields
whose components are homogeneous polynomials of degree 2 is given
by
\begin{align}
&H^2=\nonumber\\
&\text{span}\left\{\left(
\begin{array}{c}
 u_1^2  \\
 0  \\
 0
\end{array}
\right),\, \left(
\begin{array}{c}
 u_1 u_2  \\
 0  \\
 0
\end{array}
\right),\, \left(
\begin{array}{c}
 u_1 u_3  \\
 0  \\
 0
\end{array}
\right),\, \left(
\begin{array}{c}
 u_2^2  \\
 0  \\
 0
\end{array}
\right),\, \left(
\begin{array}{c}
 u_2 u_3  \\
 0  \\
 0
\end{array}
\right),\right. \nonumber \\ & \left. \ \ \ \ \  \left(
\begin{array}{c}
 u_3^2  \\
 0  \\
 0
\end{array}
\right),\,\left(
\begin{array}{c}
 0  \\
 u_1^2  \\
 0
\end{array}
\right),\, \left(
\begin{array}{c}
 0  \\
 u_1 u_2  \\
 0
\end{array}
\right),\, \left(
\begin{array}{c}
 0  \\
 u_1 u_3  \\
 0
\end{array}
\right),\, \left(
\begin{array}{c}
 0\\
 u_2^2    \\
 0
\end{array}
\right),\right. \nonumber \\ & \left. \ \ \ \ \  \left(
\begin{array}{c}
 0  \\
 u_2 u_3  \\
 0
\end{array}
\right),\, \left(
\begin{array}{c}
 0  \\
 u_3^2  \\
 0
\end{array}
\right)\,\left(
\begin{array}{c}
 0  \\
 0  \\
 u_1^2
\end{array}
\right),\, \left(
\begin{array}{c}
 0  \\
 0  \\
 u_1 u_2
\end{array}
\right),\, \left(
\begin{array}{c}
 0  \\
 0 \\
 u_1 u_3
\end{array}
\right), \right. \nonumber\\ & \left. \ \ \ \ \ \ \ \ \ \ \ \ \ \
\ \ \ \ \ \ \ \ \ \  \left(
\begin{array}{c}
 0\\
 0    \\
 u_2^2
\end{array}
\right),\, \left(
\begin{array}{c}
 0  \\
 0 \\
 u_2 u_3
\end{array}
\right),\, \left(
\begin{array}{c}
 0  \\
 0  \\
 u_3^2
\end{array}
\right)\right\}
\end{align}

By computing the action of ${\bf L_J^{(2)}}$ on each basis element
on $H^2$ we have

\begin{align}
&{\bf L_J^{(2)}}\left(H^2\right)=\nonumber\\
&\text{span}\left\{\left(
\begin{array}{c}
 u_1^2  \\
 0  \\
 0
\end{array}
\right),\, \left(
\begin{array}{c}
 u_1 u_2  \\
 0  \\
 0
\end{array}
\right),\, \left(
\begin{array}{c}
 u_1 u_3  \\
 0  \\
 0
\end{array}
\right),\, \left(
\begin{array}{c}
 u_2^2  \\
 0  \\
 0
\end{array}
\right),\, \left(
\begin{array}{c}
 u_2 u_3  \\
 0  \\
 0
\end{array}
\right), \right. \nonumber \\ & \left. \ \ \ \ \ \left(
\begin{array}{c}
 u_3^2  \\
 0  \\
 0
\end{array}
\right),\, \left(
\begin{array}{c}
 0  \\
 u_1 u_2  \\
 0
\end{array}
\right),\, \left(
\begin{array}{c}
 0\\
 u_2^2    \\
 0
\end{array}
\right),\, \left(
\begin{array}{c}
 0  \\
 u_2 u_3  \\
 0
\end{array}
\right),\,  \left(
\begin{array}{c}
 0  \\
 0  \\
 u_1 u_2
\end{array}
\right),\right. \nonumber\\ & \left. \ \ \ \ \ \ \ \ \ \ \ \ \ \ \
\ \ \ \ \ \ \ \ \ \left(
\begin{array}{c}
 0\\
 0    \\
 u_2^2
\end{array}
\right),\, \left(
\begin{array}{c}
 0  \\
 0 \\
 u_2 u_3
\end{array}
\right)\right\}\label{nonresonant2}.
\end{align}
Thus, the second order terms that are linear combinations of the
six vectors in \eqref{nonresonant2} can be eliminated
\cite{wiggins}. To determine the nature of the second order terms
that cannot be eliminated we must compute the complementary space
of \eqref{nonresonant2} which is
\begin{align}
&G^2=\nonumber\\
&\text{span}\left\{ \left(
\begin{array}{c}
 0  \\
 u_1^2  \\
 0
\end{array}
\right), \, \left(
\begin{array}{c}
 0  \\
 u_1 u_3  \\
 0
\end{array}
\right),\, \left(
\begin{array}{c}
 0\\
 u_3^2    \\
 0
\end{array}
\right),\, \left(
\begin{array}{c}
 0  \\
 0  \\
 u_1^2
\end{array}
\right),\, \left(
\begin{array}{c}
 0  \\
 0 \\
 u_1 u_3
\end{array}
\right), \right. \nonumber \\ & \left. \ \ \ \ \ \ \ \ \ \ \ \ \ \
\ \ \ \ \ \ \ \ \ \ \ \ \ \left(
\begin{array}{c}
 0  \\
 0  \\
 u_3^2
\end{array}
\right)\right\}
\end{align}
Hence, the normal form of the system \eqref{Q5secondorder} is
\begin{align}
&v_1'=v_2+{\cal O}(3),\nonumber\\
&v_2'= {\cal O}(3),\nonumber\\
   & v_3'=\frac{1}{2}
  (2-\lambda_2) \mu_2 v_1 v_3+{\cal O}(3)\label{Q5secondordernormal}
\end{align}
The solution of the truncated normal form passing through
$(v_1,v_2,v_3)=(v_{10},v_{20},v_{30})$ is
\begin{align}
& v_1=v_{10} +v_{20}\tau,\,  v_2=v_{20},\nonumber\\ & v_3=v_{30}
\exp\left[\frac{1}{4}(2-\lambda_2)\mu_2\tau(2 v_{10}
+v_{20}\tau)\right].\label{solQ5}
\end{align}
Since the phase space remains compact under the quadratic
transformation that reduces the original system
\eqref{Q5secondorder} to is normal form,
\eqref{Q5secondordernormal}, it follows that

\begin{enumerate}
\item For $v_{10}\neq0, v_{20}=v_{30}=0$ the system tends
asymptotically to the point $(v_{10},0,0);$ thus, such solutions
remains very close to the critical point $Q_5.$ \item For
$v_{20}=0,$ and $v_{10}(2-\lambda_2)\mu_2>0,$ the solution tends
asymptotically to the point $(v_{10},0,0);$ thus, such solutions
remains very close to the critical point $Q_5.$ \item If none of
the above conditions holds and if we assume that the solutions are
defined for all $\tau$-values, then, either $|v_1|\rightarrow
\infty$ or $|v_3|\rightarrow \infty$ asymptotically (i.e., for
$\tau\rightarrow\pm\infty$) in contradiction with the compactness
of the phase space. Thus, in this case, solutions should admit a
finite finite time interval of existence.
\end{enumerate}
In summary, the dynamical character $Q_5$ is very sensible to the
changes on the initial conditions. Thus, we argue that $Q_5$ is
unstable.

The basic observables \eqref{Omegas}, \eqref{wss}, \eqref{wtot}
and \eqref{decc} are given in terms of the Poincar\'e variables by

\begin{eqnarray}
&&\Omega_\phi\equiv\frac{\left(x_r-y_r\right)
   \left(x_r+y_r\right)}{x_r^2+y_r^2+z_r^2-1}=-\frac{2 \rho ^2 \cos (2 \theta ) \sin ^2(\psi )}{\cos (2 \psi ) \rho ^2-\rho ^2+2},\nonumber\\
&& w_\phi\equiv   \frac{x_r^2+y_r^2}{\left(x_r-y_r\right)
   \left(x_r+y_r\right)}=\sec (2 \theta
   ),\nonumber\\
&&w_{\text{tot}}\equiv
\frac{x_r^2+y_r^2}{x_r^2+y_r^2+z_r^2-1}=\frac{\rho ^2 (\cos (2 \psi )-1)}{\cos (2 \psi ) \rho ^2-\rho ^2+2},\nonumber\\
&&q\equiv \frac{4
   x_r^2+4 y_r^2+z_r^2-1}{2 \left(x_r^2+y_r^2+z_r^2-1\right)}=\frac{2 \cos (2 \psi ) \rho ^2-2 \rho ^2+1}{\cos (2 \psi ) \rho ^2-\rho ^2+2}\label{PoincareObsv2}.
\end{eqnarray}

\begin{table*}[ht]
\begin{center}\caption{\label{BasicObsVmassinfty2} Basic observables for the singular points  of the system
\eqref{autonomous2} (case 2). Solution types} {\small
\begin{tabular}{cccccc} \hline \hline
 Cr. P & $\Omega_\phi$ & $w_\phi$ & $w_{\text{tot}}$ & $q$ & Solution type\\
\hline $Q_5$ & 0 & 1& 0 & $\frac{1}{2}$ & dust-like \vspace{0.2cm}\\
\hline $Q_6$ & $-\infty$ & 1 & $-\infty$ & $-\infty$ & (unphys.) big-rip \\
\vspace{-0.35cm}\\
\hline $Q_7$ & $-\infty$ & 1 & $-\infty$ & $-\infty$ & (unphys.) big-rip \\
\vspace{-0.35cm}\\
\hline $Q_8$ & 0 & $-\infty$ & $-\infty$ & $-\infty$ &  big-rip \\
\vspace{-0.35cm}\\
\hline $Q_9$ &0 & $+\infty$ & $-\infty$ & $-\infty$ & big-rip \\
\vspace{-0.35cm}\\
\hline $Q_{10}$ & 0 & $-\infty$ & $-\frac{1}{18} (\la_2-2)^2$ &
$\frac{1}{12} \left(-\la_2^2+4 \la_2+2\right)$ &  Accelerated for \\
                &   &           &       &       &   $|\la_2-2|>\sqrt{6}$\\
                &   &           &       &       &   phantom for \\
                &   &           &       &       &   $|\la_2-2|>3 \sqrt{2}$\\
\vspace{-0.35cm}\\
\hline $Q_{11}$ & 0 & $+\infty$ & $-\frac{1}{18} (\la_2-2)^2$ &
$\frac{1}{12} \left(-\la_2^2+4 \la_2+2\right)$ & Accelerated for \\
                &   &           &       &       &   $|\la_2-2|>\sqrt{6}$\\
                &   &           &       &       &   phantom for \\
                &   &           &       &       &   $|\la_2-2|>3 \sqrt{2}$\\
\vspace{-0.35cm}\\
\hline $Q_{12}$ & $-\frac{1}{9} (\mu_2-2)^2$ & 1& $-\frac{1}{9}
(\mu_2-2)^2$ & $\frac{1}{6}
\left(-\mu_2^2+4 \mu_2-1\right)$ & unphysical\\
\vspace{-0.35cm}\\
\hline $Q_{13}$ & $-\frac{1}{9} (\mu_2-2)^2$ & 1& $-\frac{1}{9}
(\mu_2-2)^2$ & $\frac{1}{6}
\left(-\mu_2^2+4 \mu_2-1\right)$ & unphysical\\
\hline \hline
\end{tabular}}
\end{center}
\end{table*}

In table \ref{BasicObsVmassinfty2} are displayed the values of the
basic observables  \eqref{Omegas}, \eqref{wss}, \eqref{wtot} and
\eqref{decc} for the singular points  of the system
\eqref{autonomous2} (case 2) as well as the solution types.

In order to illustrate the above analytical results we perform
several numerical integrations.

\begin{figure}[ht]
\begin{center}
\hspace{0.4cm}
\includegraphics[scale=0.5]{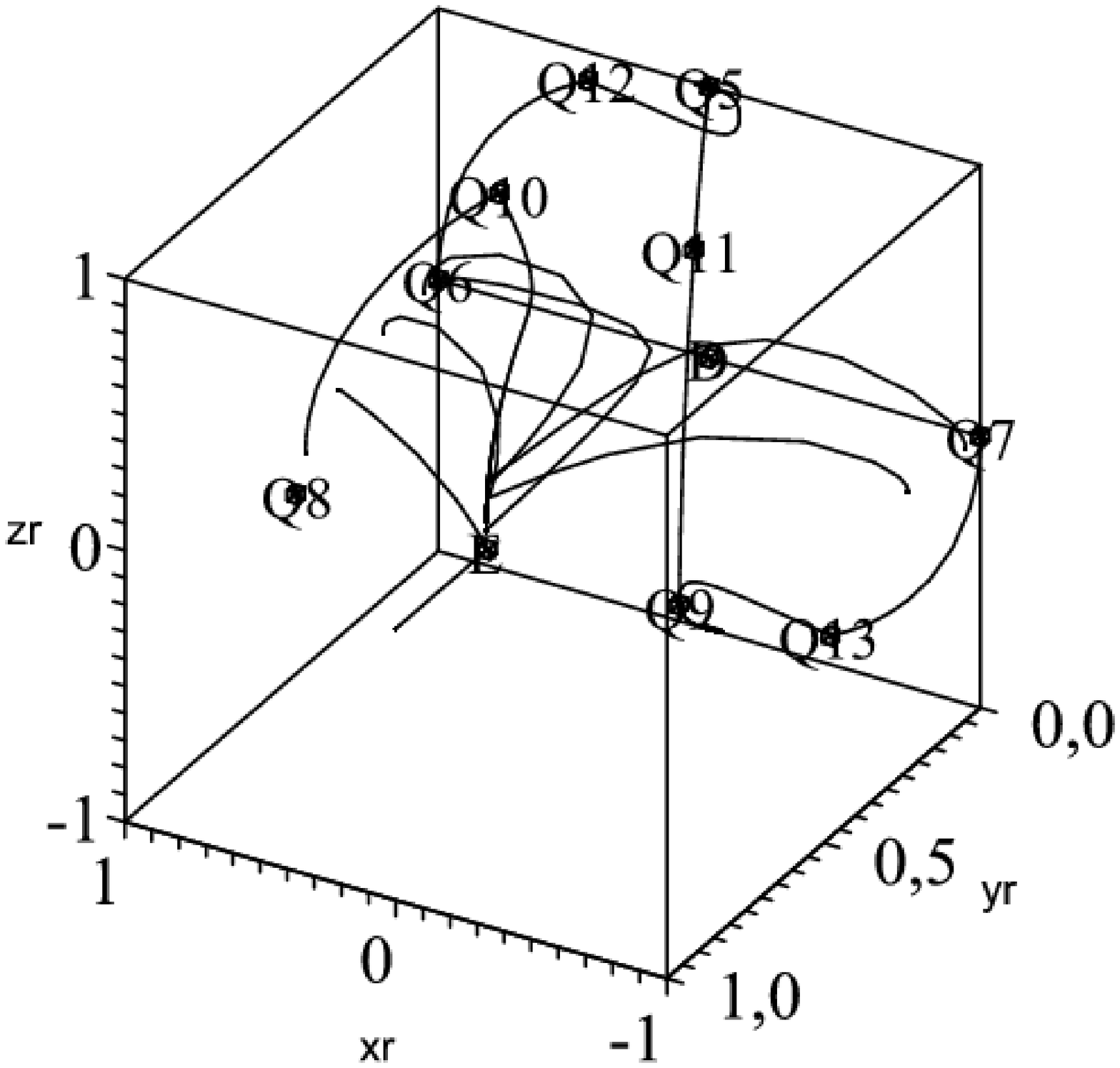}
\caption{{ Poincar\'e (global) phase-space of Model 2, for the
parameter values $\la_2=-0.5$ and $\mu_2=0.5$. The critical point
E (representing de Sitter solutions) is a local attractor for the
points at the finite region.  The points at infinity $Q_{6,7,8,9}$
are local sources; $Q_{10,11,12}$ are saddles; and $Q_{13}$ is a
local sink.}} \label{Vmass2la_-0.5_mu_0.5}
\end{center}
\end{figure}

\begin{figure}[ht]
\begin{center}
\hspace{0.4cm}
\includegraphics[scale=0.5]{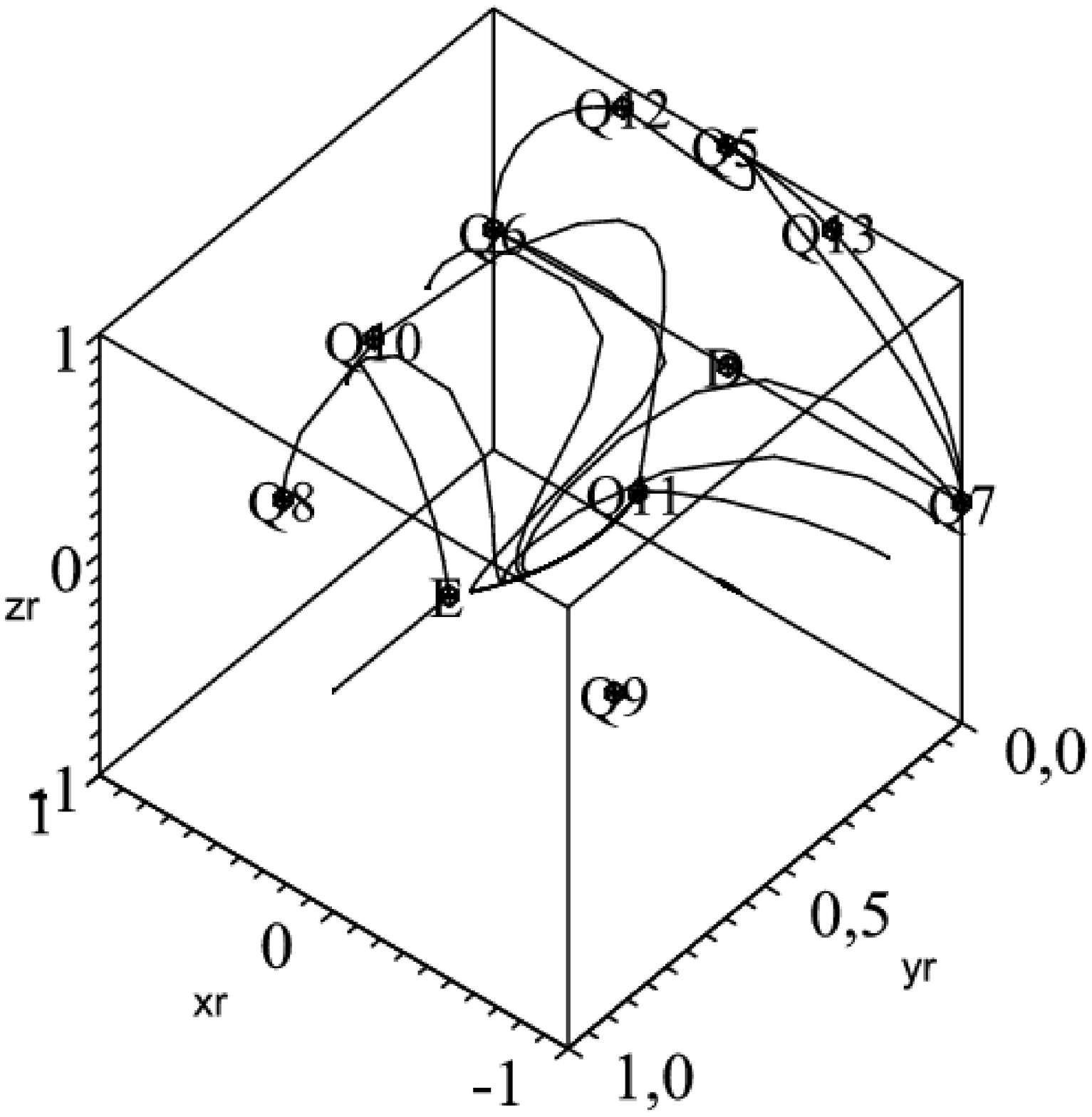}
\caption{{ Poincar\'e (global) phase-space of Model 2, for the
parameter values $\la_2=2.01+\sqrt{30}$ and $\mu_2=0.5$. The
critical point E (representing de Sitter solutions) is a saddle
point.  The points at infinity $Q_{6,7,8,9,10}$ are local sources;
$Q_{12,13}$ are saddles and $Q_{11}$ is a sink.}}
\label{la2+sqrt30+01_mu0_5}
\end{center}
\end{figure}

In the figure \ref{Vmass2la_-0.5_mu_0.5} it is showed the
Poincar\'e (global) phase-space of Model 2, for the parameter
values $\la_2=-0.5$ and $\mu_2=0.5$. The critical point E
(representing de Sitter solutions) is a local attractor for the
points at the finite region.  The singular points at infinity
$Q_{6,7,8,9}$ are local sources. In the figure are two lines
(contained in the line $y_r=z_r=0$) that connects $Q_6$ with $D$
and $Q_7$ with $D$, respectively. The points $Q_{10,11,12}$ are
saddles where as $Q_{13}$ is a local sink.

In the figure \ref{la2+sqrt30+01_mu0_5} it is showed the
Poincar\'e (global) phase-space of Model 2, for the parameter
values $\la_2=2.01+\sqrt{30}>2+\sqrt{30}$ and $\mu_2=0.5$. The
critical point E (representing de Sitter solutions) is, by theorem
\ref{centerE}, a saddle for the points at the finite region. The
singular points at infinity $Q_{6,7,8,9,10}$ are local sources;
$Q_{12,13}$ are saddles and $Q_{11}$ is a sink.

\subsubsection{Cosmological implications and discussion: Model 2}

In  this case, both real and physically meaningful critical
points, namely D and E, have a stable manifold of smaller
dimensionality than that of the phase-space. As was mentioned in
subsection \ref{mod2a} the singular point D has a  very small
probability to be the late-time attractor of the system. By using
the center manifold theory we have proved the stability of the de
Sitter solution E for Power-law potential with $\la_2\leq 0$ and
power-law dependent dark-matter particle mass. For $\la_2> 0$; E
is locally unstable (saddle type). However, even if the
cosmological evolution is managed to be attracted by these
solutions, the coincidence problem will not be solved, since D
represents a flat, non-accelerating universe dominated by dark
matter, and E correspond to de Sitter universe completely
dominated by dark energy. These critical points are located in the
region where the scalar field and the Hubble parameter diverge.
Divergencies in a cosmological scenario are represented as
asymptotic states, in particular associated with the past and
future asymptotic dynamics
\cite{Leon:2008de,Rendall:2004ic,Rendall:2006cq,Foster:1998sk,Miritzis:2003ym,Miritzis:2005hg,Hertog:2006rr,Wetterich:1994bg2}.
In the present Model 2, due to the non-compactness of the
phase-space, such a behavior can lead either to an asymptotic
state acquired at infinite time, or to a singularity reached at a
finite time. If $H\rightarrow\infty$ or
$\rho_\phi\rightarrow\infty$ at $t\rightarrow\infty$ then we
acquire an eternally expanding universe, while if
$H\rightarrow\infty$ at $t\rightarrow t_{BR}<\infty$ then the
universe results to a Big Rip \cite{Capozziello:2009hc}. In order
to give a complete picture of the physical model under
consideration we have investigate the global phase space through
Poincar\'e projection. We have obtained that the physical
solutions at infinity are $Q_{5,8,9,10,11}.$ $Q_5$ represents a
matter-dominated solution with effective equation of state
$w_{\text{tot}}=0$ (dust-like) which is unstable by our previous
analysis. $Q_8$ and $Q_9$ corresponds to initial big-rip
singularities due that $q\rightarrow -\infty$ and
$w_{\text{tot}}\rightarrow -\infty.$ That is, the points at
infinity represents supper-accelerating ($q\ll 0$) phantom
solutions ($w_{\text{tot}}\ll -1$). The solutions $Q_{10,11}$ are
accelerated for $|\la_2-2|>\sqrt{6}$ and phantom for  $|\la_2-2|>3
\sqrt{2}.$ The unphysical solutions are $Q_{6}$ and $Q_7$ that
represents unphysical big-rip singularities since $q\rightarrow
-\infty$ and $w_{\text{tot}}\rightarrow -\infty,$ and $Q_{12,13}.$
The last two singular points satisfy $\Omega_\phi<0$ for
$\mu_2\neq2.$ In the case $\mu_2=2$ they reduce to $Q_5.$ None of
these solutions allows to solve the coincidence problem.

Therefore, power-law potentials with power-law-dependent
dark-matter particle masses, cannot solve the coincidence problem.

\subsection{Model 3: Power-law potential and exponentially-dependent dark-matter particle mass}

In this case the autonomous system reads:
\begin{eqnarray}
&& x'=-
3x+\frac{3}{2}x (1-x^2-y^2)-\frac{\lambda_2y^2 z}{2}-\sqrt{\frac{3}{2}}\mu_1(1+x^2-y^2)\nonumber\\
&&y'=\frac{3}{2}y (1-x^2-y^2)-\frac{\lambda_2 xyz}{2},
\nonumber\\
&&z'=-xz^2. \label{autonomous3}
\end{eqnarray}

\subsubsection{Finite analysis}

The real and physically meaningful critical points are
\begin{eqnarray}
&&\left(x_{c6}=0,\ y_{c6}=1,\ z_{c6}=0\right), \nonumber\\
&&\left(x_{c7}=-\frac{\sqrt{\frac{3}{2}}}{\mu_1},\
y_{c7}=\sqrt{1-\frac{3}{2\mu_1^2}},\ z_{c7}=0\right),
\end{eqnarray}
and the necessary conditions for their existence are shown in
table \ref{stability3}.

In the model at hand, all critical points are non-hyperbolic and
the dimensionality of their stable manifold is presented in table
\ref{stability3}.

Additionally, we mention that there exists also an unphysical
critical point H, with coordinates
$\left(x_{c8}=-\sqrt{\frac{2}{3}}\mu_1,\ y_{c8}=0\
z_{c8}=0\right)$. Its stable manifold is 2D if $|\mu_1|>
\sqrt{\frac{3}{2}}$, and 1D if $|\mu_1|< \sqrt{\frac{3}{2}}$.

\begin{figure}[ht]
\begin{center}
\hspace{0.4cm}
\includegraphics[width=8cm,height=7cm]{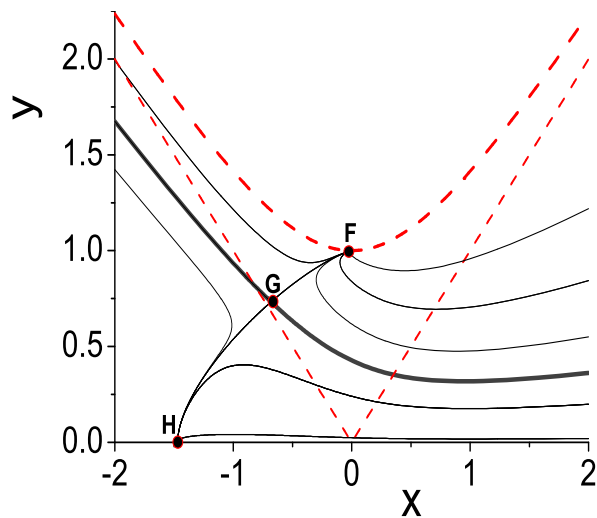}
\caption{{ \label{FFFFFFig4} xy-projection of the phase-space of
Model 3 for the parameter values $\la_2=1$ and $\mu_1=1.8$. The
stable manifold of G (thick curve) divides the physical part of
the phase space (region bounded by the dashed (red) curves) in two
regions. The orbits initially below this curve converge towards H,
while those initially above this curve converge towards F. [Taken
from \cite{PLB2010}; published with permission of Elsevier
B.V.].}}

\end{center}
\end{figure}

For the choice $|\mu_1|> \sqrt{3},$ the orbits initially below the
stable manifold of G converge towards H. The orbits initially
above this curve converge towards F. This behavior is depicted in
fig. \ref{FFFFFFig4}, which has arisen from numerical evolution
using $\la_2=1$ and $\mu_1=1.8$.

\begin{table*}[t]
\caption[The real and physically meaningful critical points of
Model 3 and their behavior.]{\label{stability3} The real and
physically meaningful critical points of  Model 3 and their
behavior.}
\begin{center}
\begin{tabular}{|c|c|c|c|c|c|c|c|c|}
\hline
 Cr. P.& $x_c$ & $y_c$ & $z_c$ & Existence & Stable   & $\Omega_\s$ &  $w_{tot}$ & Acc.   \\
 &   &  &   &  &   manifold &   &    &     \\
\hline \hline
 F&  $x_{c6}$ & $y_{c6}$ &$z_{c6}$ & Always  &2-D  &  1
 &-1
    & Always  \\
G&  $x_{c7}$ & $y_{c7}$ &$z_{c7}$ & $|\mu_1|>\sqrt{3}$  & 1-D &
$1-\frac{3}{\mu_1^2}$ & -1
& Always \\
\hline
\end{tabular}
\end{center}

\end{table*}

If we restrict ourselves in the region $|\mu_1|<
\sqrt{\frac{3}{2}}$, then the critical point G does not exists and
thus there are not scaling solutions. In this case F is indeed the
attractor for a positive-measure set of initial conditions.
Moreover, there exist exceptional orbits contained on a 1D center
manifold of F whose dynamical behavior cannot be anticipated from
the linear analysis. However, this scenario does not lead to
a solution of the coincidence problem ($\Omega_\phi=1$ always).\\

\subsubsection{Stability of de Sitter solution for Power-law
potential and exponentially-dependent dark-matter particle mass.}

The singular point $F$ represents the de Sitter solution for
Power-law potential and power-law-dependent dark-matter particle
mass. In this section we will analyze the stability of the center
manifold of $F$ for the vector field \eqref{autonomous3}.

\begin{prop}\label{centerF} For $\la_2< 0,$ the singular point $F:\left(x_{c6}=0,\ y_{c6}=1,\ z_{c6}=0\right)$ of the system
\eqref{autonomous3} is locally asymptotically stable. For
$\la_2>0,$ $F$ is locally unstable (saddle type). For $\la_2=0,$
$F$ is stable but not asymptotically stable.
\end{prop}

In order to translate $F$ to the origin and transforming the
linear part of the resulting vector field into its Jordan
canonical form, we define new variables
$(u,v_1,v_2)\equiv\mathbf{x}$, by the equations \be
u=z,\,v_1=\frac{x}{\sqrt{6} \mu_1}+\frac{z \la_2}{6 \sqrt{6}
\mu_1},\, v_2=y-1\nonumber\ee so that \be \left(\begin{array}{c}
u'\\v_1'\\v_2'
\end{array}\right)=\left(\begin{array}{ccccc} 0& 0 &0 \\
0& -3 &1
\\  0& 0 & -3
\end{array}\right)\left(\begin{array}{c}u\\v_1\\v_2
\end{array}\right)+\left(\begin{array}{c}f(u,v_1,v_2)\\g_1(u,v_1,v_2)\\g_2(u,v_1,v_2)
\end{array}\right)\label{Fcenter2}
\ee where

$f(u,v_1,v_2)=\frac{1}{6} u^2 \left(u \la_2-6 \sqrt{6} v_1
\mu_1\right),$ $g_1(u,v_1,v_2)=\frac{\left(\frac{{\la_2}^3}{144
\sqrt{6}}+\frac{{\la_2}^2}{36 \sqrt{6}}\right)
u^3}{\mu_1}+\left(\frac{1}{72} (-9 v_1-1) {\la_2}^2-\frac{v_1
{\la_2}}{6}\right)
   u^2+\left(\frac{v_1 (9 v_1+2) {\la_2} \mu_1}{2 \sqrt{6}}-\frac{v_2 (v_2+2)
   {\la_2}}{4 \sqrt{6} \mu_1}\right) u-3 v_1^2 (3 v_1+1) \mu_1^2+\frac{1}{2} v_2
   (v_2-3 v_1 (v_2+2)),$ and
$g_2(u,v_1,v_2)=\frac{1}{24} (v_2+1) \left(u^2 {\la_2}^2-216 v_1^2
\mu_1^2\right)-\frac{3}{2} v_2^2
   (v_2+3).$
\\
The system \eqref{Fcenter2} is written in diagonal form
\begin{align}
u'  &  =Cu+f\left(  u,\mathbf{v}\right) \nonumber\\
\mathbf{v}'  &  =P\mathbf{v}+\mathbf{g}\left(  u,\mathbf{v}\right)
, \label{Fcenter3}
\end{align}
where $\left(  u,\mathbf{v}\right)
\in\mathbb{R}\times\mathbb{R}^{2},$ $C$ is the zero $1\times1$
matrix, $P$ is a $2\times 2$ matrix with negative eigenvalues and
$f,\mathbf{g}$ vanish at $\mathbf{0}$ and have vanishing
derivatives at $\mathbf{0.}$ The center manifold theorem
\ref{existenceCM} asserts that there exists a 1-dimensional
invariant local center manifold $W^{c}\left( \mathbf{0}\right) $
of \eqref{Fcenter3} tangent to the center subspace (the
$\mathbf{v}=\mathbf{0}$ space) at $\mathbf{0}.$ Moreover,
$W^{c}\left( \mathbf{0}\right)  $ can be represented as
\[
W^{c}\left(  \mathbf{0}\right)  =\left\{  \left(
u,\mathbf{v}\right)
\in\mathbb{R}\times\mathbb{R}^{2}:\mathbf{v}=\mathbf{h}\left(
u\right), \left\vert u\right\vert <\delta\right\} ;
\mathbf{h}\left(  0\right) =\mathbf{0}, D\mathbf{h}\left( 0\right)
=\mathbf{0}
\]
for $\delta$ sufficiently small (see definition \ref{CMdef}). The
restriction of (\ref{Fcenter3}) to the center manifold is (see
definition \ref{vectorfieldCM})
\begin{equation}
u'=f\left( u,\mathbf{h}\left(  u\right)  \right)  . \label{Frest}
\end{equation}
According to Theorem \ref{stabilityCM}, if the origin $u=0$ of
\eqref{Frest} is stable (asymptotically stable) (unstable) then
the origin of \eqref{Fcenter3} is also stable (asymptotically
stable) (unstable). Therefore, we have to find the local center
manifold, i.e., the problem reduces to the computation of
$\mathbf{h}\left( u\right).$

Substituting $\mathbf{v}=\mathbf{h}\left(  u\right)  $ in the
second component of \eqref{Fcenter3} and using the chain rule,
$\mathbf{v }'=D\mathbf{h}\left(  u\right)  u'$, one can show that
the function $\mathbf{h}\left( u\right)  $ that defines the local
center manifold
satisfies%
\begin{equation}
D\mathbf{h}\left(  u\right)  \left[  f\left(  u,\mathbf{h}\left(
u\right) \right)  \right]  -P\mathbf{h}\left(  u\right)
-\mathbf{g}\left( u,\mathbf{h}\left(  u\right)  \right)  =0.
\label{Fh}
\end{equation}
According to Theorem \ref{approximationCM}, equation \eqref{Fh}
can be solved approximately by using an approximation of
$\mathbf{h}\left(  u\right)  $ by a Taylor series at $u=0.$ Since
$\mathbf{h}\left(  0\right)  =\mathbf{0\ } $and $D\mathbf{h}\left(
0\right)  =\mathbf{0},$ it is obvious that $\mathbf{h}\left(
u\right)  $ commences with quadratic terms. We substitute
\[
\mathbf{h}\left(  u\right)  =:\left[
\begin{array}
[c]{c}%
h_{1}\left(  u\right) \\
h_{2}\left(  u\right)
\end{array}
\right]  =\left[
\begin{array}
[c]{c}%
a_{1}u^{2}+a_{2}u^{3}+O\left(  u^{4}\right) \\
b_{1}u^{2}+b_{2}u^{3}+O\left(  u^{4}\right)
\end{array}
\right]
\]
into (\ref{Fh}) and set the coefficients of like powers of $u$
equal to zero to find the unknowns $a_{1},b_{1},...$.

We find that the non-zero coefficients are
$$a_3=\frac{{\la_2}^2}{108 \sqrt{6} \mu_1},\, b_2=\frac{{\la_2}^2}{72},$$ Therefore, \eqref{Frest}
yields \be u'=\frac{{\la_2} u^3}{6}-\frac{{\la_2}^2
u^5}{108}+O\left(u^6\right).\label{Frest1}\ee Neglecting the error
terms, this  is a gradient-like equation (i.e., $u'=-\nabla U(u)$)
with potential $U(u)=\frac{1}{648} u^4 {\la_2} \left(u^2
{\la_2}-27\right)$ for which the origin is a degenerate minimum
provided $\lambda_2<0$ and a degenerated maximum provided
$\lambda_2>0$. Thus, for $\lambda_2<0,$ the origin $u=0$ of
\eqref{Frest1} is locally asymptotically stable. Hence, the origin
$\mathbf{u}=\mathbf{0}$ of the full three-dimensional system is
asymptotically stable. For $\la_2>0$ the origin is locally
unstable (saddle type). In the same way as we proceeded in the
proof of \ref{centerE} can be proved the stability (but not
asymptotic stability) of $F$. $\blacksquare$

\subsubsection{Analysis at infinity}

Performing the transformation \eqref{TransfVmass2}, the system
\eqref{autonomous3} becomes

\begin{align}
&\rho'= -\frac{3}{2} \rho ^3 \sin ^4\psi-\frac{1}{2} \cos \theta
\left(2 \cos \psi \left(\cos ^2\psi+\lambda_2 \sin
   ^2\theta \sin ^2\psi\right) \rho ^3+\right.\nonumber\\ & \left. +\sqrt{6} \mu_1 \left(1-\rho ^2\right)^{3/2}\right) \sin \psi+\cos (2
   \theta ) \left(\frac{3}{2} \rho  \left(\rho ^2-1\right) \sin ^2\psi+\right.\nonumber\\ & \left.
   -\frac{1}{2} \mu_1 \rho ^2 \sqrt{6-6 \rho ^2}
   \cos \theta \sin ^3\psi\right),\label{Vmassinfty3a}\\
   & \theta'=3 \cos \theta \sin \theta+\frac{\mu_1 \sqrt{6-6
\rho ^2} \csc \psi \sin \theta}{2 \rho }+ \nonumber\\ &
+\frac{\rho \cos (2
   \theta ) \left(\lambda_2 \rho  \cos \psi-\mu_1 \sqrt{6-6 \rho ^2}\right) \sin \psi \sin \theta}{2
   \left(\rho ^2-1\right)}\label{Vmassinfty3b},\end{align}
\begin{align}   &   \psi'=\frac{3 \rho ^2 \cos \psi \sin ^3\psi}{2
\left(\rho ^2-1\right)}\nonumber \\ &
   +\frac{\cos \theta \cos \psi \left(
   (\la_2-2) \cos \psi \sin ^2\psi \rho ^3+\sqrt{6} \mu_1 \left(1-\rho ^2\right)^{3/2}\right)}{2 \rho  \left(\rho
   ^2-1\right)}+ \nonumber\\ &
 +\cos (2 \theta ) \left(\frac{\rho  \cos \theta \cos \psi \left(\mu_1 \sqrt{6-6 \rho
   ^2}-\lambda_2 \rho  \cos \psi\right) \sin ^2\psi}{2 \left(\rho ^2-1\right)}+ \right. \nonumber\\
   & \left.
   -\frac{3}{2} \cos \psi \sin (\psi
   )\right)\label{Vmassinfty3c}.
\end{align}

In the limit $\rho\rightarrow 1,$ the leading terms in
\eqref{Vmassinfty3a}-\eqref{Vmassinfty3c} are

\begin{align}
&\rho'\rightarrow -\frac{1}{2} \sin \psi \left(3 \sin ^3\psi+2
\cos \theta \cos \psi \left(\cos ^2\psi+\lambda_2 \sin
   ^2\theta \sin ^2\psi\right)\right),\label{Vmassinfty3d}\\
&   \theta'\rightarrow
-\frac{\lambda_2 \cos (2 \theta ) \cos \psi \sin \theta \sin \psi}{2 \left(1-\rho ^2\right)}\label{Vmassinfty3e},\\
&   \psi'\rightarrow \frac{\cos \psi (\cos \theta (\cos (2 \theta
) \lambda_2-\lambda_2+2) \cos \psi-3 \sin \psi) \sin
   ^2\psi}{2 \left(1-\rho ^2\right)}\label{Vmassinfty3f}.
\end{align}

The radial equation does not contain the radial coordinate, so the
singular points  can be obtained using the angular equations only.
Setting $\theta'=0,\psi'=0$, we obtain the singular points which
are listed in table \ref{Vmassinfty3}. The stability of these
points is studied by analyzing first the stability of the angular
coordinates and then deducing, from the sign of equation
\eqref{Vmassinfty3d}, the stability on the radial direction.

\begin{landscape}

\begin{table*}[ht]
\begin{center}\caption{\label{Vmassinfty3} Asymptotic singular points  of the system
\eqref{autonomous3} (case 3) and their stability. We use the
notations $\alpha=\frac{3\sqrt{2}}{\sqrt{22-4\la_2+\la_2^2}},$
$\epsilon=\text{sign}
(\la_2),\delta=\text{sign}(-26+4\la_2+\la_2^2).$ NH stands for
nonhyperbolic.} {\begin{tabular}{ccccc} \hline \hline
 Cr. P & Coordinates & Eigenvalues & $\rho'$ & Stability\\
    &  $\theta,\psi, x_r, y_r, z_r$ &  &  & \\
\hline $Q_{14}$ & $0, 0, 0, 0 , 1$ & $0,0$ & 0 & NH; 3D center manifold \vspace{0.2cm}\\
\hline $Q_{15}$ & $0, \frac{\pi}{2}, 1, 0, 0$ & $0,+\infty$ & $-\frac{3}{2}$& { NH; 2D unstable manifold} \\
\vspace{-0.35cm}\\
\hline $Q_{16}$ & $\pi, \frac{\pi}{2}, -1, 0, 0$ & $0,+\infty$ & $-\frac{3}{2}$& { NH; 2D unstable manifold} \\
\vspace{-0.35cm}\\
\hline $Q_{17}$ & $\frac{\pi}{4}, \frac{\pi}{2}, \frac{\sqrt{2}}{2}, \frac{\sqrt{2}}{2}, 0$ & $0,+\infty$ & $-\frac{3}{2}$& { NH; 2D unstable manifold} \\
\vspace{-0.35cm}\\
\hline $Q_{18}$ & $\frac{3\pi}{4}, \frac{\pi}{2}, -\frac{\sqrt{2}}{2}, \frac{\sqrt{2}}{2}, 0$ & $0,+\infty$ & $-\frac{3}{2}$& {NH; 2D unstable manifold}\\
\vspace{-0.35cm}\\
\hline $Q_{19}$ & $0, \cos^{-1}\left(\frac{3}{\sqrt{13}}\right), \frac{2}{\sqrt{13}}, 0, \frac{3}{\sqrt{13}}$ & $-\epsilon\infty,-\infty$ & $<0$& { saddle}\\
\vspace{-0.35cm}\\
\hline $Q_{20}$ & $\frac{\pi}{4}, \cos^{-1}(\alpha),
\frac{\alpha|\la_2-2|}{6}, \frac{\alpha|\la_2-2|}{6},
\alpha$  & $\begin{array}{c} +\infty, \delta \infty \, \text{if}\, \la_2>2\\
\epsilon\infty, -\infty \,\text{if}\, \la_2<2\end{array}  $ & $<0$
&
${\begin{array}{c} \text{source if}\, \la_2>2+\sqrt{30}\\
\text{saddle, otherwise}\end{array}} $ \\
\vspace{-0.35cm}\\
\hline $Q_{21}$ & $\frac{3\pi}{4}, \cos^{-1}(\alpha),
-\frac{\alpha|\la_2-2|}{6}, \frac{\alpha|\la_2-2|}{6},
\alpha$ & $\begin{array}{c}-\infty, - \infty \, \text{if}\, \la_2>2\\
-\epsilon\infty, \delta\infty \, \text{if}\, \la_2<2\end{array}  $
&
$\begin{array}{c} >0 \, \text{if}\, \la_2>-1\\ \text{and}\, \la_2\neq 2, \\
<0 \, \text{if}\, \la_2<-1\end{array}$ &
${\begin{array}{c} \text{sink, if}\, \la_2>-1,\\ \text{and}\, \la_2\neq 2, \\
\text{saddle, otherwise}\end{array} } $  \\
\hline \hline
\end{tabular}}
\end{center}
\end{table*}

\end{landscape}

In table \ref{BasicObsVmassinfty3} are displayed the values of the
basic observables  \eqref{Omegas}, \eqref{wss}, \eqref{wtot} and
\eqref{decc} for the singular points  of the system
\eqref{autonomous3} (case 3) as well as the solution types.

\begin{table*}[ht]
\begin{center}\caption{\label{BasicObsVmassinfty3} Basic observables for the singular points  of the system
\eqref{autonomous3} (case 3). Solution types}
{\small\begin{tabular}{cccccc} \hline \hline
 Cr. P & $\Omega_\phi$ & $w_\phi$ & $w_{\text{tot}}$ & $q$ & Solution type\\
\hline $Q_{14}$ & 0 & 1& 0 & $\frac{1}{2}$ & dust-like \vspace{0.2cm}\\
\hline $Q_{15}$ & $-\infty$ & 1 & $-\infty$ & $-\infty$ & (unphys.) big-rip \\
\vspace{-0.35cm}\\
\hline $Q_{16}$ & $-\infty$ & 1 & $-\infty$ & $-\infty$ & (unphys.) big-rip \\
\vspace{-0.35cm}\\
\hline $Q_{17}$ & 0 & $-\infty$ & $-\infty$ & $-\infty$ &  big-rip \\
\vspace{-0.35cm}\\
\hline $Q_{18}$ &0 & $+\infty$ & $-\infty$ & $-\infty$ & big-rip \\
\vspace{-0.35cm}\\
\hline $Q_{19}$ & $-\frac{4}{9}$ & 1 & $-\frac{4}{9}$ & $-\frac{1}{6}$ & unphysical \\
\vspace{-0.35cm}\\
\hline $Q_{20}$ & 0 & $-\infty$ & $-\frac{1}{18} (\la_2-2)^2$ &
$\frac{1}{12} \left(-\la_2^2+4 \la_2+2\right)$ &  Accelerated for \\
                &   &           &       &       &   $|\la_2-2|>\sqrt{6}$\\
                &   &           &       &       &   phantom for \\
                &   &           &       &       &   $|\la_2-2|>3 \sqrt{2}$\\
\vspace{-0.35cm}\\
\hline $Q_{21}$ & 0 & $+\infty$ & $-\frac{1}{18} (\la_2-2)^2$ &
$\frac{1}{12} \left(-\la_2^2+4 \la_2+2\right)$ & Accelerated for \\
                &   &           &       &       &   $|\la_2-2|>\sqrt{6}$\\
                &   &           &       &       &   phantom for \\
                &   &           &       &       &   $|\la_2-2|>3 \sqrt{2}$\\
\hline \hline
\end{tabular}}
\end{center}
\end{table*}

In order to perform the numerical experiments for the system
\eqref{Vmassinfty3a}-\eqref{Vmassinfty3c} it is useful to rewrite
the system in the cartesian coordinates $x_r, y_r, z_r.$ The
system \eqref{Vmassinfty3a}-\eqref{Vmassinfty3c} becomes

\begin{align}
& x_r'=\frac{1}{2} \left(2 x_r^2 z_r^3+3 x_r
\left(-x_r^2+y_r^2+1\right) z_r^2+\la_2 \left(2 x_r^2-1\right)
y_r^2
   z_r+\right. \nonumber\\ & \left.
   +3 x_r \left(x_r^2+y_r^2-1\right) \left(2 y_r^2+1\right)\right)-\sqrt{\frac{3}{2}} \mu_1 \left(x_r^2-1\right)
   w \left(2 y_r^2+z_r^2-1\right),
\nonumber\end{align}
\begin{align}& y_r'= \frac{1}{2} y_r
\left(\left(6 y_r^2-3 \left(z_r^2+1\right)\right)
   x_r^2+z_r \left(2 \la_2 y_r^2+2 z_r^2-\la_2\right) x_r
   \right. \nonumber\\ & \left.  +3 \left(y_r^2-1\right) \left(2
   y_r^2+z_r^2-1\right)\right)-\sqrt{\frac{3}{2}} \mu_1 x_r y_r w \left(2
   y_r^2+z_r^2-1\right),
\nonumber\end{align}
\begin{align} & z_r'= \frac{1}{2} z_r
\left(\left(6 y_r^2-3 z_r^2+3\right) x_r^2+2 z_r \left(\la_2
   y_r^2+z_r^2-1\right) x_r+\right. \nonumber\\ &
   \left.  +3 y_r^2 \left(2 y_r^2+z_r^2-1\right)\right) -\sqrt{\frac{3}{2}} \mu_1 x_r z_r
   w \left(2 y_r^2+z_r^2-1\right)\nonumber\\
   & w'=\frac{1}{2} w
   \left(\left(6 y_r^2-3 z_r^2+3\right) x_r^2+2 z_r \left(\la_2 y_r^2+z_r^2\right) x_r+\right. \nonumber\\ &
   \left.   +3 y_r^2 \left(2
   y_r^2+z_r^2-1\right)\right)-\sqrt{\frac{3}{2}} w^2 \mu_1 x_r \left(2 y_r^2+z_r^2-1\right)\label{Model3cartesian}
\end{align}
where we have used the time re-scaling $$\mathrm{d}\tau\rightarrow
\frac{\mathrm{d}\tau}{1-\rho^2}$$ which leave invariant the orbits
of the phase-space and the time direction (see theorem \ref{ext2})
and we have introduced the auxiliary variable
$w=\sqrt{1-x_r^2-y_r^2-z_r^2}$ to avoid that the numerical
procedure becomes complex-valued at the singular points.

To examine the stability of $Q_{14}$ we perform the linear
coordinate transformation $u_1=x_r,\, u_2=y_r,\, u_3=z_r-1,\,
u_4=w$ and Taylor expand the system \eqref{Model3cartesian} around
the origin up to third order in the vector norm to obtain the
approximated system
\begin{eqnarray}
&&u_1'=u_1^2+3 u_3u_1-\frac{u_2^2 \la_2}{2}+\sqrt{6} u_3u_4\mu_1+{\cal O}(3),\nonumber\\
&& u_2'=   -\frac{1}{2} u_2(6 u_3+u_1(\la_2-2))+{\cal O}(3),\nonumber\\
&& u_3'=2 u_1u_3+{\cal O}(3),\nonumber\\
&& u_4'=u_1u_4+{\cal O}(3),\label{asymptModel3}\end{eqnarray}
where  ${\cal O}(3)$ denotes ${\cal O}(\|(u_1,u_2,u_3,u_4)\|^3)$
defined in a neighborhood of the origin contained in the region
{\small \[\left\{0\leq u_1^2+u_2^2\leq 1,\,
-1-\sqrt{1-u_1^2-u_2^2}\leq u_3\leq -1+\sqrt{1-u_1^2-u_2^2}\leq
0,\, u_4>0\right\}.\]} Observe that the variables $u_i, i=1\ldots
4$ are not independent since $u_1^2+u_2^2+u_3 (u_3+2)+u_4^2=0.$
Thus, one is able to eliminate one variable. From
\eqref{asymptModel3}(c) and \eqref{asymptModel3}(d) follows that
$u_3\propto u_4^2.$ By substituting back this relation on
\eqref{asymptModel3} and neglecting the terms ${\cal
O}(\|(u_1,u_2,u_4)\|^3)$ we obtain the reduce (decoupled)
3-dimensional system
\begin{eqnarray}
&&u_1'=u_1^2-\frac{u_2^2 \la_2}{2},\nonumber\\
&& u_2'=   -\frac{1}{2}(\la_2-2) u_1 u_2,\nonumber\\
&& u_4'=u_1u_4,\label{asymptModel3b}\end{eqnarray} that represents
very accurately the nonlinear dynamics. \footnote{Another argument
in favor to neglect the contribution of $u_3$ to the nonlinear
dynamics is that $u_1^2+u_2^2+u_3 (u_3+2)+u_4^2=0$ is an exact
formula. Thus, by taking the total derivative of $u_1^2+u_2^2+u_3
(u_3+2)+u_4^2$ we obtain
$\frac{\mathrm{d}}{\mathrm{d}\tau}\left[u_1^2+u_2^2+u_3
(u_3+2)+u_4^2\right]=4 u_1 u_3+{\cal O}(\|(u_1,u_2,u_3,u_4)\|^3),$
which means that $u_1 u_3={\cal O}(\|(u_1,u_2,u_3,u_4)\|^3),$ and
then $u_3'={\cal O}(\|(u_1,u_2,u_3,u_4)\|^3).$ This result in
consistent with $u_3\propto u_4^2.$}

The system \eqref{asymptModel3b} admits the implicit solution
\begin{eqnarray}
&& u_1=\pm\sqrt{c_1 u_2^{-\frac{4}{\la_2-2}}+u_2^2},\nonumber\\
&&\tau=c_2\mp\frac{u_2^{\frac{2}{\la_2-2}} \,
_2F_1\left(\frac{1}{\la_2},\frac{1}{2};1+\frac{1}{\la_2};-\frac{u_2^{\frac{2
\la_2}{\la_2-2}}}{c_1}\right)}{\sqrt{c_1}},\nonumber\\
 && u_4= u_2^{-\frac{2}{\la_2-2}}
c_3\label{solasymptModel3b}
\end{eqnarray}
By examining the asymptotic behavior of solutions
\eqref{solasymptModel3b} we obtain that for $\la_2>2,$ the variable $u_2$
cannot approach asymptotically to zero since otherwise this would
imply the divergence of $u_1$ and $u_4$ in contradiction with the
compactness of the phase space. For $0<\la_2<2,$ the origin is
approached as $\tau\rightarrow -\infty$ in the case of $u_1(0)>0$
and it is approached as $\tau\rightarrow +\infty$ in the case of
$u_1(0)<0.$ For $\la_2<0,$
$$\frac{u_2^{\frac{2}{\la_2-2}} \,_2F_1\left(\frac{1}{\la_2},\frac{1}{2};1+\frac{1}{\la_2};-\frac{u_2^{\frac{2
\la_2}{\la_2-2}}}{c_1}\right)}{\sqrt{c_1}}\approx
\frac{u_2^{\frac{2}{\la_2-2}}}{\sqrt{c_1}}-\frac{u_2^{2+\frac{6}{\la_2-2}}}{2
   (\la_2+1) c_1^{3/2}}+\text{h.o.t.}\rightarrow +\infty$$ as $u_2\rightarrow
   0.$
This result can be interpreted as follows. For $\la_2<0$ and
$u_1(0)>0$ the origin is approached as $\tau\rightarrow -\infty$
and for $\la_2<0$ and $u_1(0)<0$ the origin is approached as
$\tau\rightarrow +\infty,$ which means that $Q_{14}$ is saddle
like. In this way we have proved the instability of $Q_{14}.$

We obtain by an explicit calculation that the center manifold of $Q_{17,18}$ is given up to fourth order by the graph
$x_r\pm y_r\mp \sqrt{2}=\mp\sqrt{2}u^2,z_r=0,w=0$ where $u\equiv y_r-\frac{\sqrt{2}}{2}$.
The dynamics on the center manifold is given by
$$u'=-6 u^3+{\cal O}(5).$$ From this  follows that the center manifold is stable since the origin is a degenerate minimum of the potential $U(u)=\frac{3 u^4}{2}.$  Hence, $Q_{17,18}$ are of saddle type.

By an explicit calculation we find that the center manifolds of $Q_{15,16}$ are $x_r=\pm \left(1-\frac{y_r^2}{2}\right)+{\cal O}(y_r)^3=\pm\sqrt{1+y^2},$  $z_r=w=0.$ The dynamics on the center manifolds of  $Q_{15,16}$ are given by $u'=\frac{3}{2}u^2+{\cal O}(u)^5,$ where $u=y_r.$ From this follows that  $Q_{15,16}$ are local sources since their center manifolds are unstable (the origin is a degenerate local maximum of the potential function $U(u)=-\frac{3 u^4}{8}$).

\begin{figure}[ht]
\begin{center}
\hspace{0.4cm}
\includegraphics[scale=1]{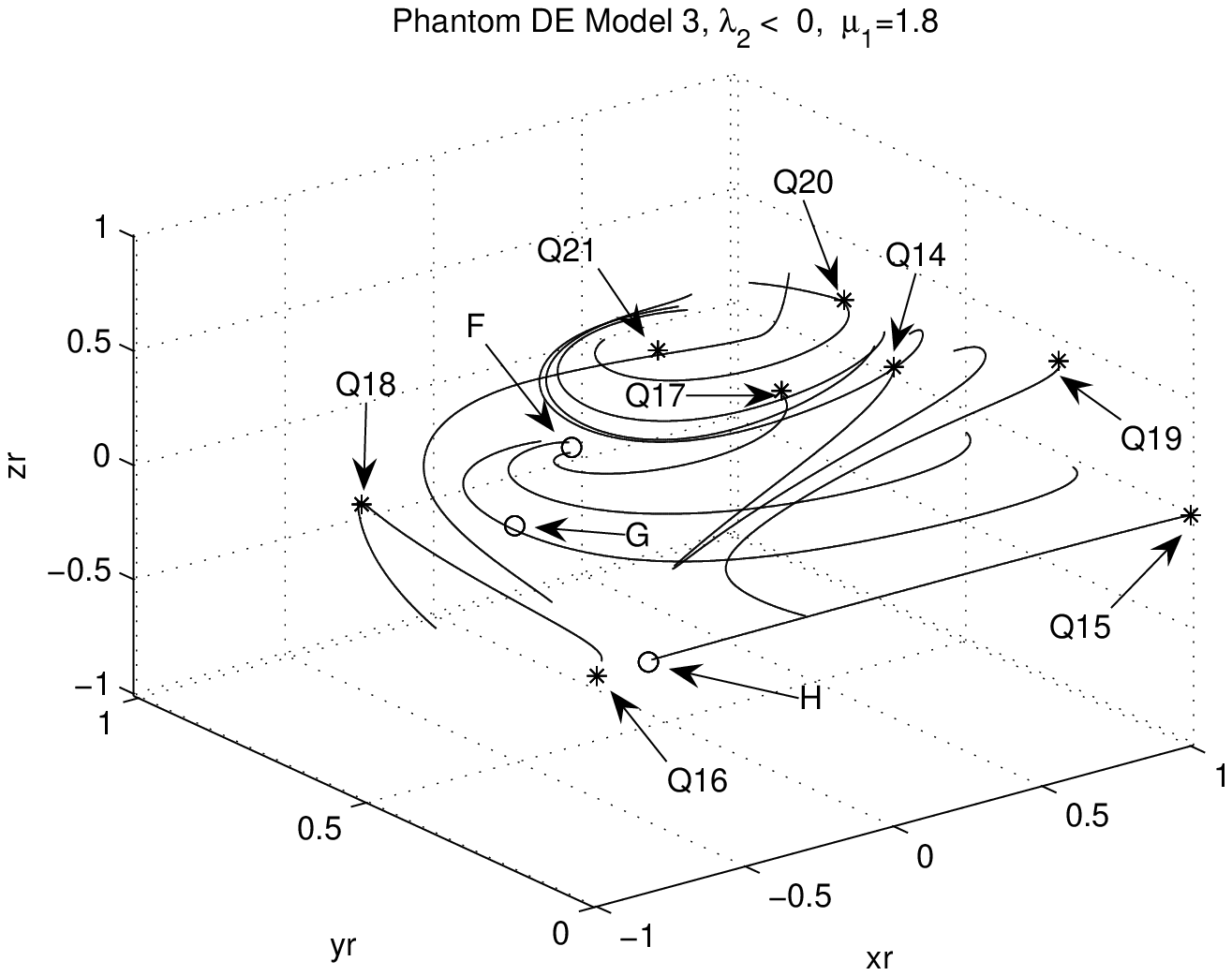}
\caption{{ Poincar\'e (global) phase-space of Model 3, for the
parameter values $\la_2=-0.01$ and $\mu_1=1.8$. For this choice of
parameters $Q_{15}$ and $Q_{16}$ are local sources; $Q_{14}$ is
unstable (of saddle type); $Q_{17,18,19,20}$ are saddles in the
infinite region; $G,H$ are saddles in the finite region; $Q_{21}$
is a sink in the infinite region and $F$ is locally asymptotically
stable.}} \label{AsymptoticPhantomDEModel3RK4B70}
\end{center}
\end{figure}

In the figure \ref{AsymptoticPhantomDEModel3RK4B70} are displayed
several trajectories in the Poincar\'e (global) phase-space of
Model 3, for the parameter values $\la_2=-0.01$ and $\mu_1=1.8$.
For this choice of parameters $Q_{15}$ and $Q_{16}$ are local
sources; $Q_{14}$ is unstable (of saddle type); $Q_{17,18,19,20}$
are saddles in the infinite region; $G,H$ are saddles in the
finite region; $Q_{21}$ is a sink in the infinite region and $F$
is locally asymptotically stable. There is one orbit joining
$Q_{15}$ and $H;$ and one orbit joining $Q_{15}$ with $Q_{19}.$

\subsubsection{Cosmological implications and discussion: Model 3}

In this model we see that the critical point F exists always,
while G exists only for $|\mu_1|>\sqrt{3}$. However, in both cases
the stable manifold is of smaller dimensionality than that of the
phase-space. However, by an explicit computation of the center
manifold we have proved  that for $\la_2\leq 0$ the singular point
F (corresponding to a de Sitter solution) is stable. Furthermore,
in order to avoid the treatment of unphysical attracting states we
have to impose the additional constraint
 $|\mu_1|< \sqrt{\frac{3}{2}}$. For this choice
of parameters, G does not exists and thus there are not scaling
solutions, while F is the attractor for a positive-measure set of
initial conditions. Point F corresponds to a dark-energy dominated
de Sitter universe, while G to a flat accelerating universe with
$\Omega_\phi=1-\frac{3}{\mu_1^2}$, that is with $0<\Omega_\phi<1$
in the region that it exists. In both points the phantom field
diverges. However, even if G possesses $0<\Omega_\phi<1$, it can
not solve the coincidence problem since it is not a relevant
late-time attractor.

In order to give a complete picture of the physical model under
consideration we have investigate the global phase space through
Poincar\'e projection. We have obtained that the physical
solutions at infinity are $Q_{14,17,18,20,21}.$ $Q_{14}$ represents a
matter-dominated solution with effective equation of state
$w_{\text{tot}}=0$ (dust-like) which is unstable by our previous
analysis. $Q_{17}$ and $Q_{18}$ corresponds to initial big-rip
singularities due that $q\rightarrow -\infty$ and
$w_{\text{tot}}\rightarrow -\infty.$ That is, the points at
infinity represents supper-accelerating ($q\ll 0$) phantom
solutions ($w_{\text{tot}}\ll -1$). The solutions $Q_{20,21}$ are
accelerated for $|\la_2-2|>\sqrt{6}$ and phantom for  $|\la_2-2|>3
\sqrt{2}.$ The unphysical solutions are $Q_{15}$ and $Q_{16}$ that
represents unphysical big-rip singularities since $q\rightarrow
-\infty$ and $w_{\text{tot}}\rightarrow -\infty,$ and $Q_{12,13}.$
The singular point $Q_{19}$ satisfy $\Omega_\phi<0;$ thus, it is unphysical.  None of
these solutions allows to solve the coincidence problem.

In summary, power-law potentials with exponentially-dependent
dark-matter particle masses cannot solve or even alleviate the
coincidence problem.

\subsection{Model 4: Exponential potential and  power-law-dependent dark-matter particle mass}

In this case the
 autonomous system writes:
\begin{eqnarray}
&&x'=-
3x+\frac{3}{2}x (1-x^2-y^2)-\sqrt{\frac{3}{2}}\la_1\, y^2-\frac{\mu_2}{2}z(1+x^2-y^2)\nonumber\\
&&y'=\frac{3}{2}y (1-x^2-y^2)- \sqrt{\frac{3}{2}}\la_1\, x y\\
&&z'=-xz^2.
\label{autonomous4}
\end{eqnarray}

\subsubsection{Finite analysis}

 The real and
physically meaningful critical points
 are
\begin{eqnarray}
&&\left(x_{c9}=0,\ y_{c9}=0,\ z_{c9}=0\right), \nonumber\\
&&\left(x_{c10}=-\frac{\la_1}{\sqrt{6}},\
y_{c10}=\sqrt{1+\frac{\la_1^2}{6}},\ z_{c10}=0\right),
\end{eqnarray}
and in table \ref{stability4} we present the necessary conditions
for their existence.
\begin{table*}[t]
\caption[The real and physically meaningful critical points of
Model 4 and their behavior.]{\label{stability4} The real and
physically meaningful critical points of Model 4 and their
behavior.}
\begin{center}
\begin{tabular}{|c|c|c|c|c|c|c|c|c|}
\hline
 Cr. P.& $x_c$ & $y_c$ & $z_c$ & Existence & Stable   & $\Omega_\s$ &  $w_{tot}$ & Acc.   \\
  &   &   &   &   & manifold &   &    &     \\
\hline \hline
 I&  $x_{c9}$ & $y_{c9}$ &$z_{c9}$ & Always  &1-D   & 0
 &0
    & Never  \\
J&  $x_{c10}$ & $y_{c10}$ &$z_{c10}$ & Always  & 2-D  & 1 &
{\small $-\frac13 (3+\la_1^2)$}
& Always \\
\hline
\end{tabular}
\end{center}
\end{table*}
The aforementioned critical points are non-hyperbolic since at
least one eigenvalue of ${\bf {Q}}$ is always zero. Linear
analysis in not conclusive in these cases, but information about
the dimensionality of the stable manifold can be obtained by
applying the center manifold theorem \cite{arrowsmith}. The
corresponding results are shown in table \ref{stability4}. Both I and J cannot solve the coincidence problem
($\Omega_\phi=1$). In the next section we examine the stability of the phantom solution for this case.

\begin{figure}[ht]
\begin{center}
\hspace{0.4cm}
\includegraphics[width=8cm,height=7cm]{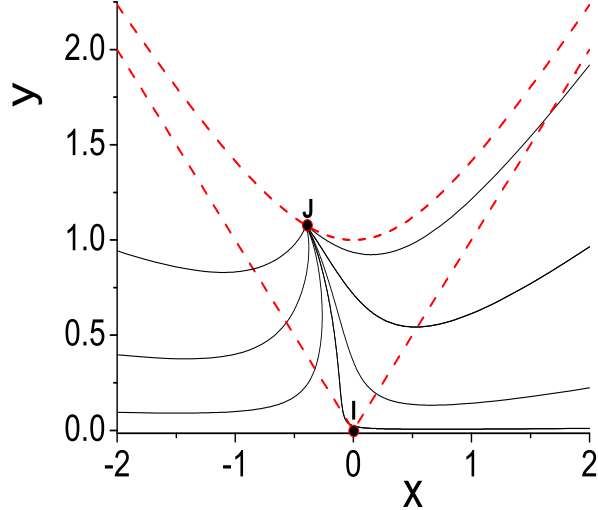}
\caption{{ \label{Vmassfig5} xy-projection of the phase-space of
Model 4 for the parameter values $\la_1=1$ and $\mu_2=1.8$. The
critical point J (corresponding to a super-accelerating universe)
attracts all the orbits in this invariant set. The dashed (red)
curves bound the physical part of the phase space, that is
corresponding to
 $0\leq \Omega_\phi\leq1$. [Taken from \cite{PLB2010}; published with permission of Elsevier
B.V.].}}
\end{center}
\end{figure}

In order to acquire a more transparent picture of the phase-space
behavior, we evolve the system numerically for $\la_1=1$ and
$\mu_2=1.8$ and we depict the results in fig. \ref{Vmassfig5}.

\subsubsection{Stability analysis of the phantom dominated solution for the exponential potential and  power-law-dependent dark-matter particle mass}

In this section we examine the stability of the phantom dominated solution for the exponential potential and  power-law-dependent dark-matter particle mass through the stability analysis of its center manifold.

\begin{prop}\label{centerJ} The singular point $J:\left(x_{c10}=-\frac{\la_1}{\sqrt{6}},\
y_{c10}=\sqrt{1+\frac{\la_1^2}{6}},\ z_{c10}=0\right)$ of the system
\eqref{autonomous4} is stable but is not asymptotically stable.
\end{prop}

{\bf Proof}.

In order to translate $J$ to the origin and transforming the
linear part of the resulting vector field into its Jordan
canonical form, we define new variables
$(u,v_1,v_2)\equiv\mathbf{x}$, by the equations \be
u=z,\,v_1=y-\frac{\sqrt{\la_1^2+6}}{\sqrt{6}},\, v_2=\frac{\sqrt{\la_1^2+6} y+x
   \la_1-\sqrt{6}}{\la_1}\nonumber\ee so that \be \left(\begin{array}{c}
u'\\v_1'\\v_2'
\end{array}\right)=\left(\begin{array}{ccccc} 0& 0 &0 \\
0& \frac{1}{2} \left(-\la_1^2-6\right) &0
\\  0& 0 & -\la_1^2-3
\end{array}\right)\left(\begin{array}{c}u\\v_1\\v_2
\end{array}\right)+\left(\begin{array}{c}f(u,v_1,v_2)\\g_1(u,v_1,v_2)\\g_2(u,v_1,v_2)
\end{array}\right)\label{Jcenter1}
\ee where

$f(u,v_1,v_2)=u^2 \left(\frac{\sqrt{\lambda_1^2+6} v_1}{\la_1}-v_2+\frac{\lambda_1}{\sqrt{6}}\right),$
$g_1(u,v_1,v_2)=-\frac{9 v_1^3}{\lambda_1^2}-3 v_1^3+\frac{3
   v_2 \sqrt{\lambda_1^2+6} v_1^2}{\lambda_1}-\frac{3 \sqrt{\frac{3}{2}} \sqrt{\lambda_1^2+6}
   v_1^2}{\lambda_1^2}-\sqrt{6} \sqrt{\lambda_1^2+6} v_1^2-\frac{3 v_2^2
   v_1}{2}+\sqrt{\frac{3}{2}} v_2 \lambda_1
   v_1+\frac{3 \sqrt{6} v_2 v_1}{\lambda_1}-\frac{1}{2} \sqrt{\frac{3}{2}} v_2^2 \sqrt{\lambda_1^2+6},$ and
$g_2(u,v_1,v_2)=-\frac{3 v_2^3}{2}+\sqrt{\frac{3}{2}} \lambda_1
   v_2^2-\frac{1}{2} u \mu_2 v_2^2+\frac{3
   v_1 \sqrt{\lambda_1^2+6} v_2^2}{\lambda_1}-\frac{3 \sqrt{\frac{3}{2}} v_2^2}{\lambda_1}-3
   v_1^2 v_2+\frac{u \lambda_1 \mu_2
   v_2}{\sqrt{6}}+\frac{u v_1 \sqrt{\lambda_1^2+6}
   \mu_2 v_2}{\lambda_1}-3 \sqrt{\frac{3}{2}}
   v_1 \sqrt{\lambda_1^2+6} v_2+\frac{3 \sqrt{6}
   v_1 \sqrt{\lambda_1^2+6} v_2}{\lambda_1^2}-\frac{9 v_1^2 v_2}{\lambda_1^2}-\frac{3
   u v_1^2 \mu_2}{\lambda_1^2}-\frac{9 \sqrt{6}
   v_1^2}{\lambda_1^3}.$
\\
The system \eqref{Jcenter1} is written in diagonal form
\begin{align}
u'  &  =Cu+f\left(  u,\mathbf{v}\right) \nonumber\\
\mathbf{v}'  &  =P\mathbf{v}+\mathbf{g}\left(  u,\mathbf{v}\right)
, \label{Jcenter2}
\end{align}
where $\left(  u,\mathbf{v}\right)
\in\mathbb{R}\times\mathbb{R}^{2},$ $C$ is the zero $1\times1$
matrix, $P$ is a $2\times 2$ matrix with negative eigenvalues and
$f,\mathbf{g}$ vanish at $\mathbf{0}$ and have vanishing
derivatives at $\mathbf{0.}$ The center manifold theorem
\ref{existenceCM} asserts that there exists a 1-dimensional
invariant local center manifold $W^{c}\left( \mathbf{0}\right) $
of \eqref{Jcenter2} tangent to the center subspace (the
$\mathbf{v}=\mathbf{0}$ space) at $\mathbf{0}.$ Moreover,
$W^{c}\left( \mathbf{0}\right)  $ can be represented as
\[
W^{c}\left(  \mathbf{0}\right)  =\left\{  \left(
u,\mathbf{v}\right)
\in\mathbb{R}\times\mathbb{R}^{2}:\mathbf{v}=\mathbf{h}\left(
u\right), \left\vert u\right\vert <\delta\right\};
\mathbf{h}\left(  0\right) =\mathbf{0}, D\mathbf{h}\left( 0\right)
=\mathbf{0}
\]
for $\delta$ sufficiently small (see definition \ref{CMdef}).

The equations for the center manifold of the origin
reduces to

\begin{align} & \frac{36 \left(\lambda_1^2+3\right)
   h_1^3}{\lambda_1}+\left(\frac{6 \sqrt{6}
   \sqrt{\lambda_1^2+6} \left(2 \lambda_1^2+3\right)}{\lambda_1}-36 \sqrt{\lambda_1^2+6}
   h_2\right) h_1^2+\nonumber \\ &+\left(18 \lambda_1
   h_2^2-6 \sqrt{6} \left(\lambda_1^2+6\right)
   h_2+6 \lambda_1 \left(\lambda_1^2+6\right)\right) h_1+\nonumber\\ &+3 \sqrt{6} \lambda_1
   \sqrt{\lambda_1^2+6} h_2^2+\nonumber\\ &+\left(2 \sqrt{6}
   \lambda_1^2 u^2+12 \sqrt{\lambda_1^2+6}
   h_1 u^2-12 \lambda_1 h_2 u^2\right)
   h_1'=0,\nonumber\\
   & 9 \lambda_1^2 h_2^3+3 \lambda_1
   \left(-\sqrt{6} \lambda_1^2+u \mu_2 \lambda_1+3 \sqrt{6}\right) h_2^2+\nonumber\\ &+\lambda_1^2 \left(6
   \lambda_1^2-\sqrt{6} u \mu_2 \lambda_1+18\right) h_2+\nonumber\\ &+h_1^2 \left(18 u \mu_2+18 \left(\lambda_1^2+3\right) h_2+\frac{54
   \sqrt{6}}{\lambda_1}\right)+\nonumber\\ &+h_1 \left(3
   \sqrt{\lambda_1^2+6} \left(3 \sqrt{6} \lambda_1^2-2 u \mu_2 \lambda_1-6 \sqrt{6}\right)
   h_2-18 \lambda_1 \sqrt{\lambda_1^2+6}
   h_2^2\right)+\nonumber\\ &+\left(\sqrt{6} u^2 \lambda_1^3-6 u^2
   h_2 \lambda_1^2+6 u^2 \sqrt{\lambda_1^2+6} h_1 \lambda_1\right) h_2'=0.\label{CcenterJ}
\end{align}
We obtain, using a Taylor series at $u=0,$  that the solution of
\eqref{centerJ} satisfying $\mathbf{h}(0)=0, D\mathbf{h}\left(
0\right) =\mathbf{0}$ is the trivial solution to arbitrary order.
This means that the center manifold of $J$ is a small segment
contained in the z-axis.

In order to examine the stability of the origin for the flow of \eqref{CcenterJ} we proceed as follows.
Using spherical coordinates
\begin{equation}u=r \cos \varphi  \sin \theta,v_1=r \sin \theta \sin \varphi ,v_2=r \cos \theta \label{sphcoordJ}
\end{equation} and
taking the limit $r\rightarrow 0$ the angular equations
$\theta',\varphi'$ become \begin{align} &  \varphi' \to
-\frac{1}{4} \left(\lambda_1^2+6\right) \sin (2 \varphi ),\nonumber\\
& \theta' \to \frac{1}{8} \left(3
   \left(\lambda_1^2+2\right)+\left(\lambda_1^2+6\right) \cos (2 \varphi )\right) \sin (2
   \theta )
   .\label{angularJ}\end{align}

Solving the approximate equations \eqref{angularJ} we obtain
\begin{align} & \theta(\tau)=\tan ^{-1}\left(e^{\left(\lambda_1^2+3\right)
\tau +2 c_2} \sqrt{1+e^{\left(-\lambda_1^2-6\right) \tau +4
   c_1}}\right),\nonumber\\
   & \varphi(\tau)=\tan ^{-1}\left(e^{2 c_1-\frac{1}{2} \left(\lambda_1^2+6\right) \tau }\right),
   \label{approxsolsangularJ}\end{align} where $c_1$ and $c_2$ are integration constants.

By Taylor expanding the radial equation around $r=0$ we obtain the
equation \be r'=-\frac{1}{2} r \left(2 \left(\lambda_1^2+3\right) \cos ^2(\theta )+\left(\lambda_1^2+6\right) \sin ^2(\theta ) \sin ^2(\varphi )\right)+O\left(r^2\right).\label{approxrJ} \ee

By substituting the first order solution
\eqref{approxsolsangularJ} into the equation
\eqref{approxrJ} and solving the resulting equation we obtain
\be r(\tau)= e^{\left(-\lambda_1^2-3\right) \tau } \sqrt{1+e^{2 \left(\lambda_1^2+3\right) \tau +4 c_2}+e^{\tau
   \lambda_1^2+4 \left(c_1+c_2\right)}} c_3\label{approxsolrJ},\ee where $c_3$ is an
integration constant. Substituting \eqref{approxsolsangularJ}
and  \eqref{approxsolrJ} in \eqref{sphcoordJ} and taking the
limit as $\tau\rightarrow +\infty$ we obtain $u\rightarrow u_0=e^{2 c_2} c_3,
v_1\rightarrow 0, v_2\rightarrow 0$ where $u_0=u(0).$ Let be
$\epsilon>0$ an arbitrary number. Then there exists a $\delta>0,$
such that $\delta<\epsilon.$ Let us consider the solution with
initial value $u(0)=u_0,v_1(0)=v_{1 0},v_2(0)=v_{2 0},$ with
$u_0^2+v_{1 0}^2+v_{20}^2<\delta^2.$ Since $u\rightarrow u_0,$
satisfying $|u_0|<\delta,$ then the solution, ${\bf x}(\tau,{\bf
x}_0)$ passing through ${\bf x}_0=\left(u_0, v_{1
0},v_{20}\right)$ at $\tau=0,$ satisfies $\|{\bf x}(\tau,{\bf
x}_0)\|<\epsilon,$  for $\tau$ arbitrarily large. In this way we
prove the stability (but not asymptotic stability) of  $J.$
$\blacksquare$

Hence, the super-accelerating cosmological solution represented by the singular point $J$ is such that nearby solutions remain close to it, but do not approach it asymptotically.

\subsubsection{Analysis at infinity}

Performing the transformation \eqref{TransfVmass2}, the system
\eqref{autonomous4} becomes

\begin{align}
&\rho'= -\frac{3}{2} \rho ^3 \sin ^4\psi+\cos (2 \theta )
\left(\frac{3}{2} \rho  \left(\rho ^2-1\right) \sin ^2\psi+\right.
\nonumber\\ & \ \ \ \ \ \ \ \ \ \ \ \ \ \ \ \ \ \ \ \ \ \left.
-\frac{1}{2}
   \mu_2 \rho ^3 \cos \theta \cos \psi \sin ^3\psi\right)+\nonumber\\ &+\cos \theta \left(-\la_1 \rho ^2
   \sqrt{6-6 \rho ^2} \sin ^2\theta \sin ^3\psi+\right. \nonumber\\ & \ \ \ \ \ \ \ \ \ \ \ \ \ \ \ \ \ \ \ \ \  \left.
   -\frac{1}{4} \rho  \left(-\mu_2 \rho ^2+\cos (2 \psi ) \rho
   ^2+\rho ^2+\mu_2\right) \sin (2 \psi )\right) ,\label{Vmassinfty4a}\\
&   \theta'=3 \cos \theta \sin \theta+\frac{1}{2} \mu_2 \cot \psi
\sin \theta+\nonumber \\ & \ \ \ \ \ \ \ \ \ \ \ \ \ \ \ \ \ \ \ \
\ +\cos (2 \theta ) \left(\frac{\mu_2 \rho ^2 \cos \psi}{2-2 \rho
^2}-\frac{\sqrt{\frac{3}{2}} \la_1 \rho }{\sqrt{1-\rho ^2}}\right)
\sin\psi
    \sin \theta\label{Vmassinfty4b},\\
&   \psi'=\frac{3 \rho ^2 \cos \psi \sin ^3\psi}{2 \left(\rho
^2-1\right)}+\cos (2 \theta ) \left(\frac{\mu_2 \rho ^2 \cos
   \theta \cos ^2\psi \sin ^2\psi}{2 \rho ^2-2}\right. \nonumber\\ & \ \ \ \ \ \ \ \ \ \ \ \ \ \ \ \ \ \ \ \ \  \left.
   -\frac{3}{2} \cos \psi \sin \psi\right)+\nonumber\\ & \ \ \ \ \ \ \ \ \ \ \ \ \ \ \ \ \ \ \ \ \ +\cos \theta
   \left(\frac{1}{2} \cos ^2\psi \left(-\frac{2 \rho ^2 \sin ^2\psi}{\rho ^2-1}-\mu_2\right)\right. \nonumber\\ & \ \ \ \ \ \ \ \ \ \ \ \ \ \ \ \ \ \ \ \ \  \left.
   -\frac{\sqrt{6}
   \lambda_1 \rho  \cos \psi \sin ^2\theta \sin ^2\psi}{\sqrt{1-\rho ^2}}\right)\label{Vmassinfty4c}.
\end{align}

\begin{landscape}

\begin{table*}[ht]
\begin{center}\caption{\label{Vmassinfty4} Asymptotic singular points  of the system
\eqref{autonomous4} (case 4) and their stability. We use the notations
$\beta=\frac{3}{\sqrt{13-4\mu_2+\mu_2^2}},$ $\epsilon=\text{sign}(\mu_2)$ and $\eta=\text{sign}(-11-4\mu_2+\mu_2^2).$ NH stands for
nonhyperbolic.} \begin{tabular}{ccccc} \hline \hline
 Cr. P & Coordinates & Eigenvalues & $\rho'$ & Stability\\
    &  $\theta,\psi, x_r, y_r, z_r$ &  &  & \\
\hline $Q_{22}$ & $0, 0, 0, 0 , 1$ & $0,0$ & 0 & NH; 3D center manifold \vspace{0.2cm}\\
\hline $Q_{23}$ & $0, \frac{\pi}{2}, 1, 0, 0$ & $0,+\infty$ & $-\frac{3}{2}$& { NH; 2D unstable manifold} \\
\vspace{-0.35cm}\\
\hline $Q_{24}$ & $\pi, \frac{\pi}{2}, -1, 0, 0$ & $0,+\infty$ & $-\frac{3}{2}$& { NH; 2D unstable manifold} \\
\vspace{-0.35cm}\\
\hline $Q_{25}$ & $\frac{\pi}{4}, \frac{\pi}{2}, \frac{\sqrt{2}}{2}, \frac{\sqrt{2}}{2}, 0$ & $0,+\infty$ & $-\frac{3}{2}$& { NH; 2D unstable manifold} \\
\vspace{-0.35cm}\\
\hline $Q_{26}$ & $\frac{3\pi}{4}, \frac{\pi}{2}, -\frac{\sqrt{2}}{2}, \frac{\sqrt{2}}{2}, 0$ & $0,+\infty$ & $-\frac{3}{2}$& { NH; 2D unstable manifold}\\
\vspace{-0.35cm}\\
\hline $Q_{27}$ & $\frac{\pi}{4}, \cos^{-1}\left(\frac{3}{\sqrt{11}}\right), \frac{1}{\sqrt{11}}, \frac{1}{\sqrt{11}}, \frac{3}{\sqrt{11}}$ & $-\epsilon\infty,-\infty$ & $\begin{array}{c} >0 \,\text{if}\, \la_1<-11\\ <0 \,\text{if}\, \la_1>-11 \end{array}$ & {$\begin{array}{c} \text{sink} \,\text{if}\, \la_1<-11 \\ \text{and}\, \mu_2>0\\ \text{saddle otherwise} \end{array}$}\\
\vspace{-0.35cm}\\
\hline $Q_{28}$ & $0, \cos^{-1}(\beta), \frac{\beta|\mu_2-2|}{3}, 0,
\beta$ &  $\begin{array}{c} \epsilon\infty, - \infty \, \text{if}\, \mu_2<2\\
+\infty, \eta\infty \, \text{if}\, \mu_2>2\end{array}  $ &
$<0$ &
${\begin{array}{c} \text{source, if}\, \mu_2 >2+\sqrt{15},\\
\text{saddle, otherwise}\end{array} } $   \\
\vspace{-0.35cm}\\
\hline $Q_{29}$ & $\pi, \cos^{-1}(\beta), -\frac{\beta|\mu_2-2|}{3},
0,
\beta$ &  $\begin{array}{c}-\epsilon\infty, \eta \infty \, \text{if}\, \mu_2<2\\
-\infty, -\infty \, \text{if}\, \mu_2>2\end{array}  $ &
$\begin{array}{c}<0 \,\text{if}\,\mu_2<2-\sqrt[3]{18}\\
\text{or}\,\mu_2>2+\sqrt[3]{18}\\>0
\,\text{if}\,2<\mu_2<2+\sqrt[3]{18}\\ \text{or}\, 2-\sqrt[3]{18}<\mu_2<2 \end{array}$ & ${
\begin{array}{c} \text{source, if}\,
\mu_2<2-\sqrt{15}\\
\text{sink, if}\,
0<\mu_2<2\\ \text{or}\, 2<\mu_2<2+\sqrt[3]{18};\\
\text{saddle, otherwise}\end{array} } $  \\
\hline \hline
\end{tabular}
\end{center}
\end{table*}

\end{landscape}

In the limit $\rho\rightarrow 1,$ the leading terms in
\eqref{Vmassinfty4a}-\eqref{Vmassinfty4c} are

\begin{align}
&\rho'\rightarrow -\frac{1}{4} \sin \psi \left(6 \sin ^3\psi+\cos
\theta \left(4 \cos ^3\psi+\mu_2 \cos (2 \theta ) \sin
   \psi \sin (2 \psi )\right)\right),\label{Vmassinfty4d}\\
&   \theta'\rightarrow \frac{\mu_2 \cos (2 \theta ) \cos \psi \sin
\theta \sin \psi}{2 \left(1-\rho ^2\right)}
\label{Vmassinfty4e},\\
&   \psi'\rightarrow -\frac{\cos \psi \sin ^2\psi (\cos \theta
(\mu_2 \cos (2 \theta )-2) \cos \psi+3 \sin \psi)}{2
   \left(1-\rho ^2\right)}\label{Vmassinfty4f}.
\end{align}

The radial equation does not contain the radial coordinate, so the
singular points  can be obtained using the angular equations only.
Setting $\theta'=0,\psi'=0$, we obtain the singular points which
are listed in table \ref{Vmassinfty4}. The stability of these
points is studied by analyzing first the stability of the angular
coordinates and then deducing, from the sign of equation
\eqref{Vmassinfty4d}, the stability on the radial direction.

In table \ref{BasicObsVmassinfty4} are displayed the values of the
basic observables  \eqref{Omegas}, \eqref{wss}, \eqref{wtot} and
\eqref{decc} for the singular points  of the system
\eqref{autonomous4} (case 4) as well as the solution types.

\begin{table*}[ht]
\begin{center}\caption{\label{BasicObsVmassinfty4} Basic observables for the singular points  of the system
\eqref{autonomous4} (case 4). Solution types. We use the notations
$M(\mu_2)=-\frac{1}{9} (\mu_2-2)^2$ and $N(\mu_2)=\frac{1}{6}
\left(-\mu_2^2+4 \mu_2-1\right).$} {\begin{tabular}{cccccc} \hline
\hline
 Cr. P & $\Omega_\phi$ & $w_\phi$ & $w_{\text{tot}}$ & $q$ & Solution type\\
\hline $Q_{22}$ & 0 & 1& 0 & $\frac{1}{2}$ & dust-like \vspace{0.2cm}\\
\hline $Q_{23}$ & $-\infty$ & 1 & $-\infty$ & $-\infty$ & (unphys.) big-rip \\
\vspace{-0.35cm}\\
\hline $Q_{24}$ & $-\infty$ & 1 & $-\infty$ & $-\infty$ & (unphys.) big-rip \\
\vspace{-0.35cm}\\
\hline $Q_{25}$ & 0 & $-\infty$ & $-\infty$ & $-\infty$ &  big-rip \\
\vspace{-0.35cm}\\
\hline $Q_{26}$ &0 & $+\infty$ & $-\infty$ & $-\infty$ & big-rip \\
\vspace{-0.35cm}\\
\hline $Q_{27}$ & 0 & Indet. & $-\frac{2}{9} $ &
$\frac{1}{6}$ &  Matter-dominated \\
\vspace{-0.35cm}\\
\hline $Q_{28}$ & $M(\mu_2)$ & 1& $M(\mu_2)$ & $N(\mu_2)$ & unphysical\\
\vspace{-0.35cm}\\
\hline $Q_{29}$ & $M(\mu_2)$ & 1& $M(\mu_2)$ & $N(\mu_2)$ & unphysical\\
\hline \hline
\end{tabular}}
\end{center}
\end{table*}

In order to perform the numerical experiments for the system
\eqref{Vmassinfty4a}-\eqref{Vmassinfty4c} it is useful to rewrite
the system in the cartesian coordinates $x_r, y_r, z_r.$ The
system \eqref{Vmassinfty4a}-\eqref{Vmassinfty4c} becomes

\begin{align}
& x_r'=\frac{1}{2} \left(\left(6 y_r^2-3 z_r^2+3\right) x_r^3+z_r
\left(-2 \mu_2 y_r^2-(\mu_2-2)
   z_r^2+\mu_2\right) x_r^2+\right. \nonumber\\ & \ \ \ \ \ \ \ \ \ \ \ \ \ \left.  +3 \left(2 y_r^4+\left(z_r^2-1\right) y_r^2+z_r^2-1\right) x_r+\sqrt{6} w \la_1
   \left(2 x_r^2-1\right) y_r^2+\right. \nonumber\\ & \ \ \ \ \ \ \ \ \ \ \ \ \
   \ \ \ \ \ \ \ \ \ \ \ \ \ \left. +\mu_2 z_r \left(2 y_r^2+z_r^2-1\right)\right),
\nonumber\end{align}
\begin{align}& y_r'= \frac{1}{2} y_r
\left(-(\mu_2-2)
   x_r z_r^3-3 \left(x_r^2-y_r^2+1\right) z_r^2+\mu_2 x_r \left(1-2 y_r^2\right) z_r+
   \right. \nonumber\\ & \ \ \ \ \ \ \ \ \ \ \ \
   \ \ \ \ \ \ \ \ \ \ \ \ \ \ \left.+\left(2 y_r^2-1\right)
   \left(\sqrt{6} w \la_1 x_r+3 \left(x_r^2+y_r^2-1\right)\right)\right),
\nonumber\end{align}
\begin{align}& z_r'= \frac{1}{2} z_r
\left(\left(6 y_r^2-3
   z_r^2+3\right) x_r^2+\right. \nonumber\\ &  \ \ \ \ \ \ \ \ \ \ \ \
   \ \ \ \ \ \ \ \ \ \ \ \ \ \ \ \left.+\left(2 y_r^2 \left(\sqrt{6} w \la_1-\mu_2 z_r\right)-(\mu_2-2) z_r
   \left(z_r^2-1\right)\right) x_r+\right. \nonumber\\ &  \ \ \ \ \ \ \ \ \ \ \ \
   \ \ \ \ \ \ \ \ \ \ \ \ \ \ \ \left.
   +3 y_r^2 \left(2 y_r^2+z_r^2-1\right)\right)\nonumber\\
   & w'=\frac{1}{2} w \left(\left(6 y_r^2-3 z_r^2+3\right)
   x_r^2+\right. \nonumber\\ &  \ \ \ \ \ \ \ \ \ \ \ \
   \ \ \ \ \ \ \ \ \ \ \ \ \ \ \ \left. +\left(-(\mu_2-2) z_r^3+\mu_2 z_r+2 y_r^2 \left(\sqrt{6} w \la_1-\mu_2
   z_r\right)\right) x_r+\right. \nonumber\\ &  \ \ \ \ \ \ \ \ \ \ \ \
   \ \ \ \ \ \ \ \ \ \ \ \ \ \ \ \left. +3 y_r^2 \left(2 y_r^2+z_r^2-1\right)\right)\label{Model4cartesian}
\end{align}
where we have used the time re-scaling $$\mathrm{d}\tau\rightarrow
\frac{\mathrm{d}\tau}{1-\rho^2}$$ which leave invariant the orbits
of the phase-space and the time direction (see theorem \ref{ext2})
and we have introduced the auxiliary variable
$w=\sqrt{1-x_r^2-y_r^2-z_r^2}$ to avoid that the numerical
procedure becomes complex-valued at the singular points.

In order to examine the stability of $Q_{22}$ we introduce the
linear coordinate transformation $$u_1=\frac{x_r}{\mu_2},\,
u_2=y_r,\, u_3=z_r-1, u_4=w$$ and Taylor expanding up to third
order the system \eqref{Model4cartesian} becomes
\begin{eqnarray}
&& u_1'=\mu_2 u_1^2+3 u_3 u_1+u_2^2+\frac{3
u_3^2}{2}+u_3+{\cal O}(3),\nonumber\\
&& u_2'=u_1 u_2 \mu_2-3 u_2
u_3+{\cal O}(3),\nonumber\\
&& u_3'=2 \mu_2(2-\mu_2)u_1 u_3+{\cal O}(3),\nonumber\\
&&u_4'=u_1 u_4 \mu_2+{\cal O}(3). \label{Model4Q22}
\end{eqnarray}

The linear part of the vector field \eqref{Model4Q22} is given by
$$\left(
\begin{array}{cccc}
 0 & 0 & 1 & 0 \\
 0 & 0 & 0 & 0 \\
 0 & 0 & 0 & 0 \\
 0 & 0 & 0 & 0
\end{array}
\right).$$

By computing the action of ${\bf L_J^{(2)}}$ on each basis element
on $H^2$ (the vector space of 4-dimensional vector fields of
second order) we have

\begin{align}
& {\bf L_J^{(2)}}(H^2)=\nonumber\\ & \text{span}\left\{\left(
\begin{array}{c}
 u_1^2 \\
 0 \\
 0 \\
 0
\end{array}
\right),\, \left(
\begin{array}{c}
 u_1 u_2 \\
 0 \\
 0 \\
 0
\end{array}
\right),\, \left(
\begin{array}{c}
 u_2^2 \\
 0 \\
 0 \\
 0
\end{array}
\right),\, \left(
\begin{array}{c}
 u_1 u_3 \\
 0 \\
 0 \\
 0
\end{array}
\right),\, \left(
\begin{array}{c}
 u_2 u_3 \\
 0 \\
 0 \\
 0
\end{array}
\right),\right. \nonumber\\ & \left. \ \ \ \ \ \left(
\begin{array}{c}
 u_3^2 \\
 0 \\
 0 \\
 0
\end{array}
\right),\,  \left(
\begin{array}{c}
 u_1 u_4 \\
 0 \\
 0 \\
 0
\end{array}
\right),\, \left(
\begin{array}{c}
 u_2 u_4 \\
 0 \\
 0 \\
 0
\end{array}
\right),\, \left(
\begin{array}{c}
 u_3 u_4 \\
 0 \\
 0 \\
 0
\end{array}
\right),\, \left(
\begin{array}{c}
 u_4^2 \\
 0 \\
 0 \\
 0
\end{array}
\right),\right. \nonumber\\ & \left. \ \ \ \ \ \left(
\begin{array}{c}
 0 \\
 u_1 u_2 \\
 0 \\
 0
\end{array}
\right),\, \left(
\begin{array}{c}
 0 \\
 u_2^2 \\
 0 \\
 0
\end{array}
\right), \,  \left(
\begin{array}{c}
 0 \\
 u_2 u_3 \\
 0 \\
 0
\end{array}
\right),\, \left(
\begin{array}{c}
 0 \\
 u_2 u_4 \\
 0 \\
 0
\end{array}
\right), \, \left(
\begin{array}{c}
 0 \\
 0 \\
 u_1 u_2 \\
 0
\end{array}
\right), \right. \nonumber\\ & \left. \ \ \ \ \ \left(
\begin{array}{c}
 0 \\
 0 \\
 u_2^2 \\
 0
\end{array}
\right),\, \left(
\begin{array}{c}
 0 \\
 0 \\
 u_2 u_3 \\
 0
\end{array}
\right),\, \left(
\begin{array}{c}
 0 \\
 0 \\
 u_2 u_4 \\
 0
\end{array}
\right),\, \left(
\begin{array}{c}
 0 \\
 0 \\
 0 \\
 u_1 u_2
\end{array}
\right),\, \left(
\begin{array}{c}
 0 \\
 0 \\
 0 \\
 u_2^2
\end{array}
\right),  \right. \nonumber\\ & \left. \ \ \ \ \ \ \ \ \ \ \ \ \ \
\ \ \ \ \ \ \ \ \ \ \ \ \left(
\begin{array}{c}
 0 \\
 0 \\
 0\\
 u_2 u_3
\end{array}
\right),\, \left(
\begin{array}{c}
 0 \\
 0 \\
 0 \\
 u_2 u_4
\end{array}
\right)\right\}. \label{nonresonant4}\end{align}

Thus, the second order terms that are linear combinations of the
twenty two vectors in \eqref{nonresonant4} can be eliminated
\cite{wiggins}. To determine the nature of the second order terms
that cannot be eliminated we must compute the complementary space
of \eqref{nonresonant4} which is
\begin{align}
& G^2=\nonumber\\ & \text{span}\left\{\left(
\begin{array}{c}
 0 \\
 u_1^2 \\
 0 \\
 0
\end{array}
\right),\, \left(
\begin{array}{c}
 0 \\
 u_1 u_3 \\
 0 \\
 0
\end{array}
\right),\, \left(
\begin{array}{c}
 0 \\
 u_3^2 \\
 0 \\
 0
\end{array}
\right),\, \left(
\begin{array}{c}
 0 \\
 u_1 u_4 \\
 0 \\
 0
\end{array}
\right),\, \left(
\begin{array}{c}
 0 \\
 u_3 u_4 \\
 0 \\
 0
\end{array}
\right),\right. \nonumber\\ & \left. \ \ \ \ \   \left(
\begin{array}{c}
 0 \\
 u_4^2 \\
 0 \\
 0
\end{array}
\right),\,\left(
\begin{array}{c}
 0 \\
 0 \\
 u_1^2 \\
 0
\end{array}
\right),\, \left(
\begin{array}{c}
 0 \\
 0 \\
 u_1 u_3 \\
 0
\end{array}
\right),\, \left(
\begin{array}{c}
 0 \\
 0 \\
 u_3^2 \\
 0
\end{array}
\right),\, \left(
\begin{array}{c}
 0 \\
 0 \\
 u_1 u_4 \\
 0
\end{array}
\right),\right. \nonumber\\ & \left. \ \ \ \ \ \left(
\begin{array}{c}
 0 \\
 0 \\
 u_3 u_4 \\
 0
\end{array}
\right),\, \left(
\begin{array}{c}
 0 \\
 0 \\
 u_4^2 \\
 0
\end{array}
\right), \,  \left(
\begin{array}{c}
 0 \\
 0 \\
 0 \\
 u_1^2
\end{array}
\right),\, \left(
\begin{array}{c}
 0 \\
 0 \\
 0 \\
 u_1 u_3
\end{array}
\right), \, \left(
\begin{array}{c}
 0 \\
 0 \\
 0 \\
 u_3^2
\end{array}
\right), \right. \nonumber\\ & \left. \ \ \ \ \ \ \ \ \ \ \ \ \ \
\ \ \ \ \ \ \ \ \ \  \left(
\begin{array}{c}
 0 \\
 0 \\
 0 \\
 u_1 u_4
\end{array}
\right), \, \left(
\begin{array}{c}
 0 \\
 0 \\
 0 \\
 u_3 u_4
\end{array}
\right),\, \left(
\begin{array}{c}
 0 \\
 0 \\
 0 \\
 u_4^2
\end{array}
\right)\right\}. \label{resonant4}\end{align}

The normal form of \eqref{Model4Q22}  is given by
\begin{eqnarray}
&& v_1'=v_3+{\cal O}(3),\nonumber\\
&& v_2'={\cal O}(3),\nonumber\\
&& v_3'=\mu_2(2-\mu_2)v_1 v_3+{\cal O}(3),\nonumber\\
&&v_4'=v_1 v_4 \mu_2+{\cal O}(3). \label{normalModel4Q22}
\end{eqnarray}
The general solution  of \eqref{normalModel4Q22} is given by
\begin{eqnarray}
&& v_1=\frac{\sqrt{2} \sqrt{c_1} \tanh \left(\frac{\sqrt{c_1}
\sqrt{\mu_2-2} \sqrt{\mu_2} \left(\tau +2
c_1\right)}{\sqrt{2}}\right)}{\sqrt{\mu_2-2} \sqrt{\mu_2}},\nonumber\\
&& v_3=c_2 \text{sech}^2\left(\frac{\sqrt{c_1} \sqrt{\mu_2-2}
\sqrt{\mu_2} \left(\tau +2
   c_1\right)}{\sqrt{2}}\right),\nonumber\\
&& v_4=c_3 \cosh ^{\frac{2}{\mu_2-2}}\left(\frac{\sqrt{c_1}
\sqrt{\mu_2-2} \sqrt{\mu_2} \left(\tau +2
c_1\right)}{\sqrt{2}}\right).\label{solModel4Q22}
\end{eqnarray}

By analyzing the qualitative behavior of solutions
\eqref{solModel4Q22} we obtain that the dynamical character
$Q_{22}$ is very sensible to the changes on the initial
conditions. Thus, we argue that $Q_{22}$ is unstable.

By an explicit calculation we find that the center manifolds of
$Q_{23,24}$ are $x_r=\pm \left(1-\frac{y_r^2}{2}\right)+{\cal
O}(y_r)^3=\pm\sqrt{1+y^2},$  $z_r=w=0.$ The dynamics on the center
manifolds of  $Q_{15,16}$ are given by $u'=\frac{3}{2}u^2+{\cal
O}(u)^5,$ where $u=y_r.$ From this follows that  $Q_{15,16}$ are
local sources since their center manifolds are unstable (the
origin is a degenerate local maximum of the potential function
$U(u)=-\frac{3 u^4}{8}$).

We obtain by an explicit calculation that the center manifold of
$Q_{25,26}$ is given up to fourth order by the graph $x_r\pm
y_r\mp \sqrt{2}=\mp\sqrt{2}u^2,z_r=0,w=0$ where $u\equiv
y_r-\frac{\sqrt{2}}{2}$. The dynamics on the center manifold is
given by
$$u'=-6 u^3+{\cal O}(5).$$ From this  follows that the center manifold is stable since the origin is a degenerate minimum of the potential $U(u)=\frac{3 u^4}{2}.$  Hence, $Q_{25,26}$ are of saddle type.

\begin{figure}[ht]
\begin{center}
\includegraphics[scale=0.9]{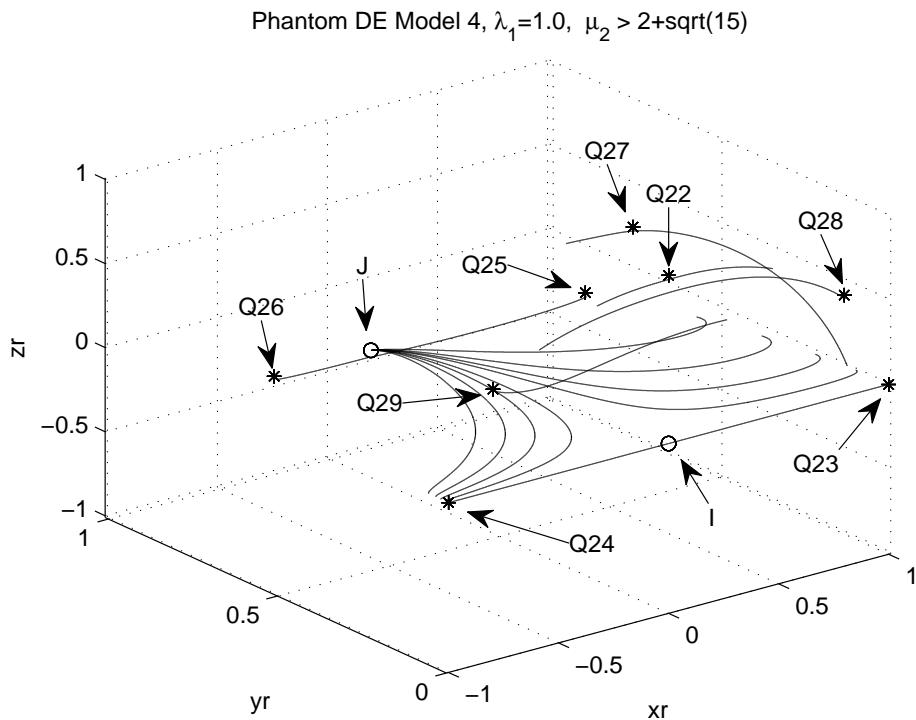}
\caption{{Poincar\'e (global) phase-space of Model 4, for the
parameter values $\la_1=1.0$ and $\mu_2=2.1+\sqrt{15}$. In the
figure the attractor in the finite region is $J.$ The points at
infinity $Q_{25,26}$  are the local sources whereas $Q_{27,28,29}$
are saddles.}} \label{AsymptoticPhantomDEModel4RK4Ba70}
\end{center}
\end{figure}

\begin{figure}[ht]
\begin{center}
\includegraphics[scale=0.9]{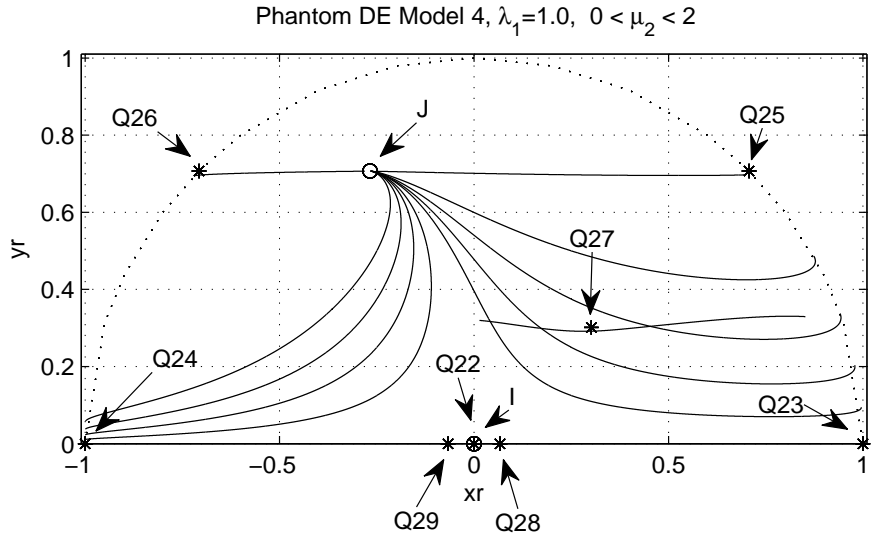}
\caption{{Projection of \ref{AsymptoticPhantomDEModel4RK4Ba70} on
the plane $x_r$-$y_r.$}} \label{AsymptoticPhantomDEModel4RK4B}
\end{center}
\end{figure}

In the figures \ref{AsymptoticPhantomDEModel4RK4Ba70} and
\ref{AsymptoticPhantomDEModel4RK4B} are displayed some orbits in
the Poincar\'e (global) phase-space of Model 4, for the parameter
values $\la_1=1.0$ and $\mu_2=2.1+\sqrt{15}$.  In the figures the
attractor in the finite region is $J.$ The points at infinity
$Q_{25,26}$  are the local sources whereas $Q_{27,28,29}$ are
saddles.

\subsubsection{Cosmological implications and discussion: Model 4}

In this case, the critical points I and J exist always. The point
I corresponds to a flat, non-accelerating, matter-dominated
universe. J corresponds to a dark-energy dominated universe, that
super-accelerates \cite{Das:2005yj}. Similarly to the previous
cases, the stable manifolds of I and J are 1D or 2D respectively,
and thus almost all orbits of the cosmological system cannot be
attracted by I at late times. We proved, however, that J is always
stable (but not asymptotically stable) having a large chance to be
the late-time attractor. Since they cannot lead to
$0<\Omega_\phi<1$, they are not relevant to solve the Coincidence
Problem. Finally, by using Poincar\'e projection, we have obtained
that the physical solutions at infinity are $Q_{22,25,26,27}.$
$Q_{22}$ represents a matter-dominated solution with effective
equation of state $w_{\text{tot}}=0$ (dust-like) which is unstable
by our previous analysis. $Q_{25}$ and $Q_{26}$ corresponds to
initial big-rip singularities due that $q\rightarrow -\infty$ and
$w_{\text{tot}}\rightarrow -\infty.$ That is, the points at
infinity represents supper-accelerating ($q\ll 0$) phantom
solutions ($w_{\text{tot}}\ll -1$). The unphysical solutions are
$Q_{23}$ and $Q_{24}$ that represents unphysical big-rip
singularities since $q\rightarrow -\infty$ and
$w_{\text{tot}}\rightarrow -\infty,$ and $Q_{28,29}.$ The last two
singular points satisfy $\Omega_\phi<0$ for $\mu_2\neq2.$ In the
case $\mu_2=2$ they reduce to $Q_{22}.$ None of these solutions
allows to solve the coincidence problem.

Therefore, an exponential potential and a power-law-dependent
dark-matter particle mass, cannot solve the coincidence problem.

\section{Conclusions}
\label{conclusions}

In this chapter we have investigated the phantom cosmological
scenario, with varying-mass dark-matter particles due to the
interaction between dark-matter and dark-energy sectors. In
particular, we performed a detailed phase-space analysis of
various models, with either exponentially or power-law dependent
dark-matter particle mass, in exponential or power-law scalar
field potentials in both finite and infinite regions. These
functions cover a wide range of the possible forms, and they
correspond to the cases that can accept a reasonable theoretical
justification
\cite{Amendola:1999er,Zhang:2005rg,Amendola:2001rc,Comelli2003,Kneller:2003xg}.
In each case we extracted the critical points in both finite and
infinite regions, we determined their stability, and we calculated
the basic cosmological observables, namely the total
equation-of-state parameter $w_{tot}$ and $\Omega_{{\text{DE}}}$
(attributed to the phantom field). Our basic goal was to examine
whether there exist late-time attractors, corresponding to
accelerating universe and possessing
$\Omega_{{\text{DE}}}/\Omega_{{\text{DM}}}\approx{\cal{O}}(1)$,
thus satisfying the basic observational requirements.

The new results of this investigation are the following:
\begin{enumerate}
\item By performing a Poncar\'e projection we find that for
exponential potential and exponentially-dependent dark-matter
particle mass there is no late-time attractors in the infinite
region. Thus, following the discussion in the section
\ref{FiniteVmass1} the relevant late-time attractor with physical
sense is the phantom-dominated super-accelerated solution, $A,$
for the choice of parameters in the range
$\la_1\left(\mu_1-\la_1\right)<3.$ \item We have proved, using the
Center Manifold theory the proposition \ref{centerE}  that states
that for $\la_2< 0,$ the de Sitter solution for Power-law
potential and power-law dependent dark-matter particle mass is
locally asymptotically stable. For $\la_2>0,$ it is locally
unstable (saddle type). For $\la_2=0,$ it is stable but not
asymptotically stable.

\item By investigating the global phase space of the above models
through Poincar\'e projection, we have obtained saddle points
corresponding to matter-dominated solutions with effective
dust-like equation of state; supper-accelerating early-time
phantom solutions and unphysical big-rip singularities.

\item For power-law potential and exponentially-dependent
dark-matter particle mass we have proved proposition \ref{centerF}
which states that for $\la_2 < 0$ the de Sitter solution is
locally asymptotically stable. For $\la_2>0$ it is locally
unstable (saddle type) whereas for $\la_2= 0$ it is stable but not
asymptotically stable.

\item By investigating the global phase space dynamics through
Poincar\'e projection of the above model-class, we have obtained
unstable matter-dominated solution with dust-like effective
equation of state; supper-accelerating early-time phantom
solutions and unphysical big-rip singularities. None of these
solutions allows to solve the coincidence problem.

\item For exponential potential and power-law-dependent
dark-matter particle mass, we have proved the proposition
\ref{centerJ} that states that the phantom-dominated solution at
finite region is stable but is not asymptotically stable.

\item By investigating the global phase space dynamics through
Poincar\'e projection of the above model-class, we have obtained
saddle points corresponding to matter-dominated solutions with
effective dust-like equation of state; supper-accelerating
early-time phantom solutions and unphysical big-rip singularities.

\end{enumerate}

Thus, we have that for power-law potential and either power-law-
or exponentially-dependent dark matter particles mass, the de
Sitter solution is locally asymptotically stable provided
$\la_2<0$ and stable, but not asymptotically stable for $\la_2=0.$
These results suggest that the dynamical character of the de
Sitter solution depends more on the potential of the scalar field
rather than on the mass-varying function. We have obtained also
that for the case of an exponential potential with an
exponentially-dependent dark-matter particle mass, the
cosmological system possesses a relevant late-time (phantom)
attractor. For the exponential potential and power-law-dependent
dark-matter particle mass, the phantom solution is a relevant
late-time attractor. However, in all the examined cases, solutions
having
$\Omega_{{\text{DE}}}/\Omega_{{\text{DM}}}\approx{\cal{O}}(1)$ are
not relevant attractors at late times. By using the Poicar\'e
projection method we have investigate the infinity region
obtaining as interesting result the existence of both early-time
phantom solutions and past big-rip singularities.

Therefore, summarizing, the coincidence problem cannot be solved
or even alleviated in varying-mass dark matter particles models in
the framework of phantom cosmology, in a radical contrast with the
corresponding quintessence case
\cite{Amendola:1999er,Zhang:2005rg,Comelli2003}. This conclusion
agrees with that of \cite{ChenSaridakis}, that interacting phantom
cosmology cannot solve the coincidence problem. It seems that
interacting phantom cosmology, either directly or through the
dependence of the dark-matter particle mass, cannot fulfill the
basic requirements that led to its construction, that is to
provide stable accelerating late-time solutions which can solve
the coincidence problem. An alternative direction could be to
consider a specially constructed potential or dark-matter particle
mass in order to solve the coincidence problem, but this would
imply significant loss of simplicity, generality, and theoretical
justification of the model.

The aforementioned conclusion has been extracted by the
negative-kinetic-energy realization of phantom, which does not
cover the whole class of phantom models. However, since it is a
qualitative statement it should intuitively be robust for general
phantom scenarios, too. Therefore, phantom cosmology with
varying-mass dark matter particles cannot easily act as a
successful candidate to describe dark energy.

\chapter{Phase-space analysis of Ho\v{r}ava-Lifshitz cosmology}

In this chapter we perform a detailed phase-space analysis of
Ho\v{r}ava-Lifshitz cosmology, with and without the
detailed-balance condition. Under detailed-balance we find that
the universe can reach a bouncing-oscillatory state at late times,
in which dark-energy, behaving as a simple cosmological constant,
is dominant. In the case where the detailed-balance condition is
relaxed, we find that the universe reaches an eternally expanding,
dark-energy-dominated solution, with the oscillatory state
preserving also a small probability. Although this analysis
indicates that Ho\v{r}ava-Lifshitz cosmology can be compatible
with observations, it does not enlighten the discussion about its
possible conceptual and theoretical problems.

\section{Introduction}

Two years ago Ho\v{r}ava proposed a power-counting renormalizable
theory with consistent ultra-violet (UV) behavior \cite{hor3}.
Although presenting an infrared (IR) fixed point, namely General
Relativity, in the  UV the theory exhibits an anisotropic,
Lifshitz scaling between time and space. Due to these novel
features, there has been a large amount of effort in examining the
properties of the theory
\cite{Cai:2009ar,Cai:2009dx,Bogdanos:2009uj,Charmousis:2009tc}.
Furthermore, application of Ho\v{r}ava-Lifshitz gravity as a
cosmological framework gives rise to Ho\v{r}ava-Lifshitz
cosmology, which proves to lead to interesting behavior
\cite{Kiritsis:2009sh}. In particular, one can examine specific
solution subclasses \cite{Lu:2009em}, the phase-space behavior
\cite{Leon:2009rc}, the gravitational wave production
\cite{Mukohyama:2009zs}, the perturbation spectrum
\cite{Mukohyama:2009gg}, the matter bounce
\cite{Brandenberger:2009yt}, the black hole properties
\cite{Danielsson:2009gi}, the dark energy phenomenology
\cite{Saridakis:2009bv}, the observational constraints on the
parameters of the theory \cite{Dutta:2009jn}, the astrophysical
phenomenology \cite{Kim:2009dq}, the thermodynamic properties
\cite{Cai:2009qs} etc. However, despite this extended research,
there are still many ambiguities if Ho\v{r}ava-Lifshitz gravity is
reliable and capable of a successful description of the
gravitational background of our world, as well as of the
cosmological behavior of the universe
\cite{Charmousis:2009tc,Bogdanos:2009uj}.

Let us briefly review the scenario where the
cosmological evolution is governed by the simple version of
Ho\v{r}ava-Lifshitz gravity \cite{Kiritsis:2009sh}. The dynamical
variables are the lapse and shift functions, $N$ and $N_i$
respectively, and the spatial metric $g_{ij}$ (roman letters
indicate spatial indices). In terms of these fields the full
metric is written as:
\begin{eqnarray}\label{metriciit}
ds^2 = - N^2 dt^2 + g_{ij} (dx^i + N^i dt ) ( dx^j + N^j dt ) ,
\end{eqnarray}
and the scaling transformation of the coordinates reads: $
 t \rightarrow l^3 t~~~{\rm and}\ \ x^i \rightarrow l x^i
$.

\subsection{Detailed Balance}

The gravitational action is decomposed into a kinetic and a
potential part as $S_g = \int dt d^3x \sqrt{g} N ({\cal L}_K+{\cal
L}_V)$. The assumption of detailed balance \cite{hor3}
  reduces the possible terms in the Lagrangian, and it allows
for a quantum inheritance principle, since the
$(D+1)$-dimensional theory acquires the renormalization properties
of the $D$-dimensional one. Under the detailed balance condition
 the full action of Ho\v{r}ava-Lifshitz gravity is given by
\begin{eqnarray}
 S_g &=&  \int dt d^3x \sqrt{g} N \left\{
\frac{2}{\kappa^2}
(K_{ij}K^{ij} - \lambda K^2) \ \ \ \ \ \ \ \ \ \ \ \ \ \ \ \ \  \right. \nonumber \\
&+&\left.\frac{\kappa^2}{2 w^4} C_{ij}C^{ij}
 -\frac{\kappa^2 \mu}{2 w^2}
\frac{\epsilon^{ijk}}{\sqrt{g}} R_{il} \nabla_j R^l_k +
\frac{\kappa^2 \mu^2}{8} R_{ij} R^{ij}
     \right. \nonumber \\
&-&\left.    \frac{\kappa^2 \mu^2}{8( 3 \lambda-1)} \left[ \frac{1
- 4 \lambda}{4} R^2 + \Lambda  R - 3 \Lambda ^2 \right] \right\},
\label{acct}
\end{eqnarray}
where $ K_{ij} =   \left( {\dot{g_{ij}}} - \nabla_i N_j - \nabla_j
N_i \right)/2N $
 is the extrinsic curvature and
$ C^{ij}  =  \epsilon^{ijk} \nabla_k \bigl( R^j_i -  R
\delta^j_i/4 \bigr)/\sqrt{g} $ the Cotton tensor, and the
covariant derivatives are defined with respect to the spatial
metric $g_{ij}$. $\epsilon^{ijk}$ is the totally antisymmetric
unit tensor, $\lambda$ is a dimensionless constant and the
variables $\kappa$, $w$ and $\mu$ are constants. Finally, we
mention that in action (\ref{acct}) we have already performed the
usual analytic continuation of the parameters $\mu$ and $w$ of the
original version of Ho\v{r}ava-Lifshitz gravity, since such a
procedure is required in order to obtain a realistic cosmology
\cite{Lu:2009em,Wang:2009rw}.

In order to add the matter component we follow the hydrodynamical
approach   of adding a cosmological stress-energy tensor to the
gravitational field equations, by demanding to recover the usual
general relativity formulation in the low-energy limit
\cite{Carloni:2009jc}. Thus, this
matter-tensor is a hydrodynamical approximation with $\rho_m$ and
$p_m$ (or $\rho_m$ and $w_m$) as parameters. Similarly, one can
additionally include the standard-model-radiation component, with
the additional parameters   $\rho_r$ and $w_r$.

In order to investigate cosmological frameworks, we impose the
projectability condition \cite{Charmousis:2009tc} and we use an
FRW metric
\begin{eqnarray}
N=1~,~~g_{ij}=a^2(t)\gamma_{ij}~,~~N^i=0~,
\end{eqnarray}
with
\begin{eqnarray}
\gamma_{ij}dx^idx^j=\frac{dr^2}{1- K r^2}+r^2d\Omega_2^2~,
\end{eqnarray}
where $ K-1,0,+1$ corresponding  to open, flat, and closed
universe respectively. By varying $N$ and $g_{ij}$, we extract the
Friedmann equations:
\begin{eqnarray}\label{Fr1fluid}
H^2 &=&
\frac{\kappa^2}{6(3\la-1)}\Big(\rho_m+\rho_r\Big)\nonumber\\
&+&\frac{\kappa^2}{6(3\la-1)}\left[ \frac{3\kappa^2\mu^2
K^2}{8(3\lambda-1)a^4} +\frac{3\kappa^2\mu^2\Lambda
^2}{8(3\lambda-1)}
 \right]\nonumber\\
 &-&\frac{\kappa^4\mu^2\Lambda  K}{8(3\lambda-1)^2a^2} \ ,
\end{eqnarray}
\begin{eqnarray}\label{Fr2fluid}
\dot{H}+\frac{3}{2}H^2 &=&
-\frac{\kappa^2}{4(3\la-1)}\Big(w_m\rho_m+w_r\rho_r\Big)\nonumber\\
&-&\frac{\kappa^2}{4(3\la-1)}\left[\frac{\kappa^2\mu^2
K^2}{8(3\lambda-1)a^4} -\frac{3\kappa^2\mu^2\Lambda
^2}{8(3\lambda-1)}
 \right]\nonumber\\
 &-&\frac{\kappa^4\mu^2\Lambda  K}{16(3\lambda-1)^2a^2}\ ,
\end{eqnarray}
where  $H\equiv\frac{\dot a}{a}$ is the Hubble parameter. As
usual, $\rho_m$ follows the standard evolution equation
$
 \dot{\rho}_m+3H(\rho_m+p_m)=0,
$ while $\rho_r$   follows $
 \dot{\rho}_r+3H(\rho_r+p_r)=0.
$
Finally, concerning the dark-energy sector we can define
\begin{equation}\label{rhoDE}
\rho_{DE}\equiv \frac{3\kappa^2\mu^2 K^2}{8(3\lambda-1)a^4}
+\frac{3\kappa^2\mu^2\Lambda ^2}{8(3\lambda-1)}
\end{equation}
\begin{equation}
\label{pDE} p_{DE}\equiv \frac{\kappa^2\mu^2
K^2}{8(3\lambda-1)a^4} -\frac{3\kappa^2\mu^2\Lambda
^2}{8(3\lambda-1)}.
\end{equation}
The term proportional to $a^{-4}$ is the usual ``dark radiation
term'', present in Ho\v{r}ava-Lifshitz cosmology
\cite{Kiritsis:2009sh}, while the constant term is just the
explicit cosmological constant. Therefore, in expressions
(\ref{rhoDE}),(\ref{pDE}) we have defined the energy density and
pressure for the effective dark energy, which incorporates the
aforementioned contributions. Note that using
(\ref{rhoDE}),(\ref{pDE}) it is straightforward to show that these
 dark energy quantities satisfy the
standard evolution equation:
$\dot{\rho}_{DE}+3H(\rho_{DE}+p_{DE})=0.
$

If we require expressions (\ref{Fr1fluid}) to coincide with the
standard Friedmann equations, in units where $c=1$  we set
\cite{Kiritsis:2009sh}:
$
G=\frac{\kappa^2}{16\pi(3\lambda-1)}$ and
$\frac{\kappa^4\mu^2\Lambda}{8(3\lambda-1)^2}=1$
where $G$ is the  Newton's constant.

\subsection{Beyond Detailed Balance}

The aforementioned formulation of Ho\v{r}ava-Lifshitz cosmology
has been performed under the imposition of the detailed-balance
condition. However, in the literature there is a discussion
whether this condition leads to reliable results or if it is able
to reveal the full information of Ho\v{r}ava-Lifshitz
 gravity \cite{Kiritsis:2009sh}. Therefore, one
 needs to investigate also the Friedman equations in the case
 where detailed balance is relaxed. In such a case one can in
 general write
 \cite{Leon:2009rc,Charmousis:2009tc}:
\begin{eqnarray}\label{Fr1c}
H^2 &=&
\frac{2\sigma_0}{(3\la-1)}\Big(\rho_m+\rho_r\Big)\nonumber\\
&+&\frac{2}{(3\la-1)}\left[ \frac{\sigma_1}{6}+\frac{\sigma_3
K^2}{6a^4} +\frac{\sigma_4 K}{6a^6}
 \right]\nonumber\\&+&\frac{\sigma_2}{3(3\la-1)}\frac{ K}{a^2}
\end{eqnarray}
\begin{eqnarray}\label{Fr2c}
\dot{H}+\frac{3}{2}H^2 &=&
-\frac{3\sigma_0}{(3\la-1)}\Big(w_m\rho_m+w_r\rho_r\Big)\nonumber\\
&-&\frac{3}{(3\la-1)}\left[ -\frac{\sigma_1}{6}+\frac{\sigma_3
K^2}{18a^4} +\frac{\sigma_4 K}{6a^6}
 \right]\nonumber\\&+&
 \frac{\sigma_2}{6(3\la-1)}\frac{ K}{a^2},
\end{eqnarray}
where $\sigma_0\equiv \kappa^2/12$, and the constants $\sigma_i$
are arbitrary (with $\sigma_2$ being negative and $\sigma_4$
positive). Furthermore, the   dark-energy quantities
 are generalized to
\begin{eqnarray}\label{rhoDEext}
&&\rho_{DE}|_{_\text{non-db}}\equiv
\frac{\sigma_1}{6}+\frac{\sigma_3 K^2}{6a^4} +\frac{\sigma_4
K}{6a^6}
\\
&&\label{pDEext} p_{DE}|_{_\text{non-db}}\equiv
-\frac{\sigma_1}{6}+\frac{\sigma_3 K^2}{18a^4} +\frac{\sigma_4
K}{6a^6}.
\end{eqnarray}
Again, it is easy to show that
\begin{eqnarray}\label{rhodotfluidnd}
\dot{\rho}_{DE}|_{_\text{non-db}}+3H(\rho_{DE}|_{_\text{non-db}}+p_{DE}|_{_\text{non-db}})=0.
\end{eqnarray}
Finally, if we force (\ref{Fr1c}),(\ref{Fr2c}) to coincide with
 the standard Friedmann equations, we obtain:
$G=\frac{6\sigma_0}{8\pi(3\lambda-1)}$
and $\sigma_2=-3(3\lambda-1)$.

 The above  basic models of   Ho\v{r}ava-Lifshitz cosmology proves to have
very interesting cosmological behavior
\cite{Leon:2009rc,Kiritsis:2009sh,Lu:2009em,Mukohyama:2009gg,Mukohyama:2009zs,Brandenberger:2009yt,
Danielsson:2009gi,Cai:2009qs,Saridakis:2009bv,Kim:2009dq,Dutta:2009jn}
. However, the gravitational sector itself proves to have
instabilities that cannot be cured by simple tricks such as
analytic continuation \cite{Bogdanos:2009uj,Charmousis:2009tc}.
Therefore, it is necessary to try to construct suitable extensions
that are free of such problems.

A quite general power-counting renormalizable action is
\cite{Kiritsis:2009vz}:
 \be S=S_{kin}+S_{1}+S_{2}+S_{new}
\label{q10},\ee with
 \be S_{kin}=\alpha
\int dtd^3x\sqrt{g}N\!\!\left[(K_{ij}K^{ij}\!-\!\l K^2)\right]
\label{q4}\nonumber\ee \be S_{1}=\int
dtd^3x\sqrt{g}N\left[\gamma_0 {\e^{ijk}\over
\sqrt{g}}R_{il}\na_j{R^l}_{k} \!+\!\zeta R_{ij}R^{ij}\!+\!\eta
R^2\!+\!\xi R\!+\!\sigma\!\right]
 \label{q5}\nonumber
 \ee
  \begin{eqnarray} S_2=\int dtd^3x\sqrt{g}N\left[ \beta_0
C_{ij}C^{ij}+\beta_1 R\square
R+\beta_2R^3\right.\nonumber\\
\left.+\beta_3RR_{ij}R^{ij}+\beta_4 R_{ij}R^{ik}{R^{j}}_k\right]
\label{q6}\nonumber
\end{eqnarray}
  \begin{eqnarray} S_{new}=\int
dtd^3x\sqrt{g}N\left[a_1 (a_ia^i)+a_2 (a_ia^i)^2+a_3R^{ij}a_ia_j
\right.\nonumber\\
\left.+a_4R\nabla_i a^i+a_5\nabla_ia_j\nabla^ia^j+ a_6\nabla^i a_i
(a_ja^j)+\cdots \right].\ \ \label{q7}\end{eqnarray} Thus, apart
from the known kinetic, detailed-balance and
beyond-detailed-balance combinations that constitute the
Ho\v{r}ava-Lifshitz gravitational action, in (\ref{q7}) we have
added a new combination, based on the term \cite{Blas:2009qj}:
 \be
 a_i\equiv {\partial_iN\over N},
 \label{q11}
 \ee
 which breaks the projectability condition,
and the ellipsis in (\ref{q7}) refers to dimension six terms
involving $a_i$ as well as curvatures.

Such a new combination of terms seems to alleviate the problems of
Ho\v{r}ava-Lifshitz gravity, although there could still be some
ambiguities. Therefore, one should repeat
all the relevant investigations of the literature for this extended
version of the theory.

\section{The cosmological equations}
\label{HLmodel}

The cosmological equations of Ho\v{r}ava-Lifshitz cosmology including a scalar field matter source, under
the imposition of the detailed-balance condition, are:
\begin{eqnarray}\label{HLFr1}
H^2 &=&
\frac{\kappa^2}{6(3\la-1)}\left[\frac{3\la-1}{4}\,\dot\phi^2
+V(\phi)\right]+\nonumber\\
&+&\frac{\kappa^2}{6(3\la-1)}\left[
-\frac{3\kappa^2\mu^2k^2}{8(3\lambda-1)a^4}
-\frac{3\kappa^2\mu^2\Lambda ^2}{8(3\lambda-1)}
 \right]+\nonumber\\
 &+&\frac{\kappa^4\mu^2\Lambda k}{8(3\lambda-1)^2a^2} \ ,
\end{eqnarray}
\begin{eqnarray}\label{HLFr2}
\dot{H}+\frac{3}{2}H^2 &=&
-\frac{\kappa^2}{4(3\la-1)}\left[\frac{3\la-1}{4}\,\dot\phi^2
-V(\phi)\right]-\nonumber\\
&-&\frac{\kappa^2}{4(3\la-1)}\left[-\frac{\kappa^2\mu^2k^2}{8(3\lambda-1)a^4}
+\frac{3\kappa^2\mu^2\Lambda ^2}{8(3\lambda-1)}
 \right]+\nonumber\\
 &+&\frac{\kappa^4\mu^2\Lambda k}{16(3\lambda-1)^2a^2}\ ,
\end{eqnarray}
where we have defined the Hubble parameter as $H\equiv\frac{\dot
a}{a}$, and we have neglected radiation from the cosmological budget.

Finally, the equation of motion for the scalar field
reads:
\begin{eqnarray}\label{HLphidott}
&&\ddot\phi+3H\dot\phi+\frac{2}{3\lambda-1}\frac{dV(\phi)}{d\phi}=0.
\end{eqnarray}

At this stage we can define the energy density and pressure for
the scalar field responsible for the matter content of the
Ho\v{r}ava-Lifshitz universe:
\begin{eqnarray}
&&\rho_M\equiv \rho_\phi=\frac{3\la-1}{4}\,\dot\phi^2
+V(\phi)\label{HLrhom}\\
&&p_M\equiv p_\phi=\frac{3\la-1}{4}\,\dot\phi^2
-V(\phi).\label{HLpressurem}
\end{eqnarray}
Concerning the dark-energy sector we can define
\begin{equation}\label{HLrhoDE}
\rho_{DE}\equiv -\frac{3\kappa^2\mu^2k^2}{8(3\lambda-1)a^4}
-\frac{3\kappa^2\mu^2\Lambda ^2}{8(3\lambda-1)}
\end{equation}
\begin{equation}
\label{HLpDE} p_{DE}\equiv
-\frac{\kappa^2\mu^2k^2}{8(3\lambda-1)a^4}
+\frac{3\kappa^2\mu^2\Lambda ^2}{8(3\lambda-1)}.
\end{equation}
The term proportional to $a^{-4}$ is the usual ``dark radiation
term'', present in Ho\v{r}ava-Lifshitz cosmology
\cite{Calcagni:2009ar,Kiritsis:2009sh}. Finally, the constant term
is just the explicit (negative) cosmological constant. Therefore,
in expressions (\ref{HLrhoDE}),(\ref{HLpDE}) we have defined the
energy density and pressure for the effective dark energy, which
incorporates the aforementioned contributions.

Using the above definitions, we can re-write the Friedmann
equations (\ref{HLFr1}),(\ref{HLFr2}) in the standard form:
\begin{eqnarray}
\label{HLFr1b} H^2 &=&
\frac{\kappa^2}{6(3\la-1)}\Big[\rho_M+\rho_{DE}\Big]+ \frac{\beta k}{a^2}\\
\label{HLFr2b} \dot{H}+\frac{3}{2}H^2 &=&
-\frac{\kappa^2}{4(3\la-1)}\Big[p_M+p_{DE}
 \Big]+ \frac{\beta k}{2a^2}.
\end{eqnarray}
In these relations we have defined
$\beta\equiv\frac{\kappa^4\mu^2\Lambda }{8(3\lambda-1)^2}$, which
is the coefficient of the curvature term. Additionally, we could
also define an effective Newton's constant and an effective light
speed \cite{Calcagni:2009ar,Kiritsis:2009sh}, but we prefer to
keep $\frac{\kappa^2}{6(3\la-1)}$ in the expressions, just to make
clear the origin of these terms in Ho\v{r}ava-Lifshitz cosmology.
Finally, note that using (\ref{HLphidott}) it is straightforward
to see that the aforementioned dark matter and dark energy
quantities verify the standard evolution equations:
\begin{eqnarray}\label{HLphidot2}
&&\dot{\rho}_M+3H(\rho_M+p_M)=0\\
\label{HLsdot2} &&\dot{\rho}_{DE}+3H(\rho_{DE}+p_{DE})=0.
\end{eqnarray}

In the literature there is a discussion whether the
detailed-balance condition leads to reliable results or if it is
able to reveal the full information of Ho\v{r}ava-Lifshitz gravity
\cite{Calcagni:2009ar,Kiritsis:2009sh}. Thus, for
 completeness, we add here the Friedmann equation in the case
 where detailed balance is relaxed. In such a case one can in
 general write
 \cite{Sotiriou:2009bx,Bogdanos:2009uj,Charmousis:2009tc}:
\begin{eqnarray}\label{HLFr1c}
H^2 &=&
\frac{2\sigma_0}{(3\la-1)}\left[\frac{3\la-1}{4}\,\dot\phi^2
+V(\phi)\right]+\nonumber\\
&+&\frac{2}{(3\la-1)}\left[
\frac{\sigma_1}{6}+\frac{\sigma_3k^2}{6a^4} +\frac{\sigma_4k}{a^6}
 \right]+\nonumber\\&+&\frac{\sigma_2}{3(3\la-1)}\frac{k}{a^2}
\end{eqnarray}
\begin{eqnarray}\label{HLFr2c}
\dot{H}+\frac{3}{2}H^2 &=&
-\frac{3\sigma_0}{(3\la-1)}\left[\frac{3\la-1}{4}\,\dot\phi^2
-V(\phi)\right]-\nonumber\\
&-&\frac{3}{(3\la-1)}\left[
-\frac{\sigma_1}{6}+\frac{\sigma_3k^2}{18a^4}
+\frac{\sigma_4k}{6a^6}
 \right]+\nonumber\\&+&
 \frac{\sigma_2}{6(3\la-1)}\frac{k}{a^2} ,
\end{eqnarray}
where $\sigma_0\equiv \kappa^2/12$, and the constants $\sigma_i$
are arbitrary (although one can set $\sigma_2$ to be positive
too). Thus, the effect of the detailed-balance relaxation is the
decoupling of the coefficients, together with the appearance of a
term proportional to $a^{-6}$. This term has a negligible impact
at large scale factors, however it could play a significant role
at small ones. Finally, in the non-detailed-balanced case, the
energy density and pressure for matter coincide with those of
detailed-balance scenario (expressions
(\ref{HLrhom}),(\ref{HLpressurem})),  since the detailed-balance
condition affects only the gravitational sector of the theory and
has nothing to do with the matter content of the universe.
However, the corresponding quantities for dark energy are
generalized to
\begin{eqnarray}\label{HLrhoDEext}
&&\rho_{DE}|_{_\text{non-db}}\equiv
\frac{\sigma_1}{6}+\frac{\sigma_3k^2}{6a^4} +\frac{\sigma_4k}{a^6}
\\
&&\label{HLpDEext} p_{DE}|_{_\text{non-db}}\equiv
-\frac{\sigma_1}{6}+\frac{\sigma_3k^2}{18a^4}
+\frac{\sigma_4k}{6a^6}.
\end{eqnarray}

\section{Detailed balance: Phase-space analysis}
\label{HLphaseanalysis}

In order to perform the phase-space and stability analysis of the
Ho\v{r}ava-Lifshitz universe, we have to transform the
cosmological equations into an autonomous dynamical system
\cite{Copeland:1997et,expon,expon1,expon2}. This will be achieved
by introducing the auxiliary variables:
\begin{eqnarray}
&&x=\frac{\kappa  \dot \phi }{2 \sqrt{6} H}, \label{HLauxilliaryx}\\
&&y=\frac{\kappa \sqrt{V(\phi)}}{\sqrt{6}H \sqrt{3 \lambda -1}} \label{HLauxilliaryy}\\
&&z=\frac{\kappa ^2 \mu }{4 (3 \lambda -1) a^2 H} \label{HLauxilliaryz}\\
&&u=\frac{\kappa ^2 \Lambda  \mu }{4 (3 \lambda -1) H},
 \label{HLauxilliaryu}
\end{eqnarray}
together with $M=\ln a$. Thus, it is easy to see that for every
quantity $F$ we acquire $\dot{F}=H\frac{dF}{dM}$.
 Using these
variables we can straightforwardly obtain the density parameters
of dark matter and dark energy (through expressions
(\ref{HLrhom}), (\ref{HLrhoDE})) as:
\begin{eqnarray}
&&
\Omega_M\equiv\frac{\kappa^{2}}{6(3\lambda-1)H^{2}}\rho_M=x^2+y^2,
 \label{HLOmegaM}\\
&&
\Omega_{DE}\equiv\frac{\kappa^{2}}{6(3\lambda-1)H^{2}}\rho_{DE}=-k^2z^2-u^2
 \label{HLOmegaDE},
\end{eqnarray}
 and in addition we can calculate the corresponding
 equation-of-state parameters:
\begin{eqnarray}
&& w_M\equiv\frac{p_M}{\rho_M}=\frac{x^2-y^2}{x^2+y^2},
 \label{HLwM}\\
 && w_{DE}\equiv\frac{p_{DE}}{\rho_{DE}}=\frac{k^2z^2-3u^2}{3k^2z^2+3u^2}
 \label{HLwDE}.
\end{eqnarray}
We mention that these relations are always valid, that is
independently of the specific state of the system (they are valid
in the whole phase-space and not only at the singular points).
Finally, for completeness, and observing (\ref{HLFr1b}), we can
define the curvature density parameter as:
\begin{eqnarray}
&& \Omega_k\equiv\frac{\beta k}{H^2a^2}=2kuz.
 \label{HLOmegaK}
\end{eqnarray}

Using the auxiliary variables
(\ref{HLauxilliaryx}),(\ref{HLauxilliaryy}),(\ref{HLauxilliaryz}),(\ref{HLauxilliaryu})
the cosmological equations of motion (\ref{HLFr1b}),
(\ref{HLFr2b}), (\ref{HLphidot2}) and (\ref{HLsdot2}), can be
transformed into an autonomous form
 $\textbf{X}'=\textbf{f(X)}, $ where $\textbf{X}$ is the column
vector constituted by the auxiliary variables, \textbf{f(X)} the
corresponding  column vector of the autonomous equations, and
prime denotes derivative with respect to $M=\ln a$.

In the following we perform a phase-space analysis of the
cosmological system at hand. As we can see from the Friedmann
equations (\ref{HLFr1}), (\ref{HLFr2}) one can have a zero or
non-zero cosmological constant, in a flat or non-flat universe.
Thus, for simplicity we investigate separately the corresponding
four cases. Finally, note that we assume $\la
>\frac{1}{3}$ as required by the consistency of the Ho\v{r}ava
gravitational background, but we do not impose any other
constraint on the model parameters (although one could do so using
the light speed and Newton's constant values) in order to remain
as general as possible.

\subsection{Case 1: Flat universe with $\Lambda=0$}\label{HLsection A}

In this scenario the variable $u$ is irrelevant, and the Friedmann
equations (\ref{HLFr1}), (\ref{HLFr2}) become:
\begin{eqnarray}
&&1=x^2+y^2 \label{HLFr1bcase1}
\\
&&\frac{H'}{H}=-3 x^2. \label{HLFr2bcase1}
\end{eqnarray}
Thus, after using the first of these relations in order to
eliminate one variable, the corresponding autonomous system
writes:
\begin{eqnarray}
x'&=&\left(3 x-\sqrt{6} s\right)
\left(x^2-1\right),\label{HLeqxcase1}\\
z'&=&  \left(3 x^2-2\right) z.\label{HLeqzcase1}
\end{eqnarray}
We mention that for simplicity we have set $s=-\frac{1}{\kappa
V(\phi)}\frac{d V(\phi)}{d\phi}$ and we have assumed it to be a
constant, that is we are investigating the usual exponential
potentials. However, as we will see this is not necessary, since
the most important results of the present work are independent of
the matter sector.

The autonomous system \eqref{HLeqxcase1}-\eqref{HLeqzcase1} is
defined in the phase space $$\Psi=\left\{(x,z): -1\leq x\leq 1,
z\in\mathbb{R}\right\}.$$ As we observe, this phase plane is not
compact since $z$ is in general unbounded. However, the system is
integrable and the orbit in the plane $\Psi$, passing initially
through $(x_0,\,z_0)$, can be obtained explicitly and it is given
by the graph
\begin{eqnarray}
z(x)=z_0\left(\frac{3 x-\sqrt{6}
   s}{3
   x_0-\sqrt{6} s}\right)^{1+\frac{1}{2 s^2-3}}
   \left(\frac{x^2-1}{x_0^2-1}\right)^{\frac{1}{6-4
   s^2}}\cdot\nonumber\\
  \cdot \exp\left\{\frac{\sqrt{6} s \left[\tanh ^{-1}(x)-\tanh ^{-1}(x_0)\right]}{6
   s^2-9}\right\}.
\end{eqnarray}

\subsubsection{Finite analysis}

The singular points $(x_c,z_c)$ of the autonomous system
(\ref{HLeqzcase1})  are obtained by setting the left hand sides of
the equations to zero. They are displayed in table \ref{HLcrit},
where we also present the necessary conditions for their
existence. In addition, for each singular point we calculate the
values of $w_{M}$ (given by relation (\ref{HLwM})), of
$\Omega_{DE}$ (given by (\ref{HLOmegaDE})), and of $w_{DE}$ (given
by (\ref{HLwDE})). Note that in this case, $w_{DE}$ remains
unspecified and the results hold independently of its value. The
cosmological solutions associated with the singular points can be
derived from the Raychaudhury equation \eqref{HLFr2bcase1}.
Concerning the type and stability of singular points, for
hyperbolic singular points (all the eigenvalues of the
linearization matrix have real parts different from zero) one can
easily extract their type (source (unstable) for positive real
parts, saddle for real parts of different sign and sink (stable)
for negative real parts). However, if at least one eigenvalue has
a zero real part (non-hyperbolic singular point) one is not able
to obtain conclusive information about the stability from
linearization and needs to resort to other tools like Normal Forms
calculations \cite{arrowsmith,wiggins}, or numerical
experimentation.

In the following we will discuss about the eigensystems
(eigenvalues and associated eigenvectors) of the linearization
evaluated at each singular points displayed in table \ref{HLcrit}.
We summarize in table \ref{HLcrit} their type and stability,
acquired by examining the sign of the real part of the eigenvalues
and some basic observables ($w_M,$ $\Omega_{DE},$and $w_{DE}$)
evaluated at the singular points. In table \ref{HLsols} we display
the corresponding cosmological solutions.
\begin{table*}[ht]
\begin{center}\caption{\label{HLcrit} Finite singular points of  a flat universe
with $\Lambda=0$ (case 1) and their behavior. NH stands for
nonhyperbolic (adapted from \cite{Leon:2009rc}).}
\begin{tabular}{cccccccc}
\hline \hline
 Cr. P &$x_c$&$z_c$&  Existence &  Stable for&  $w_M$&  $\Omega_{DE}$&  $w_{DE}$\\
\hline $P_{1,2}$& $\pm 1$ &0&  All $s$   &    unstable& 1& 0&  arbitrary  \vspace{0.2cm}\\
\hline $P_{3}$& $\sqrt{\frac{2}{3}}\,s$ &0&  $s^2<\frac{3}{2}$
  &    $s^2<1$ & $\frac{4}{3}s^2-1$& 0&arbitrary  \\
\vspace{-0.35cm}\\
\hline $P_{4}$ & $\sqrt{\frac{2}{3}} s$ &$z_c$&  $s=\pm 1$
  &   NH & $1/3$& 0&arbitrary  \\
\vspace{-0.35cm}\\
\hline \hline
\end{tabular}
\end{center}
\end{table*}

\begin{table*}[ht]
\begin{center}\caption{\label{HLsols} Finite singular points and their corresponding solutions for a flat universe
with $\Lambda=0$ (case 1).}
\begin{tabular}{ccc}
\hline \hline
 Cr. P & Solution & Energy Density\\
\hline $P_{1,2}$& $a\propto (t-t_0)^{\frac{1}{3}}$&  $\rho_M\propto (t-t_0)^{-2}$ \vspace{0.2cm}\\
\hline $P_{3}$& $a\propto (t-t_0)^{\frac{1}{2 q^2}}$&  $\rho_M\propto (t-t_0)^{-2}$  \\
\vspace{-0.35cm}\\
\hline $P_{4}$ & $a\propto (t-t_0)^{\frac{1}{2}}$&  $\rho_M\propto (t-t_0)^{-2}$   \\
\vspace{-0.35cm}\\
\hline \hline
\end{tabular}
\end{center}
\end{table*}

The singular points $P_{1,2}$ exist for all the $s$-values. The
singular points $P_{1,2}$ have eigenvalues $\{\mp 2 \sqrt{6}
s+6,\, 1\}$ with associated eigenvectors $\{1,0\},\,  \{0,1\}.$
Thus, the singular points are of saddle type with a 1-dimensional
unstable manifold tangent to the z-axis provided $\pm
s>\sqrt{\frac{3}{2}}$ (the sign $+$ for $P_1$ and the sign $-$ for
$P_2$). Otherwise they are local sources.

The singular point $P_3$ exists for $s^2<\frac{3}{2}.$ The
singular point $P_3$ has eigenvalues $\{-3+2 s^2,\,
2\left(-1+s^2\right)\}$ with associated eigenvectors $\{1,0\},\,
\{0,1\}.$ Thus, the singular point is non-hyperbolic for $s=\pm1;$
a sink provided $-1<s<1.$ Otherwise, it is a saddle with
1-dimensional unstable manifold tangent to the z-axis.

Finally, note that in the special case where $s=\pm1$, the system
admits an extra curve of singular points $P_4.$ Each point in
$P_4$ is non-hyperbolic, with center manifold tangent to the
z-axis, but the curve $P_4$ is actually ``normally hyperbolic''
\cite{Aulbach1984a}. This means that we can indeed analyze the
stability by analyzing the sign of the real parts of the non-null
eigenvalues. Therefore, since the non zero eigenvalue is negative,
$P_4$ is a local attractor.

\begin{figure}[ht]
\begin{center}
\mbox{\epsfig{figure=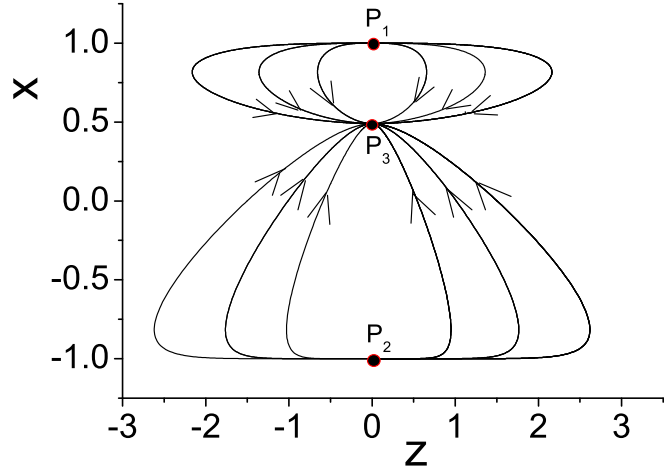,width=8.8cm,angle=0}} \caption{
{Phase plane for a flat universe with $\Lambda=0$ (case 1), for
the choice $s=0.6.$ The singular points $P_{1}$ and $P_2$ are
unstable (sources), while $P_3$ is a global attractor. [Taken from
\cite{Leon:2009rc} and published with permission of IOP Publishing
Ltd]. See the global phase space in figure
\ref{Case1FlatLequal0s0p6}.}} \label{HLFig1}
\end{center}
\end{figure}

In order to present the aforementioned behavior more
transparently, we evolve the autonomous system (\ref{HLeqzcase1})
numerically for  the choice $s=0.6$, and the results are shown in
figure \ref{HLFig1}. As we can wee, in this case the singular
point $P_3$ is the global attractor of the system.

\subsubsection{Analysis at infinity}\label{7.2.1.2}

Owing to the fact that the dynamical system
\eqref{HLeqxcase1}-\eqref{HLeqzcase1} is non-compact, there could
be features in the asymptotic regime which are non-trivial for the
global dynamics. Thus, in order to complete the analysis of the
phase space we will now extend our study using the Poincar\'e
central projection method.

Let us introduce the Poincar\'e variables
\begin{equation}\label{TransfHL1}
x_r=\rho \cos\theta,\, z_r=\rho \sin\theta,
\end{equation} where $\rho=\frac{r}{\sqrt{1+r^2}},$ $r=\sqrt{x^2+z^2}$ and $\theta\in[0,2\pi].$ Thus, the points at ``infinite''
($r\rightarrow+\infty$) are those having $\rho\rightarrow 1.$ The
region of physical interest is given by
$$2x_r^2+z_r^2\leq 1.$$

Performing the transformation \eqref{TransfHL1}, the system
\eqref{HLeqxcase1}-\eqref{HLeqzcase1} becomes
\begin{eqnarray}
& \rho'=-s \rho ^2 \sqrt{6-6 \rho ^2} \cos ^3\theta+\sqrt{6} s
\left(1-\rho ^2\right)^{3/2} \cos \theta+\nonumber\\
& +\frac{1}{2} \rho
   \left(8 \rho ^2+\left(4 \rho ^2-1\right) \cos (2 \theta )-5\right),\label{HLinfty1a}\\
&\theta'=\left(\frac{\sqrt{6} s \rho  \cos ^2(\theta
   )}{\sqrt{1-\rho ^2}}+\cos \theta-\frac{s \sqrt{6-6 \rho ^2}}{\rho }\right) \sin \theta.\label{HLinfty1b}
\end{eqnarray}

In the limit $\rho\rightarrow 1,$ the leading terms in
\eqref{HLinfty1a} and in \eqref{HLinfty1b} are
\begin{eqnarray}
&&\rho'\rightarrow 3 \cos^2\theta,\label{HLinfty1c}\\
&&\theta'\rightarrow \frac{\sqrt{6} s   \cos ^2(\theta
   ) \sin \theta}{\sqrt{1-\rho ^2}} .\label{HLinfty1d}
\end{eqnarray}

Note that the radial equation does not contain the radial
coordinate, so the singular points  can be obtained using the
angular equation only. Setting $\theta'=0$, we obtain the singular
points  which are listed in table \ref{HLinfty1}. The stability of
these points is studied by analyzing first the stability of the
angular coordinate and then deducing, from the sign of equation
\eqref{HLinfty1a}, the stability on the radial direction. We only
need to perturb the angular variable $\theta$ around the
equilibrium points $\theta_i$ via $\theta=\theta_i+\delta\theta.$
The equilibrium points will be stable if $\rho' > 0$ and the
eigenvalue $\lambda< 0$ for the linearized equation
$\delta\theta'=\lambda \delta\theta$, in the limit of
$\rho\rightarrow 1$. When both conditions are satisfied the point
is an stable node, if only one is satisfied it is a saddle and
when neither holds it is an unstable node.

In table \ref{HLinfty1} are displayed the location of the
asymptotic singular points  of the system
\eqref{HLeqxcase1}-\eqref{HLeqzcase1} (case 1) and their
stability. The point $Q_1$ is sink provided $s>0$ and a saddle
otherwise. However this solution is unphysical since
$$2x_r^2+z_r^2> 1.$$ The singular points $Q_2$ and $Q_3$ are
nonhyperbolic. Thus, we cannot anticipate its dynamical character
from linearization. In such a case we can rely on numerical
experimentation.

\begin{table*}[ht]
\begin{center}\caption{\label{HLinfty1} Asymptotic singular points  of the system
\eqref{HLeqxcase1}-\eqref{HLeqzcase1} (case 1) and their
stability. }
\begin{tabular}{ccccc}
\hline \hline
 Cr. P & Coordinates: $\theta, x_r, z_r$ & Eigenvalue & $\rho'$ & Stability\\
\hline $Q_1$ & $0, 1, 0$ & $\left\{\begin{array}{cc}
  -\infty & \text{for}\; s<0\\
  +\infty & \text{for}\; s>0\\
\end{array}\right. $  & $3$ &  $\left\{\begin{array}{c} \text{sink}\\ \text{saddle} \end{array}\right.$ \vspace{0.2cm}\\
\hline $Q_{2}$ & $\frac{\pi}{2}, 0, 1$ & $0$ & $0$ & nonhyperbolic \\
\vspace{-0.35cm}\\
\hline $Q_{3}$ & $\frac{3\pi}{2}, 0, -1$ & $0$ & $0$ &
nonhyperbolic
\\\hline \hline
\end{tabular}
\end{center}
\end{table*}

\begin{figure}[h]
    \centering
        \includegraphics[scale=0.8]{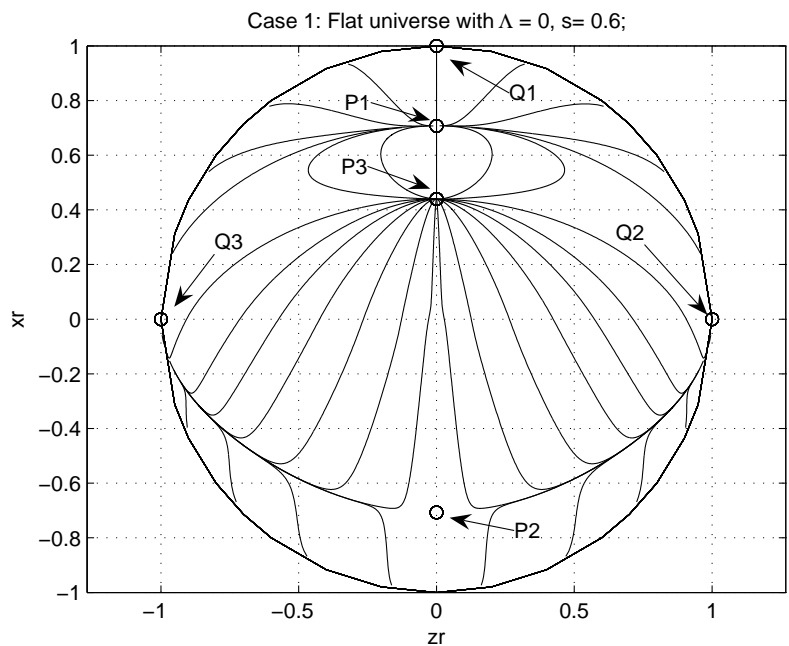}
    \caption{Poincar\'e projection (global phase space) of the system
\eqref{HLeqxcase1}-\eqref{HLeqzcase1} (case 1) for the choice
$s=0.6$. Observe that the points at infinity $Q_{1,2,3}$ are
saddles, whereas the finite point $P_3$ is the global attractor.}
    \label{Case1FlatLequal0s0p6}
\end{figure}

\begin{figure}[h]
    \centering
        \includegraphics[scale=0.8]{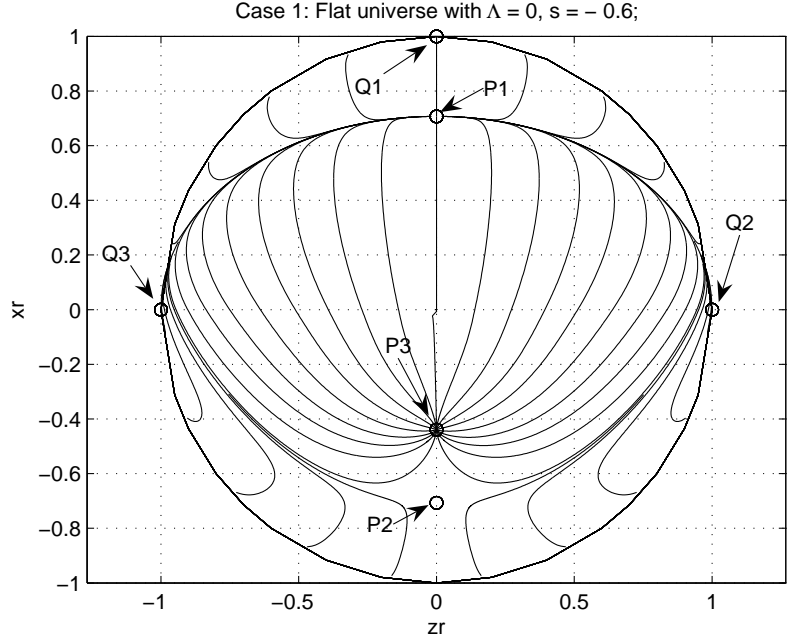}
    \caption{Global phase space of the system
\eqref{HLeqxcase1}-\eqref{HLeqzcase1} (case 1) for the choice
$s=-0.6$. The points at infinity $Q_2$ and $Q_3$ are saddles. The
point $Q_1$ is a local attractor at infinity (but it is
unphysical) and $P_3$ is a local attractor at the finite region.}
    \label{Case1FlatLequal0s0p6n}
\end{figure}

In figures \ref{Case1FlatLequal0s0p6} and
\ref{Case1FlatLequal0s0p6n} are drawn orbits in the global phase
space of the system \eqref{HLeqxcase1}-\eqref{HLeqzcase1} (case
1). That is, the projection of the Poincar\'e sphere in the plane
passing by its equator. The points at infinity $Q_{i}, i=1,2,3$
are projected on the circle $x_r^2+z_r^2=1.$ For $s>0,$ the points
at infinity $Q_{1,2,3}$ are saddles, whereas the finite point
$P_3$ is the global attractor (see figure
\ref{Case1FlatLequal0s0p6} for the choice $s=0.6$). For $s<0$ The
points at infinity $Q_2$ and $Q_3$ are saddles. The point $Q_1$ is
a local attractor at infinity and $P_3$ is a local attractor at
the finite region (see figure \ref{Case1FlatLequal0s0p6n} for the
choice $s=-0.6$).

\subsubsection{Cosmological implications for Case 1: flat universe with
$\Lambda=0$}\label{HLsection A2}

In this scenario the singular points $P_{1,2}$ are not relevant
from a cosmological point of view, since apart from being unstable
they correspond to complete dark matter domination, with the
matter equation-of-state parameter being unphysically stiff.
However, point $P_3$ is more interesting since it is stable for
$-1<q<1$ and thus it can be the late-time state of the universe.
If additionally we desire to keep the dark-matter
equation-of-state parameter in the physical range $0<w_M<1$ then
we have to restrict the parameter $q$ in the range
$\sqrt{3}/2<q<\sqrt{3/2}$. However, even in this case the universe
is finally completely dominated by dark matter. The fact that
$z_c=0$ means that in general this sub-class of universes will be
expand forever. The singular points $P_4$ consist a stable
late-time solution, with a physical dark-matter equation-of-state
parameter $w_M=1/3$, but with zero dark energy density. We mention
that the dark-matter domination of the case at hand was expected,
since in the absent of curvature and of a cosmological constant
the corresponding Ho\v{r}ava-Lifshitz universe is comprised only
by dark matter. Note however that the dark-energy
equation-of-state parameter can be arbitrary. We have obtained an
unphysical attractor at the infinity region.

\subsection{Case 2: non-flat universe with $\Lambda=0$}\label{HLsection B}

Under this scenario, and using the auxiliary variables
(\ref{HLauxilliaryx}),(\ref{HLauxilliaryy}),(\ref{HLauxilliaryz}),(\ref{HLauxilliaryu}),
the Friedmann equations (\ref{HLFr1}), (\ref{HLFr2}) become:
\begin{eqnarray}
&&1=x^2+y^2-z^2 \label{HLFr1bcase2}
\\
&&\frac{H'}{H}=-3 x^2+2 z^2, \label{HLFr2bcase2}
\end{eqnarray}
while the autonomous system writes:
\begin{eqnarray}
x'&=&x \left(3 x^2-2 z^2-3\right)+\sqrt{6} s \left(-x^2+z^2+1\right),\label{HLeqxcase2}\\
z'&=& z \left[3 x^2-2 \left(z^2+1\right)\right].\label{HLeqzcase2}
\end{eqnarray}
It is defined in the phase space $\Psi=\left\{(x,z): x^2-z^2\leq
1, z\in\mathbb{R}\right\}$ and as before the phase space is not
compact.

\subsubsection{Finite analysis}

The singular points,  the conditions for their existence and
stability, and the physical quantities are presented in  table
\ref{HLcrit2}.  Thus, $P_{1,2,3}$ are exactly the same as in case
1, while $P_{5,6}$ are  saddle points except if $s^2\rightarrow
1$, where they are nonhyperbolic because they give rise to the
eigenvalues $\left\{-\frac{1}{2}-\frac{1}{2}\mu_0,\,
-\frac{1}{2}+\frac{1}{2}\mu_0\right\}$ with associated
eigenvectors
  $\left\{\pm\mu_1,1\right\},\, \left\{\pm\mu_2,1\right\},$ where we use the notations $\mu_0=\sqrt{-15+\frac{16}{s^2}}$,
$\mu_1=\frac{-9 s^2-\sqrt{16 s^2-15 s^4}+8}{4
   \sqrt{\frac{6}{s^2}-6} s}$ and $\mu_2=\frac{-9 s^2+\sqrt{16
   s^2-15 s^4}+8}{4 \sqrt{\frac{6}{s^2}-6} s}.$ It is interesting to notice that this scenario admits two
more unstable singular points, namely $P_{7,8}$, in which
$z_c^2=-1$. The eigenvalues of the linearization are $\{4,\, -1\}$
with associated eigenvectors $\left\{\pm\frac{2}{5} i \sqrt{6}
s,1\right\},\, \{1,0\}.$ These points are of great physical
importance, as we are going to see in the next section.

\begin{table*}[t]
\begin{center}\caption{\label{HLcrit2} The singular points of  a non-flat
universe with $\Lambda=0$ (case 2) and their behavior (adapted
from \cite{Leon:2009rc}).} {
\begin{tabular}{ccccccccc} \hline \hline
 Cr. P &$x_c$&$z_c$&  Existence &  Stable for&  $w_M$&  $\Omega_{DE}$&  $w_{DE}$\\
\hline $P_{1,2}$& $\pm 1$ &0&  All $s$  &  unstable& 1& 0&  arbitrary  \vspace{0.2cm}\\
\hline $P_{3}$& $\sqrt{\frac{2}{3}}\,s$ &0&  $s^2<\frac{3}{2}$
&   $s^2<1$ & $\frac{4}{3}s^2-1$& 0&arbitrary  \\
\vspace{-0.35cm}\\
\hline $P_{5,6}$& $\sqrt{\frac{3}{2}}\frac{1}{s}\ $ &\ $\pm
\sqrt{-1+\frac{1}{s^2}}$ &  $s^2\leq 1,\, s\neq 0$  &  unstable&  $\frac{3}{s^2}-1$& 0&  arbitrary  \vspace{0.2cm}\\
\hline $P_{7,8}$& 0 &\ $\pm i$ &  always &  unstable&  arbitrary &  1&  $1/3$  \vspace{0.2cm}\\
\hline \hline
\end{tabular}}
\end{center}
\end{table*}

\begin{figure}[ht]
\begin{center}
\mbox{\epsfig{figure=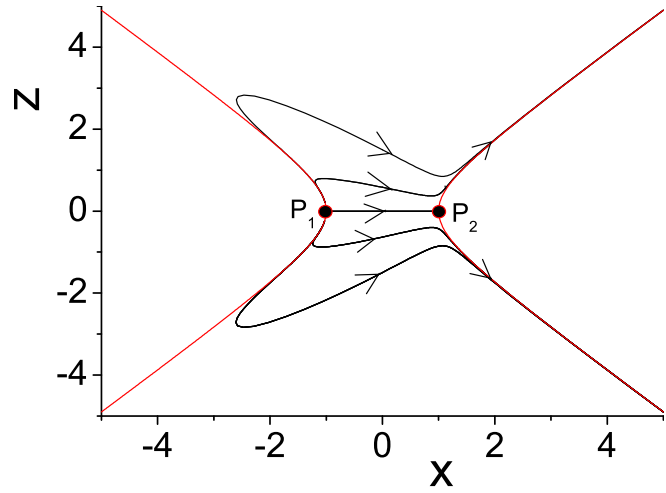,width=8.8cm,angle=0}} \caption{ {
Phase plane for a non-flat universe with $\Lambda=0$ (case 2), for
the choice $s=\sqrt{3}$. In this specific scenario the singular
points $P_3$ and $P_{5,6,7,8}$ do not exist, while
 $P_{1}$ and $P_2$ are unstable (source and saddle respectively).
 [Taken from \cite{Leon:2009rc} and published with permission of
IOP Publishing Ltd]. See the global phase space in figure
\ref{Case2NonFlatLequal0ssqrt3}.}} \label{HLFig2a}
\end{center}
\end{figure}

In order to present the results more transparently, in fig.
\ref{HLFig2a} we present the numerical evolution of the system for
the choice $s=\sqrt{3}$. In this specific realization of the
scenario the singular points $P_3$ and $P_{5,6,7,8}$ do not exist.
We find only the source $P_{1}$ and the saddle $P_2$, and we
indeed observe that there is one orbit  approaching $P_2$ (the
solution with $z\equiv 0$). Finally, note that  the divergence of
the orbits towards the future is typical and suggests that the
future attractor of the system can be located at infinite regions.

\begin{figure}[ht]
\begin{center}
\mbox{\epsfig{figure=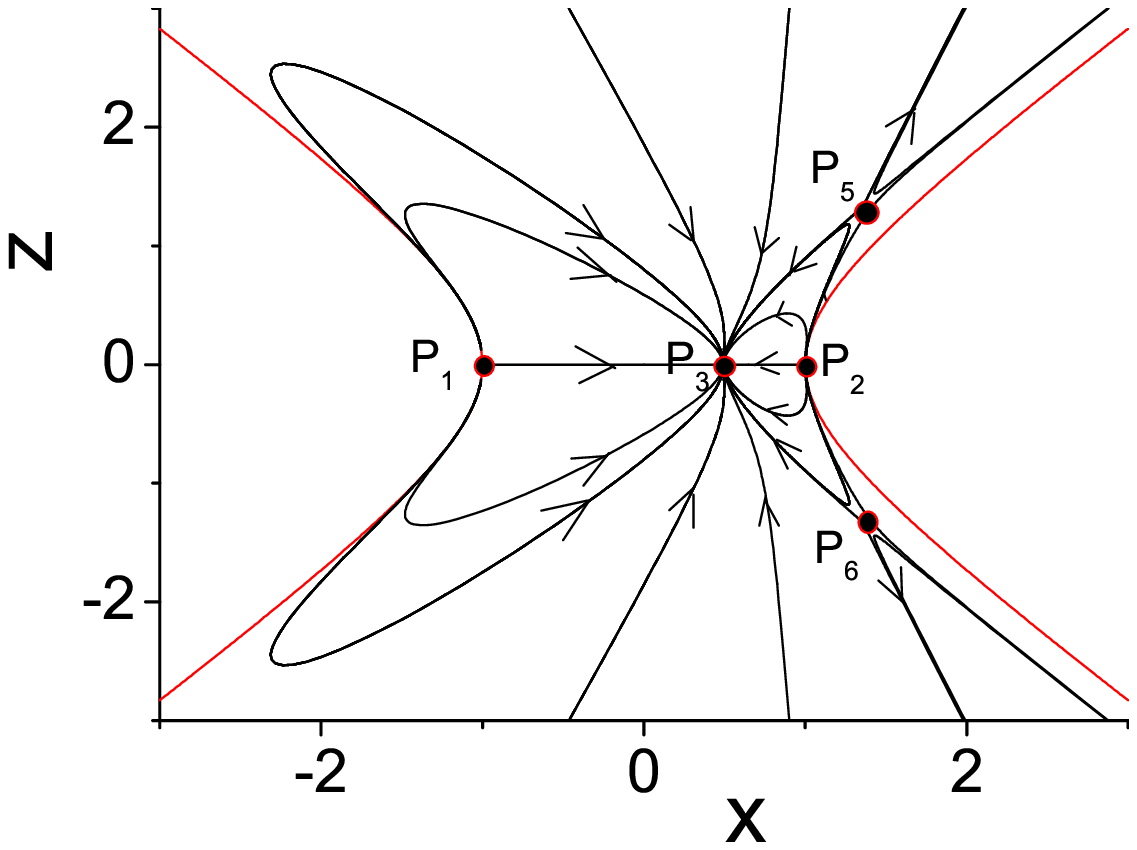,width=8.8cm,angle=0}} \caption{ {
Phase plane for a non-flat universe with $\Lambda=0$ (case 2), for
the choice $s=0.6$. In this specific scenario the singular point
$P_3$ is a local attractor, while $P_{1,2}$ are unstable (sources)
and $P_{5,6}$ are saddle ones. [Taken from \cite{Leon:2009rc} and
published with permission of IOP Publishing Ltd]. See the global
phase space in figure \ref{Case2NonFlatLequal0s0p6}.}}
\label{HLFig2b}
\end{center}
\end{figure}

In fig. \ref{HLFig2b} we depict the phase-space graph for the
choice $s=0.6$. In this case the singular points $P_{1,2}$ are
unstable (sources), while $P_3$ is a local attractor. The points
$P_{5,6}$ are saddle ones, and thus we observe that some orbits
coming from infinity spend a large amount of time near them before
diverge again in a finite time.

\subsubsection{Analysis at infinity}\label{7.2.2.2}

\begin{table*}[ht]
\begin{center}\caption{\label{HLinfty2} Asymptotic singular points  of the system
\eqref{HLeqxcase2}-\eqref{HLeqzcase2} (case 2) and their
stability.}
\begin{tabular}{ccccc}
\hline \hline
 Cr. P & Coordinates: $\theta, x_r, z_r$ & Eigenvalue & $\rho'$ & Stability\\
\hline $Q_4$ & $0, 1, 0$ & $\left\{\begin{array}{cc}
  -\infty & \text{for}\; s<0\\
  +\infty & \text{for}\; s>0\\
\end{array}\right. $  & $3$ &  $\left\{\begin{array}{c} \text{sink}\\ \text{saddle} \end{array}\right.$ \vspace{0.2cm}\\
\hline $Q_5$ & $\frac{\pi}{4}, \frac{\sqrt{2}}{2},
\frac{\sqrt{2}}{2}$ & $\left\{\begin{array}{cc}
  +\infty & \text{for}\; s<0\\
  -\infty & \text{for}\; s>0\\
\end{array}\right. $  & $\frac{1}{2}$ &  $\left\{\begin{array}{c} \text{saddle}\\ \text{sink} \end{array}\right.$ \\
\vspace{-0.35cm}\\
\hline $Q_6$ & $\frac{3\pi}{4}, -\frac{\sqrt{2}}{2},
\frac{\sqrt{2}}{2}$ & $\left\{\begin{array}{cc}
  -\infty & \text{for}\; s<0\\
  +\infty & \text{for}\; s>0\\
\end{array}\right. $  & $\frac{1}{2}$ &  $\left\{\begin{array}{c} \text{sink}\\ \text{saddle} \end{array}\right.$\\
\vspace{-0.35cm}\\
\hline $Q_7$ & $\frac{5\pi}{4}, -\frac{\sqrt{2}}{2},
-\frac{\sqrt{2}}{2}$ & $\left\{\begin{array}{cc}
  +\infty & \text{for}\; s<0\\
  -\infty & \text{for}\; s>0\\
\end{array}\right. $  & $\frac{1}{2}$ &  $\left\{\begin{array}{c} \text{saddle}\\ \text{sink}
\end{array}\right.$\\
\vspace{-0.35cm}\\
\hline $Q_8$ & $\frac{7\pi}{4}, \frac{\sqrt{2}}{2},
-\frac{\sqrt{2}}{2}$ & $\left\{\begin{array}{cc}
  +\infty & \text{for}\; s<0\\
  -\infty & \text{for}\; s>0\\
\end{array}\right. $  & $\frac{1}{2}$&  $\left\{\begin{array}{c} \text{saddle}\\ \text{sink} \end{array}\right.$\\
\\\hline \hline
\end{tabular}
\end{center}
\end{table*}

Using the same procedure as in section \ref{7.2.1.2}; that is,
performing the transformation \eqref{TransfHL1}, the system
\eqref{HLeqxcase2}-\eqref{HLeqzcase2} becomes
\begin{eqnarray}
&& \rho'=3 \rho ^3-\frac{1}{2} s \sqrt{6-6 \rho ^2} \cos (3 \theta
) \rho ^2-\frac{5 \rho }{2}+\frac{1}{2} s \sqrt{6-6 \rho ^2}
   \left(2-3 \rho ^2\right) \cos \theta+\nonumber \\&&+\left(3 \rho ^3-\frac{\rho }{2}\right) \cos (2 \theta ),\label{HLinfty2a}\\
&&\theta'=\cos \theta \sin (\theta
   )+\frac{\sqrt{6} s \rho  \cos (2 \theta ) \sin \theta}{\sqrt{1-\rho ^2}}-\frac{s \sqrt{6-6 \rho ^2} \sin \theta}{\rho
   }.\label{HLinfty2b}
\end{eqnarray}
In this case the physical region is given by
$$-\frac{\sqrt{2}}{2}\leq x_r \leq \frac{\sqrt{2}}{2},\, x_r^2+z_r^2\leq 1.$$

In the limit $\rho\rightarrow 1,$ the leading terms in
\eqref{HLinfty1a} and in \eqref{HLinfty1b} are
\begin{eqnarray}
&&\rho'\rightarrow \frac{1}{2} (5 \cos (2 \theta )+1),\label{HLinfty2c}\\
&&\theta'\rightarrow \frac{\sqrt{6} s  \cos (2 \theta ) \sin
\theta}{\sqrt{1-\rho ^2}}.\label{HLinfty2d}
\end{eqnarray}

As before, the radial equation does not contain the radial
coordinate, so the singular points  can be obtained using the
angular equation only. Setting $\theta'=0$, we obtain the singular
points  which are listed in table \ref{HLinfty2}. The stability of
these points is studied by analyzing first the stability of the
angular coordinate and then deducing, from the sign of equation
\eqref{HLinfty2a}, the stability on the radial direction. In table
\ref{HLinfty2} are displayed the asymptotic singular points  of
the system \eqref{HLeqxcase2}-\eqref{HLeqzcase2} (case 2). We
comment there on their stability. For $s<0$ the local sinks are
$Q_4$ and $Q_6;$ the rest of the points at infinity are saddles.
For $s>0$ the local sinks are $Q_{5,7,8}$ whereas the rest of the
points at infinity are saddles.

The system \eqref{HLinfty2a}-\eqref{HLinfty2b} have an apparent
singularity at $\rho=0, \sin\theta=0$ which is due to the
spherical coordinate system. Thus, for numerical integrations is
more convenient to use the cartesian coordinates $x_r,z_r.$ The
system reads \begin{eqnarray} &&x_r'=6 x_r^3-3 x_r+\frac{\sqrt{6}
s \left(2 x_r^4-3
   x_r^2+1\right)}{\sqrt{1-x_r^2-z_r^2}},\nonumber\\
  &&z_r'= \left(6 x_r^2-2\right) z_r+\frac{\sqrt{6} s x_r
   \left(2 x_r^2-1\right)
   z_r}{\sqrt{1-x_r^2-z_r^2}}.\label{543}
   \end{eqnarray}

\begin{figure}[h]
    \centering
        \includegraphics[scale=0.8]{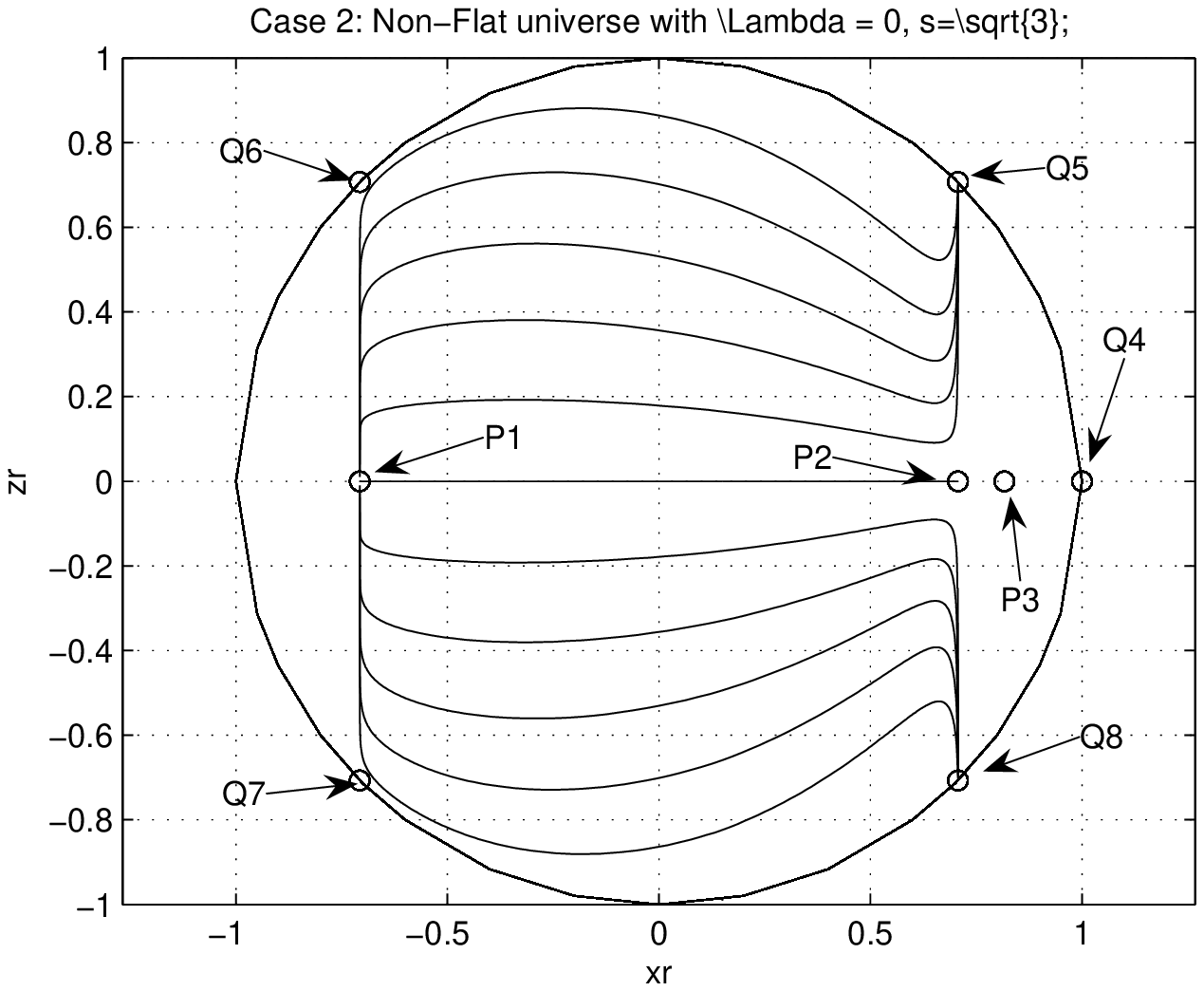}
    \caption{Global phase space of the system
\eqref{HLeqxcase2}-\eqref{HLeqzcase2} (case 2) for the choice
$s=\sqrt{3}$. The singular points $P_3$ and $P_{5,6,7,8}$ do not
exist. At finite region we find only the source $P_{1}$ and the
saddle $P_2$, and we indeed observe that there is one orbit
approaching $P_2$ (the solution with $z\equiv 0$). The future
attractors of the system are $Q_5$ and $Q_8$ located at the
infinite region. The points at infinity $Q_{4,6,7}$ are saddles.
$Q_4$ is unphysical since $x\notin \left[-\frac{\sqrt{2}}{2},
\frac{\sqrt{2}}{2}\right].$}
    \label{Case2NonFlatLequal0ssqrt3}
\end{figure}

\begin{figure}[h]
    \centering
        \includegraphics[scale=0.8]{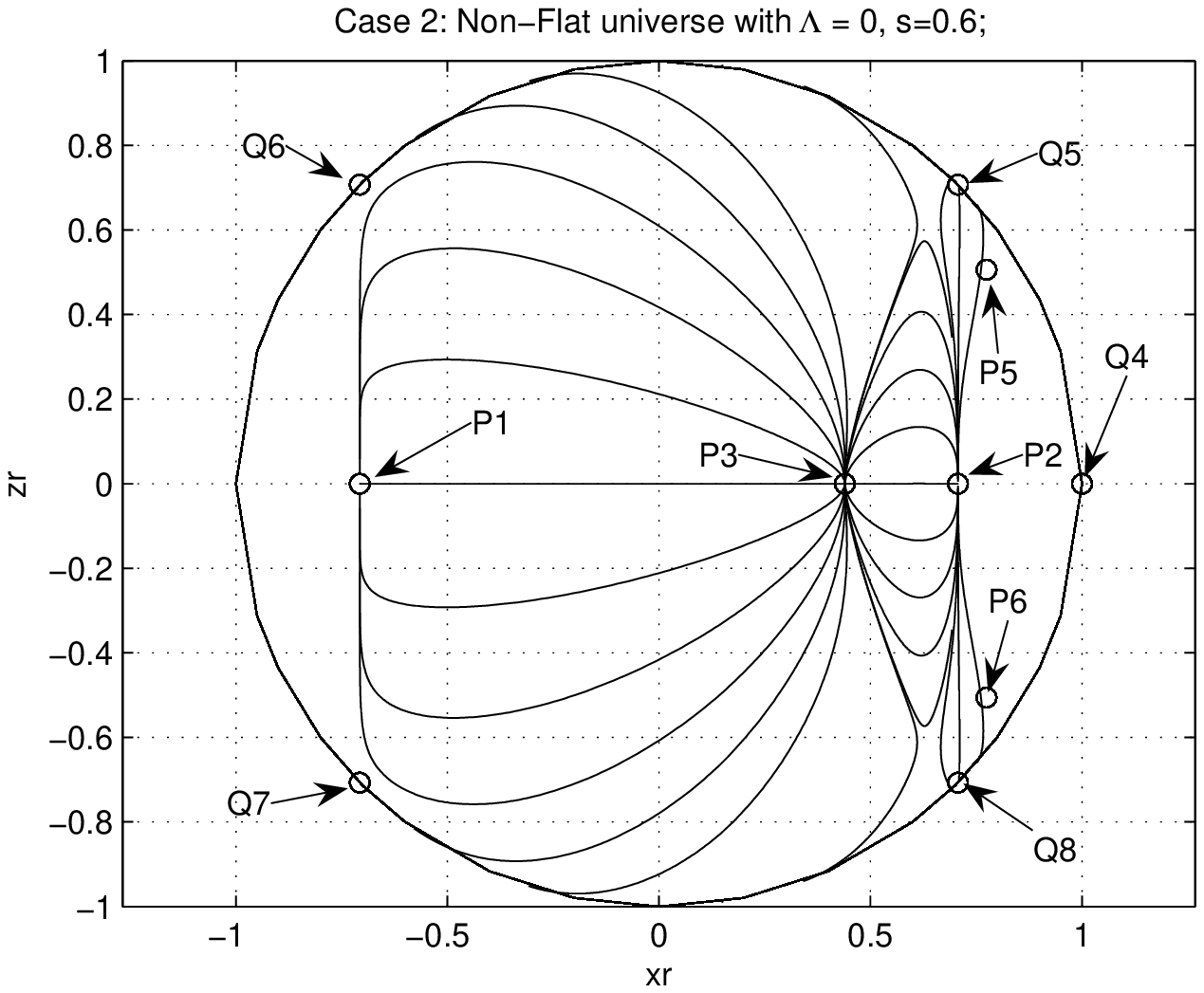}
    \caption{Global phase space of the system
\eqref{HLeqxcase2}-\eqref{HLeqzcase2} (case 2) for the choice
$s=0.6$. The finite points $P_1$ and $P_2$ are local sources.
$P_3$ is a local sink at the finite region. Observe that the
orbits spent a finite lapse of time near the saddle points $P_5$
and $P_6$ before reaching the local sinks at infinity $Q_5$ and
$Q_8$ respectively. The points at infinity $Q_{4,6,7}$ are
saddles. $Q_4$ is unphysical since $x\notin
\left[-\frac{\sqrt{2}}{2}, \frac{\sqrt{2}}{2}\right].$}
    \label{Case2NonFlatLequal0s0p6}
\end{figure}

To illustrate the global dynamics we depicts in  the figure
\ref{Case2NonFlatLequal0ssqrt3} the phase space of the system
\eqref{543} describing the flow of
\eqref{HLeqxcase2}-\eqref{HLeqzcase2} (case 2) for the choice
$s=\sqrt{3}$ in the Poincar\'e variables. As in figure
\ref{HLFig2a} the singular points $P_3$ and $P_{5,6,7,8}$ do not
exist. At finite region we find only the source $P_{1}$ and the
saddle $P_2$, and we indeed observe that there is one orbit
approaching $P_2$ (the solution with $z\equiv 0$). The future
attractors of the system are $Q_5$ and $Q_8$ located at the
infinite region. The points at infinity $Q_{4,6,7}$ are saddles.
$Q_4$ is unphysical since $x\notin \left[-\frac{\sqrt{2}}{2},
\frac{\sqrt{2}}{2}\right].$ Also, in the figure
\ref{Case2NonFlatLequal0s0p6} we depict the global phase space for
case 2 for the choice $s=0.6$. The finite points $P_1$ and $P_2$
are local sources. $P_3$ is a local sink at the finite region.
This is consistent with the drawn in figure \ref{HLFig2b}. Observe
that the orbits spent a finite lapse of time near the saddle
points $P_5$ and $P_6$ before reaching the local sinks at infinity
$Q_5$ and $Q_8$ respectively. The points at infinity $Q_{4,6,7}$
are saddles. As in the previous figure, $Q_4$ is unphysical.

\subsubsection{Cosmological implications for Case 2: non-flat universe with $\Lambda=0$}\label{HLsection B2}

In this scenario, the first three singular points are identical
with those of case 1, and thus the  physical implications are the
same. The singular points $P_{5,6}$ are unstable, corresponding to
a dark-matter dominated universe. This was expected since in the
absence of the cosmological constant $\Lambda$, the curvature role
is downgrading as the scale factor increases and thus in the end
this case tends to the case 1 above. Note however that at early
times, where the scale factor is small, the behavior of the system
will be significantly different than case 1, with the dark energy
playing an important role. This different behavior is observed in
the corresponding phase-space figures  \ref{HLFig2a},
\ref{HLFig2b} comparing with figure \ref{HLFig1}.

The case at hand admits another solution sub-class, namely points
$P_{7,8}$. In these points $z_c^2=-1$, and thus using
({\ref{HLauxilliaryz}}) we straightforwardly find the late-time
solution $a(t)=e^{i\gamma t}$, with
$\gamma=|\kappa^2\mu/[4(3\lambda-1)]|$. This solution corresponds
to an oscillatory universe \cite{cyclic.clifton,Saridakis:2007cf},
and in the context of Ho\v{r}ava-Lifshitz cosmology it has already
been studied in the literature
\cite{Brandenberger:2009yt,Brandenberger:2009ic,Cai:2009in}.
However, as we see, these singular points are unstable and thus
this solution subclass cannot be a late-time attractor in the case
of a non-flat universe with zero cosmological constant. This
situation will change in the case where the cosmological constant
is switched on.

\subsection{Case 3: flat universe with $\Lambda \neq 0$}

In this case the Friedmann equations (\ref{HLFr1}), (\ref{HLFr2})
write as
\begin{eqnarray}
&&1=x^2+y^2-u^2 \label{HLFr1bcase3}
\\
&&\frac{H'}{H}=-3 x^2, \label{HLFr2bcase3}
\end{eqnarray}
and the autonomous system becomes:
\begin{eqnarray}
x'&=&\sqrt{6} s \left(u^2-x^2+1\right)+3 x \left(x^2-1\right),\label{HLeqxcase3}\\
u'&=& 3 u x^2,\label{HLequcase3}
\end{eqnarray}
 defined
in the phase space $\Psi=\left\{(x,u): x^2-u^2\leq 1,
u\in\mathbb{R}\right\}$. As before the phase space is not compact.

\subsubsection{Finite analysis}

The singular points,  the conditions for their existence and
stability, and the physical quantities are presented in  table
\ref{HLcrit3}.
\begin{table*}[t]
\begin{center}\caption{\label{HLcrit3} The singular points of a flat universe
with $\Lambda\neq0$ (case 3) and their behavior. NH stands for
nonhyperbolic (adapted from \cite{Leon:2009rc}).}
\begin{tabular}{cccccccc} \hline \hline
 Cr. P &$x_c$&$u_c$&  Existence &  Stable for&  $w_M$&  $\Omega_{DE}$&  $w_{DE}$\\
\hline $P_{9,10}$& $\pm 1$ &0&  All $s$   &  unstable& 1& 0&  arbitrary  \vspace{0.2cm}\\
\hline $P_{11}$& $\sqrt{\frac{2}{3}}\,s$ &0&
$s^2<{\frac{3}{2}}$   & unstable & $\frac{4}{3}s^2-1$& 0&arbitrary  \\
\vspace{-0.35cm}\\
\hline $P_{12,13}$& 0 &\ $\pm i$ &  always &  NH&  arbitrary &  1&  $-1$  \vspace{0.2cm}\\
\hline \hline
\end{tabular}
\end{center}
\end{table*}

The singular points $P_{9,10}$ exist for all the $s$-values. They
have eigenvalues $\mp 2 \sqrt{6} s+6,\, 3$ with associated
eigenvectors $\{1,0\},\,  \{0,1\}.$ Thus, the singular points are
of saddle type with a 1-dimensional unstable manifold tangent to
the u-axis provided $\pm s>\sqrt{\frac{3}{2}}$ (the sign $+$ for
$P_9$ and the sign $-$ for $P_{10}$). Otherwise they are local
sources. Note that the singular point $P_{11}$ (which exists for
$s^2<{\frac{3}{2}}$) is nonhyperbolic if $s^2\in\{0,\,3/2\}$,
while it is a saddle otherwise, with stable (unstable) manifold
tangent to the x- (u-) axis because it gives rise to the
eigenvalues  $\{-3+2 s^2,\, 2s^2\}$ with associated eigenvectors $
\{1,0\},\, \{0,1\}.$ Finally, the system admits two more
nonhyperbolic singular points, namely $P_{12,13}$, in which
$u_c^2=-1$. They have eigenvalues  $\{-3, 0\}$ with associated
eigenvectors $\{1,0\},\,  \left\{2 i \sqrt{\frac{2}{3}}
s,1\right\}.$

\begin{figure}[ht]
\begin{center}
\mbox{\epsfig{figure=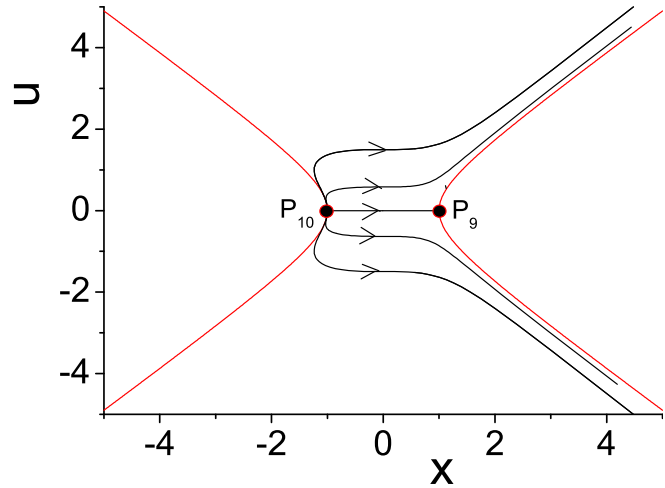,width=8.8cm,angle=0}} \caption{ {
Phase plane for a flat universe with $\Lambda\neq0$ (case 3), for
the choice $s=\sqrt{3}$. In this specific scenario the singular
point $P_{11}$ does not exists. $P_{10}$ is unstable (source),
while $P_9$ is a saddle one. [Taken from \cite{Leon:2009rc} and
published with permission of IOP Publishing Ltd]. See the global
phase space in figure \ref{Case3FlatLnonequal0ssqrt3}. }}
\label{HLFig3a}
\end{center}
\end{figure}

In fig. \ref{HLFig3a} we present the phase-space graph of the
system for the choice $s=\sqrt{3}$. In this case the singular
point $P_{11}$ does not exists, while $P_9$ and $P_{10}$ are
unstable (source and saddle respectively). The divergence of the
orbits towards the future is typical and suggests that the future
attractor of the system will be located at infinite regions.
\begin{figure}[ht]
\begin{center}
\mbox{\epsfig{figure=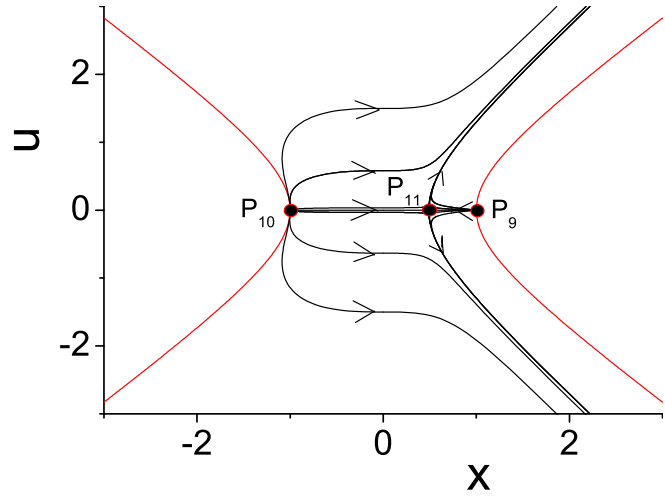,width=8.8cm,angle=0}} \caption{ {
Phase plane for a flat universe with $\Lambda\neq0$ (case 3), for
the choice $s=0.6$. In this specific scenario the singular point
$P_{11}$ is a saddle one, while $P_9$ and $P_{10}$ are unstable
(sources).  [Taken from \cite{Leon:2009rc} and published with
permission of IOP Publishing Ltd]. See the global phase space in
figure \ref{Case3FlatLnonequal0s0p6}.}} \label{HLFig3b}
\end{center}
\end{figure}
Finally, in fig. \ref{HLFig3b} we display the phase-space graph
for the choice $s=0.6$. In this case the singular point $P_{11}$
is a saddle one (with stable manifold tangent to the x-axis),
while $P_9$ and $P_{10}$ are unstable (sources). There are two
orbits, one joining $P_{10}$ with $P_{11}$ and one joining $P_9$
with $P_{11},$ both of them overlapping the $x$-axis. Note that
some orbits remain close to $P_{11}$ before finally diverge
towards the future, and this suggests that the future attractor of
the system is located at infinite regions.

\subsubsection{Analysis at infinity}

\begin{table*}[ht]
\begin{center}\caption{\label{HLinfty3} Asymptotic singular points  of the system
\eqref{HLeqxcase3}-\eqref{HLequcase3} (case 3) and their
stability.}
\begin{tabular}{ccccc}
\hline \hline
 Cr. P & Coordinates: $\theta, x_r, u_r$ & Eigenvalue & $\rho'$ & Stability\\
\hline $Q_9$ & $0, 1, 0$ & $\left\{\begin{array}{cc}
  -\infty & \text{for}\; s<0\\
  +\infty & \text{for}\; s>0\\
\end{array}\right. $  & $3$ &  $\left\{\begin{array}{c} \text{sink}\\ \text{saddle} \end{array}\right.$ \vspace{0.2cm}\\
\hline $Q_{10}$ & $\frac{\pi}{4}, \frac{\sqrt{2}}{2},
\frac{\sqrt{2}}{2}$ & $\left\{\begin{array}{cc}
  +\infty & \text{for}\; s<0\\
  -\infty & \text{for}\; s>0\\
\end{array}\right. $  & $\frac{3}{2}$ &  $\left\{\begin{array}{c} \text{saddle}\\ \text{sink} \end{array}\right.$ \\
\vspace{-0.35cm}\\
\hline $Q_{11}$ & $\frac{3\pi}{4}, -\frac{\sqrt{2}}{2},
\frac{\sqrt{2}}{2}$ & $\left\{\begin{array}{cc}
  -\infty & \text{for}\; s<0\\
  +\infty & \text{for}\; s>0\\
\end{array}\right. $  & $\frac{3}{2}$ &  $\left\{\begin{array}{c} \text{sink}\\ \text{saddle} \end{array}\right.$\\
\vspace{-0.35cm}\\
\hline $Q_{12}$ & $\frac{5\pi}{4}, -\frac{\sqrt{2}}{2},
-\frac{\sqrt{2}}{2}$ & $\left\{\begin{array}{cc}
  -\infty & \text{for}\; s<0\\
  +\infty & \text{for}\; s>0\\
\end{array}\right. $  & $\frac{3}{2}$ &  $\left\{\begin{array}{c} \text{sink}\\ \text{saddle}
\end{array}\right.$\\
\vspace{-0.35cm}\\
\hline $Q_{13}$ & $\frac{7\pi}{4}, \frac{\sqrt{2}}{2},
-\frac{\sqrt{2}}{2}$ & $\left\{\begin{array}{cc}
  +\infty & \text{for}\; s<0\\
  -\infty & \text{for}\; s>0\\
\end{array}\right. $  & $\frac{3}{2}$&  $\left\{\begin{array}{c} \text{saddle}\\ \text{sink} \end{array}\right.$\\
\\\hline \hline
\end{tabular}
\end{center}
\end{table*}

The coordinate transformation
\begin{equation}\label{TransfHL3}
x_r=\rho \cos\theta,\, u_r=\rho \sin\theta,
\end{equation} where $\rho=\frac{r}{\sqrt{1+r^2}},$ and $r=\sqrt{x^2+u^2},$  allows to
investigate the asymptotics of the system
\eqref{HLeqxcase3}-\eqref{HLequcase3} (i.e., at the region
$r\rightarrow+\infty$) by taking the limit $\rho\rightarrow 1.$ In
this case the physical region is given by
$$-\frac{\sqrt{2}}{2}\leq x_r\leq \frac{\sqrt{2}}{2},\, x_r^2+u_r^2\leq 1.$$

Performing the transformation \eqref{TransfHL3}, the system
\eqref{HLeqxcase3}-\eqref{HLequcase3} becomes
\begin{eqnarray}
\rho'=\left(6 \rho ^3-3 \rho \right) \cos ^2\theta+\left(\sqrt{6}
s \left(1-\rho ^2\right)^{3/2}-s \rho ^2 \sqrt{6-6 \rho
   ^2} \cos (2 \theta )\right) \cos \theta,\label{HLinfty3a}\\
\theta'=3 \cos \theta \sin \theta+\frac{\sqrt{6} s \rho \cos (2
\theta ) \sin
   \theta}{\sqrt{1-\rho ^2}}-\frac{s \sqrt{6-6 \rho ^2} \sin \theta}{\rho }.\label{HLinfty3b}
\end{eqnarray}
In the limit $\rho\rightarrow 1,$ the leading terms in
\eqref{HLinfty3a}-\eqref{HLinfty3b} are
\begin{eqnarray}
\rho'\rightarrow 3 \cos ^2\theta,\label{HLinfty3c}\\
\theta'=\frac{\sqrt{6} s  \cos (2 \theta ) \sin
   \theta}{\sqrt{1-\rho ^2}}.\label{HLinfty3d}
\end{eqnarray}
As before, the radial equation does not contain the radial
coordinate, so the singular points  can be obtained using the
angular equation only. Setting $\theta'=0$, we obtain the singular
points  which are listed in table \ref{HLinfty3}. The stability of
these points is studied by analyzing first the stability of the
angular coordinate and then deducing, from the sign of equation
\eqref{HLinfty3a}, the stability on the radial direction.

The system \eqref{HLinfty3a}-\eqref{HLinfty3b} have an apparent
singularity at $\rho=0, \sin\theta=0$ which is due to the
spherical coordinate system. Thus, for numerical integrations is
more convenient to use the cartesian coordinates $x_r,u_r.$ The
system reads \begin{eqnarray} &&x_r'=6 x_r^3-3 x_r+\frac{\sqrt{6}
s \left(2 x_r^4-3
x_r^2+1\right)}{\sqrt{1-u_r^2-x_r^2}},\nonumber\\
  &&u_r'= 6 u_r
   x_r^2+\frac{\sqrt{6} s u_r \left(2 x_r^2-1\right)
   x_r}{\sqrt{1-u_r^2-x_r^2}}.\label{553}
   \end{eqnarray}

   \begin{figure}[h]
    \centering
        \includegraphics[scale=0.8]{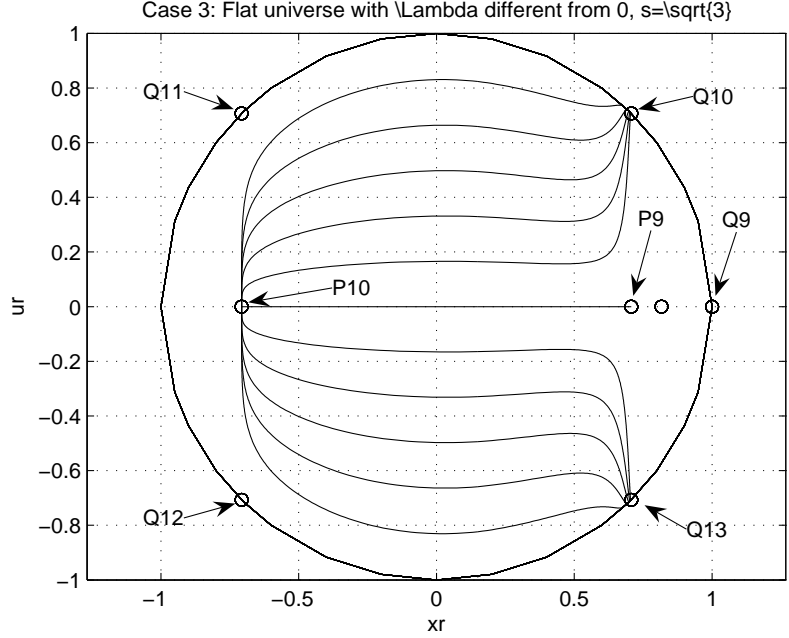}
    \caption{Global phase space of the system
\eqref{HLeqxcase1}-\eqref{HLeqzcase1} (case 3) for the choice
$s=\sqrt{3}$. As in figure \ref{HLFig3a}, the singular point
$P_{11}$ does not exists. $P_{10}$ is unstable (source), while
$P_9$ is a saddle one. The orbits passing near the saddle $P_9$
bifurcates and tends asymptotically to one of the global attractor
at infinity $Q_{10}$ or $Q_{13}$ depending on the sign of the
initial value of $u_r.$ The points at infinity $Q_{9,11,12}$ are
saddles; $Q_9$ is unphysical.}
    \label{Case3FlatLnonequal0ssqrt3}
\end{figure}

\begin{figure}[h]
    \centering
        \includegraphics[scale=0.8]{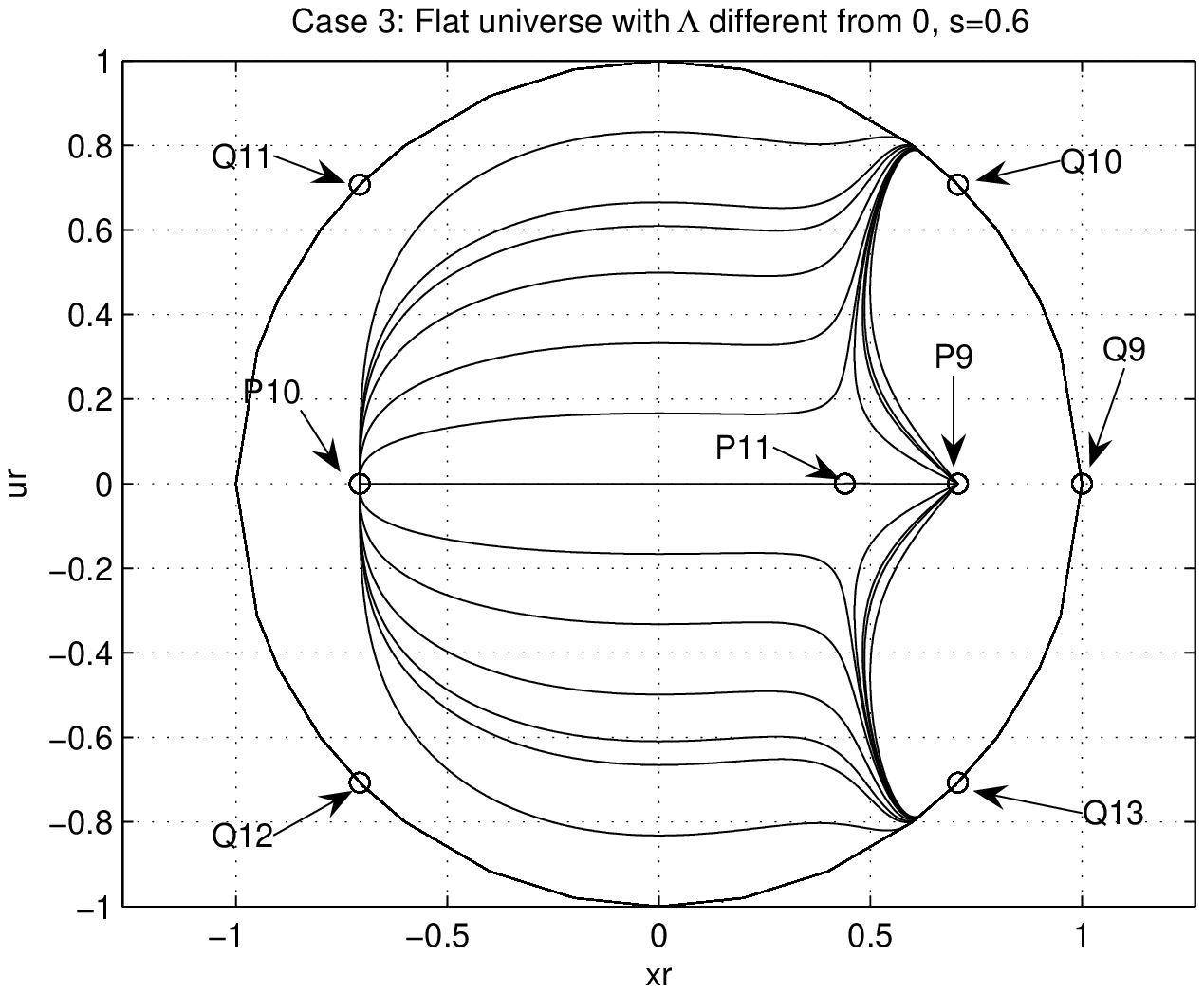}
    \caption{Global phase space of the system
\eqref{HLeqxcase3}-\eqref{HLequcase3} (case 3) for the choice
$s=0.6$. As in figure \ref{HLFig3b}, in this specific scenario the
singular point $P_{11}$ is a saddle one, while $P_9$ and $P_{10}$
are unstable (sources). The global attractors at infinity are the
points $Q_{10}$ and $Q_{13}.$ The points at infinity $Q_{9,11,12}$
are saddles; $Q_9$ is unphysical.}
    \label{Case3FlatLnonequal0s0p6}
\end{figure}

To illustrate the global dynamics we depicts in  the figure
\ref{Case3FlatLnonequal0ssqrt3} the phase space of the system
\eqref{553} describing the flow of
\eqref{HLeqxcase3}-\eqref{HLequcase3} (case 3) for the choice
$s=\sqrt{3}$ in the Poincar\'e variables. As in figure
\ref{HLFig3a}, the singular point $P_{11}$ does not exists.
$P_{10}$ is unstable (source), while $P_9$ is a saddle one. The
orbits passing near the saddle $P_9$ bifurcates and tends
asymptotically to one of the global attractor at infinity $Q_{10}$
or $Q_{13}$ depending on the sign of the initial value of $u_r.$
The points at infinity $Q_{9,11,12}$ are saddles; $Q_9$ is
unphysical. In figure \eqref{Case3FlatLnonequal0s0p6} we drawn the
global phase space for the choice $s=0.6$. As in figure
\ref{HLFig3b}, in this specific scenario the singular point
$P_{11}$ is a saddle one, while $P_9$ and $P_{10}$ are unstable
(sources). The global attractors at infinity are the points
$Q_{10}$ and $Q_{13}.$ The points at infinity $Q_{9,11,12}$ are
saddles; $Q_9$ is unphysical.

\subsubsection{Cosmological implications for Case 3: flat universe with $\Lambda \neq 0$}\label{HLsection C2}

Under this scenario, the Ho\v{r}ava-Lifshitz  universe admits two
unstable singular points ($P_{9,10}$), completely dominated by
stiff dark matter. Point $P_{11}$ exhibits a more physical dark
matter equation-of-state parameter, but still with negligible dark
energy at late times. The case at hand admits the two
nonhyperbolic points $P_{12,13}$ possessing $u_c^2=-1$, and thus
(as can be seen by ({\ref{HLauxilliaryu}})) they correspond to the
oscillatory solution $a(t)=e^{i\delta t}$, with
$\delta=|\kappa^2\mu\Lambda/[4(3\lambda-1)]|$. We mention that
these points are nonhyperbolic, with a negative eigenvalue, and
thus they have a large probability to be a late-time solution of
Ho\v{r}ava-Lifshitz universe. Additionally, they correspond to
dark-energy domination, with dark-energy equation-of-state
parameter $-1$ and an arbitrary $w_M$. These features make them
good candidates to be a realistic description of the universe. We
mention that this result is independent from the parameter $q$
which comes from the dark matter sector. Thus, we conclude that it
is valid independently of the matter-content of the universe.
Indeed, this behavior is novel, and arises purely by the extra
terms that are present in Ho\v{r}ava gravity.

\subsection{Case 4: $k\neq 0, \Lambda \neq 0$}

\begin{table*}[t]
\begin{center}\caption{\label{HLcrit4} The  singular points of a non-flat
universe with $\Lambda\neq0$ (case 4) and their behavior (adapted
from \cite{Leon:2009rc}).} {
\begin{tabular}{ccccccccc}
\hline \hline
 Cr. P &$x_c$& $z_c$&$u_c$& Existence& Stable for& $w_M$& $\Omega_{DE}$& $w_{DE}$\\
\hline $P_{14,15}$& $\pm 1$ &0&0&  All $s$  &   unstable &  1& 0&  arbitrary  \vspace{0.2cm}\\
\hline $P_{16}$& $\sqrt{\frac{2}{3}}\,s$ &0&0&
$s^2<{\frac{3}{2}}$  &   unstable &  $\frac{4}{3}s^2-1$& 0&arbitrary  \\
\vspace{-0.35cm}\\
\hline $P_{17,18}$& $\sqrt{\frac{3}{2}}\frac{1}{s}$ & $ \pm
\sqrt{-1+\frac{1}{s^2}}$ & $0$  &  $s^2\leq 1,\, s\neq 0$  &   unstable & 2 &  $1-\frac{1}{s^2}$&  $1/3$  \vspace{0.2cm}\\
\hline $P_{19,20}$& 0 & 0 & $\pm i$  & always &   NH &  arbitrary &  1  &   $-1$ \vspace{0.2cm}\\
\hline $P_{21,22}$& 0 & $\pm i$ & 0  & always &   unstable &   arbitrary &  1  &   $1/3$ \vspace{0.2cm}\\
\hline \hline
\end{tabular}}
\end{center}
\end{table*}
Under this scenario, and using the auxiliary variables
(\ref{HLauxilliaryx}),(\ref{HLauxilliaryy}),(\ref{HLauxilliaryz}),(\ref{HLauxilliaryu}),
the Friedmann equations (\ref{HLFr1}), (\ref{HLFr2}) become:
\begin{eqnarray}
&&1=x^2+y^2-(u-k z)^2 \label{HLFr1bcase4}
\\
&&\frac{H'}{H}=-3 x^2 + 2 z (-u + z). \label{HLFr2bcase4}
\end{eqnarray}
while the autonomous system writes:
\begin{eqnarray}
x'&=&\sqrt{6} s \left[-x^2+(u-z)^2+1\right]+x \left[3 x^2+2 (u-z) z-3\right],\label{HLeqxcase4}\nonumber\\
z'&=&z \left[3 x^2+2 (u-z) z-2\right],\label{HLeqzcase4}\nonumber\\
u'&=& u \left[3 x^2+2 (u-z) z\right],\label{HLequcase4}
\end{eqnarray} defined
in the phase space $\Psi=\left\{(x,z,u): x^2-(u-k z)^2\leq 1, u,
z\in\mathbb{R}\right\}$, which is not compact.

\subsubsection{Finite analysis}
 The singular points, and their corresponding information, are
presented in  table \ref{HLcrit4}.

The singular point $P_{14}$ is nonhyperbolic if $s= \sqrt{3/2},$
it is a source if $s< \sqrt{3/2}$ or a saddle otherwise, while
$P_{15}$ is nonhyperbolic $s= -\sqrt{3/2},$ it is a source if $s>
-\sqrt{3/2}$ or a saddle otherwise. $P_{16}$ is nonhyperbolic if
$s^2\in\{0, \, 1,\ 3/2\}$ and saddle otherwise, while $P_{17,18}$
are nonhyperbolic if $s^2\rightarrow 1,$ and saddle otherwise. The
points $P_{19,20}$ have the eigenvalues $\{-3, -2, 0\}$ with
associated eigenvectors $\{1,0,0\}, \{0,1,1\}, \left\{\pm 2 i
\sqrt{\frac{2}{3}} s,0,1\right\}$. Hence, they are nonhyperbolic
possessing a 2-dimensional stable manifold. Finally, $P_{21,22}$
are unstable because they give rise to the eigenvalues
$\{4,2,-1\}$, with associated eigenvectors $\left\{-\frac{2}{5} i
\sqrt{6} s,1,0\right\}, \{0,1,1\}, \{1,0,0\}$.

\subsubsection{Analysis at infinity}

\begin{table*}[ht]
\begin{center}\caption{\label{HLinfty4} Asymptotic singular points  of the system
\eqref{HLeqxcase4}-\eqref{HLequcase4} (case 4) and their
stability. NH stands for nonhyperbolic}
\begin{tabular}{ccccc}
\hline \hline
 Cr. P & Coordinates & Eigenvalues & $\rho'$ & Stability\\
    &  $\theta,\psi, x_r, x_r, u_r$ &  &  & \\
\hline $Q_{14}$ &
$0,\frac{\pi}{4},\frac{\sqrt{2}}{2},0,\frac{\sqrt{2}}{2}$ &
$\left\{\begin{array}{ccc}
 0, & -\infty & \text{for}\; s>0\\
  0, & +\infty & \text{for}\; s<0\\
\end{array}\right. $  & $-\frac{3}{2}$ &  $\left\{\begin{array}{c}
\; {\scriptsize \text{NH, 1D stable manifold}}\\ {\scriptsize \text{NH, 2D unstable manifold}} \end{array}\right.$ \vspace{0.2cm}\\
\hline $Q_{15}$ & $\frac{\pi}{2},\frac{\pi}{4},0,
\frac{\sqrt{2}}{2},\frac{\sqrt{2}}{2}$ & $0,0$  & $0$ &
NH \\
\vspace{-0.35cm}\\
\hline $Q_{16}$ &
$\pi,\frac{\pi}{4},-\frac{\sqrt{2}}{2},0,\frac{\sqrt{2}}{2}$
&$\left\{\begin{array}{ccc}
 0, & +\infty & \text{for}\; s>0\\
  0, & -\infty & \text{for}\; s<0\\
\end{array}\right. $  & $-\frac{3}{2}$ &  $\left\{\begin{array}{c}
\; {\scriptsize\text{NH, 2D unstable manifold}}\\ {\scriptsize \text{NH, 1D stable manifold}} \end{array}\right.$ \\
\vspace{-0.35cm}\\
\hline $Q_{17}$ &
$0,-\frac{\pi}{4},\frac{\sqrt{2}}{2},0,-\frac{\sqrt{2}}{2}$
&$\left\{\begin{array}{ccc}
 0, & -\infty & \text{for}\; s>0\\
  0, & +\infty & \text{for}\; s<0\\
\end{array}\right. $  & $-\frac{3}{2}$ &  $\left\{\begin{array}{c}
\; {\scriptsize \text{NH, 1D stable manifold}}\\ {\scriptsize \text{NH, 2D unstable manifold}} \end{array}\right.$ \\
\vspace{-0.35cm}\\
\hline $Q_{18}$ &
$\pi,-\frac{\pi}{4},-\frac{\sqrt{2}}{2},0,-\frac{\sqrt{2}}{2}$
&$\left\{\begin{array}{ccc}
 0, & +\infty & \text{for}\; s>0\\
  0, & -\infty & \text{for}\; s<0\\
\end{array}\right. $  & $-\frac{3}{2}$ &  $\left\{\begin{array}{c}
\; {\scriptsize\text{NH, 2D unstable manifold}}\\ {\scriptsize \text{NH, 1D stable manifold}} \end{array}\right.$ \\
\vspace{-0.35cm}\\
\hline $Q_{19}$ & $0,\frac{\pi}{2},1,0,0$
&$\left\{\begin{array}{ccc}
 +\infty, & +\infty & \text{for}\; s>0\\
  -\infty, & -\infty & \text{for}\; s<0\\
\end{array}\right. $  & $-3$ &  $\left\{\begin{array}{c} \text{source}\\ \text{saddle} \end{array}\right.$ \\
\vspace{-0.35cm}\\
\hline $Q_{20}$ & $0,-\frac{\pi}{2},-1,0,0$
&$\left\{\begin{array}{ccc}
 -\infty, & -\infty & \text{for}\; s>0\\
  +\infty, & +\infty & \text{for}\; s<0\\
\end{array}\right. $  & $-3$ &  $\left\{\begin{array}{c} \text{saddle}\\ \text{source} \end{array}\right.$ \\
\hline \hline
\end{tabular}
\end{center}
\end{table*}

The coordinate transformation
\begin{equation}\label{TransfHL4}
x_r=\rho \cos\theta\sin\psi,\, z_r=\rho \sin\theta\sin\psi,\,
u_r=\rho\cos\psi,
\end{equation} where $\rho=\frac{r}{\sqrt{1+r^2}},$ and $r=\sqrt{x^2+z^2+u^2},$
$\theta\in[0,2\pi],\,\psi\in[-\frac{\pi}{2},\frac{\pi}{2}],$
allows to investigate the asymptotics of the system
\eqref{HLeqxcase4}-\eqref{HLequcase4} (i.e., at the region
$r\rightarrow+\infty$) by taking the limit $\rho\rightarrow 1.$ In
this case the physical region is given by
$$2\left(x_r^2+k u_r z_r\right)\leq 1, u_r^2+x_r^2+z_r^2\leq 1.$$

Performing the transformation \eqref{TransfHL4}, the system
\eqref{HLeqxcase4}-\eqref{HLequcase4} becomes, as $\rho\rightarrow
1$
\begin{eqnarray}
&&\rho'\rightarrow -\frac{1}{2} \sin \psi (4 \cos \psi \sin
\theta+5 \cos (2 \theta ) \sin \psi+\sin \psi), \label{HLinfty4a}\\
&& \theta'\rightarrow \frac{\sqrt{6} s \sin \theta \left(2 \sin
\psi \cos ^2\theta-\csc \psi+2 \cos \psi \sin \theta\right)}{\sqrt{1-\rho^2}},\label{HLinfty4b}\\
&& \psi'\rightarrow \frac{\sqrt{\frac{3}{2}} s \cos \psi \left(2
\cos (2 \psi ) \cos ^3\theta+2 \sin ^2\theta \cos \theta-\sin (2
\theta )
   \sin (2 \psi )\right)}{\sqrt{1-\rho^2}}. \label{HLinfty4c}
\end{eqnarray}

As before, the radial equation does not contain the radial
coordinate, so the singular points  can be obtained using the
angular equations only. Setting $\theta'=0,\psi'=0$, we obtain the
singular points  which are listed in table \ref{HLinfty4}. The
stability of these points is studied by analyzing first the
stability of the angular coordinates and then deducing, from the
sign of equation \eqref{HLinfty4a}, the stability on the radial
direction.

\subsubsection{Cosmological implications for Case 4: non-flat universe with $\Lambda \neq 0$}\label{HLsection D2}

This case admits the unstable singular points $P_{14,15,16}$ which
correspond to a dark-matter dominated universe, and the unstable
points $P_{17,18}$ which are unphysical since they possess
$w_M=2$. As expected, the system admits also the unstable points
$P_{21,22}$ which correspond to oscillatory universes with
$a(t)=e^{i\gamma t}$ ($\gamma=|\kappa^2\mu/[4(3\lambda-1)]|$).
However, we find two more oscillatory singular points, namely
$P_{19,20}$, which correspond to $a(t)=e^{i\delta t}$, with
$\delta=|\kappa^2\mu\Lambda/[4(3\lambda-1)]|$. These points are
nonhyperbolic, with a negative eigenvalue, and thus they have a
large probability to be the late-time state of the universe, and
additionally this result is independent of the specific form of
the dark-matter content. Furthermore, they correspond to a
dark-energy dominated universe, with $w_{DE}=-1$ and arbitrary
$w_M$. Thus, they are good candidates for a realistic description
of the universe.

\section{Beyond detailed balance: phase space
analysis}\label{HLnondetbal}

In this section we extend the phase-space analysis to a universe
governed by Ho\v{r}ava gravity in which the detailed balance
condition has been relaxed. In order to transform the
corresponding cosmological equations into an autonomous dynamical
system, we use the auxiliary variables $x$ and $y$ defined in
(\ref{HLauxilliaryx}),(\ref{HLauxilliaryy}), and furthermore we
define the following four new ones:
\begin{eqnarray}
&&x_1=\frac{\sigma_1}{3 (3 \lambda -1) H^2},\nonumber\\
&&x_2=\frac{k \sigma_2}{3 (3 \lambda -1) a^2 H^2},\nonumber\\
&&x_3=\frac{\sigma_3}{3 (3 \lambda -1) a^4 H^2}\nonumber\\
&&x_4=\frac{2 k \sigma_4}{(3 \lambda -1) a^6 H^2}.
 \label{HLauxilliaryc}
\end{eqnarray}
Thus, using these variables and the definitions (\ref{HLrhoDEext})
and (\ref{HLpDEext}), we can express the dark energy density and
equation-of-state parameters respectively  as:
\begin{eqnarray}
&& \Omega_{DE}|_{_\text{non-db}}\equiv\frac{2}{(3\la-1)H^2}\left(
\frac{\sigma_1}{6}+\frac{\sigma_3k^2}{6a^4} +\frac{\sigma_4k}{a^6}
 \right)=\nonumber\\
 &&\ \ \ \ \ \ \ \ \ \ \ \ \ \ \ =x_1 + x_3 + x_4,
  \label{HLOmegaDEext}
 \end{eqnarray}
\begin{equation}
  w_{DE}|_{_\text{non-db}}\equiv \frac{-\frac{\sigma_1}{6}+\frac{\sigma_3k^2}{18a^4}
+\frac{\sigma_4k}{6a^6} }{
\frac{\sigma_1}{6}+\frac{\sigma_3k^2}{6a^4}
+\frac{\sigma_4k}{a^6}}=-\frac{6 x_1 - 2 x_3 - x_4}{6 (x_1 - x_3 +
x_4)}
 \label{HLwDEext}.
\end{equation}
Note that the corresponding quantities for dark matter coincide
with those of the detailed balance case (expressions
(\ref{HLOmegaM}) and (\ref{HLwM})).

\begin{table*}[t]
\begin{center}\caption{\label{HLcritext} The singular points of a universe
governed by Ho\v{r}ava gravity beyond detailed balance (system
\ref{HLexty})) and their behavior. NH stands for nonhyperbolic
(adapted from \cite{Leon:2009rc}).}{
\begin{tabular}{cccccccc}
\hline \hline
 Cr. P &${x_2}_c$&${x_3}_c$&${x_4}_c$&${x}_c$&${y}_c$&  Existence &  Stable for\\
\hline $P_{23}$& 0 & 0 & $1-{{x}_c}^2$ & ${x}_c$ & 0 & All $q$ & NH\vspace{0.2cm}\\
\hline $P_{24,25}$& 0 & 0 & $0$ & $\pm 1$ & 0 & All $q$ & NH\vspace{0.2cm}\\
\hline $P_{26}$& 0 & 0 & $0$ & 0 & 0 & All $q$ & stable\vspace{0.2cm}\\
\hline $P_{27,28}$& 0 & 0 & 0 & $\sqrt{\frac{2}{3}} q$ &
$\pm\sqrt{1-\frac{2 q^2}{3}}$ &  $q^2\leq \frac{3}{2}$  & unstable\vspace{0.2cm}\\
\hline $P_{29}$& 1 & 0 & 0 & 0 & 0 & All $q$ & unstable\vspace{0.2cm}\\
\hline $P_{30,31}$& $1-\frac{1}{2q^2}$ & 0 & 0 &
$\frac{1}{\sqrt{6}q}$ & $\pm\frac{1}{\sqrt{3}q}$ & $q\neq 0$ & unstable\vspace{0.2cm}\\
\hline $P_{32}$& 0 & 1 & 0 & 0 & 0 & All $q$  & unstable\vspace{0.2cm}\\
\hline $P_{33,34}$& 0 & $1-\frac{1}{q^2}$ & 0 &
$\frac{\sqrt{\frac{3}{2}}}{q}$ & $\pm\frac{1}{\sqrt{3}q}$ & $q\neq
0$ & unstable\vspace{0.2cm}\\
\hline $P_{35}$& 0 & 0 & $1-\frac{3}{2 q^2}$ &
$\frac{\sqrt{\frac{3}{2}}}{q}$ & 0 & All $q$ & NH\vspace{0.2cm}\\
\vspace{-0.35cm}\\
\hline \hline
\end{tabular}}
\end{center}
\end{table*}

\begin{table}[t]
\begin{center}\caption{\label{HLdensities} Observable cosmological quantities of
a universe governed by Ho\v{r}ava gravity beyond detailed
balance.}
\begin{tabular}{ccccccccc}
\hline \hline
 Cr. P &  $w_M$&  $\Omega_M$ &  $\Omega_{DE}$&  $w_{DE}$\\
\hline $P_{23}$ & 1 & ${x_c}^2$ & $1 - {x_c}^2$ & $1/6$ \vspace{0.2cm}\\
\hline $P_{24,25}$ &1 & 1 & 0 & arbitrary \vspace{0.2cm}\\
\hline $P_{26}$ & arbitrary& 0& 1& -1 \vspace{0.2cm}\\
\hline $P_{27,28}$ & $\frac{4 q^2}{3}-1$ &1 &$\frac{-2
q^2+\sqrt{9-6 q^2}+3}{-2
q^2+\sqrt{9-6 q^2}+6}$ &-1 \vspace{0.2cm}\\
\hline $P_{29}$ & 1 &1  &0 & arbitrary \vspace{0.2cm}\\
\hline $P_{30,31}$ & $-\frac{1}{3}$ & $\frac{1}{2 q^2}$ &
$\frac{\sqrt{3} q+1}{3 q^2+\sqrt{3} q+1}$ & -1 \vspace{0.2cm}\\
\hline $P_{32}$ & arbitrary &0 & 1& 1/3 \vspace{0.2cm}\\
\hline $P_{33,34}$& $\frac{1}{3}$ & $\frac{1}{q^2}$ &
$1+\frac{3}{\left(\sqrt{3}-3 q\right) q-1}$ &$\frac{2-q
   \left(q+\sqrt{3}\right)}{\left(\sqrt{3}-3 q\right) q+2}$\vspace{0.2cm}\\
\hline $P_{35}$ &1 & $\frac{3}{2 q^2}$ & $1-\frac{3}{2 q^2}$ &$\frac{1}{6}$\vspace{0.2cm}\\
\vspace{-0.35cm}\\
\hline \hline
\end{tabular}
\end{center}
\end{table}

  Using the aforementioned auxiliary variables,
   the Friedmann equations (\ref{HLFr1c}), (\ref{HLFr2c}) become:
\begin{eqnarray}
&&1=x_1 + x_2 + x_3 + x_4 + x^2 + y^2\label{HLFr1bextendedcase}
\\
&&\frac{H'}{H}=-3 x^2 - x_2 - 2 x_3 - 3 x_4.
\label{HLFr2bextendedcase}
\end{eqnarray}
Thus, after using the first of these relations in order to
eliminate one variable, the corresponding autonomous system
writes:
\begin{eqnarray}
&&x_2'=2 x_2 \left(3 x^2+x_2+2 x_3+3
x_4-1\right),\label{HLext2}\nonumber\\
&&x_3'=2x_3 \left(3 x^2+x_2+2 x_3+3 x_4-2\right),\label{HLext3}\nonumber\\
&&x_4'=2 x_4 \left(3x^2+x_2+2 x_3+3
   x_4-3\right),\label{HLext4}\nonumber\\
&& x'=   3 x^3+(x_2+2 x_3+3 x_4-3) x+\sqrt{6}
q y^2,\label{HLextx}\nonumber\\
&& y'=\left(3 x^2-\sqrt{6}
   q x+x_2+2 x_3+3 x_4\right) y,\label{HLexty}
\end{eqnarray}
defining a dynamical system in $\mathbb{R}^5$. Its singular points
and their properties are displayed in table \ref{HLcritext}  and
in table \ref{HLdensities} we present the corresponding observable
cosmological quantities.

The curve of nonhyperbolic singular points denoted by $P_{23}$ is
``normally hyperbolic'' \cite{Aulbach1984a} because they give rise
to the eigenvalues $\{6,\, 0,\, 2,\,4,\, 3-\sqrt{6} q x_c\}$ and
the eigenvector associated to the zero eigenvalues is tangent to
the set. \footnote{This curve contains the singular points
$P_{24,25}$ for the choice $x_c=\pm 1,$ and $P_{35}$ for the
choice $x_c=\frac{\sqrt{\frac{3}{2}}}{q}.$ In general for a
singular point located at $P_{23}$ we can obtain accurate
information about its stability by using the center manifold
theory.} Thus, examining the sign of the real parts of the
non-null eigenvalues, we find that they are always local sources
provided $q x_c>\sqrt{6}/2$. Amongst all nonhyperbolic singular
points, $P_{26},$ whose eigenvalues are $\{-6,\,-4,\,-3,\, -2,\,
0\}$, proves to be a stable one (this result was proved in
\cite{Leon:2009rc} using Normal Forms calculations). We will
present next the stability analysis of its center manifold.
 $P_{27,28}$ are saddle points
and their stable manifold can be 4-dimensional provided
$-\frac{\sqrt{2}}{2}<q<\frac{\sqrt{2}}{2}$ because they give rise
to the eigenvalues $\{4 q^2,\,-3+2 q^2,\,2(-3+2
q^2),\,4(-1+q^2),\, 2(-1+2 q^2)\}$. $P_{29}$ has eigenvalues
$\{-4,\, -2,\, -2,\, 2,\, 1\},$ thus, it admits a 2-dimensional
unstable manifold tangent to the $x_2$-$y$ plane and its stable
manifold is always 3-dimensional. The eigenvalues of the
linearization around $P_{30,31}$ are $\left\{-4,\,-2,\,2,\,
-1-\sqrt{-3+\frac{2}{q^2}},\,
 -1+\sqrt{-3+\frac{2}{q^2}}\right\},$ thus, they have a 4-dimensional
stable manifold provided $q^2>\frac{2}{3}$ or
$-\sqrt{\frac{2}{3}}\leq q\leq -\frac{\sqrt{2}}{2}$ or
$\frac{\sqrt{2}}{2}< q\leq \sqrt{\frac{2}{3}}$. The point $P_{32}$
is a saddle point since its eigenvalues are  $\{-4,\,-2,\,2,\,2,\,
 -1\}.$ Finally, $P_{33,34}$
has a 3-dimensional stable manifold if $q^2>\frac{16}{15}$ or
$-{\frac{4}{\sqrt{15}}}\leq q< -1$ or $1< q\leq
{\frac{4}{\sqrt{15}}}$ because they gives rise to the eigenvalues
$\left\{4,\,-2,\,2,\,
-\frac{1}{2}\left(1-\sqrt{-15+\frac{16}{q^2}}\right),\,
 -\frac{1}{2}\left(1+\sqrt{-15+\frac{16}{q^2}}\right)\right\}.$

\subsection{Stability Analysis of the \emph{de Sitter} Solution
in Ho\v{r}ava-Lifshitz cosmology} \label{stabilityP26}

In order to analyze the stability of \emph{de Sitter} solution we
can use center manifold theorem.

\begin{prop}\label{centerP26} The origin for the system
\eqref{HLexty} is locally asymptotically stable.
\end{prop}

In order to determine the local center manifold of \eqref{HLexty}
at the origin we have to transform the system into a form suitable
for the application of the center manifold theorem (see section
\ref{sectionCM} for a summary of the techniques involved in the
proof).

{\bf Proof}. In order to transform the linear part of the vector
field into its Jordan canonical form, we define new variables
$(u,v_1,v_2,v_3,v_4)\equiv\mathbf{x}$, by the equations \be
u=y,\,v_1=x_4,\, v_2=x_3,\,v_3=x,\,v_4=x_2\nonumber\ee so that \be
\left(\begin{array}{c} u'\\v_1'\\v_2'\\v_3'\\v_4'
\end{array}\right)=\left(\begin{array}{ccccc} 0& 0 &0 &0 &0 \\
0& -6 &0 &0 &0
\\  0& 0 & -4 &0 &0
\\ 0& 0 & 0 &-3 &0\\0& 0 &0 &0 & -2
\end{array}\right)\left(\begin{array}{c}u\\v_1\\v_2\\v_3\\v_4
\end{array}\right)+\left(\begin{array}{c}f(u,v_1,v_2,v_3,v_4)\\g_1(u,v_1,v_2,v_3,v_4)\\g_2(u,v_1,v_2,v_3,v_4)
\\g_3(u,v_1,v_2,v_3,v_4)\\g_3(u,v_1,v_2,v_3,v_4)\end{array}\right)\label{HLcenter2}
\ee where

$f(u,v_1,v_2,v_3,v_4)=u \left(3 v_1+2 v_2-\sqrt{6} q v_3+4
   v_4\right),$
\\
$g_1(u,v_1,v_2,v_3,v_4)=2 v_1 \left(3 v_3^2+3 v_1+2
   v_2+v_4\right),$
\\
$g_2(u,v_1,v_2,v_3,v_4)=2 v_2 \left(3 v_3^2+3 v_1+2
   v_2+v_4\right),$
\\
$g_3(u,v_1,v_2,v_3,v_4)=\sqrt{6} q u^2+v_3 \left(3 v_3^2+3
   v_1+2 v_2+v_4\right),$ and
 $g_4(x,y_1,y_2,y_3)=2 v_4 \left(3 v_3^2+3
   v_1+2 v_2+v_4\right).$
\\
The system \eqref{HLcenter2} is written in diagonal form
\begin{align}
u'  &  =Cu+f\left(  u,\mathbf{v}\right) \nonumber\\
\mathbf{v}'  &  =P\mathbf{v}+\mathbf{g}\left(  u,\mathbf{v}\right)
, \label{HLcenter3}
\end{align}
where $\left(  u,\mathbf{v}\right)
\in\mathbb{R}\times\mathbb{R}^{4},$ $C$ is the zero $1\times1$
matrix, $P$ is a $4\times 4$ matrix with negative eigenvalues and
$f,\mathbf{g}$ vanish at $\mathbf{0}$ and have vanishing
derivatives at $\mathbf{0.}$ The center manifold theorem
\ref{existenceCM} asserts that there exists a 1-dimensional
invariant local center manifold $W^{c}\left( \mathbf{0}\right) $
of \eqref{HLcenter3} tangent to the center subspace (the
$\mathbf{v}=\mathbf{0}$ space) at $\mathbf{0}.$ Moreover,
$W^{c}\left( \mathbf{0}\right)  $ can be represented as
\[
W^{c}\left(  \mathbf{0}\right)  =\left\{  \left(
u,\mathbf{v}\right)
\in\mathbb{R}\times\mathbb{R}^{4}:\mathbf{v}=\mathbf{h}\left(
u\right) ,\;\left\vert u\right\vert <\delta\right\}
;\;\;\;\mathbf{h}\left(  0\right) =\mathbf{0},\;D\mathbf{h}\left(
0\right)  =\mathbf{0}\}
\]
for $\delta$ sufficiently small (see definition \ref{CMdef}). The
restriction of (\ref{HLcenter3}) to the center manifold is (see
definition \ref{vectorfieldCM})
\begin{equation}
u'=f\left( u,\mathbf{h}\left(  u\right)  \right)  . \label{HLrest}
\end{equation}
According to Theorem \ref{stabilityCM}, if the origin $x=0$ of
\eqref{HLrest} is stable (asymptotically stable) (unstable) then
the origin of \eqref{HLcenter3} is also stable (asymptotically
stable) (unstable). Therefore, we have to find the local center
manifold, i.e., the problem reduces to the computation of
$\mathbf{h}\left( u\right).$

Substituting $\mathbf{v}=\mathbf{h}\left(  u\right)  $ in the
second component of \eqref{HLcenter3} and using the chain rule,
$\mathbf{v }'=D\mathbf{h}\left(  u\right)  u'$, one can show that
the function $\mathbf{h}\left( u\right)  $ that defines the local
center manifold
satisfies%
\begin{equation}
D\mathbf{h}\left(  u\right)  \left[  f\left(  u,\mathbf{h}\left(
u\right) \right)  \right]  -P\mathbf{h}\left(  u\right)
-\mathbf{g}\left( u,\mathbf{h}\left(  u\right)  \right)  =0.
\label{HLh}
\end{equation}
According to Theorem \ref{approximationCM}, equation \eqref{HLh}
can be solved approximately by using an approximation of
$\mathbf{h}\left(  u\right)  $ by a Taylor series at $u=0.$ Since
$\mathbf{h}\left(  0\right)  =\mathbf{0\ } $and $D\mathbf{h}\left(
0\right)  =\mathbf{0},$ it is obvious that $\mathbf{h}\left(
u\right)  $ commences with quadratic terms. We substitute
\[
\mathbf{h}\left(  x\right)  =:\left[
\begin{array}
[c]{c}%
h_{1}\left(  x\right) \\
h_{2}\left(  x\right) \\
h_{3}\left(  x\right)\\
h_{4}\left(  x\right)
\end{array}
\right]  =\left[
\begin{array}
[c]{c}%
a_{1}u^{2}+a_{2}u^{3}+a_{3}u^{4}+O\left(  u^{5}\right) \\
b_{1}u^{2}+b_{2}u^{3}+b_{3}u^{4}+O\left(  u^{5}\right) \\
c_{1}u^{2}+c_{2}u^{3}+c_{3}u^{4}+O\left(  u^{5}\right) \\
d_{1}u^{2}+d_{2}u^{3}+d_{3}u^{4}+O\left(  u^{5}\right)
\end{array}
\right]
\]
into (\ref{HLh}) and set the coefficients of like powers of $u$
equal to zero to find the unknowns $a_{1},b_{1},c_{1},d_{1},...$.

We find that the non-zero coefficients are $$c_1=
   \sqrt{\frac{2}{3}} q,\, c_3=
\frac{4}{3} \sqrt{\frac{2}{3}} q^3,$$ Therefore, \eqref{HLrest}
yields \be u'=-2 q^2 u^3+\left(2 q^2-\frac{8 q^4}{3}\right) u^5+O\left(u^6\right).\label{HLrest1}\ee Neglecting the error
terms, this  is a gradient-like equation (i.e., $u'=-\nabla U(u)$)
with potential $U(u)=\frac{1}{9} q^2 \left(4 q^2-3\right) u^6+\frac{q^2 u^4}{2}$ for
which the origin is a degenerate minimum.  Thus, the origin $u=0$
of \eqref{HLrest1} is locally asymptotically stable. Hence, the
origin $\mathbf{u}=\mathbf{0}$ of the full five-dimensional system
is asymptotically stable. $\blacksquare$

\subsection{Cosmological implications: Beyond detailed balance}\label{HLsection O2}

Let us now discuss about the cosmological behavior of a
Ho\v{r}ava-Lifshitz universe, in the case where the detailed
balance condition is abandoned. In this case the system admits the
unstable singular points $P_{27,28,29}$ which correspond to dark
matter domination, the unstable point $P_{32}$ corresponding to an
unphysical dark-energy dominated universe, and the unstable
$P_{30,31,33,34}$ which have physical $w_M$, $w_{DE}$ but
dependent on the specific dark-matter form. The system admits also
the singular points $P_{23}$, $P_{35}$ which are nonhyperbolic
with positive non-null eigenvalues, thus unstable, with
furthermore unphysical cosmological quantities. Additionally,
points $P_{24,25}$ are also dark-matter dominated, unstable
nonhyperbolic ones.

It is interesting to notice that since $\sigma_3$ has an arbitrary
sign, $P_{33,34}$ could also correspond to an oscillatory
universe, for a wide region of the parameters $\sigma_3$ and $q$.
However, this oscillatory behavior has a small probability to be
the late-time state of the universe because it is not stable (with
at least two positive eigenvalues). Additionally, the fact that it
depends on $q$ means that this solution depends on the matter form
of the universe.

The scenario at hand admits a final singular point, namely
$P_{26}$. As we showed in detail in section \ref{HLnondetbal}
using Center Manifold Theory, it is indeed asymptotically stable
and thus it can be a late-time attractor of Ho\v{r}ava-Lifshitz
universe beyond detailed balance. Using the definition of the
auxiliary variables, we can straightforwardly show that it
corresponds to an eternally expanding solution. Additionally, it
is characterized by complete dark energy domination, with
dark-energy equation-of-state parameter $-1$ and arbitrary $w_M$.
Note also that this result is independent of the specific form of
the dark-matter content. These feature make it a very good
candidate for the description of our universe. We mention that
according to the initial conditions, this universe on its way
towards this late-time attractor can be just an expanding universe
with a non-negligible dark matter content, which is in agreement
with observations, and this can be verified also by numerical
investigation. This fact makes the aforementioned result more
concrete.

\section{Conclusions}
\label{HLconclusions}

In this work we performed a detailed phase-space analysis of
Ho\v{r}ava-Lifshitz cosmology, with and without the
detailed-balance condition. In particular, we examined if a
universe governed by Ho\v{r}ava gravity can have late-time
solutions compatible with observations.

In the case where the detailed-balance condition is imposed, we
find that the universe can reach a bouncing-oscillatory state at
late times, in which dark-energy, behaving as a simple
cosmological constant, will be dominant. Such solutions were
already investigated in the context of Ho\v{r}ava-Lifshitz
cosmology
\cite{Brandenberger:2009yt,Brandenberger:2009ic,Cai:2009in} as
possible ones, but now we see that they can indeed be the
late-time attractor for the universe. They arise purely from the
novel terms of Ho\v{r}ava-Lifshitz cosmology, and in particular
the dark-radiation term proportional to $a^{-4}$ is responsible
for the bounce, while the cosmological constant term is
responsible for the turnaround.

In the case where the detailed-balance condition is abandoned, we
find that the universe reaches an eternally expanding solution at
late times, in which dark-energy, behaving like a cosmological
constant, dominates completely. Note that according to the initial
conditions, the universe on its way to this late-time attractor
can be an expanding one with non-negligible matter content. We
mention that this behavior is independent of the specific form of
the dark-matter content. Thus, the aforementioned features make
this scenario a good candidate for the description of our
universe, in consistency with observations. Finally, in this case
the universe has also a probability to reach an oscillatory
solution at late times, if the initial conditions lie in its basin
of attraction (in this case the eternally expanding solution will
not be reached).

Although this analysis indicates that Ho\v{r}ava-Lifshitz
cosmology can be compatible with observations, it does not
enlighten the discussion about possible conceptual and
phenomenological problems and instabilities of Ho\v{r}ava-Lifshitz
gravity,  nor it can interfere with the questions concerning the
validity of its theoretical background, which is the subject of
interest of other studies. It just faces the problem from the
cosmological point of view, and thus its results can been taken
into account only if  Ho\v{r}ava gravity passes successfully the
aforementioned theoretical tests.

\chapter{Cardassian Cosmologies}

In this chapter we analyze the asymptotic behavior of Cardassian
cosmological models filled with a perfect fluid and a scalar field
with an exponential potential. Cardassian cosmologies arise from
modifications of the Friedmann equation, and among the different
proposals within that framework we will choose those of the form
$3 H^2-\rho \propto \rho^n$ with $n < 1$. We construct a three
dimensional dynamical systems arising from the evolution
equations. Using standard dynamical systems techniques we find the
fixed points and characterize the solutions they represent. We pay
especial attention to the properties inherent to the modifications
and compare with the (standard) unmodified scenario. Among other
interesting results, we find there are no late-time scaling
attractors.

\section{Introduction}

Cardassian cosmologies are non-relativistic phenomenological
models without a covariant formulation as those based in General
Relativity and they have a different justification than
quintessence. However, with regard to the observational tests that
depend only on the scale or the Hubble factor, in the late-time
regime Cardassian models filled with just matter ($ \rho\propto
a^{-3}$) are indistinguishable from perfect fluid models with a
$p=(\gamma-1)\rho$ equation of state under the identification
$n\equiv \gamma$. These perfect fluid models are in turn
kinematically equivalent to scalar field (quintessence) models
with an exponential potential. In this way, the Cardassian model
can make contact with quintessence with regard to observational
tests. Interestingly, observational tests seem to favor $n<0$, so
that asymptotically one would get a phantom equation of state
\cite{Caldwell:1999ew,Nojiri:2003vn,Carroll:2003st,Onemli:2002hr,Elizalde:2004mq,Singh:2003vx,Gonzalez-Diaz:2004vq,Johri:2003rh,Hannestad:2002ur,Gonzalez-Diaz:2004eu,Chimento:2003qy,Stefancic:2003rc,Stefancic:2003bj,Onemli:2004mb,Brunier:2004sb,Nojiri:2003ag}.
However, the equivalence between Cardassian and perfect fluid
models is not extensible to the dynamical realm, the evolution of
perturbations may differ significantly, and this can lead to
discrepancies for instance in observational tests associated with
the cosmic microwave bakground. Nevertheless, as stated in
\cite{Freese:2002sq}, questions of interpretation remain open,
because  in the Cardassian model matter alone is responsible for
the accelerated behavior, and yet the universe can be flat. The
condition for acceleration is $n<2/3.$

Dynamical systems techniques have been using for exploring
Cardassian models filled with baryonic matter in
\cite{Szydlowski:2004np}. However, in \cite{Lazkoz:2005bf} was
given a step further by allowing as well for a scalar field
component (non-baryonic matter). As well as in
\cite{Lazkoz:2005bf} in this chapter we investigate for early and
late-time tracking (scaling) solutions; i.e., such that both
baryonic and non-baryonic matter contributes with non-negligible
and proportional fractions to the critical density. Tracking
solutions are particularly interesting because their dynamical
effects mimics a decaying cosmological constant (see the classical
references
\cite{Amendola:1999qq,Chimento:2000kq,Copeland:1997et,vandenHoogen:1999qq,Nunes:2000yc,Steinhardt:1999nw,Uzan:1999ch,Billyard:1998hv}).
Such solutions would be devoid of the fine-tuning problems posed
by a cosmological constant precisely because of the independence
on the initial conditions.

For this study we consider a self-interacting exponential
potential
\cite{Amendola:1999qq,Chimento:2000kq,Copeland:1997et,vandenHoogen:1999qq,Nunes:2000yc,Steinhardt:1999nw,Uzan:1999ch,Billyard:1998hv,Chimento:1998ju,Liddle:1998xm,Kehagias:2004bd}
for allowing the reduction of the dimension of the resulting
autonomous dynamical systems \cite{Copeland:2004hq}.

For the case $0<n<2/3$ it is possible to make comparisons with
\cite{Copeland:1997et}. However, since the recent observations
favors the case $n<0$
\cite{Wang:2003cs,Zhu:2004ij,Lazkoz:2005sp,Szydlowski:2003fg,Nesseris:2004wj}
we also consider these values in our numerical simulations.

\goodbreak

\section{Field equations}
The evolution equations for a flat Friedmann-Robertson-Walker
(FRW) Cardassian cosmological model filled with a scalar field
$\phi$ with self-interaction potential $V(\phi)=V_0\exp(-s\phi)$
and a barotropic perfect fluid with equation of state
$p_{\gamma}=(\gamma-1)\rho_{\gamma}$ are
\begin{eqnarray}
&&2\dot H+\left(\gamma\rho_{\gamma}+\dot\phi^2\right)
\left(1+n\sigma \rho_{\rm tot}^{n-1}\right)=0\label{cardF2}
\\
&&\ddot\phi+3H\dot\phi+\frac{d V(\phi)}{d\phi}=0,\label{cardKG}\\
&&\dot\rho_{\gamma}+3\gamma H\rho_{\gamma}=0 ,\label{cardCE}
\end{eqnarray}
where for the total energy density $\rho_{\rm tot}$ we have
\begin{equation}
    \rho_{\rm tot}=\frac{1}{2}\dot\phi^2+V(\phi)+\rho_{\gamma}.
\end{equation}

The evolution equations (\ref{cardF2}-\ref{cardCE}) are in turn
subject to the constraint
\begin{equation}
    H^2=\frac{1}{3}\rho_{\rm tot}\left(1+\sigma \rho_{\rm tot}^{n-1}\right).\label{carF1}\\
\end{equation}
Here and throughout $\sigma$, $V_0$  and $s$ will be free
parameters, and we will restrict ourselves to the $n<1$ case.

\section{Phase-space analysis}

Experience has demonstrated that dynamical systems methods can be
used to describe the evolution of cosmological models  by means of
past and future attractors. In order to cast our set of equations
as a dynamical system, it is convenient to normalize the
variables,  because in the vicinity of an hypothetical initial
singularity physical variables would typically diverge, whereas at
late times they commonly tend to zero \cite{Wainwright:2004cd}.
Due to physical considerations normalization with the Hubble
factor is an appropriate choice in cosmology. Besides, all
available mathematical evidence suggests that Hubble-normalized
variables are bounded into the past (that is, as the initial
singularity is approached), and if there is a cosmological
constant (or something that mimics it) it seems those variables
will also be bounded into the future. Thus, even though the
Hubble-normalized state space is unbounded, it is sensible to
expect that the evolution equations will admit a past attractor
and a future attractor.

Let us introduce the normalized variables \be w=\frac{\sigma
\rho_{\rm tot}^n}{3H^2}, x=\frac{\dot\phi}{\sqrt{6}H},\,
y=\frac{\sqrt{V}}{\sqrt{3} H}, \,
z=\frac{\sqrt{\rho_\gamma}}{\sqrt{3} H}\label{cardvars}\ee

This coordinates will allow us analyzing the solutions of
(\ref{cardF2}-\ref{cardKG}), and the cosmological models
associated with them. In addition, the variables will be related
among them through
\begin{equation}
w+x^2+y^2+z^2=1\label{cardlig}.
\end{equation}

The constraint (\ref{cardlig}) lets us ``forget'' about the
evolution of one of the coordinates. Here we will choose the
discarded coordinate to be $w$. Using the variables
\eqref{cardvars}, equation (\ref{cardlig}), and the conservation
equations (\ref{cardKG}) and (\ref{cardCE}) we get the equations

\begin{eqnarray}
 x'&=&\frac{3 n x \left(z^2 (\gamma -2)-2 y^2\right)}{2
\left(x^2+y^2+z^2\right)}+\frac{1}{2} \left(\sqrt{6} s y^2-3 (n-1)
x
   \left(2 x^2+z^2 \gamma -2\right)\right)\label{cardx},\\
y'&=& \frac{3 n y \left(2 x^2+z^2 \gamma \right)}{2
\left(x^2+y^2+z^2\right)}-\frac{1}{2} y
   \left(\sqrt{6} s x+3 (n-1) \left(2 x^2+z^2 \gamma \right)\right),\label{cardy}\\
z'&=&\frac{3 n z \left(2 x^2+z^2 \gamma \right)}{2
   \left(x^2+y^2+z^2\right)}-\frac{3}{2} z \left(2 (n-1) x^2+\left((n-1) z^2+1\right) \gamma \right). \label{cardz}
 \end{eqnarray}

Observe that the equations are invariant under the variable
changes $y\to -y$, and $z\to -z$, but not under $x\to -x$, and so
in our numerical examples we will concentrate on the region
$\{x^2+y^2+z^2\leq1,  -1\leq x\leq 1,y\geq 0,\,z\geq 0\}.$ This is
equivalent to saying we are just considering expanding universes
($H>0$). However, an analytic description of all singular points
is presented in the lines below and in table \ref{eigencard}. We
will also set restrictions $0<\gamma<2$ and $s^2<6$ so that
neither the barotropic fluid nor the scalar field have
supraluminical sound speeds and the fluid satisfies the weak
energy condition. We will also assume $n<1$, as this is the case
of interest.

\begin{figure}[ht]
\begin{center}
\includegraphics[width=9cm, height=8.0cm,angle=0]{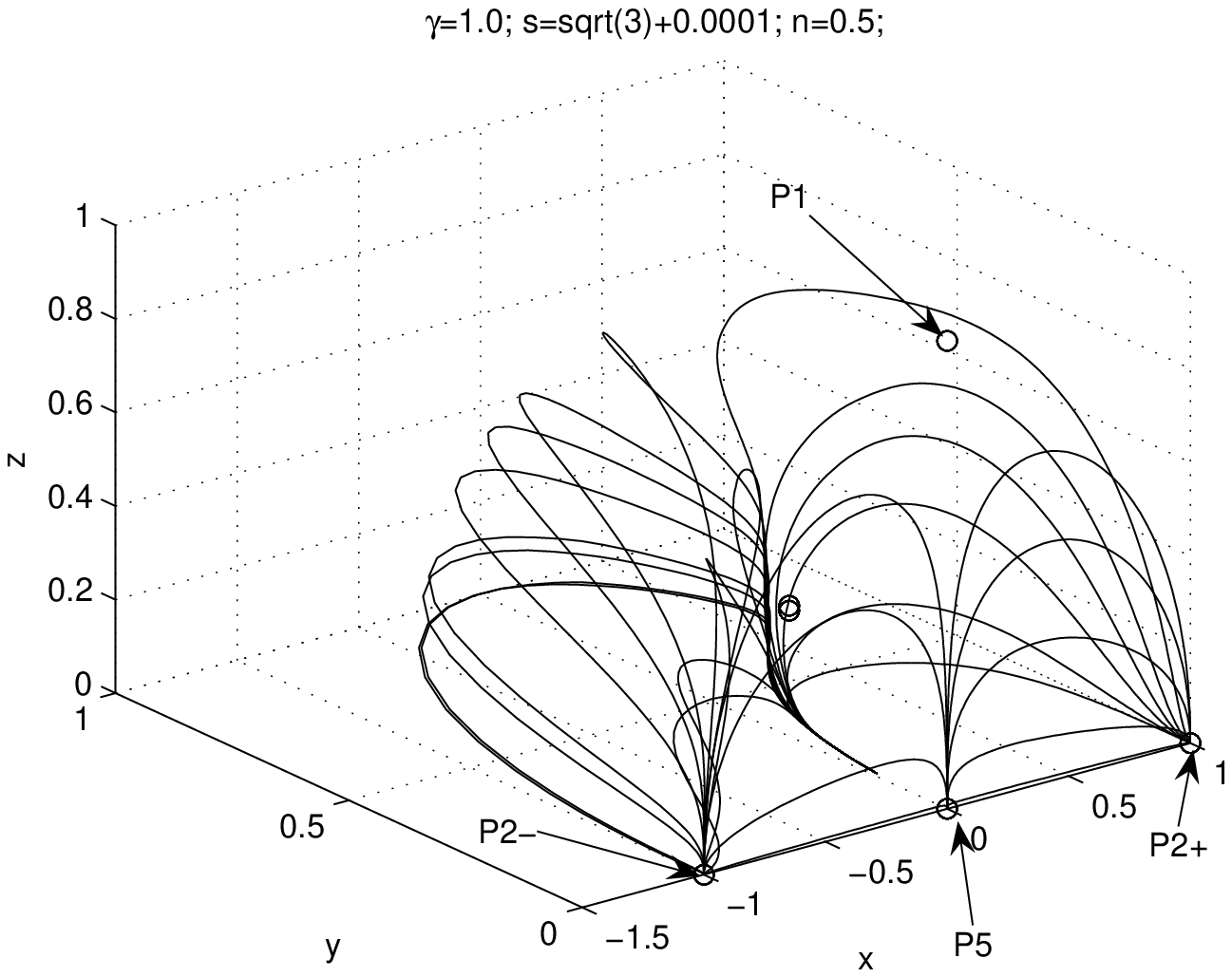}

\begin{center} (a) \end{center}
\includegraphics[width=9cm, height=8.0cm,angle=0]{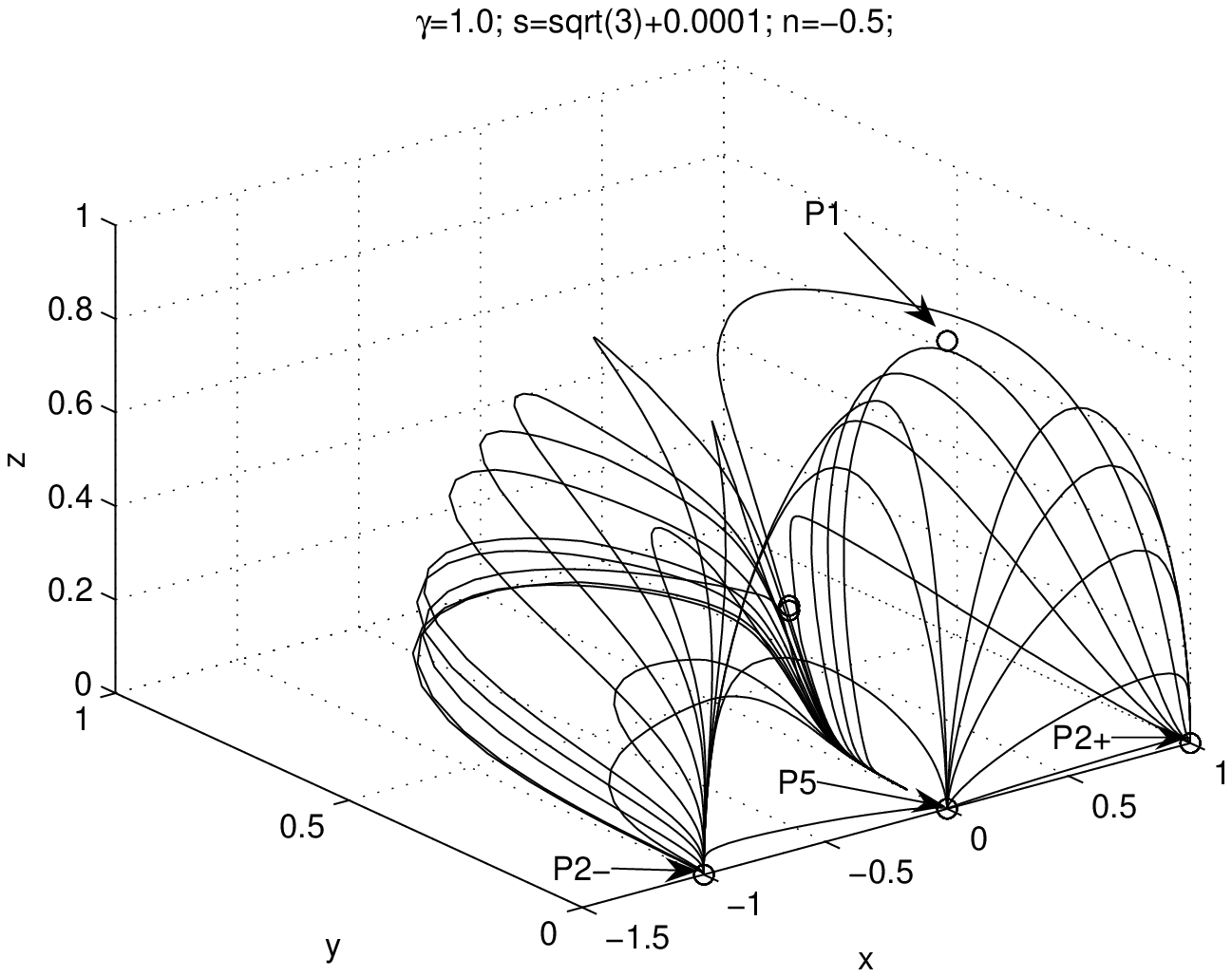}
\begin{center} (b) \end{center}
\caption{{\label{CardassianModel1}} Phase space of  Cardassian
models for the choices: (a) $\gamma=1.0, s=\sqrt{3}+0.0001,
n=0.5.$ and (b) $\gamma=1.0, s=\sqrt{3}+0.0001, n=-0.5.$}
\end{center}
\end{figure}

\begin{figure}[ht]
\begin{center}
\includegraphics[width=9.5cm, height=8.0cm,angle=0]{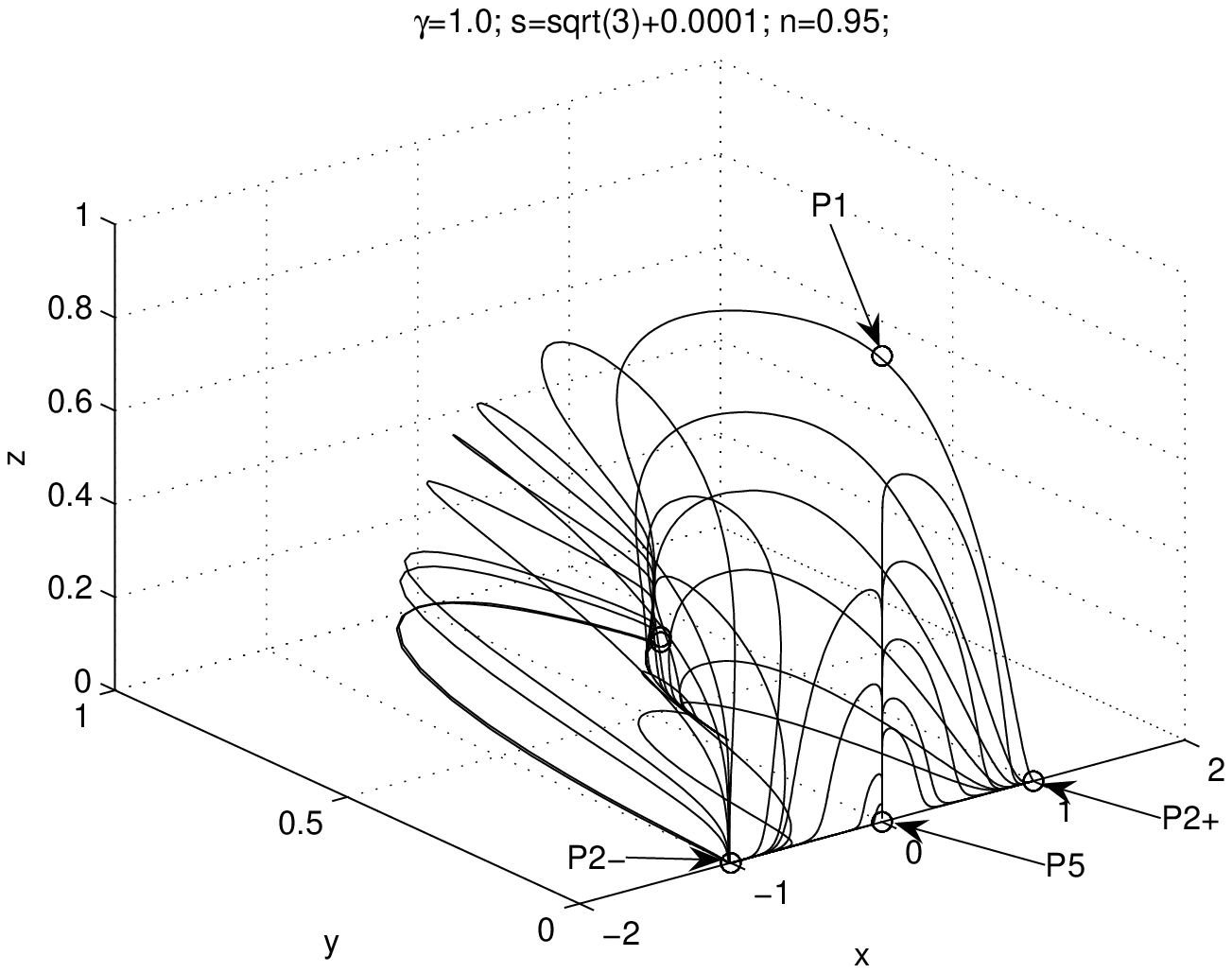}
\begin{center} (a) \end{center}
\includegraphics[width=9.5cm, height=8.0cm,angle=0]{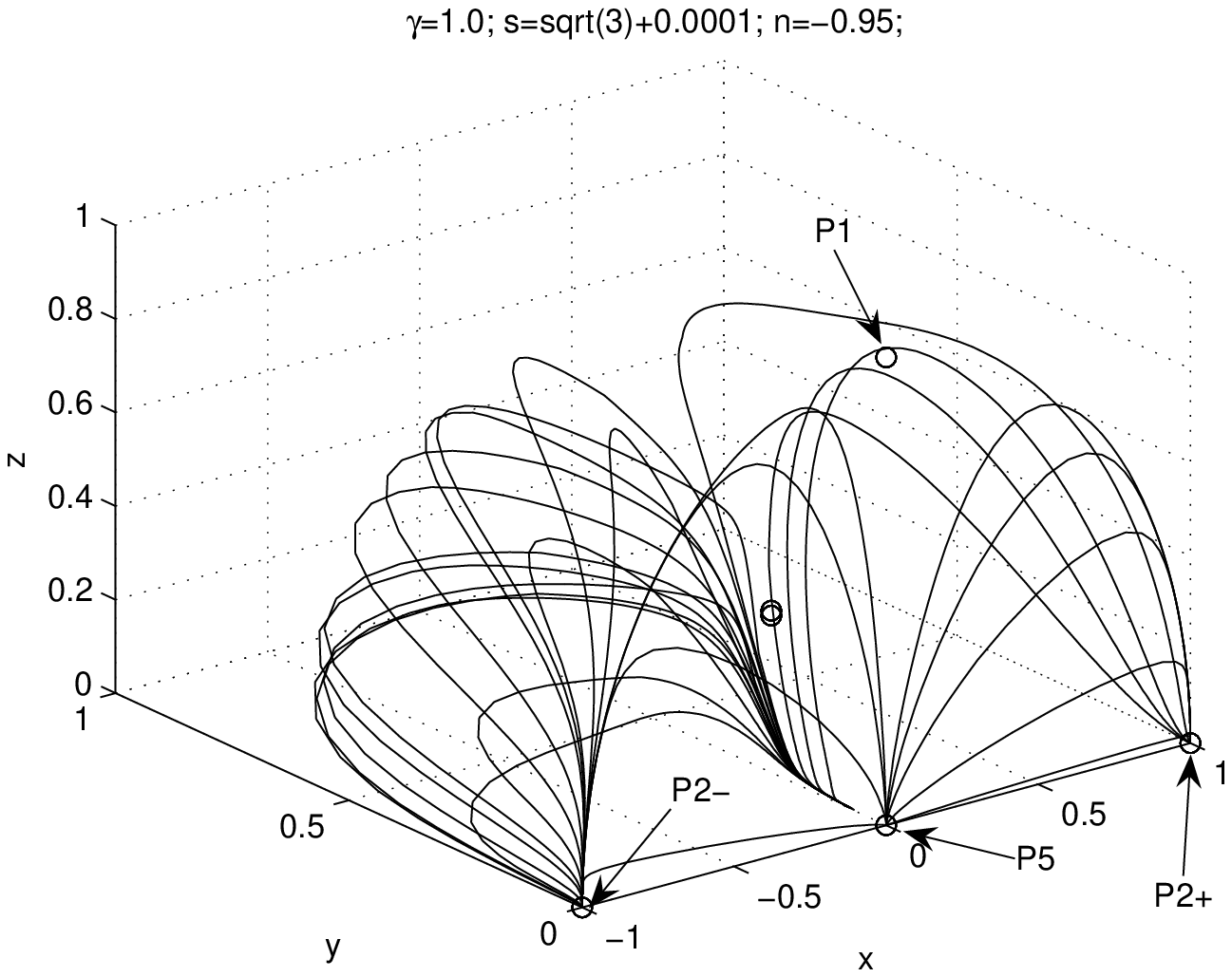}
\begin{center} (b) \end{center}
\caption{{\label{CardassianModel2}} Phase space of Cardassian
models for the choices: (a) $\gamma=1.0, s=\sqrt{3}+0.0001,
n=0.95.$ and (b) $\gamma=1.0, s=\sqrt{3}+0.0001, n=-0.95.$}
\end{center}
\end{figure}

In table \ref{eigencard} are displayed the coordinates, existence
conditions and stability conditions of the critical points of the
dynamical system \eqref{cardx}-\eqref{cardz}. Although the
location of the critical points of this dynamical system does not
depend  on $n$ \footnote{This is confirmed by numerical tests.},
the same is not true for their dynamical character. In table
\ref{eigencard} are displayed the eigenvalues of the Jacobian
matrix evaluated at each critical points.Due to the symmetries of
the equations \eqref{cardx}-\eqref{cardz}, the system can be
simplified by performing the coordinate transformation to
spherical coordinates $\{r,\theta ,\varphi \},$ i.e.,
$$x=r\sin\theta \cos \varphi, \, y=r\sin\theta \sin \varphi,\,
z=r\cos\theta.$$ Our region of interest is $$0\leq r\leq 1,\,
0\leq\theta\leq\frac{\pi}{2},\, 0<\varphi<\pi.$$ Equations
\eqref{cardx}-\eqref{cardz} becomes in spherical coordinates to
\begin{eqnarray}
& r'=-\frac{3}{4} (n-1) r \left(r^2-1\right) \left(2 \cos (2 \varphi ) \sin ^2\theta+\gamma +(\gamma -1) \cos (2 \theta )+1\right)\label{cardr},\\
& \theta'= -\frac{3}{2} \cos \theta (-\gamma +\cos (2 \varphi )+1) \sin \theta,\label{cardtheta}\\
& \varphi'= \frac{1}{2} \left(6 \cos (\varphi )-\sqrt{6} r s \sin
\theta\right) \sin (\varphi ). \label{cardvarphi}
 \end{eqnarray}
The origin of coordinates ($P_5$) corresponds to $r=0.$ Its
dynamical character cannot be determined using the linearization.
In fact, for $r=0,\theta=0,\varphi=0,$ the eigenvalues of the
linearization of \eqref{cardr}-\eqref{cardvarphi} are $3,\frac{3
(\gamma -2)}{2},\frac{3}{2} (n-1) \gamma,$ whereas for
$r=0,\theta=\pi/2,\varphi=\pi/2,$ they are
$-3,0,-\frac{3\gamma}{2}.$ Since both are different
representations for the origin, its dynamical character cannot be
anticipated from the linearization. This is due to the fact that
the dynamical system is not of class $C^1$ at the origin. Based in
numerical investigation (see figures \ref{CardassianModel1},
\ref{CardassianModel2}) we can support the claim that the origin
is a local sink for \eqref{cardx}-\eqref{cardz}.

\begin{landscape}

\begin{center}
\begin{table}[h!]
\caption{\label{eigencard} Location and existence conditions,
eigenvalues and dynamical character of the critical points of the
dynamical system \eqref{cardx}-\eqref{cardz}. We assume $n<1$,
$0<\gamma<2$ and $s^2< 6.$ We use the notation
$\beta_\pm=-\frac{3}{4}\left((2-\gamma)\pm\sqrt{(2-\gamma)(24\gamma^2/s^2+(2-9\gamma))}\right).$}
\vspace{0.3cm}\begin{center} {
\begin{tabular}{cccccc}
\hline \hline
Name & Coordinates: $x,y,z$  & Existence &  Eigenvalues & Dynamical Character\\\\
\hline\\
$P_1$  &$0,0,1$ & All $\gamma$ and $s$ & $\frac{3\gamma}{2}, \,
3(1-n)\gamma,-\frac{3(2-\gamma)}{2}$ & saddle. \\\vspace{5pt}
$P_2^\pm$ & $\pm 1,0,0$ & All $\gamma$ and $s$ & $6(1-n),\, 3\mp
\sqrt{\frac{3}{2}}s, \, \frac{3(2-\gamma)}{2}$ & source for $\pm
s<\sqrt{6};$ \\\vspace{5pt} & &  &  & saddle
otherwise.\\\vspace{5pt} $P_3$ &
$\displaystyle\frac{s}{\sqrt{6}},\displaystyle\sqrt{1-\frac{s^2}{6}},0$
&  $\displaystyle s^2<6$ & $-\frac{(6-s^2)}{2},\,
\frac{(s^2-3\gamma)}{2}, \, (1-n)s^2$ & non-hyp. for $s^2=3\gamma$
or $s=0;$
\\\vspace{5pt} & &  saddle otherwise.
\\\vspace{5pt}
$P_4$ &
$\,\displaystyle\sqrt{\frac{3}{2}}\frac{\gamma}{s},\,\displaystyle\sqrt{\frac{3(2-\gamma)\gamma}{2s^2}},\,\displaystyle\sqrt{1-\frac{3
\gamma }{s^2}}$ & $\displaystyle s^2>3\gamma\,$& $3(1-n)\gamma,\,
\beta_+,\,\beta_-$ & non-hyp. for $\gamma=0,2;$
\\\vspace{5pt} & &  saddle otherwise.
\\\vspace{5pt}
$P_5$ & $0,0,0$& All $\gamma$ and $s$ & undefined & local sink
(see text).
\\\vspace{-0.30cm}\\
\hline \hline
\end{tabular}}
\end{center}
\end{table}
\end{center}

\end{landscape}

\begin{prop}\label{PropCardassian}
The origin of coordinates of the dynamical system
\eqref{cardx}-\eqref{cardz} is:
\begin{enumerate}
\item stable for initial solutions outside the invariant set
$y=0;$ \item asymptotically stable for solutions in the invariant
set $y=0.$
\end{enumerate}
\end{prop}

{\bf Proof}.

In the limit $r\rightarrow 0,$ equations \eqref{cardtheta},
\eqref{cardvarphi} reduce to
\begin{eqnarray}
&& \theta'= \frac{3}{4} (\gamma -\cos (2 \varphi )-1) \sin (2 \theta ),\label{cardthetaapprox}\\
&& \varphi'= -\frac{3}{2} \sin (2 \varphi ).
\label{cardvarphiapprox}
 \end{eqnarray}
The approximated system
\eqref{cardthetaapprox}-\eqref{cardvarphiapprox} is completely
integrable with solution
\begin{eqnarray*}
\theta (\tau )= \tan ^{-1}\left(e^{\frac{3}{2} (\gamma -2)
   \tau +2 c_2} \sqrt{1+e^{6 \tau +4 c_1}}\right),\\
   \varphi (\tau )= \tan ^{-1}\left(e^{2 c_1-3 \tau }\right),
\end{eqnarray*}
where $c_1$ and $c_2$ are integration constants.

Taylor expanding \eqref{cardr} around $r=0$ we obtain
\begin{equation}\label{approxr}
r'=\frac{3}{4} (n-1) \left(2 \cos (2 \varphi ) \sin
^2\theta+\gamma +(\gamma -1) \cos (2 \theta )+1\right)
r+O\left(r^3\right)
\end{equation}
By substituting the previous first order solutions for $\theta
(\tau ), \varphi(\tau)$ in \eqref{approxr} and integrating the
resulting equation we obtain
\begin{equation*}
\rho(\tau)=e^{\frac{3}{2} (n-1) \gamma  \tau } \left(1+e^{3
(\gamma -2) \tau +4 c_2}+e^{3 \gamma  \tau +4
   \left(c_1+c_2\right)}\right){}^{\frac{1-n}{2}} c_3.
\end{equation*}

Passing to Cartesian coordinates we have that
\begin{eqnarray*}
&& x= a_2 e^{\frac{3}{2} (n \gamma -2) \tau } \left(e^{3 \gamma
\tau } a_1^2+a_2^2 e^{3 (\gamma -2) \tau
   }+1\right)^{-n/2} c_3,\\
&& y= a_1 e^{\frac{3 n \gamma  \tau }{2}} \left(e^{3 \gamma  \tau
} a_1^2+a_2^2 e^{3
   (\gamma -2) \tau }+1\right)^{-n/2} c_3,\\
&& z= e^{\frac{3}{2} (n-1) \gamma  \tau } \left(e^{3 \gamma  \tau
}
   a_1^2+a_2^2 e^{3 (\gamma -2) \tau }+1\right)^{-n/2} c_3,
\end{eqnarray*}
where we have introduced the reescaling $$c_1\to \frac{\log
(|a_1|)}{2}-\frac{\log (|a_2|)}{2},c_2\to \frac{\log
(|a_2|)}{2}.$$ Assuming when $n<1,$ and taking the limit as
$\tau\rightarrow +\infty$  in the above expressions, we obtain
that $x$ and $z$ approaches to zero at an exponential rate,
whereas $y\rightarrow \bar{y}\equiv y_0^{1-n}
\left(x_0^2+y_0^2+z_0^2\right)^{n/2},$ where
$x(0)=x_0,y(0)=y_0,z(0)=z_0.$ Let us assume that $n<1$ and let be
$\epsilon>0$ an arbitrary number. Then there exists a $\delta>0,$
such that $\delta<\epsilon.$ Let us consider the solution with
initial value $x(0)=x_0,y(0)=y_0,z(0)=z_0,$ with
$x_0^2+y_0^2+z_0^2<\delta^2.$ Since $y\rightarrow \bar{y},$
satisfying $|\bar{y}|<\delta,$ then the solution, ${\bf
x}(\tau,{\bf x}_0)$ passing through ${\bf
x}_0=\left(x_0,y_0,z_0\right)$ at $\tau=0,$ satisfies $\|{\bf
x}(\tau,{\bf x}_0)\|<\epsilon,$  for $\tau$ arbitrarily large. In
this way we prove the stability of $P_5.$ For solutions passing
through ${\bf x}_0=\left(x_0,y_0,z_0\right)$ at $\tau=0,$ with
$y_0=0,$ $P_5$ is asymptotically stable. $\blacksquare$

The result of proposition  \ref{PropCardassian} is illustrated
numerically in the figures \ref{CardassianModel1},
\ref{CardassianModel2}.

\section{Basic observables}

In this section we evaluate the deceleration parameter $$q\equiv
-\frac{a\ddot a}{\dot a^2},$$ and the effective EoS parameter
$$w_{\rm eff}\equiv\frac{\frac{1}{2}\dot\phi^2-V(\phi)+(\gamma-1)\rho_\gamma}{\rho_{\rm
tot}},$$ at the singular points.

\begin{center}
\begin{table}[h!]
\caption{\label{obscard} Deceleration parameter, $q,$ and
effective EoS, $w_{\rm eff}$ at the critical points of the
dynamical system \eqref{cardx}-\eqref{cardz}.}
\vspace{0.3cm}\begin{center} {
\begin{tabular}{ccc}
\hline \hline
Name & $q,\,w_{\rm eff}$  & Type of solution\\\\
\hline\\
$P_1$ & $\frac{1}{2} (3 \gamma -2),\, \gamma -1$ & accelerating
for $\gamma<\frac{2}{3};$ matter-dominated.
\\\vspace{5pt} $P_2^\pm$ & $2,\,1$ & decelerating;
kinetic-dominated.\\\vspace{5pt} $P_3$ & $\frac{1}{2}
\left(s^2-2\right),\,\frac{1}{3} \left(s^2-3\right)$ &
accelerating for $s^2<2;$ quintessence.
\\\vspace{5pt}
$P_4$ & $\frac{3 \gamma }{2}-1,\,\gamma -1$ & accelerating for
$\gamma<\frac{2}{3};$ matter-scalar scaling.
\\\vspace{5pt}
$P_5$ & undefined & Cardassian corrections
\\\vspace{-0.30cm}\\
\hline \hline
\end{tabular}}
\end{center}
\end{table}
\end{center}

Using the variables \eqref{cardvars}, and equation (\ref{cardlig})
we obtain
$$q=\frac{3 n \left(2 x^2+z^2 \gamma \right)}{2 \left(x^2+y^2+z^2\right)}+\frac{1}{2} \left(-6 (n-1) x^2-3 (n-1) z^2 \gamma
   -2\right)$$ and
$$w_{\rm eff}=\frac{x^2-y^2+z^2 (\gamma -1)}{x^2+y^2+z^2}.$$

In table \ref{obscard} are displayed the deceleration parameter,
$q,$ and the effective EoS, $w_{\rm eff}$ at the critical points
of the dynamical system \eqref{cardx}-\eqref{cardz}.

\section{Physical interpretation}

In what follows, and in order to complete the information provided
in the tables, we will characterize the cosmological models
represented by the singular point living in the above mentioned
phase-space.

The first point, called $P_1$ (equivalent to $W_+$ in the notation
of \cite{Lazkoz:2005bf}), represents a solution completely
dominated by the fluid. The unstable character of these solutions
agrees with what one might have anticipated, are they are only
expected to be relevant at early times.

The second point, $P_2^\pm $ (equivalent to $X_\pm$ in the
notation of \cite{Lazkoz:2005bf}),  represents a solution
completely dominated by the scalar field, more specifically by its
kinetic energy. As discussed in table \ref{eigencard} they are
either saddles or sources.

The third point, called $P_3$ (equivalent to $XY_+$in the notation
of \cite{Lazkoz:2005bf}), represents a scalar field dominated
solution, which is inflationary if $s^2<2$ \cite{Kitada-Maeda}.
They are unstable saddle in the asymptotic future.

The fourth point, called $P_4$ (equivalent to $XY_+W$ in the
notation of \cite{Lazkoz:2005bf}), depicts a tracking  solution,
neither the fluid nor the scalar field dominate completely
\cite{Holden:1999hm,Copeland:1997et}.

The fifth point, called $P_5$ (equivalent to $\rm O$ in the
notation of \cite{Lazkoz:2005bf}), represents a regime where the
Cardassian corrections dominates. As a difference with the
analysis in \cite{Lazkoz:2005bf} where this point is
non-hyperbolic, in the present study it is possible to completely
characterize this point using spherical coordinates. The critical
point $P_5$ is stable for initial solutions outside the invariant
set $y=0,$ whereas, for solutions in the invariant set $y=0,$ it
is asymptotically stable.  Thus, for a massless scalar field,
$P_5$ is the local sink. We have presented several numerical
experiments to show this feature.

\section{Cosmological consequences}

The cosmological consequences of this analysis are simple but
important. As compared to the situation in standard cosmology, for
the description of Cardassian models we find that the  first
complication stems from the necessity of
introducing an additional  variable, which we call $w$. \\

Our numerical analysis tell us that  the past attractors
correspond to $ x^2+y^2+z^2 \equiv 1$, and because of the
constraint the latter enforces $w=0$ which  with in turn implies
the recovery of the usual form of the Friedmann equation. In the
case of  models expanding from an initial singularity will and for
($n<1$) we then conclude that the past attractors corresponds more
specifically to $w=0$, and from the definition of $w$ we see that
those are solutions with an initial singularity. Summarizing, from
the perspective of dynamical systems Cardassian models with a
fluid and a scalar field with an exponential potential will
preferably have a big bang.

More specifically, the early-time attractor is a solution
completely dominated by the kinetic energy of the scalar field and
satisfying $ \rho\propto a^{-6}$, and its evolution is
indistinguishable from that of perfect fluid models with a
$p=(\gamma-1)\rho$ equation of state under the identification
$n\equiv \gamma/2$, and the condition for inflation is simply
$n<1/3$.

Interestingly, there are no tracking late-time attractors neither
de Sitter attractors. The past attractor for a massles scalar
field is given by a point where cardassian corrections dominates.
This is an important difference with respect to the behavior in
standard (non-Cardassian) models. If the potential of the scalar
field is important in the dynamics, then the orbits near the
origin do not approach asymptotically the solution dominated by
Cardassian corrections. This fact is due to the
non-differentiability of the system at the origin.

\section{Conclusions}

Cardassian models have been proposed as yet one more possible
explanation for late-time acceleration. The main interest of the
proposal is it involves only matter and radiation and does not
invoke either vacuum energy or a cosmological constant. The idea
consists in introducing a modification to the Friedmann equation,
so that the effects of the modification become important at low
redshift.

We have concentrated here on modifications of the form
$3H^2-\rho\propto \rho^n$ with $n<1$, and we have studied its
asymptotic behavior assuming $\rho$ is made up of two
contributions: the energy density of a perfect fluid with a
$p=(\gamma-1)\rho$ equation of state and a self-interacting scalar
field with an exponential potential.

Our analysis falls mainly on the analytical side, but we have also
carried out some numerical investigations. We constructed a
dynamical system arising from the evolution equations.

Our analysis allows us to say that for $n<0,$ the late-time
solution attractor is a solution completely dominated by the
Cardassian corrections which is accelerating for $n<\frac{2}{3
\gamma },$ and that there are not tracking late-time attractor.

\chapter*{Conclusions}
\addcontentsline{toc}{chapter}{Conclusions}

Dynamical systems provide one of the best ways to study the
stability of cosmological models, and crucially, the fine-tuning
of initial conditions required to match with observations. The
main purpose of the analysis of the flow of a dynamical system is
to describe the global dynamics (phase portrait) and the effect of
small perturbations of the initial conditions on the dynamics
(stability). After an exhaustive bibliographical review on the
more recent results on the subject of dynamical studies, in this
book we have provided an overview of the applications of these
methods to Cosmology. We have used several of the well-known
mathematical tools for the stability analysis such that the
calculation of Normal forms and the application of Center Manifold
Theory for the investigation of non-hyperbolic singular points for
which the classical linearization techniques fails to be applied.

The standard procedure to analyze the properties of the flow of a
cosmological dynamical system determined by an ordinary
differential equation (and some algebraic restrictions) has
several steps. To determine whether the state space is compact,
since this would imply the existence of both past and future
attractors. A nice tool for obtaining a compact phase space is
through Poicar\'e Projection. Identify the lower-dimensional
invariant sets, which contain the orbits of more special classes
of models with additional symmetries. In order to do that, the
devising of monotonic functions is very helpful. Find all the
singular points and analyze their local stability. Where possible
identify the stable and unstable manifolds of the singular points,
which may coincide with some of the invariant sets previously
identified. Find Dulac's functions or monotone functions in
various invariant sets where possible. Investigate any
bifurcations that occur as the equation of state parameter (or any
other parameters) varies. The bifurcations are associated with
changes in the local stability of the singular points. Having all
the information in the previous points one can hope to formulate
precise conjectures about the asymptotic evolution, by identifying
the past and the future attractors. The past attractor will
describe the evolution of a typical universe near the initial
singularity while the future attractor will play the same role at
late times. The monotone functions, in conjunction with theorems
of dynamical systems theory, may enable some of the conjectures to
be proved. Knowing the stable and unstable manifolds of the
singular points it is possible to construct all possible
heteroclinic sequences that join the past attractor, thereby
gaining insight into the intermediate evolution of cosmological
models. Having at hand all the available dynamical information of
a cosmological model, and using the observational available
evidence, one crucially can establish the correspondence between
stability conditions, observational evidence and the cosmological
structure that we observe.

In this book we have presented some progresses in the theoretical
and/or phenomenological modelling of the Universe on the basis of
an increasing number of observational data that inform to us into
how it is the Universal kinematics on great scales, and on the
other hand, in the deepening in the understanding of the
fundamental theory that it describes the gravitational
interaction.

In order to give a phase space description of several cosmological
models we have combined topological, analytical and numerical
techniques for obtaining all possible asymptotic behaviors for
coupled quintessence dark energy models including or not
radiation, based on Scalar-Tensor theories. We have considered
mass-varying dark matter-particles in the framework of phantom
cosmologies for analyzing the viability of them in order to solve
the coincidence problem (why the energy densities of dark matter
and of dark energy are comparable in order the magnitude today?)
obtaining that this problem cannot be solved or even alleviated
for our cases of study. We have obtained all possible asymptotic
behaviors for Ho\v{r}ava-Lifshitz cosmologies with and without
detailed-balance and finally, we have obtained all possible
asymptotic behaviors for the so-called Cardassian cosmologies.

As coupled quintessence models we have considered scalar fields
with arbitrary (positive) potentials and arbitrary coupling
functions. Then, we have straightforwardly introduced mild
assumptions under such functions (differentiable class, number of
singular points, asymptotes, etc.) in order to clarify the
structure of the phase space of the dynamical system. We have
obtained several analytical results. Also, we have presented
several numerical evidences that confirm some of these results. We
have proved the if the potential is nonnegative and has a local
zero minimum at the origin; its derivative and the potential are
simultaneously bounded; and provided that the coupling function is
of exponential order, then the energy density of the background as
well as the kinetic term tend to zero when the time goes forward,
meaning the stability of the de Sitter solution. We have devised a
monotonic function for the flow of the dynamical system which
allow for the identification of some invariant sets. We have
provided approximated center manifolds for the vector field around
the inflection points and the strict degenerate local minimum of
the potential. It is proved that the scalar field typically
diverges into the past, generalizing previous results. By assuming
some regularity conditions on the potential and on the coupling
function in that regime, we have obtained radiation-dominated
cosmological solutions; power-law scalar-field dominated
inflationary cosmological solutions; matter-kinetic-potential
scaling solutions and radiation-kinetic-potential scaling
solutions. Scaling attractors are relevant to give an answer to
the Coincidence Problem. Also we have investigated, for the
general model including radiation, the important examples of
higher order gravity theories $F(R) = R + \alpha R^2$ (quadratic
gravity) and $F(R) = R^n.$

In the literature it is reported that non-minimally coupled
quintessence models are well suited to give an affirmative answer
to the Coincidence problem; thus the natural question would be
that if it is possible for phantom cosmological models to
alleviate the coincidence problem. In this book, we have
investigated phantom cosmological models with dark-matter
particles. The dark matter particles particles acquire masses due
to the interaction with the dark-energy sector. We have performed
a detailed phase-space analysis of various models, with either
exponentially or power-law dependent dark-matter particle mass, in
exponential or power-law scalar field potentials. In all the
examined cases, solutions having Energy densities of the same
orders that might solve the coincidence problem are not relevant
attractors at late times. Thus, the coincidence problem cannot be
solved or even alleviated in varying-mass dark matter particles
models in the framework of phantom cosmology, in a radical
contrast with the corresponding quintessence models. Therefore,
phantom cosmology with varying-mass dark matter particles cannot
easily act as a successful candidate to describe dark energy.

For the case of Ho\v{r}ava gravity we have obtained late-time
solutions compatible with observations. In the case where the
detailed-balance condition is imposed, we find that the universe
can reach a bouncing-oscillatory state at late times, in which
dark-energy, behaving as a simple cosmological constant, will be
dominant. Such solutions were already investigated in the context
of Ho\v{r}ava-Lifshitz cosmology as possible ones, but now we see
that they can indeed be the late-time attractor for the universe.
They arise purely from the novel terms of Ho\v{r}ava-Lifshitz
cosmology, and in particular the dark-radiation term proportional
to $a^4$ is responsible for the bounce, while the cosmological
constant term is responsible for the turnaround. In the case where
the detailed-balance condition is abandoned, we find that the
universe reaches an eternally expanding solution at late times, in
which dark-energy, behaving like a cosmological constant,
dominates completely. Note that according to the initial
conditions, the universe on its way to this late-time attractor
can be an expanding one with noneligible matter content. We
mention that this behavior is independent of the specific form of
the dark-matter content. Thus, the aforementioned features make
this scenario a good candidate for the description of our
universe, in consistency with observations. Finally, in this case
the universe has also a probability to reach an oscillatory
solution at late time; if the initial conditions lie in its basin
of attraction (in this case the eternally expanding solution will
not be reached). Although this analysis indicates that
Ho\v{r}rava-Lifshitz cosmology can be compatible with
observations, it does not enlighten the discussion about possible
conceptual and phenomenological problems and instabilities of
Ho\v{r}rava-Lifshitz gravity, nor can it interfere with the
questions concerning the validity of its theoretical background,
which is the subject of interest of other studies. It just faces
the problem from the cosmological point of view, and thus its
results can been taken into account only if Ho\v{r}ava gravity
passes successfully the aforementioned theoretical tests.

Finally we have investigated Cardassian models. These have been
proposed as yet one more possible explanation for late-time
acceleration. The main interest of the proposal is it involves
only matter and radiation and does not invoke either vacuum energy
or a cosmological constant. The idea consists in introducing a
modification to the Friedmann equation, so that the effects of the
modification become important at low redshift. Our analysis allows
us to say that for $n < 0$; the late-time solution attractor is a
solution completely dominated by the Cardassian corrections which
can be acceleration; and that there are not tracking late-time
attractor.

\label{lastpage-01}

\end{document}